\documentclass[a4paper,12pt]{report}
\usepackage{epsfig}
\usepackage[koi8-r]{inputenc}
\usepackage[russian]{babel}
\usepackage{mathptmx}
\usepackage{amsmath}
\DeclareMathOperator{\sgn}{sgn}
\usepackage{indentfirst}
\usepackage{tocloft}
\usepackage{cite}
\usepackage{geometry}
\usepackage{textcomp}

\addtocontents{toc}{\cftpagenumbersoff{subsection}}

\DeclareMathOperator{\arch}{arch}

\begin{document}

\begin{titlepage}

\vspace*{93mm}
\centerline{{\usefont{T2A}{cmr}{b}{n} \large З. К. Силагадзе}}
\vspace*{4mm}
\centerline{\usefont{T2A}{cmr}{b}{n} \Large МЕХАНИКА И ТЕОРИЯ ОТНОСИТЕЛЬНОСТИ}
\vspace*{2mm}
\centerline{\usefont{T2A}{cmr}{b}{n} \Large Задачи семинарских занятий с решениями}
\vspace*{8mm}
\centerline{}
\vspace*{68mm}
\centerline{Новосибирск}
\vspace*{3mm}
\centerline{2017}
\end{titlepage}

\begin{titlepage}

\vspace*{60mm}
\begin{center}
{\usefont{T2A}{cmr}{b}{n}\selectfont\large 85-летию Карло Силагадзе, 
моего отца, посвящается.}
\end{center}

\vspace*{30mm}
В учебном пособии приведено решение 230  задач,  которые использовались автором  в  
течение  ряда  лет на семинарских  занятиях по курсу <<Механика и теория 
относительности>>.  Учебное пособие состоит из 36 глав - по числу семинаров по 
данному курсу.  В~каждой  главе разбираются от 5 до 10 задач. Где это возможно, 
приведены несколько решений одной и той же задачи, чтобы студенты привыкли 
рассматривать физические явления с разных точек  зрения. Большинство задач взято из 
задачника  Бельченко Ю.~И., Гилев Е.~А., Силагадзе З.~К., Соколов В.~Г. Сборник 
задач по механике частиц и тел. Новосибирск: НГУ, 2000. Для таких задач в скобках 
указан номер задачи в задачнике.

\vspace*{25mm}

\hspace*{110mm}\copyright 
\hspace*{1mm} Силагадзе З. К., 2017

\end{titlepage}

\clearpage
\setcounter{page}{3}

\tableofcontents
\newpage

\section[Семинар 1]
{\centerline{Семинар 1}}

\subsection*{Задача 1 {\usefont{T2A}{cmr}{b}{n}(1.5)}: Максимальное время 
наблюдения Венеры после захода Солнца}
Определить максимальное время наблюдения Венеры после захода Солнца. Радиус орбиты 
Венеры 0,72 а. е.

\subsection*{Задача 2 {\usefont{T2A}{cmr}{b}{n}(\hspace*{-1.5mm}\cite{2})}: 
Время торможения}
На частицу с массой $m$ и начальной скоростью $V$ действует зависящая от скорости 
сила сопротивления воздуха вида $F=bV^n$. Для $n=0,1,2,\ldots$ оценить, как время 
торможения зависит от $m,V$ и $b$.

\subsection*{Задача 3 {\usefont{T2A}{cmr}{b}{n}(\hspace*{-1.5mm}\cite{3})}: 
Исчезнование тени}
Источник света движется горизонтально на высоте $H$ с большой скоростью $V$ и 
приближается к стене высоты $h<H$. На каком расстоянии от стены начнет исчезать тень, 
если $V/c>1-h/H$ ($c$ --- скорость света)?  

\subsection*{Задача 4 {\usefont{T2A}{cmr}{b}{n}(\hspace*{-1.5mm}\cite{2})}: 
Скорости частиц 
после столкновения}
Найти скорости частиц после упругого лобового столкновения частицы с массой $m$, 
скоростью $V$ с первоначально неподвижной частицей с массой $M$. Что будет при 
$m\ll M$ и $m\gg M$?

\subsection*{Задача 5 {\usefont{T2A}{cmr}{b}{n}(\hspace*{-1.5mm}\cite{4,5,6})}: 
Как быстро доползет жук?}
К стенке прикреплена резиновая нить длиной $L=1~\mbox{м}$. На ней с постоянной 
скоростью начинает ползти жук и за время $\tau=1~\mbox{с}$ проползает расстояние 
$l=1~\mbox{см}$. В конце промежутка $\tau$ нить быстро удлиняют на $L=1~\mbox{м}$
и так продолжается после каждого промежутка $\tau$. За сколько времени жук доползет
до конца нити? 

\subsection*{Задача 6 {\usefont{T2A}{cmr}{b}{n}(\hspace*{-1.5mm}\cite{6A})}: 
Электрон на краю наблюдаемой Вселенной}
Допустим, у нас есть сосуд с газом и суперкомпьютер, который может точно предсказать
траекторию молекул газа, если в нем заложены координаты и характеристики всех тел
во Вселенной. Оценить насколько быстро мы потеряем предсказательную способность, если
забудем учесть гравитационное взаимодействие одного-единственного электрона на краю 
наблюдаемой Вселеной на расстоянии $10^{10}$ световых лет. 

\section[Семинар 2]
{\centerline{Семинар 2}}

\subsection*{Задача 1 {\usefont{T2A}{cmr}{b}{n}(1.30)}: Форма шнура из 
взрывчатого вещества}
Имеется однородный шнур из взрывчатого вещества. Скорость распространения реакции 
взрыва вдоль шнура $V$, скорость распространения взрывной волны по воздуху $c$. Найти 
форму линии, по которой надо расположить шнур, чтобы волна от всех точек шнура
пришла в заданную точку одновременно. 

\subsection*{Задача 2 {\usefont{T2A}{cmr}{b}{n}(\hspace*{-1.5mm}\cite{7}, 
2.4.12)}: Скорость нижнего цилиндра}
Два гладких одинаковых цилиндра радиуса $R$ прислонены к стенке. Из-за того, что 
нижний цилиндр чуть-чуть стронулся вправо по горизонтальной плоскости, верхный стал 
опускаться по вертикали, и система пришла в движение. Найдите конечную скорость 
нижнего цилиндра.
\begin{figure}[htb]
\centerline{\epsfig{figure=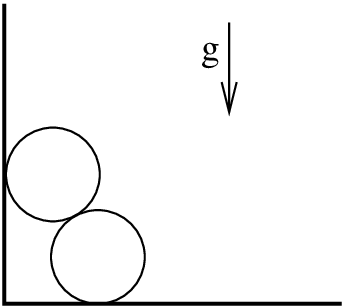,height=4cm}}
\end{figure}

\subsection*{Задача 3 {\usefont{T2A}{cmr}{b}{n}(1.31)}: Угол между векторами}
Сферические координаты векторов $\vec{r}_1=(r_1,\theta_1,\phi_1)$ и 
$\vec{r}_2=(r_2,\theta_2,\phi_2)$. Определить угол между векторами $\vec{r}_1$ и 
$\vec{r}_2$.

\subsection*{Задача 4: Периметр эллипса}
Найти периметр эллипса с маленьким эксцентриситетом $e\ll 1$.

\subsection*{Задача 5: Площадь эллипса}
Найти площадь эллипса, если его полуоси равны $a$ и $b$.

\subsection*{Задача 6 {\usefont{T2A}{cmr}{b}{n}(5.33)}:
Установившаяся скорость монеты}
На наклонной плоскости с углом наклона $\alpha$ поле тяжести лежит монета.
Монете сообщается скорость $V_0$ вдоль горизонтальной оси (см. рисунок). Найти
установившуюся скорость движения монеты $u$, если коэффициент трения $\mu=
\tg{\alpha}$ не зависит от скорости.
\begin{figure}[htb]
\centerline{\epsfig{figure=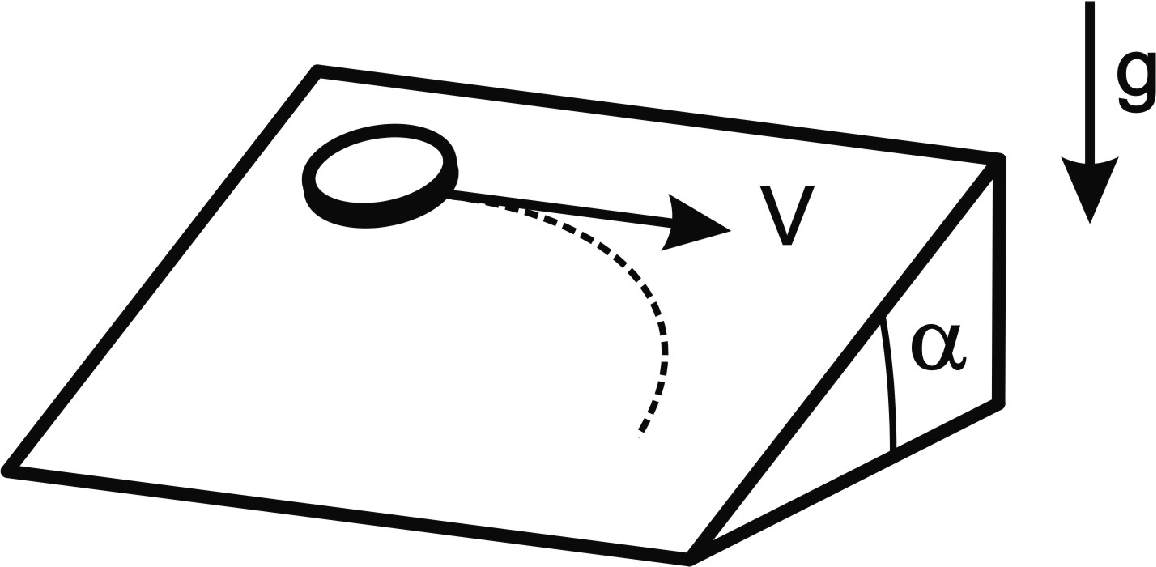,height=4cm}}
\end{figure}

\subsection*{Задача 7 {\usefont{T2A}{cmr}{b}{n}(1.9)}:
Траектория конца тени от вертикально стоящей палочки}
Нарисуйте траекторию конца тени от вертикально стоящей палочки в солнечный день 
22 июня в Новосибирске. Оцените долготу дня. Проследите эволюцию траектории со 
временем. Что будет на других широтах?

\section[Семинар 3]
{\centerline{Семинар 3}}

\subsection*{Задача 1 {\usefont{T2A}{cmr}{b}{n}(1.33)}:
Орты сферической и цилиндрической систем координат}
Выразить орты сферической и цилиндрической систем координат через орты 
декартовой системы координат.

\subsection*{Задача 2 {\usefont{T2A}{cmr}{b}{n}(1.24)}:
Скорость и ускорение в цилиндрической системе координат}
Получить выражение для компонент радиус-вектора, скорости и ускорения точки 
в цилиндрической системе координат.

\subsection*{Задача 3 {\usefont{T2A}{cmr}{b}{n}(1.36)}: расстояние 
наибольшего сближения}
Для двух кораблей, движущихся неизменными пересекающимися курсами, выразить 
расстояние наибольшего сближения и время до сближения через векторы скоростей 
и начальных положений.

\subsection*{Задача 4 {\usefont{T2A}{cmr}{b}{n}(1.29)}: Заяц и собака} 
Заяц бежит по прямой линии со скоростью $u$. В начальный момент времени из
положения, показанного на рисунке, его начинает преследовать собака со скоростью 
$V$. В ходе погони собака всегда бежит в направлении зайца. Через какое время 
собака настигнет зайца? Начальное расстояние между ними $L$.
\begin{figure}[htb]
\centerline{\epsfig{figure=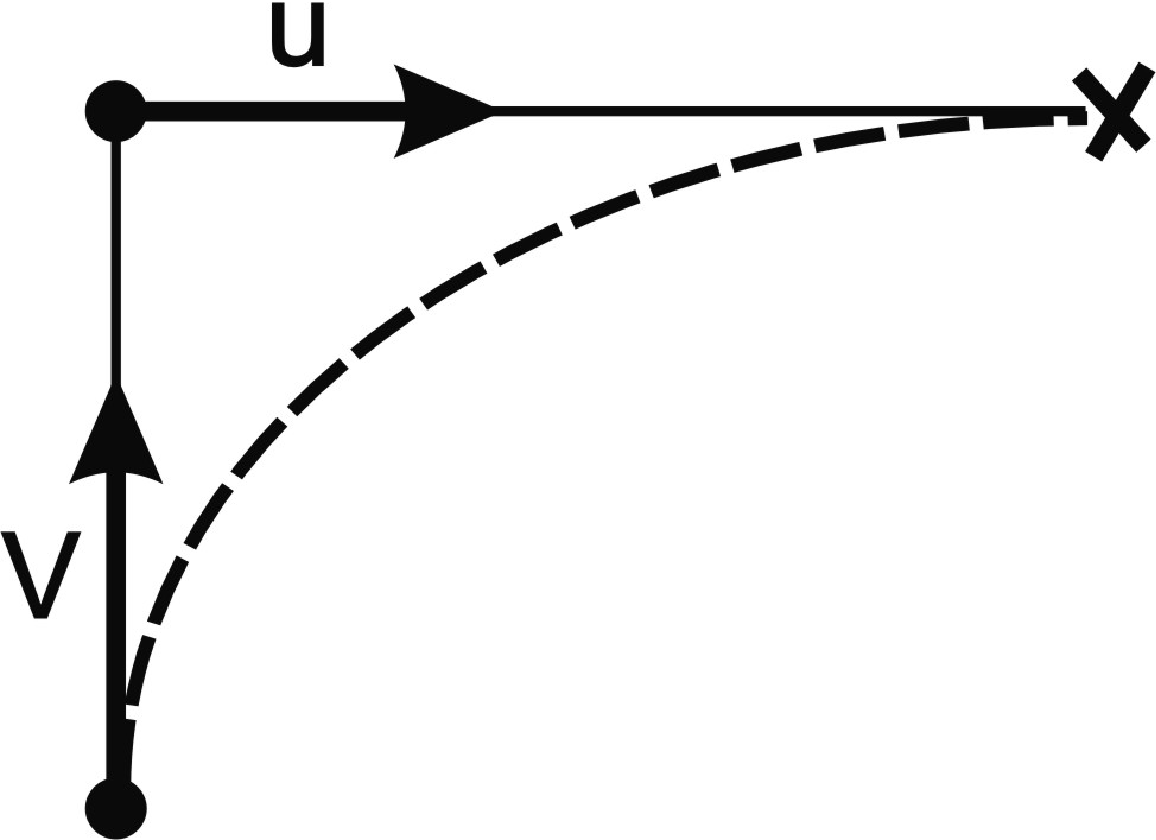,height=4cm}}
\end{figure}

\subsection*{Задача 5 {\usefont{T2A}{cmr}{b}{n}(1.26)}: Радиус кривизны 
траектории}
Точка движется по закону $x=2t,\;y=t^2$ ($x,y$ -- в м, $t$ -- в с). 
Определить радиус кривизны траектории в начале движения и через 2 с. 

\subsection*{Задача 6 {\usefont{T2A}{cmr}{b}{n}(1.38)}: Облет треугольника}
Самолет облетел стороны треугольника с длинами $A$, $B$ и $C$ за
время $t_1$, $t_2$, $t_3$ соответственно. Найти скорость ветра и самолета в
случае, когда скорость ветра параллельна плоскости треугольника.

\section[Семинар 4]
{\centerline{Семинар 4}}

\subsection*{Задача 1 {\usefont{T2A}{cmr}{b}{n}(1.23)}: Траектория точки}
Нарисовать траекторию точки, движущейся по закону
$$r=\frac{b}{t},\;\;\;\;\phi=\gamma t\;\;\;(b>0). $$
Найти закон движения и уравнение траектории в декартовых координатах.

\subsection*{Задача 2 {\usefont{T2A}{cmr}{b}{n}(1.25)}: Траектория, скорость, 
ускорение и радиус кривизны траектории}
Tочка движется по закону
$$\rho=ae^{kt},\;\;\;\phi=kt.$$
Найти траекторию, скорость, ускорение и радиус кривизны траектории
в зависимости от радиус-вектора точки.

\subsection*{Задача 3 {\usefont{T2A}{cmr}{b}{n}(1.40)}: Преобразования Галилея}
Как изменяются импульс и кинетическая энергия системы частиц при преобразованиях 
Галилея? В какой системе отсчета кинетическая энергия частиц минимальна?

\subsection*{Задача 4 {\usefont{T2A}{cmr}{b}{n}(1.41)}: Кинетическая энергия 
гусеницы}
Какова кинетическая энергия гусеницы трактора в системе дороги и в системе трактора, 
если скорость трактора $V$?

\subsection*{Задача 5 {\usefont{T2A}{cmr}{b}{n}(1.28)}: Через какое время 
собаки догонят друг друга?}
Четыре собаки преследуют друг друга, так что скорость догоняющей собаки $\vec{V}$
всегда направлена на убегающую собаку (см. рисунок). Через какое время собаки 
догонят друг друга, если сначала они находились в углах квадрата со
стороной а? Какова траектория собак? Какой путь проходит до встречи каждая собака?
\begin{figure}[htb]
\centerline{\epsfig{figure=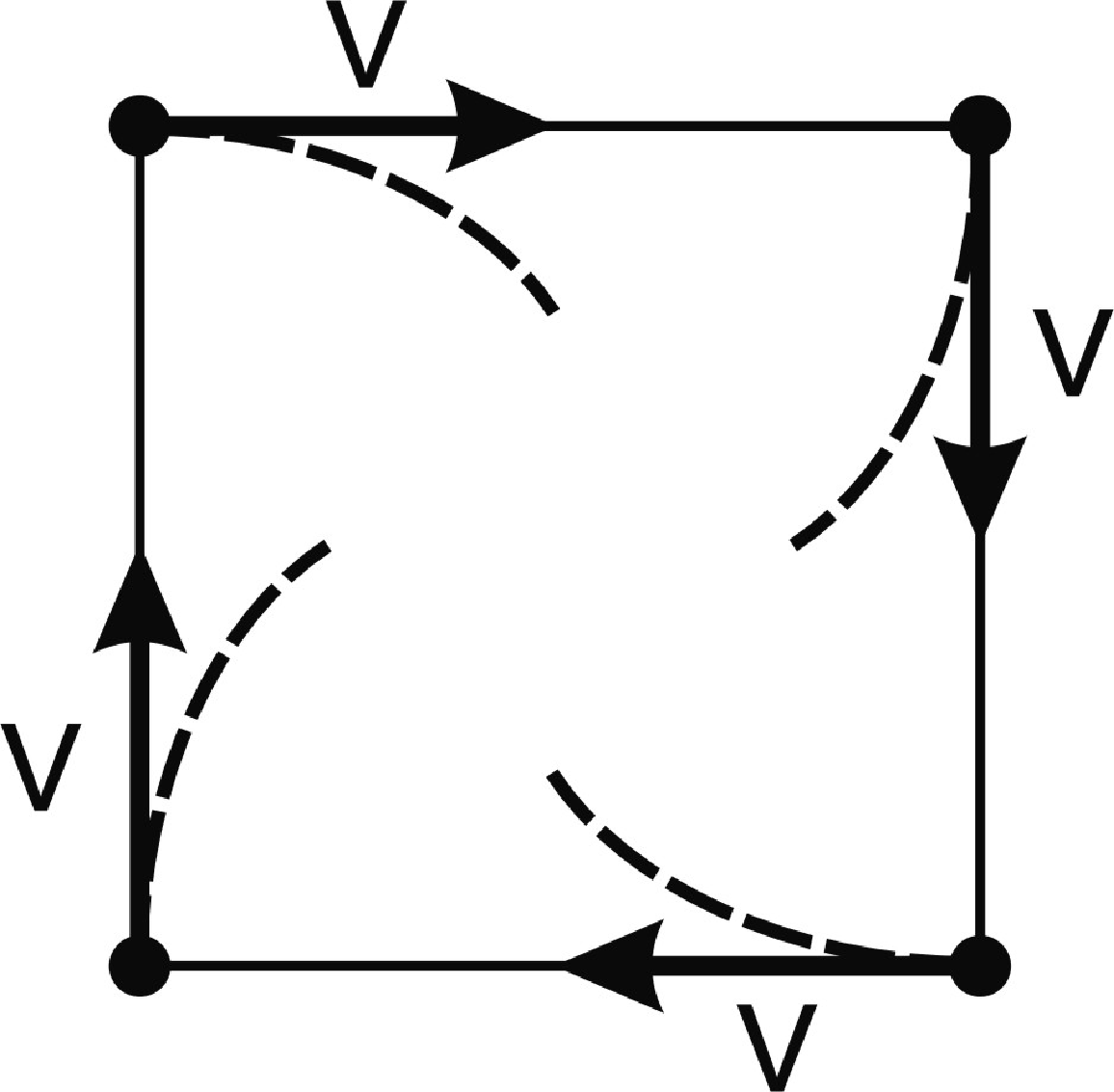,height=4cm}}
\end{figure}

\subsection*{Задача 6:
Сферический треугольник}
Какое расстояние пролетел самолет при полете по маршруту Новосибирск -- Рио-де-Жанейро
-- Нью-Йорк -- Новосибирск? Географические координаты городов: Новосибирск ($\varphi=
83^\circ$ восточной долготы, $\theta=55^\circ$ северной широты), 
Рио-де-Жанейро ($\varphi= 43^\circ$ западной долготы,  $\theta=23^\circ$ южной 
широты), Нью-Йорк ($\varphi=74^\circ$ западной долготы, $\theta=40^\circ$ северной 
широты). Найти сумму углов сферического треугольника, образованного участками пути 
самолета между городами (считая, что самолет летит на одной и той же высоте над 
землей). 

\section[Семинар 5]
{\centerline{Семинар 5}}

\subsection*{Задача 1 {\usefont{T2A}{cmr}{b}{n}(1.42)}: Скорость поезда}
Определите скорость поезда, если при приближении к неподвижному наблюдателю гудок 
поезда имел частоту в $\alpha$ раз большую, чем при удалении от наблюдателя.

\subsection*{Задача 2 {\usefont{T2A}{cmr}{b}{n}(1.42)}: Два встречных поезда}
Машинисты двух сближающихся поездов сигнализируют друг другу гудками. Определите 
скорость поездов, если частоты принимаемых машинистами сигналов превышают ``собственную'' 
частоту гудка в $\alpha$ и $\beta$ раз соответственно. Сигнальные 
устройства ло\-ко\-мотивов одинаковы.

\subsection*{Задача 3 {\usefont{T2A}{cmr}{b}{n}(1.53)}: Поступательное и 
угловое ускорение шестеренки}
Между двумя зубчатыми рейками зажата шестеренка радиусом $R$= 0,5 м. Ускорения реек 
$a_1$ = 1,5 м/с$^2$ и $a_2$ = 2,5 м/с$^2$ (см. рисунок). Найти поступательное и 
угловое ускорение шестеренки.
\begin{figure}[htb]
\centerline{\epsfig{figure=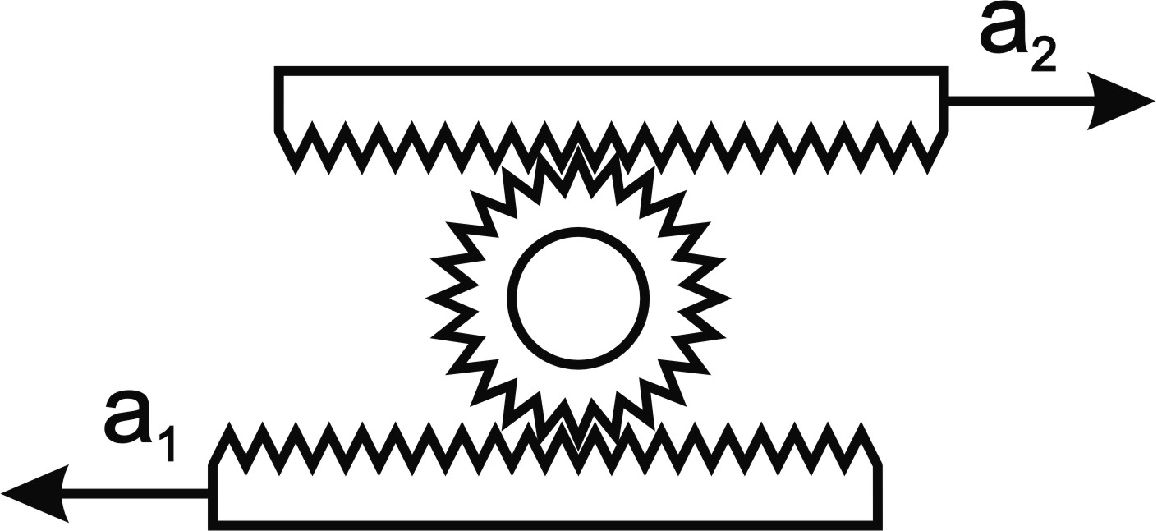,height=2cm}}
\end{figure}

\subsection*{Задача 4 {\usefont{T2A}{cmr}{b}{n}(2.6)}: Реальная скорость объекта}
За пять лет наблюдения с Земли светящийся объект, находящийся на расстоянии $10^5$ 
св. лет, совершил видимое угловое перемещение $10^{-4}$ рад, т. е. его кажущаяся 
скорость перемещения равна удвоенной скорости света. Найдите, под каким углом к линии 
наблюдения может двигаться объект, чтобы его реальная скорость была меньше скорости
света. Какова минимально возможная скорость объекта?

\subsection*{Задача 5 {\usefont{T2A}{cmr}{b}{n}(2.7)}: Время обращения
спутника Ио}
Каким будет казаться земному наблюдателю время обращения спутника Ио вокруг Юпитера? 
Как меняется это время в течение года? Истинный период обращения Ио 42 часа.

\subsection*{Задача 6: Преобразования Лоренца}
Исходя из физических предположений о свойствах масштабов и часов, как надо 
модифицировать преобразования Галилея, чтобы скорость света не зависела от выбора 
инерциальной системы отсчета? 

\section[Семинар 6]
{\centerline{Семинар 6}}

\subsection*{Задача 1 {\usefont{T2A}{cmr}{b}{n}(2.10)}: Период движения фотона}
Квадратная (в собственной системе отсчета) платформа со стороной $L$ движется вдоль 
своей диагонали со скоростью $V$. В углах платформы установлены зеркала. Отражаясь 
от них, по периметру платформы движется фотон. Найти период его движения в Л-системе 
отсчета.

\subsection*{Задача 2 {\usefont{T2A}{cmr}{b}{n}(2.8)}: Какая лампочка загорится 
раньше?}
На концах стержня с собственной длиной $L_0$, движущегося со скоростью $V$, 
одновременно в системе стержня зажигаются две лампочки. Какая из них загорится 
раньше (и насколько) в Л-системе отсчета? Какую вспышку увидит раньше (и насколько) 
неподвижный наблюдатель, находящийся в точке $O$ (см. рисунок)?
\begin{figure}[htb]
\centerline{\epsfig{figure=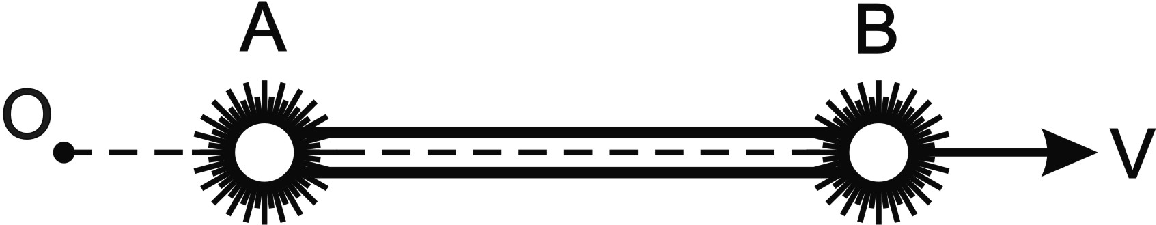,height=1cm}}
\end{figure}

\subsection*{Задача 3 {\usefont{T2A}{cmr}{b}{n}(2.11)}: Время движения фотона}
Луч света движется через систему зеркал, расположенных в вершинах квадрата со стороной 
$L$. Найти время движения фотона через систему с точки зрения наблюдателей, 
движущихся со скоростью $V$ = 0,8 с в направлениях, указанных на рисунке.
\begin{figure}[htb]
\centerline{\epsfig{figure=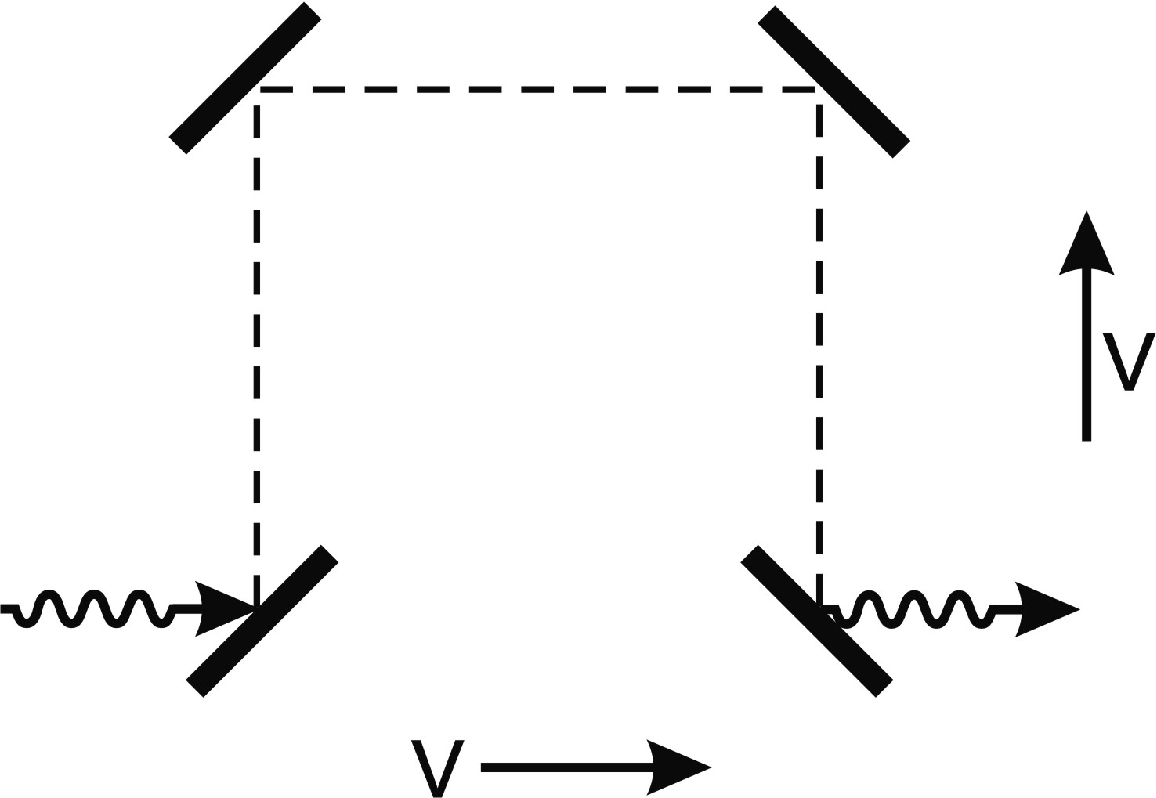,height=4cm}}
\end{figure}

\subsection*{Задача 4 {\usefont{T2A}{cmr}{b}{n}(2.15)}: Время пролета галактики}
В системе галактики фотон пролетает ее диаметр за время $T$ = $10^5$ лет. Сколько 
времени потребуется фотону на это путешествие в системе отсчета протона 
с релятивистским фактором $\gamma=10^{10}$, летящего следом за фотоном? Как 
изменится результат, если протон летит навстречу фотону?

\subsection*{Задача 5 {\usefont{T2A}{cmr}{b}{n}(2.17)}: Стеклянный брусок с  
посеребренной гранью}
Стеклянный брусок длиной $L$ движется со скоростью $V$ параллельно своей грани. Одна 
из сторон бруска, перпендикулярная к скорости, посеребрена. Сколько времени по часам 
неподвижного наблюдателя потребуется свету, летящему параллельно $V$, чтобы пройти 
сквозь брусок, отразиться от серебряной грани и выйти из бруска? Скорость света в
неподвижном бруске $c/n$ ($n$ --- показатель преломления).

\subsection*{Задача 6 {\usefont{T2A}{cmr}{b}{n}(2.16)}: Доля оставшихся нейтронов}
Из-за распада количество покоящихся нейтронов уменьшается экспоненциально с постоянной 
времени $10^3$ с. Какая доля нейтронов с релятивистским фактором $\gamma=10^{10}$,
стартовавших вместе с фотоном, ``останется в живых'' к моменту, когда фотон достигнет 
края галактики с размером $10^5$ световых лет? Рассмотреть задачу с точки зрения
наблюдателя, движущегося вместе с нейтронами, и с точки зрения Л-системы отсчета.

\section[Семинар 7]
{\centerline{Семинар 7}}

\subsection*{Задача 1 {\usefont{T2A}{cmr}{b}{n}(2.23)}: Релятивистский трактор}
Релятивистский трактор движется по полю с постоянной скоростью $V$. Сидящий в кабине 
тракторист насчитал на каждой половине гусеницы по $N$ траков. Сколько траков 
насчитает на верхней и нижней половинах гусеницы неподвижный наблюдатель?

\subsection*{Задача 2 {\usefont{T2A}{cmr}{b}{n}(2.13)}: Как далеко улетел 
корабль?}
Космический корабль половину времени (по часам корабля) двигался с релятивистской 
скоростью $V_1$, а вторую половину ---  со скоростью $V_2$. Как далеко улетел 
корабль, если его путешествие по часам на Земле длилось время $T$?

\subsection*{Задача 3 {\usefont{T2A}{cmr}{b}{n}(2.19)}: Вспышка света}
В вершине $O$ прямоугольного треугольника $ABO$ происходит вспышка света (см.
рисунок). С какой скоростью и в каких направлениях может двигаться наблюдатель, 
чтобы в его системе отсчета свет достиг точки $B$ раньше, чем точки $A$? 
$AO=AB=L$.
\begin{figure}[htb]
\centerline{\epsfig{figure=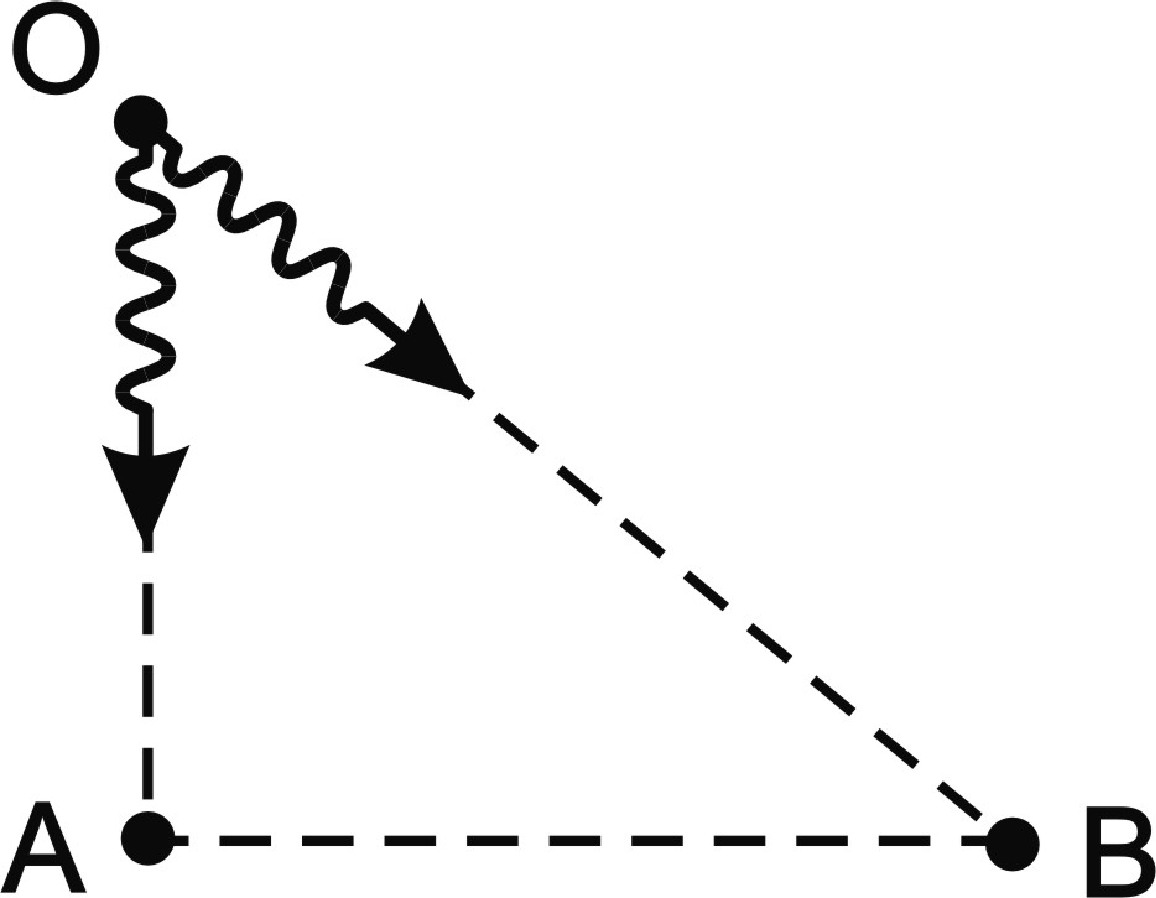,height=3cm}}
\end{figure}

\subsection*{Задача 4 {\usefont{T2A}{cmr}{b}{n}(2.26)}: Стабилизированный 
электронный пучок}
В пучок релятивистских электронов, движущихся со скоростью $V$ и имеющих плотность 
частиц $n_-$ , добавлено некоторое количество неподвижных однозарядных ионов 
с концентрацией $n_+<n_-$ , так что в Л-системе отсчета пучок заряжен отрицательно. 
Найти плотность частиц каждого сорта в системе отсчета, связанной с электронами. 
При каком условии суммарная плотность заряда в этой СО будет положительной, т. е.
электроны будут стягиваться к оси пучка?

\subsection*{Задача 5 {\usefont{T2A}{cmr}{b}{n}(2.22)}: Пенал и карандаш}
Вдоль оси пенала длиной $L$, закрывающегося с торцов крышками $A$ и $B$ 
(см. левый рисунок), со скоростью $V$ движется карандаш. Собственная длина карандаша 
$L_0$ удовлетворяет условию $L_0>L>L_0/\gamma$ (где $\gamma$ --- релятивистский 
фактор карандаша). Сначала крышка $A$ пенала открыта, а крышка $B$ закрыта. Когда 
карандаш влетает в пенал, крышка $A$ закрывается, так что в течение некоторого 
времени карандаш находится в закрытом пенале (правый рисунок). Затем открывается 
крышка $B$, и карандаш свободно вылетает из пенала. Опишите явление в системе отсчета 
карандаша. 
\begin{figure}[htb]
\begin{center}
\epsfig{figure=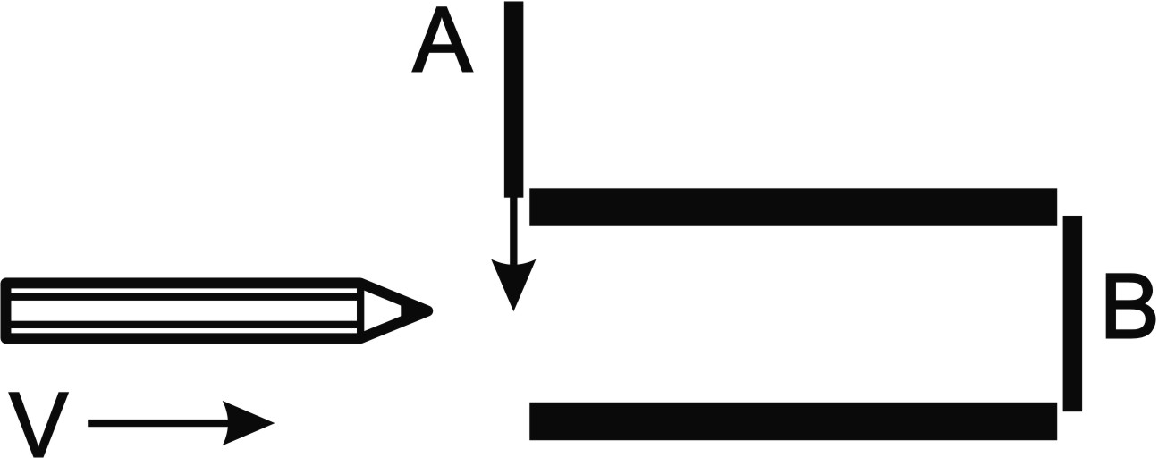,width=4cm}
\hspace*{15mm}
\epsfig{figure=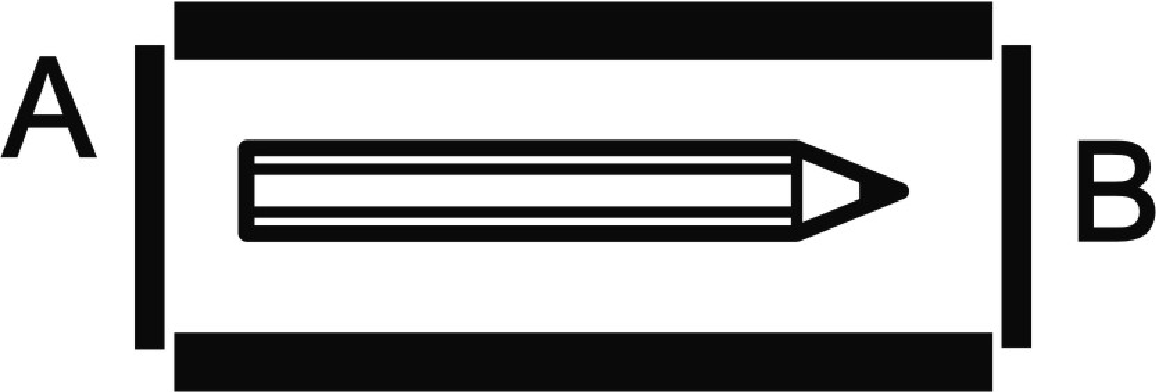,width=2.5cm}
\end{center}
\end{figure}

\subsection*{Задача 6 {\usefont{T2A}{cmr}{b}{n}(2.20)}: Сколько таких систем 
отсчета существует?}
Координаты двух событий в Л-системе отсчета $\vec{r}_1,t_1$ и $\vec{r}_2,t_2$.
В какой системе отсчета эти события одновременны? В какой системе они одноместны? 
Сколько таких систем отсчета существует?
 
\section[Семинар 8]
{\centerline{Семинар 8}}

\subsection*{Задача 1 {\usefont{T2A}{cmr}{b}{n}(2.24)}: Релятивистский танк}
Релятивистский танк движется по направлению к крепости со скоростью $V$. Он выпускает 
$n$ снарядов в секунду (по часам стрелка). Скорость снарядов относительно танка $u$. 
Сколько снарядов в секунду попадает в крепость (по часам гарнизона в крепости)?

\subsection*{Задача 2 {\usefont{T2A}{cmr}{b}{n}(2.25)}: Какова скорость реки?}
Лодочник, отплывая от моста, уронил в реку багор. Когда он заметил потерю и повернул
обратно, по его часам прошло время $\tau$. Он догнал багор напротив дерева, растущего
на расстоянии $L$ от моста. Скорость лодки относительно реки $V$. Какова скорость 
реки?

\subsection*{Задача 3 {\usefont{T2A}{cmr}{b}{n}(2.30)}: Фотографирование 
быстролетящих параллелепипеда и шара}
Что получится при моментальном фотографировании быстролетящих параллелепипеда, шара? 
Фотографирование производится в параллельных лучах света, падающих перпендикулярно 
фотопластинке.

\subsection*{Задача 4 {\usefont{T2A}{cmr}{b}{n}(2.27)}: Инвариантность площади 
поперечного сечения пучка света}
Показать, что площадь поперечного сечения, проведенного перпендикулярно направлению 
движения параллельного пучка света, является релятивистским инвариантом.

\subsection*{Задача 5 {\usefont{T2A}{cmr}{b}{n}(2.28)}: Перекрытие света диском}
Между двумя линзами сформирован пучок света, имеющий круглое сечение радиусом $R$ 
и движущийся вертикально вниз (см. рисунок). Перпендикулярно пучку вдоль оси 
$x$ со скоростью $V$ движется диск такого же радиуса (в собственной системе отсчета). 
Плоскость диска перпендикулярна пучку. С точки зрения лабораторной системы отсчета 
диск испытывает лоренцево сокращение и не может перекрыть пучок света. Для наблюдателя
на диске сокращается сечение пучка и должен наступить момент полного перекрытия света 
в фокусе второй линзы. Объясните парадокс.
\begin{figure}[htb]
\centerline{\epsfig{figure=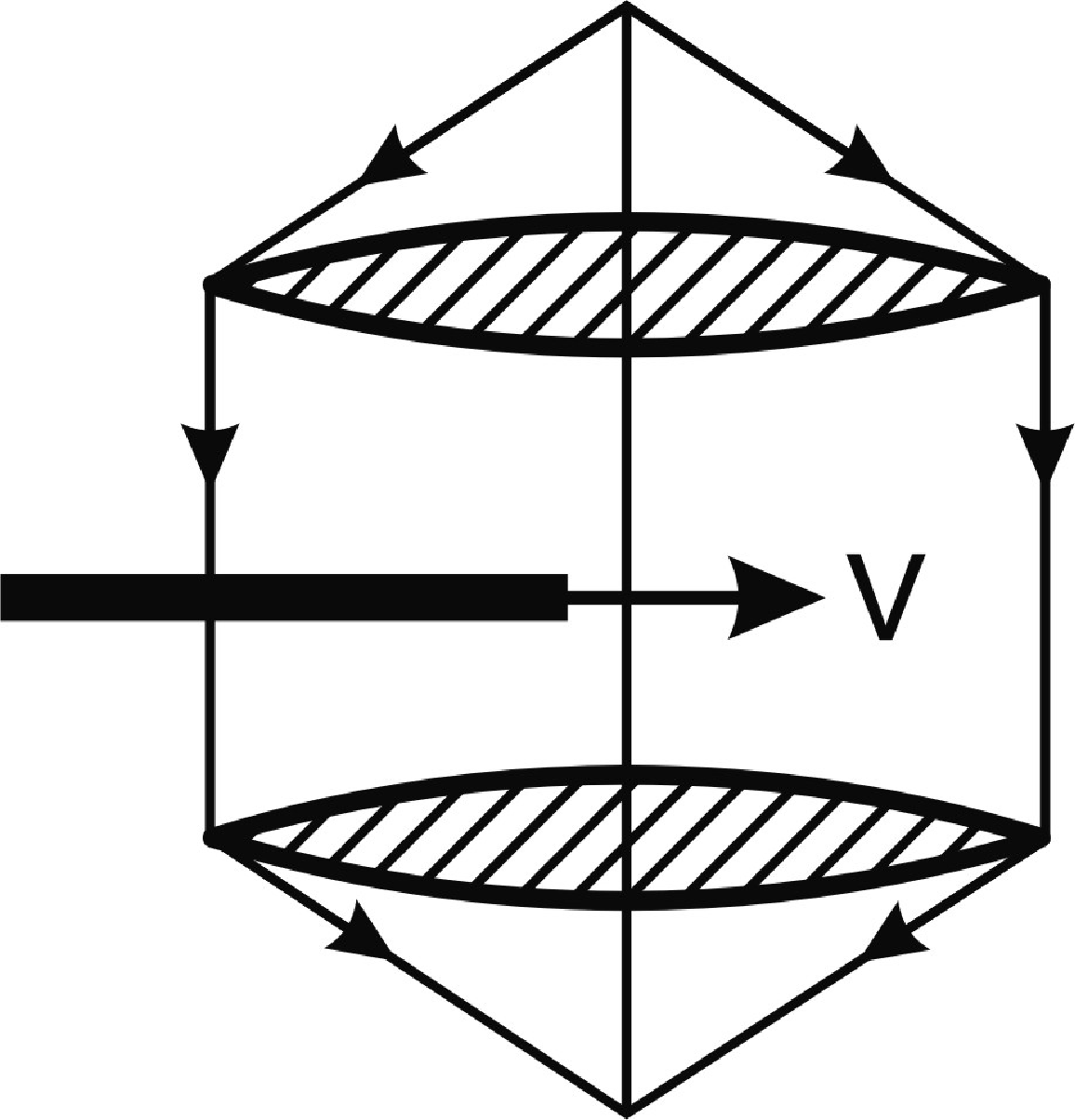,height=4cm}}
\end{figure}

\subsection*{Задача 6: Сколько 
ступеней должен иметь релятивистская ракета?}
Сколько ступеней должен иметь релятивистская ракета, чтобы достичь скорости $0,9c$, 
где $c$ --- скорость света, если каждая ступень увеличивает скорость ракеты на 
$0,1c$ относительно предыдущей?

\section[Семинар 9]
{\centerline{Семинар 9}}

\subsection*{Задача 1 {\usefont{T2A}{cmr}{b}{n}(2.32)}: В какой системе отсчета 
длины стержней будут равными?}
Один из двух одинаковых стержней покоится, а другой движется вдоль него со скоростью 
$V$. В какой системе отсчета длины стержней будут равными?

\subsection*{Задача 2 {\usefont{T2A}{cmr}{b}{n}(2.40)}: Найти минимальное 
расстояние между частицами}
Из двух точек, разделенных расстоянием $L$, одновременно вылетают две частицы 
с перпендикулярными друг другу одинаковыми по величине скоростями $V$ (см. рисунок). 
Найти минимальное расстояние между частицами: а) в Л-системе отсчета; 
б) в системе одной из частиц.
\begin{figure}[htb]
\centerline{\epsfig{figure=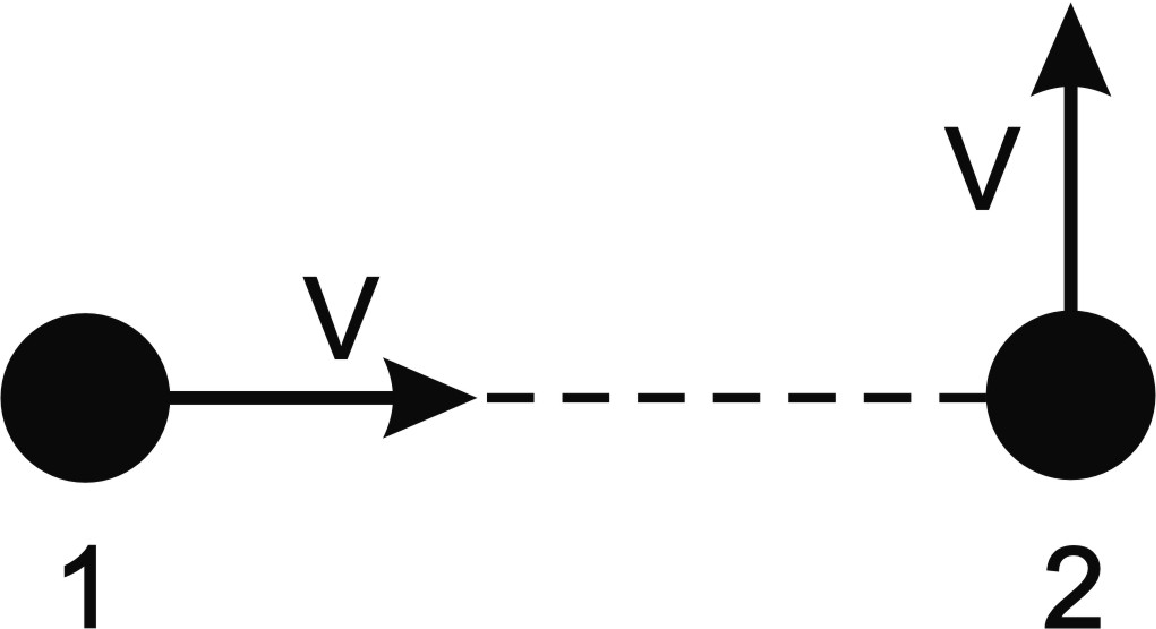,height=2cm}}
\end{figure}

\subsection*{Задача 3 {\usefont{T2A}{cmr}{b}{n}(2.46)}: Скорость движения
звездочета}
При быстром движении наблюдателя относительно небосвода в передней полусфере 
насчитывается в $N$ раз больше звезд, чем в задней. Определите скорость этого 
движения, если для неподвижного наблюдателя звезды распределены по небу изотропно.

\subsection*{Задача 4 {\usefont{T2A}{cmr}{b}{n}(2.52)}: Длина и угол наклона
падающего стержня}
Стержень длины $L$ падает на пол со скоростью $u$ (ускорением пренебречь), так что 
его концы достигают пола одновременно. Какой будет длина стержня и под каким углом 
к полу он будет расположен с точки зрения наблюдателя, движущегося по полу со 
скоростью $V$ параллельно плоскости падения стержня?

\subsection*{Задача 5 {\usefont{T2A}{cmr}{b}{n}(2.44)}: В какой точке выйдет
свет раньше всего?}
Прозрачная пластинка с показателем преломления $n$, толщиной $d$ движется параллельно 
своей плоскости с релятивистской скоростью $V$. В точке $A$, расположенной снаружи от
пластинки, произошла вспышка света. В какую точку $X$ Л-системы отсчета, 
расположенную на другой стороне пластинки (см. рисунок), свет дойдет раньше всего? 
Свет в пластинке распространяется со скоростью $c/n$.
\begin{figure}[htb]
\centerline{\epsfig{figure=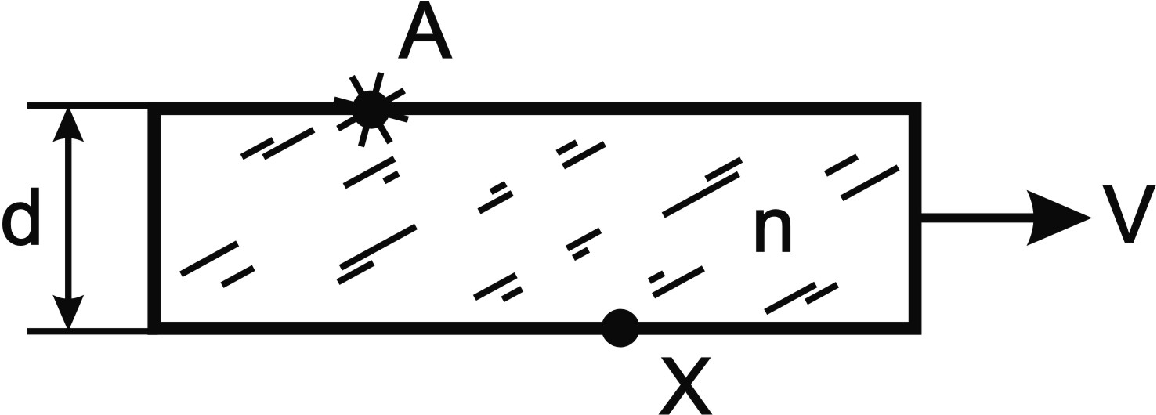,height=2.0cm}}
\end{figure}

\section[Семинар 10]
{\centerline{Семинар 10}}

\subsection*{Задача 1 {\usefont{T2A}{cmr}{b}{n}(2.37)}: Релятивистский 
эскалатор}
Релятивистский эскалатор, движущийся со скоростью $V=c/3$, имеет $N$ ступенек. 
Пассажир, опаздывающий на поезд, сбегает по эскалатору со скоростью $u=c/2$ 
(относительно эскалатора), не пропуская ни одной ступеньки. Сколько шагов 
сделает пассажир по эскалатору?

\subsection*{Задача 2 {\usefont{T2A}{cmr}{b}{n}(2.29)}: Коэффициент пропускания 
решетки}
Параллельный пучок света падает на решетку, состоящую из брусков сечением $a\times d$
с расстоянием между брусками $b$ (см. рисунок). Какая часть падающего света сможет
пройти через решетку, если ее двигать перпендикулярно пучку с релятивистской 
скоростью $V$?
\begin{figure}[htb]
\centerline{\epsfig{figure=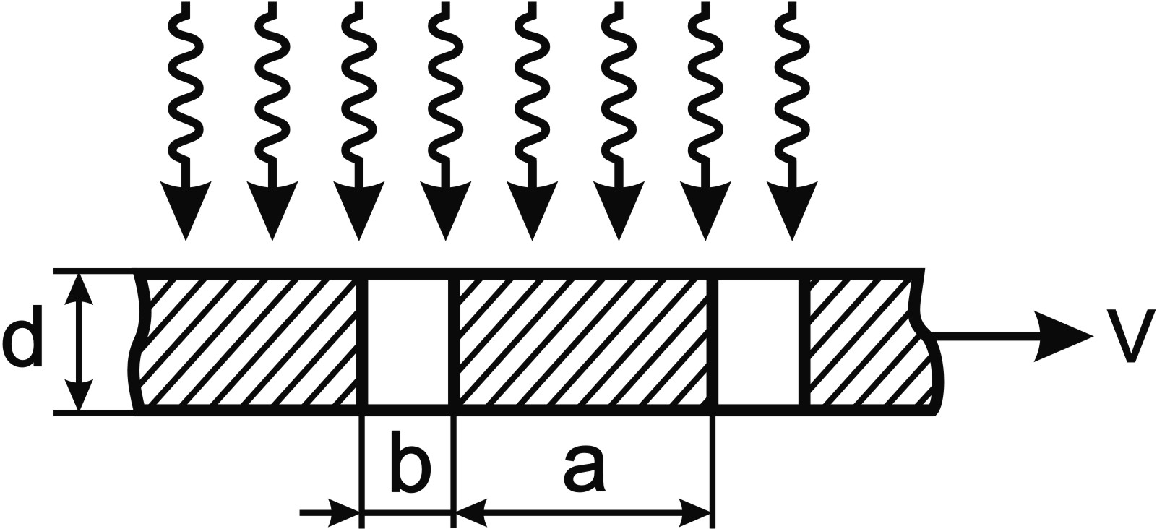,height=3.0cm}}
\end{figure}

\subsection*{Задача 3 {\usefont{T2A}{cmr}{b}{n}(3.8)}: Какова масса зеркала?}
Давлением лазерного луча зеркало удерживается в поле тяжести. Какова масса зеркала, 
если мощность лазера 200 кВт?

\subsection*{Задача 4 {\usefont{T2A}{cmr}{b}{n}(3.9)}: Энергия и длительность 
светового импульса после отражения}
Лазер испускает импульс света длительностью $T$ и полной энергией $E$, который 
отражается от идеального зеркала, приближающегося к лазеру со скоростью $V$. Каковы 
энергия и длительность светового импульса после отражения? Свет падает по нормали к
поверхности зеркала.

\subsection*{Задача 5 {\usefont{T2A}{cmr}{b}{n}(3.16)}: Минимальная кинетическая 
энергия}
Докажите, что суммарная кинетическая энергия ансамбля невзаимодействующих частиц имеет 
минимальную величину в системе центра масс.

\section[Семинар 11]
{\centerline{Семинар 11}}

\subsection*{Задача 1 {\usefont{T2A}{cmr}{b}{n}(3.11)}: Какова должна быть
мощность лазера?}
На космическом корабле, удаляющемся от Земли со скоростью $V=c/2$, вышла из строя 
энергетическая установка. Чтобы обеспечить корабль энергией, с Земли посылают лазерный 
луч. Какова должна быть мощность лазера, если на борту корабля потребляется мощность 
$N$?

\subsection*{Задача 2 {\usefont{T2A}{cmr}{b}{n}(3.13)}:  Какова максимальная 
кинетическая энергия электронов?}
Пучок электронов со средней энергией частиц 50~ГэВ и относительным разбросом энергий
1\,\% движется вдоль оси $x$ Л-системы отсчета. Какова максимальная кинетическая 
энергия электронов в сопровождающей пучок системе отсчета?

\subsection*{Задача 3 {\usefont{T2A}{cmr}{b}{n}(3.28)}:  Ускорение идеального 
зеркала лучом лазера}
Идеальное зеркало массой 1 кг ускоряется лучом лазера, расположенного на Земле. Какой 
должна быть мощность лазера, чтобы ускорить зеркало до скорости 0,8 c за один год?

\subsection*{Задача 4 {\usefont{T2A}{cmr}{b}{n}(3.21)}:  Эффект Доплера в среде}
Источник света с частотой $\nu_0$ движется со скоростью $V$ в неподвижной среде 
с показателем преломления $n$ мимо неподвижного наблюдателя. Какую частоту будет 
регистрировать наблюдатель при приближении источника света и при его удалении?

\subsection*{Задача 5 {\usefont{T2A}{cmr}{b}{n}(4.12)}: Найти энергию 
$\pi^0$-мезона }
$\pi^0$-мезон распадается на лету на два $\gamma$-кванта, углы вылета которых 
составляют соответственно $\theta_1$ и $\theta_1$  начальным направлением
движения $\pi^0$-мезона. Найти энергию $\pi^0$-мезона, если его масса равна $M$.  

\section[Семинар 12]
{\centerline{Семинар 12}}

\subsection*{Задача 1 {\usefont{T2A}{cmr}{b}{n}(4.15)}: Найти отношение 
суммарных энергий}
Пучок $\pi^0$-мезонов, летящих со скоростью $V=0,5c$, распадается по схеме $\pi^0\to
\gamma +\gamma$. Во сколько раз суммарная энергия образовавшихся фотонов, летящих 
в переднюю полусферу (по направлению движения пионов), больше, чем суммарная энергия 
фотонов, летящих назад? В системе отсчета $\pi^0$-мезонов распад изотропен.

\subsection*{Задача 2 {\usefont{T2A}{cmr}{b}{n}(4.6)}: Найти энергию 
$\pi$-мезонов}
Летящий $\pi$-мезон распадается на мюон и нейтрино: $\pi^+\to\mu^++\nu_\mu$.
Найти энергию $\pi$-мезонов, если известно, что максимальная энергия рождающихся при 
распаде нейтрино в $\alpha=100$ раз больше минимальной. Масса $\pi^+$-мезона 
140~МэВ. Массу нейтрино считать равной нулю.

\subsection*{Задача 3 {\usefont{T2A}{cmr}{b}{n}(4.22)}: Определите массу каона}
При распаде каонов с энергией 510~МэВ по схеме $K^0\to\pi^+\pi^-$ минимальный угол 
разлета $\pi$-мезонов равен $150^\circ$. Определите массу каона.

\subsection*{Задача 4 {\usefont{T2A}{cmr}{b}{n}(4.41)}: Преймущество встречных 
пучков}
При какой энергии протонов становится возможным рождение $J/\psi$-мезонов с массой 
$3,1\cdot 10^9$~эВ на мишени из неподвижных протонов по схеме $p+p\to p+p+J/\psi$?
При какой энергии электронов и позитронов наблюдается рождение $J/\psi$-мезона в
экспериментах на встречных электрон-позитронных пучках?

\subsection*{Задача 5 {\usefont{T2A}{cmr}{b}{n}(4.37)}: Излучение и поглощение 
света свободным электроном}
Доказать, что излучение и поглощение света свободным электроном в вакууме невозможно. 
Возможна ли однофотонная аннигиляция электрон-позитронной пары по схеме $e^-+e^+\to
\gamma$?

\subsection*{Задача 6 {\usefont{T2A}{cmr}{b}{n}(4.42)}: Пороговая энергия 
рождения электрон-позит\-ронной пары}
Найти пороговую энергию рождения электрон-позитронной пары при столкновении 
$\gamma$-квантов с разными энергиями: $\gamma+\gamma\to e^++e^-$. Энергия 
``холодного'' $\gamma$-кванта 1~эВ.

\section[Семинар 13]
{\centerline{Семинар 13}}

\subsection*{Задача 1 {\usefont{T2A}{cmr}{b}{n}(4.38)}: Симметричный разлет
фотонов}
Найти угол симметричного разлета фотонов $\alpha$ (см. рисунок), получающихся при
аннигиляции покоящегося электрона с движущимся позитроном.
\begin{figure}[htb]
\centerline{\epsfig{figure=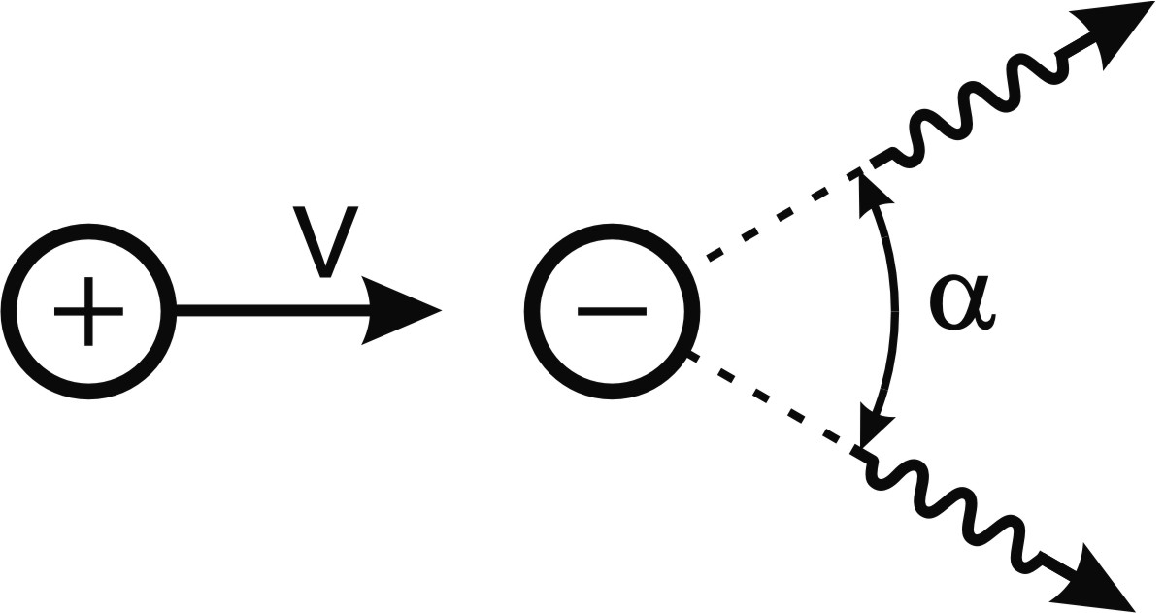,height=2cm}}
\end{figure}

\subsection*{Задача 2 {\usefont{T2A}{cmr}{b}{n}(4.43)}: Найти суммарную 
кинетическую энергию нуклонов}
Какова суммарная кинетическая энергия нуклонов, образующихся в реакции 
$\gamma+d\to n+p$ при пороговой для этой реакции энергии \mbox{$\gamma$-квантов?} 
Энергия связи дейтрона 2~МэВ.

\subsection*{Задача 3 {\usefont{T2A}{cmr}{b}{n}(4.32)}: Максимальная энергия 
электрона}
Найти максимальную энергию электрона, возникающего при распаде неподвижного мюона:
$\mu\to e+\nu+\bar\mu$.

\subsection*{Задача 4 {\usefont{T2A}{cmr}{b}{n}(4.35)}: Максимальная энергия 
вторичных $\gamma$-квантов}
При столкновении встречных пучков электронов и позитронов с энергиями 300~МэВ возможна 
реакция $e^++e^-\to 2\pi^0$. Каждый из образующихся $\pi^0$-мезонов затем 
распадается на два $\gamma$-кванта. Найти максимальную энергию образующихся  
$\gamma$-квантов.

\subsection*{Задача 5 {\usefont{T2A}{cmr}{b}{n}(4.51)}: Масса и скорость
составной частицы}
Две частицы массами $m_1$ и $m_2$, летящие со скоростями $\vec{V}_1$ и 
$\vec{V}_2$, слипаются в одну. Найти массу и скорость образовавшейся частицы.

\subsection*{Задача 6 {\usefont{T2A}{cmr}{b}{n}(4.71)}: Эффект Комптона}
Фотон с энергией $E$ сталкивается с покоящимся электроном. Найти энергию фотона, 
рассеянного на угол $\theta$. Насколько изменится длина волны фотона при таком 
столкновении?

\section[Семинар 14]
{\centerline{Семинар 14}}

\subsection*{Задача 1 {\usefont{T2A}{cmr}{b}{n}(4.61)}: Передача энергии при 
упругом столкновении}
Найти минимальную кинетическую энергию, которая может остаться у частицы массой $m_1$
после упругого столкновения с покоящейся частицей массой $m_2$ в а) нерелятивистском 
и б) релятивистском случаях.

\subsection*{Задача 2 {\usefont{T2A}{cmr}{b}{n}(4.69)}: Лобовое столкновение
электрона с протоном.}
Электрон испытывает лобовое столкновение с неподвижным протоном и передает ему 
половину своей энергии. Найти начальную энергию электрона.

\subsection*{Задача 3 {\usefont{T2A}{cmr}{b}{n}(4.73)}:  Максимальный угол 
отклонения}
На какой максимальный угол может отклониться релятивистский протон при столкновении 
с первоначально неподвижным электроном?

\subsection*{Задача 4 {\usefont{T2A}{cmr}{b}{n}(4.72)}:  Энергия рассеянного 
фотона}
Фотон с энергией 10~эВ рассеивается на угол $90^\circ$ на электроне, летящем 
навстречу. Найти энергию рассеянного фотона, если кинетическая энергия электрона была: 
a) 100~эВ; б) 10~ГэВ.

\subsection*{Задача 5 {\usefont{T2A}{cmr}{b}{n}(5.12)}: Мюоны в магнитном поле}
Мюоны движутся по круговой траектории радиусом 10~м в поперечном магнитном поле
величиной 1~Т. За какое время мюонный ток уменьшится в 20 раз? Собственное время 
жизни мюона $2 \cdot 10^{-6}\mbox{с}$, масса 105~МэВ.

\subsection*{Задача 6 {\usefont{T2A}{cmr}{b}{n}(5.7)}: Скорость 
$\Upsilon$-частицы}
При столкновении ультрарелятивистских электрон-позитронных пучков из двух одинаковых 
по размерам кольцевых накопителей наблюдается рождение $\Upsilon$-частицы ($e^++e^-
\to\Upsilon$). Определите скорость $\Upsilon$-частицы, если напряженности магнитного 
поля на дорожках накопителей отличаются в три раза. Масса $\Upsilon$-частицы 10~ГэВ.

\section[Семинар 15]
{\centerline{Семинар 15}}

\subsection*{Задача 1 {\usefont{T2A}{cmr}{b}{n}(5.10)}: Ширина пучка на выходе
из магнитного поля}
Пучок протонов со средней энергией $E=2$~ГэВ и энергоразбросом $2\,\%$ инжектируется 
в полупространство с однородным поперечным магнитным полем $B=1$~Т (см. рисунок). 
Каким будет ширина пучка на выходе из магнитного поля?
\begin{figure}[htb]
\centerline{\epsfig{figure=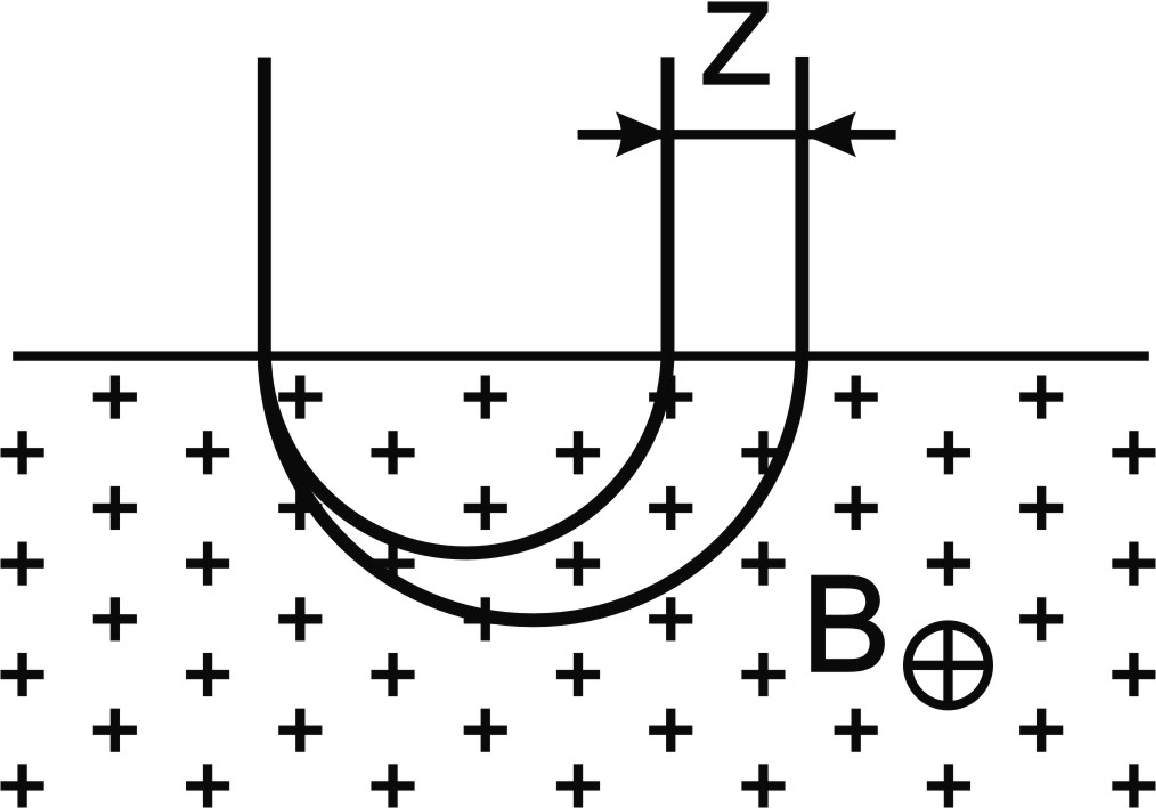,height=4cm}}
\end{figure}

\subsection*{Задача 2 {\usefont{T2A}{cmr}{b}{n}(5.8)}: Равенство углов падения и
отражения}
Показать, что угол падения равен углу отражения для электрона, влетающего под углом
$\alpha$ в полупространство с неоднородным магнитным полем (см. рисунок). Поле 
перпендикулярно плоскости рисунка и изменяется лишь в направлении, перпендикулярном 
границе.
\begin{figure}[htb]
\centerline{\epsfig{figure=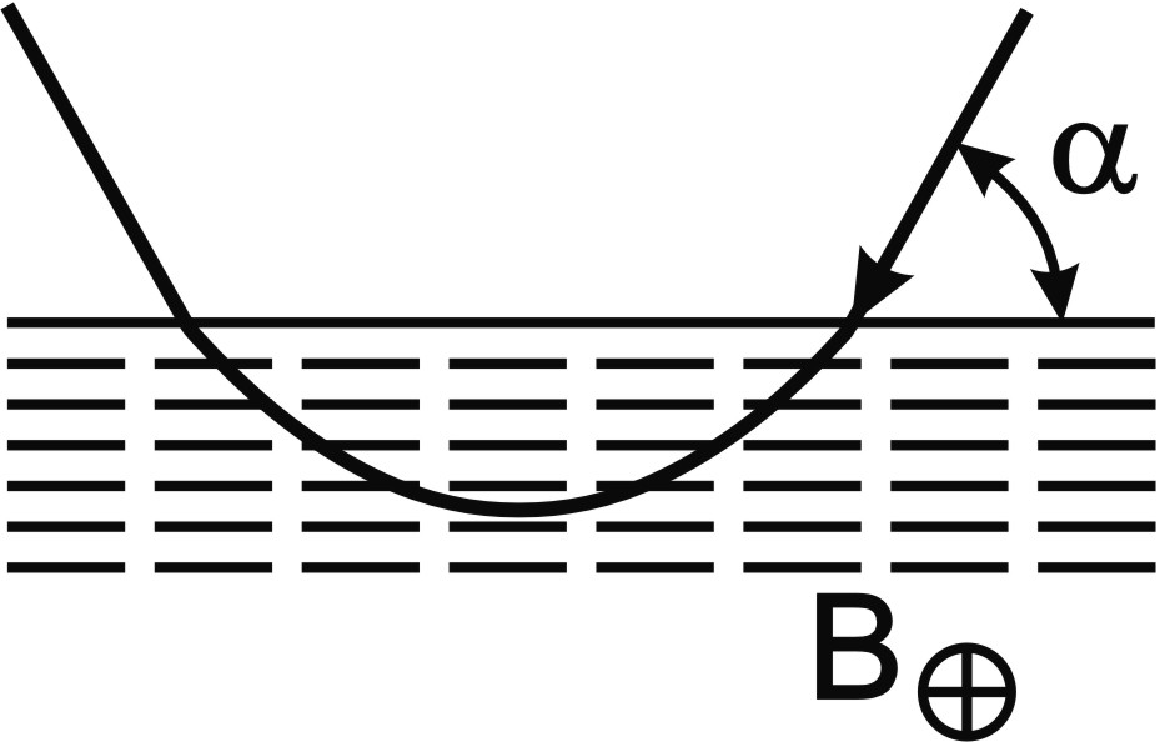,height=2cm}}
\end{figure}

\subsection*{Задача 3 {\usefont{T2A}{cmr}{b}{n}(5.21)}: Торможение в электрическом
поле}
Электрон влетает в тормозящее постоянное однородное электрическое поле напряженностью
${\cal{E}}$ с начальной скоростью $\vec{V}_0 \parallel {\cal{E}}$. Через какое 
время электрон вернется в начальную точку? Какой путь он пройдет за это время?

\subsection*{Задача 4 {\usefont{T2A}{cmr}{b}{n}(5.23)}: За какое время  сигнал 
связи догонит ракету?}
Ракета удаляется от Земли с постоянным ускорением $a^\prime$ в сопутствующей системе 
отсчета. Через время $T$ после старта ей вдогонку посылается сигнал связи. За какое 
время он догонит ракету? При каком $T$ сигнал уже не сможет догнать ракету?

\subsection*{Задача 5 {\usefont{T2A}{cmr}{b}{n}(5.11)}: Магнитный фильтр}
Пучок протонов и электронов с одинаковыми энергиями 2~ГэВ проходит через магнитный 
фильтр, который состоит из двух участков длины 0,2~м с однородным и перпендикулярным 
пучку магнитным полем величиной 1~T разной полярности (см. рисунок). Найти расстояние 
между пучками протонов и электронов после прохождения фильтра.
\begin{figure}[htb]
\centerline{\epsfig{figure=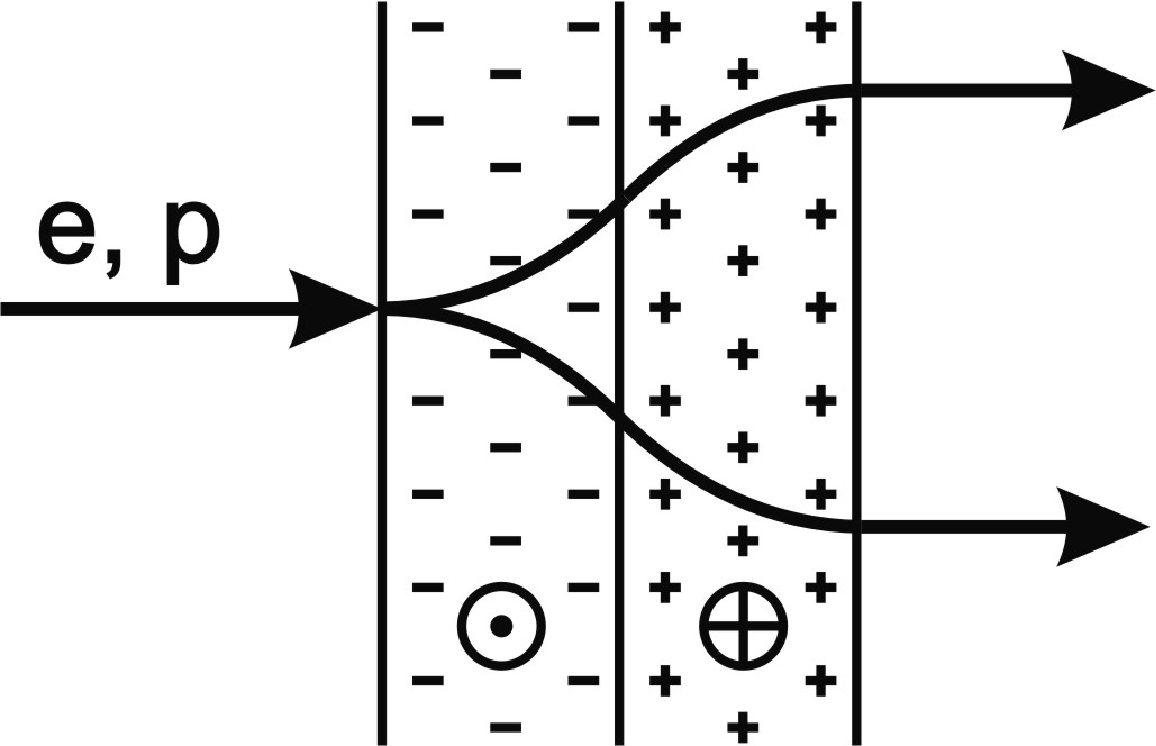,height=4cm}}
\end{figure}

\subsection*{Задача 6 {\usefont{T2A}{cmr}{b}{n}(5.19)}:  Радиус 
кривизны траектории электрона в верхней точке}
Электрон со скоростью $V$ влетает по нормали в область поперечного магнитного поля, 
увеличивающегося в направлении $z$ по закону $B=B_0(1+z/a)$. Определите радиус 
кривизны траектории электрона в верхней точке траектории $O$ (см. рисунок).
\begin{figure}[htb]
\centerline{\epsfig{figure=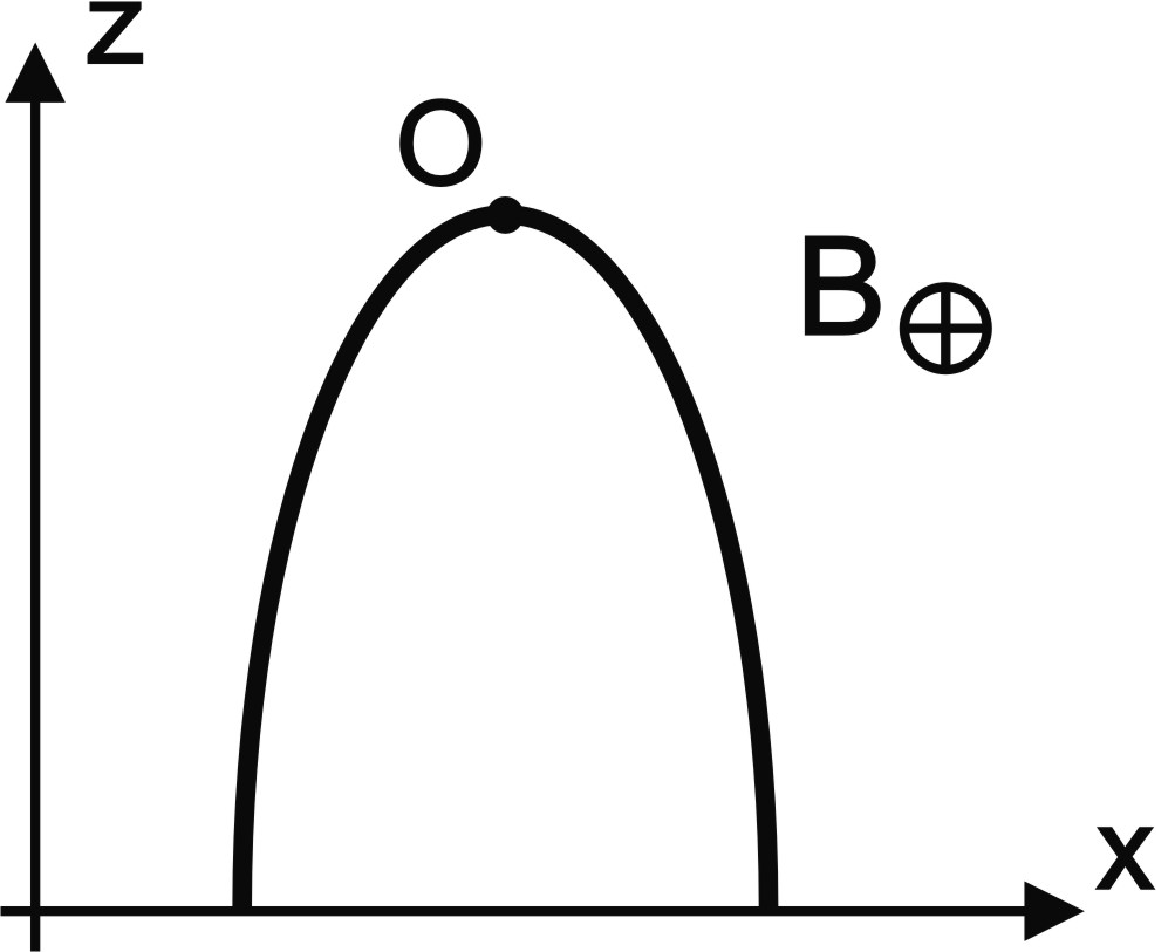,height=4cm}}
\end{figure}

\subsection*{Задача 7 {\usefont{T2A}{cmr}{b}{n}(5.9)}: Какая часть мюонов выйдет 
из области магнитного поля, не распавшись? }
Мюоны влетают под углом $\alpha$ в полупространство с однородным поперечным магнитным 
полем величиной $B=10^{-2}$~Т. Какая часть мюонов выйдет из области магнитного поля, 
не распавшись? Масса мюона 105~МэВ, собственное время жизни $2\cdot 10^{-6}$~с.

\section[Семинар 16]
{\centerline{Семинар 16}}

\subsection*{Задача 1 {\usefont{T2A}{cmr}{b}{n}(5.27)}: Столкновение протона и
электрона в электрическом поле}
Протон и электрон ускоряются из состояния покоя навстречу друг другу однородным 
электрическим полем напряженностью ${\cal{E}}=10^7$~В/м. При каком минимальном 
расстоянии $L$ между точками старта протона и электрона возможно образование пионов 
при столкновении в реакции $p+e^-\to p+e^-\pi^-+\pi^+$? Каким будет это расстояние, 
если протон стартует раньше электрона на время $T=10^{-7}$~с? $M_\pi=140$~МэВ.

\subsection*{Задача 2 {\usefont{T2A}{cmr}{b}{n}(5.29)}: Предельные значения 
компонент импульса и скорости}
Частица массой $m$ влетает с начальной скоростью $V_0$ в область, где на нее 
действует постоянная сила $\vec{F}\perp\vec{V}_0$. К каким значениям стремятся 
составляющие импульса $p_\parallel,\;p_\perp$  и скорости $V_\parallel,\;
V_\perp$, направленные соответственно вдоль и поперек начальной скорости?

\subsection*{Задача 3 {\usefont{T2A}{cmr}{b}{n}(5.30)}: Высота траектории и 
минимальная скорость электрона}
Электрон с кинетической энергией 1~МэВ влетает в тормозящее однородное электрическое 
поле напряженностью ${\cal{E}}=10^6$~В/м под углом $\alpha=45^\circ$ 
(см. рисунок). Какова высота траектории и минимальная скорость электрона? Найти 
расстояние между точками влета и вылета электрона и время его пролета через 
конденсатор.
\begin{figure}[htb]
\centerline{\epsfig{figure=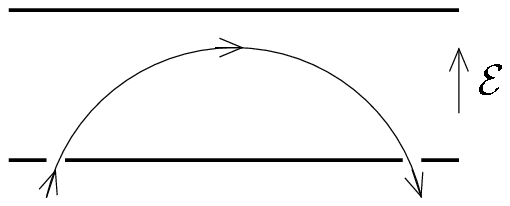,height=2cm}}
\end{figure}

\subsection*{Задача 4: Минимальный угол разлета $\gamma$-квантов}
Найти минимальный угол разлета $\gamma$-квантов, образующихся при аннигиляции 
покоящегося электрона с движущимся позитроном энергией $E$.

\subsection*{Задача 5 {\usefont{T2A}{cmr}{b}{n}(3.19)}: Скорость второго корабля 
относительно Земли}
Космический корабль, летящий  относительно Земли со скоростью $V_1$, посылает сигнал 
частотой $\omega_1$ и после его отражения от летящего навстречу другого космического 
корабля принимает сигнал частотой $\omega_2$ (частоты даны в системе отсчета первого 
корабля). Найти скорость второго корабля относительно Земли.

\section[Семинар 17]
{\centerline{Семинар 17}}

\subsection*{Задача 1 {\usefont{T2A}{cmr}{b}{n}(\hspace*{-1.5mm}
\cite{8}, 2.14)}: Какое время покажут часы?}
Два космических корабля летят встречными курсами со скоростями $V=0,5c$ каждый. При 
пролете мимо друг друга (в точке $O$ рисунка) их часы были синхронизованы. Через час
после встречи (по своим часам) один из кораблей посылает радиосигнал вдогонку другому. 
Какое время покажут часы, установленные на втором корабле, в момент приема посланного 
радиосигнала?
\begin{figure}[htb]
\centerline{\epsfig{figure=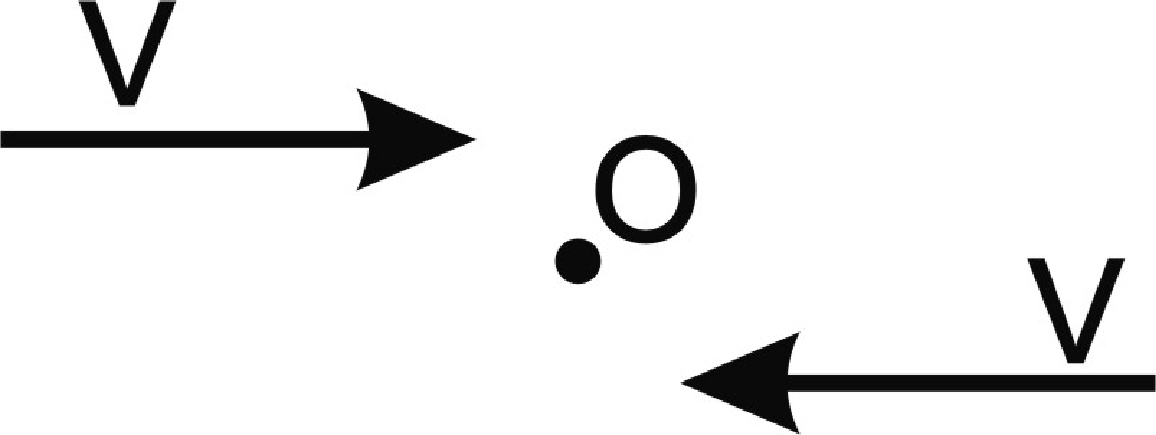,height=1.5cm}}
\end{figure}

\subsection*{Задача 2: Частота принимаемого сигнала при постоянном угле пеленга}
Космический корабль движется со скоростью $V=0,8c$, сохраняя постоянный угол пеленга
$\theta$ между скоростью корабля и направлением на маяк (в СО маяка). Частота сигнала
маяка $\nu_0$. При каких значениях угла пеленга частота принимаемого на корабле
сигнала будет меньше $\nu_0$?  

\subsection*{Задача 3 {\usefont{T2A}{cmr}{b}{n}(\hspace*{-1.5mm}\cite{8}, 
2.53)}: Под каким углом выйдет луч света через боковой торец?}
Луч света падает вертикально к поверхности стеклянного прямоугольного бруска 
(см. рисунок), движущегося вдоль горизонтальной оси со скоростью $V$. Найти угол 
к вертикали, под которым луч выйдет через боковой торец бруска после преломления, 
и условие, при котором свет выйдет из бруска. Показатель преломления стекла $n$. 
Закон преломления света на границе двух сред в системе бруска $n_1\sin{\alpha_1}=
n_2\sin{\alpha_2}$.
\begin{figure}[htb]
\centerline{\epsfig{figure=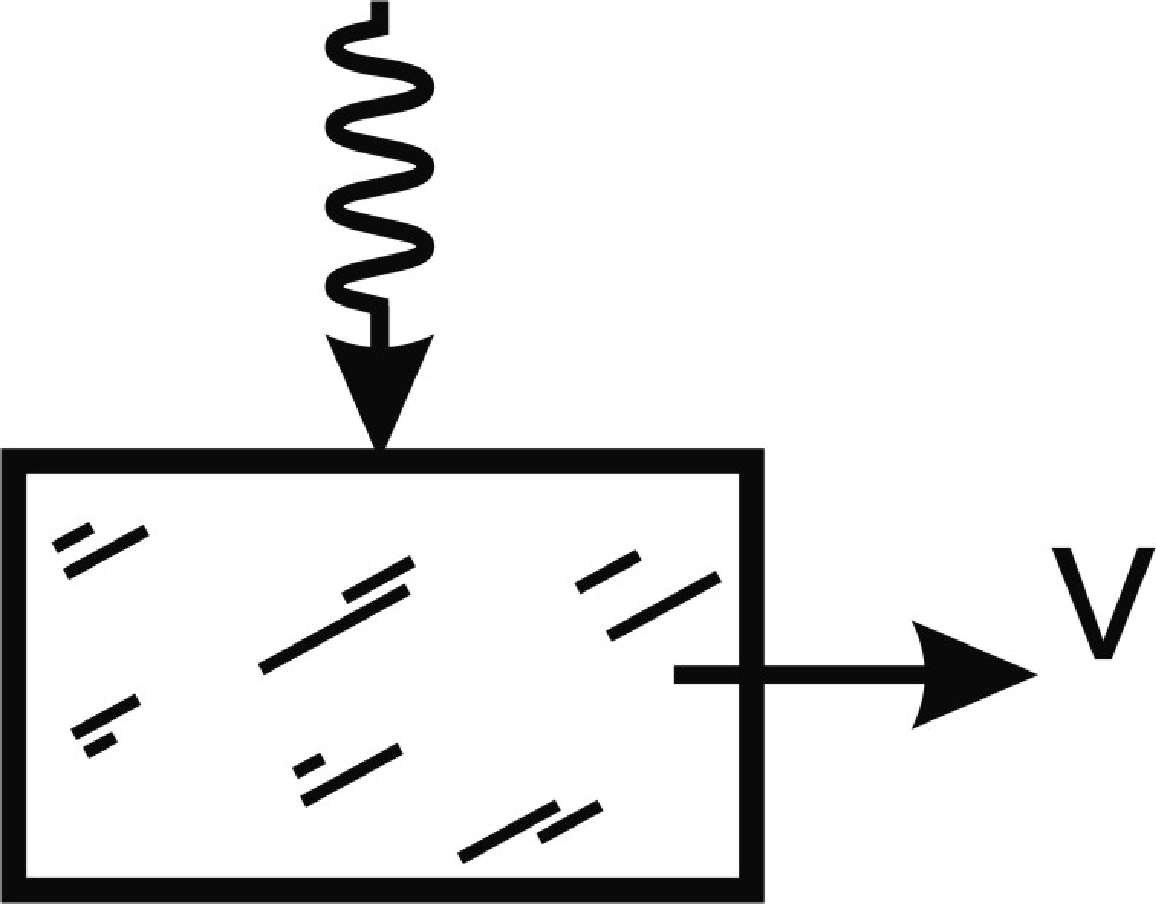,height=2cm}}
\end{figure}

\subsection*{Задача 4: Энергия $\gamma$-кванта в реакции $e^++e^-\to 
\omega+\gamma$}
При столкновении встречных электрон-позитронных пучков с энергией 450~МэВ образуется
$\omega$-мезон и $\gamma$-квант по схеме $e^++e^-\to \omega+\gamma$. Найти
энергия $\gamma$-кванта. Масса $\omega$-мезона 782~МэВ.

\subsection*{Задача 5 {\usefont{T2A}{cmr}{b}{n}(\hspace*{-1.5mm}\cite{8}, 
5.16)}: Распад каона в пузырьковой камере}
В пузырьковую камеру, которая находится в поперечном однородном магнитном поле, 
влетает поток каонов $K^-$. Трек каона в камере имеет вид дуги окружности (см. 
рисунок). При распаде каона по схеме $K^-\to\mu^-+\nu_\mu$ отношение максимального 
радиуса трека образующегося мюона к минимальному радиусу равно $N=3$ (для треков, 
лежащих в плоскости, перпендикулярной полю). Найти отношение масс каона и мюона 
$M_K/M_\mu$. Время жизни каона $\tau=1,2\cdot 10^{-8}$~с, средняя длина трека 
каона до распада в камере 1,85~м.
\begin{figure}[htb]
\centerline{\epsfig{figure=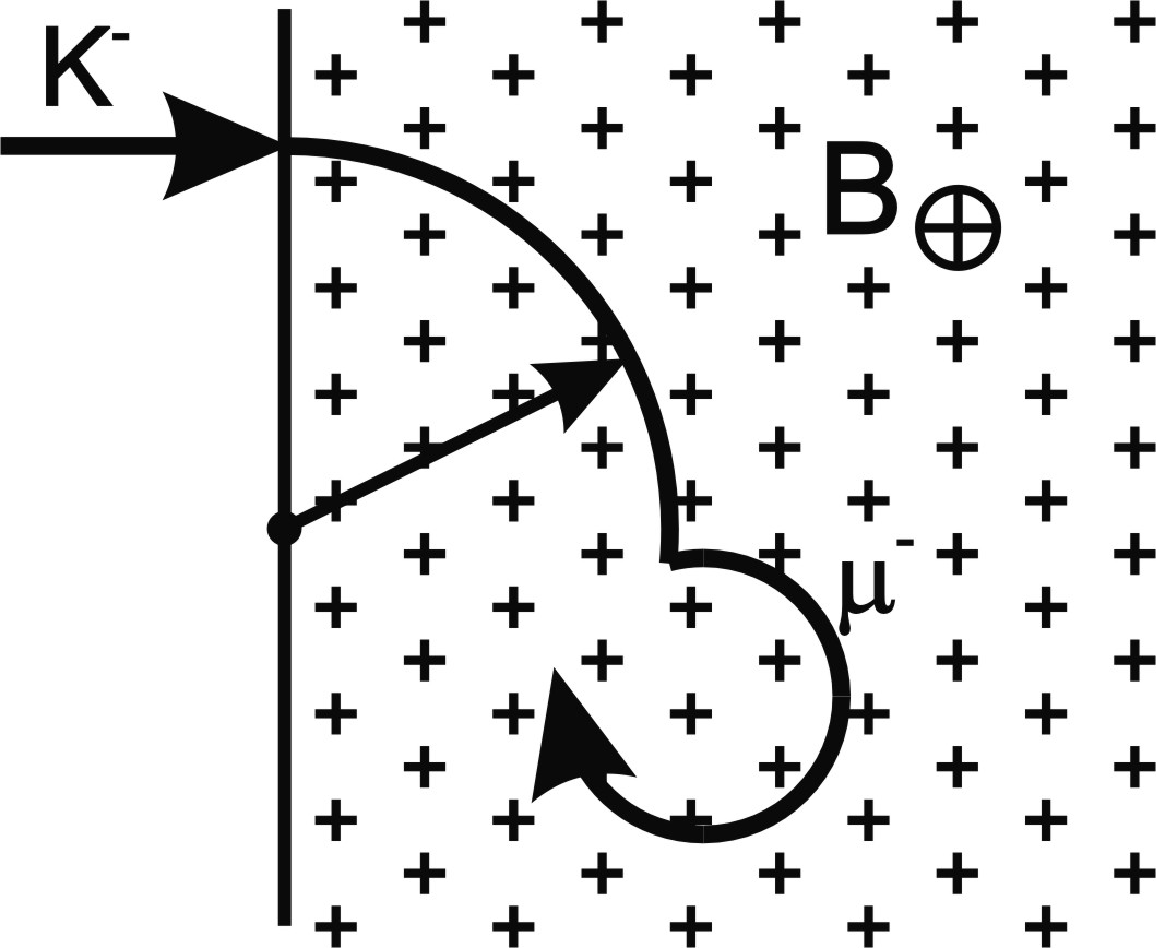,height=4cm}}
\end{figure}

\subsection*{Задача 6 {\usefont{T2A}{cmr}{b}{n}(\hspace*{-1.5mm}\cite{8}, 
5.41)}: Какую часть топлива израсходует ракета?}
Какую часть топлива израсходует ракета с фотонным двигателем при доускорении от 
скорости 0,9 с до 0,99 с?

\section[Семинар 18]
{\centerline{Семинар 18}}

\subsection*{Задача 1: За какое время волна от камня достигнет берега?}
В реку со скоростью воды $V\sim c$ перпендикулярно линии берега из точки $A$ на 
берегу бросили камень, упавший в воду на расстоянии $a$ от берега. За какое время 
волна от камня достигнет берега? Через какое время волна придет к точке $A$? Скорость
волны в стоячей воде $u>V$. 

\subsection*{Задача 2 {\usefont{T2A}{cmr}{b}{n}(\hspace*{-1.5mm}\cite{8}, 
2.47)}: Сколько вещества окажется внутри пенала?}
Цилиндрический пенал сечением $S$ и собственной длиной $L_0$ движется 
с релятивистской скоростью $V$ навстречу неподвижному облаку пыли плотностью $\rho$. 
Когда первые пылинки прилипают к дну пенала $B$ (см. рисунок), по его стенкам 
начинает распространяться упругая волна, движущаяся относительно пенала со скоростью 
$u$. Когда волна достигает переднего конца пенала, его крышка $A$ закрывается. 
Сколько вещества окажется внутри пенала?
\begin{figure}[htb]
\centerline{\epsfig{figure=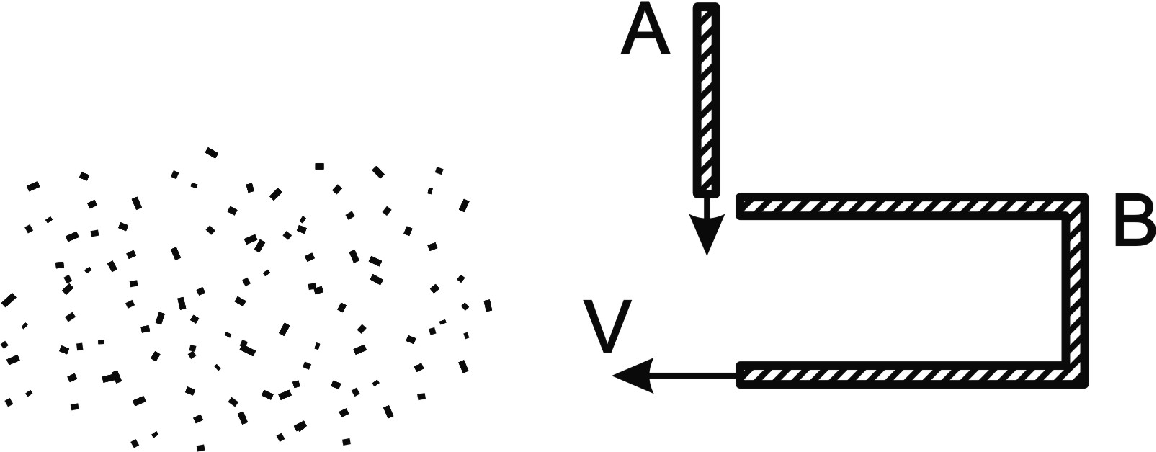,height=3cm}}
\end{figure}

\subsection*{Задача 3 {\usefont{T2A}{cmr}{b}{n}(\hspace*{-1.5mm}\cite{8}, 
3.23)}: Какова частота сигналов, принимаемых на подводных лодках?}
Две подводные лодки движутся навстречу друг другу с релятивистскими скоростями $V_1$ 
и $V_2$ относительно воды. Радары лодок имеют одинаковую собственную частоту $\nu_0$, 
скорость сигнала в воде равна $c/n$. Какова частота сигналов, принимаемых на лодках?

\subsection*{Задача 4: Минимальный угол разлета $\gamma$-квантов}
При столкновении протона с первоначально неподвижным протоном в реакции $p+p\to p+p+
\pi^0$ рождается $\pi^0$-мезон, который на лету распадается на два $\gamma$-кванта.
Найти минимальный угол разлета квантов при пороговой энергии налетающего протона. Масса
протона $M=938$~МэВ, $\pi^0$-мезона --- $m=135$~МэВ.

\subsection*{Задача 5: Скорость дрейфа в магнитном поле}
Найти среднюю скорость перемещения электрона вдоль оси $x$ при движении в поперечном 
магнитном поле, направленном вдоль оси $z$ и имеющем величину $B_1$ при $y<0$ и
величину $B_2$ при $y>0$. В начальный момент скорость электрона $V$ была направлена 
вдоль оси $y$.

\subsection*{Задача 6: Найти первоначальную энергию частицы}
В область поперечного электрического поля напряженностью ${\cal{E}}$ влетает частица
с импульсом $p_0$, зарядом $e$ и пересекает ее за время $T$. Ширина области $d$.
Найти первоначальную энергию частицы.

\section[Семинар 19]
{\centerline{Семинар 19}}

\subsection*{Задача 1: Какую задержку между вспышками зарегистрирует наблюдатель?}
Точка $A$ имеет декартовы координаты $(0,R)$, точка $B$ --- $(R,0)$. Вдоль осей
координат с одинаковыми по величине скоростями $V=0,6c$ движутся два наблюдателя.
В момент, когда оба наблюдателя находились в начале координат $O$, в точке $A$
произошла вспышка света. Вторая вспышка произошла в точке $B$, причем наблюдатель, 
движущийся вдоль оси $x$, увидел свет от обоих вспышек одновременно. Какую задержку
между вспышками зарегистрирует наблюдатель, движущийся вдоль оси $y$?

\subsection*{Задача 2: Распад $K_S$-мезона}
Короткоживущий $K_S$-мезон распадается на лету на два $\pi$-мезона по схеме 
$K_S\to \pi^++\pi^-$. Найти энергию исходного $K_S$-мезона, если в Л-системе
регистрируется $\pi^+$-мезон с кинетической энергией 26,66~МэВ, летящий 
перпендикулярно направлению движения $K_S$-мезона. Найти импульс второго $\pi$-мезона
в Л-системе отсчета. Масса $K_S$-мезона $M=500$~МэВ, масса $\pi$-мезона
$m=140$~МэВ.

\subsection*{Задача 3: Пучок мюонов в магнитном поле}
Пучок мюонов движется в поперечном магнитном поле по траектории радиусом 50~м. После 
115 оборотов пучка количество мюонов уменьшилось из-за распада в 3 раза. Определите
величину магнитного поля и начальную плотность пучка в собственной системе отсчета.
Время жизни мюона равно $2,2\cdot 10^{-6}$~с, масса $m_\mu=105$~МэВ, а начальная
плотность в лабораторной системе была $n_0=10^{10}$~см$^{-3}$.

\subsection*{Задача 4 {\usefont{T2A}{cmr}{b}{n}(5.33)}: Монета на наклонной 
плоскости}
На наклонной плоскости с углом наклона $\alpha$ в поле тяжести лежит монета. Монете 
сообщается релятивистская скорость $V_0$ вдоль горизонтальной оси (см. рисунок). Найти
установившуюся скорость движения монеты $u$, если коэффициент трения 
$\mu=\tg{\alpha}$ не зависит от скорости.
\begin{figure}[htb]
\centerline{\epsfig{figure=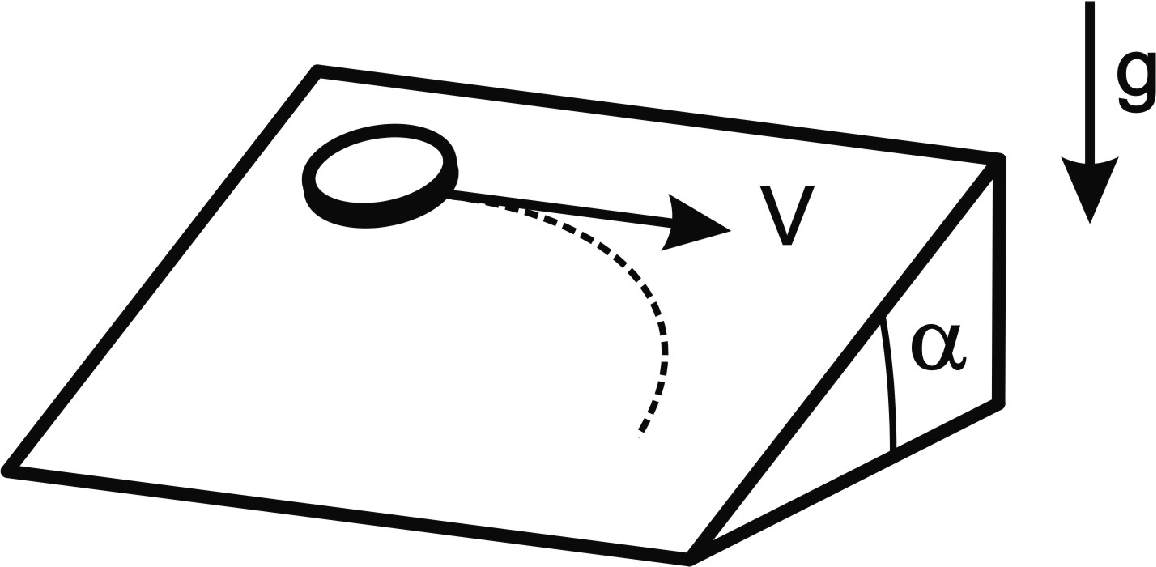,height=3cm}}
\end{figure}

\subsection*{Задача 5 {\usefont{T2A}{cmr}{b}{n}(2.39)}: Скорость частиц в системе 
центра масс}
Найти скорость двух одинаковых частиц в системе центра масс, если в Л-системе отсчета 
их скорости $\vec{V}_1$ и $\vec{V}_2$.

\subsection*{Задача 6 {\usefont{T2A}{cmr}{b}{n}(2.51)}: Ориентация ракеты связи}
Космический корабль летит со скоростью $V$. В системе отсчета корабля под углом 
$\alpha^\prime$ к направлению движения запускают ракету связи со скоростью $u^\prime$
(см. рисунок). Найти в Л-системе отсчета углы между: а) векторами скорости ракеты и 
корабля, б) осью ракеты и скоростью корабля.
\begin{figure}[htb]
\centerline{\epsfig{figure=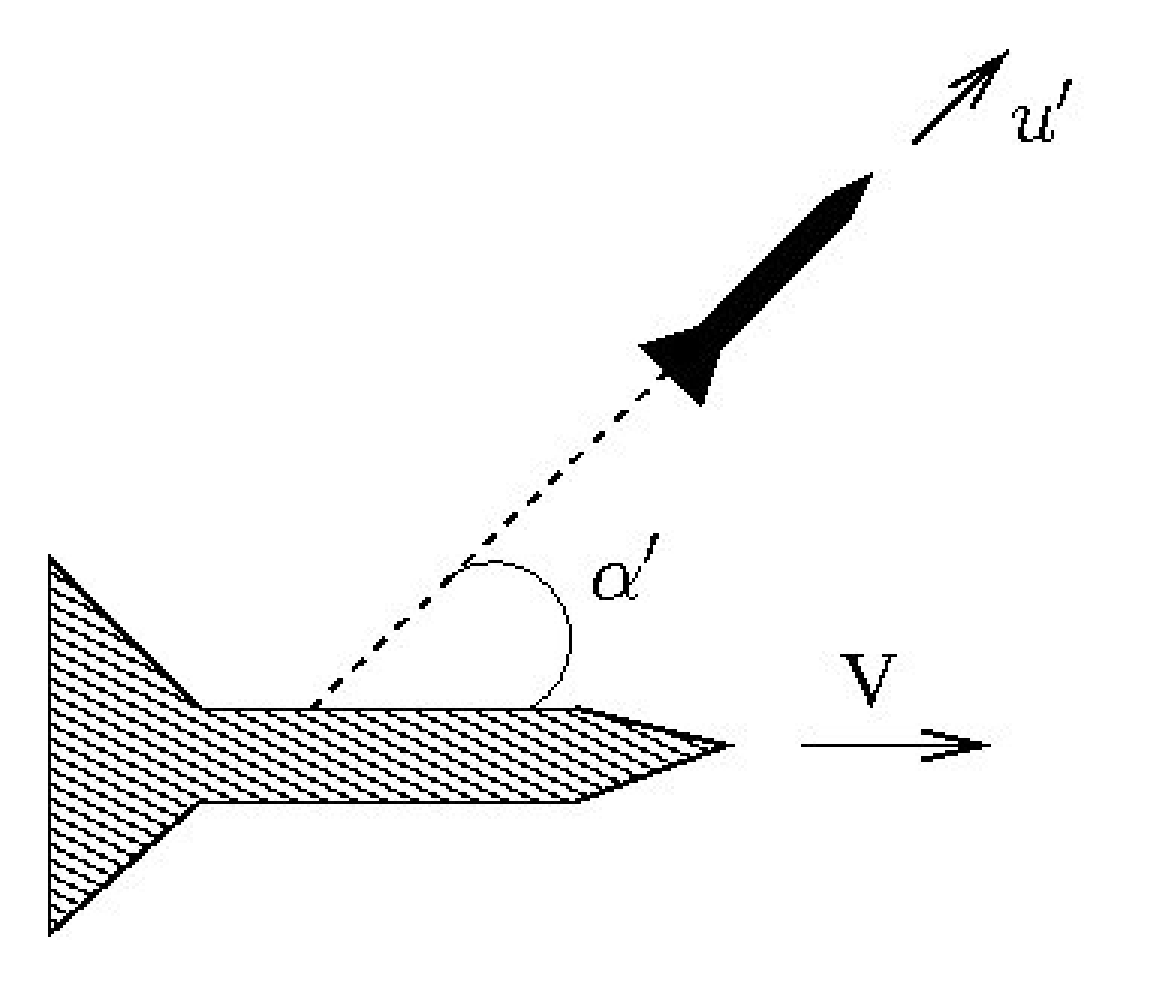,height=3.5cm}}
\end{figure}

\section[Семинар 20]
{\centerline{Семинар 20}}

\subsection*{Задача 1 {\usefont{T2A}{cmr}{b}{n}(6.5)}: Зависимость силы и 
потенциала от координаты}
\begin{figure}[htb]
\centerline{\epsfig{figure=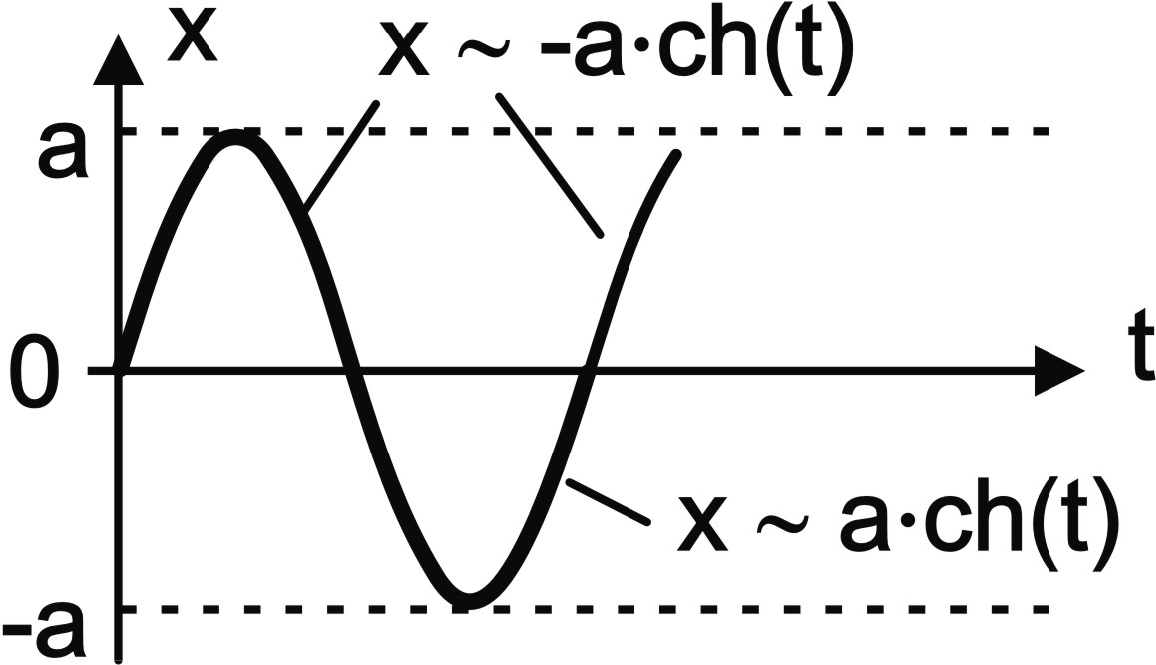,height=2.5cm}}
\end{figure}
Частица движется в статическом силовом поле по закону, показанному на рисунке. Найти 
зависимость силы и потенциала от координаты.

\subsection*{Задача 2 {\usefont{T2A}{cmr}{b}{n}(6.6)}: Закон движения частицы
при нулевой полной энергии}
Найти закон движения частицы в поле $U=-\alpha x^4$ в случае, когда ее полная
энергия равна нулю. Нарисовать траекторию частицы на фазовой плоскости.

\subsection*{Задача 3 {\usefont{T2A}{cmr}{b}{n}(6.8)}: Траектории частицы на 
фазовой плоскости}
Нарисовать траектории частицы на фазовой плоскости для следующих одномерных полей:
\begin{eqnarray} &&
1)\;\; U(x)=\alpha^2(x^2-b^2)^2,\;\;\;\; 2)\;\; U(x)=-\alpha^2(x^2-b^2)^2,
\nonumber \\ &&
3)\;\; U(x)=U_0\sin{(kx)},\;\;\;\; 4)\;\; U(x)=\alpha^2\left (\frac{b^2}
{x^2}-\frac{1}{x}\right ). 
\nonumber
\end{eqnarray}

\subsection*{Задача 4 {\usefont{T2A}{cmr}{b}{n}(6.9)}: Как изменится  форма и 
объем области в фазовом пространстве?}
Как изменится со временем форма и объем области в фазовом пространстве, занимаемой 
группой движущихся вдоль оси $x$ невзаимодействующих друг с другом частиц, помещенных 
в одномерный ящик с координатами стенок $x_1=0,\; x_2= L$? Столкновения со стенками 
упругие. В начальный момент частицы занимали область $x_0,\;x_0+\Delta x$ и
$p_0,\;p_0+\Delta p$.

\subsection*{Задача 5 {\usefont{T2A}{cmr}{b}{n}(6.11)}: Зависимость периода от
энергии}
Как зависит период движения частицы в поле $U=\alpha |x|^\beta$ от ее
энергии? $\alpha>0$.

\subsection*{Задача 6 {\usefont{T2A}{cmr}{b}{n}(6.19)}: Закон движения тела при
наличии силы трения}
Найти закон движения тела при падении в поле тяжести. Сила сопротивления воздуха
$F=-\alpha V^2$. Начальная скорость тела равна нулю.

\subsection*{Задача 7: Отношение периодов движения частицы в левой и 
правой потенциальных ямах}
Найти отношение периодов движения частицы в левой и правой потенциальных ямах 
(см. рисунок) с потенциалом $$U(x)=k(b-x)^2\left (1-\frac{x^2}{a^2}\right ),\;\;
\;b>a,\;\;k>0,$$ 
при положительной энергии частицы $E\approx 0$
\begin{figure}[htb]
\centerline{\epsfig{figure=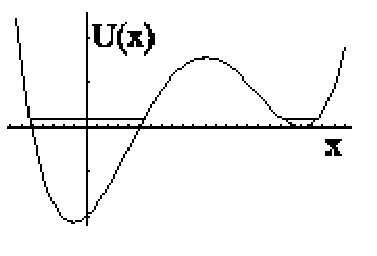,height=4cm}}
\end{figure}

\section[Семинар 21]
{\centerline{Семинар 21}}

\subsection*{Задача 1 {\usefont{T2A}{cmr}{b}{n}(6.40)}: Падение капли при наличии
тумана}
Найти закон движения сферической капли жидкости через неподвижный туман в поле 
тяжести. В начальный момент масса капли мала, а скорость равна нулю.

\subsection*{Задача 2 {\usefont{T2A}{cmr}{b}{n}(6.34)}:  Закон движения цепи}
Свернутая в клубок тяжелая однородная цепь лежит на краю горизонтального стола, причем 
вначале одно звено цепи свешивается со стола. Под действием силы тяжести цепь начинает 
соскальзывать. Принимая нулевые начальные условия, определить закон движения цепи.
Считать, что звенья цепи поочередно приобретают только вертикальную скорость.
\begin{figure}[htb]
\centerline{\epsfig{figure=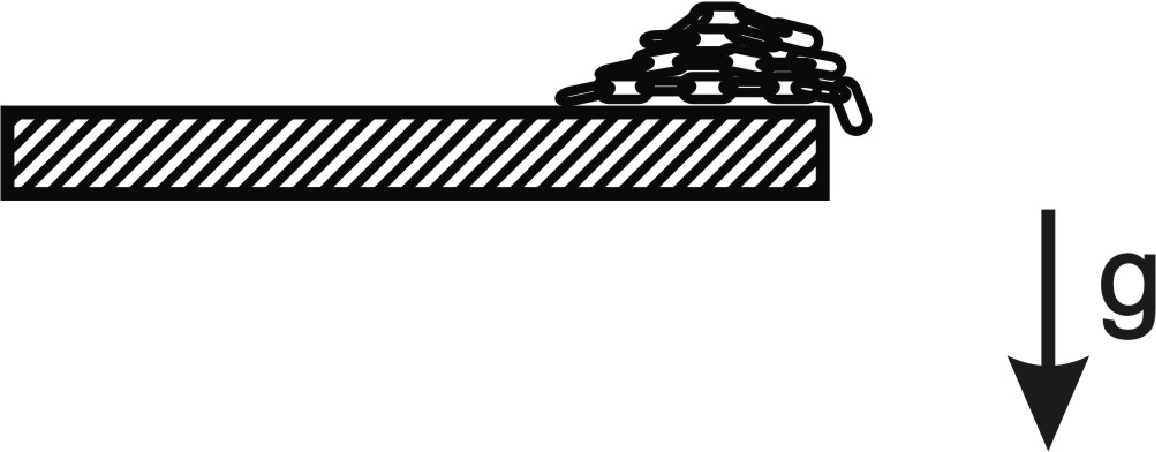,height=2.5cm}}
\end{figure}

\subsection*{Задача 3 {\usefont{T2A}{cmr}{b}{n}(6.31)}:  Падение каната на чашку
весов}
Однородный гибкий канат, висевший вертикально, падает на площадку весов. Найти 
зависимость показаний весов от времени.

\subsection*{Задача 4 {\usefont{T2A}{cmr}{b}{n}(6.13)}: Столкновение металлических 
шаров}
Изучая столкновения металлических шаров, студент обнаружил, что время соприкосновения 
шаров $T=\alpha V^{-1/5}$, где $V$ --- скорость шаров перед столкновением. 
Восстановите зависимость силы сопротивления шара деформации от величины деформации. 
Нарисуйте траектории сталкивающихся шаров на фазовой плоскости в системе центра масс.

\subsection*{Задача 5 {\usefont{T2A}{cmr}{b}{n}(6.14)}: Зависимость периода 
колебаний частицы от энергии}
Найдите зависимость периода колебаний частицы от энергии в одномерном поле 
с потенциалом
$$U(x)=\left\{\begin{array}{c} kx^2/2\;\;\;\mbox{при}\;|x\le a, \\
\infty \;\;\;\;\;\;\;\;\;\;\;\mbox{при}\;|x>a.\end{array}\right . $$ 

\subsection*{Задача 6 {\usefont{T2A}{cmr}{b}{n}(6.15)}:  По какому закону 
обращается в бесконечность период движения частицы?}
Определите, по какому закону обращается в бесконечность период движения частицы 
в поле, изображенном на рисунке, при приближении полной энергии частицы $E$ к $U(a)$,
если производная потенциала $U^\prime(a)=0$, причем $U^{\prime\prime}(a)\ne 0$.
\begin{figure}[htb]
\centerline{\epsfig{figure=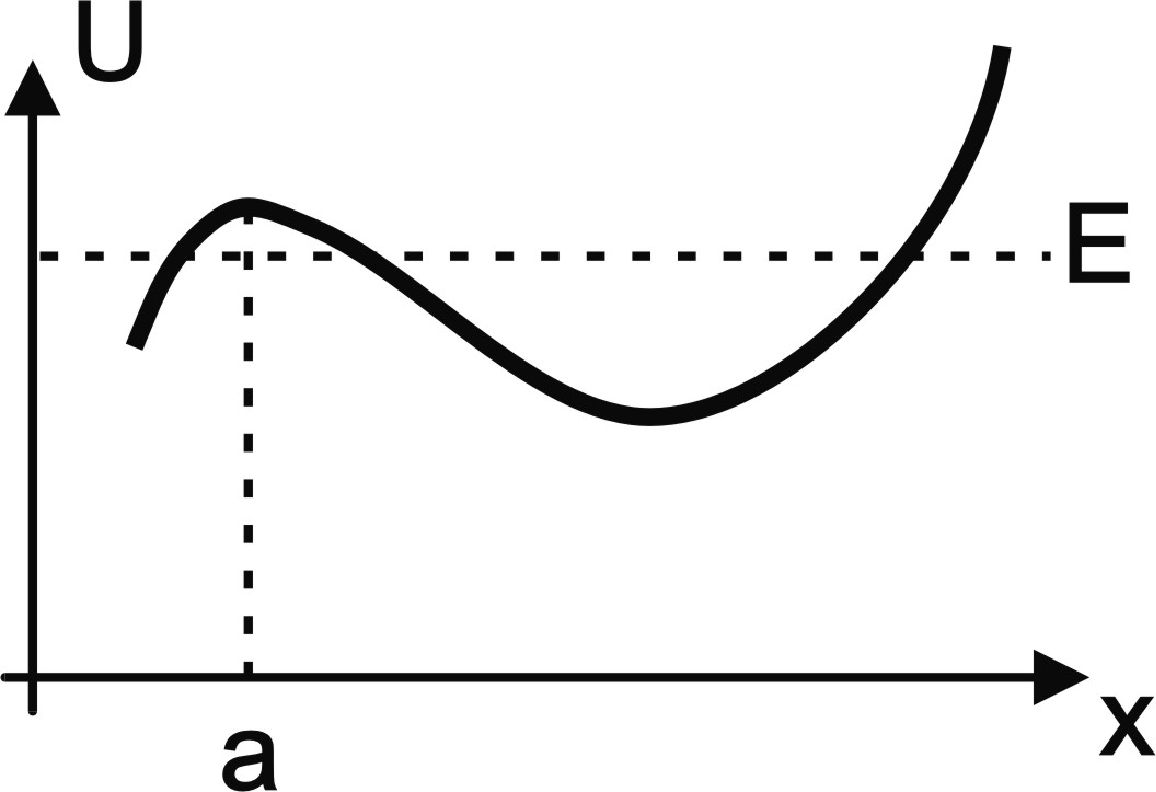,height=3cm}}
\end{figure}

\subsection*{Задача 7 {\usefont{T2A}{cmr}{b}{n}(6.24)}: Скорость движения 
кальмара}
Найти зависимость скорости движения кальмара от времени, если он затрачивает мощность 
$N$ и выбрасывает воду со скоростью $u$. Стартовая скорость кальмара равна нулю. Сила 
трения $F=-\alpha V$.

\subsection*{Задача 8 {\usefont{T2A}{cmr}{b}{n}(6.27)}: Закон движения пробки в 
горлышке бутылки с подогретым шампанским}
Исследовать теоретически движение пробки в горлышке бутылки с подогретым шампанским. 
Нарисовать графики зависимости от времени скорости и смещения пробки.

\section[Семинар 22]
{\centerline{Семинар 22}}

\subsection*{Задача 1 {\usefont{T2A}{cmr}{b}{n}(6.25)}: Сколько времени тело 
находилось в полете?}
Тело массой $m$, подброшенное вертикально вверх с малой скоростью $V_1$, вернулось 
обратно со скоростью $V_2$. Сила сопротивления воздуха $F=-\alpha V$, ускорение 
свободного падения $g$. Сколько времени тело находилось в полете?

\subsection*{Задача 2 {\usefont{T2A}{cmr}{b}{n}(7.1)}: Частота колебаний доски}
Определите частоту колебаний доски, положенной на два быстро вращающихся 
в противоположные стороны валика (см. рисунок), если расстояние между их осями $L$, 
коэффициент трения $\mu$.
\begin{figure}[htb]
\centerline{\epsfig{figure=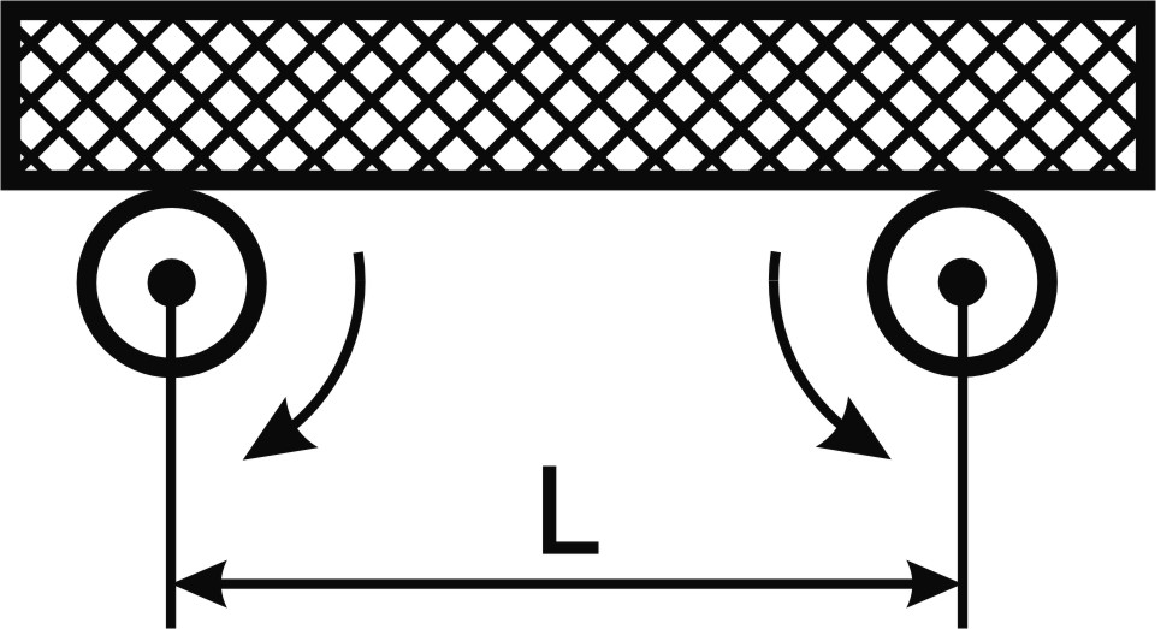,height=1.5cm}}
\end{figure}

\subsection*{Задача 3 {\usefont{T2A}{cmr}{b}{n}(7.2)}: Частота малых колебаний 
жидкости в трубке} 
Найти частоту малых колебаний жидкости в трубке, показанной на рисунке. Высота
трубки существенно больше радиуса закругления. Капиллярными эффектами пренебречь.
\begin{figure}[htb]
\centerline{\epsfig{figure=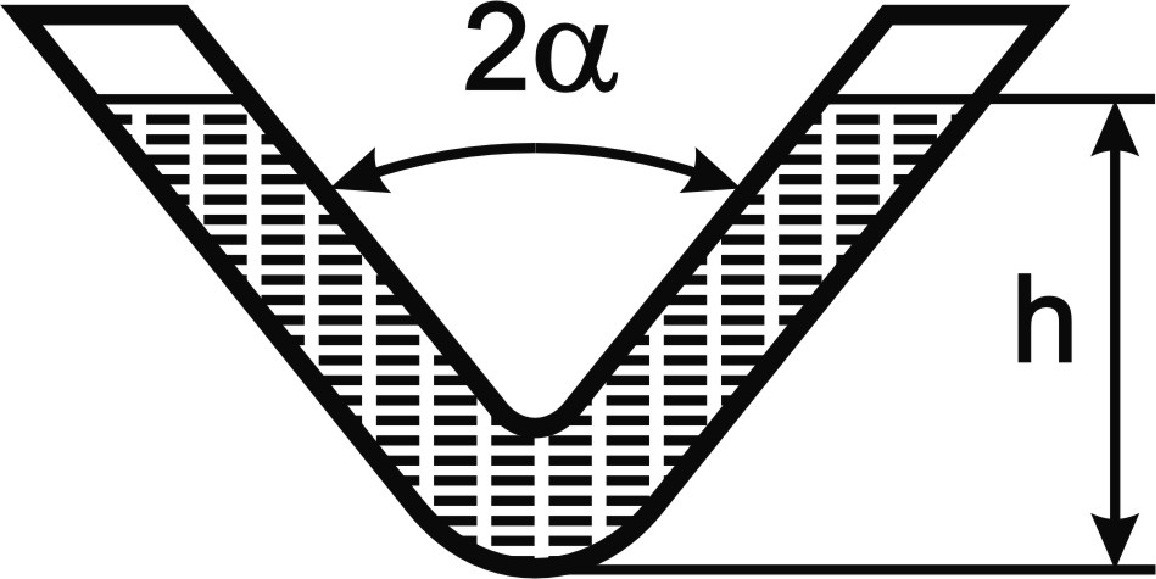,height=1.5cm}}
\end{figure}

\subsection*{Задача 4 {\usefont{T2A}{cmr}{b}{n}(7.3)}: При каком условии 
колебания устойчивы?} 
Через невесомый блок перекинута нерастяжимая нить массой $M$, длиной $L$, концы
которой натянуты пружинами жесткостью $k$ (см. рисунок). Найти собственную частоту 
колебаний нити в поле тяжести. При каком условии колебания устойчивы? Трения нет.
\begin{figure}[htb]
\centerline{\epsfig{figure=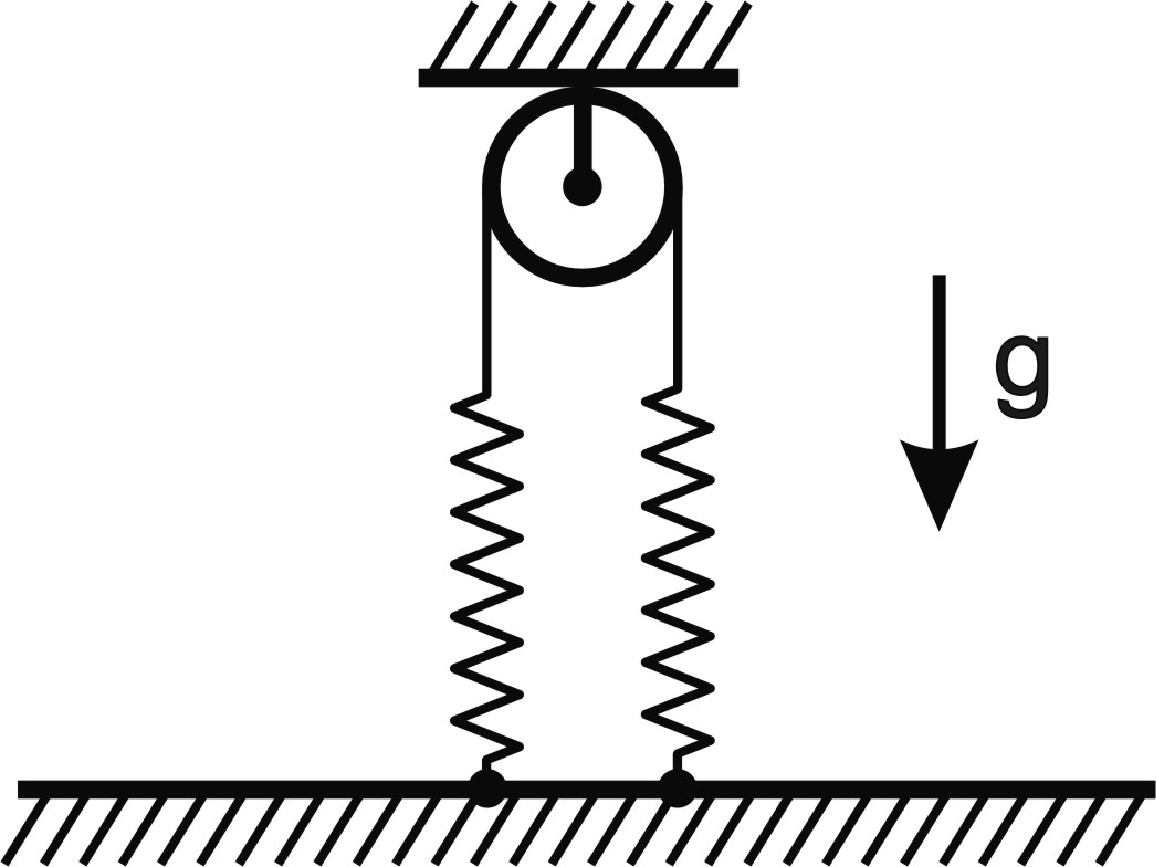,height=3cm}}
\end{figure}

\subsection*{Задача 5 {\usefont{T2A}{cmr}{b}{n}(6.35)}: За какое время цепь 
соскользнет со стола?}
Свернутая в клубок тяжелая однородная цепь полной длиной $3H$ лежит на краю стола 
высотой $H$, так что ее один конец свешивается, касаясь пола (см. рисунок). Под 
действием силы тяжести цепь начинает соскальзывать. Найти зависимость скорости
движущегося участка цепи от времени. За какое время цепь соскользнет со стола?
\begin{figure}[!h]
\centerline{\epsfig{figure=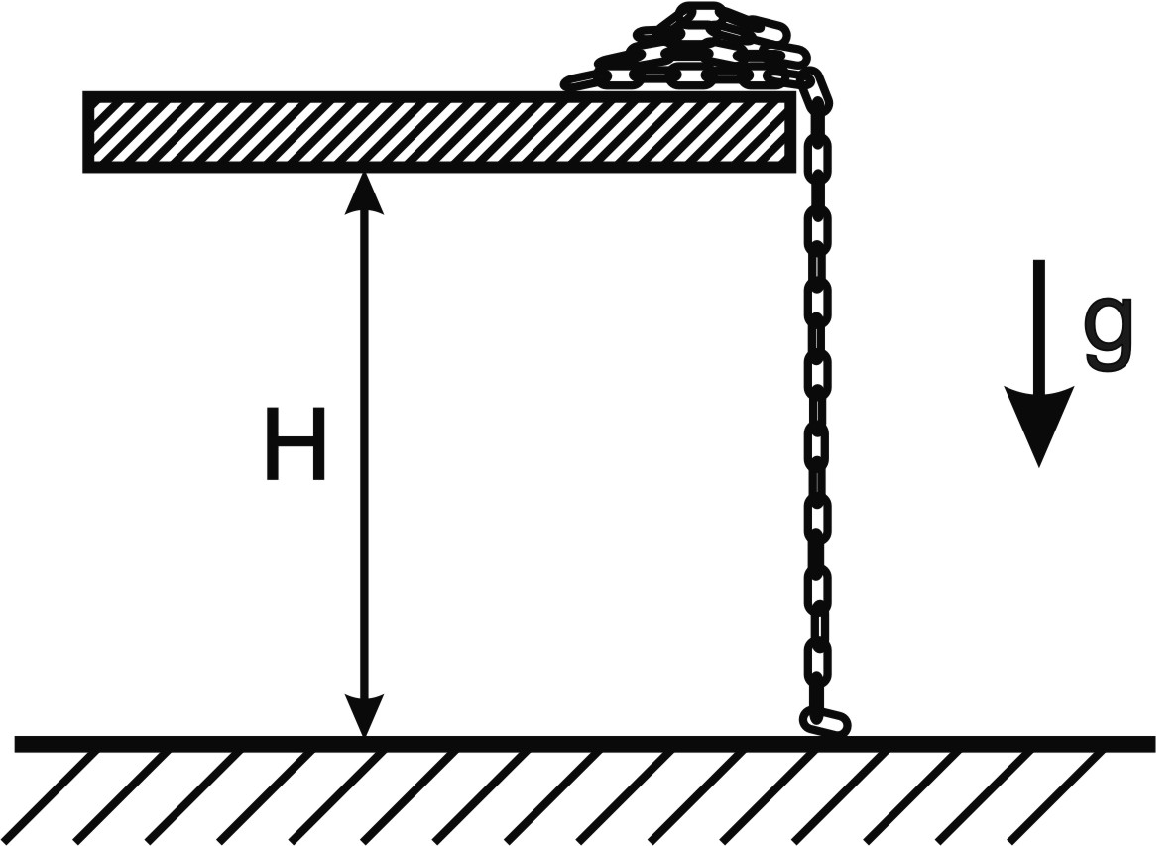,height=4cm}}
\end{figure}

\subsection*{Задача 6 {\usefont{T2A}{cmr}{b}{n}(6.42)}: Закон движения ледяного
метеорита}
Ледяной метеорит сферической формы тормозится в атмосфере Земли. Сила трения о воздух 
пропорциональна его площади и скорости: $F=-\alpha SV$. Скорость испарения вещества 
метеорита пропорциональна его площади: $dM/dt=-\beta S$,  причем в СО метеорита 
испарение изотропно. Найти зависимость скорости метеорита от времени, если его 
плотность $\rho$, а начальный радиус равен $R_0$. Силой тяжести пренебречь. 
Начальная скорость метеорита $V_0$.

\subsection*{Задача 7: Когда исчезнет сила давления на блок?}
Гладкая однородная веревка длиной $2L$ перекинута через небольшой блок так, что 
вначале находится в равновесии. Веревку немного смещают, и она под действием силы 
тяжесты начинает соскальзывать с блока. При каком перемещении конца веревки она 
перестанет давить на блок? Чему равна сила натяжения веревки около блока в этот момент?

\section[Семинар 23]
{\centerline{Семинар 23}}

\subsection*{Задача 1 {\usefont{T2A}{cmr}{b}{n}(7.19)}: Какой путь пройдет шайба 
до остановки?}
Лежащая на плоскости шайба массой $M$ прикреплена к пружинке жесткости $k$ 
(см. рисунок). Шайбу сместили из положения равновесия на расстояние $A$ и отпустили. 
Какой путь пройдет шайба до остановки? Сила трения мала и пропорциональна скорости
$F=-\alpha V\;(\alpha^2\ll kM)$.
\begin{figure}[htb]
\centerline{\epsfig{figure=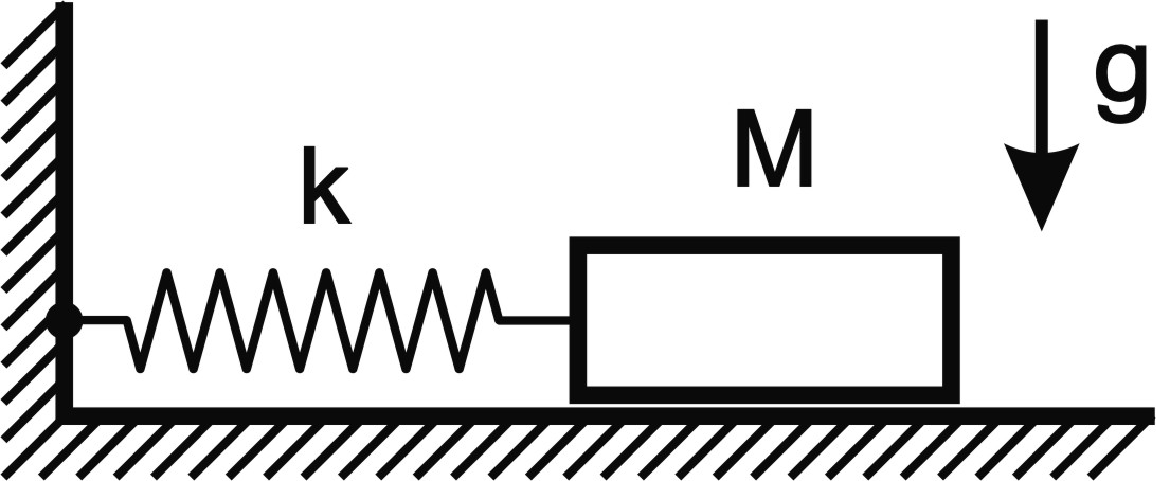,height=3cm}}
\end{figure}

\subsection*{Задача 2 {\usefont{T2A}{cmr}{b}{n}(6.44)}: Закон движения нити}
На гладком столе лежит нить, сложенная пополам. К одному из ее концов приложена 
постоянная сила $F$ (см. рисунок). Описать движение нити.
\begin{figure}[htb]
\centerline{\epsfig{figure=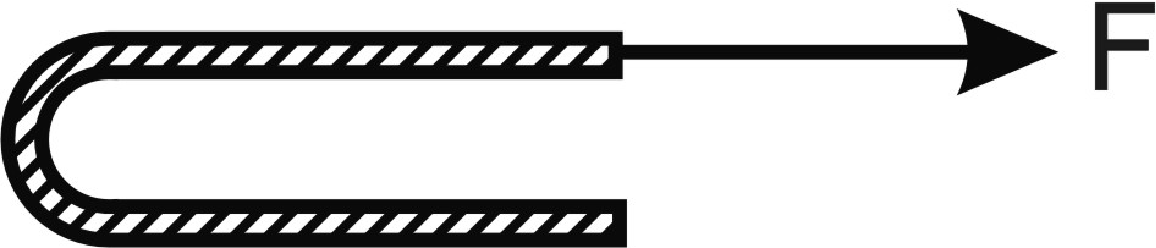,height=1cm}}
\end{figure}

\subsection*{Задача 3 {\usefont{T2A}{cmr}{b}{n}(7.4)}: Время столкновения сильно 
накачанного мяча со стенкой}
Определите время столкновения сильно накачанного мяча со стенкой. Масса мяча 
$m=0,5$~кг, радиус $R=0,1$~м. Избыточное давление $P=7,5\cdot 10^4$~Па в мяче 
в процессе удара меняется незначительно. Скорость мяча перпендикулярна стенке.

\subsection*{Задача 4 {\usefont{T2A}{cmr}{b}{n}(7.11)}: Период колебаний грузика}
Грузик на нити подвешен к стенке с малым углом наклона к вертикали (см. рисунок). 
Определить зависимость периода колебаний грузика от угла отклонения от вертикали для 
случаев: а) удар о стенку упругий; б) удар неупругий.
\begin{figure}[htb]
\centerline{\epsfig{figure=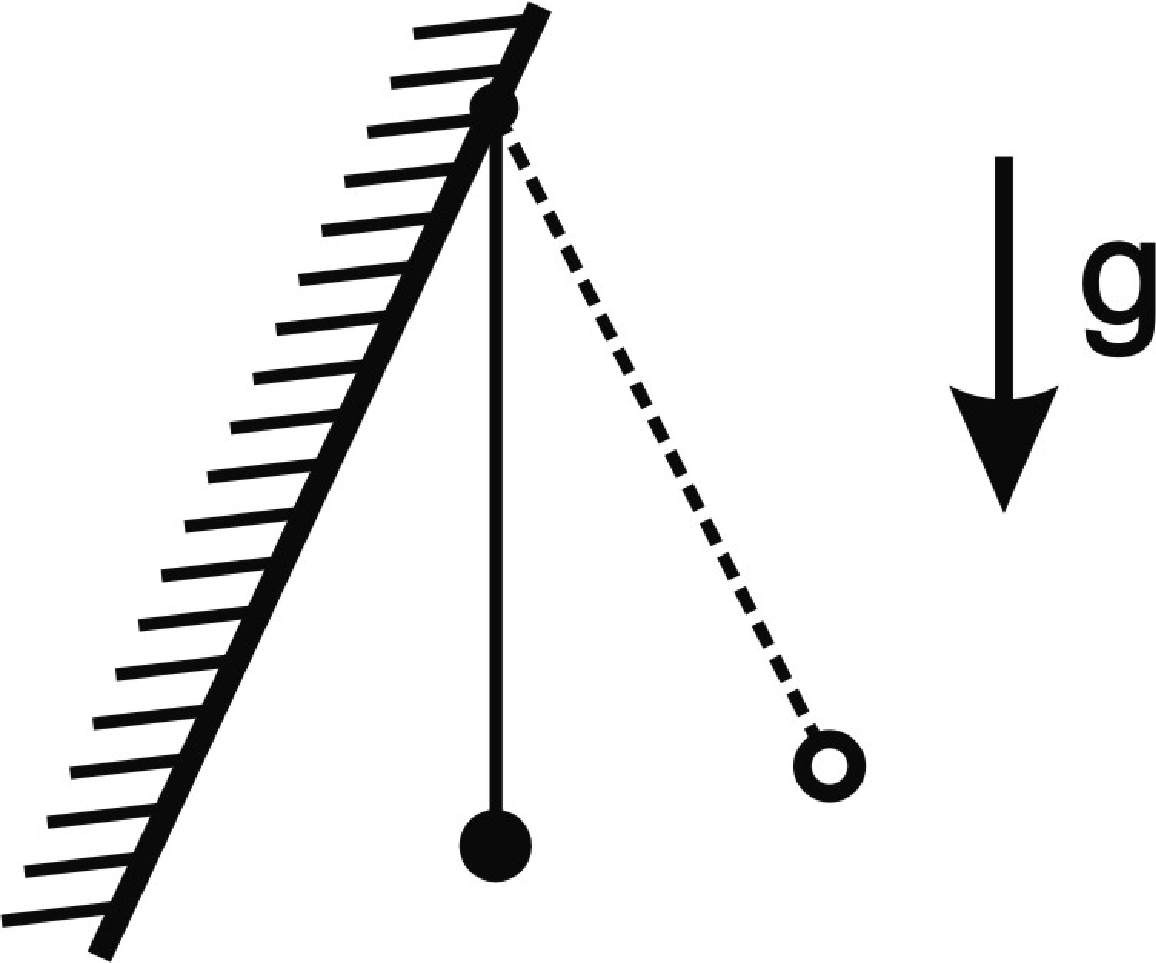,height=3.5cm}}
\end{figure}

\subsection*{Задача 5 {\usefont{T2A}{cmr}{b}{n}(7.13)}: Частота малых 
колебаний бусинки}
Бусинка надета на невесомую гладкую нить длиной $L$, концы которой закреплены на 
одинаковой высоте на расстоянии $d$ друг от друга (см. рисунок). Найти частоту малых 
колебаний бусинки вдоль нити.
\begin{figure}[htb]
\centerline{\epsfig{figure=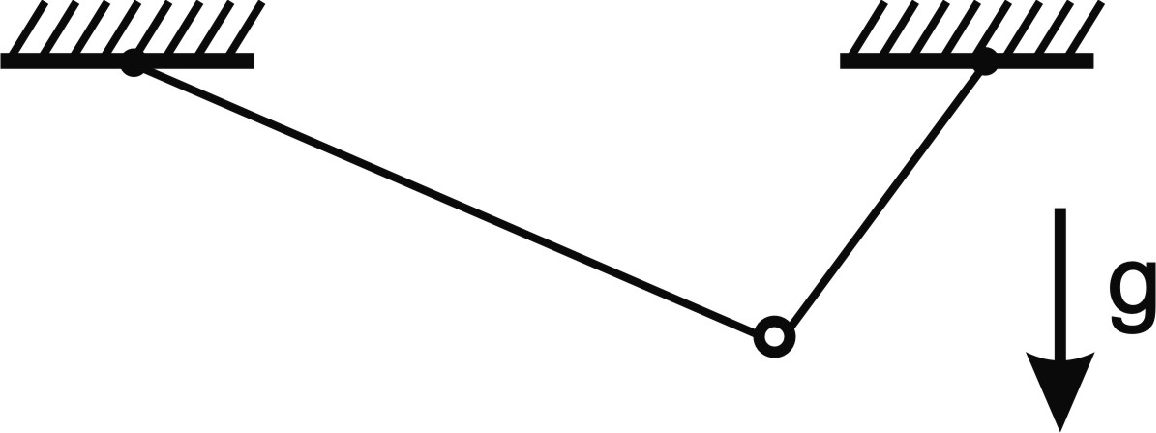,height=3.0cm}}
\end{figure}

\subsection*{Задача 6 {\usefont{T2A}{cmr}{b}{n}(7.14)}: Период колебаний точки
в циклоидальной чашке}
Найти период колебаний точки массой $m$, движущейся в поле тяжести в гладкой 
циклоидальной чашке $x=R(\phi+\sin{\phi})$, $y=R(1-\cos{\phi})$.

\section[Семинар 24]
{\centerline{Семинар 24}}

\subsection*{Задача 1 {\usefont{T2A}{cmr}{b}{n}(7.18)}:  Движение стрелки 
амперметра после разрыва цепи}
Исследовать движение стрелки амперметра с нулем в центре шкалы после быстрого разрыва 
цепи постоянного тока, протекавшего через него. Учесть момент сил сухого трения 
в подшипниках оси. Нарисовать траекторию на фазовой плоскости для угла отклонения
стрелки $\phi$ и угловой скорости $\dot\phi$.

\subsection*{Задача 2 {\usefont{T2A}{cmr}{b}{n}(7.20)}: Оптимальный режим
демпфирования колебаний}
При каком соотношении между сопротивлением $R$, индуктивностью $L$ и емкостью $C$ 
контура в цепи гальванометра осуществляется наиболее оптимальный апериодический режим
демпфирования колебаний рамки гальванометра?

\subsection*{Задача 3 {\usefont{T2A}{cmr}{b}{n}(7.22)}: Добротность и затухание 
колебаний стрелки амперметра}
Собственная частота стрелки амперметра с нулем в центре шкалы 1~Гц, добротность при 
затухании колебаний $Q=20$. За какое время амплитуда колебаний стрелки уменьшится 
в 20 раз, если выключить протекавший через амперметр постоянный ток?

\subsection*{Задача 4 {\usefont{T2A}{cmr}{b}{n}(7.34)}: Амплитуда 
установившихся колебаний при осциллирующей точке подвеса}
Шарик массой $m$ подвешен в поле тяжести на пружине жесткости $k$. Точка подвеса 
пружины движется по вертикали по закону $z=a\;\cos{\Omega t}$. Найти амплитуду 
установившихся малых колебаний шарика.

\subsection*{Задача 5 {\usefont{T2A}{cmr}{b}{n}(7.29)}: Амплитуда 
колебаний после воздействия внешней силы}
Определить амплитуду колебаний массы $m$, закрепленной на пружинке жесткостью $k$, 
оставшихся после воздействия внешней силы, график которой (один полупериод синусоиды) 
показан на рисунке. Период собственных колебаний осциллятора $T$ совпадает с периодом 
внешней силы. До включения силы осциллятор покоился. Каким будет результат, если
внешняя сила действовала в течение $N$ полупериодов синусоиды?
\begin{figure}[htb]
\centerline{\epsfig{figure=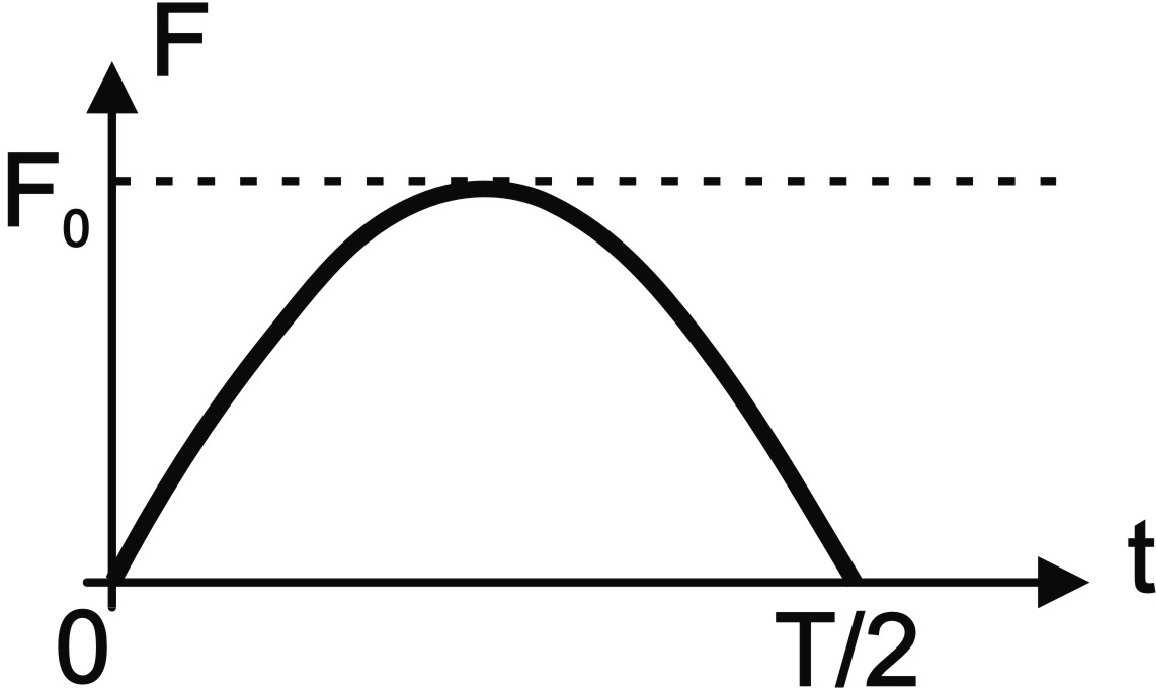,height=4cm}}
\end{figure}

\subsection*{Задача 6 {\usefont{T2A}{cmr}{b}{n}(7.24)}: Амплитуда 
колебаний после действия прямоугольного импульса}
Груз массой $m$ подвешен на пружине жесткости $k$. Найти амплитуду его колебаний, 
оставшихся после действия прямоугольного импульса силы амплитудой $F$, длительностью 
$\tau$. Сила направлена вдоль пружины. Начальная скорость груза была равна нулю.

\section[Семинар 25]
{\centerline{Семинар 25}}

\subsection*{Задача 1 {\usefont{T2A}{cmr}{b}{n}(6.43)}: Обезьянка на веревке}
На невесомый блок намотана тонкая веревка массой $m$, длиной $L$. По веревке начинает 
подниматься обезьянка массой $M$, при этом расстояние от нее до блока в процессе 
подъема остается постоянным и равным $l$. Найти зависимость от времени скорости 
обезьянки относительно веревки, если начальная длина свешивающейся части веревки 
$l_0>l$. 

\subsection*{Задача 2 {\usefont{T2A}{cmr}{b}{n}(7.35)}: Амплитуда установившихся 
колебаний маятника при горизонтальных осцилляциях точки подвеса}
Точка подвеса математического маятника длиной $L$ движется по горизонтали по закону
$x=b\;\cos{\Omega t}$. Найти угловую амплитуду установившихся малых колебаний.

\subsection*{Задача 3 {\usefont{T2A}{cmr}{b}{n}(7.36)}: Резонанс при движении 
поезда}
Рессоры железнодорожного вагона прогибаются под его тяжестью на 4~см, расстояние между 
стыками железнодорожного полотна 25~м. При какой скорости поезда амплитуда 
вертикальных колебаний вагона будет максимальной?

\subsection*{Задача 4 {\usefont{T2A}{cmr}{b}{n}(7.30)}: Амплитуда колебаний при 
резонансе}
Период собственных колебаний рамки амперметра магнито-электрической системы равен 1~с, 
добротность $Q=10$. Найти амплитуду установившихся колебаний стрелки (в единицах 
показываемого тока), если через амперметр пропускается синусоидальный ток с амплитудой 
1~А и частотой 10~Гц. Какой будет амплитуда колебаний при резонансе?

\subsection*{Задача 5 {\usefont{T2A}{cmr}{b}{n}(7.26)}: Найти энергию, 
приобретенную осциллятором}
Найти энергию, приобретенную осциллятором за все время действия силы $F=F_0-
F_0e^{-t/\tau}$. В начальный момент времени $t=0$ энергия осциллятора была равна 
$E_0$ и он проходил через положение равновесия.

\subsection*{Задача 6 {\usefont{T2A}{cmr}{b}{n}(7.39)}: Как меняется высота 
подскока шарика над плитой?}
Упругий шарик подпрыгивает в поле тяжести над горизонтальной плитой. Поле тяжести 
медленно изменяется. Как меняется высота подскока шарика над плитой?

\section[Семинар 26]
{\centerline{Семинар 26}}

\subsection*{Задача 1 {\usefont{T2A}{cmr}{b}{n}(8.3)}: Мягкая посадка космического 
аппарата}
На какой высоте от поверхности планеты нужно включить тормозной двигатель космического 
аппарата, чтобы обеспечить мягкую посадку на поверхность? Спуск происходит по прямой, 
проходящей через центр планеты. Сила торможения $F$ постоянна. Сопротивлением воздуха и
изменением массы аппарата пренебречь. Масса аппарата $m$, скорость вдали от Земли 
$V_\infty$.

\subsection*{Задача 2 {\usefont{T2A}{cmr}{b}{n}(8.10)}: Период движения частицы
в центральном поле}
Найти период движения частицы массой $m$ в центральном поле с потенциалом 
$U=\alpha\;r^2$ $(\alpha>0)$.

\subsection*{Задача 3 {\usefont{T2A}{cmr}{b}{n}(8.2)}: Давление в центре жидкой 
планеты}
Найти давление в центре жидкой планеты шаровой формы. Плотность жидкости считать 
однородной и равной 5,5~г/см$^3$. Радиус планеты 6400~км. Как изменится результат, 
если не пренебрегать сжимаемостью жидкости при увеличении давления?

\subsection*{Задача 4 {\usefont{T2A}{cmr}{b}{n}(8.12)}: Покинет ли планета 
звезду?}
По круговой орбите вокруг звезды массой $M$ движется планета массой $m\ll M$.  
В результате взрыва звезда сбрасывает массу $\alpha M$. Найти, при каком значении
$\alpha$  планета покинет звезду. Считать, что сбрасываемая масса выходит за орбиту 
планеты сферически симметрично и мгновенно.

\subsection*{Задача 5 {\usefont{T2A}{cmr}{b}{n}(8.23)}: При какой скорости
частица не провалится в воронку?}
Частица скользит без трения по стенке воронки (см. рисунок). В начальный момент 
частица находилась на высоте $h$ и двигалась горизонтально со скоростью $V$. При 
какой минимальной скорости $V$ частица не провалится в воронку, отверстие которой 
имеет радиус 
$\rho_0$?
\begin{figure}[htb]
\centerline{\epsfig{figure=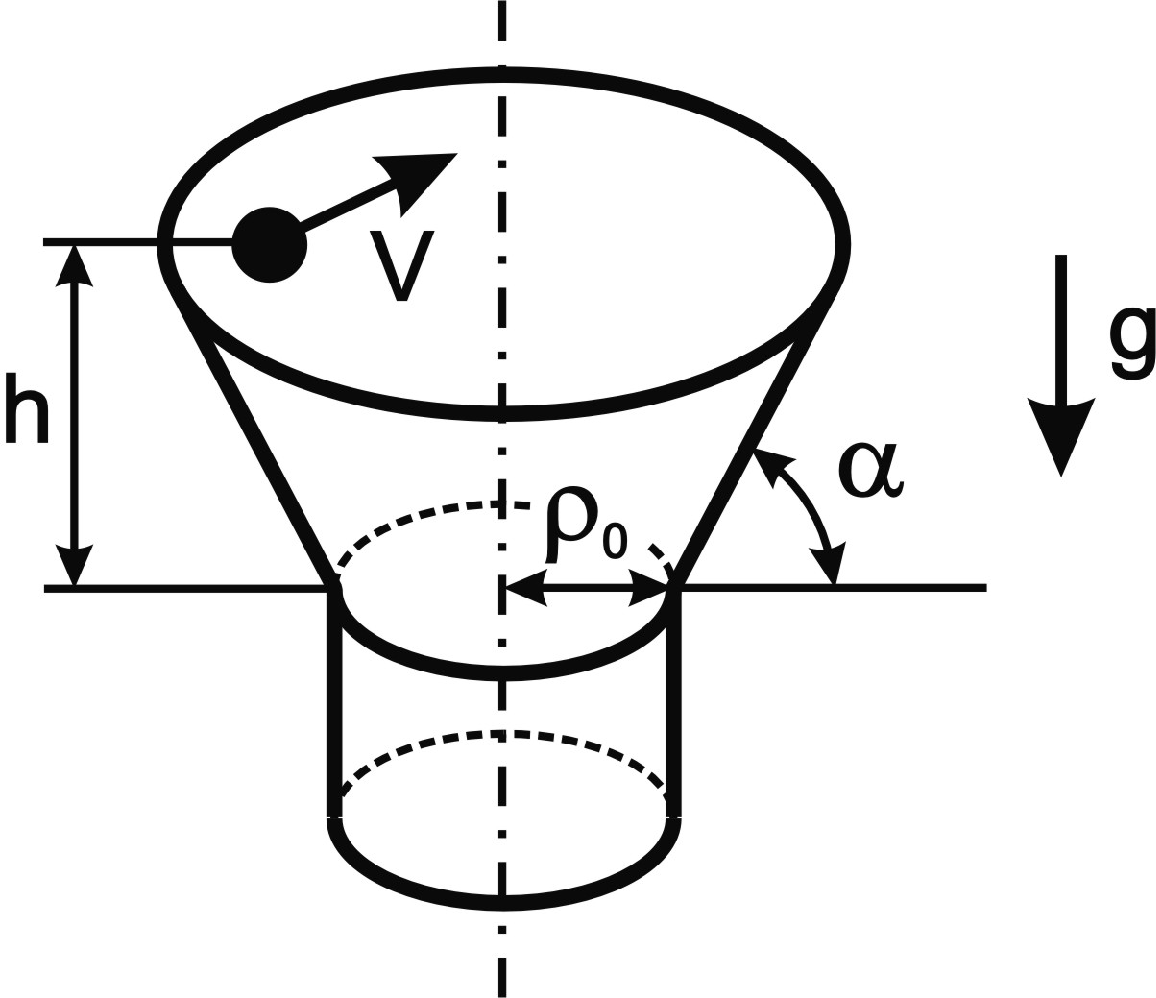,height=4cm}}
\end{figure}

\subsection*{Задача 6 {\usefont{T2A}{cmr}{b}{n}(8.26)}: Сечение падения потока 
метеоритов на Землю}
Найти сечение падения потока метеоритов на Землю. Скорость метеоритов вдали от Земли
$V_\infty$.

\subsection*{Задача 7 {\usefont{T2A}{cmr}{b}{n}(7.42)}: Адиабатическое изменение 
радиуса круговой орбиты}
Звезда теряет за счет излучения $10^{-9}$ часть своей массы в год. За какое время 
радиус круговой орбиты планеты, вращающейся вокруг звезды, изменится вдвое? Влиянием 
излучения на планету пренебречь.

\subsection*{Задача 8 {\usefont{T2A}{cmr}{b}{n}(7.40)}: Осциллятор с медленно 
возрастающей массой}
Масса осциллятора медленно возрастает. Как при этом изменяются амплитуда и период его 
колебаний? Рассмотреть оба случая, показанных на рисунке.
\begin{figure}[htb]
\centerline{\epsfig{figure=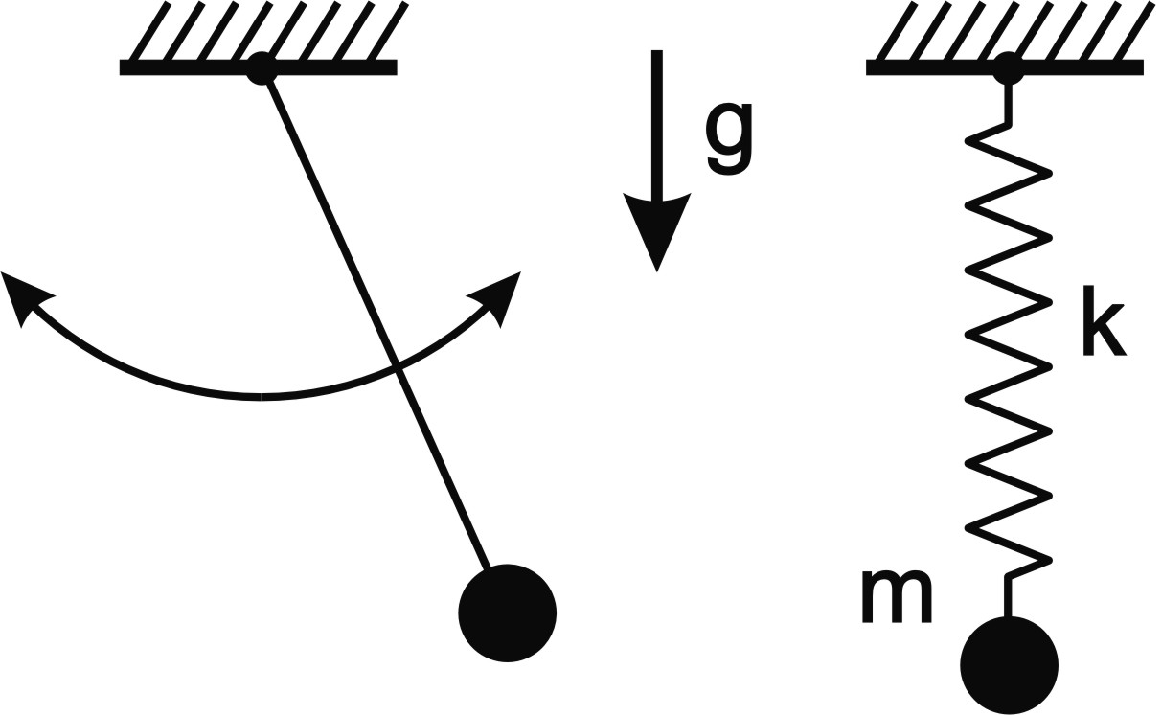,height=3.5cm}}
\end{figure}

\subsection*{Задача 9 {\usefont{T2A}{cmr}{b}{n}(7.41)}: Осциллятор в процессе 
таяния}
Масса осциллятора медленно уменьшается (например, из-за таяния). Как при этом 
изменяются амплитуда и период его колебаний? Рассмотреть оба случая, показанных на 
рисунке к предыдущей задаче.

\subsection*{Задача 10 {\usefont{T2A}{cmr}{b}{n}(7.43)}: Частица в ящике, 
который медленно поднимают за один конец}
Частица движется со скоростью $V$ вдоль стороны $L$ в расположенном горизонтально 
прямоугольном ящике (см. рисунок), упруго отражаясь от его стенок. Ящик медленно 
поднимают за один конец, поворачивая вокруг ребра, перпендикулярного $L$. При каком 
угле наклона $\alpha$  дна ящика частица не будет достигать его верхней стенки?
\begin{figure}[htb]
\centerline{\epsfig{figure=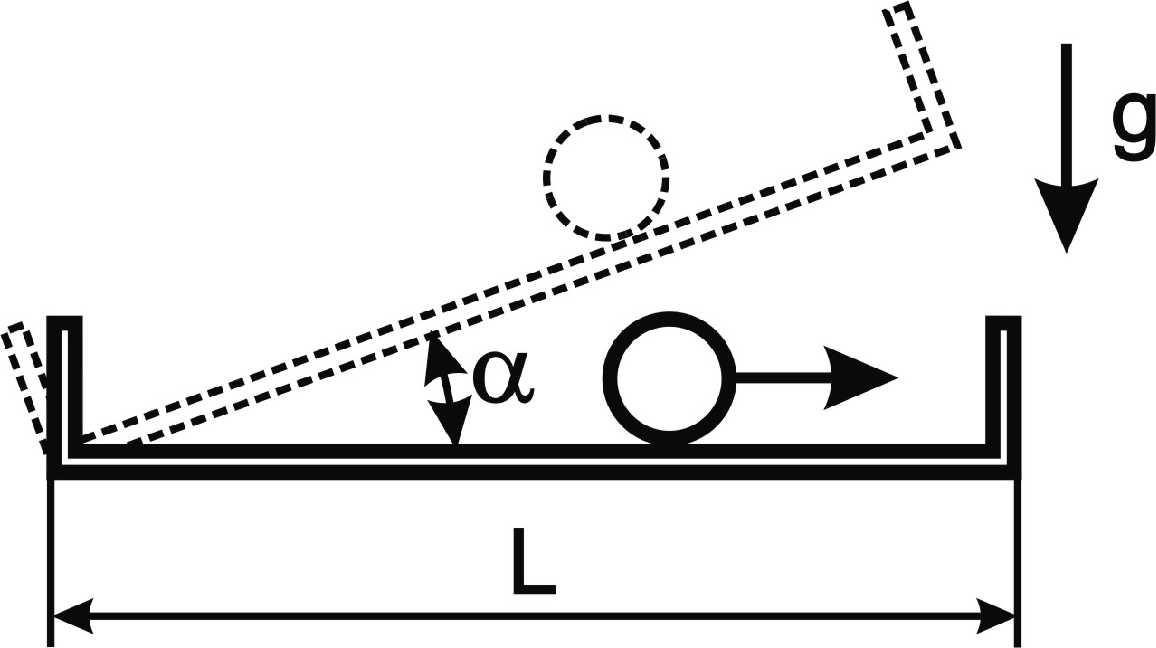,height=4cm}}
\end{figure}

\section[Семинар 27]
{\centerline{Семинар 27}}

\subsection*{Задача 1: Уравнение орбиты в задаче Кеплера}
Получить уравнение орбиты в задаче Кеплера.

\subsection*{Задача 2 {\usefont{T2A}{cmr}{b}{n}(8.39)}: Третья космическая 
скорость}
Какой должна быть минимальная скорость ракеты при выходе из атмосферы Земли, чтобы она 
смогла покинуть Солнечную систему без дополнительного ускорения?

\subsection*{Задача 3 {\usefont{T2A}{cmr}{b}{n}(8.60)}: Комета Галлея}
Сколько лет нужно ожидать возвращения кометы, удаляющейся от Солнца на 35~а.~е.? 
Перигелий кометы 0,6~а.~е.

\subsection*{Задача 4 {\usefont{T2A}{cmr}{b}{n}(8.52)}: Изменение орбиты спутника}
Спутник движется по околоземной круговой орбите радиусом $r$. Какую радиальную добавку 
скорости ему нужно сообщить, чтобы его орбита стала эллиптической с перигеем $r_1$?

\subsection*{Задача 5 {\usefont{T2A}{cmr}{b}{n}(8.4)}: Поле тяготения внутри 
полости}
Внутри шара плотностью $\rho$ имеется сферическая полость, центр которой находится на 
расстоянии $\vec{a}$ от центра шара. Найти напряженность поля тяготения внутри 
полости.

\subsection*{Задача 6 {\usefont{T2A}{cmr}{b}{n}(8.28)}: Сечение падения частиц 
на сферу} 
Найти сечение падения частиц энергией $E$ на сферу радиусом $R$, находящуюся в центре 
поля с потенциалом отталкивания $U=\alpha/r$. 

\subsection*{Задача 7 {\usefont{T2A}{cmr}{b}{n}(8.24)}: Границы движения частицы}
Частица движется без трения по поверхности параболической чашки, описываемой в 
цилиндрической системе координат уравнением $z=\alpha\rho^2$. Поле тяжести 
направлено вдоль оси $z$. На высоте $H$ скорость частицы $V$ была горизонтальна 
(см. рисунок). Найти границы движения частицы.
\begin{figure}[htb]
\centerline{\epsfig{figure=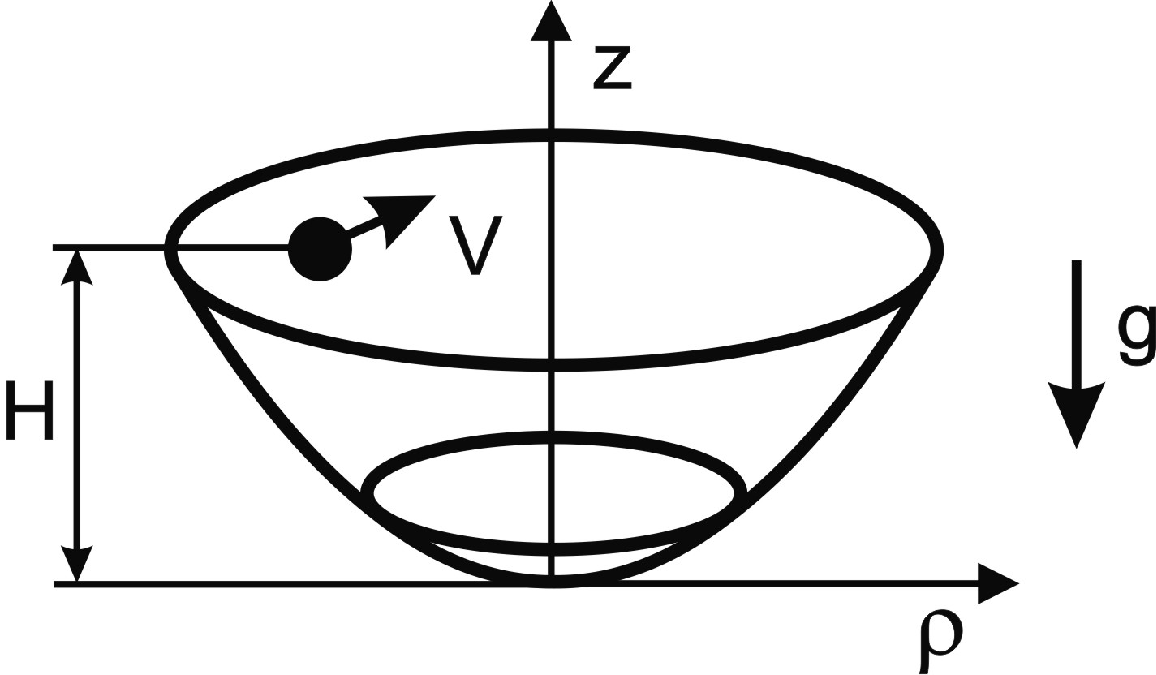,height=4cm}}
\end{figure}

\subsection*{Задача 8 {\usefont{T2A}{cmr}{b}{n}(8.15)}: Сечение отрыва планеты 
быстрой звездой}
Оценить сечение ``ионизации'' (отрыва планеты) Солнечной системы быстрой звездой. 
Скорость звезды $u$ много больше орбитальной скорости планеты $V_0$. 

\subsection*{Задача 9 {\usefont{T2A}{cmr}{b}{n}(8.17)}: Время жизни атома 
водорода}
Оцените время жизни атома водорода с точки зрения классической физики, считая, что 
электрон вращается по круговой орбите радиусом $r_0=5\cdot 10^{-9}$~см и в единицу 
времени излучает энергию $$\frac{2e^2a^2}{3c^3},$$ где $a$ --- ускорение электрона,
$e$ --- его заряд, $c$ --- скорость света (в системе CGSE).

\section[Семинар 28]
{\centerline{Семинар 28}}

\subsection*{Задача 1: Вектор Рунге--Ленца}
Доказать, что в задаче Кеплера сохраняется вектор Рунге--Ленца
$$\vec{e}=\frac{1}{\alpha}\,\dot{\vec{r}}\times\vec{L}-\frac{\vec{r}}{r},$$
где $\vec{L}$ --- момент импульса. Используя сохранение вектора Рунге--Ленца, получить
уравнение траектории.

\subsection*{Задача 2 {\usefont{T2A}{cmr}{b}{n}(8.65)}: Падение Земли на Солнце}
За какое время Земля упадет на Солнце, если остановить ее движение по орбите?

\subsection*{Задача 3 {\usefont{T2A}{cmr}{b}{n}(8.43)}: Возвращение с Луны}
Оцените, с какой минимальной скоростью нужно стартовать с поверхности Луны, чтобы 
вернуться на Землю? Ускорение свободного падения на Луне $g/6$, скорость движения 
Луны по орбите 1~км/с. Радиус Луны~1740 км.

\subsection*{Задача 4 {\usefont{T2A}{cmr}{b}{n}(8.51)}: Минимальная начальная 
скорость при полете на Марс и на Венеру}
С какой минимальной скоростью должен покинуть атмосферу Земли космический корабль, 
направляющийся к Марсу и стартующий по касательной к орбите Земли? Каким будет 
расстояние от Земли до Марса при посадке корабля на Марс? Радиус орбиты Марса 
1,52~а.~е. Какова минимальная начальная скорость при полете на Венеру? Радиус орбиты
Венеры 0,72~а.~е.

\subsection*{Задача 5 {\usefont{T2A}{cmr}{b}{n}(8.42)}: Радиус оптимальной орбиты}
Находящийся на круговой орбите космический корабль тангенциальной добавкой скорости 
переводят на гиперболическую орбиту со скоростью на бесконечности $V_\infty$. При 
каком радиусе начальной круговой орбиты эта добавка скорости минимальна?

\subsection*{Задача 6 {\usefont{T2A}{cmr}{b}{n}(8.66)}: Падение баллистической 
ракеты}
Оценить время, через которое возвратится баллистическая ракета, запущенная 
с поверхности Земли со скоростью 10~км/с. Сопротивлением атмосферы пренебречь.

\section[Семинар 29]
{\centerline{Семинар 29}}

\subsection*{Задача 1 {\usefont{T2A}{cmr}{b}{n}(8.18)}: Эффект 
Пойнтинга--Робертсона}
Сферическая частичка радиусом 1~мм, массой $10^{-2}$~г движется по круговой орбите 
радиусом 500~св.~с вокруг Солнца. Оцените силу торможения, обусловленную 
взаимодействием частички с излучением Солнца. Частичка разогревается и переизлучает 
тепло изотропно в своей системе отсчета. Мощность излучения Солнца 
$4\cdot 10^{26}$~Вт.

\subsection*{Задача 2 {\usefont{T2A}{cmr}{b}{n}(8.50)}: Изменение орбиты 
космической станции}
Орбитальная станция движется по круговой траектории на расстоянии 200~км от 
поверхности Земли. Какую наименьшую дополнительную скорость надо сообщить станции, 
чтобы ее максимальное удаление от Земли достигло 210~км?

\subsection*{Задача 3 {\usefont{T2A}{cmr}{b}{n}(8.49)}: Прирост скорости 
в перигее и высота апогея}
В перигее величиной $r_{min}$ скорость спутника $V$. При каком касательном приросте 
скорости в перигее высота апогея увеличится на $1\,\%$? 

\subsection*{Задача 4 {\usefont{T2A}{cmr}{b}{n}(8.53)}: Баллистическая ракета}
Баллистическую ракету запускают с Северного полюса, так что после выхода из атмосферы 
и выключения двигателей она имеет скорость $V_0$ и угол вылета $\theta$ по отношению 
к горизонту. При каком соотношении между $V_0$ и $\theta$ ракета достигнет Южного 
полюса?

\subsection*{Задача 5 {\usefont{T2A}{cmr}{b}{n}(8.74)}: Частота малых колебаний 
грузиков}
Через невесомый блок перекинута нерастяжимая нить, к концам которой через пружины 
жесткостью $k$ подвешены грузики массой $m$ и $M$. Найти частоту малых колебаний 
грузиков в поле тяжести. Трения нет.

\section[Семинар 30]
{\centerline{Семинар 30}}

\subsection*{Задача 1 {\usefont{T2A}{cmr}{b}{n}(8.73)}: Как изменится скорость 
хода часов?}
Как изменится скорость хода часов с крутильным маятником, если их снять с пульта 
космического корабля и оставить свободно парить в кабине? От каких параметров и как 
будет зависеть это изменение?

\subsection*{Задача 2 {\usefont{T2A}{cmr}{b}{n}(8.72)}: Период малых продольных 
колебаний осциллятора}
Найти период малых продольных колебаний осциллятора, состоящего из двух масс $m$ и 
$M$, закрепленных на концах пружины жесткостью $k$.

\subsection*{Задача 3 {\usefont{T2A}{cmr}{b}{n}(8.72)}: Через какое время 
столкнутся две точки?}
Через какое время столкнутся две точки с разными массами $M$ и $m$, начавшие 
двигаться из состояния покоя под действием силы взаимного гравитационного притяжения?

\subsection*{Задача 4 {\usefont{T2A}{cmr}{b}{n}(8.85)}: Полная масса системы 
``двойная звезда''}
Найти полную массу системы ``двойная звезда'' по периоду обращения, минимальному и 
максимальному расстояниями между составляющими ее звездами.

\subsection*{Задача 5 {\usefont{T2A}{cmr}{b}{n}(8.96)}: Сечение рассеяния 
электрона на протоне}
Найти сечение рассеяния на угол, больший $90^\circ$, при столкновении электрона 
с энергией $T=10$~кэВ с неподвижным протоном. Как изменится результат, если протон 
не закреплен?

\subsection*{Задача 6 {\usefont{T2A}{cmr}{b}{n}(8.95)}: Рассеяние быстрых 
электронов}
Найти зависимость угла рассеяния от прицельного параметра для быстрых электронов, 
пролетающих мимо тонкой заряженной проволочки перпендикулярно ее оси (напряженность 
электрического поля проволочки обратно пропорциональна расстоянию от нее).

\section[Семинар 31]
{\centerline{Семинар 31}}

\subsection*{Задача 1 {\usefont{T2A}{cmr}{b}{n}(8.90)}: Рассеяние на абсолютно 
упругой сфере}
Найти зависимость угла рассеяния точечных частиц на абсолютно упругой сфере радиусом 
$R$ от прицельного параметра $\rho$.

\subsection*{Задача 2 {\usefont{T2A}{cmr}{b}{n}(8.91)}: Сечение рассеяния при 
упругом столкновении частицы со сферой}
Найти сечение рассеяния на угол, больший $90^\circ$, при упругом столкновении 
точечной частицы массой $m$ с первоначально неподвижной сферой радиусом $R$, массой 
$2m$.

\subsection*{Задача 3 {\usefont{T2A}{cmr}{b}{n}(8.102)}: Зависимость угла 
рассеяния от прицельного параметра}
Найти зависимость угла рассеяния от прицельного параметра в поле $U=\alpha/r+
\beta/r^2$, где $\alpha\;,\beta>0$.

\subsection*{Задача 4 {\usefont{T2A}{cmr}{b}{n}(8.98)}: Теневая область при
кулоновском рассеянии}
Плоский поток частиц рассеивается на отталкивающем кулоновском потенциале, который имеет 
вид $U(r)=\alpha/r$. Найти область, в которую частицы попасть не могут.
 
\subsection*{Задача 5 {\usefont{T2A}{cmr}{b}{n}(9.4)}: Сила упругого сжатия 
обруча}
Найти силу упругого сжатия обруча радиусом $r$, весом $P$, лежащего горизонтально 
внутри гладкой сферической чашки радиуса $R$.

\section[Семинар 32]
{\centerline{Семинар 32}}

\subsection*{Задача 1 {\usefont{T2A}{cmr}{b}{n}(9.1)}: Натяжение цепочки, надетой 
на гладкий конус}
Найти натяжение кольцевой цепочки, надетой на гладкий конус с углом при вершине 
$\alpha$.

\subsection*{Задача 2 {\usefont{T2A}{cmr}{b}{n}(9.2)}: Натяжение цепочки, надетой 
на гладкую сферу}
Найти натяжение кольцевой цепочки радиусом $r$, весом $P$, надетой на гладкую сферу
радиусом $R$ (см. рисунок).
\begin{figure}[htb]
\centerline{\epsfig{figure=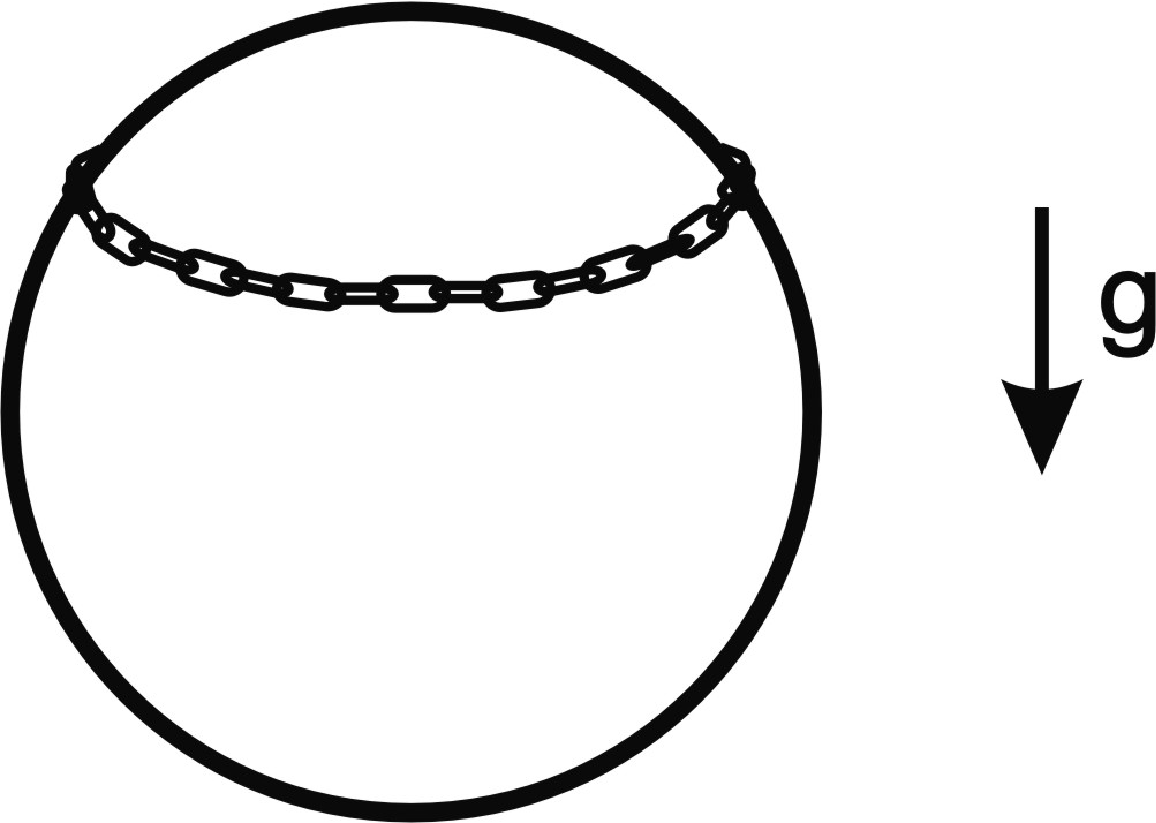,height=2cm}}
\end{figure}

\subsection*{Задача 3 {\usefont{T2A}{cmr}{b}{n}(9.8)}: Найти натяжение нити}
\begin{figure}[!h]
\centerline{\epsfig{figure=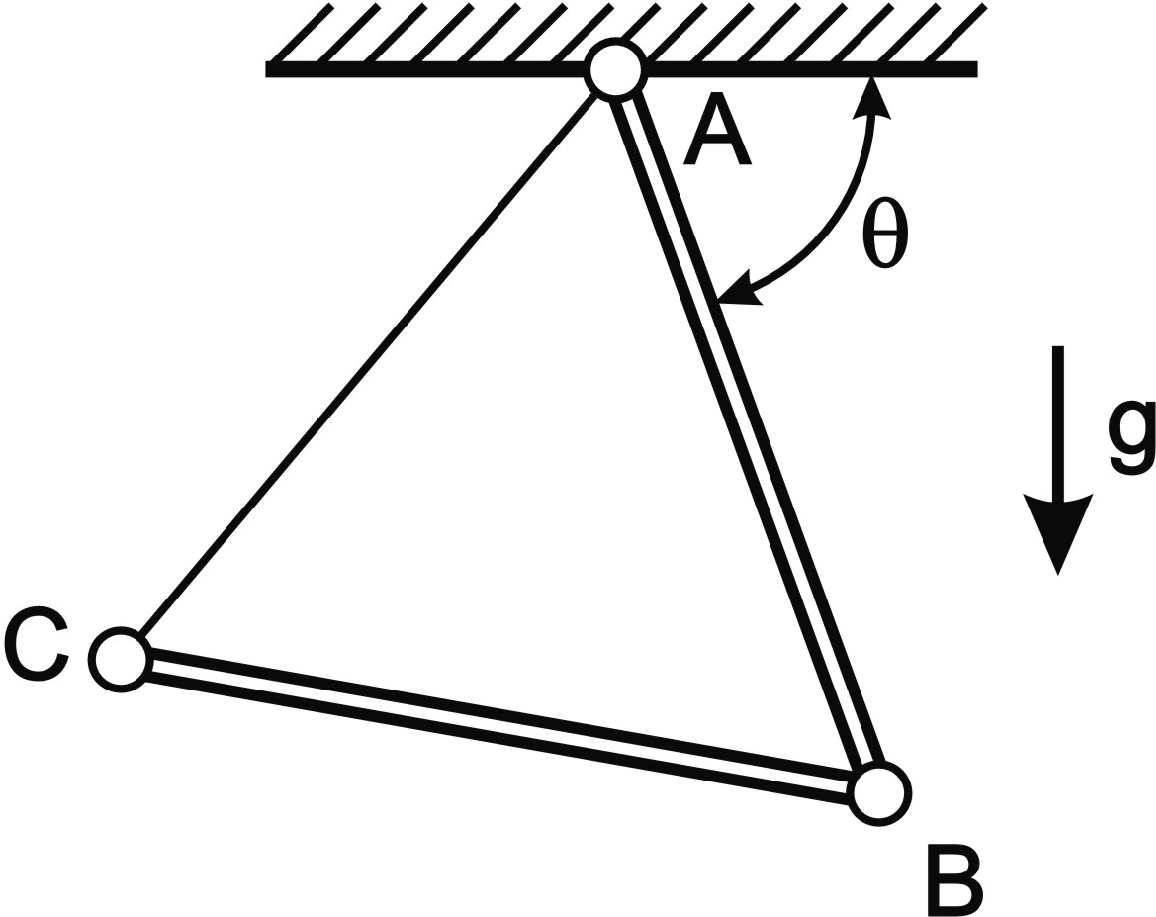,height=3cm}}
\end{figure}
Одинаковые стержни $AB$ и $BC$ длиной $L$, весом $P$ соединены шарнирно. Точки $A$ и 
$C$ соединены невесомой нитью длины $L$. Найти натяжение нити и угол $\theta$, 
образуемый стержнем $AB$ с горизонталью в положении равновесия.

\subsection*{Задача 4 {\usefont{T2A}{cmr}{b}{n}(9.19)}: Период вращения
нейтронной звезды}
Оценить период вращения Солнца, если бы оно превратилось в нейтронную звезду 
с плотностью $10^{14}$~г/см$^3$. Средняя плотность Солнца 1,4~г/см$^3$, период 
вращения $2\cdot 10^6$~с.

\subsection*{Задача 5 {\usefont{T2A}{cmr}{b}{n}(9.25)}: Сколько оборотов сделает 
цилиндр?}
Сплошной цилиндр радиусом $R$, вращающийся с угловой скоростью $\omega$, ставят 
вертикально на шероховатую горизонтальную плоскость. Коэффициент трения $\mu$. Сколько 
оборотов сделает цилиндр?

\subsection*{Задача 6 {\usefont{T2A}{cmr}{b}{n}(9.28)}: Цилиндрическая банка 
с жидкостью} 
Цилиндрическая банка с жидкостью раскручена вокруг оси симметрии так, что жидкость не 
успела закрутиться. Как изменится угловая скорость вращения к моменту, когда угловые 
скорости банки и жидкости уравняются? Сколько энергии перейдет в тепло? Моменты
инерции жидкости и пустой банки равны.

\subsection*{Задача 7 {\usefont{T2A}{cmr}{b}{n}(9.26)}: Установившаяся скорость 
вращения дисков}
Два одинаковых диска, насаженных на гладкие оси, один из которых вращался с угловой
скоростью $\omega$, привели в соприкосновение (см. рисунок). Найти установившуюся 
скорость вращения дисков. Какая часть энергии перейдет в тепло?
\begin{figure}[htb]
\centerline{\epsfig{figure=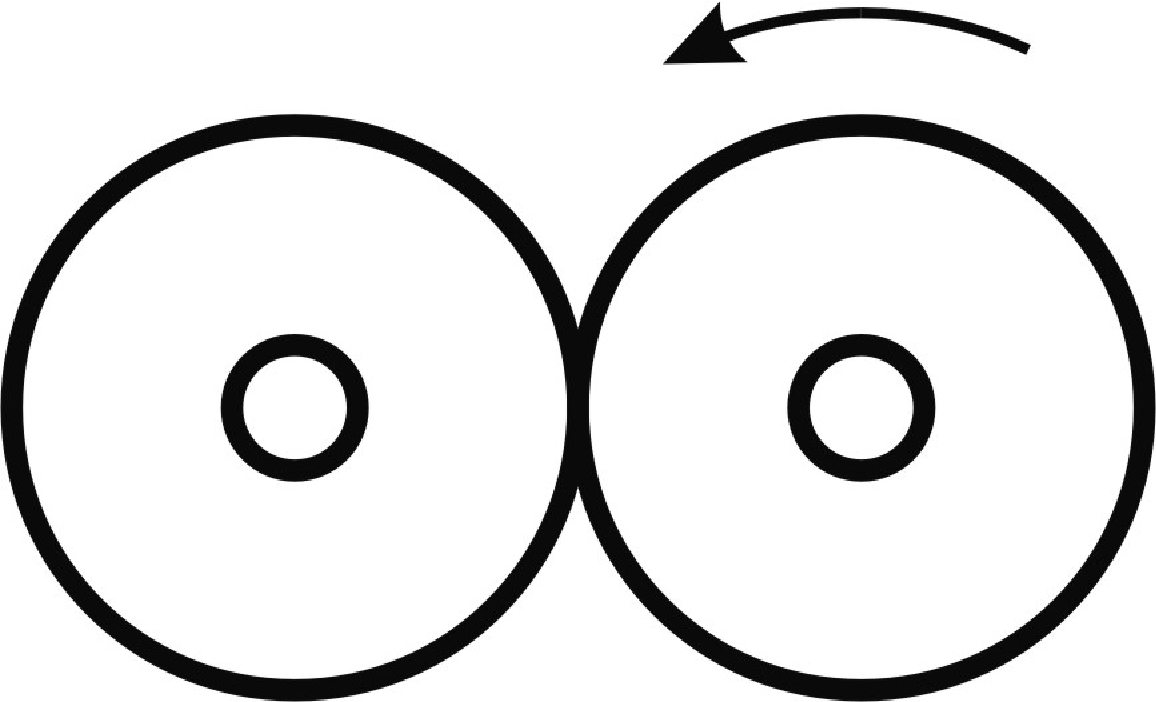,height=1.5cm}}
\end{figure}

\subsection*{Задача 8 {\usefont{T2A}{cmr}{b}{n}(9.32)}: Частота колебаний обруча}
Обруч подвешен за верхнюю точку и может колебаться в вертикальной плоскости. Найти 
частоту малых колебаний: а) в плоскости обруча; б) перпендикулярно плоскости обруча.

\subsection*{Задача 9 {\usefont{T2A}{cmr}{b}{n}(9.40)}: Частота колебаний обруча
внутри неподвижной гладкой сферы}
Обруч радиусом $R$ лежит горизонтально внутри неподвижной гладкой сферы радиусом 
$2R$. Найти частоту малых колебаний обруча под действием силы тяжести.

\section[Семинар 33]
{\centerline{Семинар 33}}

\subsection*{Задача 1 {\usefont{T2A}{cmr}{b}{n}(9.13)}: Моменты инерции 
однородных тел}
Найдите моменты инерции однородных тел: кольца, диска, цилиндра, палочки, сферы, шара.

\subsection*{Задача 2 {\usefont{T2A}{cmr}{b}{n}(9.18)}: Метеорная пыль и 
продолжительность суток}
На поверхность Земли выпадает метеорная пыль. Поток ее изотропен, плотность потока 
$\mu$. Найти зависимость продолжительности суток от времени.

\subsection*{Задача 3 {\usefont{T2A}{cmr}{b}{n}(9.27)}: Раскручивание неподвижных 
дисков}
К диску, раскрученному вокруг оси, проходящей через центр симметрии перпендикулярно 
плоскости, поднесли два таких же неподвижных диска, так что они начали раскручиваться 
вокруг своих осей, тормозя первый диск. Какой будет угловая скорость диска к моменту,
когда скорости дисков уравняются?

\subsection*{Задача 4 {\usefont{T2A}{cmr}{b}{n}(9.35)}: Частота колебаний 
проволочного равностороннего треугольника}
Найти частоту малых колебаний проволочного равностороннего треугольника, подвешенного
на шарнире за вершину в поле тяжести (см. рисунок). Треугольник колеблется 
в плоскости рисунка.
\begin{figure}[htb]
\centerline{\epsfig{figure=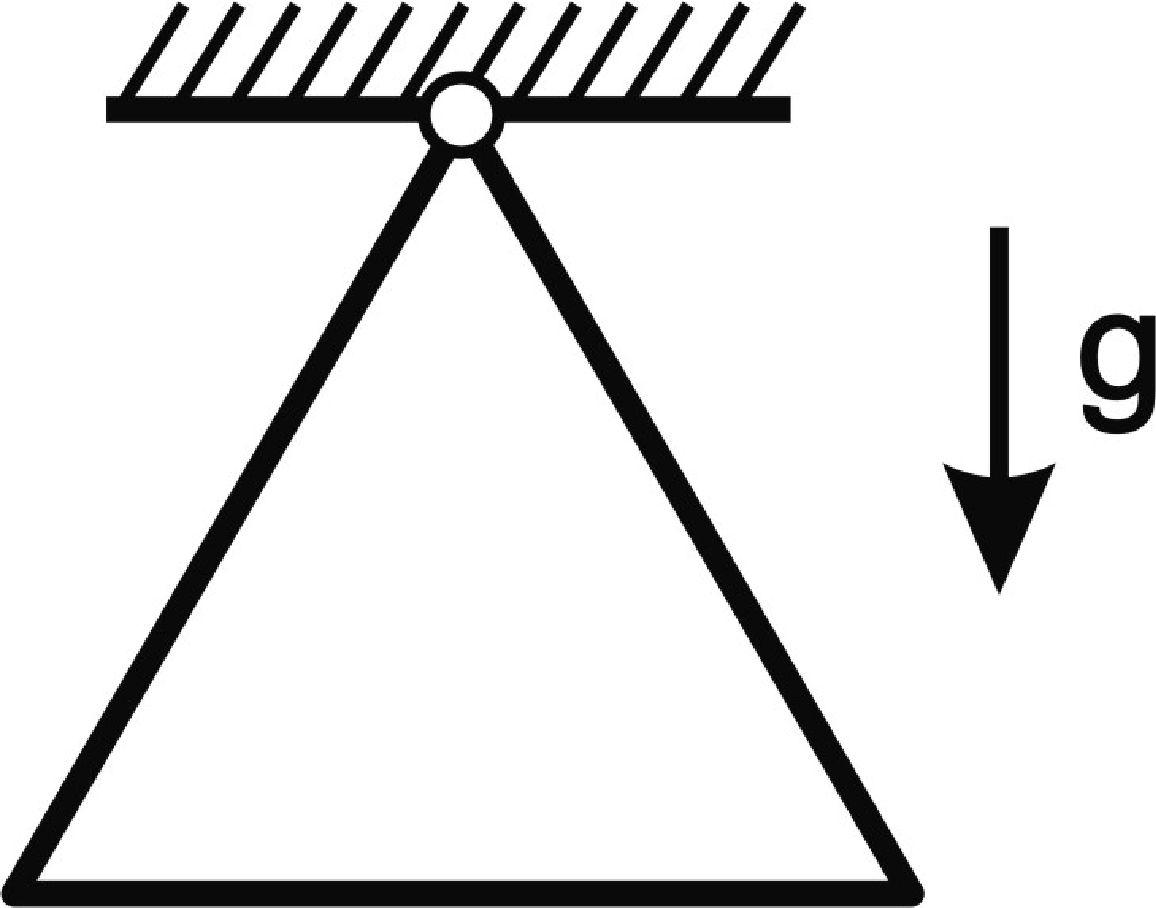,height=2.5cm}}
\end{figure}

\subsection*{Задача 5 {\usefont{T2A}{cmr}{b}{n}(9.38)}: Частота колебаний 
номерка из гардероба}
Номерок из гардероба представляет собой диск радиусом $R$, на краю которого имеется 
отверстие радиусом $r$. Номерок висит на тонком гвозде. Найти частоту его малых 
колебаний в своей плоскости.

\subsection*{Задача 6 {\usefont{T2A}{cmr}{b}{n}(9.43)}: Частота малых колебаний 
полушара}
Найти частоту малых колебаний полушара радиусом $R$ на горизонтальной плоскости в поле 
тяжести. Проскальзывания нет.

\subsection*{Задача 7 {\usefont{T2A}{cmr}{b}{n}(8.97)}: Рассеяние протона на 
протоне}
Найти сечение рассеяния на угол, больший $90^\circ$, при упругом столкновении протона 
с энергией $T=10$~эВ с летящим навстречу протоном такой же энергии.

\subsection*{Задача 8 {\usefont{T2A}{cmr}{b}{n}(8.107)}: Интенсивность пучка 
$\alpha$-частиц}
Пучок $\alpha$-частиц с энергией 10~МэВ проходит через золотую фольгу толщиной 
10~мк. За час происходит в среднем одно рассеяние на угол, больший $90^\circ$. Найти 
интенсивность пучка $\alpha$-частиц.

\section[Семинар 34]
{\centerline{Семинар 34}}

\subsection*{Задача 1 {\usefont{T2A}{cmr}{b}{n}(9.55)}: Каким участком сабли 
следует рубить лозу?}
Каким участком сабли следует рубить лозу, чтобы рука не чувствовала удара? 
Саблю считать однородным стержнем длиной $L$.

\subsection*{Задача 2 {\usefont{T2A}{cmr}{b}{n}(9.80)}: С каким ускорением надо 
тянуть вверх нить?}
С каким ускорением надо тянуть вверх нить, намотанную на катушку, чтобы катушка не 
падала? Плотность материала катушки $\rho$. Размеры катушки указаны на рисунке.
 \begin{figure}[htb]
\centerline{\epsfig{figure=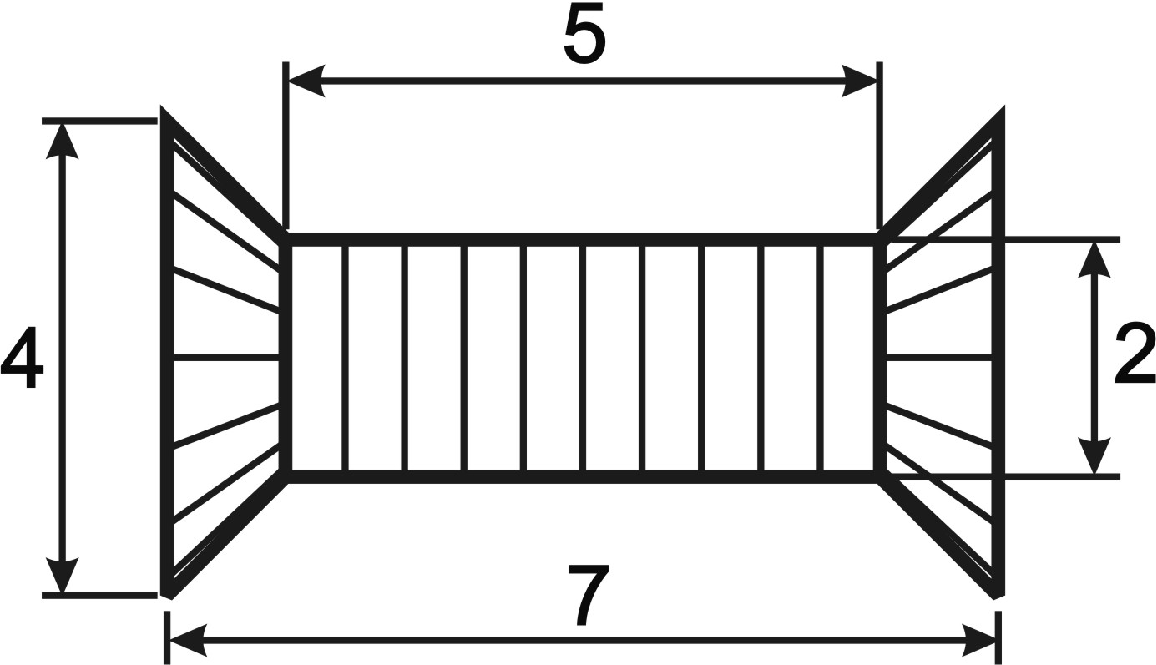,height=3.5cm}}
\end{figure}

\subsection*{Задача 3 {\usefont{T2A}{cmr}{b}{n}(9.76)}: На какой высоте цилиндр
оторвется от горки?}
С верхней точки цилиндрической горки радиусом $R$ скатывается без проскальзывания и 
без начальной скорости цилиндр радиусом $r$ (см. рисунок). На какой высоте цилиндр 
оторвется от горки?
\begin{figure}[htb]
\centerline{\epsfig{figure=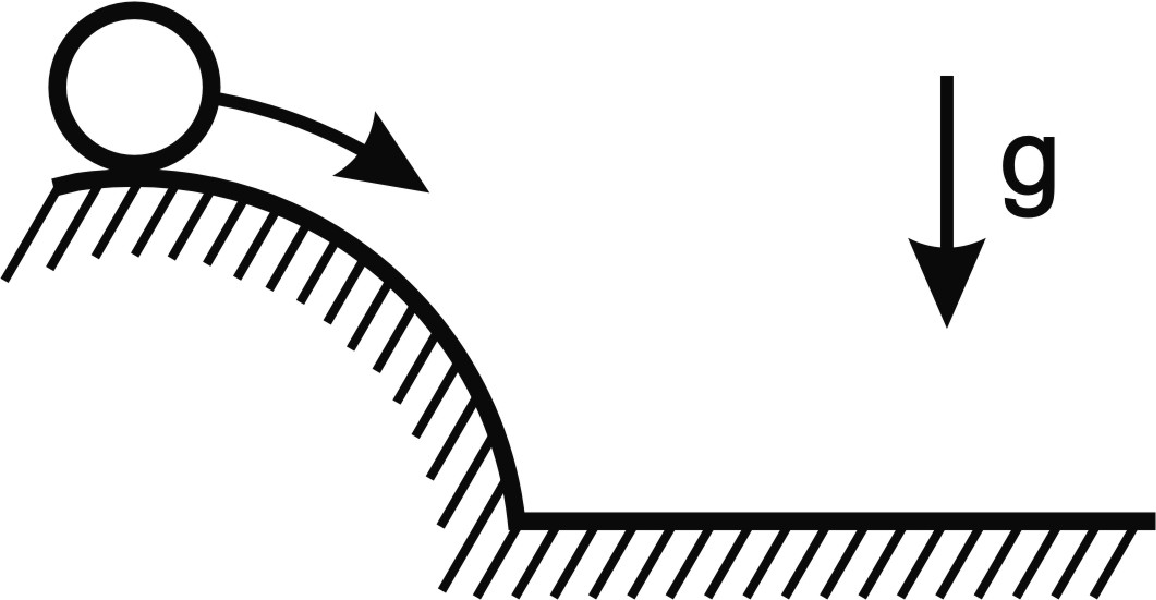,height=3cm}}
\end{figure}

\subsection*{Задача 4 {\usefont{T2A}{cmr}{b}{n}(9.62)}: Максимальное число 
оборотов футбольного мяча}
Какое максимальное число оборотов вокруг своей оси может сделать футбольный мяч после 
одиннадцатиметрового удара? Радиус мяча 0,11~м. Начальная скорость мяча направлена 
под малым углом к горизонту. Сопротивлением воздуха пренебречь.

\subsection*{Задача 5 {\usefont{T2A}{cmr}{b}{n}(9.53)}: Столкновение шарика со
стержнем}
На льду лежит стержень длиной $L$, массой $M$, с которым упруго сталкивается шарик 
массой $m$. Скорость шарика $V$ направлена по нормали к стержню. Точка удара близка 
к концу стержня. Найти скорость шарика после столкновения.

\subsection*{Задача 6 {\usefont{T2A}{cmr}{b}{n}(9.68)}: С каким ускорением
скатывается бочка?}
Определить ускорение, с которым цилиндрическая бочка массой $m$, целиком заполненная 
жидкостью массой $M$, скатывается без проскальзывания с наклонной плоскости с углом 
наклона $\alpha$. Вязкостью жидкости и моментом инерции днищ бочки пренебречь.

\subsection*{Задача 7: Частота малых колебаний стержня}
Однородный стержень длиной $L$ может скользить своими концами без трения по параболе 
$y=\alpha x^2$. Найти частоту малых колебаний стержня при различных значениях $L$ и 
$\alpha$.

\section[Семинар 35]
{\centerline{Семинар 35}}

\subsection*{Задача 1 {\usefont{T2A}{cmr}{b}{n}(9.66)}: Минимальное значение 
угловой скорости}
Определить минимальное значение угловой скорости, при которой обруч, брошенный вперед 
с закруткой, сможет покатиться назад. Определить его установившуюся скорость, если 
начальная угловая скорость превышает минимальную.

\subsection*{Задача 2 {\usefont{T2A}{cmr}{b}{n}(9.67)}:  Ускорение скатывания 
с наклонной плоскости}
Определить ускорение скатывания с наклонной плоскости с углом наклона к горизонту
$\alpha$:  а) полого и сплошного цилиндров; б) шара.

\subsection*{Задача 3 {\usefont{T2A}{cmr}{b}{n}(10.16)}: Устойчивость 
равновесия пузырька}
Тонкая кольцевая стеклянная трубка заполнена водой и вращается в поле тяжести вокруг 
своего вертикального диаметра. В трубке имеется пузырек воздуха. При какой скорости 
вращения трубки равновесие пузырька в ее нижней точке будет устойчивым? Диаметр кольца
$D$.

\subsection*{Задача 4 {\usefont{T2A}{cmr}{b}{n}(9.88)}: Устойчивость свободного 
вращения спичечного коробка}
Исследовать устойчивость свободного вращения спичечного коробка вокруг оси, проходящей 
через его центр и параллельной одной из его сторон.
\begin{figure}[htb]
\centerline{\epsfig{figure=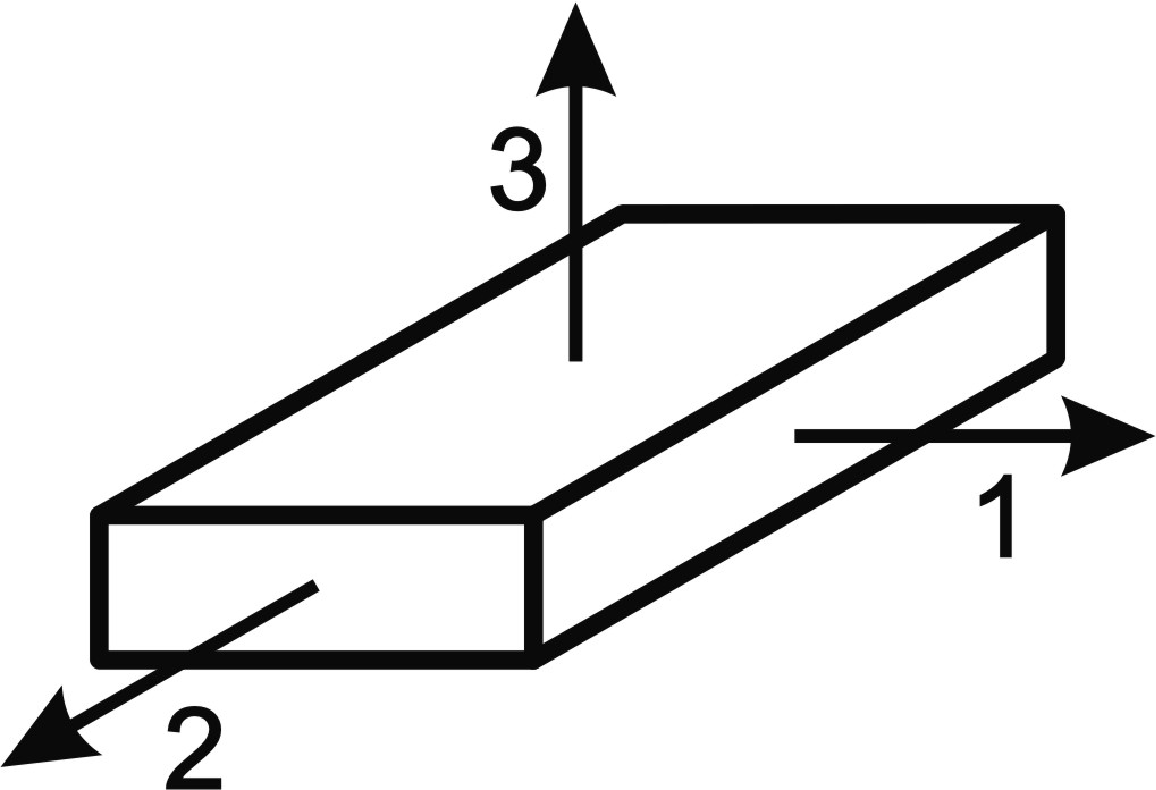,height=3cm}}
\end{figure}

\subsection*{Задача 5: Длина висящего каната}
Однородный и гибкий канат имеет массу $m$, длину $l$ и жесткость $k$. Найти длину
каната, если его повесить на верхний конец вертикально в поле тяжести.

\subsection*{Задача 6 {\usefont{T2A}{cmr}{b}{n}(\hspace*{-1.5mm}\cite{32}, 
9.6)}: Колебания полушара на гладкой горизонтальной плоскости}
Найти период малых колебаний полушара радиусом $R$ на гладкой горизонтальной плоскости 
в поле тяжести.

\section[Семинар 36]
{\centerline{Семинар 36}}

\subsection*{Задача 1 {\usefont{T2A}{cmr}{b}{n}(10.4)}: Отклонение тела при 
падении}
Тело свободно падает с высоты 500~м на землю. Принимая во внимание вращение Земли и 
пренебрегая сопротивлением воздуха, определить, насколько отклонится тело при падении. 
Географическая широта места $60^\circ$.

\subsection*{Задача 2 {\usefont{T2A}{cmr}{b}{n}(10.5)}: Наклон поверхности воды 
в реке}
Вращение Земли вызывает наклон поверхности воды в реке. Оценить угол наклона 
поверхности воды для реки, текущей с севера на юг на широте $\phi$.

\subsection*{Задача 3 {\usefont{T2A}{cmr}{b}{n}(10.12)}: Скорость пробки 
относительно трубки в момент вылета}
Горизонтальная трубка длиной $L$ равномерно вращается вокруг вертикальной оси 
с угловой скоростью $\omega$. Внутри трубки свободно скользит пробка. Определить 
скорость пробки относительно трубки в момент вылета и время движения пробки в трубке. 
В начальный момент пробка покоилась на расстоянии $x_0$ от оси.

\subsection*{Задача 4 {\usefont{T2A}{cmr}{b}{n}(10.13)}: Уравнение движения
бусинки}
На палочку длиной $L$ надета бусинка массой $m$. Коэффициент ее трения о палочку 
$\mu$.  Палочка вращается по конусу с углом раствора $2\alpha$ с угловой скоростью
$\omega$  относительно вертикальной оси, проходящей через конец. Пренебрегая весом 
бусинки, написать уравнение ее движения во вращающейся системе координат.

\subsection*{Задача 5 {\usefont{T2A}{cmr}{b}{n}(9.22)}: Угловая скорость вращения 
желоба после соскальзывания тела}
Тело массой $m$ соскальзывает с высоты $H$ по винтовому желобу с радиусом $R$ и 
углом наклона $30^\circ$ (см. рисунок). Желоб массой $M$ может свободно вращаться 
вокруг своей вертикальной оси. Найти угловую скорость вращения желоба после 
соскальзывания тела. Трения нет.
\begin{figure}[htb]
\centerline{\epsfig{figure=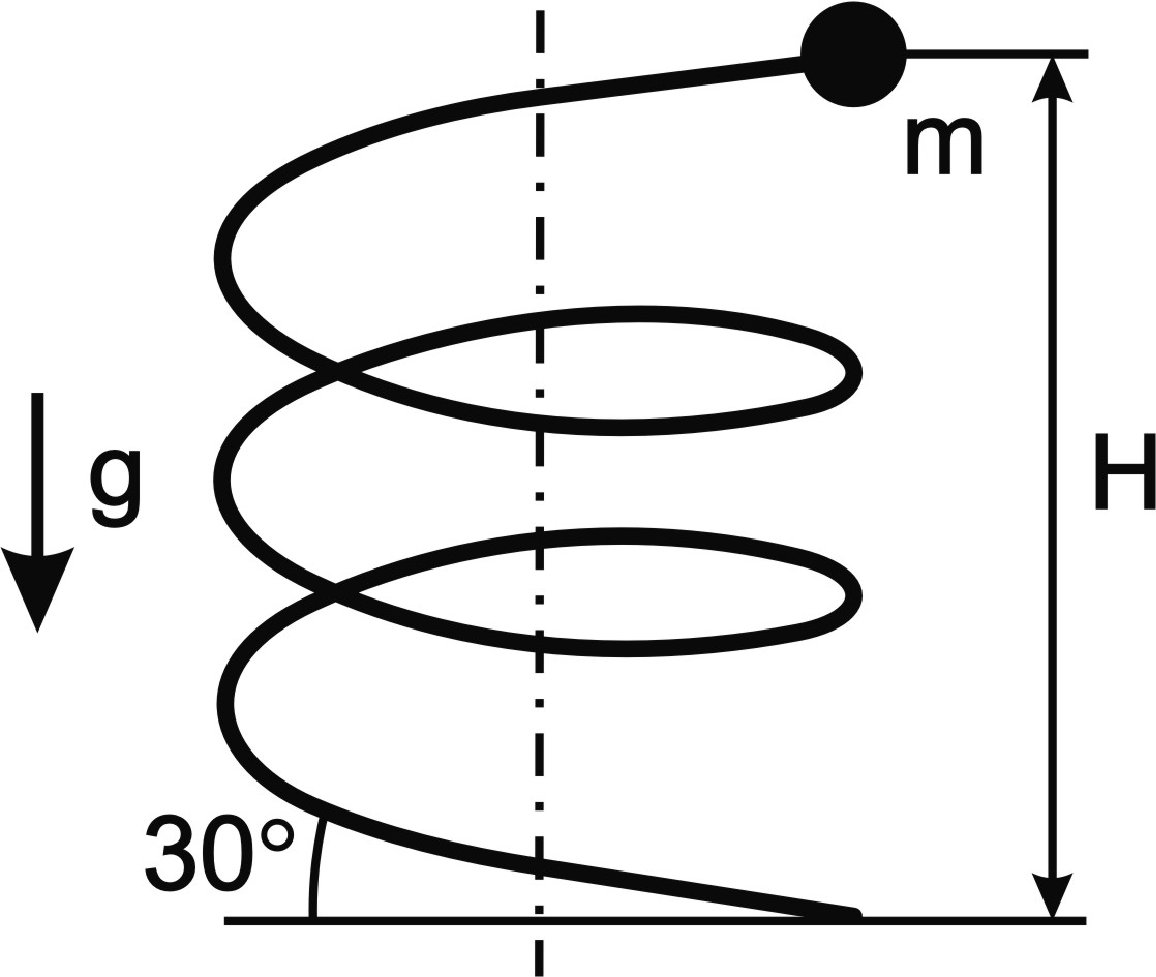,height=4cm}}
\end{figure}

\subsection*{Задача 6 {\usefont{T2A}{cmr}{b}{n} (\hspace*{-1.5mm}\cite{36})}: 
Отклонение отвеса от линии к центру Земли}
Найти отклонение отвеса от линии к центру Земли на широте $\theta$. Землю считать 
эллипсоидом вращения, гравитационный потенциал на поверхности которого зависит от 
широты $\theta$ и расстояния до центра $r$ как 
$$U=-\frac{GM}{r}+\frac{a_2}{2}\,\frac{GM}{r}\left [\frac{R^2}{r^2}\,
(\left (3\cos^2{\theta}-1\right )\right ],$$
где $M$ --- масса Земли, $R$ --- ее экваториальный радиус. Численный коэффициент
$a_2=1,1\cdot 10^{-3}$.

\subsection*{Задача 7 {\usefont{T2A}{cmr}{b}{n}(\hspace*{-1.5mm}\cite{37}, 
487)}: Орбита геостационарного спутника после радиального возмущения}
Спутник, круговая орбита которого расположена в экваториальной плоскости, ``висит'' 
неподвижно над некоторой точкой земной поверхности на расстоянии $R$ от центра Земли.
Спутник получает возмущающий импульс, сообщающий ему малую радиальную добавку 
к скорости $\Delta V$. Найти траекторию спутника по отношению к земному наблюдателю 
после возмущения.

\section[\LARGE Решения]{\centerline{\LARGE Решения}}

\section[Семинар 1]
{\centerline{Семинар 1}}

\subsection*
{Задача 1 {\usefont{T2A}{cmr}{b}{n}(1.5)}: Максимальное время наблюдения Венеры 
после захода Солнца}
\begin{figure}[htb]
\centerline{\epsfig{figure=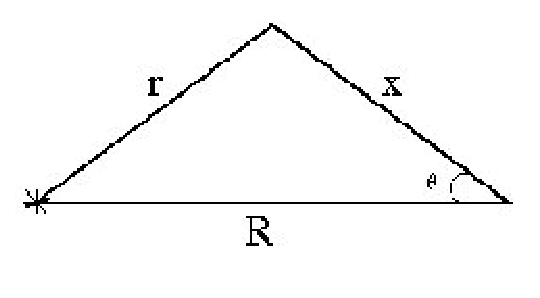,height=4cm}}
\end{figure}
{\usefont{T2A}{cmr}{m}{it} Первое решение.} Пусть $x$ --- расстояние Земля--Венера,
а $\theta$ --- угловое удаление Венеры от Солнца для земного наблюдателя. По теореме 
косинусов (см. рисунок)
\begin{equation}
r^2=R^2+x^2-2Rx\cos{\theta},
\label{eq1_1}
\end{equation}
где $r$ --- радиус орбиты Венеры, $R$ --- радиус орбиты Земли. Функция $\theta(x)$
достигает максимального значения $\theta=\theta_m$ в точке $x=x_m$, где 
$$\dot\theta(x_m)=\left . \frac{d\theta}{dx}\right |_{x=x_m}=0.$$ 
Поэтому, дифференцируя (\ref{eq1_1}) по $x$, получаем при $x=x_m$:
$$0=2x-2R\cos{\theta_m}.$$
Следовательно, $x_m=R\cos{\theta_m}$ и тогда (\ref{eq1_1}) дает 
$r=R\sin{\theta_m}$. Отсюда $\sin{\theta_m}=r/R\approx 0,72$ и
$\theta_m \approx\arcsin{\,0,72}\approx 46^\circ$. Для земного наблюдателя
Венера на небосводе движется с угловой скоростью $360^\circ/24~\mbox{ч}$. Поэтому
максимальное время наблюдения Венеры после захода Солнца составит
$$t=46^\circ:\frac{360^\circ}{24~\mbox{ч}}\approx 3,1~\mbox{ч}.$$

{\usefont{T2A}{cmr}{m}{it} Второе решение.}
Перейдем во вращающуюся систему, в которой Земля неподвижна относительно Солнца
(т. е. которая вращается с такой же угловой скоростью, как Земля). В этой системе 
Венера движется по окружности радиуса $r$ и ясно (см. рисунок), что максимальному
угловому удалению Венеры от Солнца соответствует ситуация, когда линия Земля--Венера
касательна к орбите Венеры. Следовательно, $\sin{\theta_m}=r/R$.
\begin{figure}[htb]
\centerline{\epsfig{figure=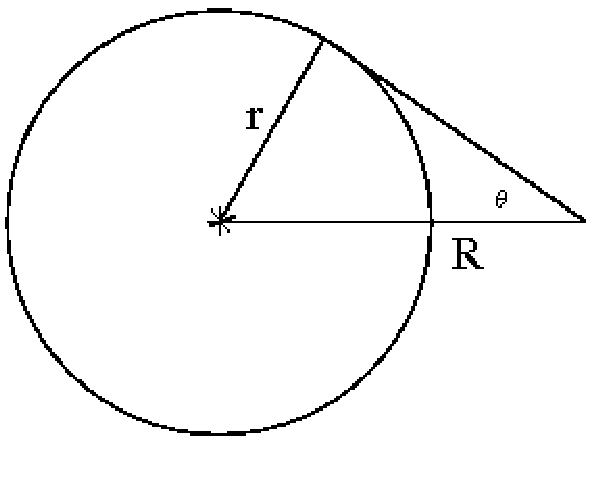,height=4cm}}
\end{figure}
 
\subsection*{Задача 2 {\usefont{T2A}{cmr}{b}{n}(\hspace*{-1.5mm}\cite{2})}: 
Время торможения}
Сила имеет размерность $[M][L]/[T]^2$, скорость --- $[L]/[T]$. Поэтому соотношение
$F=bV^n$ показывает, что коэффициент $b$ имеет размерность $[b]=[M][T]^{n-2}
[L]^{1-n}$. Пусть время торможения $t\sim b^\alpha m^\beta V^\gamma$. Приравняв
размерности, имеем
$$[T]=[M]^{\alpha+\beta} [T]^{\alpha(n-2)-\gamma} [L]^{\alpha(n-1)+\gamma}.$$
Следовательно,
$$\alpha=-\beta,\;\;\alpha(n-1)=-\gamma,\;\;\alpha(n-2)-\gamma=1.$$
Решением этой системы является $\alpha=-1$, $\beta=1$, $\gamma=-(n-1)$. Таким
образом, 
\begin{equation}
t=f(n)\frac{m}{b}V^{1-n}.
\label{eq1_2}
\end{equation}
Если $n=0$, это уравнение дает $t=f(0)mV/b$. С другой стороны, в этом случае
тормозящая сила $F=b$ постоянна и сообщает частице постоянное отрицательное ускорение
$a=-b/m$. Время торможения $t=V/|a|=mV/b$, что согласуется с (\ref{eq1_2}), и 
показывает, что $f(0)=1$.

Но рассмотрим случай $n=1$. Формула (\ref{eq1_2}) дает
$$t=f(1)\frac{m}{b},$$
что не зависит от начальной скорости $V$. Это довольно странно. Попробуем разобраться, 
в чем дело. При $n=1$ уравнение движения имеет вид
$$m\frac{dV}{dt}=-bV.$$
Разделяем переменные и интегрируем
$$\int\limits_V^{V_1}\frac{dV}{V}=-\frac{b}{m}\int\limits_0^t dt,$$
где $V_1$ --- конечная скорость при времени $t$. Это дает
$$t=\frac{m}{b}\ln{\frac{V}{V_1}}\to\infty,\;\;\mbox{когда}\;\;V_1\to 0,$$
что опять согласуется с формулой (\ref{eq1_2}), но делает ее бесполезной, так как 
в данном случае оказывается, что $f(1)=\infty$.

Как видим, требуется некоторая осторожность при использовании метода размерностей.

\subsection*{Задача 3 {\usefont{T2A}{cmr}{b}{n}(\hspace*{-1.5mm}\cite{3})}: 
Исчезнование тени}
Пусть $t=0$ соответствует моменту времени, когда источник света находится
точно над стеной, и пусть в момент времени $t$ в точке $P$ на расстоянии $x$
от стены наблюдается свет от источника и исчезнование тени. Этот свет был
излучен в более ранний момент момент времени $t_S=-y/V$, когда источник
света находился на расстоянии $y$ от стены. Ясно, что
\begin{equation}
t=t_S+\frac{SP}{c},
\label{eq1_3a}
\end{equation}
где $c$ --- скорость света. 
\begin{figure}[htb]
\centerline{\epsfig{figure=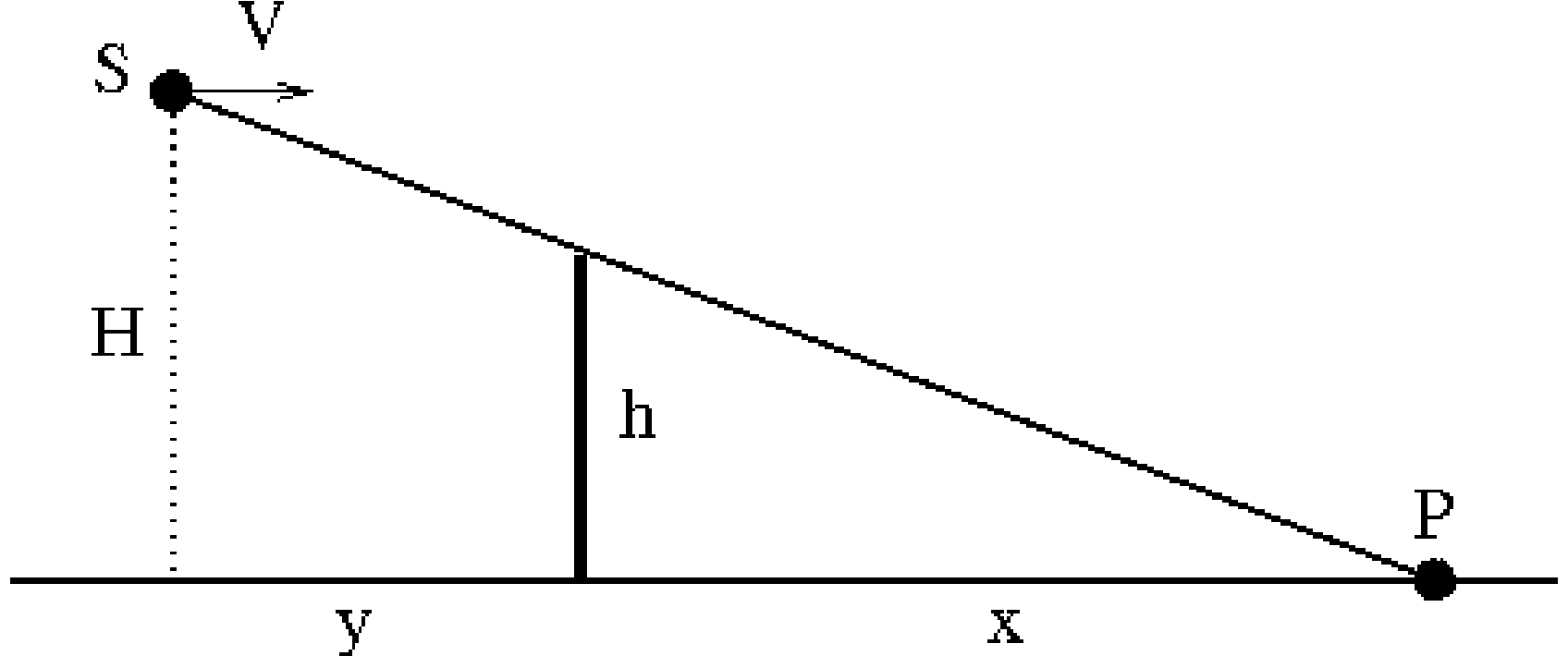,height=4cm}}
\end{figure}
Из подобия соответствующих треугольников имеем
$$\frac{x}{x+y}=\frac{h}{H},$$
следовательно
\begin{equation}
x=\frac{h}{H-h}\,y,\;\;\;x+y=\frac{H}{H-h}\,y,
\label{eq1_3b}
\end{equation}
и
$$SP=\sqrt{H^2+(x+y)^2}=H\sqrt{1+\frac{y^2}{(H-h)^2}}.$$
Тогда (\ref{eq1_3a}) принимает вид
\begin{equation}
t=-\frac{y}{V}+\frac{H}{c}\sqrt{1+\frac{y^2}{(H-h)^2}}.
\label{eq1_3c}
\end{equation}
Найдем такой $y=y_m$, который соответствует минимуму этой функции. Равенство
нулю производной
$$\frac{dt}{dy}=-\frac{1}{V}+\frac{H}{c}\,\frac{1}{\sqrt{1+\frac{y^2}
{(H-h)^2}}}\,\frac{y}{(H-h)^2}=0$$
дает уравнение для $y$:
\begin{equation}
\frac{y}{\sqrt{1+\frac{y^2}{(H-h)^2}}}=\beta^{-1}\;\frac{(H-h)^2}{H},
\label{eq1_3d}
\end{equation}
где $\beta=V/c$. Возводя (\ref{eq1_3d}) в квадрат и решая полученное уравнение для
$y^2$, находим
$$y_m=\beta^{-1}\;\frac{(H-h)^2}{H}\frac{1}{\sqrt{1-\beta^{-2}\frac{(H-h)^2}
{H^2}}}.$$
Вспоминая (\ref{eq1_3b}), находим соответствующее значение $x=x_m$:
\begin{equation}
x_m=\frac{h}{H-h}\,y_m=h\left(1-\frac{h}{H}\right)\frac{1}
{\sqrt{\beta^2-\left(1-\frac{h}{H}\right)^2}}.
\label{eq1_3e}
\end{equation}
Это и есть то расстояние от стены, где быстрее всего исчезнет тень. Из 
(\ref{eq1_3e}) видно, что такой сценарий предполагает
$$\beta>1-\frac{h}{H}.$$
Если же $\beta<1-h/H$, то $t(y)$ убывающая функция и при $y\to\infty$
$$t\to\frac{-yH}{V(H-h)}\left (1-\frac{h}{H}-\beta\right)\to -\infty.$$
В этом случае тень ведет себя обычно: светлая полоса приближается к стенке из 
бесконечности справа и тень полностью исчезает в момент времени $t=H/c$ при $y=0$.

{\usefont{T2A}{cmr}{m}{it} Вариант решения.} Несколько более удобно использовать
в качестве независимой переменной не $y$, a $z=\cot{\theta}$, где $\theta$ --- это
угол между прямой $SP$ и горизонталью. Тогда $X=hz$, $x+y=Hz$, $y=(H-h)z$ и
$$t=-\frac{(H-h)z}{V}+\frac{H\sqrt{1+z^2}}{c}.$$
Условие
$$\frac{dt}{dz}=-\frac{H-h}{V}+\frac{H}{c}\,\frac{z}{\sqrt{1+z^2}}=0$$
дает
$$\frac{z}{\sqrt{1+z^2}}=\beta^{-1}\;\left(1-\frac{h}{H}\right),$$
что легко решается:
$$z_m=\left(1-\frac{h}{H}\right)\frac{1}
{\sqrt{\beta^2-\left(1-\frac{h}{H}\right)^2}}.$$
Окончательно
$$x_m=hz_m=h\left(1-\frac{h}{H}\right)\frac{1}
{\sqrt{\beta^2-\left(1-\frac{h}{H}\right)^2}}.$$

\subsection*{Задача 4 {\usefont{T2A}{cmr}{b}{n}(\hspace*{-1.5mm}\cite{2})}: 
Скорости частиц после столкновения}
Пусть $r=M/m$. Законы сохранения импульса и энергии приводят к системе
$$V=V_1+rV+2,\;\;\;V^2=V_1^2+rV_2^2,$$
где $V_1,V_2$ --- скорости частиц после столкновения. Замечая, что $V^2-V_1^2=
(V+V_1)(V-V_1)=(V+V_1)rV_2$, эту систему можно переписать и так:
$$V-V_1=rV_2,\;\;\;V+V_1=V_2.$$
Решением ее является
$$V_1=\frac{1-r}{1+r}V,\;\;\;V_2=\frac{2V}{1+r}.$$
Если $r\gg 1$ (т. е. $M\gg m$), будем иметь $V_1=-V,V_2=0$, что соответствует
упругому столкновению частицы со стенкой. Если $r\ll 1$ (т. е. $M\ll m$), то
$V_1=V,V_2=2V$, что тоже ожидаемо, так как в системе первой частицы, вторая налетает
на нее со скоростью $V$ и упруго отражается с такой же по величине скоростью, как от 
стенки. Поэтому в лабораторной системе скорость второй частицы будет $V+V=2V$.

\subsection*{Задача 5 {\usefont{T2A}{cmr}{b}{n}(\hspace*{-1.5mm}\cite{4,5,6})}: 
Как быстро доползет жук?}
{\usefont{T2A}{cmr}{m}{it} Первое решение.}
В начале $n$-й секунды (или в общем случае в начале $n$-го промежутка $\tau$),
длина нити будет $nL$. Пусть при этом жук находится на расстоянии $S_n$ от стены 
(так что $S_1=0$). В течении следующей секунды жук проползет еще расстояние $l$,
а потом нить растянут до длины $(n+)L$, и расстояние от жука до стены пропорционально
увеличится. Поэтому
$$S_{n+1}=(S_n+l)\frac{(n+1)L}{nL}.$$
Отсюда получаем
\begin{equation}
\frac{S_{n+1}}{n+1}=\frac{S_n}{n}+\frac{l}{n}.
\label{eq1_5a}
\end{equation}
Повторные применения рекуррентного соотношения (\ref{eq1_5a}) приводят к результату
\begin{equation}
\frac{S_{n}}{n}=l\left(\frac{1}{n-1}+\frac{1}{n-2}+\ldots +\frac{1}{n-1}
\right)=lH_{n-1},
\label{eq1_5b}
\end{equation}
где
$$H_n=1+\frac{1}{2}+\frac{1}{3}+\ldots +\frac{1}{n}$$
является $n$-м гармоническим числом. В течений $n$-й секунды жук доползет до 
конца, если $S_n+l\ge nL$, что с учетом (\ref{eq1_5b}) дает условие
\begin{equation}
H_{n-1}\ge \frac{L}{l}-\frac{1}{n},\;\;\mbox{что означает}\;\;
H_n\ge \frac{L}{l}.
\label{eq1_5c}
\end{equation}
Интуитивно ясно, что $n$ будет очень большим. Поэтому можно использовать 
асимптотическую формулу для гармонических чисел $H_n\approx \ln{n}$. Это дает 
$\ln{n}\ge L/l$. Как видим, жук доползет до конца нити за чудовищно большое время
$$T\approx e^{L/l}\tau=e^{100}~\mbox{с}.$$
Интересно, что это время можно вычислить с точностью до секунды (хотя, конечно, глупо 
это делать в контексте данной задачи). Пусть $A$ некоторое число и $n=n_A$ обозначает
наименьшее целое число, такое что $H_n>A$. Можно показать \cite{9}, что 
$$n_A=\left[e^{A-\gamma}+\frac{1}{2}\right],$$
кроме тех случаев, когда $e^{A-\gamma}+1/2$ подходит слишком близко к целому 
значению. Здесь скобки обозначают целое значение, а $\gamma\approx 0,5772$ --- 
константа Эйлера. В частности \cite{9},
$$n_{100}=1509 26886 22113 78832 36935  63264 53810 14498 59497.$$

{\usefont{T2A}{cmr}{m}{it} Второе решение} \cite{6}.
В течение первой секунды жук проползает $l/L$ часть нити. Ключевым моментом является
наблюдение, что это соотношение не изменится после растяжения нити, так как расстояние
от жука до стены и длина нити увеличатся в одинаковое число раз. Следовательно, на второй
секунде жук проползает $l/2L$ часть нити, на третьей --- $l/3L$, и так далее. На
$n$-й секунды жук достигнет конца нити, если
$$\frac{l}{L}+\frac{l}{2L}+\frac{l}{3L}+\ldots+\frac{l}{nL}=\frac{l}{L}
H_n\ge 1.$$

{\usefont{T2A}{cmr}{m}{it} Третье решение} (это, в сущности, решение Сахарова 
\cite{4}).
Будем считать, что нить растягивают непрерывно со скоростью $u=L/\tau$, а жук ползет
по нити со скоростью $V=l/\tau$. Метрика на нити дается выражением $dl=a(t)dx$,
где $x$ --- сопутствующая координата жука, которая может меняться в пределах от нуля
(начало нити) до $L$ (конец нити), а $a(t)$ --- масштабный множитель, учитывающий 
тот факт, что нить постоянно растягивают. Его можно найти так: с одной стороны, длина
нити в момент времени $t$ равна $a(t)L$, а с другой стороны, эта длина должна 
быть равна $L+ut=L(1+t/\tau)$. Поэтому $a(t)=1+t/\tau$, и
\begin{equation}
dl=\left( 1+\frac{t}{\tau}\right) dx
\label{eq1_5d}
\end{equation}
Относительно стены скорость жука равна
$$V=\frac{dl}{dt}=l/\tau.$$
Поэтому (\ref{eq1_5d}) дает
$$\frac{l}{\tau}=\left( 1+\frac{t}{\tau}\right)\frac{dx}{dt}.$$
Разделяем переменные и интегрируем
$$\int\limits_0^L dx=\frac{l}{\tau}\int\limits_0^T\frac{dt}
{1+\frac{t}{\tau}},$$
или
$$L=l\ln{\left( 1+\frac{T}{\tau}\right)}\approx  l\ln{\frac{T}{\tau}}.$$
Следовательно, $T\approx e^{L/l}\tau$.

{\usefont{T2A}{cmr}{m}{it} Четвертое решение} \cite{10}.
Пусть в момент времени $t$ расстояние от жука до стены есть $l(t)$. Длина нити при
этом будет $L(t)=L+ut=L(1+t/\tau)$. Посмотрим, как меняется отношение $F(t)=
l(t)/L(t)$. За инфинитезимальное время $dt$ координата жука $l(t)$ увеличится до
$$l(t)\,\frac{L(t)+udt}{L(t)}=l(t)\left(1+\frac{udt}{L(t)}\right)$$
за счет растяжения нити и увеличится на $Vdt$ из-за того, что жук ползет вперед.
Следовательно,
$$F(t+dt)=\frac{l(t)\left(1+\frac{udt}{L(t)}\right)+Vdt}{L(t)+udt}\approx
F(t)+\frac{Vdt}{L(t)}.$$
Отсюда получаем
\begin{equation}
\frac{dF}{dt}=\frac{V}{L(t)}=\frac{V}{L+ut}.
\label{eq1_5e}
\end{equation}
Интегрируем с начальным условием $F(0)=0$ и получаем
\begin{equation}
F(t)=\frac{V}{u}\ln{\frac{L+ut}{L}}.
\label{eq1_5f}
\end{equation}
При $t=T$ имеем $F(T)=1$. Поэтому (\ref{eq1_5f}) дает
$$T=\frac{L}{u}\left(e^{u/V}-1\right )=\tau\left(e^{L/l}-1\right )\approx
e^{L/l}\tau.$$
Заметим, что расстояние от жука до свободного конца нити (которого тянут) сначала 
увеличивается, а потом уменшается. Найдем, на каком максимальном расстоянии жук окажется
от свободного конца. Это расстояние равно
$$s(t)=[1-F(t)]L(t)=\left [1-\frac{V}{u}\ln{\left (1+\frac{ut}{L}\right )}
\right ](L+ut).$$
Условие, что производная этой функции равна нулю,
$$\frac{ds}{dt}=u\left [1-\frac{V}{u}\ln{\left (1+\frac{ut}{L}\right )}
\right ]-V=0,$$
дает
$$1-\frac{V}{u}\ln{\left (1+\frac{ut_m}{L}\right )}=\frac{V}{u},\;\;
L+ut_m=L\exp{\left (\frac{u}{V}-1\right)}.$$
Следовательно, максимальное расстояние равно
$$s_m=L\frac{V}{u}\exp{\left(\frac{u}{V}-1\right)}=l\exp{\left(\frac{L}{l}
-1\right)}\approx e^{99}~\mbox{см}\approx 10^{25}~\mbox{св. лет}.$$ 

\subsection*{Задача 6 {\usefont{T2A}{cmr}{b}{n}(\hspace*{-1.5mm}\cite{6A})}: 
Электрон на краю наблюдаемой Вселенной}
Пусть скорость молекулы $V$, а длина свободного пробега $l$. На молекулу действует 
сила гравитационного притяжения электрона, равная $F=GmM/L^2$, где $G$ --- константа 
Ньютона, $m$ --- масса электрона, $M$ --- масса молекулы. Будем считать, что эта 
сила сообщает молекуле поперечное ускорение $a=F/M$. Тогда за  время $t=l/V$ 
молекула отклонится от прямой в поперечном направлении на $\Delta y=at^2/2$, 
т. е. на угол
\begin{equation}
\Delta \theta=\frac{\Delta y}{l}=G\frac{m}{L^2}\frac{l}{2V^2}\approx
G\frac{mM}{L^2}\frac{l}{6kT}\approx 1,5\cdot 10^{-105},
\label{eq1_6}
\end{equation}
где заменили $MV^2\approx 3kT$, а для численной оценки подставили $T=300^\circ K$,
$m=9,1\cdot 10^{-31}$~кг, $M=32\cdot 1,7\cdot 10^{-27}$~кг (молекула кислорода)
и $l\approx 10^{-7}$~м (что для данной температуры соответствует атмосферному давлению).

Как и ожидалось, $\Delta \theta$ очень маленькая величина. Но посмотрим, во сколько 
раз эта неопределенность увеличивается при столкновении молекул.
\begin{figure}[htb]
\centerline{\epsfig{figure=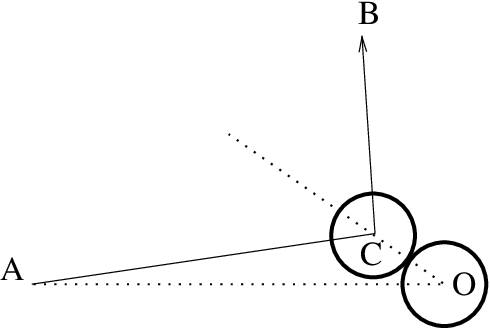,height=4cm}}
\end{figure}
Для оценки будем считать, что молекула имеет сферическую форму радиусом $R$. 
Пусть $\angle CAO=\Delta \theta$, $\angle COA=\Delta \theta_1$. Так как оба
угла маленькие, имеем $l\Delta \theta\approx 2R \Delta \theta_1$. Следовательно,
$$\Delta \theta_1=\frac{l}{2R}\Delta \theta.$$
После отражения молекула летит назад под углом $2\Delta \theta_1-\Delta \theta
\approx 2\Delta \theta_1$ по отношению к той траектории, которую она бы имела, если
бы не было электрона на краю Вселенной. Следовательно, при каждом столкновении 
неопределенность в угле увеличивается $l/R\approx 0,7\cdot 10^3$ раз, так как 
$R\approx 1,5\cdot 10^{-10}$~м. Достаточно всего лишь около сорока столкновений,
чтобы очень маленькая первоначальная неопределенность (\ref{eq1_6}) превратилась в 
неопределенность порядка единицы, и мы полностью потеряем предсказательную способность.
Это случится меньше чем за микросекунду. 

\section[Семинар 2]
{\centerline{Семинар 2}}

\subsection*{Задача 1 {\usefont{T2A}{cmr}{b}{n}(1.30)}: Форма шнура из 
взрывчатого вещества}
{\usefont{T2A}{cmr}{m}{it} Первое решение.}
Пусть $r_0,\phi_0$ --- полярные координаты начала шнура $A$, которую поджигают 
(с которого начинается взрыв). Длина шнура до точки $B$ с координатами $r,\phi$ есть
$$L=\int\limits_{\phi_0}^\phi\sqrt{r^2+\left(\frac{dr}{d\phi}\right)^2}\,
d\phi.$$
В точке $B$ взрыв произойдет на $L/V$ позже по времени. Поэтому условие, что из точек 
$A$ и $B$ взрывная волна одновременно приходит в начало координат, имеет вид
\begin{equation}
\frac{r_0}{c}=\frac{r}{c}+\frac{1}{V}\int\limits_{\phi_0}^\phi\sqrt{r^2+
\left(\frac{dr}{d\phi}\right)^2}\,d\phi.
\label{eq2_1a}
\end{equation}
Рассматривая $r=r(\phi)$ как функцию от $\phi$, продифференцируем (\ref{eq2_1a}) по
$\phi$, используя формулу
$$\frac{d}{d\phi}\int\limits_{\phi_0}^\phi f(\varphi)\,d\varphi=f(\phi).$$
В результате получим
\begin{equation}
0=\frac{1}{c}\frac{dr}{d\phi}+\frac{1}{V}\sqrt{r^2+\left(\frac{dr}{d\phi}
\right)^2}.
\label{eq2_1b}
\end{equation}
Отсюда получаем дифференциальное уравнение
\begin{equation}
\frac{dr}{d\phi}=-\frac{r}{\sqrt{\frac{V^2}{c^2}-1}}.
\label{eq2_1c}
\end{equation}
При извлечении корня мы выбрали знак минус в (\ref{eq2_1c}), так как ясно, что если 
шнур закручивается влево при увеличении $\phi$, расстояние $r$ должно уменьшаться 
(см. рисунок). Если шнур закручивается вправо, надо выбрать знак плюс.
\begin{figure}[htb]
\centerline{\epsfig{figure=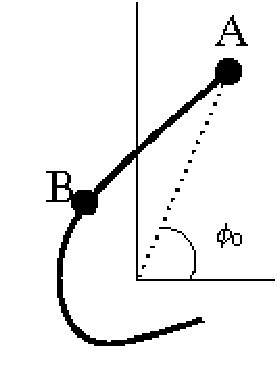,height=4cm}}
\end{figure}
Разделяем переменные в (\ref{eq2_1c}),
$$\frac{dr}{r}=-\frac{d\phi}{\sqrt{\frac{V^2}{c^2}-1}}$$
и интегрируем:
$$\ln{\frac{r}{r_0}}=-\frac{\phi-\phi_0}{\sqrt{\frac{V^2}{c^2}-1}}.$$
Следовательно, шнур должен иметь форму спирали
$$r=r_0\exp{\left(-\frac{\phi-\phi_0}{\sqrt{\frac{V^2}{c^2}-1}}\right)},\;\;
\phi\ge \phi_0.$$

{\usefont{T2A}{cmr}{m}{it} Второе решение.}
Пусть точки $A(r,\phi)$ и $B(r+dr,\phi+d\phi)$ на шнуре инфинитазимально близки и
пусть взрыв сначала происходит в точке $A$. Так как расстояние между этими точками равно
$AB=\sqrt{(dr)^2+r^2(d\phi)^2}$, то взрыв в точке $B$ произойдет позже на время 
$\Delta t_1=AB/V$. Но точка $B$ находится ближе на $-dr$ (рассматриваем левую 
закрутку шнура, так что $dr<0$ при $d\phi>0$), поэтому волна от взрыва в точке $B$ 
придет в начало координат на $\Delta t_2=-dr/c$ быстрее. Требуется равенство времен 
$\Delta t_1=\Delta t_2$. Это дает $$\frac{\sqrt{(dr)^2+r^2(d\phi)^2}}{V}=
-\frac{dr}{c}.$$ Это эквивалентно уравнению (\ref{eq2_1b}).

\subsection*{Задача 2 {\usefont{T2A}{cmr}{b}{n}(\hspace*{-1.5mm}\cite{7} 
2.4.12)}:  Скорость нижнего цилиндра}
Важно осознать, что цилиндры расцепятся раньше, чем верхний цилиндр успеет полностью
упасть. Посмотрим, когда это случится. Пусть $x$ --- горизонтальная $x$-координата
центра нижнего цилиндра, $y$ --- высота центра верхнего цилиндра (его 
$y$-координата).  Тогда (см. рисунок)
\begin{figure}[htb]
\centerline{\epsfig{figure=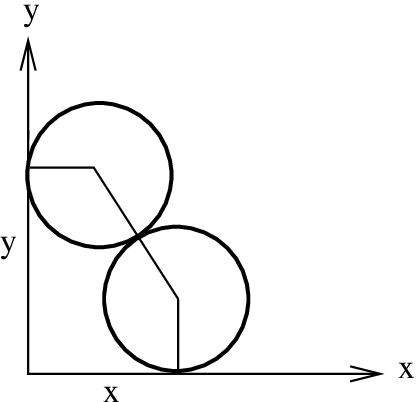,height=4cm}}
\end{figure}
\begin{equation}
(x-R)^2+(y-R)^2=(2R)^2.
\label{eq2_2a}
\end{equation}
Продифференцируем (\ref{eq2_2a}) по времени,
\begin{equation}
\dot x (x-R)+\dot y (y-R)=0.
\label{eq2_2b}
\end{equation}
Продифференцируем еще раз:
\begin{equation}
\dot x^2+\dot y^2+\ddot x (x-R)+\ddot y (y-R)=0.
\label{eq2_2c}
\end{equation}
В момент расцепления цилиндров $\ddot x=0$ (так как на нижний цилиндр горизонтально
никто не давит) и $\ddot y=-g$ (так как на  верхний цилиндр действует только сила
тяжести). Поэтому из (\ref{eq2_2c}) получаем, что в момент расцепления
\begin{equation}
\dot x^2+\dot y^2=g(y-R).
\label{eq2_2d}
\end{equation}
Центр верхнего цилиндра упал с высоты $3R$ до высоты $y$. Поэтому сохранение энергии,
с учетом (\ref{eq2_2d}), дает
$$mg(3R-y)=\frac{m}{2}(\dot x^2+\dot y^2)=\frac{mg}{2}(y-R).$$
Следовательно, $6R-2y=y-R$, и это определяет высоту 
\begin{equation}
y=\frac{7}{3}R,
\label{eq2_2e}
\end{equation} 
при которой происходит
расцепление цилиндров. Подставляя это значение $y$ в (\ref{eq2_2a}), получаем 
уравнение для $x$, из которого следует
\begin{equation}
x-R=\frac{2}{3}\sqrt{5}\,R.
\label{eq2_2f}
\end{equation}
Подставим (\ref{eq2_2e}) и (\ref{eq2_2f}) в (\ref{eq2_2b}). В результате получим
\begin{equation}
\dot y=-\frac{\sqrt{5}}{2}\;\dot x.
\label{eq2_2g}
\end{equation}
Наконец, (\ref{eq2_2e}) и  (\ref{eq2_2g}) можно подставить в (\ref{eq2_2d}) и 
получить уравнения для определения $\dot x$ в момент расцепления цилиндров. Это дает
\begin{equation}
\dot x =\frac{4}{3}\sqrt{\frac{gR}{3}}.
\label{eq2_2h}
\end{equation} 
Так как после расцепление цилиндров скорость нижнего цилиндра не меняется, 
(\ref{eq2_2h}) и есть конечная скорость нижнего цилиндра.    

\subsection*{Задача 3 {\usefont{T2A}{cmr}{b}{n}(1.31)}: Угол между векторами}
Декартовы координаты векторов равны
$$\vec{r}_1=(r_1\,\sin{\theta_1}\,\cos{\phi_1},r_1\,\sin{\theta_1}\,
\sin{\phi_1},r_1\cos{\theta_1})$$
и
$$\vec{r}_2=(r_2\,\sin{\theta_2}\,
\cos{\phi_2},r_2\,\sin{\theta_2}\,\sin{\phi_2},r_2\cos{\theta_2}).$$
Поэтому их скалярное произведение, с одной стороны, равно
$$\vec{r}_1\cdot \vec{r}_2=r_1r_2\cos{\alpha},$$
где $\alpha$ есть угол между векторами, а с другой стороны,
$$\vec{r}_1\cdot \vec{r}_2=r_1r_2[\sin{\theta_1}
\sin{\theta_2}\cos{\phi_1}\cos{\phi_2}+\sin{\theta_1}\sin{\theta_2}
\sin{\phi_1}\sin{\phi_2}+\cos{\theta_1}\cos{\theta_2}],$$
Отсюда получаем уравнение для определения
$\alpha$:
$$\cos{\alpha}=\cos{\theta_1}\cos{\theta_2}+\sin{\theta_1}\sin{\theta_2}
\,\cos{(\phi_1-\phi_2)}.$$

\subsection*{Задача 4: Периметр эллипса}
Уравнение эллипса в декартовых координатах имеет вид
$$\frac{x^2}{a^2}+\frac{y^2}{b^2}.$$
Удобно ввести параметризацию
$$x=a\sin{\tau},\;\;\;y=b\cos{\tau},$$
где параметр $\tau$ меняется в пределах от $0$ до $2\pi$. Тогда
$$dx=a\cos{\tau}\;d\tau,\;\;\;dy=-b\sin{\tau}\;d\tau$$
и для элемента дуги эллипса будем иметь
$$dl=\sqrt{a^2\cos^2{\tau}+b^2\sin^2{\tau}}\,d\tau=
a\sqrt{1-e^2\sin^2{\tau}}\,d\tau,$$
где мы учли, что $b^2=a^2(1-e^2)$. Так как $e\ll 1$, для периметра эллипса будем
иметь
$$L=a\int\limits_0^{2\pi}\sqrt{1-e^2\sin^2{\tau}}\,d\tau\approx
a\int\limits_0^{2\pi}\left(1-\frac{e^2}{2}\sin^2{\tau}\right)\,d\tau=
2\pi a\left(1-\frac{e^2}{4}\right ).$$
Интеграл удобно вычислить, используя
$$\sin^2{\tau}=\frac{1}{2}(1-\cos{2\tau}).$$

\subsection*{Задача 5: Площадь эллипса}
{\usefont{T2A}{cmr}{m}{it} Первое решение.}
Имеем
\begin{equation}
S=\iint dx\,dy=4\int\limits_0^a dx\int\limits_0^{b\sqrt{1-x^2/a^2}}dy=
4b\int\limits_0^a\sqrt{1-\frac{x^2}{a^2}}\;dx,
\label{eq2_5}
\end{equation}
где мы воспользовались уравнением эллипса
$$\frac{x^2}{a^2}+\frac{y^2}{b^2},$$
и тем фактом, что его площадь в четыре раза больше, чем площадь его части (четвертинки)
в первом квадранте. Интеграл в (\ref{eq2_5}) можно вычислить подстановкой 
$x=a\sin{\tau}$. Тогда
$$S=4ab\int\limits_0^{\pi/2}\cos^2{\tau}\,d\tau=
2ab\int\limits_0^{\pi/2}(1+\cos{2\tau})\,d\tau=\pi ab.$$

{\usefont{T2A}{cmr}{m}{it} Второе решение.}
От декартовых координат $x,y$ перейдем к координатам $\tau,\varphi$ с помощью
соотношений 
$$x=a\tau\cos{\varphi},\;\; y=b\tau\sin{\varphi},\;\;0\le\tau\le 1,\;\;
0\le\varphi<2\pi.$$ 
Якобиан этого преобразования равен
$$\frac{\partial(x,y)}{\partial(\tau,\varphi)}=\left | \begin{array}{cc} 
a\cos{\varphi} & b\sin{\varphi} \\ -a\tau\sin{\varphi} & b\tau\cos{\varphi}
\end{array}\right |=ab\tau.$$
Поэтому площадь эллипса
$$S=\iint ab\tau\;d\tau\,d\varphi=ab\int\limits_0^1\tau\,d\tau
\int\limits_0^{2\pi}d\varphi=\pi ab.$$ 

\subsection*{Задача 6 {\usefont{T2A}{cmr}{b}{n}(5.33)}:
Установившаяся скорость монеты}
Пусть $\vec{F}$ есть компонента силы тяжести вдоль наклонной плоскости. Направим ось
$y$ вдоль $\vec{F}$. Условие задачи $\mu=\tg{\alpha}$ означает, что сила трения
по величине равна $F$. Так как сила трения направлена всегда против скорости
монеты $\vec{V}$, уравнение движения монеты имеет вид
\begin{equation}
m\frac{d\vec{V}}{dt}=\vec{F}-F\frac{\vec{V}}{V}.
\label{eq2_6a}
\end{equation}
Так как $\vec{V}\cdot \vec{V}=V^2$, имеем с учетом (\ref{eq2_6a})
$$V\frac{dV}{dt}=\vec{V}\cdot\frac{d\vec{V}}{dt}=\frac{1}{m}(\vec{F}\cdot
\vec{V}-FV)=\frac{F}{m}(V_y-V).$$
Следовательно,
\begin{equation}
\frac{dV}{dt}=\frac{F}{m}\left (\frac{V_y}{V}-1\right ).
\label{eq2_6b}
\end{equation}
С другой стороны, из $y$ компоненты векторного уравнения (\ref{eq2_6a}) получаем
\begin{equation}
\frac{dV_y}{dt}=\frac{F}{m}\left (1-\frac{V_y}{V}\right ).
\label{eq2_6c}
\end{equation}
Вместе (\ref{eq2_6b}) и (\ref{eq2_6c}) означают, что
$$\frac{d}{dt}(V+V_y)=0,$$
т. е. величина $V+V_y$ сохраняется. В начале $V=V_0,\, V_y=0$, в конце монета 
движется вдоль оси $y$ и $V=V_y=u$. Поэтому $V_0=2u$ и конечная скорость монеты
$u=V_0/2$.

\subsection*{Задача 7 {\usefont{T2A}{cmr}{b}{n}(1.9)}:
Траектория конца тени от вертикально стоящей палочки}
Угол наклона земного экватора к эплиптике равен $23^\circ$. Поэтому 22 июня солнечный 
свет падает на северное полушарие под углом $\theta_0=90^\circ-23^\circ=67^\circ$
к земной оси. Вертикально стоящая палочка в Новосибирске составляет угол $\theta_1=
90^\circ-55^\circ=35^\circ$ к земной оси, так как Новосибирск находится на $55^\circ$
северной широты. Солнечные лучи, проходящие через верхний конец палочки, лежат на 
поверхности конуса ось которого параллельна к земной оси, а угол раствора составляет 
$2\theta_0$. Конец тени получается в местах пересечения этих лучей с земной 
поверхностью, которую в малых масштабах можно считать плоской. Следовательно, 
траектория конца тени соответствует коническому сечению. Для Новосибирска эта 
гипербола.

Проследим движение конца тени на траектории с течением  времени. Выберем систему 
отсчета так, что ось $z$ направлена вдоль палочки, а земная ось лежит в плоскости 
$yOz$. Если направление на солнце соответствует сферическим углам $\theta$ и $\phi$,
то конец тени имеет координаты $x=-\tg{\theta}\sin{\phi},\;y=-\tg{\theta}
\cos{\phi}$, где мы считаем, что длина палочки равна единице.

Если систему отсчета повернем вокруг оси $x$ на угол $\theta_1$, то направление оси
$z^\prime$ совпадет с направлением земной оси. Поэтому в новой системе направление на
Солнце задается сферическими углами $\theta^\prime=\theta_0$, $\phi^\prime=
\omega \,t$, где  $\omega$ --- угловая скорость вращения Земли вокруг своей оси. С 
другой стороны, при таком повороте
$$y^\prime=\cos{\theta_1}\,y+\sin{\theta_1}\,z,\;\;
z^\prime=-\sin{\theta_1}\,y+\cos{\theta_1}\,z.$$
Отсюда получаем связь между сферическими углами $\theta^\prime,\phi^\prime$ и 
$\theta,\phi$:
$$\sin{\theta^\prime}\sin{\phi^\prime}=\cos{\theta_1}\sin{\theta}\sin{\phi}+
\sin{\theta_1}\cos{\theta},$$
$$\cos{\theta^\prime}=-\sin{\theta_1}\sin{\theta}\sin{\phi}+
\cos{\theta_1}\cos{\theta}.$$
Из этой системы получаем
$$\sin{\theta}\sin{\phi}=\cos{\theta_1}\sin{\theta_0}\sin{\omega\, t}-
\sin{\theta_1}\cos{\theta_0},$$
$$\cos{\theta}=\sin{\theta_1}\sin{\theta_0}\sin{\omega \, t}+
\cos{\theta_1}\cos{\theta_0},$$
где мы учли, что $\theta^\prime=\theta_0$, $\phi^\prime=\omega \,t$. Возводя эти
равенства в квадрат и складывая, получаем после некоторой алгебры $\sin{\theta}
\cos{\phi}=\sin{\theta_0}\cos{\omega t}$. Следовательно,
$$x=-\frac{\sin{\theta_0}\cos{\omega t}}{\sin{\theta_1}\sin{\theta_0}
\sin{\omega \, t}+\cos{\theta_1}\cos{\theta_0}},$$
$$y=-\frac{\cos{\theta_1}\sin{\theta_0}\sin{\omega\, t}-
\sin{\theta_1}\cos{\theta_0}}{\sin{\theta_1}\sin{\theta_0}
\sin{\omega \, t}+\cos{\theta_1}\cos{\theta_0}}.$$
Эти формулы определяют движение конца тени на траектории с течением  времени. Чтобы 
убедиться, что действительно имеем коническое сечение, выразим из второго равенства
$$\sin{\omega \, t}=\frac{\cos{\theta_0}(\sin{\theta_1}-y\cos{\theta_1})}
{\sin{\theta_0}(\cos{\theta_1}+y\sin{\theta_1})}.$$
Тогда
$$\sin{\theta_1}\sin{\theta_0}\sin{\omega \, t}+\cos{\theta_1}\cos{\theta_0}
=\frac{\cos{\theta_0}}{\cos{\theta_1}+y\sin{\theta_1}},$$
и из выражения для $x$-координаты получаем
$$\cos{\omega \, t}=-\frac{x\cos{\theta_0}}{\sin{\theta_0}(\cos{\theta_1}+
y\sin{\theta_1})}.$$
Следовательно,
$$1=\sin^2{\omega \, t}+\cos^2{\omega \, t}=\frac{\cos^2{\theta_0}}
{\sin^2{\theta_0}}\,\frac{(\sin{\theta_1}-y\cos{\theta_1})^2+x^2}
{(\cos{\theta_1}+y\sin{\theta_1})^2},$$
или, после некоторых преобразований,
$$\cos^2{\theta_0}\,x^2+\cos{(\theta_0+\theta_1)}\cos{(\theta_0-\theta_1)}\,
y^2-\sin{2\theta_1}\, y=\sin{(\theta_0+\theta_1)}\sin{(\theta_0-\theta_1)}.$$
На северном полюсе ($\theta_1=0$) будем иметь окружность. До северного полярного круга
($\theta_1<23^\circ$, или $\theta_1+\theta_0<90^\circ$) -- эллипс. На северном
полярном круге, коэффициент при $y^2$ зануляется, т.е. будем иметь параболу. Южнее
полярного круга (для северного полушария) $\cos{(\theta_0+\theta_1)}
\cos{(\theta_0-\theta_1)}<0$, т.е. будем иметь гиперболу.

В течение дня, $x$- и $y$-координаты конца тени меняются для Новосибирска от 
$-\infty$ до $+\infty$. Поэтому долгота дня определяется разностью между двумя 
последовательными решениями уравнения 
$$\sin{\theta_1}\sin{\theta_0}\sin{\omega \, t}+\cos{\theta_1}\cos{\theta_0}
=0,$$
такими, что $t_1<0$ и $t_2>0$ (так как $t=0$ отвечает промежуточному положению конца
тени между $\mp\infty$). Следовательно, долгота дня
$$T=t_2-t_1=\frac{2\pi}{\omega}\left [\frac{1}{2}+\frac{1}{\pi}\arcsin{\left 
(\ctg{\theta_1}\ctg{\theta_0}\right )}\right ]\approx 17~\mbox{ч}.$$

\section[Семинар 3]
{\centerline{Семинар 3}}

\subsection*{Задача 1 {\usefont{T2A}{cmr}{b}{n}(1.33)}:
Орты сферической и цилиндрической систем координат}

Используя сферические координаты $(r,\theta,\phi)$, радиус-вектор $\vec{r}$
можно выразить как $$\vec{r}=r\sin{\theta}\cos{\phi}\,\vec{i}+
r\sin{\theta}\sin{\phi}\,\vec{j}+r\cos{\theta}\,\vec{k}.$$
Рассмотрим его изменение $(d\vec{r})_r$ по направлению $r$-координаты, т. е. когда
$r$ получает инфинитезимальную добавку $dr$, а координаты $\theta$ и $\phi$ не
меняются:
$$(d\vec{r})_r=dr(\sin{\theta}\cos{\phi}\,\vec{i}+
\sin{\theta}\sin{\phi}\,\vec{j}+\cos{\theta}\,\vec{k}).$$
Орт $\vec{e}_r$ отличается от$(d\vec{r})_r$ только тем, что имеет единичную норму.
Поэтому
$$\vec{e}_r=\frac{(d\vec{r})_r}{|(d\vec{r})_r|}=\sin{\theta}\cos{\phi}\,
\vec{i}+\sin{\theta}\sin{\phi}\,\vec{j}+\cos{\theta}\,\vec{k}.$$
Аналогично находим
$$(d\vec{r})_\theta=rd\theta(\cos{\theta}\cos{\phi}\,\vec{i}+
\cos{\theta}\sin{\phi}\,\vec{j}-\sin{\theta}\,\vec{k})$$
и
$$(d\vec{r})_\phi=rd\phi(-\sin{\theta}\sin{\phi}\,\vec{i}+
\sin{\theta}\cos{\phi}\,\vec{j}).$$
Следовательно,
$$\vec{e}_\theta=\frac{(d\vec{r})_\theta}{|(d\vec{r})_\theta|}=
\cos{\theta}\cos{\phi}\,\vec{i}+
\cos{\theta}\sin{\phi}\,\vec{j}-\sin{\theta}\,\vec{k},$$
и
$$\vec{e}_\phi=\frac{(d\vec{r})_\phi}{|(d\vec{r})_\phi|}=
-\sin{\phi}\,\vec{i}+\cos{\phi}\,\vec{j}.$$
В случае цилиндрических координат $\rho,\varphi,z$ имеем
$$\vec{r}=\rho\cos{\varphi}\,\vec{i}+\rho\sin{\varphi}\,\vec{j}
+z\,\vec{k}.$$
Поэтому
$$(d\vec{r})_\rho=d\rho(\cos{\varphi}\,\vec{i}+\sin{\varphi}\,\vec{j}),$$
$$(d\vec{r})_\varphi=\rho\, d\varphi(-\sin{\varphi}\,\vec{i}+
\cos{\varphi}\,\vec{j}),$$
и $(d\vec{r})_z=dz\,\vec{k}$. Следовательно,
$$\vec{e}_\rho=\frac{(d\vec{r})_\rho}{|(d\vec{r})_\rho|}=
\cos{\varphi}\,\vec{i}+\sin{\varphi}\,\vec{j},$$
$$\vec{e}_\varphi=\frac{(d\vec{r})_\varphi}{|(d\vec{r})_\varphi|}=
-\sin{\varphi}\,\vec{i}+\cos{\varphi}\,\vec{j},$$
и $\vec{e_z}=\vec{k}$.

\subsection*{Задача 2 {\usefont{T2A}{cmr}{b}{n}(1.24)}:
Скорость и ускорение в цилиндрической системе координат}
Так как связь между декартовыми и цилиндрическими координатами имеет вид 
$x=\rho\cos{\varphi},\,y=\rho\sin{\varphi},\,z=z$, а орты цилиндрической
системы выражаются через декартовы орты так
\begin{equation}
\vec{e}_\rho=\cos{\varphi}\,\vec{i}+\sin{\varphi}\,\vec{j},\;\;
\vec{e}_\varphi=-\sin{\varphi}\,\vec{i}+\cos{\varphi}\,\vec{j},\;\;
\vec{e}_z=\vec{k},
\label{eq3_2a}
\end{equation}
то для компонент радиус-вектора в цилиндрической системе координат получаем
$$r_\rho=\vec{r}\cdot\vec{e}_\rho=x\,\cos{\varphi}+y\,\sin{\varphi}=\rho,$$
$$r_\varphi=\vec{r}\cdot\vec{e}_\varphi=-x\,\sin{\varphi}+,\cos{\varphi}=0,$$
и $r_z=z$. Следовательно, радиус-вектор в цилиндрической системе координат имеет вид
\begin{equation}
\vec{r}=\rho\,\vec{e}_\rho+z\,\vec{e}_z.
\label{eq3_2b}
\end{equation}
Чтобы найти скорость, продифференцируем (\ref{eq3_2b}) и учтем, что, как следует из
(\ref{eq3_2a}),
\begin{equation}
\dot{\vec{e}}_\rho=\dot\varphi\,\vec{e}_\varphi,\;\;
\dot{\vec{e}}_\varphi=-\dot\varphi\,\vec{e}_\rho,\;\;
\dot{\vec{e}}_z=0.
\label{eq3_2c}
\end{equation}
В результате получим
\begin{equation}
\vec{V}=\dot{\vec{r}}=\dot\rho\,\vec{e}_\rho+\rho\,\dot{\vec{e}}_\rho+
\dot z\,\vec{e}_z=\dot\rho\,\vec{e}_\rho+\rho\dot\varphi\,\vec{e}_\varphi+
\dot z\,\vec{e}_z.
\label{eq3_2d}
\end{equation}
Чтобы найти ускорение, продифференцируем (\ref{eq3_2d}) 
$$\vec{a}=\dot{\vec{V}}=\ddot\rho\,\vec{e}_\rho+
\dot\rho\dot\varphi\,\vec{e}_\varphi+\rho\ddot\varphi\,\vec{e}_\varphi+
\ddot z\,\vec{e}_z+\dot\rho\,\dot{\vec{e}}_\rho+
\rho\dot\varphi\,\dot{\vec{e}}_\varphi$$
и учтем (\ref{eq3_2c})
$$\vec{a}=(\ddot\rho-\rho\dot\varphi^2)\,
\vec{e}_\rho+(\rho\ddot\varphi+2\dot\rho\dot\varphi)\,\vec{e}_\varphi+
\ddot z\,\vec{e}_z.$$

\subsection*{Задача 3 {\usefont{T2A}{cmr}{b}{n}(1.36)}: Расстояние 
наибольшего сближения}
{\usefont{T2A}{cmr}{m}{it} Первое решение.}
Радиус-векторы кораблей со временем меняются как
$$\vec{r}_1(t)=\vec{r}_1+\vec{V}_1t,\;\;\;\vec{r}_2(t)=\vec{r}_2+\vec{V}_2t.$$
Поэтому расстояние между кораблями $l(t)$ удовлетворяет
\begin{equation}
l^2(t)=[\vec{r}_1-\vec{r}_2+(\vec{V}_1-\vec{V}_2)t]^2.
\label{eq3_3a}
\end{equation}
Нам надо найти минимум функции $l(t)$, но при этом функция $l^2(t)$ тоже будет иметь
минимум. Следовательно, требуем
$$\frac{dl^2}{dt}=2[\vec{r}_1-\vec{r}_2+(\vec{V}_1-\vec{V}_2)t]\cdot
(\vec{V}_1-\vec{V}_2)=0.$$
Это определяет время $t=t_m$ наибольшего сближения:
\begin{equation}
t_m=-\frac{(\vec{r}_1-\vec{r}_2)\cdot (\vec{V}_1-\vec{V}_2)}
{(\vec{V}_1-\vec{V}_2)^2}.
\label{eq3_3b}
\end{equation}
Подставляя (\ref{eq3_3b}) в (\ref{eq3_3a}), находим
$$l_m^2=(\vec{r}_1-\vec{r}_2)^2-\frac{[(\vec{r}_1-\vec{r}_2)\cdot 
(\vec{V}_1-\vec{V}_2)]^2}{(\vec{V}_1-\vec{V}_2)^2}=
\frac{[(\vec{r}_1-\vec{r}_2)\times (\vec{V}_1-\vec{V}_2)]^2}
{(\vec{V}_1-\vec{V}_2)^2}.$$
Следовательно,
$$l_m=\frac{|(\vec{r}_1-\vec{r}_2)\times (\vec{V}_1-\vec{V}_2)|}
{|\vec{V}_1-\vec{V}_2|}.$$

{\usefont{T2A}{cmr}{m}{it} Второе решение.}
Перейдем в систему первого корабля. Тогда второй корабль в начале имеет радиус-вектор
$\vec{r}=\vec{r_2}-\vec{r_1}$ и движется со скоростью 
$\vec{V}=\vec{V_2}-\vec{V_1}$. Из рисунка время сближения
$$t_m=\frac{AB}{V}=\frac{r\cos{\alpha}}{V}=-\frac{\vec{r}\cdot\vec{V}}{V^2}=
-\frac{(\vec{r}_2-\vec{r}_1)\cdot (\vec{V}_2-\vec{V}_1)}
{(\vec{V}_2-\vec{V}_1)^2}.$$
Знак минус, так как $\alpha$ --- это угол между векторами $\vec{V}$ и $-\vec{r}$.
Расстояние наибольшего сближения
$$L_m=r\sin{\alpha}=\frac{|\vec{r}\times\vec{V}|}{V}=
\frac{|(\vec{r}_2-\vec{r}_1)\times (\vec{V}_2-\vec{V}_1)|}
{|\vec{V}_2-\vec{V}_1|}.$$
\begin{figure}[htb]
\centerline{\epsfig{figure=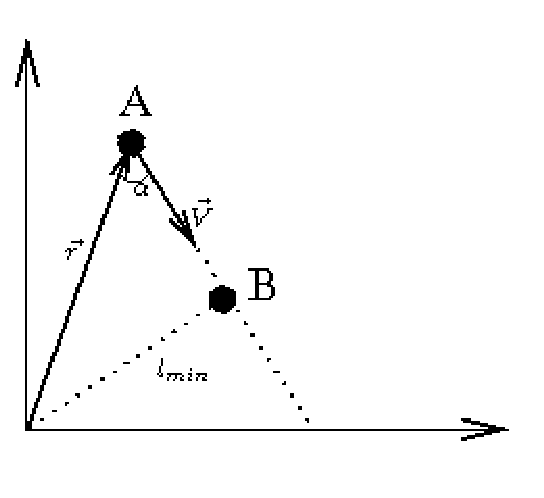,height=5cm}}
\end{figure}

\subsection*{Задача 4 {\usefont{T2A}{cmr}{b}{n}(1.29)}: Заяц и собака}
{\usefont{T2A}{cmr}{m}{it} Первое решение.}
Перейдем в систему зайца; пусть $(\rho,\varphi)$ --- полярные координаты собаки в 
этой системе. Так как декартовый орт $\vec{i}$ так выражается через полярные орты 
$\vec{i}=\cos{\varphi}\,\vec{e}_\rho-\sin{\varphi}\,\vec{e}_\varphi$, скорость 
собаки в этой системе равна
\begin{equation}
\vec{V}_C=-V\vec{e}_\rho -u\vec{i}=-(V+u\cos{\varphi})\vec{e}_\rho+
u\sin{\varphi}\,\vec{e}_\varphi.
\label{eq3_4a}
\end{equation}
С другой стороны, скорость в полярной системе
\begin{equation}
\vec{V}_C=\dot{\rho}\,\vec{e}_\rho+\rho\dot{\varphi}\,\vec{e}_\varphi.
\label{eq3_4b}
\end{equation}
Сравнивая (\ref{eq3_4a}) и (\ref{eq3_4b}), получаем систему
\begin{equation}
\dot{\rho}=-(V+u\cos{\varphi}),\;\;\;\rho\dot{\varphi}=u\sin{\varphi},
\label{eq3_4c}
\end{equation}
из которой следует, что
$$\ddot{\rho}=u\sin{\varphi}\,\dot{\varphi}=\frac{u^2\sin^2{\varphi}}{\rho}.$$
Таким образом,
$$\rho\,\ddot{\rho}=u^2\sin^2{\varphi},\;\;\;(\dot{\rho}+V)^2=
u^2\cos^2{\varphi}.$$
Складывая эти два уравнения, получаем после перегруппировки слагаемых
$$u^2-V^2=\rho\,\ddot{\rho}+\dot{\rho}^2+2\dot{\rho}V=\frac{d}{dt}[\rho\,
\dot{\rho}+2\rho V].$$
Следовательно,
\begin{equation}
\rho[\dot{\rho}+2V]=(u^2-V^2)t+C,
\label{eq3_4d}
\end{equation}
где константа интегрирования $C$ определяется из начальных условий. В начале 
при $t=0$ имеем $\rho=L,\,\varphi=-\pi/2$ и, согласно первому соотношению из 
(\ref{eq3_4c}), $\dot{\rho}=-(V+u\cos{\varphi})=-V$. Подставляя эти начальные 
значения в (\ref{eq3_4d}), определяем константу интегрирования: $C=LV$. 
Следовательно,
\begin{equation}
\rho[\dot{\rho}+2V]=(u^2-V^2)t+LV.
\label{eq3_4e}
\end{equation}
Когда собака поймает зайца, будем иметь $\rho=0$, и (\ref{eq3_4e}) дает нам 
соответствующее время
$$T=\frac{LV}{V^2-u^2}.$$

{\usefont{T2A}{cmr}{m}{it} Второе решение.}
Пусть $\vec{r}_1$ --- радиус-вектор собаки, $\vec{r}_2$ --- зайца. Тогда
$$\vec{r}_1(t)=\vec{r}_0+\int\limits_0^t\vec{V}(\tau)\,d\tau$$
и $\vec{r}_2=\vec{u}t$, где $\vec{r}_0=(0,-L)$ и $\vec{u}=(u,0)$. По условию
задачи, скорость собаки $\vec{V}$ всегда параллельна вектору
$\vec{r}_2-\vec{r}_1$. Поэтому
\begin{equation}
\vec{r}_1-\vec{r}_2=\vec{r}_0-\vec{u}t+\int\limits_0^t\vec{V}(\tau)\,d\tau=
k(t)\,\vec{V}(t),
\label{eq3_4f}
\end{equation}
где коэффициент пропорциональности $k$ тоже зависит от времени. Продифференцируем 
(\ref{eq3_4f}) по времени, что дает
$$\dot{k}\vec{V}+k\dot{\vec{V}}=\vec{V}-\vec{u}.$$
Умножим это равенство скалярно на $\vec{V}$ и учтем, что
$$0=\frac{d\vec{V}^2}{dt}=2\vec{V}\cdot\dot{\vec{V}}.$$
В результате получим
$$\dot{k}V^2=V^2-\vec{u}\cdot\vec{V}=V^2-uV_x.$$
Интегрируем по времени от $t=0$ до времени встречи $t=T$  и учтем, что
$$\int\limits_0^T V_x(\tau)\,d\tau=uT.$$
Получаем
\begin{equation}
V^2[k(T)-k(0)]=(V^2-u^2)T.
\label{eq3_4g}
\end{equation}
Но $k(t)\vec{V}(t)=\vec{r}_1(t)-\vec{r}_2(t)$ показывает, что $k(T)=0$, так как
$\vec{V}(T)\ne 0$ и $\vec{r}_1(T)-\vec{r}_2(T)=0$. Кроме того, при $t=0$ имеем
$k(0)(0,V)=(0,-L)-(0,0)=(0,-L)$. Следовательно,
$$k(0)=-\frac{L}{V},\;\;\;k(T)=0.$$
Подставляя это в (\ref{eq3_4g}), получаем окончательно
$$T=\frac{LV}{V^2-u^2}.$$

{\usefont{T2A}{cmr}{m}{it} Третье решение} \cite{11,11A}.
Пусть $\vec{r}=\vec{r}_1-\vec{r}_2$ относительный радиус-вектор. Имеем
$$\frac{d}{dt}\left [\vec{r}\cdot (\vec{V}+\vec{u})\right ]=
(\vec{V}-\vec{u})\cdot (\vec{V}+\vec{u})+\vec{r}\cdot\dot{\vec{V}}=
V^2-u^2+\vec{r}\cdot\dot{\vec{V}}.$$
Но по условию задачи $\vec{r}=k\vec{V}$ и, следовательно, $\vec{r}\cdot
\dot{\vec{V}}=k\vec{V}\cdot\dot{\vec{V}}=0$, так как скорость собаки не меняется
по величине. Таким образом,
$$\frac{d}{dt}\left [\vec{r}\cdot (\vec{V}+\vec{u})\right ]=V^2-u^2.$$
Правая часть этого равенства не зависит от времени. Поэтому после интегрирования  
от $t=0$ до момента встречи $t=T$ получаем
\begin{equation}
T=\frac{\left .\left [\vec{r}\cdot (\vec{V}+\vec{u})\right]\right |_{t=0}}
{u^2-V^2}. 
\label{eq3_4h}
\end{equation}
Но 
$$\left .\left [\vec{r}\cdot (\vec{V}+\vec{u})\right]\right |_{t=0}=
(0,-L)\cdot ((0,V)+(u,0))=-LV$$
и (\ref{eq3_4h}) дает прежний ответ.

\subsection*{Задача 5 {\usefont{T2A}{cmr}{b}{n}(1.26)}: Радиус кривизны 
траектории}
{\usefont{T2A}{cmr}{m}{it} Первое решение.}
Так как $x=2t,\,y=t^2$, компоненты скорости и ускорения равны $V_x=2,\,
V_y=2t$ и $a_x=0,\,a_y=2$. Находим величины скорости и ускорения 
$V=\sqrt{V_x^2+V_y^2}=2\sqrt{1+t^2}$ и $a=\sqrt{a_x^2+a_y^2}=2$. 
Тангенциальное ускорение $$a_\tau=\frac{dV}{dt}=
\frac{2t}{\sqrt{1+t^2}}.$$ Нормальное  ускорение $$a_n=\sqrt{a^2-a_\tau^2}=
\frac{2}{\sqrt{1+t^2}}=\frac{V^2}{R},$$
где $R$ --- радиус кривизны траектории. Отсюда
$$R(t)=\frac{V^2}{a_n}=2(1+t^2)\sqrt{1+t^2}.$$
В частности, $R(0)=2$, и $R(2)=10\sqrt{5}$.

{\usefont{T2A}{cmr}{m}{it} Второе решение.}
\begin{figure}[htb]
\centerline{\epsfig{figure=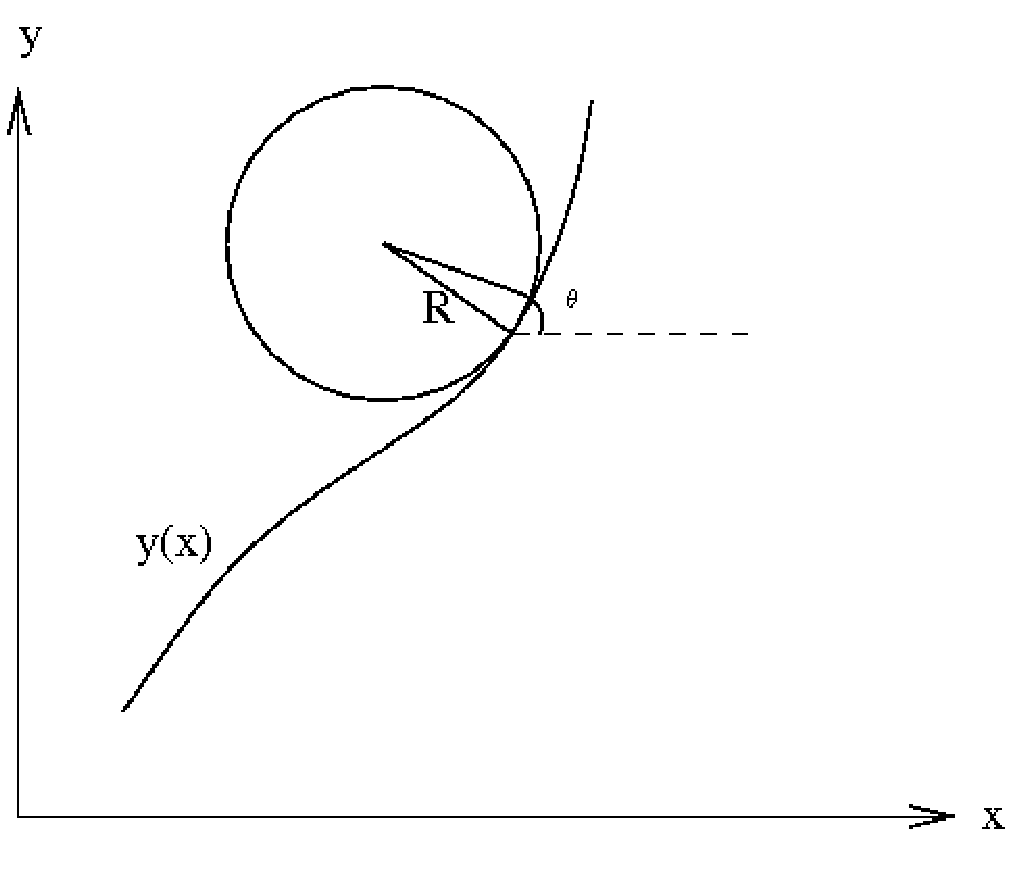,height=6cm}}
\end{figure}
Пусть плоская кривая задана в декартовых координатах уравнением $y=y(x)$.
Выведем формулу для радиуса кривизны этой кривой \cite{12}. Пусть 
$\theta$ --- угол наклона касательной. При инфинитизимальном сдвиге вдоль
кривой будем иметь (см. рисунок) $ds=R\,d\theta,$ где $ds=\sqrt{dx^2+dy^2}=
\sqrt{1+y^{\prime\,2}}\,dx$ --- элемент дуги вдоль кривой. Следовательно,
\begin{equation}
R=\frac{ds}{d\theta}=\frac{\sqrt{1+y^{\prime\,2}}}{\theta^\prime},
\label{eq3_5a}
\end{equation}
где штрих означает производную по $x$. Но $\tg{\theta}=y^\prime$, и поэтому
$$\theta^\prime=\frac{d}{dx}\left (\arctan{(y^\prime)}\right)=
\frac{y^{\prime\prime}}{1+y^{\prime\,2}}.$$
Подставляя это в (\ref{eq3_5a}), получаем окончательно
\begin{equation}
R=\frac{(1+y^{\prime\,2})\,\sqrt{1+y^{\prime\,2}}}{y^{\prime\prime}}.
\label{eq3_5b}
\end{equation}
В нашем случае $x=2t,\,y=t^2$ означает $y=x^2/4$. Поэтому $y^\prime=x/2$,
$y^{\prime\prime}=1/2$ и формула (\ref{eq3_5b}) дает 
$$R=\frac{\left(1+\frac{x^2}{4}\right )\sqrt{1+\frac{x^2}{4}}}{1/2}=
2(1+t^2)\sqrt{1+t^2}.$$

Формулу (\ref{eq3_5b}), с точностью до знака, можно получить еще следующим образом. 
Пусть уравнение соприкасающейся окружности имеет вид
\begin{equation}
(x-x_0)^2+(y-y_0)^2=R^2.
\label{eq3_5c}
\end{equation}
Дифференцируя по $x$, получаем в точке соприкосновения
\begin{equation}
x-x_0+y^\prime \,(y-y_0)=0.
\label{eq3_5d}
\end{equation}
Дифференцирование по $x$ еще раз дает
$$1+y^{\prime\, 2}+y^{\prime\prime}(y-y_0)=0.$$
Следовательно, в точке соприкосновения
\begin{equation}
y-y_0=-\frac{1+y^{\prime\, 2}}{y^{\prime\prime}}.
\label{eq3_5e}
\end{equation}
Подставляя это в  (\ref{eq3_5d}), получаем
\begin{equation}
x-x_0=\frac{y^{\prime}(1+y^{\prime\, 2})}{y^{\prime\prime}}.
\label{eq3_5f}
\end{equation}
Наконец, подставляя (\ref{eq3_5e}) и (\ref{eq3_5f}) в (\ref{eq3_5c}), будем иметь
$$R^2=\frac{(1+y^{\prime\, 2})^3}{y^{\prime\prime\,2}}.$$ 

\subsection*{Задача 6 {\usefont{T2A}{cmr}{b}{n}(1.38)}: Облет треугольника}
{\usefont{T2A}{cmr}{m}{it} Первое  решение.}
Вектора $\vec{A}$, $\vec{B}$ и $\vec{C}$ вдоль сторон треугольника удовлетворяют
равенству 
\begin{equation}
\vec{A}+\vec{B}+\vec{C}=0.
\label{eq3_6a}
\end{equation}
Поэтому
\begin{equation}
\vec{A}\cdot\vec{B}=\frac{1}{2}(C^2-A^2-B^2).
\label{eq3_6b}
\end{equation}
Скорости самолета при полете вдоль сторон треугольника равны
\begin{equation}
\vec{V_1}=\frac{\vec{A}}{t_1},\;\;\;\vec{V_2}=\frac{\vec{B}}{t_2},\;\;\;
\vec{V_3}=\frac{\vec{C}}{t_3}.
\label{eq3_6}
\end{equation}
Но $\vec{V_i}=\vec{u}+V\vec{n_i}$, где $\vec{u}$ --- скорость ветра, $V$ ---
величина скорости самолета относительно воздуха, и $\vec{n_i},\,i=1,2,3$ --- 
единичные вектора. Имеем
$$V^2=(\vec{V_1}-\vec{u})^2=V_1^2-2\vec{V_1}\cdot\vec{u}+u^2=
\left(\frac{A}{t_1}\right)^2-2\frac{\vec{A}\cdot\vec{u}}{t_1}+u^2,$$
и аналогично для $V_2$ и $V_3$. Эти равенства можно переписать следующим образом
\begin{eqnarray} &&
t_1(V^2-u^2)=\frac{A^2}{t_1}-2\vec{A}\cdot\vec{u},\nonumber \\ &&
t_2(V^2-u^2)=\frac{B^2}{t_2}-2\vec{B}\cdot\vec{u},\nonumber \\ &&
t_3(V^2-u^2)=\frac{C^2}{t_3}-2\vec{C}\cdot\vec{u}.
\label{eq3_6c}
\end{eqnarray}
Сложим их и учтем (\ref{eq3_6a}). В результате получим
\begin{equation}
V^2-u^2=\frac{1}{t_1+t_2+t_3}\left(\frac{A^2}{t_1}+\frac{B^2}{t_2}+
\frac{C^2}{t_3}\right).
\label{eq3_6d}
\end{equation}
С другой стороны, переписывая (\ref{eq3_6c}) в виде
\begin{eqnarray} &&
V^2=\frac{A^2}{t^2_1}-\frac{2}{t_1}\vec{A}\cdot\vec{u}+u^2,\nonumber \\ &&
V^2=\frac{B^2}{t^2_2}-\frac{2}{t_2}\vec{B}\cdot\vec{u}+u^2,\nonumber \\ &&
V^2=\frac{C^2}{t^2_3}-\frac{2}{t_3}\vec{C}\cdot\vec{u}+u^2,
\nonumber
\end{eqnarray}
получаем систему соотношений для определения $\vec{A}\cdot\vec{u}$ и 
$\vec{B}\cdot\vec{u}$: 
\begin{eqnarray} &&
\frac{A^2}{t^2_1}-\frac{B^2}{t^2_2}=2\left(\frac{\vec{A}\cdot\vec{u}}{t_1}-
\frac{\vec{B}\cdot\vec{u}}{t_1}\right),\nonumber \\ &&
\frac{A^2}{t^2_1}-\frac{C^2}{t^2_3}=2\left(\frac{\vec{A}\cdot\vec{u}}{t_1}-
\frac{\vec{C}\cdot\vec{u}}{t_3}\right)=
2\left(\frac{\vec{A}\cdot\vec{u}}{t_1}+\frac{\vec{A}\cdot\vec{u}}{t_3}+
\frac{\vec{B}\cdot\vec{u}}{t_3}\right ).
\label{eq3_6e}
\end{eqnarray}
Решением системы (\ref{eq3_6e}) является
\begin{eqnarray} &&
\vec{A}\cdot\vec{u}=\frac{1}{2(t_1+t_2+t_3)}\left (\frac{t_2+t_3}{t_1}\,A^2-
\frac{t_1}{t_2}\,B^2-\frac{t_1}{t_3}\,C^2\right ),\nonumber \\ &&
\vec{B}\cdot\vec{u}=\frac{1}{2(t_1+t_2+t_3)}\left (\frac{t_1+t_3}{t_2}\,B^2-
\frac{t_2}{t_1}\,A^2-\frac{t_2}{t_3}\,C^2\right ).
\label{eq3_6f}
\end{eqnarray}
Знание $\vec{A}\cdot\vec{u}$ и $\vec{B}\cdot\vec{u}$ позволяет определить величину
скорости ветра $u$. Действительно, $\vec{u}$ можно разложить по взаимно 
перпендикулярным векторам
$$\vec{e}_1=\vec{A},\;\;\;\vec{e}_2=\vec{B}-\frac{\vec{A}\cdot\vec{B}}{A^2}\,
\vec{A}$$
и получить
\begin{equation}
u^2=\frac{(\vec{u}\cdot\vec{e}_1)^2}{\vec{e}_1\cdot\vec{e}_1}+
\frac{(\vec{u}\cdot\vec{e}_2)^2}{\vec{e}_2\cdot\vec{e}_2}=
\frac{A^2(\vec{u}\cdot\vec{B})^2+B^2(\vec{u}\cdot\vec{A})^2-
2\vec{A}\cdot\vec{B}\,\vec{u}\cdot\vec{A}\,\vec{u}\cdot\vec{B}}
{A^2B^2-(\vec{A}\cdot\vec{B})^2}.
\label{eq3_6g}
\end{equation}
Зная $u^2$, из (\ref{eq3_6d}) можно найти скорость самолета $V$ относительно 
воздуха. 

{\usefont{T2A}{cmr}{m}{it} Второе решение} \cite{13}.
\begin{figure}[htb]
\centerline{\epsfig{figure=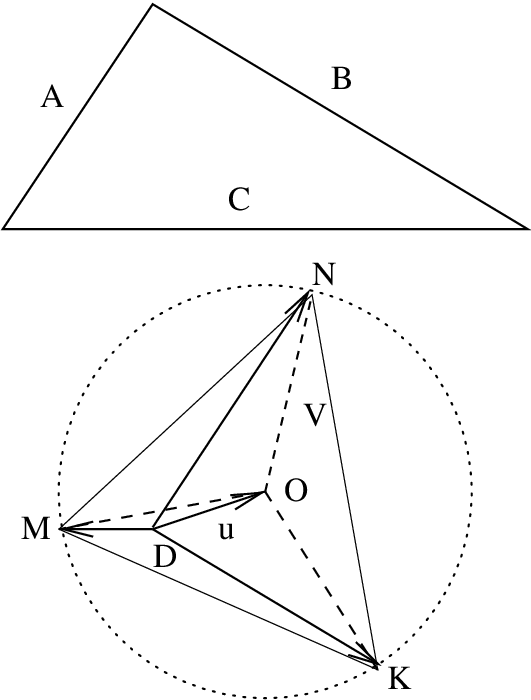,height=6cm}}
\end{figure}
Рассмотрим векторную диаграмму скоростей самолета при полете вдоль сторон треугольника.
На рисунке $\vec{DO}=\vec{u}$ --- скорость ветра, $\vec{DN}=\vec{V}_1$,
$\vec{DK}=\vec{V}_2$ и $\vec{DM}=\vec{V}_3$ --- скорости самолета при полете 
вдоль сторон треугольника относительно земли,  эти скорости параллельны сторонам 
треугольника $A$, $B$ и $C$. $\vec{ON}=V\vec{n}_1$, $\vec{OK}=V\vec{n}_2$ 
и $\vec{OM}=V\vec{n}_3$ --- скорости самолета при полете вдоль сторон треугольника 
относительно воздуха. Как видно на рисунке, величина скорости самолета относительно 
воздуха $V$ равна радиусу окружности, описанной вокруг треугольника $MNK$. 
Следовательно, по известной формуле из геометрии,
\begin{equation}  
V=\frac{abc}{\sqrt{(a+b+c)(a+b-c)(b+c-a)(c+a-b)}},
\label{eq3_6h}
\end{equation}
где $a=MN$, $b=NK$, $c=MK$ --- стороны треугольника $MNK$. Найдем их. Так как
угол между векторами $\vec{V}_1$ и $\vec{V}_3$ равен $\pi-\alpha$, где $\alpha$
--- это угол между сторонами $A$ и $C$ исходного треугольника, то
\begin{equation} 
a^2=V_1^2+V_3^2-2V_1V_3\cos{(\pi-\alpha)}=V_1^2+V_3^2+2V_1V_3\cos{\alpha}.
\label{eq3_6k}
\end{equation}
Но из $\vec{A}+\vec{C}=-\vec{B}$ следует, что $A^2+C^2-2AC\cos{\alpha}=B^2$.
Определяя из этого соотношения $\cos{\alpha}$ и вспоминая (\ref{eq3_6}), получаем
из (\ref{eq3_6k}):
\begin{equation}
a^2=\frac{A^2}{t_1^2}+\frac{C^2}{t_3^2}-\frac{B^2-A^2-C^2}{t_1t_3}.
\label{eq3_6l}
\end{equation}
Это определяет $a$. Аналогично находим
\begin{eqnarray} && 
b^2=\frac{A^2}{t_1^2}+\frac{B^2}{t_2^2}-\frac{C^2-A^2-B^2}{t_1t_2},\nonumber 
\\ &&
c^2=\frac{B^2}{t_2^2}+\frac{C^2}{t_3^2}-\frac{A^2-B^2-C^2}{t_2t_3}.
\label{eq3_6m}
\end{eqnarray}
Зная $a,b,c$, по формуле (\ref{eq3_6h}) находим скорость самолета $V$, а потом по 
формуле (\ref{eq3_6d}) --- скорость ветра $u$.

\section[Семинар 4]
{\centerline{Семинар 4}}

\subsection*{Задача 1 {\usefont{T2A}{cmr}{b}{n}(1.23)}: Траектория точки}
Декартовы координаты точки имеют вид $$x(t)=r(t)\cos{\varphi(t)}=\frac{b}{t}
\cos{(\gamma t)},\;\;\;y(t)=r(t)\sin{\varphi(t)}=\frac{b}{t}
\sin{(\gamma t)}.$$
Отсюда $$x^2+y^2=\frac{b^2}{t^2}\;\;\mbox{и}\;\;\frac{y}{x}=
\tg{(\gamma t)}.$$
Из второго уравнения $$t=\frac{1}{\gamma}\arctan{\left(\frac{y}{x}\right)}$$
подставим в первое
$$x^2+y^2=\frac{b^2\gamma^2}{\left(\arctan{\left(\frac{y}{x}\right)}
\right)^2}.$$ Отсюда
$$\arctan{\left(\frac{y}{x}\right)}=\sqrt{\frac{b^2\gamma^2}{x^2+y^2}}.$$
Это и есть уравнение траектории в декартовых координатах. Заметим, что
$$x=\frac{b}{t}\cos{(\gamma t)}\to\infty,\;\;\;
y=\frac{b}{t}\sin{(\gamma t)}\to b\gamma$$
при $t\to 0$. Поэтому частица приходит справа из бесконечности по
горизонтальной асимптоте и падает на центр по спирали, делая при этом
бесконечно много поворотов вокруг него. Примерный вид траектории показан на
рисунке.
\begin{figure}[htb]
\centerline{\epsfig{figure=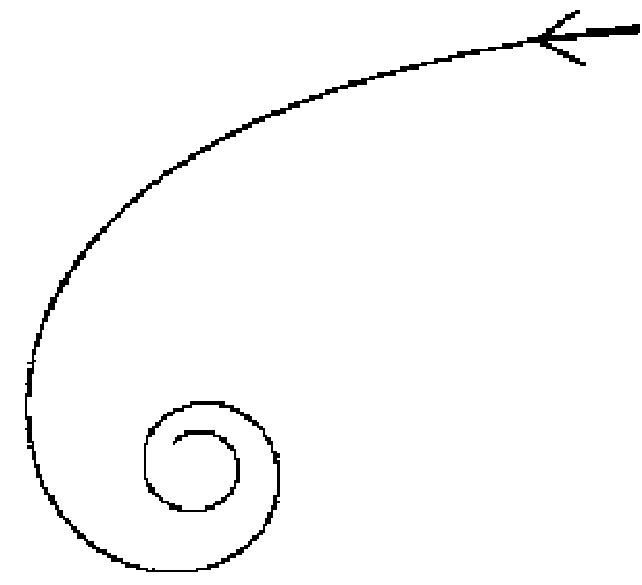,height=4cm}}
\end{figure}

\subsection*{Задача 2 {\usefont{T2A}{cmr}{b}{n}(1.25)}: Траектория, скорость, 
ускорение и радиус кривизны траектории}
{\usefont{T2A}{cmr}{m}{it} Первое решение.}
Так как $r=ae^{kt},\,\varphi=kt$, то уравнение траектории в полярной системе
координат имеет вид $r=ae^\varphi$. В этой системе
$\vec{V}=\dot{r}\,\vec{e}_r+r\dot{\varphi}\,\vec{e}_\varphi$ и
$\vec{a}=(\ddot{r}-r\dot{\varphi}^2)\,\vec{e}_r+(r\ddot{\varphi}+
2\dot{r}\dot{\varphi})\,\vec{e}_\varphi$. Поэтому
$$V^2=\dot{r}^2+(r\dot{\varphi})^2=a^2k^2e^{2kt}=2k^2r^2.$$
Следовательно, $V=\sqrt{2}kr$. Затем, используя $\dot{r}=kr$,
$\ddot{r}=k^2r$, $\dot{\varphi}=k$, $\ddot{\varphi}=0$, находим
$$a_r=\ddot{r}-r\dot{\varphi}^2=k^2r-k^2r=0,\;\;
a_\varphi=r\ddot{\varphi}+2\dot{r}\dot{\varphi}=2k^2r.$$
Следовательно, $a=2k^2r$. Находим тангегциальное ускорение
$a_\tau=\dot{V}=\sqrt{2}k\dot{r}=\sqrt{2}k^2r$, нормальное ускорение
$a_n=\sqrt{a^2-a_\tau^2}=\sqrt{2}k^2r$ и радиус кривизны
$$R=\frac{V^2}{a_n}=\sqrt{2}r.$$

{\usefont{T2A}{cmr}{m}{it} Второе решение.}
Несколько проще $V$ и $a$ можно вычислить следующим образом.  Пусть 
$z=x+iy$. Тогда $\dot{z}=V_x+iV_y$ и $\ddot{z}=a_x+ia_y$. Поэтому
$V=|\dot{z}|$ и $a=|\ddot{z}|$. Но 
$$x+iy=re^{i\varphi}=ae^{kt}\,e^{ikt}=ae^{(1+i)kt}.$$
Поэтому $\dot{z}=(1+i)kz$ и $V=k|1+i|\,|z|=\sqrt{2}kr$, так как
$|1+i|=\sqrt{2}$ и $|z|=r$. Аналогично, $\ddot{z}=(1+i)^2k^2z$ и 
$a=k^2|1+i|^2|z|=2k^2r$.

\subsection*{Задача 3 {\usefont{T2A}{cmr}{b}{n}(1.40)}: Преобразования Галилея}
Пусть система $S^\prime$ движется со скоростью $\vec{V}$ относительно системы 
$S$. Преобразования Галилея 
\begin{eqnarray} &&
\vec{r}_i^{\,\prime}=\vec{r}_i-\vec{V}t,\nonumber \\ &&
\vec{v}_i^{\,\prime}=\vec{v}_i-\vec{V} \nonumber
\end{eqnarray}
показывают, что импульс системы частиц $\vec{p}=\sum_i m_i\vec{v}_i$
преобразуется следующим образом:
$$\vec{p}^{\,\prime}=\sum_i m_i\vec{v}_i^{\,\prime}=
\sum_i m_i(\vec{v}_i-\vec{V})=
\vec{p}-\vec{V}\sum_im_i=\vec{p}-M\vec{V},$$
где $M=\sum_i m_i$. Аналогично получаем закон преобразования кинетической 
энергии,
$$E^\prime=\sum_i\frac{1}{2}m_i\vec{v}_i^{\,\prime\,2}=
\sum_i\frac{1}{2}m_i(\vec{v}_i^2-2\vec{v}_i\cdot\vec{V}+\vec{V}^2)=
E+\frac{1}{2}MV^2-\vec{p}\cdot\vec{V}.$$
Если в системе отсчета $S$ импульс системы частиц равен нулю, $\vec{p}=0$, 
то $$E^\prime=E+\frac{1}{2}MV^2>E.$$
Т. е. кинетическая энергия системы частиц минимальна в системе покоя, где
$\vec{p}=\sum_i\vec{p}_i=0$.

\subsection*{Задача 4 {\usefont{T2A}{cmr}{b}{n}(1.41)}: Кинетическая энергия 
гусеницы}
В системе трактора обе половинки гусеницы имеют одинаковую скорость $V$,
поэтому в этой системе кинетическая энергия гусеницы равна
$E^\prime=\frac{1}{2}MV^2$, где $M$ --- масса гусеницы. В системе Земли
верхняя половинка гусеницы движется со скоростью $2V$, а нижняя покоится.
Поэтому в этой системе кинетическая энергия гусеницы равна $E=\frac{1}{2}
(\frac{M}{2})(2V)^2=MV^2$.

\subsection*{Задача 5 {\usefont{T2A}{cmr}{b}{n}(1.28)}: Через какое время 
собаки догонят друг друга?}
Из симметрии условия задачи ясно, что собаки всегда находятся в вершинах 
некоторого квадрата. При этом каждая собака приближается к центру со скоростью
$V\cos{45^\circ}=V/\sqrt{2}$. Первоначально расстояние до центра равно
$a/\sqrt{2}$. Поэтому собаки встретятся через время
$$t=\frac{a/\sqrt{2}}{V/\sqrt{2}}=\frac{a}{V}.$$
До встречи каждая собака проходит путь $s=Vt=a$.

Траекторию собак можно найти так. В полярной системе координат имеем (для
указанного на рисунке к задаче направления движения собак)
$$V_\rho=\dot{\rho}=-V\cos{45^\circ}=-V/\sqrt{2},\;\;
V_\varphi=\rho\dot{\varphi}=-V\sin{45^\circ}=-V/\sqrt{2}.$$
Деля первое уравнение на второе, получаем дифференциальное уравнение 
траектории $$\frac{1}{\rho}\frac{d\rho}{d\varphi}=1.$$
Можно разделить переменные и интегрировать, что дает $\ln{\rho/\rho_0}=
\varphi-\varphi_0$, или
$$\rho=\rho_0e^{\varphi-\varphi_0},$$
где $\rho_0=a/\sqrt{2}$, а $\varphi_0$ есть начальное значение
$\varphi$-координаты для данной собаки. Заметим, что $\varphi\le\varphi_0$.

Поучительно получить пройденную собаками путь как длину бесконечной спирали:
$$s=\int\limits_{-\infty}^{\varphi_0}\sqrt{\rho^2+\left(\frac{d\rho}
{d\varphi}\right )^2}\,d\varphi=\sqrt{2}\rho_0\int\limits_{-\infty}^{\varphi_0}
e^{\varphi-\varphi_0}\,d\varphi=\sqrt{2}\rho_0=a.$$

\subsection*{Задача 6: Сферический треугольник}
Сферические координаты городов:
Новосибирск -- $\theta=35^\circ$, $\phi=83^\circ$; Рио-де-Жанейро -- 
$\theta=113^\circ$, $\phi=317^\circ$; Нью-Йорк -- $\theta=50^\circ$, 
$\phi=286^\circ$. Угол между векторами $\vec{r}_1$ и  $\vec{r}_2$ определяется 
выражением  $$\cos{\theta_{12}}=\cos{\theta_1}\cos{\theta_2}+
\sin{\theta_1}\sin{\theta_2}\cos{(\phi_1-\phi_2)}.$$
Пусть Новосибирск -- 1, Рио-де-Жанейро -- 2, Нью-Йорк -- 3. Находим 
$$\cos{\theta_{12}}\approx -0,6304,\;\;\cos{\theta_{23}}\approx 0,3533,
\;\;\cos{\theta_{13}}\approx 0,1221.$$
Или $\theta_{12}\approx 2,253, \;\;\theta_{23}\approx 1,210,\;\;
\theta_{13}\approx 1,448$. Радиус Земли $R=6400~\mbox{км}$. Находим расстояния 
$d_{12}=R\theta_{12}\approx 14420~\mbox{км},\;\; d_{23}\approx 7740~\mbox{км},
\;\;d_{13}\approx 9270 ~\mbox{км}$ и их сумму $d=d_{12}+d_{23}+d_{13}
\approx 31430 ~\mbox{км}$.

Углы сферического треугольника удобно найти с помощью теоремы косинусов сферической 
тригонометрии, которую мы сейчас докажем. Пусть вершины треугольника $A$,$B$ и $C$.
Сферические углы при этих вершинах обозначим также, а угловые величины (как дуг 
окружностей) противоположных сторон -- $a$,$b$ и $c$. Выберем систему координат с 
центром в центре сферы так, чтобы вершина $A$ лежала на оси $z$, а вершина $B$ --
в плоскости $xOz$. Тогда азимутальный угол для вершины $C$ как раз равен углу $A$
сферического треугольника. Радиус-векторы точек $B$ и $C$ будут
$\vec{r}_B=R\sin{c}\;\vec{i}+R\cos{c}\;\vec{k}$ и  $\vec{r}_C=R\sin{b}\cos{A}
\;\vec{i}+R\sin{b}\sin{A}\;\vec{j}+R\cos{b}\;\vec{k}$. Отсюда
$$R^2\cos{a}=\vec{r}_B\cdot\vec{r}_C=R^2[\cos{b}\cos{c}+\sin{b}\sin{c}
\cos{A}]$$
и получаем теорему косинусов сферической тригонометрии:
$$\cos{a}=\cos{b}\cos{c}+\sin{b}\sin{c}\cos{A}.$$ 
С ее помощью находим косинусы углов сферического треугольника
$$\cos{\alpha_1}=\frac{\cos{\theta_{23}}-\cos{\theta_{12}}\cos{\theta_{13}}}
{\sin{\theta_{12}}\sin{\theta_{13}}}\approx 0,5584,\;\;
\cos{\alpha_2}=\frac{\cos{\theta_{13}}-\cos{\theta_{12}}\cos{\theta_{23}}}
{\sin{\theta_{12}}\sin{\theta_{23}}}\approx 0,4748,$$ $$
\cos{\alpha_3}=\frac{\cos{\theta_{12}}-\cos{\theta_{13}}\cos{\theta_{23}}}
{\sin{\theta_{13}}\sin{\theta_{23}}}\approx -0,7254,$$
сами углы $\alpha_1\approx 0,978,\;\; \alpha_2\approx 1,076,\;\;
\alpha_3\approx 2,3824$ и их сумму $\alpha_1+\alpha_2+\alpha_3\approx
4,437\approx 254^\circ$.

\section[Семинар 5]
{\centerline{Семинар 5}}

\subsection*{Задача 1 {\usefont{T2A}{cmr}{b}{n}(1.42)}: Скорость поезда}
По эффекту Доплера, при приближении принимаемая частота будет ($c$ --- скорость
звука)
$$\nu^\prime=\frac{\nu}{1-\frac{V}{c}}.$$
При удалении
$$\nu^{\prime\prime}=\frac{\nu}{1+\frac{V}{c}}.$$
По условию задачи
$$\alpha=\frac{\nu^\prime}{\nu^{\prime\prime}}=\frac{1+\frac{V}{c}}
{1-\frac{V}{c}}.$$
Решая относительно $V$, находим
$$V=\frac{\alpha-1}{\alpha+1}\,c.$$

\subsection*{Задача 2 {\usefont{T2A}{cmr}{b}{n}(1.42)}: Два встречных поезда}
Перейдем в систему первого поезда. Тогда источник звука приближается к нему со 
скоростью $V_1+V_2$. Поэтому, по эффекту Доплера, частота звука, принимаемого первым 
машинистом, будет $$\nu_1=\frac{\nu}{1-\frac{V_1+V+2}{c_1}},$$
где $c_1=c+V_1$ --- скорость звука в системе первого поезда, когда звук движется
навстречу. Следовательно,
$$\nu_1=\frac{c+V_1}{c-V_2}\,\nu\;\;\mbox{и}\;\; \alpha=\frac{\nu_1}{\nu}=
\frac{c+V_1}{c-V_2}.$$
Аналогично
$$\beta=\frac{c+V_2}{c-V_1}.$$
Решая систему 
\begin{eqnarray}&&
V_1+\alpha V_2=(\alpha-1)c,\nonumber\\ &&
\beta V_1+V_2=(\beta-1)c,\nonumber
\end{eqnarray}
находим скорости поездов:
$$V_1=\frac{\alpha \beta+1-2\alpha}{\alpha \beta-1}\,c,\;\;
V_2=\frac{\alpha \beta+1-2\beta}{\alpha \beta-1}\,c.$$

\subsection*{Задача 3 {\usefont{T2A}{cmr}{b}{n}(1.53)}: Поступательное и 
угловое ускорение шестеренки}
В точке контакта шестеренка и рейка имеют все время одинаковую скорость.
Следовательно, их ускорения в этой точке тоже совпадают. Пусть поступательное
ускорение шестеренки $a$, а угловое ускорение $\epsilon$. Тогда должны иметь
(положительным направлением для вращения считаем по часовой стрелке, а для
поступательного движения --- вправо)
\begin{eqnarray} &&
a+\epsilon R=a_2,\nonumber \\ &&
a-\epsilon R=-a_1. \nonumber
\end{eqnarray}
Решая эту систему, находим
$$a=\frac{a_2-a_1}{2}=0,5~\mbox{м/с}^2,\;\;\;\epsilon=\frac{a_1+a_2}{2R}=
4~\mbox{с}^{-2}.$$

\subsection*{Задача 4 {\usefont{T2A}{cmr}{b}{n}(2.6)}: Реальная скорость объекта}
\begin{figure}[htb]
\centerline{\epsfig{figure=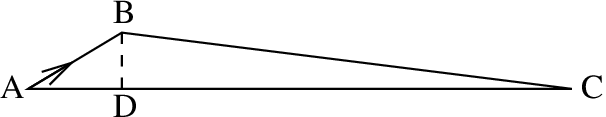,height=1.5cm}}
\end{figure}
Пусть объект движется со скоростью $V$ под углом $\angle BAC=\alpha$ к линии
наблюдения. Видимое с Земли перемещение за $T=5$~лет будет $L=BD=10$~св.
лет. Пусть свет от точки $A$ приходит на Землю (точка $C$) в момент времени
$t_A=0$. Тогда это означает, что он излучился в момент времени
$t^\prime_A=-AC/c$, $c$ --- скорость света. Объект в точке $B$ окажется в
момент $t^\prime_B=t^\prime_A+AB/V=t^\prime_A+L/(V\sin{\alpha})$, так как
$AB=BL/sin{\alpha}=L/\sin{\alpha}$. Свет излученный объектом в точке $B$ в
момент времени $t^\prime_B$ на Землю придет в момент
$t_B=t^\prime_B+BC/c\approx t^\prime_B+DC/c$, так как $BC\approx DC$.
Следовательно, $$t_B\approx -\frac{AC}{c}+\frac{L}{V\sin{\alpha}}+
\frac{DC}{c}=\frac{L}{V\sin{\alpha}}-\frac{AD}{c}.$$
Но $AD=BD\,\cot{\alpha}=L\,\cot{\alpha}$, т. е.
$$t_B\approx \frac{L}{V\sin{\alpha}}-\frac{L\,\cot{\alpha}}{c}.$$
По условию задачи, $T=t_B-t_A$. Поэтому получаем уравнение
$$\frac{L}{V\sin{\alpha}}-\frac{L\,\cot{\alpha}}{c}=T,$$
решением которого является
\begin{equation}
V=\frac{c}{\cos{\alpha}+\frac{cT}{L}\sin{\alpha}}.
\label{eq5_4}
\end{equation}
Требуется $V<c$, т. е. 
$$\cos{\alpha}+\frac{cT}{L}\sin{\alpha}>1.$$
Используя $\sin{\alpha}=2\sin{\frac{\alpha}{2}}\cos{\frac{\alpha}{2}}$ и 
$1-\cos{\alpha}=2\sin^2{\frac{\alpha}{2}}$, можно решить неравенство и
получить $$\tg{\frac{\alpha}{2}}<\frac{cT}{L}=\frac{5}{10}=\frac{1}{2}.$$
Следовательно, $\alpha<2\,\arctan{0,5}\approx 53,1^\circ$.

$V(\alpha)$ функция минимальна в точке, где $\frac{dV}{d\alpha}=0$. Но 
$$\frac{dV}{d\alpha}=-\frac{c}{\left (\cos{\alpha}+\frac{cT}{L}\sin{\alpha}
\right)^2}\,\left (-\sin{\alpha}+\frac{cT}{L}\cos{\alpha}\right ).$$
Поэтому точка минимума удовлетворяет
$$-\sin{\alpha_m}+\frac{cT}{L}\cos{\alpha_m}=0,\;\;\mbox{или}\;\;
\tg{\alpha_m}=\frac{cT}{L}=\frac{1}{2}.$$
Подставляя это в (\ref{eq5_4}), получаем минимальное значение скорости
$$V_m=\frac{c}{\cos{\alpha_m}}\,\frac{1}{1+\frac{cT}{L}\tg{\alpha_m}}=
\frac{\sqrt{1+\tg^2{\alpha_m}}}{1+\frac{cT}{L}\tg{\alpha_m}}\,c=
\frac{2c}{\sqrt{5}}\approx 0,89\,c.$$

\subsection*{Задача 5 {\usefont{T2A}{cmr}{b}{n}(2.7)}: Время обращения
спутника Ио}
{\usefont{T2A}{cmr}{m}{it} Первое решение.}
Пусть $\omega_1$ и $\omega_2$ --- угловые скорости обращения Земли и Юпитера
вокруг Солнца, $r_1$ и $r_2$ --- радиусы их орбит. Свяжем с Солнцем систему
координат, которая вращается с угловой скоростью $\omega_2$. В этой системе
Юпитер стоит на месте, а Земля  вращается с угловой скоростью
$\omega_1-\omega_2$ по окружности радиуса $r_1$ (см. рисунок). 
\begin{figure}[htb]
\centerline{\epsfig{figure=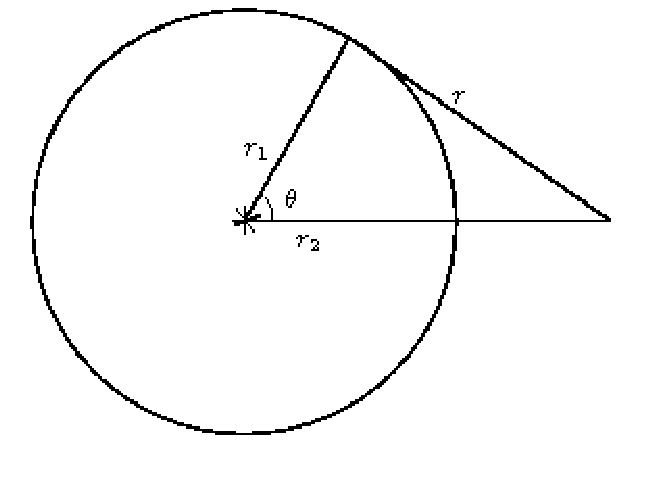,height=4cm}}
\end{figure}

\noindent Расстояние между  Юпитером и Землей
$$r=\sqrt{r_1^2+r_2^2-2r_1r_2\cos{\theta}}=
\sqrt{r_1^2+r_2^2-2r_1r_2\cos{(\omega_1-\omega_2)t}}.$$
Но $r_2\approx 5r_1$. Поэтому в первом порядке по точности по малому
параметру $r_1/r_2$ будем иметь 
\begin{equation}
r\approx r_2\left (1-\frac{r_1}{r_2}\cos{\theta}\right)=
r_2-r_1\,\cos{(\omega_1-\omega_2)t}.
\label{eq5_5a}
\end{equation}
Т. е. расстояние меняется по гармоническому закону. Как повляет это
обстоятельство на видимый период обращения Ио?
\begin{figure}[htb]
\centerline{\epsfig{figure=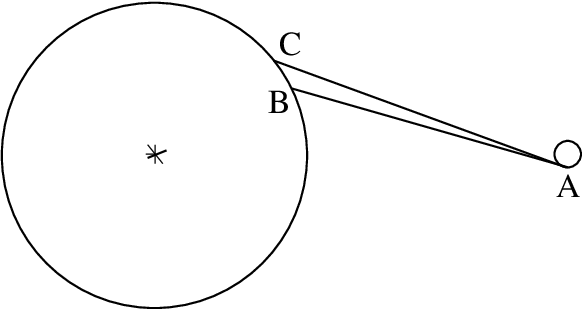,height=4cm}}
\end{figure}

Пусть в момент времени $t=t_B$ мы видим Ио около края диска Юпитера (в точке $A$). 
Этот свет излучился в момент времени $t_B^\prime=t_B-AB/c$, $c$ --- скорость
света. После одного оборота в момент времени $t_C^\prime=t_B^\prime+T_0$,
где $T_0$ --- истинный период обращения Ио, Ио опять окажется в том же
месте, но свет, испущенный им, дойдет до Земли в момент $t_C=t_C^\prime+AC/c$.
Поэтому видимый период обращения Ио будет
\begin{equation}
T=t_C-t_B=\frac{AC}{c}-\frac{AB}{c}+T_0=T_0+\frac{r(t_C)-r(t_B)}{c}\approx
T_0+\frac{T}{c}\left .\frac{dr}{dt}\right |_{t_B},
\label{eq5_5b}
\end{equation}
так как $$r(t_C)-r(t_B)\approx \left .\frac{dr}{dt}\right |_{t_B}(t_C-t_B).$$
Но, cогласно (\ref{eq5_5a}),  $$\frac{dr}{dt}=r_1(\omega_1-\omega_2)\,
\sin{(\omega_1-\omega_2)t}\approx V_1\,\sin{(\omega_1-\omega_2)t},$$
где $V_1=r_1\omega_1$ --- орбитальная скорость Земли, и мы учли, что
$\omega_1\approx 12\omega_2\gg \omega_2$. Следовательно, (\ref{eq5_5b})
дает $$T\left(1-\frac{V_1}{c}\,\sin{(\omega_1-\omega_2)t}\right )=T_0.$$
Отсюда $$T=\frac{T_0}{1-\frac{V_1}{c}\,\sin{(\omega_1-\omega_2)t}}\approx
T_0\left (1+\frac{V_1}{c}\,\sin{(\omega_1-\omega_2)t}\right ).$$
Заметим, что $T_0\frac{V_1}{c}\approx 15$ секунд.

{\usefont{T2A}{cmr}{m}{it} Второе решение.}
Наблюдаемый период обращения спутника Ио отличается от истинного периода из-за 
эффекта Доплера и дается формулой
$$T=T_0\left(1-\frac{V}{c}\cos{\alpha}\right),$$
где $V=(\omega_1-\omega_2)r_2$ --- скорость Юпитера по отношению к земному 
наблюдателю, $\alpha$ --- угол между скоростью  Юпитера и линией наблюдения (см.
рисунок).  
\begin{figure}[htb]
\centerline{\epsfig{figure=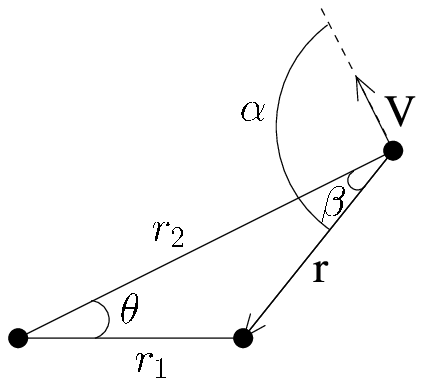,height=3cm}}
\end{figure}

\noindent Но $\alpha=\frac{\pi}{2}+\beta$ и из теоремы синусов
$$\frac{r_1}{\sin{\beta}}=\frac{r}{\sin{\theta}},$$
где $\theta=(\omega_1-\omega_2)t$, а $r$ дается выражением (\ref{eq5_5a}).
Поэтому 
$$\cos{\alpha}=-\sin{\beta}=-\frac{r_1}{r_2}\,\sin{\theta},$$
и
$$T=T_0\left(1+\frac{V}{c}\,\frac{r_1}{r_2}\,\sin{\theta}\right)=
T_0\left(1+\frac{(\omega_1-\omega_2)r_1}{c}\,\sin{(\omega_1-\omega_2)t}
\right )\approx T_0\left (1+\frac{V_1}{c}\,\sin{(\omega_1-\omega_2)t}
\right ).$$

\subsection*{Задача 6: Преобразования Лоренца \cite{14}}
Пусть инерциальная системы отсчета $S^\prime$ движется вдоль оси  $x$ со скоростью 
$V$ относительно ``неподвижной'' системы $S$.
\begin{figure}[!h]
\centerline{\epsfig{figure=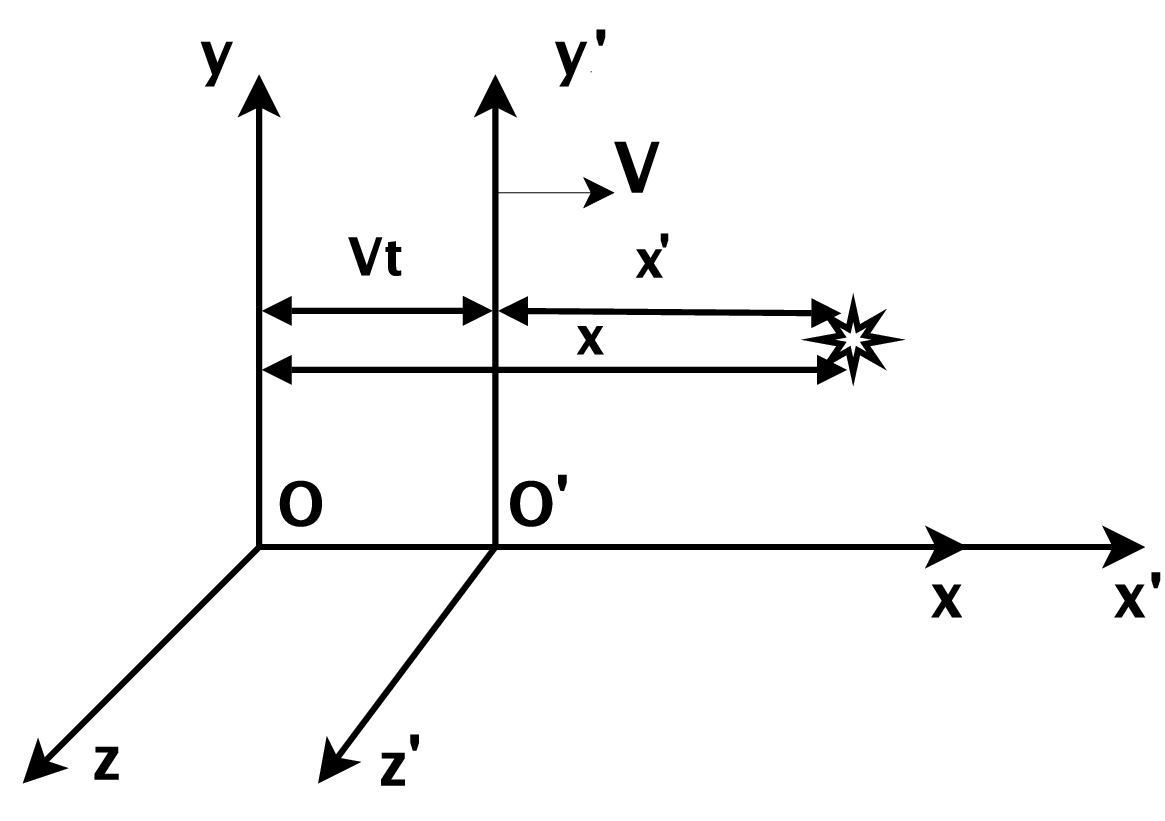,height=5cm}}
\end{figure}

Простого взгляда на приведенный рисунок достаточно, чтобы написать преобразование 
Галилея, связывающее $x$-координаты какого-либо события (например, вспышки света) 
в системах $S$ и $S^\prime$:
\begin{equation}
x=Vt+x^\prime.
\label{eq5_6a}
\end{equation}
Но в (\ref{eq5_6a}) мы неявно предполагали, что измерительная линейка не меняет свою 
длину, если ее плавно привести в равномерном движении. Хотя это весьма привлекательное
предположение, его правильность не совсем очевидна. Например, в соответствии с 
уравнениями Максвелла, заряды в покое взаимодействуют только через кулоновское поле, 
тогда как во время движения они испытывают также магнитное взаимодействие. Кроме того, 
когда источник движется очень быстро, его электрическое поле больше не сферически
симметрично. Поэтому разумно ожидать, что совсем не ислючено, что измерительная 
линейка при быстром движении изменит форму, если электромагнитные силы играют важную 
роль в обеспечении внутреннего равновесия материи \cite{15}. Так или иначе, мы просто 
примем эту более общую возможность и измененим (\ref{eq5_6a}) следующим образом:
\begin{equation}
x=Vt+k(V^2)\,x^\prime,
\label{eq5_6b}
\end{equation}
где масштабный коэффициент $k(V^2)$ учитывает возможное изменение в длине 
измерительной линейки.

Принцип относительности и изотропность пространства присутствуют  в (\ref{eq5_6b}) 
в неявной форме, потому что подразумевается, что масштабный коэффициент зависит только 
от величины относительной скорости $V$.

Уравнение (\ref{eq5_6b}) позволяет выразить штрихованные  координаты через 
нештрихованные:
\begin{equation}
x^\prime=\frac{1}{k(V^2)}\left (x-Vt\right ).
\label{eq5_6c}
\end{equation}
Тогда принцип относительности говорит нам, что такое же соотношение имеет место, если
нештрихованные координаты выражаются через штрихованные, с заменой $V$ на $-V$,
потому что скорость $S$ относительно $S^\prime$ есть $-V$. Поэтому
$$x=\frac{1}{k(V^2)}\left (x^\prime+Vt^\prime\right )=
\frac{1}{k(V^2)}\left [\frac{1}{k(V^2)}\left (x-Vt\right )+ Vt^\prime
\right ].$$
Решая относительно $t^\prime$, получаем
\begin{equation}
t^\prime =\frac{1}{k(V^2)}\left [t-\frac{1-k^2(V^2)}{V}\,x\right ].
\label{eq5_6d}
\end{equation}
Сразу видно, что время не является абсолютным, если масштабный коэффициент $k(V^2)$ 
не тождественно единице.

Из (\ref{eq5_6c}) и (\ref{eq5_6d}) можно вывести правило сложения скоростей:
$$v^\prime_x=\frac{dx^\prime}{dt^\prime}=\frac{dx-Vdt}{dt-\frac{1-k^2}{V}
\,dx}=\frac{v_x-V}{1-\frac{1-k^2}{V}\,v_x}.$$
В дальнейшем будет удобнее работать с правилом сложения скоростей, которое выражает 
нештрихованные скорости через штрихованные:
\begin{equation}
v_x=\frac{v^\prime_x+V}{1+\frac{1-k^2}{V}\,v^\prime_x}\equiv F(v^\prime_x,V).
\label{eq5_6e}
\end{equation}
Если изменим знаки обеих скоростей $v^\prime_x$ и $V$, очевидно, что знак 
результирующей скорости $v_x $ также поменяется. Поэтому $F$ должна быть нечетной 
функцией своих аргументов
\begin{equation}
F(-x,-y)=-F(x,y).
\label{eq5_6f}
\end{equation}
Рассмотрим теперь три тела $A$, $B$ и $C$ в относительном движении. Пусть $V_{AB}$
oбозначает скорость $A$ относительно $B$. Тогда
\begin{equation}
V_{BA}=-V_{AB}.
\label{eq5_6g}
\end{equation}
Этот принцип взаимности уже использовали выше. Используя  (\ref{eq5_6f}) и 
(\ref{eq5_6g}), получаем \cite{16}
$$F(V_{CB},V_{BA})=V_{CA}=-V_{AC} $$
$$=-F(V_{AB},V_{BC})=-F(-V_{BA},-V_{CB})=
F(V_{BA},V_{CB}). $$
Поэтому $F$ является симметричной функцией своих аргументов. Тогда $F(v^\prime_x,V)
=F(V,v^\prime_x)$ сразу дает, согласно (\ref{eq5_6e}), 
 $$\frac{1-k^2(V^2)}{V}\,v^\prime_x=\frac{1-k^2(v^{\prime\,2}_x)}
{v^\prime_x}\,V,$$
или
\begin{equation}
\frac{1-k^2(V^2)}{V^2}=\frac{1-k^2(v^{\prime\,2}_x)}{v^{\prime\,2}_x}
\equiv K,
\label{eq5_6h}
\end{equation}
где последний шаг следует из того, что единственный способ удовлетворить уравнению 
(\ref{eq5_6h}) для всех значений $V$ и $v^\prime_x$ --- это предположить, что
его левая и правая части равны некоторой постоянной $K$. Тогда правило сложения 
скоростей будет иметь вид
\begin{equation}
v_x=\frac{v^\prime_x+V}{1+Kv^\prime_x V}.
\label{eq5_6i}
\end{equation}
Если $K=0$, получаем преобразование Галилея и правило сложения скоростей
$v_x=v^\prime_x+V$.

Если $K<0$, можно положить $K=-\frac{1}{c^2}$ и ввести безразмерный
параметр $\beta=\frac{V}{c}$. Тогда 
\begin{eqnarray} & &
x^\prime=\frac{1}{\sqrt{1+\beta^2}}(x-Vt), \nonumber \\ & &
t^\prime=\frac{1}{\sqrt{1+\beta^2}}\left (t+\frac{V}{c^2}\,x\right ),
\label{eq5_6k}
\end{eqnarray}
в то время как правило сложения скоростей имеет вид
\begin{equation}
v_x=\frac{v^\prime_x+V}{1-\frac{v^\prime_x V}{c^2}}.
\label{eq5_6l}
\end{equation}
Если $v^\prime_x=V=\frac{c}{2}$, тогда (\ref{eq5_6l}) дает $v_x=\frac{4}{3}c$.
Поэтому скорости больше чем $c$ легко получаются в случае $K<0$. Но если
$v^\prime_x=V=\frac{4}{3}c$, тогда
$$v_x=\frac{\frac{8}{3}c}{1-\frac{16}{9}}=-\frac{24}{7}c<0.$$
Сумма двух положительных скоростей может быть отрицательной! Это не единственная 
странность случая $K<0$. Например, если $v^\prime_x V =c^2$, то сумма $v^\prime_x$
и $V$ будет бесконечной.

Если мы выполним два последовательных преобразований $(x,t)\to(x^\prime,t^\prime)
\to(x^{\prime\prime},t^{\prime\prime})$, согласно (\ref{eq5_6k}) с $\beta=4/3$,
получим в итоге, $$x^{\prime\prime}=-\frac{7}{25}x-\frac{24}{25}ct.$$
Это не может быть выражено как результат преобразования $(x,t)\to
(x^{\prime\prime},t^{\prime\prime})$ типа (\ref{eq5_6k}), так как коэффициент 
$1/\sqrt{1+\beta^2}$ перед $x$ в (\ref{eq5_6k}) всегда положительный. Поэтому 
преобразования (\ref{eq5_6k}) не составляют группу. Таким образом, случай $K<0$
не может считаться удовлетворительной реализацией принципа относительности.

Наконец, если $K=\frac{1}{c^2}>0$, получаем преобразования Лоренца
\begin{eqnarray} & &
x^\prime=\frac{1}{\sqrt{1-\beta^2}}(x-Vt),
\nonumber \\ & &
t^\prime=\frac{1}{\sqrt{1-\beta^2}}\left (t-\frac{V}{c^2}\,x\right ),
\label{eq5_6m}
\end{eqnarray}
с правилом сложения скоростей
\begin{equation}
v_x=\frac{v^\prime_x+V}{1+\frac{v^\prime_x V}{c^2}}.
\label{eq5_6n}
\end{equation}
Теперь $c$ является инвариантной скоростью, как показывает (\ref{eq5_6n}). Однако 
в вышеприведенном выводе ничего не указывает, что $c$ является скоростью света. 
Только тогда, когда мы привлекаем  уравнения Максвелла, становится ясно, что скорость 
электромагнитных волн, следующая из этих уравнений, должна совпадать с $c$, если мы 
хотим, чтобы уравнения Максвелла были инвариантны при преобразованиях Лоренца
(\ref{eq5_6m}).

Заметим, что подобный вывод преобразований Лоренца восходит к Игнатовскому \cite{17}.

\section[Семинар 6]
{\centerline{Семинар 6}}

\subsection*{Задача 1 {\usefont{T2A}{cmr}{b}{n}(2.10)}: Период движения фотона}
{\usefont{T2A}{cmr}{m}{it} Первое решение.}
В системе платформы
$$t^\prime=\frac{4L}{c},\;\;\;x^\prime=0.$$
Поэтому в Л-системе
$$t=\gamma\left (t^\prime+\frac{V}{c^2}\,x^\prime\right )=\frac{4L}{c}\,
\gamma.$$

{\usefont{T2A}{cmr}{m}{it} Второе решение.}
Поучительно проследить, как получается ответ, вычисляя отдельные времена при облете 
сторон платформы. В Л-системе платформа выглядит так:
\begin{figure}[htb]
\centerline{\epsfig{figure=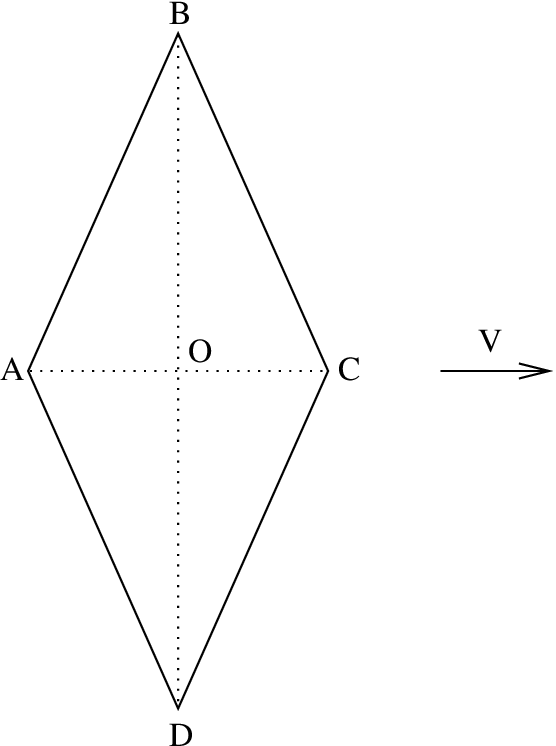,height=6cm}}
\end{figure}

При этом $$AO=OC=\frac{L}{\sqrt{2}\gamma}\;\;\;\mbox{и}\;\;\;
BO=OD=\frac{L}{\sqrt{2}},$$
т. е. размеры вдоль движения сокращены в $\gamma$ раз. Какое время 
(в Л-системе) требуется фотону, чтобы дойти от $A$ до $B$? Пока
фотон летит, зеркало $B$ передвинется в точку $B^\prime$. 
\begin{figure}[htb]
\centerline{\epsfig{figure=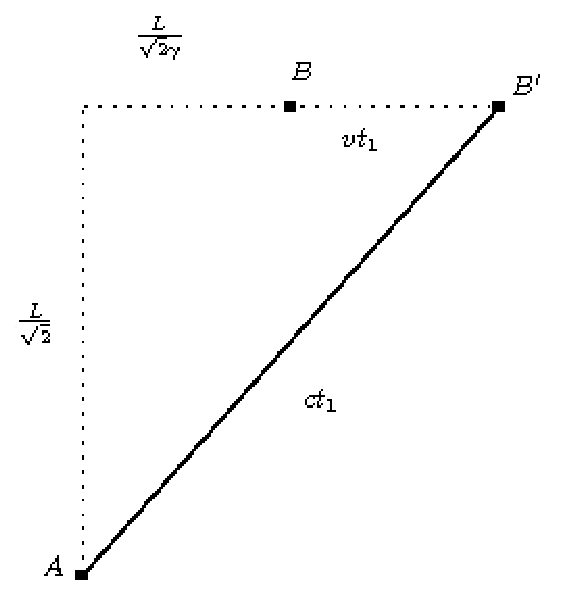,height=5cm}}
\end{figure}
Из рисунка очевидно, что 
\begin{equation}
\frac{L^2}{2}=c^2t_1^2-\left(\frac{L}{\sqrt{2}\gamma}+Vt_1\right)^2,
\label{eq6_1a}
\end{equation}
или
$$t_1^2-\frac{\sqrt{2}L}{c}\beta\gamma\,t_1-\frac{L^2}{2c^2}(\gamma^2+1)=0.$$
Используя $\beta^2\gamma^2=\gamma^2-1$, решение этого квадратного уравнения
можно представить так:
\begin{equation}
t_1=\frac{\sqrt{2}\,L\gamma}{2c}(\beta\pm\sqrt{2}).
\label{eq6_1b}
\end{equation}
Так как $t_1>0$ и $\beta<1<\sqrt{2}$, в уравнении (\ref{eq6_1b}) надо
выбрать знак плюс:
$$t_1=\frac{\sqrt{2}\,L\gamma}{2c}(\beta+\sqrt{2}).$$
Аналогично легко показать, что фотон тратит на прохождение от $B$ до $C$
такое же время $t_2=t_1$.

Теперь проследим за фотоном от $C$ до $D$.
\begin{figure}[htb]
\centerline{\epsfig{figure=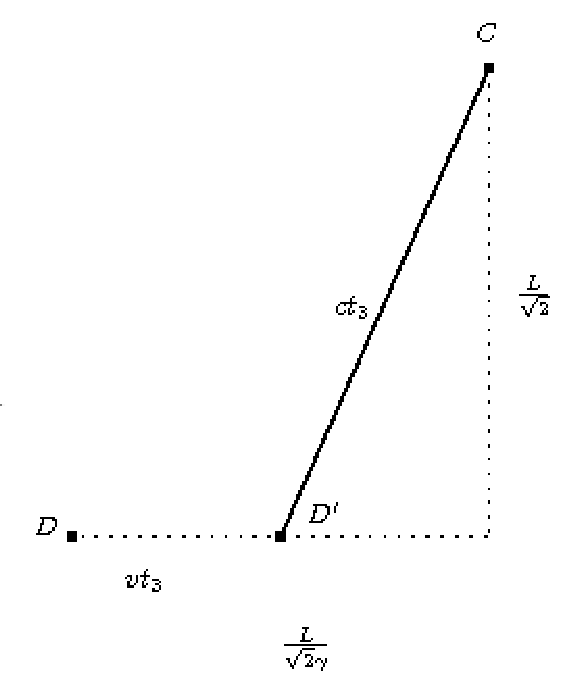,height=6cm}}
\end{figure}
Из рисунка ясно, что
\begin{equation}
\frac{L^2}{2}=c^2t_3^2-\left(\frac{L}{\sqrt{2}\gamma}-Vt_3\right)^2.
\label{eq6_1aa}
\end{equation}
Это уравнение отличается от уравнения для $t_1$ (\ref{eq6_1a}) только знаком
$V$. Поэтому сразу можем написать решение
$$t_3=\frac{\sqrt{2}\,L\gamma}{2c}(-\beta+\sqrt{2}).$$
Легко видеть, что от $D$ до $A$ фотону понадобится такое же время $t_4=t_3$.
Окончательно
$$T=t_1+t_2+t_3+t_4=\frac{\sqrt{2}\,L\gamma}{c}(\beta+\sqrt{2})+
\frac{\sqrt{2}\,L\gamma}{c}(-\beta+\sqrt{2})=\frac{4L}{c}\,\gamma.$$

{\usefont{T2A}{cmr}{m}{it} Вариант решения.}
Можно не решать квадратные уравнения для определения $t_1=t_2$ и $t_3=t_4$, 
а поступить следующим образом. Если из (\ref{eq6_1a}) отнять (\ref{eq6_1aa}), 
получим
$$c^2(t_1-t_3)(t_1+t_3)=V(t_1+t_3)\left(\frac{\sqrt{2}L}{\gamma}+V(t_1-t_3)
\right).$$
Отсюда
$$t_1-t_3=\frac{\sqrt{2}LV}{\gamma(c^2-V^2)}=\frac{\sqrt{2}L}{c}\,\gamma
\beta.$$
С другой стороны, если сложить (\ref{eq6_1a}) и (\ref{eq6_1aa}), будем иметь
$$L^2=(c^2-V^2)(t_1^2+t_3^2)-\frac{L^2}{\gamma^2}-\frac{\sqrt{2}LV}{\gamma}\,
(t_1-t_3).$$
Отсюда
$$(c^2-V^2)(t_1^2+t_3^2)=L^2(2-\beta^2)+\frac{\sqrt{2}LV}{\gamma}\,(t_1-t_3)=
L^2(2-\beta^2)++2L^2\beta^2=L^2(2+\beta^2)$$
и
$$t_1^2+t_3^2=\frac{L^2}{c^2}\,\gamma^2\,(2+\beta^2).$$
Тогда
$$(t_1+t_3)^2=2(t_1^2+t_3^2)-(t_1-t_3)^2=2\frac{L^2}{c^2}\,\gamma^2\,
(2+\beta^2)-2\frac{L^2}{c^2}\,\gamma^2\beta^2=4\frac{L^2}{c^2}\,\gamma^2$$
и
$$T=2(t_1+t_3)=4\,\frac{L}{c}\,\gamma.$$

\subsection*{Задача 2 {\usefont{T2A}{cmr}{b}{n}(2.8)}: Какая лампочка загорится
раньше?}
Пусть начало отсчета совпадает с точкой $A$, и лампочка $A$ загорается в
момент времени $t_A=t_A^\prime=0$, когда начала систем $S$ и $S^\prime$
совпадают. Тогда в системе $S^\prime$ событие ``загорание лампочки $B$''
имеет координаты $t_B^\prime=t_A^\prime=0$, $x_B^\prime=L_0$. По
преобразованию Лоренца находим время зажигания лампочки $B$ в лабораторной
$S$-системе $$t_B=\gamma\left (t_B^\prime+\frac{V}{c^2}x_B^\prime\right )=
\beta\gamma\,\frac{L_0}{c}>0.$$
Т. е. в Л-системе лампочка $B$ зажигается позже. 

К наблюдателю свет от лампочки $B$ проходит на $x_B-x_A$ больше путь, чем
свет от лампочки $A$. Поэтому наблюдатель увидит вспышку  $A$ раньше на
$$\Delta t=\beta\gamma\,\frac{L_0}{c}+\frac{x_B-x_a}{c}.$$
Но $x$-координаты лампочек в Л-системе в момент зажигания равны $x_A=0$ и
$x_B=\gamma(x_B^\prime+Vt_B^\prime)=\gamma\,L_0$. По-другому $x_B$ можно
получить так: в момент $t=0$ координата лампочки $B$ была $L_0/\gamma$ (так
как в $S$ системе стержень $\gamma$ раз сокращен), поэтому через время
$t_B=\beta\gamma\,L_0/c$, когда лампочка вспыхнет, координата будет
$$x_B=\frac{L_0}{\gamma}+\beta\gamma\frac{L_0}{c}V=\frac{L_0}{\gamma}
(1+\beta^2\gamma^2)=\gamma\,L_0.$$
Следовательно,
$$\Delta t=\beta\gamma\frac{L_0}{c}+\frac{\gamma L_0}{c}=\frac{L_0}{c}
\gamma (1+\beta)=\frac{L_0}{c}\sqrt{\frac{1+\beta}{1-\beta}}.$$

{\usefont{T2A}{cmr}{m}{it} Вариант решения.}
Пусть первое событие --- ``наблюдатель $O$ увидел вспышку $A$'', второе ---
``наблюдатель $O$ увидел вспышку $B$''. Временной интервал между этими событиями
в Л-системе 
\begin{equation}
t_2-t_1=\gamma\left [t_2^\prime-t_1^\prime+\frac{V}{c^2}(x_2^\prime-
x_1^\prime)\right ].
\label{eq6_2a}
\end{equation}
Но в системе $S^\prime$ (в системе стержня) наблюдатель $O$ убегает от стержня 
со скоростью $V$. Следовательно,
$$t_1^\prime=\frac{L}{c-V},\;\;\;\mbox{и}\;\;\;t_2^\prime=
\frac{L+L_0}{c-V},$$
где $L$ --- первоначальное расстояние (в момент вспышки $t^\prime=0$) от $A$ до $O$
в системе $S^\prime$. Таким образом,
\begin{equation}
t_2^\prime-t_1^\prime=\frac{L_0}{c-V}=\frac{L_0}{c}\,\frac{1}{1-\beta}.
\label{eq6_2b}
\end{equation}
Пусть начало отсчета в системе $S^\prime$ находится около точки $A$. Тогда
$x_1^\prime=-(L+Vt_1^\prime)$ и $x_2^\prime=-(L+Vt_2^\prime)$. Следовательно,
\begin{equation}
x_2^\prime-x_1^\prime=-V(t_2^\prime-t_1^\prime).
\label{eq6_2c}
\end{equation}
Подставим (\ref{eq6_2b}) и (\ref{eq6_2c}) в (\ref{eq6_2a}), получим
$$\Delta t=t_2-t_1=\gamma(1-\beta^2)(t_2^\prime-t_1^\prime)=
\frac{t_2^\prime-t_1^\prime}{\gamma}=\frac{L_0}{c}\sqrt{\frac{1+\beta}
{1-\beta}}.$$

\subsection*{Задача 3 {\usefont{T2A}{cmr}{b}{n}(2.11)}: Время движения фотона}
Пусть наблюдатель движется горизонтально. При использовании стандартного
пеобразования Лоренца оси $x,x^\prime$ надо направить вдоль направления 
движения наблюдателя. Позтому в $S$-системе событие прихода фотона на зеркало 
$D$ имеет координаты $X_D=L$, $t_D=3L/c$. По преобразованию Лоренца, находим
время прихода фотона на зеркало $D$ в системе $S^\prime$:
$$t_D^\prime=\gamma\left (t_D-\frac{V}{c^2}x_D\right)=
\gamma\left (\frac{3L}{c}-\frac{V}{c^2}L\right)=
\gamma\,\frac{L}{c}(3-\beta).$$
\begin{figure}[htb]
\centerline{\epsfig{figure=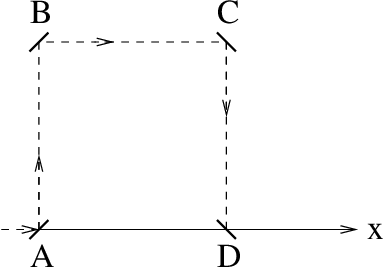,height=3cm}}
\end{figure}

Получим тот же ответ, проследив за движением фотона в системе $S^\prime$.
В этой системе платформа движется слева со скоростью $V$. Поэтому пока фотон
летит от зеркала $A$ к $B$, это последнее сместится горизонтально на
$Vt_1^\prime$, где $t_1^\prime$ --- время движения фотона от зеркала $A$ до 
$B$ в системе $S^\prime$. Так как поперечный размер платформы не меняется
при переходе в систему $S^\prime$, имеем
$c^2t_1^{\prime\,2}-V^2t_1^{\prime\,2}=L^2$. Отсюда
$$t_1^{\prime}=\frac{L}{\sqrt{c^2-V^2}}=\frac{L}{c}\gamma.$$
Продольный размер платформы $BC$ в системе $S^\prime$ сокращается в $\gamma$
раз. Когда фотон летит от $B$ к $C$, зеркало $C$ движется к нему навстречу
со скоростью $V$. Поэтому
$$t_2^{\prime}=\frac{L}{\gamma}\frac{1}{c+V}=\frac{L}{c}\frac{1}
{\gamma(1+\beta)}.$$
Наконец, легко показать, что время движения фотона от $C$ к $D$ такое же, как
время движения от $A$ к $B$: $t_3^{\prime}=t_1^{\prime}$. Следовательно,
$$t^\prime_D=2t_1^{\prime}+t_2^{\prime}=\frac{2L}{c}\gamma+
\frac{L}{c}\frac{1}{\gamma(1+\beta)}=\frac{L}{c}\gamma\left(2+
\frac{1-\beta^2}{1+\beta}\right)=\gamma\,\frac{L}{c}(3-\beta).$$

Теперь рассмотрим второй случай, когда наблюдатель движется вверх и туда же
направлены оси $x,x^\prime$. 
\begin{figure}[htb]
\centerline{\epsfig{figure=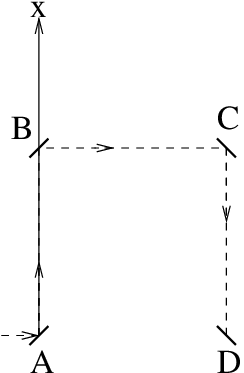,height=5cm}}
\end{figure}
В системе $S$ событие прихода фотона на зеркало $D$ имеет координаты 
$X_D=0$, $t_D=3L/c$. Поэтому
$$t^\prime_D=\gamma\left (t_D-\frac{V}{c^2}x_D\right)=\frac{3L}{c}\gamma.$$
Проследим за фотоном и в этом случае. В системе $S^\prime$ платформа движется
со скоростью $V$ вниз. Так как в этом случае в роли продольного размера
выступает $AB$ и, следовательно, $AB=L/\gamma$, для соответствующего времени
будем иметь $$t_1^{\prime}=\frac{L}{\gamma}\frac{1}{c+V}=\frac{L}{c}\frac{1}
{\gamma(1+\beta)}.$$
Поперечный размер $BC$ не меняется, $BC=L$, но при полете фотона от зеркала 
$B$ к $C$ зеркало $C$ опускается вниз на $Vt_2^\prime$. Теорема Пифагора дает
$$t_2^{\prime}=\frac{L}{\sqrt{c^2-V^2}}=\frac{L}{c}\gamma.$$
Наконец, при полете от $C$ к $D$ зеркало $D$ убегает от фотона со скоростью
$V$. При этом $CD=L/\gamma$ из за сокращения продольных размеров. Поэтому
$$t_3^{\prime}=\frac{L}{\gamma}\frac{1}{c-V}=\frac{L}{c}\frac{1}
{\gamma(1-\beta)}.$$
Следовательно, 
$$t_D^\prime=t_1^{\prime}+t_2^{\prime}+t_3^{\prime}=\frac{L}{c}\gamma
\left(1+\frac{1-\beta^2}{1+\beta}+\frac{1-\beta^2}{1-\beta}\right)=
\frac{3L}{c}\gamma.$$

\subsection*{Задача 4 {\usefont{T2A}{cmr}{b}{n}(2.15)}: Время пролета галактики}
{\usefont{T2A}{cmr}{m}{it} Первое решение.}
Пусть $S$ --- система галактики, $S^\prime$ --- система протона и пусть фотон стартует
в момент $t_1=t_1^\prime=0$, имея $x$-координату $x_1=x_1^\prime=0$. В системе 
$S$ событие ``встреча фотона с концом галактики'' имеет координаты $t_2=T=10^5$~лет,
$x_2=L=10^5$~св. лет.  В системе $S^\prime$ будем иметь
$$t_2^\prime=\gamma\left (t_2-\frac{V}{c^2}x_2\right)=\gamma\left (T-
\frac{V}{c^2}L\right).$$
Но $L=cT$, поэтому
$$t_2^\prime=\gamma T(1-\beta)=\gamma T\frac{1-\beta^2}{1+\beta}=
\frac{T}{\gamma}\,\frac{1}{1+\beta}\approx \frac{T}{2\gamma}=0,5\cdot 10^{-5}
~\mbox{лет}.$$
Если протон летит в другую сторону, знак $\beta$ поменяется на противоположный, и мы
будем иметь $t_2^\prime=\gamma T(1+\beta)\approx 2 \gamma\, 
T=2\cdot 10^{15}$~лет.

{\usefont{T2A}{cmr}{m}{it} Второе решение.}
В системе протона галактика имеет продольный размер $L/\gamma$, и дальний край 
галактики приближается навстречу фотону со скоростью $V$. Поэтому
$$t_2^\prime=\frac{L}{\gamma}\frac{1}{c+V}=\frac{T}{\gamma(1+\beta)}\approx
\frac{T}{2\gamma}.$$
Если протон летит в другую сторону, в его системе галактика по-прежнему имеет 
$L/\gamma$ продольный размер, но дальний край галактики убегает от фотона  со 
скоростью $V$. Поэтому
$$t_2^\prime=\frac{L}{\gamma}\frac{1}{c-V}=\frac{T}{\gamma(1-\beta)}=
\gamma\,T(1+\beta)\approx 2\gamma \,T.$$

\subsection*{Задача 5 {\usefont{T2A}{cmr}{b}{n}(2.17)}: Стеклянный брусок с
посеребренной гранью}
{\usefont{T2A}{cmr}{m}{it} Первое решение.}
В $S^\prime$-системе (в системе бруска) 
$$\Delta t^\prime=\frac{2L}{c/n}=\frac{2L}{c}n,\;\;\mbox{и}\;\;
\Delta x^\prime=0,$$
так как свет и входит и выходит в брусок в точке $x^\prime=0$. Перейдем в Л-систему 
по преобразованию Лоренца
$$\Delta t=\gamma\left [\Delta t^\prime+\frac{V}{c^2}\Delta x^\prime\right ]=
\frac{2L}{c}n\gamma.$$

{\usefont{T2A}{cmr}{m}{it} Второе решение.}
В Л-системе брусок имеет длину $L/\gamma$. Когда свет входит в брусок, он движется 
относительно бруска со скоростью $c/n$. Относительно Л-системы скорость света
находим из закона сложения скоростей:
$$c_1=\frac{\frac{c}{n}+V}{1+\frac{cV}{nc^2}}=\frac{1+\beta n}{n+\beta}\,c.$$  
Посеребренный конец убегает от света со скоростью $V$. Поэтому свет догонит его за 
время $$t_1=\frac{L/\gamma}{c_1-V}=\frac{L}{c}\gamma (n+\beta),$$
так как $$c_1-V=c\left (\frac{1+\beta n}{n+\beta}-\beta\right )=
\frac{1-\beta^2}{n+\beta}c=\frac{c}{\gamma^2(n+\beta)}.$$
После отражения $x$-компоненту скорости света относительно лабораторной системы тоже
находим по формуле сложения скоростей:
$$c_{2x}=\frac{-\frac{c}{n}+V}{1-\frac{cV}{nc^2}}=
\frac{-1+\beta n}{n-\beta}\,c<0.$$
Величина скорости будет
$$c_2=\frac{1-\beta n}{n-\beta}\,c.$$
Теперь грань бруска, из которого свет в конце концов выйдет, летит навстречу свету
со скоростью $V$. Поэтому, чтобы выйти из бруска после отражения, свету понадобится 
время $$t_2=\frac{L/\gamma}{c_2+V}=\frac{L}{c}\gamma (n-\beta),$$ 
так как $$c_2+V=c\left (\frac{1-\beta n}{n-\beta}+\beta\right )=
\frac{1-\beta^2}{n-\beta}c=\frac{c}{\gamma^2(n-\beta)}.$$
Полное время равно
$$\Delta t=t_1+t_2=\frac{2L}{c}n\gamma.$$

\subsection*{Задача 6 {\usefont{T2A}{cmr}{b}{n}(2.16)}: Доля оставшихся нейтронов}
{\usefont{T2A}{cmr}{m}{it} Первое решение.}
В системе нейтрона размер галактики $L/\gamma$, и край галактики движется
навстречу фотона со скоростью $V$. Поэтому когда фотон достигнет края
галактики, часы нейтрона будут показывать время 
$$t_1^\prime=\frac{L}{\gamma}\frac{1}{c+V}\approx \frac{L}{2\gamma \,c}.$$
Доля оставшихся нейтронов 
$$\frac{N_1}{N_0}=\exp{\left(-\frac{t_1^\prime}{\tau}\right )}=\exp{\left (
-\frac{L}{2c\gamma\tau}\right)}=
\exp{\left(-\frac{10^5\cdot 9,46\cdot 10^{15}}{2\cdot 3\cdot 10^8\,10^{10}
\cdot 10^3}\right )}\approx 0,85.$$
Мы учли, что 1~св.г.$\approx 9,46\cdot 10^{15}$~м.

В Л-системе фотон пролетает галактику за время $t=L/c=10^5$~лет. В
собственной системе нейтрона пройдет $\gamma$ раз меньше времени, 
$t_2^\prime=t/\gamma=L/(c\gamma)$. 
Поэтому доля оставшихся нейтронов
$$\frac{N_2}{N_0}=\exp{\left(-\frac{t_2^\prime}{\tau}\right)}=\exp{\left(
-\frac{L}{c\gamma\tau}\right)}=\left(\frac{N_1}{N_0}\right)^2\approx 0.73.$$
Объясним в Л-системе величину времени $t_2^\prime$ по-другому. Когда
фотон долетит до края галактики за время $t=L/c$, у нейтрона будет координата 
$x=Vt$. Поэтому событию ``нахождение нейтрона в точке пространства-времени 
$(t,Vt)$'' соответствует время в $S^\prime$-системе (в системе нейтрона) 
$$t_2^\prime=\gamma\left (t-\frac{V}{c^2}x\right )=\gamma t\left (1-\frac{V^2}
{c^2}\right )=\frac{t}{\gamma}.$$

{\usefont{T2A}{cmr}{m}{it} Второе решение.}
Мы проверяем количество оставшихся нейтронов {\usefont{T2A}{cmr}{m}{it}
одновременно} с достижением фотоном края галактики. Но одновременность
относительна. Пусть первое событие ``проверка количества нейтронов'', второе
событие --- ``достижение фотоном края галактики''.  Доля оставшихся нейтронов
есть всегда 
$$\frac{N}{N_0}=\exp{\left(-\frac{t_1^\prime}{\tau}\right )},$$
где время $t_1^\prime$ соответствует первому событию в системе $S^\prime$ (в
системе нейтрона). В Л-системе первое событие имеет координаты $t_1=
\frac{L}{c}$, $x_1=V\frac{L}{c}$. Поэтому 
$$t_1^\prime=\gamma\left(t_1-\frac{V}{c^2}x_1\right)=\gamma\frac{L}{c}\left
(1-\frac{V^2}{c^2}\right)=\frac{L}{\gamma c}$$ и
$$\frac{N}{N_0}=\exp{\left(-\frac{L}{\gamma c\tau}\right)}.$$
Но второе событие в Л-системе  имеет координаты $t_2=
\frac{L}{c}$, $x_2=L$ и, следовательно,
$$t_2^\prime=\gamma\left(\frac{L}{c}-\frac{V}{c^2}L\right)=\frac{\gamma L}
{c}(1-\beta)\approx\frac{L}{2\gamma c}.$$
Как видим, хотя $t_1=t_2$, тем не менее $t^\prime_1\ne t^\prime _2$ --- 
одновременность относительна. Та проверка, которую мы произвели в системе
$S$ (в лабораторной системе) одновременно с наступлением второго события, не
является одновременным для второго события с точки зрения наблюдателя из 
системы $S^\prime$.  С его точки зрения, проверку надо произвести в момент 
времени $t_1^{\prime\prime}=t^\prime_2$, и после проверки он обнаружит, что
доля оставшихся нейтронов равна 
$$\frac{\tilde N}{N_0}=\exp{\left(-\frac{t_1^{\prime\prime}}{\tau}\right )}
=\exp{\left(-\frac{L}{2\gamma c\tau}\right)}.$$

\section[Семинар 7]
{\centerline{Семинар 7}}

\subsection*{Задача 1 {\usefont{T2A}{cmr}{b}{n}(2.23)}: Релятивистский трактор}
{\usefont{T2A}{cmr}{m}{it} Первое решение.}
Тракторист всегда видит $2N$ траков на гусенице (траки не могут исчезать или 
появляться). Когда гусеница движется относительно тракториста со скоростью $V$, все 
траки одинаково сокращены. Поэтому длина одного трака в системе тракториста 
$$l=\frac{2L}{2N}=\frac{L}{N},$$ 
где $L$ --- длина рамы в этой системе. Но рама всегда неподвижна 
относительно тракториста и, следовательно, не меняет свою длину. Выходит, для 
тракториста длина трака всегда одна и та же независимо от скорости. Как это возможно? 
Ведь траки должны Лоренц-сокращаться? Дело в том, что деформация трака компенсирует 
Лоренцево сжатие \cite{15,18}. Т. е. траки ровно $\gamma$ раз растягиваются: если
$l_0$ --- собственная длина трака, когда трактор не движется и деформации нет, то при 
движении трактора собственная длина трака станет $l_V=\gamma l_0$ и $l=l_V/\gamma=
l_0=\mathrm{const}$. В системе неподвижного наблюдателя рама сокращена --- имеет 
длину $L/\gamma$. Значит, такую же длину имеют половинки гусеницы. Но на нижней 
гусенице траки неподвижны, т. е. имеют длину $l_V=\gamma l_0$. Поэтому их число 
$$N_1=\frac{L}{\gamma}{\gamma l_0}=\frac{1}{\gamma^2}\frac{L}{l_0}=
\frac{1}{\gamma^2}N=(1-\beta^2)N.$$
Но $N_1+N_2=2N$, где $N_2$ --- количество траков в Л-системе на верхней половинке
гусеницы $N_2=2N-N_1=(1+\beta^2)N$.   

{\usefont{T2A}{cmr}{m}{it} Второе решение.}
В Л-системе нижняя половинка гусеницы неподвижна, а верхняя движется со скоростью
$$u=\frac{V+V}{1+\frac{V^2}{c^2}}=\frac{2V}{1+\beta^2}.$$
Длина одного трака на нижней половинке $l_V$, на верхней --- $l_v/\gamma_u$. Но
$$\gamma_u^{-2}=1-\frac{4\beta^2}{(1+\beta^2)^2}=\frac{(1-\beta^2)^2}
{(1+\beta^2)^2}=\gamma^{-4}(1+\beta^2)^{-2},$$
т. е. $$\gamma_u=\gamma^2(1+\beta^2)=\frac{1+\beta^2}{1-\beta^2}.$$ Если длина 
рамы $L_V$ (в Л-системе), то количество траков на нижней и верхней половинках будет
$$N_1=\frac{L_V}{l_V},\;\;\;\mbox{и}\;\;\;N_2=\frac{L_V}{l_v/\gamma_u}=
N_1\frac{1+\beta^2}{1-\beta^2}.$$
Но $N_1+N_2=2N$, т. е. $$N_1\left( 1+\frac{1+\beta^2}{1-\beta^2}\right )=2N,$$
и $N_1=(1-\beta^2)N$. Тогда $N_2=2N-N_1=(1+\beta^2)N$.

\subsection*{Задача 2 {\usefont{T2A}{cmr}{b}{n}(2.13)}: Как далеко улетел
корабль?}
Пусть в Л-системе корабль двигался со скоростью $V_1$ в течение времени $t_1$ 
и со скоростью $V_2$ в течение времени $t_2$. Тогда $t_1+t_2=T$ и 
$$\frac{t_1}{\gamma_1}=\frac{t_2}{\gamma_2},$$ так как в системе корабля 
времена одинаковы. Отсюда 
$$t_1=\frac{T\gamma_1}{\gamma_1+\gamma_2},\;\;\;\mbox{и}\;\;\;
t_2=\frac{T\gamma_2}{\gamma_1+\gamma_2}.$$ Корабль улетит (в Л-системе) на 
расстояние
$$L=V_1t_1+V_2t_2=\frac{V_1\gamma_1+V_2\gamma_2}{\gamma_1+\gamma_2}T.$$

\subsection*{Задача 3 {\usefont{T2A}{cmr}{b}{n}(2.19)}: Вспышка света}
\begin{figure}[htb]
\centerline{\epsfig{figure=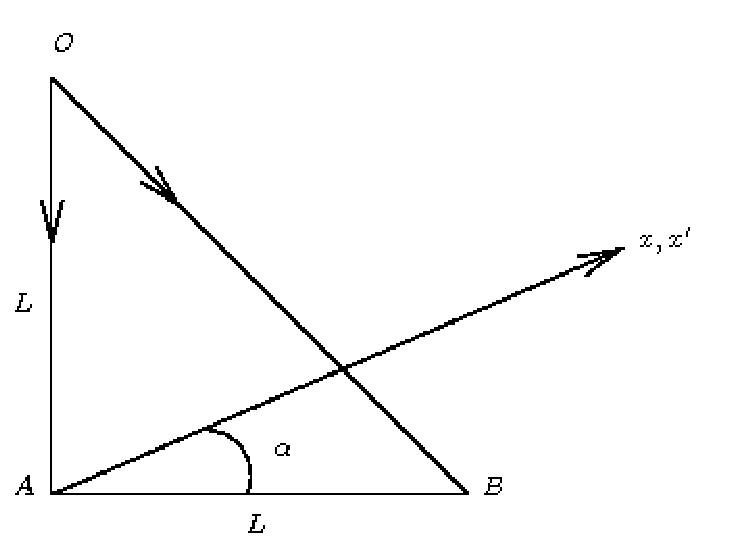,height=4cm}}
\end{figure}
Пусть наблюдатель движется под углом $\alpha$, как показано на рисунке, и пусть 
вспышка происходит в момент времени $t=t^\prime=0$. В Л-системе свет достигнет
точки $A$ в момент времени $t_A=L/c$. При этом $x$-координата этого события 
будет $x_A=0$. Аналогично для точки $B$ будем иметь $t_B=\sqrt{2}L/c$, 
$x_B=L\cos{\alpha}$. По преобразованию Лоренца получаем в системе $S^\prime$:
$$t_B^\prime-t_A^\prime=\gamma\left [(t_B-t_A)-\frac{V}{c^2}(x_B-x_A)\right]=
\gamma\left[(\sqrt{2}-1)\frac{L}{c}-\frac{V}{c^2} L\cos{\alpha}\right].$$ Нам 
нужно $t_B^\prime-t_A^\prime<0$, т. е. 
$$(\sqrt{2}-1)\frac{L}{c}<\frac{V}{c^2} L\cos{\alpha}.$$ Отсюда
$$V>\frac{\sqrt{2}-1}{\cos{\alpha}}c.$$ Минимальная скорость требуется, когда 
$\alpha=0$: $V>(\sqrt{2}-1)c\approx 0,41c$. Условие $V<c$ ограничивает угол 
$\alpha$. В частности, из $$\frac{\sqrt{2}-1}{\cos{\alpha}}c<c$$ получаем 
$\cos{\alpha}>\sqrt{2}-1$ и $\alpha<\arccos{(\sqrt{2}-1)}$.
 
\subsection*{Задача 4 {\usefont{T2A}{cmr}{b}{n}(2.26)}: Стабилизированный
электронный пучок}
Плотность частиц в сгустке $$n=\frac{N}{V}=\frac{N}{S\frac{L}{\gamma}}=
\gamma\frac{N}{SL}=\gamma n_0,$$
где $L,S$ --- длина и поперечное сечение сгустка в собственной системе (где
он покоится). Следовательно, в системе, где электроны неподвижны, их
плотность будет в $\gamma$ раз меньше, т. е. $n_-/\gamma$. В этой системе
положительные ионы движутся со скоростью $V$, и их плотность будет в $\gamma$ раз 
больше, т. е. $\gamma n_+$, чем в их собственной системе (в лабораторной).
Требуется  $\gamma n_+>n_-/\gamma$, т. е. $\gamma^2>n_-/n_+$, что дает
$$\beta=\frac{V}{c}>\sqrt{1-\frac{n_+}{n_-}}.$$

\subsection*{Задача 5 {\usefont{T2A}{cmr}{b}{n}(2.22)}: Пенал и карандаш}
\begin{figure}[htb]
\centerline{\epsfig{figure=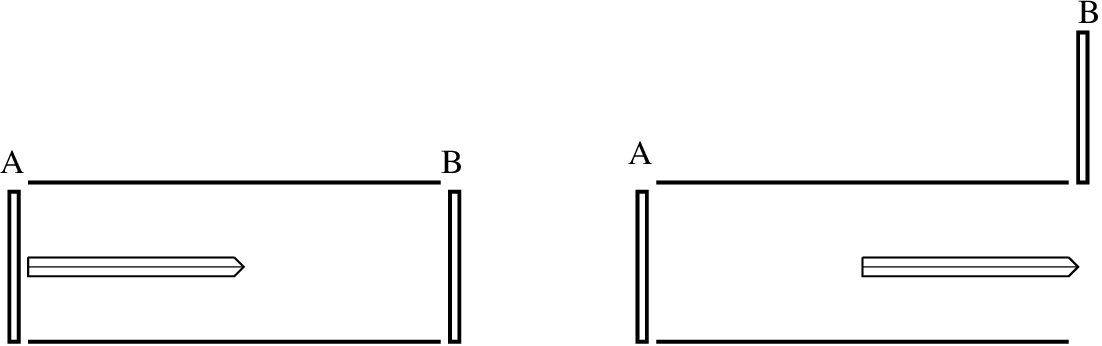,height=3cm}}
\end{figure}
Пусть крышка $A$ закрывается в момент времени $t_A=0$ (в Л-системе), а
крышка $B$ открывается как раз перед вылетом карандаша, т. е. при
$$t_B=\frac{1}{V}\left (L-\frac{L_0}{\gamma}\right ).$$
В лабораторной системе $x$-координата первого события $x_A=0$, а второго
$x_B=L$. В системе карандаша будем иметь
$$t_A^\prime=\gamma\left (t_A-\frac{V}{c^2}x_A\right )=0$$
и
$$t_B^\prime=\gamma\left [\frac{1}{V}\left (L-\frac{L_0}{\gamma}\right )-
\frac{V}{c^2}L\right ]=\frac{\gamma}{V}\left [L(1-\beta^2)-\frac{L_0}{\gamma}
\right ]=\frac{1}{V}\left[\frac{L}{\gamma}-L_0\right ]<0,$$
т. е. в системе карандаша крышка $B$ открывается раньше, чем закрывается
крышка $A$, и короткий пенал пролетает мимо карандаша.
\begin{figure}[htb]
\centerline{\epsfig{figure=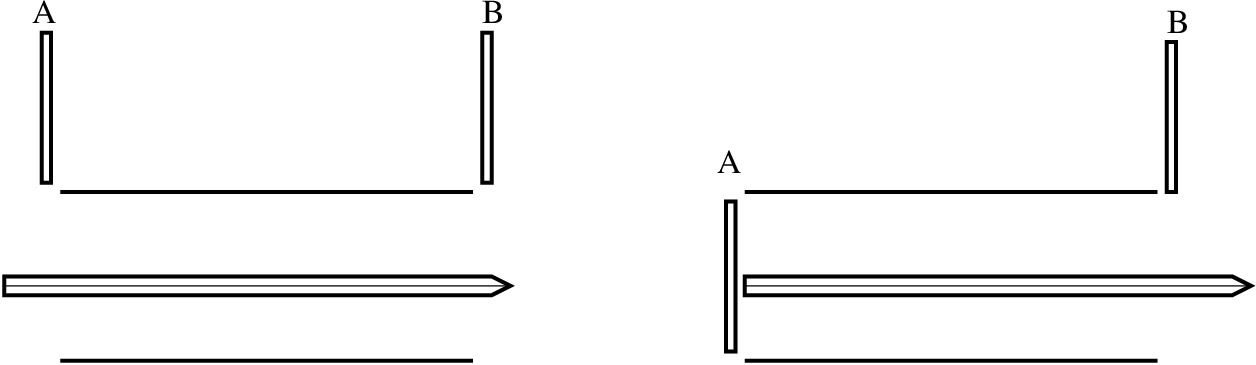,height=3cm}}
\end{figure}

\subsection*{Задача 6 {\usefont{T2A}{cmr}{b}{n}(2.20)}: Сколько таких систем
отсчета существует?}
Пусть система $S^\prime$ движется движется со скоростью $\vec{V}$
относительно лабораторной системы. Разложим $\vec{r}_1$ и $\vec{r}_2$ на
параллельные и перпендикулярные компоненты относительно $V$:
$$\vec{r}_{1\parallel}=\frac{\vec{r}_1\cdot\vec{V}}{V^2}\,\vec{V},\;\;\;
\vec{r}_{1\perp}=\vec{r}_1-\vec{r}_{1\parallel},$$ и аналогично для
$\vec{r}_2$. Тогда преобразования Лоренца будут иметь вид
\begin{eqnarray}&&
r_{1\parallel}^\prime=\gamma (r_{1\parallel}-Vt_1),\;\;\;r_{2\parallel}^\prime
=\gamma (r_{2\parallel}-Vt_2), \nonumber \\ &&
\vec{r}_{1\perp}^\prime=\vec{r}_{1\perp},\hspace*{4cm}\vec{r}_{2\perp}^\prime
=\vec{r}_{2\perp}, \nonumber \\ &&
t_1^\prime=\gamma\left (t_1-\frac{V}{c^2}r_{1\parallel}\right),\;\;
t_2^\prime=\gamma\left (t_2-\frac{V}{c^2}r_{2\parallel}\right).
\label{eq7_6a}
\end{eqnarray}
События одновременны в системе $S^\prime$, если $t_1^\prime=t_2^\prime$, или,
согласно (\ref{eq7_6a}),
\begin{equation}
t_1-t_2=\frac{V}{c^2}(r_{1\parallel}-r_{2\parallel})=\frac{(\vec{r}_1-
\vec{r}_2)\cdot\vec{V}}{c^2}.
\label{eq7_6b}
\end{equation}
Пусть $\alpha$ --- угол между векторами $\vec{r}_1-\vec{r}_2$ и $\vec{V}$.
Тогда (\ref{eq7_6b}) перепишется как
$c(t_1-t_2)=\beta |\vec{r}_1-\vec{r}_2|\cos{\alpha}$. Т. е. надо выбрать такую
$S^\prime$-систему, которая движется под углом $\alpha$ к вектору
$\vec{r}_1-\vec{r}_2$ со скоростью $$V=\frac{c(t_1-t_2)}{|\vec{r}_1-
\vec{r}_2|\cos{\alpha}}c.$$ Таких систем бесконечно много, но угол $\alpha$ не
может быть произвольным: из $V<c$ следует условие
$$\frac{c(t_1-t_2)}{|\vec{r}_1-\vec{r}_2|\cos{\alpha}}<1,$$
или
$$\alpha<\arccos{\left (\frac{c(t_1-t_2)}{|\vec{r}_1-\vec{r}_2|}\right )}.$$
Такой угол всегда существует, если
$$\frac{c(t_1-t_2)}{|\vec{r}_1-\vec{r}_2|}<1,$$
или $c^2(t_1-t_2)^2-(\vec{r}_1-\vec{r}_2)^2<0$ (интервал между событиями
пространственноподобен). В противном случае такой системы не существует.

События одноместны, если $\vec{r}_1^\prime=\vec{r}_2^\prime$, т. е. если 
$\vec{r}_{1\perp}^\prime=\vec{r}_{2\perp}^\prime$ и 
$\vec{r}_{1\parallel}^\prime=\vec{r}_{2\parallel}^\prime$. Первое из этих
условий дает $(\vec{r}_1-\vec{r}_2)_\perp=0$, или
\begin{equation}
\vec{V}=k(\vec{r}_1-\vec{r}_2).
\label{eq7_6c}
\end{equation}
Второе условие (равенство параллельных компонент) дает
$$(\vec{r}_1-\vec{r}_2)\cdot\frac{\vec{V}}{V}=V(t_1-t_2).$$
Подставляя сюда $\vec{r}_1-\vec{r}_2=k^{-1}\vec{V}$ из (\ref{eq7_6c}),
определяем коэффициент $k$ и получаем окончательно
$$\vec{V}=\frac{\vec{r}_1-\vec{r}_2}{t_1-t_2}.$$
Следовательно, если такая система существует, то она единственная: движется
вдоль $$\frac{\vec{r}_1-\vec{r}_2}{t_1-t_2}$$ со скоростью 
$$V=\left|\frac{\vec{r}_1-\vec{r}_2}{t_1-t_2}\right |.$$
Из $V<c$ следует условие $|\vec{r}_1-\vec{r}_2|<c(t_1-t_2)$, т. е. такая
система существует, если только $c^2(t_1-t_2)^2-(\vec{r}_1-\vec{r}_2)^2>0$
(интервал между событиями временноподобен).

\section[Семинар 8]
{\centerline{Семинар 8}}

\subsection*{Задача 1 {\usefont{T2A}{cmr}{b}{n}(2.24)}: Релятивистский танк}
{\usefont{T2A}{cmr}{m}{it} Первое решение.}
Интервал времени между выстрелами (по часам стрелка) есть $\tau^\prime=1/n$.
Поэтому расстояние между последующими снарядами будет $l^\prime=u\tau^\prime=
u/n$. В системе стрелка крепость надвигается со скоростью $V$, т. е.
снаряды и крепость сближаются со скоростью $u+V$. Поэтому интервал времени
между последующими попаданиями снарядов будет $$\Delta t^\prime=
\frac{l^\prime}{u+V}=\frac{1}{n}\frac{u}{u+V}.$$ За это время крепость
передвинется на $\Delta x^\prime=-V\Delta t^\prime$. Следовательно,
релятивистский интервал между попаданиями снарядов в системе $S^\prime$ (в
системе стрелка) равен $s^2=c^2(\Delta t^\prime)^2-(\Delta x^\prime)^2=
(c^2-V^2)(\Delta t^\prime)^2$. Пусть в системе $S$ (в системе крепости) в
крепость попадает $N$ снарядов в секунду. Тогда интервал времени между
попаданиями $\Delta t=1/N$. При этом крепость стоит на месте, т. е. $\Delta
x=0$ и $s^2=c^2(\Delta t)^2-(\Delta x)^2=c^2/N^2$. Следовательно,
$$\frac{c^2}{N^2}=(c^2-V^2)\left(\frac{1}{n}\frac{u}{u+V}\right)^2$$
и
$$N=\gamma_V\,n\left(1+\frac{V}{u}\right ),\;\;\;\mbox{где}\;\;\;
\gamma_V=\frac{1}{\sqrt{1-\frac{V^2}{c^2}}}.$$

{\usefont{T2A}{cmr}{m}{it} Второе решение.}
Скорость снаряда в системе крепости равна 
$$u_K=\frac{u+V}{1+\frac{uV}{c^2}}.$$
Интервал времени между выстрелами в системе стрелка (система $S$) есть 
$\tau^\prime=1/n$, а в системе крепости $\tau=\gamma_V\tau^\prime=
\gamma_V/n$. За это время первый снаряд в системе $S$ передвинется на 
$u_K\tau$, а танк на $V\tau$. Поэтому расстояние между последующими снарядами 
в системе $S$ будет $l=(u_K-V)\tau$. Эти снаряды летят к крепости со
скоростью $u_K$, поэтому их попадание в крепость произойдет с интервалом 
$$t=\frac{l}{u_K}=\frac{u_K-V}{u_K}\frac{\gamma_V}{n}.$$ С другой стороны,
$t=1/N$. Следовательно, $$N=\frac{n}{\gamma_V}\frac{u_K}{u_K-V}.$$
Заметим, что
$$u_K-V=u\frac{1-\beta_V^2}{1+\frac{uV}{c^2}}=\frac{u}{\gamma_V^2}
\frac{1+\frac{uV}{c^2}}.$$
Поэтому 
$$N=\frac{n}{\gamma_V}\frac{\gamma_V^2}{u}\left (1+\frac{uV}{c^2}\right )
\frac{u+V}{1+\frac{uV}{c^2}}=\gamma_V\,n\left(1+\frac{V}{u}\right ).$$

\subsection*{Задача 2 {\usefont{T2A}{cmr}{b}{n}(2.25)}: Какова скорость реки?}
{\usefont{T2A}{cmr}{m}{it} Первое решение.}
Решим задачу в системе реки. В этой системе лодка всегда движется по
величине с одной и той же скоростью $V$. Сначала она отплывает от багра и
через время $\gamma_V\tau$ (в системе реки) поворачивает назад. При этом в
системе лодочника проходит время $\tau$ --- движущиеся часы идут медленнее.
Так как лодка приплывает к багру с такой же по величине скоростью $V$, она
потратит на это такое же время $\gamma_V\tau$. Следовательно, в системе
реки багор оставался бесхозным на время $t=2\gamma_V\tau$. С другой
стороны, в системе реки берег движется со скоростью $u$, где $u$ --- искомая
величина скорости реки (в системе берега). Поэтому расстояние между мостом и
деревом в этой системе будет $L/\gamma_u$ и $t=/(\gamma_u\,u)$ (все время,
пока багор был бесхозным, дерево приближалось к багру со скоростью $u$, а
первоначально около багра ``проплывал'' мост). Следовательно,
\begin{equation}
2\gamma_V\tau=\frac{L}{\gamma_u\,u},
\label{eq8_2}
\end{equation}
или $$\gamma_u\beta_u=\frac{L}{2\gamma_Vc\tau}.$$
Возводя обе части этого равенства в квадрат и используя $\gamma^2_u\beta^2_u=
\gamma_u^2-1$, получим простое уравнение для $\gamma_u$, т. е. для $\beta_u$.
Из этого уравнения находим 
$$u=\beta_u\,c=\frac{c}{\sqrt{1+\left(\frac{2c\tau\gamma_V}
{L}\right)^2}}.$$

{\usefont{T2A}{cmr}{m}{it} Второе решение.}
Решим задачу в системе берега. Пусть лодка первоначально плывет против
течения. Относительно берега ее скорость будет
$$V_1=\frac{V-u}{1-\frac{uV}{c^2}}.$$
Следовательно, когда по часам лодочника пройдет время $\tau$, в системе
берега пройдет время $t_1=\gamma_{V_1}\tau$. За это время лодка проплывет
расстояние $l_1=V_1t_1$, а багор --- $l_2=ut_1$ в противоположную сторону. 
Поэтому когда лодка повернет назад и ее скорость относительно берега станет
$$V_2=\frac{V+u}{1+\frac{uV}{c^2}},$$
расстояние между лодкой и багром будет $l=l_1+l_2=\gamma_{V_1}\tau (V_1+u)$.
Следовательно, лодка догонит бугор за время
$$t_2=\frac{l}{V_2-u}=\gamma_{V_1}\tau\frac{V_1+u}{V_2-u}=\gamma_{V_1}\tau
\frac{1+\beta_u\beta_V}{1-\beta_u\beta_V}.$$
Таким образом, в системе берега багор оставался бесхозным на время 
$$t=t_1+t_2=\frac{2\gamma_{V_1}\tau}{1-\beta_u\beta_V}.$$
По условию задачи, он проплыл за это время расстояние $L$. Следовательно,
$ut=L$. Но
$$\gamma_{V_1}^{-2}=1-\left(\frac{\beta_V-\beta_u}{1-\beta_u\beta_V}\right
)^2=\frac{(1-\beta_V^2)(1-\beta_u^2)}{(1-\beta_u\beta_V)^2}=
\left[\gamma_V\gamma_u (1-\beta_u\beta_V)\right]^{-2},$$
т. е. $\gamma_{V_1}=\gamma_V\gamma_u (1-\beta_u\beta_V)$ и
$t=2\gamma_V\gamma_u\tau$. Тогда $ut=L$ дает уравнение (\ref{eq8_2}).

\subsection*{Задача 3 {\usefont{T2A}{cmr}{b}{n}(2.30)}: Фотографирование
быстролетящих параллелепипеда и шара}
Рассмотрим фотографирование прямоугольника. Пусть фотографирование
производится в момент времени $t=0$. 
Тогда из точки $B$ придет фотон, который излучился в момент $t_B=-l/c$, когда точка
$B$ занимала положение $B^\prime$, а из точки $A$ --- который излучился в момент 
$t_A=-(l+H)/c=t_B-H/c$, когда точка $A$ занимала положение  $A^\prime$. 
\begin{figure}[!h]
\centerline{\epsfig{figure=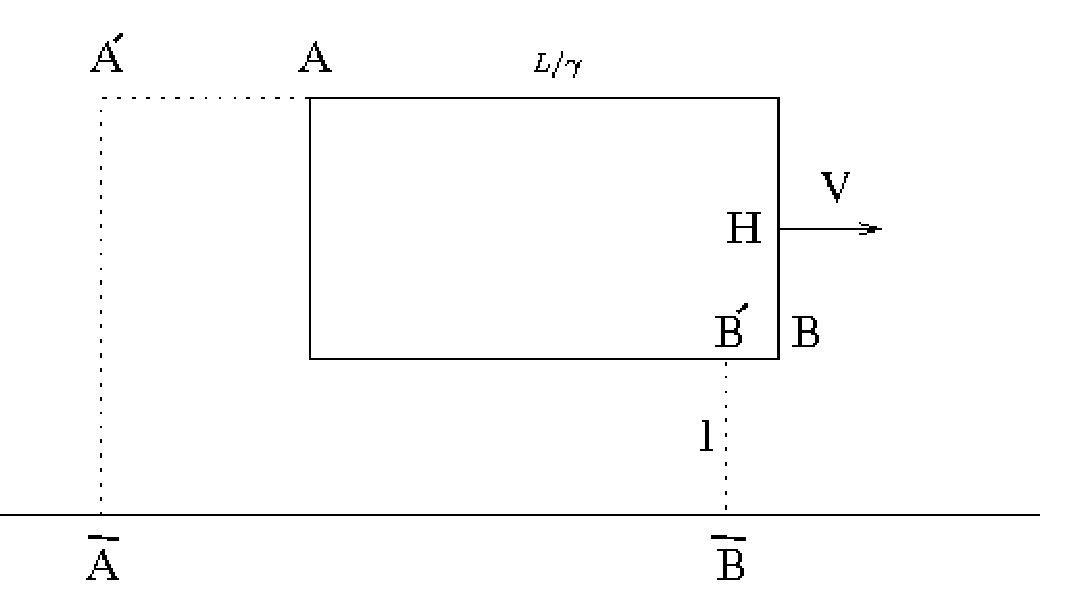,height=4cm}}
\end{figure}

\noindent Заметим,
что $A^\prime A-B^\prime B=(H/c)V=\beta H$. Следовательно, на фотографии длина 
прямоугольника будет $$\bar{A}\bar{B}=A^\prime A+\frac{L}{\gamma}-B^\prime B=
\frac{L}{\gamma}+\beta H,$$ так как в системе фотопластинки прямоугольник имеет 
длину $L/\gamma$. Это то же самое, что получится при фотографировании с длительной 
экспозицией неподвижного прямоугольника, повернутого на угол $\alpha=\arcsin{\beta}$, 
как показано на рисунке.
\begin{figure}[htb]
\centerline{\epsfig{figure=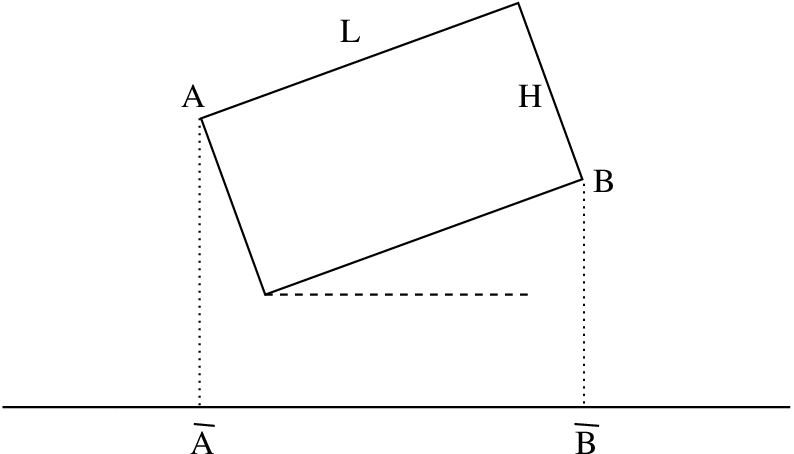,height=4cm}}
\end{figure}

Рассмотрим фотографирование шара. Быстролетящий шар в продольном направлении
сокращен в $\gamma$ раз, т. е. уравнение вертикального поперечного сечения
шара будет $$\frac{x^2}{(R/\gamma)^2}+\frac{y^2}{R^2}=1.$$
Будет удобнее, если введем параметр $\phi$, $0\le\phi<2\pi$ и запишем
уравнение вертикального поперечного сечения в следующем виде
$$x=\frac{R}{\gamma}\cos{\phi},\;\;\;y=R\sin{\phi}.$$
\begin{figure}[htb]
\centerline{\epsfig{figure=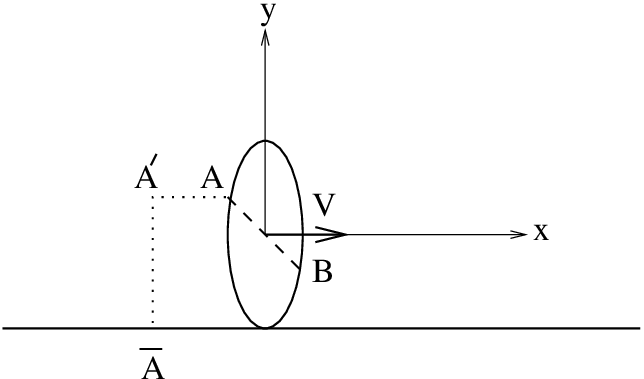,height=4cm}}
\end{figure}

Для простоты будем считать, что фотопластинка расположена на уровне нижней
точки шара, и возьмем точку $A$ на шаре, которая соответствует какому-то
конкретному значению параметра $\phi$. Если в момент времени $t=0$ на
фотопластинке регистрируется фотон, излученный в точке $A$, то это означает,
что фотон был излучен в момент $t_A=-(R+y_A)/c$ (считаем, что начало системы
отсчета совпадает с мгновенным положением центра шара при $t=0$). В момент 
$t_A$ точка $A$ находился левее на 
$A^\prime A=|t_A|V=\beta(R+y_A)=\beta\,R(1+\sin{\phi})$. Поэтому
изображение точки $A$ на фотопластинке будет иметь $x$-координату
\begin{equation}
\bar{x}_A=x-A^\prime A=\frac{R}{\gamma}\cos{\phi}-\beta\,R(1+\sin{\phi}).
\label{eq8_3a}
\end{equation}
Границы изображения шара на фотопластинке определяются минимальным и
максимальным значениями $\bar{x}_A$. Соответствующие значения параметра
$\phi$ определяются из уравнения
$$\frac{d\bar{x}}{d\phi}=-\beta\,R\cos{\phi}-\frac{R}{\gamma}\sin{\phi}=0,$$
т. е. $\tg{\phi}=-\beta\gamma$. Это означает, что левый край изображения
дается точкой $A$, которому соответствует $\sin{\phi}=\beta$, $\cos{\phi}=
-1/\gamma$ и, согласно (\ref{eq8_3a}), изображение этого края имеет
$x$-координату 
\begin{equation}
\bar{x}_A=-\beta\,R(1+\beta)-\frac{R}{\gamma^2}=-R-\beta\,R.
\label{eq8_3b}
\end{equation}
Правый  край изображения дается точкой $B$, которому соответствует 
$\sin{\phi}=-\beta$, $\cos{\phi}=1/\gamma$ и изображение этого края имеет
$x$-координату
\begin{equation}
\bar{x}_B=-\beta\,R(1-\beta)+\frac{R}{\gamma^2}=R-\beta\,R.
\label{eq8_3c}
\end{equation}
Из (\ref{eq8_3b}) и (\ref{eq8_3c}) следует, что изображение шара на
фотопластинке имеет ширину $x_B-x_A=2R$. Так как $z$-координаты точек шара
не меняются при переходе из системы шара в систему фотопластинки, приходим к
выводу, что изображение быстролетящего шара будет иметь форму круга радиуса $R$.

Заметим, что точка $A$ находится на задней стороне шара. Действительно, при
$t=0$ точка $A$ имеет координаты $x_A=-R/\gamma^2$ и $y_A=R\beta$. Поэтому в
собственной системе шара эта точка имеет координаты $x_A^\prime=\gamma(x_A+
Vt)=-R/\gamma$ и $y_A^\prime=y_A=R\beta$. Следовательно, при фотографировании 
с длительной экспозицией неподвижного шара, чтобы точка $A$ дала изображение 
левого края, шар следует повернуть на угол 
$$\alpha=\arcsin{\frac{y_A}{\sqrt{x_F^2+y_A^2}}}=\arcsin{\beta}.$$

Ненаблюдаемость Лоренцева сокращения в подобных ситуациях была замечена
австрийским физиком Антоном Лампа еще в 1924 году \cite{19}. Но этот 
интересный результат был предан забвению, пока его не переоткрыли Джеймс 
Террелл \cite{20} и Роджер Пенроуз \cite{21} в 1959 году.

\subsection*{Задача 4 {\usefont{T2A}{cmr}{b}{n}(2.27)}: Инвариантность площади
поперечного сечения пучка света}
{\usefont{T2A}{cmr}{m}{it} Первое решение} \cite{22}.
Пусть система $S^\prime$ движется вдоль оси $x$, а ось $y$ выбрана так, что скорость
фотонов $\vec{c}=(c\,\cos{\alpha},\,c\,\sin{\alpha},\,0)$ параллельна плоскости 
$xy$. Рассмотрим маленький прямоугольный участок волнового фронта $ABCD$ (см. рисунок).
\begin{figure}[!h]
\centerline{\epsfig{figure=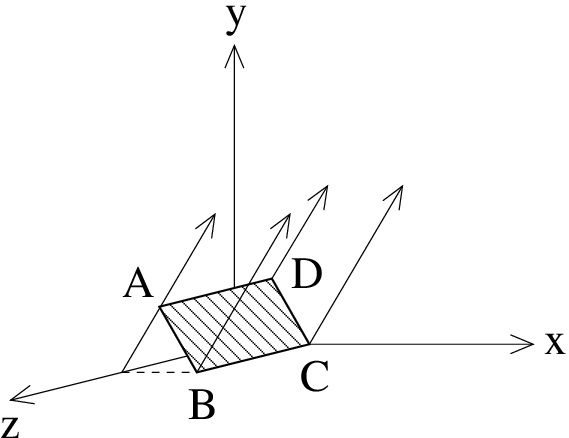,height=4cm}}
\end{figure}
Так как при преобразованиях Лоренца поперечные размеры $AD=BC$ не меняются, достаточно 
выяснить, что происходит с размерамы $AB=DC$. Пусть в системе $S$ события $C$ и $D$ 
(нахождение фотонов в точках $C$ и $D$ в момент $t=0$) имеют пространственные координаты
$C(\Delta x,\,0,\,0)$ и $D(\Delta x\,\cos^2{\alpha},\, \Delta x\,\cos{\alpha}
\sin{\alpha},\,0)$ соответственно. По преобразованиям Лоренца находим координаты этих 
событий в системе $S^\prime$ (нулевую $z$-координату опускаем):
$$x^\prime_C=\gamma\Delta x,\;\; y^\prime_C=0,\;\;t^\prime_C=-\gamma\beta\,
\frac{\Delta x}{c}$$
и
$$x^\prime_D=\gamma\Delta\,\cos^2{\alpha},\;\;y^\prime_D=\Delta x\,\cos{\alpha}
\sin{\alpha},\;\;t^\prime_D=-\gamma\beta\cos^2{\alpha}\,\frac{\Delta x}{c}.$$
В этой системе эти события уже не одновременны и соответственно не лежат на одном и том же
волновом фронте. Так как 
$$\Delta t^\prime=t^\prime_D-t^\prime_C=\gamma\beta(1-\cos^2{\alpha})
\frac{\Delta x}{c}=\gamma\beta\sin^2{\alpha}\,\frac{\Delta x}{c},$$
то надо найти координаты фотона $D$ на время $\Delta t^\prime$ раньше, чем 
$t^\prime=t^\prime_D$, чтобы узнать, где он находился на волновом фронте относительно
фотона $C$  при времени $t^\prime=t^\prime_C$. По формуле сложения скоростей находим 
скорость фотона в системе $S^\prime$:
$$\vec{c}^{\,\prime}=\left(\frac{c\,\cos{\alpha}-V}{1-\beta\,\cos{\alpha}},\;
\frac{c\,\sin{\alpha}}{\gamma(1-\beta\,\cos{\alpha})},\, 0\right ).$$
Поэтому в момент времени $t^\prime=t^\prime_C$ координаты фотона $D$ будут
$$x^\prime_{D^\prime}=\gamma\Delta\,\cos^2{\alpha}-c^\prime_x\Delta t^\prime=
\frac{\gamma\Delta x}{1-\beta\cos{\alpha}}\left [\cos^2{\alpha}+\beta^2
\sin^2{\alpha}-\beta\cos{\alpha}\right ]=\gamma\Delta x-\frac{\Delta x\,
\sin^2{\alpha}}{\gamma(1-\beta\cos{\alpha})}$$
и
$$y^\prime_{D^\prime}=\Delta x\,\cos{\alpha}\sin{\alpha}-c^\prime_y\Delta 
t^\prime=\Delta x\,\sin{\alpha}\,\frac{\cos{\alpha}-\beta}{1-\beta
\cos{\alpha}}.$$
Заметим, что вектор
$$\overrightarrow{CD^\prime}=\left (-\frac{\Delta x\,\sin^2{\alpha}}
{\gamma(1-\beta\cos{\alpha})},\;\Delta x\,\sin{\alpha}\,\frac{\cos{\alpha}-
\beta}{1-\beta\cos{\alpha}},\,0\right)$$
перпендикулярен вектору скорости фотонов $\vec{c}^{\,\prime}$ в системе $S^\prime$:
$\overrightarrow{CD^\prime}\cdot\vec{c}^{\,\prime}=0$. Следовательно, точки $C$ и 
$D^\prime$ действительно лежат на одном и том же волновом фронте в системе $S^\prime$.
При этом, так как
$$\frac{\sin^2{\alpha}}{\gamma^2}+(\cos{\alpha}-\beta)^2=(1-\beta
\cos{\alpha})^2,$$
для длины вектора $\overrightarrow{CD^\prime}$  в системе $S^\prime$ имеем
$$CD^\prime=\frac{\Delta x\,\sin{\alpha}}{1-\beta\cos{\alpha}}\sqrt{
\frac{\sin^2{\alpha}}{\gamma^2}+(\cos{\alpha}-\beta)^2}=\Delta x\,
\sin{\alpha},$$
что то же самое, что длина вектора $CD$ в системе $S$. Таким образом, элементу волнового
фронта $ABCD$ в системе $S$  при преобразованиях Лоренца соответствует элемент волнового
фронта такой же площади  в системе $S^\prime$. Следовательно, и полная площадь 
поперечнего сечения параллельного пучка света при преобразованиях Лоренца не меняется.

{\usefont{T2A}{cmr}{m}{it} Второе решение} \cite{23}.
Пусть $x_A$, $x_B$, и $x_C$ --- 4-радиус-векторы событий $A,B,C$. Тогда площадь
прямоугольного участка волнового фронта $ABCD$ равна $dS=|dx_{AB}||dx_{CB}|$, где
$dx_{AB}=x_A-x_B$ и $dx_{CB}=x_C-x_B$ --- два пространственно подобных 4-вектора,
лежащих на волновом фронте, и их длины определены так:
$$|dx_{AB}|=\sqrt{-dx_{AB}\cdot dx_{AB}},\;\;
|dx_{CB}|=\sqrt{-dx_{CB}\cdot dx_{CB}}.$$
То, что $dx_{AB}$ и $dx_{CB}$ лежат на одном и том же волновом фронте, подразумевает
выполнение двух условий. Первое условие состоит в том, что $dx_{AB}$ и $dx_{CB}$ 
перпендикулярны (в четырехмерном пространстве Минковского) 4-импульсу фотона $k$:
$dx_{AB}\cdot k=dx_{AB}\cdot k=0$. Действительно, в системе $S$ имеем
$dx_{AB}=(0,\overrightarrow{AB})$, $dx_{CB}=(0,\overrightarrow{CB})$ и оба
вектора  $\overrightarrow{AB}$ и $\overrightarrow{CB}$, находясь целиком на волновом
фронте, перпендикулярны импульсу фотона $\vec{k}$. Второе условие состоит в том, что 
события $A$, $B$ и $C$ одновременны в системе $S$. Это условие эквивалентно утверждению,
что 4-вектора $dx_{AB}$ и $dx_{CB}$ перпендикулярны 4-скорости наблюдателя $S$.

Если скорость системы $S^\prime$, $\vec{V}$, параллельна  импульсу фотона $\vec{k}$, 
утверждение задачи будет тривиально верно, так как поперечные размеры при преобразованиях 
Лоренца не меняются. Если же вектора $\vec{V}$ и $\vec{k}$ не параллельны, в системе 
$S^\prime$ второе условие не будет выполняться, так как  4-скорость наблюдателя 
$S^\prime$, $u=(c\gamma,\vec{V}\gamma)$, не будет перпендикулярна 4-векторам 
$dx_{AB}$ и $dx_{CB}$. В этом случае на мировых линиях фотонов $A$ и $C$ надо найти 
такие события $A^\prime$ и $C^\prime$, которые будут одновременны с событием $B$ 
в системе $S^\prime$. Для этой цели параметры $\lambda$ и $\mu$ в 4-векторах 
$x_{A^\prime}=x_A+\lambda k$ и $x_{C^\prime}=x_C+\mu k$ 
надо подобрать таким образом, чтобы 4-вектора $dx_{A^\prime B}=x_{A^\prime}-x_B=
dx_{AB}+\lambda k$ и $dx_{C^\prime B}=x_{C^\prime}-x_B=dx_{CB}+\mu k$ были 
перпендикулярны 4-скорости u. Это однозначно определяет $\lambda$ и $\mu$.
Окончательно будем иметь (заметим, что $k\cdot u\ne 0$, в чем легко можно убедиться,
вычислив этот инвариант в системе покоя наблюдателя $S^\prime$, где 
$u=(c,\,\vec{0})$):
$$dx_{A^\prime B}=dx_{AB}-\frac{dx_{AB}\cdot u}{k\cdot u}\,k,\;\;
dx_{C^\prime B}=dx_{CB}-\frac{dx_{CB}\cdot u}{k\cdot u}\,k.$$ 
Эти два пространственно подобных 4-вектора, $dx_{A^\prime B}$ и $dx_{C^\prime B}$,
лежат на одном и том же волновом фронте в системе $S^\prime$ и определяют элементарную 
площадь $dS^\prime$ в этой системе как $dS^\prime=|dx_{A^\prime B}|
|dx_{C^\prime B}|$. Но, так как $k\cdot k=k\cdot dx_{AB}= k\cdot dx_{CB}=0$, 
легко видеть, что $|dx_{A^\prime B}|= |dx_{AB}|,\;|dx_{C^\prime B}|= |dx_{CB}|$ 
и, следовательно, $dS^\prime=dS$.

{\usefont{T2A}{cmr}{m}{it} Третье решение} \cite{23A}.
Рассмотрим два фотона $A$ и $B$ на волновом фронте (т.~е. $k\cdot dx_{AB}=0$). 
Эти фотоны определяют два луча, расстояние между которыми в системе $S$ равно 
$AB=|dx_{AB}|$. Чтобы найти расстояние между этими лучами в системе $S^\prime$, надо 
найти такое событие $A^\prime$ на мировой линии фотона $A$, которое будет одновременным 
по отношению к событию $B$  в системе $S^\prime$ (т.~е. требуется условие
$u\cdot dx_{A^\prime B}=0$. Легко видеть, что условие $k\cdot dx_{A^\prime B}=0$
будет автоматически выполняться, так как  $dx_{A^\prime B}=dx_{AB}+\lambda k$).  
Тогда расстояние между лучами в системе $S^\prime$ будет $A^\prime B=|dx_{A^\prime 
B}|$. Но даже для любого значения $\lambda$, а не только для того значения, которое 
определяется условием одновременности $u\cdot dx_{A^\prime B}=0$, релятивистский 
интервал между событиями $A^\prime$ и $B$ будет один и тот же:
$$dx_{A^\prime B}\cdot dx_{A^\prime B}=(dx_{AB}+\lambda k)\cdot
(dx_{AB}+\lambda k)=dx_{A B}\cdot dx_{A B},$$
так как $k\cdot dx_{AB}=0$ и $k\cdot k=0$. 

Таким образом, растояние между любыми лучами в параллельном пучке света не меняется при
преобразованиях Лоренца. Следовательно, площадь поперечного сечения тоже не поменяется.
Более того, параллельный пучок света с круговым сечением будет иметь круговое сечение
одного и того же радиуса во всех инерциальных системах отсчета. 

\subsection*{Задача 5 {\usefont{T2A}{cmr}{b}{n}(2.28)}: Перекрытие света диском}
{\usefont{T2A}{cmr}{m}{it} Первое решение.}
Согласно предыдущей задаче, поперечный размер пучка не меняется при преобразовании 
Лоренца. Поэтому крайние точки волнового фронта $A$ и $B$ в системе заслонки отстоят 
друг от друга на расстоянии $2R$, которое равно длине заслонки в ее системе покоя. 
Но тогда очевидно, что заслонка не может перекрыть путь фотону $B$, так как в этой 
системе фотоны падают под углом $\alpha=\arcsin{\beta}$ к вертикали (см. рисунок). 
\begin{figure}[htb]
\centerline{\epsfig{figure=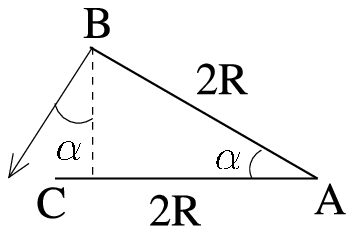,height=2.5cm}}
\end{figure}

{\usefont{T2A}{cmr}{m}{it} Второе решение.}
В фокусе линзы в каждый момент времени собираются те фотоны, которые находятся на
соответствующем волновом фронте одновременно. Пусть фотон, который выходит из правого 
края линзы $A$, блокируется правым краем заслонки $D$ (см. рисунок). 
Рассмотрим другой фотон, который выходит из другого края линзы и находится на том же 
волновом фронте, которому принадлежит первый фотон. Тогда в системе покоя $S^\prime$ 
заслонки эти фотоны покидают край линзы одновременно, так как они принадлежат одному 
и тому же волновому фронту, а в этой системе волновые фронты параллельны линии $AB$ 
линзы. Следовательно, в лабораторной системе $S$ эти фотоны покидают линзу 
неодновременно: левый край линзы фотон покидает на
$$\Delta t=\gamma\frac{V}{c^2}L=\gamma\beta\frac{L}{c}$$
время позже, где $L=2R$. Следовательно, он излучается в точке $B^\prime$ и
$$B^\prime A=AB+V\Delta t=\frac{L}{\gamma}+\gamma\beta^2 L=\gamma L.$$
Так как $B^\prime A>CD=L$, то левый край заслонки $C$ не может перекрыть путь 
фотону, который выходит из левого края линзы в точке $B^\prime$ (см. рисунок).
\begin{figure}[htb]
\centerline{\epsfig{figure=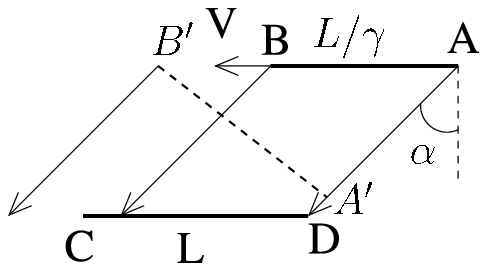,height=4cm}}
\end{figure}

\noindent Принадлежность фотонов одному и тому же волновому фронту является инвариантным 
свойством (не зависит от инерциальной системы отсчета). Проверим это в нашем случае. 
Когда фотон в лабораторной системе излучается из левого края линзы в точке $B^\prime$, 
второй фотон, который излучился из правого края линзы на $\Delta t$ раньше, находится 
в точке $A^\prime$, притом $AA^\prime=c\Delta t=\gamma\beta L$. По теореме 
косинусов,
$$A^\prime B^\prime=\sqrt{\gamma^2\beta^2 L^2+\gamma^2 L^2-2\gamma^2\beta L^2
\cos{\left(\frac{\pi}{2}-\alpha\right )}}=L\sqrt{\gamma^2-\gamma^2\beta^2}=
L,$$
так как $\sin{\alpha}=\beta$. Следовательно, как и ожидалось, поперечная ширина пучка
не изменилась и, как можно легко проверить, 
$(AB^\prime)^2=(AA^\prime)^2+(A^\prime B^\prime)^2$. Это
означает, что $A^\prime B^\prime$ перпендикулярен лучам света в этой системе и точки 
$A^\prime$ и $B^\prime$ лежат на одном волновом фронте.

\subsection*{Задача 6 {\usefont{T2A}{cmr}{b}{n}(2.28)}: Сколько 
ступеней должен иметь релятивистская ракета?}
Пусть $x_+=ct+x$. Тогда
$$x_+^\prime=ct^\prime+x^\prime=\gamma(1-\beta)(ct+x)=\sqrt{\frac{1-\beta}
{1+\beta}}x_+.$$
Пусть теперь система $S_1$ движется со скоростью $V_1$ относительно $S$, а система 
$S_2$ со скоростью $V_2$ относительно $S_1$. Тогда
$$x_+^{(1)}=\sqrt{\frac{1-\beta_1}{1+\beta_1}}x_+\;\;\mbox{и}\;\;
x_+^{(2)}=\sqrt{\frac{1-\beta_2}{1+\beta_2}}x_+^{(1)}.$$
Поэтому
$$x_+^{(2)}=\sqrt{\frac{(1-\beta_1)(1-\beta_2)}{(1+\beta_1)(1+\beta_2)}}x_+.$$
С другой стороны,
$$x_+^{(2)}=\sqrt{\frac{1-\beta}{1+\beta}}x_+,$$
где $\beta$ есть скорость $S_2$ относительно $S$. В случае $n$ коллинеарных 
скоростей получим
$$\frac{1-\beta}{1+\beta}=\frac{(1-\beta_1)(1-\beta_2)\cdots(1-\beta_n)}{
(1+\beta_1)(1+\beta_2)\cdots(1+\beta_n)}.$$
Если $\beta_1=\beta_2=\cdots=\beta_n$, то отсюда
$$\ln{\frac{1-\beta}{1+\beta}}=n\ln{\frac{1-\beta_1}{1+\beta_1}}$$
и для $\beta=0,9,\,\beta_1=0,1$ будем иметь $n\approx 14,7$. Следовательно, нужно
$n=15$ ступеней, чтобы достичь скорости $\beta=0,9$.

\section[Семинар 9]
{\centerline{Семинар 9}}

\subsection*{Задача 1 {\usefont{T2A}{cmr}{b}{n}(2.32)}: В какой системе отсчета
длины стержней будут равными?}
{\usefont{T2A}{cmr}{m}{it} Первое решение.}
Пусть система отсчета $S^\prime$ движется вдоль стержней со скоростью $u$. Тогда 
относительно $S^\prime$ первый стержень движется со скоростью $V_1=-u$, а второй 
стержень --- со скоростью $$V_2=\frac{V-u}{1-\frac{Vu}{c^2}}.$$ Длины стержней 
будут одинаковыми в системе $S^\prime$, если величины их скоростей равны, т.~е.
$$u=\frac{V-u}{1-\frac{Vu}{c^2}}.$$ 
Отсюда $$u^2-2\frac{c^2}{V}\,u+c^2=0$$ и $$u=\frac{c^2}{V}\pm\sqrt{
\frac{c^4}{V^2}-c^2}=\frac{c}{\beta_V\gamma_V}(\gamma_V\pm 1).$$
Так как $\beta\gamma=\sqrt{\gamma^2-1}$, если возьмем знак плюс, будем иметь
$$u=c\frac{\gamma_V+1}{\sqrt{\gamma_V^2-1}}=c\sqrt{\frac{\gamma_V+1}
{\gamma_V-1}}>c,$$
что недопустимо. Следовательно, подходит только знак минус и окончательно
$$u=c\frac{\gamma_V-1}{\sqrt{\gamma_V^2-1}}=c\sqrt{\frac{\gamma_V-1}
{\gamma_V+1}}.$$

{\usefont{T2A}{cmr}{m}{it} Второе решение.}
Длины стержней будут одинаковыми в системе $S^\prime$, если их гамма-факторы 
равны. Но заметим, что $$\gamma_2^{-2}=1-\left(\frac{\beta_V-\beta_u}{1-
\beta_V\beta_u}\right)^2=\frac{(1-\beta_V^2)(1-\beta_u^2)}
{(1-\beta_V\beta_u)^2},$$
и, следовательно, $\gamma_2=\gamma_V\gamma_u (1-\beta_V\beta_u)$.
Тогда условие $\gamma_1=\gamma_2$, с учетом $\gamma_1=\gamma_u$, дает уравнение
$\gamma_V (1-\beta_V\beta_u)=1$, решением которого является 
$$\beta_u=\frac{\gamma_V- 1}{\beta_V\gamma_V}=\sqrt{\frac{\gamma_V-1}
{\gamma_V+1}}.$$
Заметим, что этот способ не требует решения квадратного уравнения с выбором 
правильного знака.

\subsection*{Задача 2 {\usefont{T2A}{cmr}{b}{n}(2.40)}: Найти минимальное
расстояние между частицами}
{\usefont{T2A}{cmr}{m}{it} Первое решение.}
а) В Л-системе координаты частиц меняются так:
\begin{equation}
x_1(t)=Vt,\;\; y_1(t)=0,\;\; x_2(t)=L,\;\; y_2(t)=Vt.
\label{eq9_2a}
\end{equation}
Поэтому расстояние между частицами $l(t)=\sqrt{(Vt-L)^2+V^2t^2}$. Надо найти
минимум $l(t)$. Удобнее минимизировать $l^2(t)=(Vt-L)^2+V^2t^2$. Приравнивая
производную этой функции к нулю, получаем уравнение $V(Vt-L)+V^2t=0$,
решением которого является $t_m=L/(2V)$. Следовательно, 
$$l_m^2=\left(V\frac{L}{2V}-L\right)^2+V^2\frac{L^2}{4V^2}=\frac{L^2}{2},
\;\;\mbox{и}\;\;l_m=\frac{L}{\sqrt{2}}.$$

б) В системе первой частицы ($S^\prime$) скорость второй частицы имеет 
компоненты
$$V_{2x}^\prime=\frac{V_{2x}-V_1}{1-\frac{V_{2x}V_1}{c^2}}=-V,\;\;\;
V_{2y}^\prime=\frac{V_{2y}}{\gamma\left(1-\frac{V_{2x}V_1}{c^2}\right)}=
\frac{V}{\gamma}.$$
В Л-системе при $t_0=0$ координаты второй частицы были $x_{20}=L,\,y_{20}=0$. 
По преобразованиям Лоренца находим соответствующие координаты в системе
$S^\prime$:
$$t_0^\prime=\gamma\left(t_0-\frac{V}{c^2}x_{20}\right)=-\gamma\frac{V}{c^2}L,
\;\;x_{20}^\prime=\gamma(x_{20}-Vt_0)=\gamma L,\;\;y_{20}^\prime=y_{20}=0.$$
Поэтому уравнения движения второй частицы в системе $S^\prime$ имеют вид
\begin{equation}
x_2^\prime(t)=x_{20}^\prime+V_{2x}^\prime (t^\prime-t_0^\prime)=
\gamma L-V\left(t^\prime+\gamma\frac{V}{c^2}L\right)=\frac{L}{\gamma}-
Vt^\prime
\label{eq9_2b}
\end{equation}
и
\begin{equation}
y_2^\prime(t)=y_{20}^\prime+V_{2y}^\prime (t^\prime-t_0^\prime)=
\frac{V}{\gamma}\left(t^\prime+\gamma\frac{V}{c^2}L\right)=\beta^2L+
\frac{V}{\gamma}t^\prime.
\label{eq9_2c}
\end{equation}
Следовательно, расстояние между частицами $l^\prime(t^\prime)$ в системе 
$S^\prime$ удовлетворяет уравнению 
$$l^{\prime\,2}(t^\prime)=\left(\frac{L}{\gamma}-Vt^\prime\right)^2+
\left(\beta^2L+\frac{V}{\gamma}t^\prime\right)^2.$$
Приравнивая производную этой функции к нулю, получаем уравнение 
$$-V\left(\frac{L}{\gamma}-Vt_m^\prime\right)+\frac{V}{\gamma}
\left(\beta^2L+\frac{V}{\gamma}t_m^\prime\right)=0.$$
Это определяет 
$$Vt_m^\prime=\frac{L}{\gamma (\gamma^2+1)}.$$
Следовательно,
$$l_m^{\prime\,2}=\left(\frac{L}{\gamma}-\frac{L}{\gamma(\gamma^2+1)}
\right)^2+\left(\beta^2L+\frac{1}{\gamma}\frac{L}{\gamma(\gamma^2+1)}
\right)^2=\frac{L^2\gamma^2}{\gamma^2+1},$$
и
$$l_m^{\prime}=\frac{L\gamma}{\sqrt{\gamma^2+1}}.$$

{\usefont{T2A}{cmr}{m}{it} Второе решение.}
а) Введем относительные координаты $x=x_2-x_1$ и $y=y_2-y_1$. Из 
(\ref{eq9_2a}) видно, что $y=L-x$ --- эта прямая линия (см. рисунок). 
\begin{figure}[htb]
\centerline{\epsfig{figure=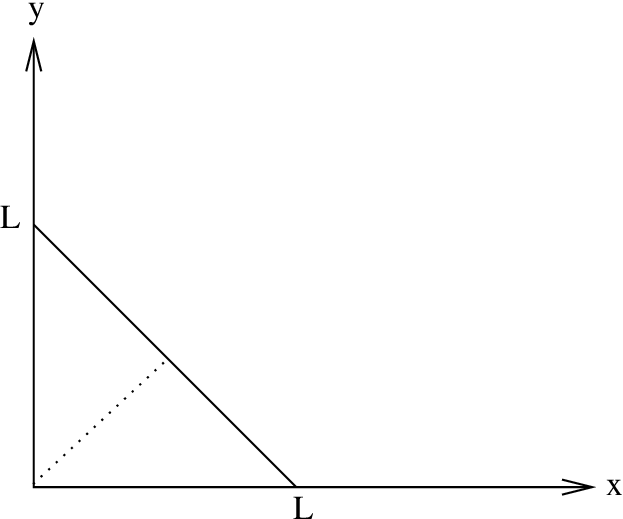,height=4cm}}
\end{figure}
Причем расстоянию между частицами $l=\sqrt{x^2+y^2}$ геометрически отвечает
расстояние от начала координат до этой линии. Из рисунка ясно, что
$$l_m=L\sin{45^\circ}=\frac{L}{\sqrt{2}}.$$

б) Из (\ref{eq9_2b}) и (\ref{eq9_2c}) следует, что
$$y_2^\prime=\beta^2 L+\frac{1}{\gamma}\left(\frac{L}{\gamma}-x_2^\prime
\right)=L-\frac{1}{\gamma}x_2^\prime.$$
\begin{figure}[htb]
\centerline{\epsfig{figure=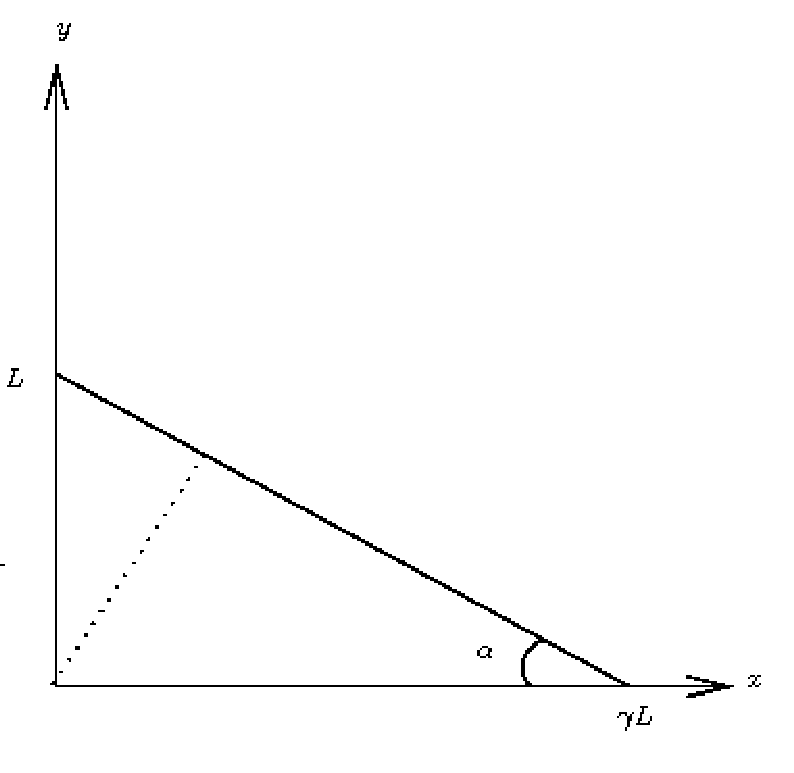,height=6cm}}
\end{figure}
Расстояние между частицами в системе $S^\prime$ то же самое, что расстояние
от начала координат до этой линии (так как $x_1^\prime=0$, $y_1^\prime=0$ 
--- первая частица в системе $S^\prime$ неподвижна).

Из рисунка ясно, что с одной стороны $l^\prime_m=L\gamma\sin{\alpha}$, 
а с другой $\cot{\alpha}=(L\gamma)/L=\gamma$. Поэтому
$$\sin{\alpha}=\frac{1}{\sqrt{1+\cot^2{\alpha}}}=\frac{1}
{\sqrt{\gamma^2+1}},$$
и $$l_m^{\prime}=\frac{L\gamma}{\sqrt{\gamma^2+1}}.$$

\subsection*{Задача 3 {\usefont{T2A}{cmr}{b}{n}(2.46)}: Скорость движения
звездочета}
Рассмотрим луч света, который в Л-системе падает под углом $\alpha$ к
вертикали, как показано на рисунке. 
\begin{figure}[htb]
\centerline{\epsfig{figure=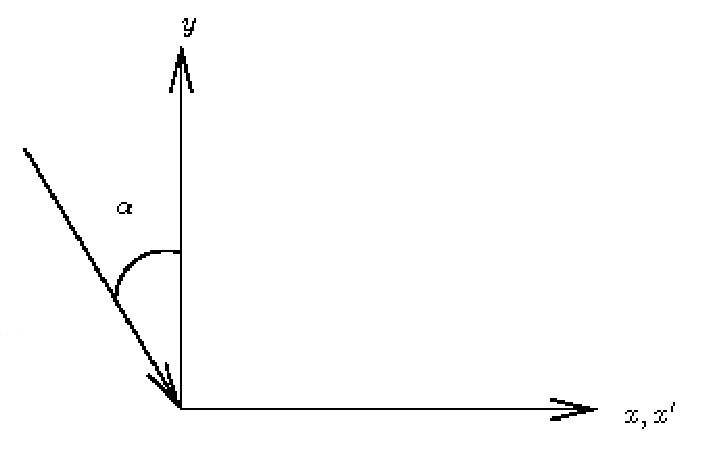,height=6cm}}
\end{figure}
Тогда $c_x=c\sin{\alpha}$, $c_y=-c\cos{\alpha}$ и в системе $S^\prime$ будем
иметь $$c_x^\prime=\frac{c_x-V}{\left (1-\frac{c_xV}{c^2}\right)}=
\frac{c\sin{\alpha}-V}{1-\beta\sin{\alpha}},\;\;c_y^\prime=
\frac{c_y}{\gamma\left(1-\frac{c_xV}{c^2}\right)}=-\frac{c\cos{\alpha}}
{\gamma(1-\beta\sin{\alpha})}.$$
Потребуем, чтобы свет в системе $S^\prime$ падал вертикально вниз. Тогда
$c^\prime_x=0$, что дает $c\sin{\alpha}-V=0$, или $\sin{\alpha}=\beta$.
Заметим, что тогда 
$$c_y^\prime=-\frac{c\sqrt{1-\beta^2}}{\gamma(1-\beta^2)}=-c\frac{1}
{\gamma\sqrt{1-\beta^2}}=-c,$$
т. е., как и положено свету, $c^\prime =c$.

Следовательно, все звезды, которые неподвижный наблюдатель видел в телесном
угле, показанным на рисунке внизу,
\begin{figure}[htb]
\centerline{\epsfig{figure=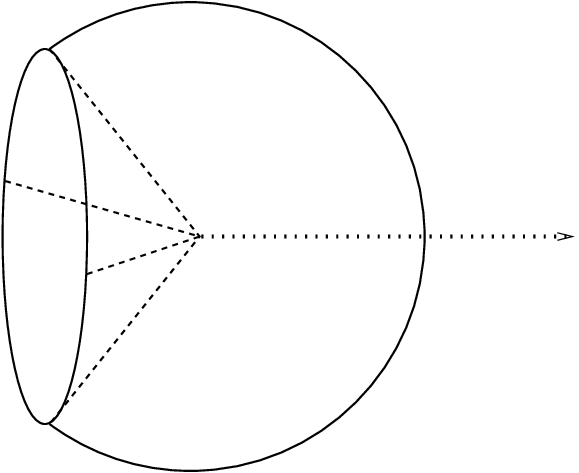,height=4cm}}
\end{figure}
движущийся наблюдатель увидит в передней полусфере. Этот телесный угол равен
$$\Omega=\int\limits_0^{\pi/2+\alpha}\sin{\theta}\,d\theta\int\limits_0^
{2\pi}d\varphi=2\pi\left(1-\cos{\left(\frac{\pi}{2}+\alpha\right)}\right)=
2\pi(1+\sin{\alpha})=2\pi(1+\beta).$$
Полный телесный угол равен $4\pi$. Следовательно, движущийся наблюдатель в
передней полусфере увидит $$\frac{2\pi(1+\beta)}{4\pi}=\frac{1}{2}(1+\beta)$$
часть всех звезд, а в задней $$1-\frac{1}{2}(1+\beta)=\frac{1}{2}(1-\beta).$$
По условию задачи, $$\frac{\frac{1}{2}(1+\beta)}{\frac{1}{2}(1-\beta)}=N.$$
Отсюда
$$\beta=\frac{N-1}{N+1}.$$

\subsection*{Задача 4 {\usefont{T2A}{cmr}{b}{n}(2.52)}: Длина и угол наклона
падающего стержня}
{\usefont{T2A}{cmr}{m}{it} Первое решение.}
\begin{figure}[htb]
\centerline{\epsfig{figure=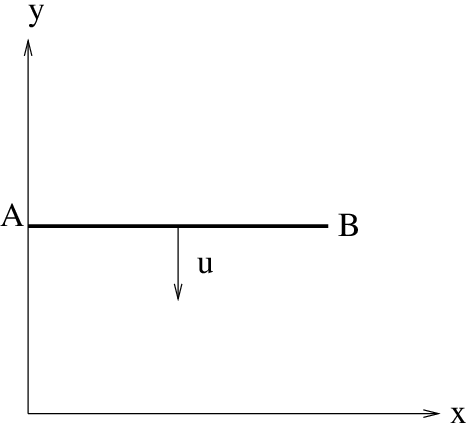,height=6cm}}
\end{figure}
Пусть точка $A$ падает на пол в начало отсчета. Т.~е.~событие ``касание точки
$A$ к полу'' имеет координаты $t_A=0,x_A=0$. Координаты события ``касание точки
$B$ к полу'' будут $t_B=0,x_B=L$. Перейдем в систему $S^\prime$ наблюдателя: 
$t_A^\prime=0,x_A^\prime=0$, но 
$$t_B^\prime=\gamma_V\left (t_B-\frac{V}{c^2}x_B\right)=-\gamma_V\beta_V
\frac{L}{c},\;\;\;x_B^\prime=\gamma_V(x_B-Vt_B)=\gamma_VL.$$
Т. е. в системе $S^\prime$ точка $B$ падает на пол раньше, чем точка $A$.
Найдем скорость стержня в системе $S^\prime$:
$$u^\prime_x=-V,\;\;\;u^\prime_y=\frac{u_y}{\gamma_V\left(1-\frac{u_xV}{c^2}
\right)}=-\frac{u}{\gamma_V},\;\;\;u^\prime=\sqrt{V^2+\frac{u^2}
{\gamma_V^2}}.$$ В момент $t^\prime=0$  в системе $S^\prime$ будем иметь
картину, показанную на рисунке.
\begin{figure}[htb]
\centerline{\epsfig{figure=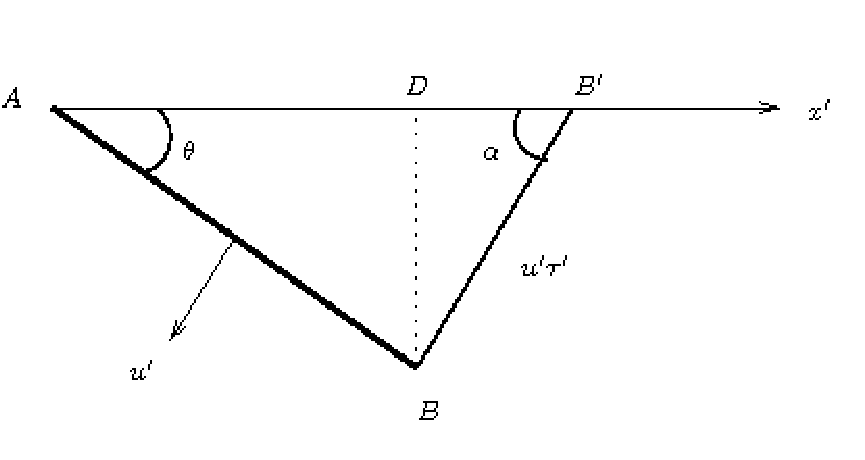,height=6cm}}
\end{figure}

\noindent При этом $$\tg{\alpha}=\left |\frac{u^\prime_y}{u^\prime_x}\right|=
\frac{u}{V\gamma_V},$$
а точка $B$ пересекает пол раньше на время $\tau^\prime=\gamma_V\beta_V 
(L/c)$ в точке $B^\prime$. Причем $AB^\prime=\gamma_VL$. Заметим, что
$$\frac{1}{\cos^2{\alpha}}=1+\tg^2{\alpha}=1+\frac{u^2}{\gamma_V^2V^2}=
\frac{1}{V^2}\left (V^2+\frac{u^2}{\gamma_V^2}\right)=\frac{u^{\prime\,2}}
{V^2}.$$
Следовательно, $u^\prime\cos{\alpha}=V$ и
$$u^\prime\sin{\alpha}=u^\prime\sqrt{1-\frac{V^2}{u^{\prime\,2}}}=
\sqrt{{u^{\prime\,2}}-V^2}=\frac{u}{\gamma_V}.$$
Из рисунка
$$\tg{\theta}=\frac{DB}{AD}=\frac{u^\prime\tau^\prime\sin{\alpha}}
{\gamma_V L-u^\prime\tau^\prime\cos{\alpha}}=\frac{u\tau^\prime/\gamma_V}
{\gamma_V L-\tau^\prime V}=\frac{\beta_V\beta_u}{\gamma_V(1-\beta_V^2)}=
\gamma_V\beta_V\beta_u.$$
Следовательно, $\theta=\arctan{(\gamma_V\beta_V\beta_u)}$. Длина стержня в
системе $S^\prime$ равна $$L^\prime=\sqrt{(AD)^2+(DB)^2}=L\sqrt{(\gamma_V-
\gamma_V\beta_V^2)^2+\beta_V^2\beta_u^2}=L\sqrt{1-\beta_V^2+\beta_V^2
\beta_u^2}.$$

{\usefont{T2A}{cmr}{m}{it} Второе решение.}
В системе $S$ $x_B=L$ всегда, так как стержень параллелен полу. Событие
``точка $A$ касается пола'' в системе $S$ имеет координаты $t_A=0,x_A=0$,
а в системе $S^\prime$ --- $t_A^\prime=0,x_A^\prime=0$. Рассмотрим положение
конца $B$ в системе $S^\prime$ в момент $t_B^\prime=0$. Тогда из
$L=x_B=\gamma_V(x_B^\prime+Vt_B^\prime)=\gamma_V x_B^\prime$ получим
$x_B^\prime=L/\gamma_V$. Но тогда 
$$t_B=\gamma_V\left(t_B^\prime+\frac{V}{c^2}x_B^\prime\right)=\gamma_V
\frac{V}{c^2}x_B^\prime=\beta_V\frac{L}{c}.$$
Поэтому $y_B^\prime=y_B=-ut_B=-\beta_V\beta_u L$.
\begin{figure}[htb]
\centerline{\epsfig{figure=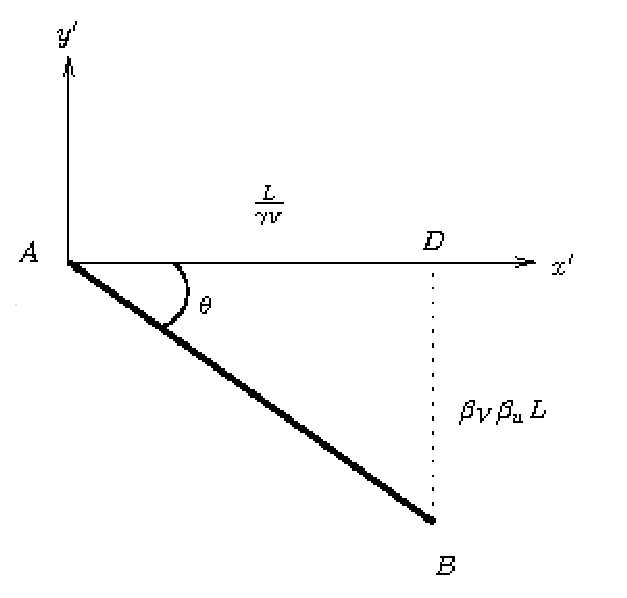,height=6cm}}
\end{figure}

Следовательно (см. рисунок),
$$L^\prime=\sqrt{\frac{L^2}{\gamma_V^2}+\beta_u^2\beta_V^2L^2}=
=L\sqrt{1-\beta_V^2+\beta_V^2\beta_u^2}$$
и $$\tg{\theta}=\frac{\beta_V\beta_u L}{L/\gamma_V}=\gamma_V\beta_V
\beta_u.$$

{\usefont{T2A}{cmr}{m}{it} Третье решение.}
В системе падающего стержня $S^{\prime\prime}$ скорость наблюдателя имеет компоненты
(ось $x^{\prime\prime}$ направлена по движению стержня, т.е. вертикально вниз. Ось 
$y^{\prime\prime}$ направлена слева направо, т.е. по движению наблюдателя 
в лабораторной системе):
$$V^{\prime\prime}_x=-u,\;\;\;V^{\prime\prime}_y=\frac{V}{\gamma_u}.$$  
Таким образом, направление движения наблюдателя составляет со стержнем угол $\beta$
такой, что
\begin{equation}
\tg{\beta}=\frac{|V^{\prime\prime}_x|}{|V^{\prime\prime}_y}=\frac{\beta_u
\gamma_u}{\beta_V}.
\label{eq9_4a}
\end{equation}
Следовательно, при переходе в систему наблюдателя $S^\prime$, где стержень движется
со скоростью $\vec{u}\ominus\vec{V}$, продольный размер стержня $L\cos{\beta}$
сокращается $\gamma_{\vec{u}\ominus\vec{V}}$ раз, а поперечный размер
$L\sin{\beta}$ не меняется. Но
$$\gamma_{\vec{u}\ominus\vec{V}}=\gamma_u\gamma_V\left(1-\frac{\vec{u}\cdot
\vec{V}}{c^2}\right)=\gamma_u\gamma_V.$$
Поэтому получаем
\begin{equation}
L^{\prime\,2}=\left(\frac{L\cos{\beta}}{\gamma_u\gamma_V}\right)^2+L^2
\sin^2{\beta}.
\label{eq9_4b}
\end{equation}
С учетом $\beta^2\gamma^2=\gamma^2-1$, из (\ref{eq9_4a}) следует, что
$$\cos^2{\beta}=\frac{1}{1+\tg^2{\alpha}}=\frac{\beta_V^2}{\beta_V^2+
\gamma_u^2-1}=\frac{\beta_V^2}{\gamma_u^2-\frac{1}{\gamma_V^2}}=
\frac{\gamma_V^2-1}{\gamma_u^2\gamma_V^2-1}.$$
и
$$\sin^2{\beta}=1-\cos^2{\alpha}=\frac{\gamma_V^2(\gamma_u^2-1)}
{\gamma_u^2\gamma_V^2-1}.$$
Поэтому (\ref{eq9_4b}) принимает вид 
\begin{equation}
L^{\prime\,2}=\frac{L^2}{\gamma_u^2\gamma_V^2-1}\left [\frac{\gamma_V^2-1}
{\gamma_u^2\gamma_V^2}+\gamma_u^2\gamma_V^2-\gamma_V^2\right].
\label{eq9_4c}
\end{equation}
Но выражение в квадратных скобках можно переписать так
$$\frac{\gamma_V^2-1}{\gamma_u^2\gamma_V^2}-(\gamma_V^2-1)+
\gamma_u^2\gamma_V^2-1=(\gamma_u^2\gamma_V^2-1)\left[1-\frac{\gamma_V^2-1}
{\gamma_u^2\gamma_V^2}\right]=(\gamma_u^2\gamma_V^2-1)\left[1-\frac{\beta_V^2}
{\gamma_u^2}\right].$$
Следовательно, длина стержня в системе $S^\prime$ равна
$$L^\prime=L\sqrt{1-\frac{\beta_V^2}{\gamma_u^2}}=
L\sqrt{1-\beta_V^2(1-\beta_u^2)}=L\sqrt{1-\beta_V^2+\beta_V^2\beta_u^2}.$$

Вычисление угла наклона стержня к горизонту этим способом содежит один тонкий момент.
Наивно было бы считать, что ось $x^\prime$ составляет с направлением 
относительного движения стержня и наблюдателя тот же угол $\beta$, который мы 
вычислили в системе $S^{\prime\prime}$, так как оси $x^\prime$ и 
$y^{\prime\prime}$ по отдельности параллельны оси $x$. Это не так. При 
решении задачи первым способом мы уже нашли, что направление относительного движения 
составляет с осью $x^\prime$ угол $\alpha$ такой, что
$$\tg{\alpha}=\frac{\beta_u}{\gamma_V\beta_V}.$$
Сравнивая с (\ref{eq9_4a}), мы видим, что $\tg{\beta}>\tg{\alpha}$ и, 
следовательно, $\beta>\alpha$. Заметим, что разность $\epsilon=\beta-\alpha$
есть пресловутый поворот Вигнера (или Вигнера--Томаса, так как он приводит к прецессии
Томаса) \cite{24,25,26}.  

Таким образом, имеем следующую картину (на рисунке $AC$ соответствует линии 
относительного движения стержня $AB$ и наблюдателя) 
\begin{figure}[htb]
\centerline{\epsfig{figure=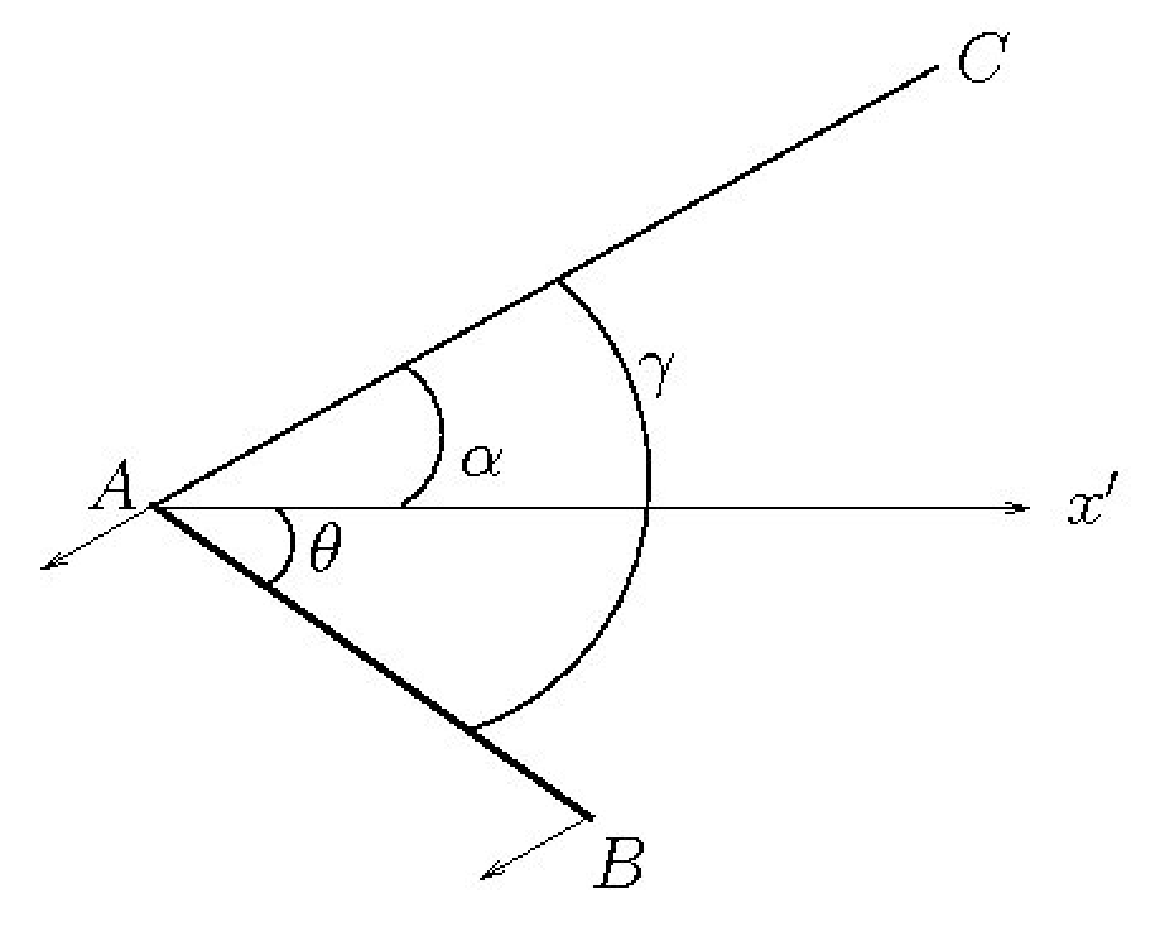,height=6cm}}
\end{figure}

При этом тангенс угла $\gamma$ равен отношению поперечного размера стержня 
к продольному:
$$\tg{\gamma}=\frac{L\sin{\beta}}{L\cos{\beta}/\gamma_u\gamma_V}=
\gamma_u\gamma_V\tg{\beta}=\frac{\gamma_u^2\gamma_V\beta_u}{\beta_V}.$$
Тогда, так как $\theta=\gamma-\alpha$, имеем
$$\tg{\theta}=\frac{\tg{\gamma}-\tg{\alpha}}{1+\tg{\gamma}\,\tg{\alpha}}=
\frac{\beta_V\beta_u}{\gamma_V}\,\frac{\gamma_u^2\gamma_V^2-1}{\beta_V^2+
\beta_u^2\gamma_u^2}.$$
Но
$$\beta_V^2+\beta_u^2\gamma_u^2=\beta_V^2+\gamma_u^2-1=\gamma_u^2-\frac{1}
{\gamma_V^2}=\frac{\gamma_u^2\gamma_V^2-1}{\gamma_V^2}$$
и окончательно получаем $\tg{\theta}=\gamma_V\beta_V\beta_u$.
 
\subsection*{Задача 5 {\usefont{T2A}{cmr}{b}{n}(2.44)}: В какой точке выйдет
свет раньше всего?}
{\usefont{T2A}{cmr}{m}{it} Первое решение.}
Пусть начало системы отсчета находится в точке вспышки $A$. Если какой-то
луч в системе $S^\prime$ (в системе пластинки) выходит в точке с координатой
$x^\prime$, то этому будет соответствовать момент времени
$$t^\prime=\frac{n}{c}\sqrt{x^{\prime\,2}+d^2}.$$
Тогда в Л-системе $$t=\gamma\left (t^\prime+\frac{V}{c^2}x^\prime\right)=
\gamma\left [\frac{n}{c}\sqrt{x^{\prime\,2}+d^2}+\frac{V}{c^2}x^\prime
\right ]$$ Найдем минимум $t$ как функции от $x^\prime$. Уравнение
$$\frac{dt}{dx^\prime}=\gamma\left
[\frac{n}{c}\,\frac{x^\prime}{\sqrt{x^{\prime\,2}+d^2}}+\frac{V}{c^2}
\right ]=0$$
дает 
\begin{equation}
\frac{x^\prime}{\sqrt{x^{\prime\,2}+d^2}}=-\frac{\beta}{n}.
\label{eq9_5a}
\end{equation}
Отсюда следует, что минимуму $t$ соответствует тот луч, который в системе
пластинки выходит в точке с координатой (заметим, что из (\ref{eq9_5a}) 
следует, что $x^\prime<0$) $$x^\prime=-\frac{\beta d}{\sqrt{n^2-\beta^2}}.$$
Тогда в системе $S^\prime$ время выхода луча из пластинки будет
$$t^\prime=\frac{n}{c}\sqrt{\frac{\beta^2 d^2}{n^2-\beta^2}+d^2}=
\frac{n^2d}{c}\frac{1}{\sqrt{n^2-\beta^2}}.$$
В лабораторной системе точка выхода света из пластинки имеет $x$-координату
$$x=\gamma(x^\prime+Vt^\prime)=\gamma\left(-\frac{\beta d}
{\sqrt{n^2-\beta^2}}+\beta n^2d\frac{1}{\sqrt{n^2-\beta^2}}\right)=
\gamma\beta d\frac{n^2-1}{\sqrt{n^2-\beta^2}}.$$

{\usefont{T2A}{cmr}{m}{it} Второе решение.}
Так как в Л-системе пластинка имеет тот же самый поперечный размер $d$, что и в своей 
собственной системе,  первым из нее выйдет тот луч, который обладает максимальной 
поперечной скоростю $c_{ym}$, и время выхода $$t_m=\frac{d}{c_{ym}}.$$
Согласно релятивистскому закону сложения скоростей,
$$c_y=\frac{c_y^\prime}{\gamma(1+\frac{c_x^\prime V}{c^2})}=
\frac{c\,\cos{\alpha}}{\gamma (n+\beta\sin{\alpha})},$$
где $\alpha$ --- угол, который луч составляет с вертикалью в системе пластинки, и мы 
учли, что $$c_x^\prime=\frac{c}{n}\sin{\alpha},\;\;\;
c_y^\prime=\frac{c}{n}\cos{\alpha}.$$
Находим минимум функции $c_y(\alpha)$. Условие
$$\frac{dc_y}{d\alpha}=-\frac{c}{\gamma}\,\frac{n\sin{\alpha}+\beta}
{(n+\beta\sin{\alpha})^2}=0$$
дает $$\sin{\alpha_m}=-\frac{\beta}{n},\;\;\mbox{и, следовательно,}\;\;
\cos{\alpha_m}=\sqrt{1-\frac{\beta^2}{n^2}}.$$
Поэтому $$c_{ym}=\frac{c\sqrt{1-\frac{\beta^2}{n^2}}}{\gamma\left(n-
\frac{\beta^2}{n^2}\right )}=\frac{c}{\gamma\sqrt{n^2-\beta^2}}$$
и $$t_m=\frac{d}{c}\gamma\sqrt{n^2-\beta^2}.$$
Найдем горизонтальную скорость луча:
$$c_{xm}=\frac{c_{xm}^\prime+V}{1+\frac{c_{xm}^\prime V}{c^2}}=
\frac{\frac{c}{n}\sin{\alpha_m}+V}{1+\frac{\beta}{n}\sin{\alpha_m}}=
\frac{n^2-1}{n^2-\beta^2}\,\beta c.$$
Следовательно, в Л-системе точка выхода света из пластинки имеет $x$-координату
$$x=c_{xm}t_m=\frac{n^2-1}{n^2-\beta^2}\;\beta c\,\frac{d}{c}\gamma
\sqrt{n^2-\beta^2}=\gamma\beta d\frac{n^2-1}{\sqrt{n^2-\beta^2}}.$$

\section[Семинар 10]
{\centerline{Семинар 10}}

\subsection*{Задача 1 {\usefont{T2A}{cmr}{b}{n}(2.37)}: Релятивистский
эскалатор}
{\usefont{T2A}{cmr}{m}{it} Первое решение.}
Пусть в лабораторной системе (в системе эскалатора) длина эскалатора равна
$L$. Тогда длина одной ступеньки в Л-системе будет $l=L/N$, а собственная
длина ступеньки --- $l_0=\gamma_V\,l=(L/N)\gamma_V$. Перейдем в систему
$S^\prime$, которая движется со скоростью $V$ вдоль ленты эскалатора, так
что в этой системе лента неподвижна, пассажир движется со скоростью $u$, а
край эскалатора движется со скоростью $V$ навстречу пассажиру. Так как в
системе $S^\prime$ длина эскалатора равна $L/\gamma_V$, пассажир сбегает по
нему (встретится с противоположным краем  эскалатора) за время 
$$\tau=\frac{L}{\gamma_V(u+V)}.$$
За это время пассажир пройдет по неподвижному эскалатору участок длиной
$$s=u\tau=\frac{Lu}{\gamma_V(u+V)}.$$  Так как в системе $S^\prime$ лента
эскалатора неподвижна, длина одной ступеньки будет $l_0=(L/N)\gamma_V$.
Следовательно, искомое число ступенек равно
$$n=\frac{s}{l_0}=\frac{Nu}{\gamma_V^2(u+V)}=N\frac{1-\frac{V^2}{c^2}}
{1+\frac{V}{u}}=N\frac{1-\frac{1}{9}}{1+\frac{2}{3}}=\frac{8}{15}N.$$

{\usefont{T2A}{cmr}{m}{it} Второе решение.}
Перейдем в систему пассажира. В этой системе лента под ногами пассажира 
движется со скоростью $u$, край  эскалатора приближается к нему со скоростью
\begin{equation}
V^\prime=\frac{u+V}{1+\frac{uV}{c^2}},
\label{eq10_1a}
\end{equation}
а сам эскалатор имеет длину $L^\prime=L/\gamma_{V^\prime}$. Следовательно,
пассажир сбегает по эскалатору (точнее, край эскалатора придет к пассажиру) за
время $t^\prime=L^\prime/V^\prime$. При этом мимо пассажира пройдет участок
ленты длиной $$s^\prime=ut^\prime=\frac{uL}{\gamma_{V^\prime}\,V^\prime}=
\frac{Lu}{\gamma_u\gamma_V(u+V)},$$
где мы на последнем этапе учли, что, как следует из (\ref{eq10_1a}),
$$\gamma_{V^\prime}=\gamma_u\gamma_V\left (1+\frac{uV}{c^2}\right).$$
В системе пассажира ступенька имеет длину $$l^\prime=\frac{l_0}{\gamma_u}=
\frac{L}{N}\,\frac{\gamma_V}{\gamma_u}.$$
Поэтому искомое число ступенек равно
$$n=\frac{s^\prime}{l^\prime}=\frac{Nu}{\gamma_V^2(u+V)}.$$

{\usefont{T2A}{cmr}{m}{it} Третье решение.}
Останемся в лабораторной системе отсчета. В этой системе пассажир движется со
скоростью $V^\prime$, которая дается формулой (\ref{eq10_1a}). Поэтому он
сбегает по эскалатору за время $t=L/V^\prime$. За это время начальная
ступенька эскалатора пройдет расстояние $\Delta S=Vt=L(V/V^\prime)$, поэтому
общая длина ступенек, пройденных пассажиром, равна
$$S=L-\Delta S=L\left (1-\frac{V}{V^\prime}\right).$$
В Л-системе ступенька имеет длину $l=L/N$, поэтому искомое число ступенек равно
$$n=\frac{S}{l}=N\left (1-\frac{V}{V^\prime}\right)=\frac{Nu}{u+V}\left (
1-\frac{V^2}{c^2}\right )=N\frac{1-\frac{V^2}{c^2}}{1+\frac{V}{u}}.$$

\subsection*{Задача 2 {\usefont{T2A}{cmr}{b}{n}(2.29)}: Коэффициент пропускания
решетки}
{\usefont{T2A}{cmr}{m}{it} Первое решение.}
В Л-системе продольные размеры решетки сокращены в $\gamma$ раз.
\begin{figure}[htb]
\centerline{\epsfig{figure=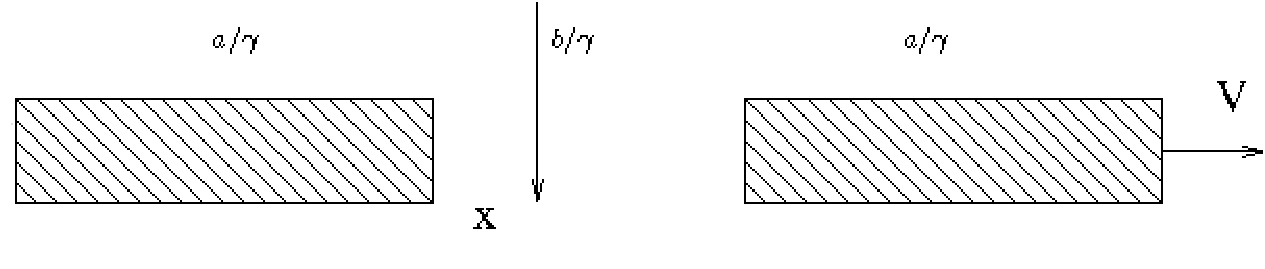,height=3cm}}
\end{figure}

\noindent Из-за движения решетки все лучи, которые на верхнем уровне решетки падают 
от правого края отверстия (см. рисунок) на расстояние меньше, чем
$x=(d/c)V=\beta d$, будут перекрыты вертикальной стенкой отверстия ($x$ ---
это расстояние, на которое смещается справа нижний край отверстия, пока фотон
летит сквозь отверстия толщиной $d$). Следовательно, если взять
периодическую ячейку решетки длиной (в Л-системе) $(a+b)/\gamma$, свет
должен попасть в участок длиной $b/\gamma-x$, чтобы он прошел сквозь
решетку. Поэтому коэффициент пропускания решетки равен
$$k=\frac{\frac{b}{\gamma}-x}{\frac{a+b}{\gamma}}=\frac{b-\beta\gamma\, d}
{a+b}.$$

{\usefont{T2A}{cmr}{m}{it} Второе решение.}
В системе решетки она неподвижна и, следовательно, не сокращена. Но в этой
системе свет падает под углом. В частности, так как в Л-системе свет падает
вертикально вниз и $c_x=0,\;c_y=c$ (ось y направлена вниз), то в системе
решетки $$c_x^\prime=\frac{c_x-V}{1-\frac{c_x V}{c^2}}=-V,\;\;\;
c_y^\prime=\frac{c_y}{\gamma \left (1-\frac{c_x V}{c^2}\right )}=\frac{c}
{\gamma}$$  и $$\tg{\alpha}=\left |\frac{c_y^\prime}{c_x^\prime}\right |=
\frac{1}{\beta\gamma}.$$
Свет должен попасть на участок $AB$ периодической ячейки решетки (см.
рисунок), чтобы пройти сквозь решетку. Но $AB=b-d\cot{\alpha}=b-\beta\gamma\,
d$, и получаем $$k=\frac{AB}{a+b}==\frac{b-\beta\gamma\, d}{a+b}.$$
\begin{figure}[htb]
\centerline{\epsfig{figure=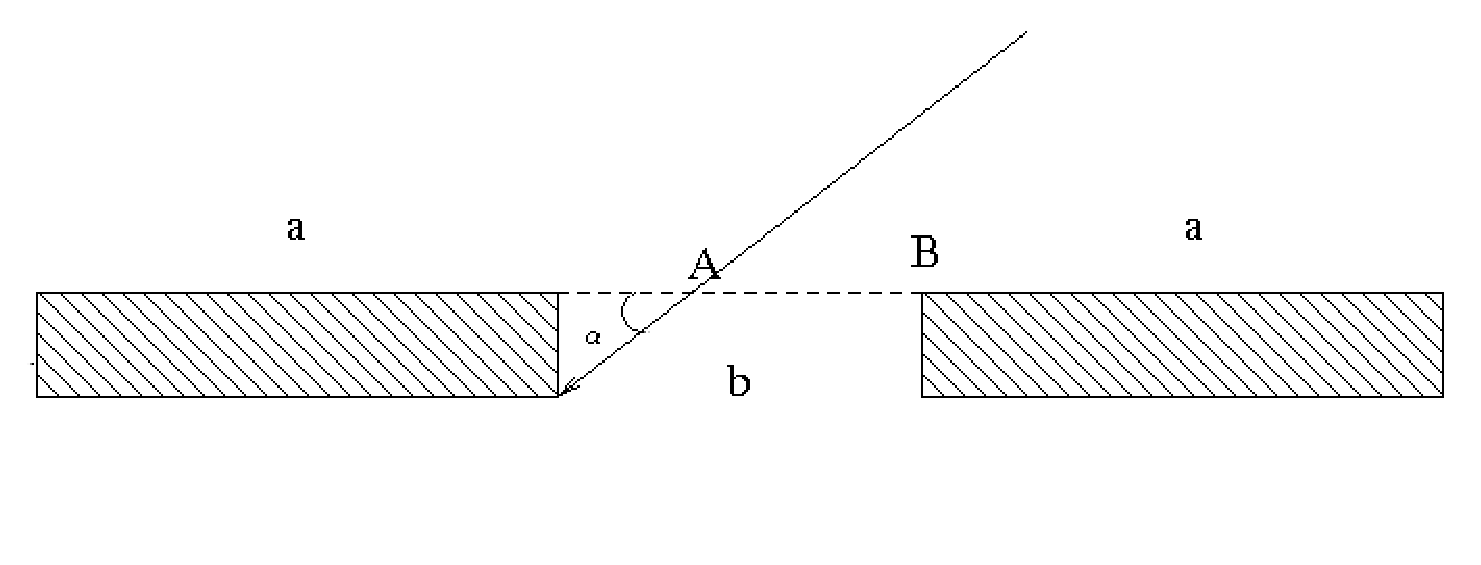,height=4cm}}
\end{figure}

\subsection*{Задача 3 {\usefont{T2A}{cmr}{b}{n}(3.8)}: Какова масса зеркала?}
За время $\Delta t$ лазер излучает энергию $E=N\Delta t$. Суммарный импульс
фотонов в этом пучке $p=(E/c)=(N/c)\Delta t$. При отражении от зеркала
импульс меняет направление на противоположное, т.~е. зеркалу передается
импульс $\Delta p=2p=(2N/c)\Delta t$. Сила давления света на зеркало 
$$F=\frac{\Delta p}{\Delta t}=2\frac{N}{c}.$$ По условию задачи $F=mg$.
Следовательно, $$m=2\frac{N}{cg}\approx 0,14\cdot 10^{-3}~\mbox{кг}.$$

\subsection*{Задача 4 {\usefont{T2A}{cmr}{b}{n}(3.9)}: Энергия и длительность
светового импульса после отражения}
Длина лазерного импульса $L=cT$. Этот импульс и зеркало сближаются со
скоростью $c+V$, поэтому процесс отражения в Л-системе длится в течение
времени $$\tau=\frac{cT}{c+V}=\frac{T}{1+\beta}.$$
За это время передний край отраженного импульса передвинется на $c\tau$, а
зеркало на $V\tau$. Следовательно, длина отраженного импульса будет
$$L_{\mbox{отр}}=(c-V)\tau=(c-V)\frac{T}{1+\beta},$$
а его длительность $$T_{\mbox{отр}}=\frac{L_{\mbox{отр}}}{c}=\frac{1-\beta}
{1+\beta}\,T.$$
Энергия и импульс падающего лазерного импульса в системе зеркала $S^\prime$
равны: 
\begin{equation}
E^\prime=\gamma(E-Vp_x),\;\;\;p_x^\prime=\gamma\left(p_x-
\frac{V}{c^2}E\right).
\label{eq10_4a}
\end{equation}
Но $p_x=-E/c$, и (\ref{eq10_4a}) принимает вид
\begin{equation}
E^\prime=\gamma(1+\beta)E,\;\;\;p_x^\prime=-\gamma\frac{E}{c}(1+\beta)=
-\frac{E^\prime}{c}.
\label{eq10_4b}
\end{equation}
В системе $S^\prime$ после отражения будем иметь
$$p_x^{\prime\prime}=-p_x^\prime=\frac{E^\prime}{c},\;\;\mbox{и}\;\;
E^{\prime\prime}=E^\prime.$$
Поэтому в Л-системе энергия отраженного импульса будет
$$E_{\mbox{отр}}=\gamma(E^{\prime\prime}+Vp_x^{\prime\prime})=
\gamma\frac{E^\prime}{c}(1+\beta),$$
и с учетом (\ref{eq10_4b}) получим окончательно
$$E_{\mbox{отр}}=\gamma^2(1+\beta)^2E=\frac{1+\beta}{1-\beta}E.$$

\subsection*{Задача 5 {\usefont{T2A}{cmr}{b}{n}(3.16)}: Минимальная кинетическая
энергия}
Пусть $S$ --- система центра масс, а система $S^\prime$ движется
относительно нее со скоростью $V$ вдоль оси $x$. Тогда кинетическая энергия
в системе $S^\prime$ равна (предполагаем, что скорость света $c=1$)
$$T^\prime=\sum_{i}(E^\prime_i-m_i)=\sum_{i} [\gamma(E_i-Vp_{ix})-m_i].$$
Но в системе центра масс $\sum_{i}p_{ix}=0$, и получаем
$$T^\prime=\sum_{i}(\gamma E_i-m_i)=\sum_{i}[(\gamma-1)E_i+(E_i-m_i)]=
(\gamma-1)E+T>T,$$
так как $\gamma>1$ и полная энергия системы $E=\sum_{i}E_i>0$.

\section[Семинар 11]
{\centerline{Семинар 11}}

\subsection*{Задача 1 {\usefont{T2A}{cmr}{b}{n}(3.11)}: Какова должна быть
мощность лазера?}
Пусть импульс лазера длится время $T$ и имеет энергию $E$. В системе корабля
энергия импульса будет $E^\prime=\gamma(E-Vp_x)=\gamma(1-\beta)E$, так как
$p_x=E/c$. Пусть длительность принимаемого импульса $T^\prime$;
$T^\prime=\Delta t^\prime$ --- это время между приходом на корабле переднего 
фронта импульса и приходом заднего фронта. В системе Земли импульс лазера 
имеет длину $cT$ и догоняет корабль со скоростью света. Поэтому интервал 
времени между этими двумя событиями есть $$\Delta t=\frac{cT}{c-V}
=\frac{T}{1-\beta},$$ а пространственный интервал --- $\Delta x=V\Delta t$, 
так как корабль движется со скоростью $V$. Следовательно,
$$T^\prime=\gamma\left (\Delta t-\frac{V}{c^2}\Delta x\right )=
\gamma(1-\beta^2)\Delta t=\frac{\Delta t}{\gamma}=\frac{T}{\gamma(1-\beta)}.$$
Принимаемая мощность в системе корабля равна
$$N=\frac{E^\prime}{T^\prime}=\gamma^2(1-\beta^2)\frac{E}{T}=
\frac{1-\beta}{1+\beta}\frac{E}{T}.$$
Поэтому мощность лазера в системе Земли должна быть
$$N_0=\frac{E}{T}=\frac{1+\beta}{1-\beta}N=\frac{1+0,5}{1-0,5}N=3N.$$

\subsection*{Задача 2 {\usefont{T2A}{cmr}{b}{n}(3.13)}:  Какова максимальная
кинетическая энергия электронов?}
Пусть средняя энергия электронов в пучке $E$, а импульс $p$. Тогда
$E=mc^2\gamma,\;p=mV\gamma$ показывают, что скорость сопутствующей пучку
системы отсчета есть
\begin{equation}
V=\frac{pc^2}{E}.
\label{eq11_2a}
\end{equation}
Если энергия какого нибудь конкретного электрона в пучке есть $E+\Delta E$,
а импульс $p+\Delta p$, то мы должны иметь
\begin{equation}
\frac{(E+\Delta E)^2}{c^2}-(p+\Delta p)^2=m^2c^2.
\label{eq11_2b}
\end{equation}
Но $$\frac{E^2}{c^2}-p^2=m^2c^2$$ и (\ref{eq11_2b}) дает с точностью до
линейных по $\Delta E$ членов 
\begin{equation}
\Delta p\approx \frac{E\Delta E}{pc^2}=
\frac{\Delta E}{V}.
\label{eq11_2c}
\end{equation}
Импульс электрона в сопутствующей пучку системе находим по преобразованию
Лоренца $$p_x^\prime=\gamma\left (p_x-\beta\frac{E+\Delta E}{c}\right )=
\gamma\left (p+\Delta p-\beta\frac{E+\Delta E}{c}\right ).$$
Подставляя $V$ из (\ref{eq11_2a}), находим
$$\gamma\left (p-\beta\frac{E}{c}\right )=0$$
(это выражение дает импульс ``среднего'' электрона в сопутствующей системе,
где этот электрон покоится). Следовательно, с учетом (\ref{eq11_2c}), 
$$p_x^\prime=\gamma\left (\Delta p-\beta\frac{\Delta E}{c}\right )=
\gamma\frac{\Delta E}{V}(1-\beta^2)=\frac{\Delta E}{\gamma V}.$$
Но $$\gamma=\frac{E}{mc^2}\gg 1.$$ Поэтому $V\approx c$, и
$$p_x^\prime\approx \frac{\Delta E}{E}mc\ll mc.$$
Кинетическая энергия этого нерелятивистского электрона равна
$$T^\prime=\frac{p^{\prime\,2}}{2m}=\frac{1}{2}\left(\frac{\Delta E}{E}
\right )^2 mc^2\approx 25~\mbox{эВ}.$$
Заметим, что $p^\prime$ линейно зависит от $\Delta E$, а $T^\prime$ ---
квадратично. Поэтому если бы воспользовались формулой преобразования
энергии, а не импульса, пришлось бы сохранить члены порядка  $(\Delta E)^2$,
и было бы сложнее получить правильный ответ. Тем не менее это возможно, и
для полноты картины  приведем такой вывод тоже.

Из (\ref{eq11_2b}) получим следующее уравнение для $\Delta p$ (для простоты
положим $c=1$ в промежуточных выражениях)
$$(\Delta p)^2+2p\Delta p-2E\Delta E-(\Delta E)^2=0,$$
решением которого является $$\Delta p=-p\pm\sqrt{p^2+2E\Delta E+
(\Delta E)^2}.$$ Ясно, что подходит только знак плюс, так как $\Delta p$ ---
инфинитезимальная величина. Используя $$\sqrt{1+x}\approx 1+\frac{x}{2}-
\frac{x^2}{8},\;\;\mbox{для}\;\;x\ll 1,$$ получаем
\begin{equation}
\Delta p\approx \frac{E\Delta E}{p}-\frac{(\Delta E)^2}{2p}\,\frac{m^2}{p^2}.
\label{eq11_2d}
\end{equation}
Находим энергию электрона в сопутствующей пучку системе отсчета:
$$E^\prime=\gamma [E+\Delta E-V(p+\Delta p)].$$
Но из (\ref{eq11_2a}) следует равенство $\gamma (E-Vp)=m$ (что естественно,
так как ``средний'' электрон в сопутствующей системе покоится). Поэтому
$E^\prime=m+\gamma(\Delta E-V\Delta p)$ и с учетом (\ref{eq11_2d}) получаем
$$T^\prime=\gamma(\Delta E-V\Delta p)=\frac{1}{2}\,\frac{(\Delta E)^2}{p^2}
m\approx \frac{1}{2}\left(\frac{\Delta E}{E} \right )^2 mc^2,$$
где на последнем этапе восстановили скорость света $c$.

\subsection*{Задача 3 {\usefont{T2A}{cmr}{b}{n}(3.28)}:  Ускорение идеального
зеркала лучом лазера}
Пусть в какой-то момент зеркало имеет скорость $V$. За время $\Delta t$ на
зеркало успеют упасть те фотоны, которые сначала находились на расстоянии
$\Delta l=(c-V)\Delta t$ от зеркала, так как расстояние до зеркала для
фотонов сокращается со скоростью $c-V$ (зеркало убегает со скоростью $V$, а
фотоны догоняют его, имея скорость $c$). Участок луча такой длины лазер
излучает за время $\tau=\Delta l/c=(1-\beta)\Delta t$ и, следовательно, в
нем находится энергия ${\cal{E}}=N\tau=(1-\beta)N\Delta t$, где $N$ ---
мощность лазера. Таким образом, за время  $\Delta t$ на зеркало падает
световой энергии ${\cal{E}}=(1-\beta)N\Delta t$. Отразится меньше,  
${\cal{E}}^\prime$, так как часть энергии перейдет в кинетическую энергию 
зеркала. Если начальная энергия и импульс зеркала были $E,p$, а стали 
$E+\Delta E,p+\Delta p$, то законы сохранения энергии (зеркало идеальное) и
импульса дают $${\cal{E}}+E={\cal{E}}^\prime+E+\Delta E,\;\;\;
\frac{{\cal{E}}}{c}+p=-\frac{{\cal{E}}^\prime}{c}+p+\Delta p,$$
или
$${\cal{E}}-{\cal{E}}^\prime=\Delta E,\;\;\;\frac{{\cal{E}}+{\cal{E}}^\prime}
{c}=\Delta p.$$
Из первого уравнения выразим ${\cal{E}}^\prime$ и подставим во второе
уравнение, получим $$\Delta p=\frac{2{\cal{E}}}{c}-\frac{\Delta E}{c},$$
или, если вспомнить ${\cal{E}}=(1-\beta)N\Delta t$ и перейти от конечных
разностей к дифференциалам,
\begin{equation}
2Ndt=\frac{cdp}{1-\beta}+\frac{dE}{1-\beta}.
\label{eq11_3a}
\end{equation}
Но из $$\frac{E^2}{c^2}-p^2=m^2c^2$$ следует $$\frac{EdE}{c^2}=pdp.$$ Кроме
того, $\beta=pc/E$, поэтому (\ref{eq11_3a}) можно переписать как
$$\frac{2N}{c}dt=\frac{E+cp}{E-cp}dp=\frac{(E+cp)^2}{m^2c^4}dp.$$
Но $(E+cp)^2=(c^2p^2+m^2c^4)+c^2p^2+2cpE$. Поэтому 
$$\frac{2N}{c}dt=\left (1+2\frac{c^2p^2}{m^2c^4}+2E\frac{cp}{m^2c^4}\right )
dp.$$ В последнем члене  заменим $pdp$ на $\frac{EdE}{c^2}$, и получим
окончательно
\begin{equation}
\frac{2N}{c}dt=\left (1+2\frac{c^2p^2}{m^2c^4}\right )dp+2\frac{E^2}{m^2c^5}
dE.
\label{eq11_3b}
\end{equation}
Пусть конечная энергия зеркала равна $E_1$, а импульс $p_1$. В начале при
$t=0$ зеркало имело энергию $E_0=mc^2$ и  импульс $p_0=0$. Поэтому,
интегрируя (\ref{eq11_3b}), 
$$\frac{2N}{c}\int\limits_0^T dt=\int\limits_0^{p_1}\left
(1+2\frac{c^2p^2}{m^2c^4}\right )dp+
2\int\limits_{mc^2}^{E_1}\frac{E^2}{m^2c^5}dE,$$
получаем
\begin{equation}
\frac{2NT}{c}=p_1+\frac{2}{3}\,\frac{p_1^3}{m^2c^2}+\frac{2}{3}\left (
\frac{E_1^3}{m^2c^5}-mc\right ).
\label{eq11_3c}
\end{equation}
Чтобы получить окончательный ответ для мощности лазера $N$, подставим 
$p_1=mc\beta\gamma$  и $E_1=mc^2\gamma$ в (\ref{eq11_3c}), где 
$\beta=0,8$, $\gamma=1/0,6=5/3$. В результате получим
$$N=\frac{mc^2}{2T}\left
[\beta\gamma-\frac{2}{3}+\frac{2}{3}\gamma^3(1+\beta^3)\right ]=
\frac{8mc^2}{3T}\approx 7,6\cdot 10^9~\mbox{Вт}.$$ 

\subsection*{Задача 4 {\usefont{T2A}{cmr}{b}{n}(3.21)}:  Эффект Доплера в среде}
За время одного колебания по часам источника пройдет время $T_0=1/\nu_0$, а в 
Л-системе --- $\gamma T_0$. За это время начало светового импульса переместится на
$l=(c/n)\gamma T_0$, а источник света --- на $V\gamma T_0$. Поэтому в Л-системе
длина светового импульса, который соответствует одному колебанию, будет 
$$L=l-V\gamma T_0=\frac{\gamma}{\nu_0}\left (\frac{c}{n}-V\right ).$$
Около неподвижного наблюдателя этот импульс пройдет за время
$$T=\frac{L}{c/n}=\frac{\gamma}{\nu_0}(1-n\beta).$$
С другой стороны, $T=1/\nu_{\mbox{приб}}$. Следовательно,
$$\nu_{\mbox{приб}}=\frac{\nu_0}{\gamma(1-n\beta)}.$$
Если источник удаляется, знак $\beta$ поменяется и
$$\nu_{\mbox{уд}}=\frac{\nu_0}{\gamma(1+n\beta)}.$$  

\subsection*{Задача 5 {\usefont{T2A}{cmr}{b}{n}(4.12)}: Найти энергию
$\pi^0$-мезона }
Из закона сохранения 4-импульса $p=k_1+k_2$ следует $p^2=k_1^2+k_2^2+
2k_1\cdot k_2$. Но (в промежуточных вычислениях будем предполагать $c=1$)
$p^2=M^2$, $k_1^2=k_2^2=0$ и 
$$2k_1\cdot k_2=2(E_1E_2-\vec{k}_1\cdot\vec{K}_2)=
2E_1E_2(1-\cos{(\theta_1+\theta_2)})=4E_1E_2\sin^2{\frac{\theta_1+\theta_2}
{2}}.$$ Следовательно,
\begin{equation}
M^2=4E_1E_2\sin^2{\frac{\theta_1+\theta_2}{2}}.
\label{eq11_5}
\end{equation}
С другой стороны, сохранение поперечной компоненты импульса дает
$E_1\sin{\theta_1}=E_2\sin{\theta_2}$. Поэтому (\ref{eq11_5}) можно
переписать как в виде
$$M^2=4E_1^2\,\frac{\sin{\theta_1}}{\sin{\theta_2}}\,
\sin^2{\frac{\theta_1+\theta_2}{2}},$$
так и в виде
$$M^2=4E_2^2\,\frac{\sin{\theta_2}}{\sin{\theta_1}}\,
\sin^2{\frac{\theta_1+\theta_2}{2}}.$$
Поэтому для энергии $E_1$ и $E_2$ получаем
$$E_1=\sqrt{\frac{\sin{\theta_2}}{\sin{\theta_1}}}\,\frac{M}
{2\sin{\frac{\theta_1+\theta_2}{2}}},\;\;
E_2=\sqrt{\frac{\sin{\theta_1}}{\sin{\theta_2}}}\,\frac{M}
{2\sin{\frac{\theta_1+\theta_2}{2}}}.$$
Но $E=E_1+E_2$ и окончательно (востановили скорость света $C$)
$$E=\frac{Mc^2}{2\sin{\frac{\theta_1+\theta_2}{2}}}\left [
\sqrt{\frac{\sin{\theta_2}}{\sin{\theta_1}}}+
\sqrt{\frac{\sin{\theta_1}}{\sin{\theta_2}}} \right ].$$

\section[Семинар 12]
{\centerline{Семинар 12}}

\subsection*{Задача 1 {\usefont{T2A}{cmr}{b}{n}(4.15)}: Найти отношение
суммарных энергий}
\begin{figure}[htb]
\centerline{\epsfig{figure=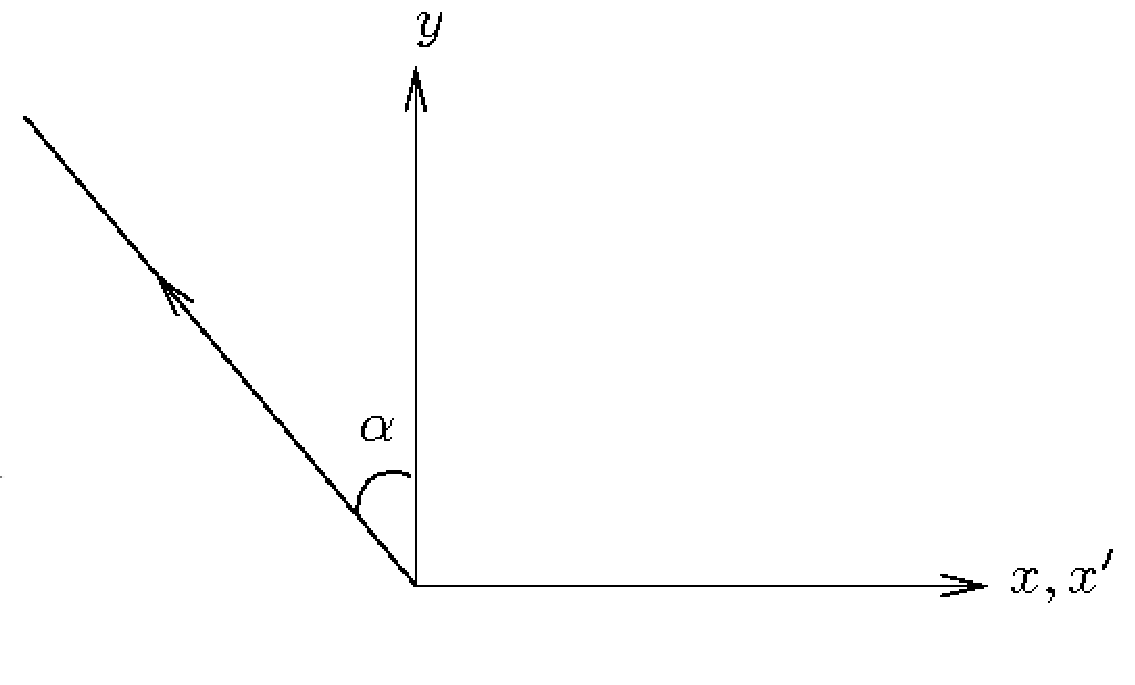,height=4cm}}
\end{figure}
Пусть в системе покоя $\pi^0$-мезона $\gamma$-квант летит под углом
$\frac{\pi}{2}+\alpha$ к направлению движения $\pi^0$-мезона. При каком
значении угла $\alpha$ этот $\gamma$-квант полетит вертикально вверх в
Л-системе? Если $c_x=0$, то формула сложения скоростей дает
$$0=c_x=\frac{c_x^\prime+V}{1+\frac{c_x^\prime V}{c^2}},$$
где $V$ --- скорость $\pi^0$-мезона. Но $c_x^\prime=-c\sin{\alpha}$.
Следовательно, должны иметь $\sin{\alpha}=\beta$. Все фотоны, которые в
системе покоя $\pi^0$-мезона летят в телесный угол
$$\Omega_1=\int\limits_0^{\pi/2+\alpha}\sin{\theta}\,d\theta\int\limits_0^
{2\pi}d\varphi=2\pi(1+\sin{\alpha})=2\pi(1+\beta),$$
в Л-системе полетят в переднюю полусферу. Пусть общее количество фотонов
$N$. Каждый фотон в системе покоя $\pi^0$-мезона имеет энергию $mc^2/2$, где
$m$ --- масса $\pi^0$-мезона. Следовательно, суммарная энергия фотонов,
которые в Л-системе летят в переднюю полусферу, в системе покоя
$\pi^0$-мезона равна $$E_1^\prime=\frac{1}{2}mc^2N\frac{2\pi(1+\beta)}{4\pi}=
\frac{1}{4}Nmc^2(1+\beta)$$ (так как в системе покоя $\pi^0$-мезона
$\gamma$-кванты вылетают изотропно, в телесный угол $\Omega_1=2\pi(1+\beta)$
попадет $N\frac{\Omega_1}{4\pi}$ фотона). Найдем суммарную
$x^\prime$-компоненту импульса этих фотонов в системе покоя $\pi^0$-мезона.
В телесный угол $d\Omega=\sin{\theta}\,d\theta\,d\varphi$ в этой системе
летит $N\frac{d\Omega}{4\pi}$ $\gamma$-квантов, и каждый имеет
$x^\prime$-компоненту импульса $\frac{1}{2}mc\,\cos{\theta}$ (так как фотон
с энергией $\frac{1}{2}mc^2$ имеет импульс $\frac{1}{2}mc$). Следовательно, 
$$p^\prime_{1x}=\int\frac{1}{2}mc\,\cos{\theta}\,N\frac{d\Omega}{4\pi}=
\frac{N}{4}mc\int\limits_0^{\pi/2+\alpha},\cos{\theta}\,\sin{\theta}\,d\theta=
\frac{N}{8}mc(1-\sin^2{\alpha})=\frac{N}{8}mc(1-\beta^2).$$
По преобразованию Лоренца находим суммарную энергию $\gamma$-квантов в
Л-системе: $$E_1=\gamma (E_1^\prime+\beta p_{1x}^\prime c)=
\frac{\gamma}{8}Nmc^2(2+3\beta-\beta^3).$$
Эта энергия летит в переднюю полусферу. Если  $\gamma$-квантов $N$, то число
распавшихся  $\pi^0$-мезонов равно $N/2$. Значит, если в заднюю полусферу летит
$E_2$ энергия, то должны иметь $\frac{N}{2}mc^2\gamma=E_1+E_2$. Отсюда
$$E_2=\frac{N}{2}mc^2\gamma-E_1=\frac{\gamma}{8}Nmc^2(2-3\beta+\beta^3).$$
Следовательно,
$$\frac{E_1}{E_2}=\frac{2+3\beta-\beta^3}{2-3\beta+\beta^3}=\frac{27}{5}
\approx 5,4.$$

\subsection*{Задача 2 {\usefont{T2A}{cmr}{b}{n}(4.6)}: Найти энергию
$\pi$-мезонов}
Для простоты выкладок будем предполагать $c=1$ в промежуточных выражениях. 
Пусть энергия $\pi$-мезона равна $E$. Тогда его $\gamma$-фактор будет
$\gamma=E/m$. В системе покоя $\pi$-мезона распад изотропный, т. е. нейтрино 
с одинаковой вероятностью может вылететь в любом направлении. Пусть вылетает 
под углом $\theta^\prime$ к скорости $\pi$-мезона (т. е. к оси $x^\prime$). 
Закон сохранения 4-импульса $p_\pi=p_\nu+p_\mu$, с учетом того, что для 
нейтрино $p_\nu^2=0$, дает $p_\mu^2=(p_\pi-p_\nu)^2=p_\pi^2-2p_\pi\cdot p_nu$. 
Но в системе покоя $\pi$-мезона $p_\pi=(m_\pi,\vec{0})$, и $p_\pi\cdot p_nu=
m_\pi E_\nu^\prime$, где $E_\nu^\prime$ --- энергия нейтрино в этой системе. 
Следовательно, $m_\mu^2=m_\pi^2-2m_\pi E_\nu^\prime$ и
$$E_\nu^\prime=\frac{m_\pi^2-m_\mu^2}{2m_\pi}.$$
Так как нейтрино не имеет массы, его импульс в системе покоя $\pi$-мезона
будет $p_\nu^\prime=E_\nu^\prime$. По преобразованию Лоренца, находим
энергию нейтрино в Л-системе:
$$E=\gamma ( E_\nu^\prime+\beta p_{\nu x}^\prime )=\gamma  E_\nu^\prime
(1+\beta \cos{\theta^\prime}),$$
так как $ p_{\nu x}^\prime =p_\nu^\prime \cos{\theta^\prime}$. Как видим,
$E_{max}=\gamma  E_\nu^\prime (1+\beta )$ и $E_{min}=\gamma  E_\nu^\prime
(1-\beta )$. По условию $$\frac{E_{max}}{E_{min}}=\frac{1+\beta}{1-\beta}
\alpha,$$ что дает
$$\beta=\frac{\alpha -1}{\alpha +1}\;\;\mbox{и}\;\;\gamma=\frac{1}
{\sqrt{1-\beta^2}}=\frac{1+\alpha}{2\sqrt{\alpha}}.$$
Следовательно, энергия $\pi$-мезона равна (восстановили скорость света $c$)
$$E_\pi=m_\pi c^2 \gamma=m_\pi c^2\frac{1+\alpha}{2\sqrt{\alpha}}\approx
707~\mbox{МэВ}.$$

\subsection*{Задача 3 {\usefont{T2A}{cmr}{b}{n}(4.22)}: Определите массу каона}
{\usefont{T2A}{cmr}{m}{it} Первое решение.}
Из сохранения 4-импульса $p_K=p_{\pi^+}+p_{\pi^-}$ следует соотношение 
(положили $c=1$)
$$m_K^2=p_K^2=(p_{\pi^+}+p_{\pi^-})^2=2m_\pi^2+2p_{\pi^+}\cdot p_{\pi^-}=
2m_\pi^2+2(E_+E_--p_+p_-\cos{\theta}).$$
Отсюда для угла разлета, с учетом $E_+=E-E_-$, получаем 
$$\cos{\theta}=\frac{2E_+E_--m_K^2+2m_\pi^2}{2p_+p_-}=
\frac{2(E-E_-)E_--m_K^2+2m_\pi^2}{2\sqrt{(E-E_-)^2-m_\pi^2}\,
\sqrt{E_-^2-m_\pi^2}}.$$
Так как $|\cos{\theta}\le 1$, мы должны иметь
$$[2(E-E_-)E_--m_K^2+2m_\pi^2]^2\le 4(E_-^2-m_\pi^2)[(E-E_-)^2-m_\pi^2],$$
или
\begin{equation}
E_-^2-EE_-+\frac{m_\pi^2}{m_K^2}\,(2E^2-m_\pi^2)+\frac{m_K^2}{4}\le 0.
\label{eq12_3a}
\end{equation}
Чтобы найти минимальный угол разлета, вычислим
$$\frac{d\cos{\theta}}{dE_-}=\frac{(2E_--E)(2m_\pi^2E^2-m_K^2EE_-+m_K^2E_-^2-
m_K^2m_\pi^2)}{2(E_-^2-m_\pi^2)^{3/2}((E-E_-)^2-m_\pi^2)^{3/2}}.$$
Следовательно, минимальному углу разлета соответствует или
\begin{equation}
E_-^2-EE_-+\frac{m_\pi^2}{m_K^2}\,(2E^2-m_K^2)=0,
\label{eq12_3b}
\end{equation}
или
\begin{equation} 
2E_--E=0.
\label{eq12_3c}
\end{equation}
Но (\ref{eq12_3a}) и (\ref{eq12_3b}) вместе означают, что мы должны иметь
$$\frac{m_K^2}{4}-\frac{E^2m_\pi^2}{m_K^2}\le 0,$$
что не выполняется, так как $m_K^2>2Em_\pi$. Следовательно, минимальному углу 
разлета соответствует (\ref{eq12_3c}), т. е. симметричный разлет с
$$E_-=E_+=\frac{E}{2}.$$
Тогда
$$\cos{\theta_m}=\frac{\frac{E^2}{2}-m_K^2+2m_\pi^2}{2\left (\frac{E^2}{4}-
m_\pi^2\right )},$$
и масса $K$-мезона определяется из соотношения
$$m_K^2=\frac{E^2}{2}(1-\cos{\theta_m})+2m_\pi^2(1+\cos{\theta_m}).$$
Окончательно (восстановили скорость света $c$)
$$m_K=\sqrt{\left (2m_\pi \cos{\frac{\theta_m}{2}}\right )^2+
\left (\frac{E}{c^2}\,\sin{\frac{\theta_m}{2}}\right )^2}\approx
498~\mbox{МэВ}.$$

\subsection*{Задача 4 {\usefont{T2A}{cmr}{b}{n}(4.41)}: Преимущество встречных
пучков}
Пусть $p_1,p_2$ --- 4-импульсы первоначальных протонов, $q_1,q_2,q$ --- 
4-импульсы конечных протонов и $J/\Psi$, соответственно. Закон сохранения 
4-импульса дает $(p_1+p_2)^2=(q_1+q_2+q)^2$. Левую часть считаем в
лабораторной системе: $(p_1+p_2)^2=2m_p^2+2m_pE$, так как $p_1\cdot p_2=
m_pE$, $E$ --- энергия налетающего протона. Правую часть считаем в системе
центра инерции, где при пороговой энергии все частицы неподвижны. Тогда
$q_1+q_2+q=(2m_p+M,\vec{0})$, $M$ --- масса $J/\Psi$, и $(q_1+q_2+q)^2=
(2m_p+M)^2$. Следовательно, $2m_p^2+2m_pE=(2m_p+M)^2$ и (восстановили
скорость света $c$) $$E=\frac{(2m_p+M)^2-2m_p^2}{2m_p}c^2=
\left (m_p+2M+\frac{M^2}{2m_p}\right )c^2\approx 11,5~\mbox{ГэВ}.$$
На встречных электрон-позитронных пучках имеем реакцию $e^++e^-\to J/\Psi$,
и закон сохранения 4-импульса дает $M^2=q^2=(p_++p_-)^2=4E^2$, так как  
$p_++p_-=(2E,\vec{0})$. Следовательно, в этом случае (снова восстановили $c$)
$E=\frac{Mc^2}{2}\approx 1,5~\mbox{ГэВ}.$

\subsection*{Задача 5 {\usefont{T2A}{cmr}{b}{n}(4.37)}: Излучение и поглощение
света свободным электроном}
Сохранение 4-импульса в процессе $e^-\to e^-+\gamma$ дает $q_1=q_2+k$, где $q_1$ 
--- 4-импульс электрона до излучения фотона, $q_1$ --- после излучения, $k$ ---
4-импульс фотона. Так как $k^2=0$, то получаем $q_1^2=(q_2+k)^2=q_2^2+2q_2
\cdot k$. Но $q_1^2=q_2^2=m_e^2$ (полагаем $c=1$). Т. е. мы должны иметь 
$q_2\cdot k=0$. Посчитаем этот инвариант в системе, где электрон после излучения 
фотона покоится, т. е. $q_2=(m_e,\vec{0})$. Тогда $q_2\cdot k=m_e E_\gamma=0$ 
показывает, что в этой системе энергия фотона $E_\gamma$ равна нулю, что невозможно.

Аналогично, для процесса $e^-+e^+\to \gamma$ имеем $q_-+q_+=k$, и из соотношения
$q_+^2=(k-q_-)^2$ следует $k\cdot q_-=0$, что невозможно, так как в системе покоя
электрона энергия фотона не может равняться нулю. 

\subsection*{Задача 6 {\usefont{T2A}{cmr}{b}{n}(4.42)}: Пороговая энергия 
рождения электрон-позит\-ронной пары}
Сохранение 4-импульса приводит к соотношению $(k_1+k_2)^2=(q_-+q_+)^2$, где
$k_1,k2$ --- 4-импульсы фотонов, $q_-,q_+$ --- 4-импульсы электрона и позитрона
после их рождения. Правую часть вычислим в системе центра инерции, где $e^-$ и $e^+$
покоятся и, следовательно, $(q_-+q_+)^2=(2m)^2=4m^2$, $m$ --- масса электрона.
Левую часть вычислим в лабораторной системе. Так как поперечная компонента импульса не
меняется при преобразованиях Лоренца, в этой системе и электрон и позитрон должны 
летать вдоль оси $x$. Если они летят в одну сторону, то $k_1=(E_1,E_1,0,0)$, 
$k_2=(E_2,E_2,0,0)$ и $(k_1+k_2)^2=0$, что противоречит закону сохранения 
4-импульса. Поэтому $k_1=(E_1,E_1,0,0)$  и $k_2=(E_2,-E_2,0,0)$, где 
$E_1=1~\mbox{МэВ}$. Тогда  $(k_1+k_2)^2=(E_1+E_2)^2-(E_1-E_2)^2=4E_1E_2$. 
Следовательно, $4E_1E_2=4m^2$ и (восстановили скорость света $c$)
$$E_2=\frac{mc^2}{E_1}mc^2=2,5\cdot 10^{11}~\mbox{эВ}.$$

\section[Семинар 13]
{\centerline{Семинар 13}}

\subsection*{Задача 1 {\usefont{T2A}{cmr}{b}{n}(4.38)}: Симметричный разлет
фотонов}
Сохранение 4-импульса дает $(p_++p_-)^2=(k_1+k_2)^2$. Но
$(p_++p_-)^2=2m^2+2p_+\cdot p_-=2m^2+2mE$, где $E$ --- энергия налетающего
позитрона. Далее имеем $(k_1+k_2)^2=2k_1\cdot k_2=2E_1E_2(1-\cos{\alpha})=
4E_1E_2\sin^2{(\alpha/2)}$. Следовательно,
$$m(m+E)=2E_1E_2\sin^2{\frac{\alpha}{2}}.$$
Так как фотоны разлетаются симметрично, то $E_1=E_2$ (более формально это
следует из сохранения поперечного импульса: $0=E_1\sin{(\alpha/2)}-
E_2\sin{(\alpha/2)}$). Следовательно, $E_1=E_2=(E+m)/2$ и 
$$\sin^2{\frac{\alpha}{2}}=\frac{2m}{m+E}=\frac{2}{1+\frac{E}{m}}.$$
Окончательно (восстановили скорость света $c$)
$$\alpha=2\arcsin{\sqrt{\frac{2}{1+\frac{E}{mc^2}}}}.$$

\subsection*{Задача 2 {\usefont{T2A}{cmr}{b}{n}(4.43)}: Найти суммарную
кинетическую энергию нуклонов}
Если масса нуклона $m$, а суммарная кинетическая энергия нуклонов $T$, то
$E+M_d=T+2m$, где $E$ ---  энергия налетающего $\gamma$-кванта. Но $M_d=2m-
E_{\mbox{св}}$. Следовательно, получаем $T=E-E_{\mbox{св}}$. Пороговую энергию 
$E$ найдем обычным образом. Имеем $(k+p_d)^2=(p_n+p_p)^2$. Правую часть
вычислим в системе центра масс, где при пороговой энергии нуклоны покоятся:
$(p_n+p_p)^2=(2m)^2=4m^2$. Правую часть вычислим в лабораторной системе:
$(k+p_d)^2=M_d^2+2k\cdot p_d=M_d^2+2EM_d$. Следовательно, $M_d^2+2EM_d=4m^2$
и $$E=\frac{(2m)^2-M_d^2}{2M_d}=\frac{(M_d+E_{\mbox{св}})^2-M_d^2}{2M_d}=
E_{\mbox{св}}+\frac{E_{\mbox{св}}^2}{2M_d}.$$ Поэтому
$T=E-E_{\mbox{св}}=\frac{E_{\mbox{св}}^2}{2M_d}$. Если восстановить скорость света
$c$, $$T=\frac{E_{\mbox{св}}^2}{2M_dc^2}\approx 10^3~\mbox{эВ}.$$

\subsection*{Задача 3 {\usefont{T2A}{cmr}{b}{n}(4.32)}: Максимальная энергия
электрона}
Пусть система $\nu+\bar{\nu}$ имеет инвариантную массу $m^*$, суммарную
энергию $E^*$ и суммарный импульс $p^*$. Тогда в системе покоя мюона будем
иметь $m_\mu=E_e+E^*$ и $p_e=p^*$. Но $p_e=\sqrt{E_e^2-m_e^2}$ и 
$p^*=\sqrt{E^{*\,2}-m^{*\,2}}$. Поэтому получаем
$E_e^2-m_e^2=E^{*\,2}-m^{*\,2}$  и $E^{*\,2}=E_e^2+m^{*\,2}-m_e^2$.
Подставляя это в равенство $E^{*\,2}=(m_\mu-E_e)^2=m_\mu^2-2m_\mu E_e+E_e^2$,
получим $$E_e=\frac{m_\mu^2+m_e^2-m^{*\,2}}{2m_\mu}.$$
Видно, что энергия $E_e$ максимальна, когда $m^*$ минимальна. Но $m^{*\,2}=
(k_1+k_2)^2=2k_1\cdot k_2=2E_1E_2(1-\cos{\alpha})$, где $\alpha$ --- угол
между импульсами нейтрино $\vec{k}_1$  и $\vec{k}_2$. Минимальному $m^*_m=0$
соответствует $\alpha=0$. Следовательно,
$$E_{e\,max}=\frac{m_\mu^2+m_e^2}{2m_\mu},$$
или, если восстановить скорость света $c$,
$$E_{e\,max}=\frac{m_\mu^2+m_e^2}{2m_\mu}c^2\approx 53~\mbox{МэВ}.$$

\subsection*{Задача 4 {\usefont{T2A}{cmr}{b}{n}(4.35)}: Максимальная энергия
вторичных $\gamma$-квантов}
Законы сохранения энергии и импульса дают $2E=E_1+E_2$, $p_1=p_2$. Но 
$p_1=\sqrt{E_1^2-m^2}$ и $p_2=\sqrt{E_2^2-m^2}$. Поэтому из равенства
величин импульса $p_1=p_2$ следует равенство энергий. Т. е. $E_1=E_2=E$ и
$\pi$-мезон имеет энергию 300~МэВ. В системе покоя $\pi$-мезона
$\gamma$-квант от распада $\pi^0\to 2\gamma$ имеет энергию $m_\pi/2$. В
Л-системе его энергия будет $E_\gamma=\gamma(E_\gamma^\prime+\beta 
p_{\gamma x}^\prime)=\gamma E_\gamma^\prime (1+\beta\cos{\alpha})$, где
$\alpha$ --- угол вылета $\gamma$-кванта в системе покоя $\pi$-мезона
относительно оси $x^\prime$. Поэтому $E_{\gamma\, max}=\gamma
E_\gamma^\prime (1+\beta)=\gamma (m_\pi/2)(1+\beta)=(E/2)(1+\beta)$.
Так как $$\gamma=\frac{1}{\sqrt{1-\beta^2}}=\frac{E}{m_\pi},$$
то скорость $\pi$-мезона $$\beta=\sqrt{1-\left(\frac{m_\pi}{E}\right)^2}$$
и окончательно (по размерности восстановили скорость света $c$):
$$E_{\gamma\, max}=\frac{E}{2}\left (1+\sqrt{1-\left(\frac{m_\pi c^2}{E}
\right)^2}\right )\approx 270~\mbox{МэВ}.$$

\subsection*{Задача 5 {\usefont{T2A}{cmr}{b}{n}(4.51)}: Масса и скорость
составной частицы}
4-импульсы частиц имеют вид
$$p_1=(m_1c\gamma_1,\,m_1\vec{V_1}\gamma_1),\;\;\;
p_2=(m_2c\gamma_2,\,m_2\vec{V_2}\gamma_2).$$
Поэтому 4-импульс составной частицы будет
$$p=p_1+p_2=((m_1\gamma_1+m_2\gamma_2)c,\,m_1\vec{V_1}\gamma_1+
m_2\vec{V_2}\gamma_2).$$
Масса этой частицы определяется из 
\begin{eqnarray} &
M^2c^2=p^2= & \\ &
=c^2\left [m_1^2\gamma_1^2+m_2^2\gamma_2^2+2m_1m_2\gamma_1\gamma_2-
m_1^2\beta_1^2\gamma_1^2-m_2^2\beta_2^2\gamma_2^2-2m_1m_2\gamma_1\gamma_2
\vec{\beta}_1\cdot\vec{\beta}_2\right ].& \nonumber
\label{eq13_5}
\end{eqnarray}
Но $\beta^2\gamma^2=\gamma^2-1$ и (\ref{eq13_5}) дает 
$$M=\sqrt{m_1^2+m_2^2+2m_1m_2\gamma_1\gamma_2
(1-\vec{\beta}_1\cdot\vec{\beta}_2)}.$$
Скорость частицы $\vec{V}$ найдем из 
$$\vec{p}=M\vec{V}\gamma_V=\frac{\vec{V}}{c^2}\,Mc^2\gamma_V=
\frac{\vec{V}}{c^2}\,E,$$
что дает
$$\vec{V}=\frac{\vec{p}c^2}{E}=\frac{m_1\vec{V_1}\gamma_1+
m_2\vec{V_2}\gamma_2}{m_1\gamma_1+m_2\gamma_2}.$$

\subsection*{Задача 6 {\usefont{T2A}{cmr}{b}{n}(4.71)}: Эффект Комптона}
Сохранение 4-импульса дает $k+p_e=k^\prime+p_e^\prime$, или $p_e^\prime=
k+p_e-k^\prime$. Поэтому $m_e^2=p_e^{\prime\,2}=(k+p_e-k^\prime)^2=m_e^2-
2k\cdot k^\prime+2k\cdot p_e-2k^\prime \cdot p_e$. Но $p_e=(m_e,\vec{0})$,
поэтому $k\cdot p_e=m_e E$ и $k^\prime\cdot p_e=m_e E^\prime$. Кроме того
$k\cdot k^\prime=E E^\prime (1-\cos{\theta})$. Следовательно, $E E^\prime
(1-\cos{\theta})=m_e (E-E^\prime)$ и (восстановили скорость света $c$)
$$E^\prime=\frac{m_e c^2}{m_e c^2+E(1-\cos{\theta})}\,E.$$
Длина волны фотона $\lambda=hc/E$. Поэтому
$$\lambda^\prime=\frac{hc}{E^\prime}=\left( 1+\frac{E}{m_e
c^2}(1-\cos{\theta})\right )\lambda$$
и
$$\Delta \lambda=\lambda^\prime-\lambda=\frac{\lambda E}{m_e
c^2}(1-\cos{\theta})=\frac{h}{m_e c}(1-\cos{\theta}).$$

\section[Семинар 14]
{\centerline{Семинар 14}}

\subsection*{Задача 1 {\usefont{T2A}{cmr}{b}{n}(4.61)}: Передача энергии при
упругом столкновении}
{\usefont{T2A}{cmr}{m}{it} Первое решение.}
Сразу рассмотрим релятивистский случай, а нерелятивистский получим предельным переходом
при $T_0/(m_1c^2)\to 0$, $T_0$ --- кинетическая энергия налетающей частицы, 
$m_1$ --- ее масса. Первая частица потеряет наибольшую энергию при лобовом 
столкновении. При этом в системе центра масс ее скорость просто поменяет знак. Пусть 
$V$ --- скорость системы  центра масс, $u$ --- величина скорости первой частицы 
в этой системе. В Л-системе величина скорости первой частицы до столкновения будет 
$$v_1=\frac{u+V}{1+\frac{uV}{c^2}},$$
а после столкновения
$$v_1^\prime=\frac{u-V}{1-\frac{uV}{c^2}}.$$
Поэтому, как легко проверить,
$$\gamma_1=\frac{1}{\sqrt{1-\frac{v_1^2}{c^2}}}=\gamma_u\gamma_V(1+\beta_u
\beta_V),\;\;\mbox{и}\;\;\gamma_1^{\,\prime}=
\frac{1}{\sqrt{1-\frac{v_1^{\prime 2}}
{c^2}}}=\gamma_u\gamma_V(1-\beta_u \beta_V).$$
Для кинетических энергий первой частицы $T_0$ и $T$, до и после столкновения, будем
иметь 
\begin{equation}
T_0-T=m_1c^2(\gamma_1-\gamma_1^{\,\prime})=2m_1c^2\gamma_u\gamma_V
\beta_u\beta_V
\label{eq14_1a}
\end{equation}
 и 
\begin{equation}
T_0+T+2m_1c^2=m_1c^2(\gamma_1+\gamma_1^{\,\prime})=
2m_1c^2\gamma_u\gamma_V.
\label{eq14_1b}
\end{equation}
Величина скорости второй частицы в системе центра масс есть $V$. Поэтому из равенствa 
величин импульсов сталкивающихся частиц в этой системе следует соотношение 
\begin{equation}
m_1\beta_u\gamma_u=m_2\beta_V\gamma_V.
\label{eq14_1c}
\end{equation} 
С учетом этого соотношения, (\ref{eq14_1a}) перепишется как
$T_0-T=2m_2c^2\beta_V^2\gamma_V^2=2m_2c^2(\gamma_V^2-1)$. Отсюда
\begin{equation}
\gamma_V^2=1+\frac{T_0-T}{2m_2c^2} 
\label{eq14_1d}
\end{equation}
С другой стороны, из (\ref{eq14_1c}) с учетом $\beta^2\gamma^2=\gamma^2-1$ и
соотношения (\ref{eq14_1d}) следует равенство
\begin{equation}
\gamma_u^2==1+\left(\frac{m_2}{m_1}\right)^2(\gamma_V^2-1)=
1+\frac{m_2}{m_1}\,\frac{T_0-T}{2m_1c^2}.
\label{eq14_1e}
\end{equation} 
Возведя равенство (\ref{eq14_1b}) в квадрат и подставляя $\gamma_V^2$ и 
$\gamma_u^2$ из (\ref{eq14_1d}) и (\ref{eq14_1e}), получаем уравнение
$$[2T_0-(T_0-T)+2m_1c^2]^2=4m_1^2c^4\left(1+\frac{T_0-T}{2m_2c^2}\right)
\left(1+\frac{m_2}{m_1}\,\frac{T_0-T}{2m_1c^2}\right).$$
Из этого уравнения находим 
\begin{equation}
T_0-T=2T_0\,\frac{m_2(T_0+2m_1c^2)}{2T_0m_2+(m_1+m_2)^2c^2},
\label{eq14_1ee}
\end{equation}
и
$$T=T_0-(T_0-T)=T_0\,\frac{(m_1-m_2)^2}{(m_1+m_2)^2+2\frac{T_0m_2}{c^2}}.$$
Отсюда видно, что в нерелятивистском пределе будем иметь
$$T=T_0\,\frac{(m_1-m_2)^2}{(m_1+m_2)^2}.$$

{\usefont{T2A}{cmr}{m}{it} Второе решение.}
Пусть $V$ --- скорость системы  центра масс. До столкновения вторая частица движется
со скоростью $-V$, а после столкновения --- со скоростью $V$. Следовательно, в 
Л-системе скорость второй частицы после столкновения будет
$$V_2^\prime=\frac{V+V}{1+\frac{V^2}{c^2}},$$
а его $\gamma$-фактор
$$\gamma_2^{\,\prime}=\frac{1}{\sqrt{1-\frac{V_2^{\prime\,2}}{c^2}}}=
\gamma_V^2(1+\beta_V^2)=\frac{1+\beta_V^2}{1-\beta_V^2}.$$
Тогда закон сохранения энергии $E_1+m_2c^2=E_1^\prime+m_2c^2\gamma_2^{\,\prime}$
дает
\begin{equation}
T_0-T=E_1-E_1^\prime=m_2c^2(\gamma_2^{\,\prime}-1)=2m_2c^2\frac{\beta_V^2}
{1-\beta_V^2}.
\label{eq14_1f}
\end{equation}
С другой стороны, скорость центра инерции, деленная на скорость света, равна
$$\beta_V=\frac{p_1c}{E_1+m_2c^2}=\frac{\sqrt{E_1^2-m_1^2c^4}}{E_1+m_2c^2},$$
где $p_1$ --- импульс первой частицы до столкновения. Подставляя это в 
(\ref{eq14_1f}), получаем уравнение (\ref{eq14_1ee}):
$$T_0-T=2m_2c^2\frac{E_1^2-m_1^2c^4}{(m_1^2+m_2^2)c^4+2m_2E_1c^2}=
2m_2T_0\frac{T_0+2m_1c^2}{(m_1+m_2)^2c^2+2T_0m_2},$$
так как $E_1=T_0+m_1c^2$. 

{\usefont{T2A}{cmr}{m}{it} Третье решение.}
Пусть $E^\prime,p^\prime$ --- энергия и величина импульса первой частицы в системе 
центра масс, а $\theta^\prime$ --- угол рассеяния. Тогда в Л-системе после рассеяния
будем иметь (мы положили $c=1$ в этой и в последующих формулах)
$E=\gamma_u(E^\prime+\beta_u p^\prime\cos{\theta^\prime})$, где
$u$ --- скорость системы центра масс. Ясно, что минимальной энергии $E_{min}$ 
соответствует $\cos{\theta^\prime}=-1$, т.~е. 
$E_{min}=\gamma_u(E^\prime-\beta_u p^\prime)$.
Если $E_1,p_1$ --- энергия и величина импульса первой частицы до рассеяния 
в Л-системе, то по преобразованию Лоренца $E^\prime=\gamma_u(E_1-\beta_u p_1)$ 
(заметим, что в системе центра масс энергия частицы в процессе рассеяния не меняется).
Но скорость системы центра масс равна $$\beta_u=\frac{p_1}{E_1+m_2}.$$ 
Следовательно, $p_1=\beta_u(E_1+m_2)$ и
$$E^\prime=\gamma_u[E_1-\beta_u^2(E_1+m_2)]=\frac{E_1}{\gamma_u}-
\gamma_u\beta_u^2m_2.$$ Кроме того, $p^\prime=m_2\beta_u\gamma_u$, так как в 
системе центра масс импульсы частиц равны по величине а вторая частица движется со 
скоростью $-\vec{u}$. Поэтому $E_{min}=\gamma_u(E^\prime-\beta_u p^\prime)=
E_1-2\gamma_u^2\beta_u^2m_2$ и $T_{min}=E_{min}-m_1=T_0-2\gamma_u^2\beta_u^2
m_2$, где $T_0=E_1-m_1$. Но
$$\gamma_u^2\beta_u^2=\frac{\beta_u^2}{1-\beta_u^2}=\frac{(p_1/(E_1+m_2))^2}
{1-(p_1/(E_1+m_2))^2}=\frac{p_1^2}{E_1^2-2E_1m_2+m_2^2-p_1^2}= $$ $$
=\frac{p_1^2}{m_1^2+2(T_0+m_1)m_2+m_2^2}=\frac{p_1^2}{(m_1+m_2)^2+2T_0m_2}.$$
Следовательно, $$T_{min}=T_0\left [1-2\,\frac{m_2(T_0+2m_1)}{(m_1+m_2)^2+
2T_0m_2}\right ]=T_0\,\frac{(m_1-m_2)^2}{(m_1+m_2)^2+2T_0m_2},$$
где мы учли, что $p_1^2=E_1^2-m_1^2=(E_1-m_1)(E_1+m_1)=T_0(T_0+2m_1)$.
Если восстановить скорость света $c$, 
$$T_{min}=T_0\,\frac{(m_1-m_2)^2}{(m_1+m_2)^2+2T_0m_2/c^2}.$$

{\usefont{T2A}{cmr}{m}{it} Четвертое решение} \cite{27}.
Пусть $p_1,p_2$ --- 4-импульсы частиц до столкновения, $p^\prime_1,p^\prime_2$ ---
после столкновения. Из закона сохранения 4-импульса $p_1+p_2=p^\prime_1+p^\prime_2$,
получаем $(p_1-p^\prime_2)^2=(p^\prime_1-p_2)^2$, или $p_1\cdot p^\prime_2=
p^\prime_1\cdot p_2$, так как $p_1^2=p_1^{\prime\, 2}=m_1^2$  и $p_2^2=
p_2^{\prime\, 2}=m_2^2$ (как обычно, положили $c=1$). Но $$p^\prime_1\cdot p_2=
m_2E_1^\prime=m_2(E_1+m_2-E_2^\prime)=m_2(E_1-T_2),$$ где $T_2=E_2^\prime-m_2$ 
--- кинетическая энергия второй (первоначально неподвижной) частицы после 
столкновения. Кроме того, имеем
$$p_1\cdot p^\prime_2=E_1(m_2+T_2)-|\vec{p}_1||\vec{p}_2^\prime|
\cos{\theta_2},$$
где $\theta_2$ --- угол рассеяния второй частицы. Следовательно,
$$E_1(m_2+T_2)-|\vec{p}_1||\vec{p}_2^\prime|\cos{\theta_2}=m_2(E_1-T_2).$$
Отсюда получаем
$$T_2(E_1+m_2)=|\vec{p}_1||\vec{p}_2^\prime|\cos{\theta_2}.$$
Возведем обе части этого  уравнения в квадрат и учтем, что
$$|\vec{p}_2^\prime|^2=E_2^{\prime\, 2}-m_2^2=(m_2+T_2)^2-m_2^2=
T_2(T_2+2m_2).$$
В результате получим
$$T_2(E_1+m_2)^2=|\vec{p}_1|^2(T_2+2m_2)\cos^2{\theta_2}.$$
Решаяя это уравнение относительно $T_2$, получаем
$$T_2=\frac{2m_2|\vec{p}_1|^2\cos^2{\theta_2}}{(E_1+m_2)^2-|\vec{p}_1|^2
\cos^2{\theta_2}}=2m_2\left [\frac{(E_1+m_2)^2}{(E_1+m_2)^2-|\vec{p}_1|^2
\cos^2{\theta_2}}-1\right ].$$
Отсюда видно, что максимальной передаче энергий ко второй частице соответствует 
$\theta_2=0$ (лобовое столкновение, при этом $\cos{\theta_2}=1$) и
\begin{equation}
T_{2,max}=\frac{2m_2|\vec{p}_1|^2}{(E_1+m_2)^2-|\vec{p}_1|^2}.
\label{eq14_1g}
\end{equation}
Но $|\vec{p}_1|^2=(m_1+T_0)^2-m_1^2=T_0(T_0+2m_1)$ и $(E_1+m_2)^2-
|\vec{p}_1|^2=(T_0+m_1+m_2)^2-T_0(T_0+2m_1)=(m_1+m_2)^2+2T_0m_2$. Поэтому
(\ref{eq14_1g}) перепишется так
$$T_{2,max}=\frac{2m_2T_0(T_0+2m_1)}{(m_1+m_2)^2+2T_0m_2}.$$
Тогда
$$T_{min}=T_0-T_{2,max}=T_0\left[1-\frac{2m_2(T_0+2m_1)}{(m_1+m_2)^2+2T_0m_2}
\right]=T_0\,\frac{(m_1-m_2)^2}{(m_1+m_2)^2+2T_0m_2}.$$

\subsection*{Задача 2 {\usefont{T2A}{cmr}{b}{n}(4.69)}: Лобовое столкновение
электрона с протоном}
{\usefont{T2A}{cmr}{m}{it} Первое решение.}
Закон сохранения 4-импульса $p_e+p_p=p_e^\prime+p_p^\prime$ означает, что
\begin{equation}
M^2=p_p^{\prime\,2}=(p_e+p_p-p_e^\prime)^2=2m^2+M^2+2p_e\cdot p_p-
2p_e\cdot p_e^\prime-2p_p\cdot p_e^\prime,
\label{eq14_2a}
\end{equation}
где $M$ --- масса протона, $m$ --- масса электрона. Но $p_p=(M,0,0,0)$. 
Поэтому $p_e\cdot p_p=EM$ и $p_e^\prime\cdot p_p=\frac{E}{2}M$, где 
$E$ --- начальная энергия электрона. Кроме того,
$$p_e\cdot p_e^\prime=E\frac{E}{2}-\vec{p}_e\cdot \vec{p}_e^\prime=
\frac{E^2}{2}+|\vec{p}_e||\vec{p}_e^\prime|=\frac{E^2}{2}+
\sqrt{E^2-m^2}\sqrt{\frac{E^2}{4}-m^2}.$$
Следовательно, из (\ref{eq14_2a}) получаем
$$0=m^2+EM-\frac{E}{2}M-\frac{E^2}{2}-\sqrt{E^2-m^2}
\sqrt{\frac{E^2}{4}-m^2},$$
или
\begin{equation}
(E^2-m^2)\left (\frac{E^2}{4}-m^2\right )=\left(\frac{E^2}{2}-\frac{E}{2}M-
m^2\right)^2.
\label{eq14_2b}
\end{equation}
Раскрывая скобки и упрощая, (\ref{eq14_2b}) можно привести к виду
\begin{equation}
2ME^2-(m^2+M^2)E-4Mm^2=0.
\label{eq14_2bb}
\end{equation}
Решение имеет вид
$$E=\frac{m^2+M^2\pm\sqrt{(m^2+M^2)^2+32M^2m^2}}{4M}.$$
Знак ``-'', очевидно, не годится, так как дает $E<0$. Окончательно 
(восстановили скорость света $c$),
$$E=\frac{m^2+M^2+\sqrt{m^4+M^4+34M^2m^2}}{4M}\,c^2.$$
Так как $M\gg m$, то
$$E\approx\frac{1}{2}Mc^2\approx 0,5~\mbox{ГэВ}.$$

{\usefont{T2A}{cmr}{m}{it} Второе решение.}
Пусть $V$ --- скорость системы  центра масс, $u$ --- скорость электрона до 
столкновения в этой системе. Тогда в Л-системе $\gamma$-факторы электрона до и
после столкновения будут (см. предыдущую задачу) $\gamma_1=\gamma_u\gamma_V(1+
\beta_u\beta_V)$ и $\gamma_1^{\,\prime}=\gamma_u\gamma_V(1-\beta_u \beta_V)$.
Кроме того,
\begin{equation}
m\beta_u\gamma_u=M\beta_V\gamma_V,
\label{eq14_2c}
\end{equation}
где $m$ --- масса электрона, $M$ --- масса протона. По условию задачи
$$\frac{E^\prime}{E}=\frac{\gamma_1^{\,\prime}}{\gamma_1}=
\frac{1-\beta_u \beta_V}{1+\beta_u \beta_V}=\frac{1}{2}.$$
Отсюда находим $\beta_u \beta_V=1/3$. Возводя это равенство в квадрат и используя
$\beta^2=(\gamma^2-1)/\gamma^2$, получаем
\begin{equation}
9(\gamma_u^2-1)(\gamma_V^2-1)=\gamma_u^2\gamma_V^2.
\label{eq14_2d}
\end{equation}
Из (\ref{eq14_2c}) следует
$$\gamma_u^2-1=\left (\frac{M}{m}\right)^2(\gamma_V^2-1),$$
так как $\beta^2\gamma^2=\gamma^2-1$. Поэтому (\ref{eq14_2d}) можно переписать
как уравнение для $\gamma_V^2-1$:
\begin{equation}
9\left (\frac{M}{m}\right)^2(\gamma_V^2-1)^2=[1+(\gamma_V^2-1)]
\left[1+\left (\frac{M}{m}\right)^2(\gamma_V^2-1)\right ].
\label{eq14_2e}
\end{equation}
Энергия электрона до столкновения $$E=m\gamma_1=m\gamma_u\gamma_V
(1+\beta_u\beta_V)=\frac{4}{3}m\gamma_u\gamma_V.$$
С другой стороны, если умножить обе стороны равенства (\ref{eq14_2c}) на $\beta_V
\gamma_V$, то можно получить соотношение
$$\frac{1}{3}m\gamma_u\gamma_V=M(\gamma_V^2-1).$$
Следовательно, $E=4M(\gamma_V^2-1)$ и (\ref{eq14_2e}), как уравнение для 
определения $E$, принимает форму
$$\frac{9}{16}\,\frac{E^2}{m^2}=\left (1+\frac{ME}{4m^2}\right)
\left (1+\frac{E}{4M}\right).$$
Раскрывая скобки и приводя подобные члены, получаем
$$\frac{1}{2}\,\frac{E^2}{m^2}=1+\frac{m^2+M^2}{4m^2M}E.$$
Легко видеть, что это то же самое уравнение (\ref{eq14_2bb}). 

{\usefont{T2A}{cmr}{m}{it} Третье решение.}
Пусть $V$ --- скорость системы  центра масс. В этой системе протон до столкновения
движется со скоростью $-V$, а после столкновения --- со скоростью $V$. Следовательно,
в Л-системе протон после столкновения движется со скоростью
$$V_p^\prime=\frac{V+V}{1+\frac{V^2}{c^2}},$$
и его $\gamma$-фактор будет 
$$\gamma_p^{\,\prime}=\frac{1}{\sqrt{1-\frac{V_p^{\prime\,2}}{c^2}}}=
\gamma_V^2(1+\beta_V^2)=\frac{1+\beta_V^2}{1-\beta_V^2}.$$
Из закона сохранения энергии следует, что
$$E+Mc^2=\frac{E}{2}+Mc^2\gamma_p^{\,\prime},$$
где $E$ --- энергия электрона до столкновения, $M$ --- масса протона. Поэтому,
\begin{equation}
\frac{E}{2}=Mc^2\left [\frac{1+\beta_V^2}{1-\beta_V^2}-1\right]=
2Mc^2\,\frac{\beta_V^2}{1-\beta_V^2}.
\label{eq14_2f}
\end{equation}
С другой стороны, для скорости центра инерции имеем
$$\beta_V=\frac{pc}{E+Mc^2}=\frac{\sqrt{E^2-m^2c^4}}{E+Mc^2},$$
где $p$ есть импульс электрона до столкновения, $m$ --- его масса. Подставляя это
в (\ref{eq14_2f}), получаем
$$\frac{E}{2}=2Mc^2\frac{E^2-m^2c^4}{(m^2+M^2)c^4+2EMc^2}.$$
Отсюда следует
$$2ME^2-(m^2+M^2)c^2E-4Mm^2C^4=0,$$
что есть уравнение (\ref{eq14_2bb}) (на этот раз мы не полагали $c=1$ в промежуточных
вычислениях). 

\subsection*{Задача 3 {\usefont{T2A}{cmr}{b}{n}(4.73)}:  Максимальный угол
отклонения}
{\usefont{T2A}{cmr}{m}{it} Первое решение.}
Из закона сохранения 4-импульса $p_p+p_e=p_p^\prime+p_e^\prime$, получаем
\begin{equation}
m^2=p_e^{\prime\,2}=(p_p+p_e-p_p^\prime)^2=2M^2+m^2+2p_p\cdot p_e-
2p_p\cdot p_p^\prime-2p_e\cdot p_p^\prime,
\label{eq14_3a}
\end{equation}
где $m$ --- масса электрона, $M$ --- масса протона. Но $p_e=(m,0,0,0)$ и, 
следовательно, $p_p\cdot p_e=mE$, $p_e\cdot p_p^\prime=mE^\prime$, где $E$ 
и $E^\prime$ --- это энергии протона до и после рассеяния на электроне. Кроме того, 
$p_p\cdot p_p^\prime=EE^\prime-\vec{p}\cdot\vec{p^\prime}=
EE^\prime-\sqrt{E^2-M^2}\sqrt{E^{\prime\,2}-M^2}\cos{\theta}$, где 
$\theta$ --- угол рассеяния протона. Поэтому из (\ref{eq14_3a}) получаем
\begin{equation}
\cos{\theta}=\frac{E^\prime(E+m)-Em-M^2}{\sqrt{E^2-M^2}
\sqrt{E^{\prime\,2}-M^2}}.
\label{eq14_3b}
\end{equation}
Максимальному углу рассеяния соответствует минимальное значение $\cos{\theta}$ как
функции $E^\prime$. Но
$$\frac{d\cos{\theta}}{dE^\prime}=\frac{1}{\sqrt{E^2-M^2}
\sqrt{(E^{\prime\,2}-M^2)^3}}[(M^2+Em)E^\prime-(E+m)M^2].$$
Следовательно, минимуму $\cos{\theta}$ отвечает
$$E_m^\prime=\frac{(E+m)M^2}{M^2+Em}.$$
Подставляя это в (\ref{eq14_3b}), находим 
\begin{equation}
\cos{\theta_m}=\frac{\sqrt{(E+m)^2M^2-(M^2+Em)^2}}{M\sqrt{E^2-M^2}}.
\label{eq14_3c}
\end{equation}
Но $$(E+m)^2M^2-(M^2+Em)^2=(M^2-m^2)(E^2-M^2),$$
и (\ref{eq14_3c}) принимает вид
$$\cos{\theta_m}=\sqrt{1-\frac{m^2}{M^2}}.$$
Следовательно, максимальный угол рассеяния не зависит от энергии и равен
$$\theta_m=\arccos{\sqrt{1-\frac{m^2}{M^2}}}=\arcsin{\frac{m}{M}}.$$

{\usefont{T2A}{cmr}{m}{it} Второе решение.}
Пусть $V$ --- скорость системы центра масс, $u$ --- скорость протона до 
столкновения в этой системе, $\theta^\prime$ --- угол рассеяния протона
в системе центра масс. Величина скорости протона в системе центра масс не
меняется в процессе рассеяния. Поэтому после рассеяния компоненты скорости
протона в системе центра масс будут $u^\prime_x=u\cos{\theta^\prime}$
и $u^\prime_y=u\sin{\theta^\prime}$. В Л-системе компоненты скорости
протона после рассеяния, $V_{p,x}^\prime$ и $V_{p,y}^\prime$ находим по формулам
релятивистского сложения скоростей:
$$V_{p,x}^\prime=\frac{u\cos{\theta^\prime}+V}{1+\beta_V\beta_u
\cos{\theta^\prime}},\;\;\;V_{p,y}^\prime=\frac{u\sin{\theta^\prime}}
{\gamma_V(1+\beta_V\beta_u\cos{\theta^\prime})}.$$
Угол рассеяния протона в Л-системе $\theta$ определяется уравнением
\begin{equation}
\tg{\theta}=\frac{V_{p,y}^\prime}{V_{p,x}^\prime}=\frac{u\sin{\theta^\prime}}
{\gamma_V(V+u\cos{\theta^\prime})}.
\label{eq14_3d}
\end{equation}
Максимальному углу $\theta$ соответствует максимум $\tg{\theta}$ как функции 
$\theta^\prime$. Но
$$\frac{d\tg{\theta}}{d\theta^\prime}=\frac{u+V\cos{\theta^\prime}}
{\gamma_V(V+u\cos{\theta^\prime})^2}.$$
Поэтому максимуму отвечает 
$$\cos{\theta_m^\prime}=-\frac{u}{V}.$$
Подставляя это в (\ref{eq14_3d}), находим
$$\tg{\theta_m}=\frac{u}{\gamma_V\sqrt{V^2-u^2}}.$$
Тогда
\begin{equation}
\frac{1}{\sin^2{\theta}}=1+\ctg^2{\theta}=\frac{\beta_V^2(1-\beta_u^2)}
{\beta_u^2(1-\beta_V^2)}=\frac{\beta_V^2\gamma_V^2}{\beta_u^2\gamma_u^2}.
\label{eq14_3e}
\end{equation}
Но в системе центра масс величины импульсов протона и электрона равны, что означает
равенство $M\beta_u\gamma_u=m\beta_V\gamma_V$, так как электрон, который до
столкновения покоится в Л-системе, в системе центра масс движется навстречу протону 
со скоростью $-\vec{V}$. С учетом этого соотношения, (\ref{eq14_3e}) дает
$$\sin{\theta}=\frac{m}{M}.$$   

{\usefont{T2A}{cmr}{m}{it} Третье решение.}
Пусть $\vec{V}$ --- скорость системы центра масс, $p^\prime=mV\gamma_V$ ---
величина импульса протона (и электрона) в этой системе. После рассеяния компоненты 
импульса протона в системе центра масс будут $p^\prime_x=p^\prime\cos{\theta^
\prime}$ и $p^\prime_y=p^\prime\sin{\theta^\prime}$. В Л-системе компоненты 
импульса протона после рассеяния находим по формулам преобразования Лоренца:
$$p_x=\gamma_V\left(p^\prime\cos{\theta^\prime}+\beta_V\frac{E^\prime}{c}
\right),\;\;\;p_y=p^\prime\sin{\theta^\prime}.$$
Из этих формул следует, что  
$$\frac{1}{\gamma_V^2}\left(p_x-\frac{p^\prime E^\prime}{mc^2}\right)^2+
p_y^2=p^{\prime\,2}.$$
Это эллипс и максимальному углу рассеяния соответствует касательная к этому эллипсу
(см. левый рисунок).
\begin{figure}[htb]
\begin{center}
\epsfig{figure=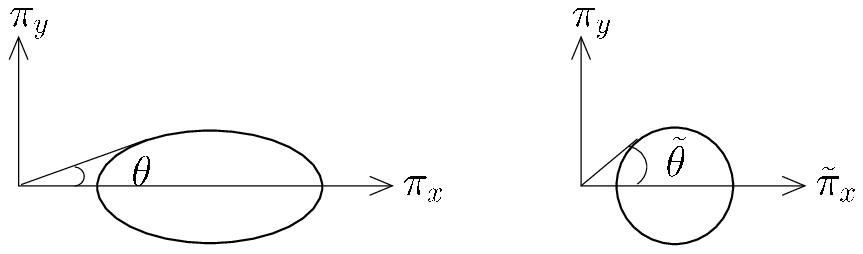,height=4cm}
\end{center}
\end{figure}

\noindent Вводя безразмерные величины 
$$\pi_x=\frac{p_x}{p^\prime},\;\;\;\pi_y=\frac{p_y}{p^\prime},$$
уравнение эллипса можно переписать как
\begin{equation}
\frac{1}{\gamma_V^2}\left (\pi_x-\sqrt{a+\gamma_V^2}\right )^2+\pi_y^2=1,
\label{eq14_3f}
\end{equation}
где $$a=\frac{M^2}{m^2}-1,$$
и мы учли, что
$$\frac{E^\prime}{mc^2}=\frac{1}{mc^2}\sqrt{p^{\prime\,2}c^2+M^2c^4}
=\sqrt{\frac{M^2}{m^2}+\beta_V^2\gamma_V^2}=\sqrt{a+\gamma_V^2}.$$
Чтобы найти соответствующий максимальный угол рассеяния $\theta$, мы сначала перейдем
к новой переменной $\tilde{\pi}_x=\pi_x/\gamma_V$ (т.е. сожмем горизонтальную
ось $\gamma_V$-раз). При этом эллипс превратится в окружность единичного радиуса,
центр которой смещен от начала координат вправо на  $\sqrt{1+\frac{a}{\gamma_V^2}}$
(см. правый рисунок):
$$\left(\tilde{\pi}_x-\sqrt{1+\frac{a}{\gamma_V^2}}\right)^2+\pi_y^2=1.$$
Угол $\theta$ при этом изменится --- станет $\tilde{\theta}$, такой, что 
$\tg{\tilde{\theta}}=\gamma_V\,\tg{\theta}$. На рисунке видно, что
$$\tg{\tilde{\theta}}=\frac{1}{\sqrt{1+\frac{a}{\gamma_V^2}-1}}=\frac{\gamma_V}
{\sqrt{a}}.$$
Следовательно, $\tg{\theta}=1/\sqrt{a}$, и
$$\frac{1}{\sin^2{\theta}}=1+a=\frac{M^2}{m^2}.$$
 
\subsection*{Задача 4 {\usefont{T2A}{cmr}{b}{n}(4.72)}:  Энергия рассеянного
фотона}
{\usefont{T2A}{cmr}{m}{it} Первое решение.}
Пусть $k,k^\prime$ 4-импульсы фотона до и после рассеяния, $p,p^\prime$ --- 
4-импульсы электрона. Из закона сохранения 4-импульса $k+p=k^\prime+p^\prime$ 
получаем ($m$ --- масса электрона) 
\begin{equation}
m^2=p^{\prime\,2}=(k+p-k^\prime)^2=
m^2+2k\cdot p-2k^\prime\cdot p-2k\cdot k^\prime.
\label{eq14_4a}
\end{equation}
Но, так как фотон рассеивается на угол $90^\circ$, $k^\prime\cdot p=
{\cal{E}}^\prime E$ и $ k\cdot k^\prime=
{\cal{E}}{\cal{E}}^\prime$, где ${\cal{E}}, {\cal{E}}^\prime$ --- энергии
фотона до и после расеяния, $E$ --- энергия электрона до рассеяния. Кроме того,
$k\cdot p={\cal{E}}E-\vec{k}\cdot\vec{p}={\cal{E}}E+|\vec{k}|\,|\vec{p}|=
{\cal{E}}(E+\sqrt{E^2-m^2})$. Поэтому (\ref{eq14_4a}) принимает вид
${\cal{E}}(E+\sqrt{E^2-m^2})={\cal{E}}^\prime (E+{\cal{E}})$ и, следовательно,
$${\cal{E}}^\prime={\cal{E}}\,\frac{E+\sqrt{E^2-m^2}}{E+{\cal{E}}}.$$
Но $E=m+T$ и $E^2-m^2=T^2+2mT$. Поэтому окончательно (восстановили скорость света 
$c$)
$${\cal{E}}^\prime={\cal{E}}\,\frac{mc^2+T+\sqrt{T^2+2mc^2T}}{mc^2+T+
{\cal{E}}}.$$
Если $T=100~\mbox{эВ}$, то $mc^2\gg {\cal{E}},T$ и ${\cal{E}}^\prime\approx
{\cal{E}}=10~\mbox{эВ}$. Если $T=10~\mbox{ГэВ}$, то $T\gg {\cal{E}},mc^2$ и
${\cal{E}}^\prime\approx 2{\cal{E}}=20~\mbox{эВ}$.

{\usefont{T2A}{cmr}{m}{it} Второе решение.}
Пусть $V$ --- скорость системы центра масс. В этой системе энергия налетающего фотона
будет $\tilde{\cal{E}}=\gamma_V{\cal {E}}(1+\beta_V)$. В системе центра масс
энергия фотона после рассеяния не изменится. С другой стороны, известно, что рассеянный
фотон летит в Л-системе под углом $90^\circ$ и, следовательно, имеет нулевую 
$x$-компоненту импульса. Поэтому энергии этого фотона в Л-системе 
${\cal{E}}^\prime$ и в системе центра масс $\tilde{\cal{E}}$ связаны так: 
$\tilde{\cal{E}}=\gamma_V{\cal{E}}^\prime$. Сравнивая два выражения для 
$\tilde{\cal{E}}$, приходим к выводу, что ${\cal{E}}^\prime={\cal{E}}(1+
\beta_V)$. Но скорость центра масс
$$\beta_V=\frac{p-{\cal{E}}}{E+{\cal{E}}}=\frac{\sqrt{E^2-m^2}-{\cal{E}}}
{E+{\cal{E}}},$$
где $p=\sqrt{E^2-m^2}$ есть первоначальный импульс электрона в Л-системе. С учетом
этого окончательно получаем
$${\cal{E}}^\prime={\cal{E}}\left (1+\frac{\sqrt{E^2-m^2}-{\cal{E}}}
{E+{\cal{E}}}\right )={\cal{E}}\,\frac{E+\sqrt{E^2-m^2}}{E+{\cal{E}}}.$$

\subsection*{Задача 5 {\usefont{T2A}{cmr}{b}{n}(5.12)}: Мюоны в магнитном поле}
Так как скорость мюонов не меняется, время в Л-системе $t$ и собственное время для
мюонов $t^\prime$ связаны как $t^\prime=t/\gamma$ и количество мюонов меняется
по закону $$N=N_0e^{-\frac{t^\prime}{\tau}}=N_0e^{-\frac{t}{\gamma\tau}}.$$
Ток тоже будет уменьшаться по этому закону:
$$I=I_0e^{-\frac{t}{\gamma\tau}}.$$
По условию задачи $N/N_0=1/20$. Поэтому получаем $t=\gamma\tau\ln{20}$. Надо
найти $\gamma$. Для радиуса круговой траектории мюона имеем $R=p/eB$, где $p=mv
\gamma=mc\beta\gamma$ --- импульс мюона. Поэтому
$$\gamma^2-1=\beta^2\gamma^2=\left(\frac{ReB}{mc}\right)^2.$$
Из этого соотношения можно определить $\gamma$ и окончательно
$$t=\sqrt{1+\left(\frac{cRB}{(mc^2/e)}\right)^2}\tau\ln{20}\approx 1,8\cdot 
10^{-4}~\mbox{с}.$$
Заметим, что в системе СИ, $mc^2/e$, где $m$ --- масса мюона, можно заменить на 
$105\cdot 10^6~\mbox{В}$.

\subsection*{Задача 6 {\usefont{T2A}{cmr}{b}{n}(5.7)}: Скорость 
$\Upsilon$-частицы}
Законы сохранения энергии $E_++E_-=E_\Upsilon$ и импульса $p_+-p_-=p_\Upsilon$
(допустим $p_+>p_-$) определяют энергию и импульс $\Upsilon$-частицы. Тогда ее 
скорость будет
$$v=\frac{p_\Upsilon c^2}{E_\Upsilon}=c^2\,\frac{p_+-p_-}{E_++E_-}.$$
Электрон и позитрон ультрарелятивистские. Следовательно, $E_+\approx p_+c$ и 
$E_-\approx p_-c$. Поэтому
$$v=c\,\frac{p_+-p_-}{p_++p_-}.$$
Но, согласно условию задачи, мы можем написать 
$$R=\frac{p_-}{eB},\;\;\;\;\mbox{и}\;\;\;\;R=\frac{p_+}{e3B}.$$
Отсюда $p_+=3p_-$ и окончательно $$v\approx c\,\frac{3-1}{3+1}=\frac{c}{2}.$$

Можно решить задачу точно, не пренебрегая массой электрона $m$ по сравнению с массой
$\Upsilon$-частицы $M_\Upsilon$. Пусть в системе центра масс, которая движется со 
скоростью $v$ и является системой покоя для $\Upsilon$-частицы,
позитрон имеет энергию $E=M_\Upsilon/2$ и импульс $p=\sqrt{E^2-m^2}$ 
(в промежуточных формулах будем полагать, что скорость света $c=1$). В Л-системе 
будем иметь 
$$p_{+,x}=\gamma(p_x+\beta E),\;\;\;p_{-,x}=\gamma(-p_x+\beta E).$$
Ось $x$ направлена по движению позитрона (мы полагаем, что $E_+>E_-$). Поэтому
$p_x=p$, $p_{+,x}=p_+$, и $p_{-,x}=-p_-$, где $p_+,\,p_-$ --- величины импульсов
позитрона и электрона в Л-системе и, как уже знаем, $p_+=3p_-$. Следовательно, получаем
$$3=\frac{p_+}{p_-}=\frac{p+\beta E}{p-\beta E}.$$
Отсюда $$\beta=\frac{p}{2E}=\frac{1}{2}\sqrt{1-\frac{m^2}{E^2}},$$
и
$$v=\frac{c}{2}\sqrt{1-\frac{4m^2}{M_\Upsilon^2}}.$$

\section[Семинар 15]
{\centerline{Семинар 15}}

\subsection*{Задача 1 {\usefont{T2A}{cmr}{b}{n}(5.10)}: Ширина пучка на выходе
из магнитного поля}
Из рисунка видно, что $z=4\Delta R$. Но
$$R=\frac{p}{eB}\;\;\;\mbox{и}\;\;\;\Delta R=\frac{\Delta p}{eB}.$$
Из $(E/c)^2-p^2=m^2c^2$ получаем дифференцированием $E\Delta E/c^2=p\Delta p$, или
$$\Delta p=\frac{E\Delta E}{pc^2}=\frac{E\Delta E}{c\sqrt{E^2-m^2c^4}}.$$
Подставляя это в формуле для $\Delta R$, получаем
$$z=2\frac{\Delta E}{E}\frac{E}{eBc}\frac{1}{\sqrt{1-\left (\frac{mc2}{E}
\right )^2}}\approx 0,6~\mbox{м}.$$
Заметим, что в системе СИ
$$\frac{E}{eBc}=\frac{2\cdot 10^9~\mbox{эВ}}{e\cdot 1~\mbox{Тл}\cdot
3\cdot 10^8~\mbox{м/с}}=\frac{20}{3}\,\frac{\mbox{В}}{\mbox{Тл}\cdot~
\mbox{м/с}}=\frac{20}{3}~\mbox{м},$$
так как $\mbox{эВ}/e=\mbox{В}$.

\subsection*{Задача 2 {\usefont{T2A}{cmr}{b}{n}(5.8)}: Равенство углов падения и
отражения}
Рассмотрим электрон в точке $A$ в поле, на глубине $z$ (см. рисунок).
\begin{figure}[htb]
\centerline{\epsfig{figure=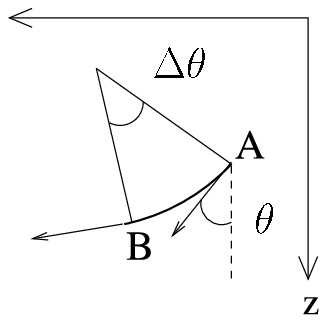,height=4cm}}
\end{figure}
Пусть его скорость составляет угол $\theta$ с осью $z$. Через время $\Delta t$
электрон окажется в точке $B$, и его скорость будет составлять угол $\theta+\Delta 
\theta$ с осью $z$. Если $\Delta t$ --- маленький промежуток времени, электрон от 
$A$ к $B$ будет двигаться на окружности радиуса $R=\frac{p}{eB(z)}$. Величина 
импульса электрона $p$ не меняется, так как магнитное поле не совершает работы. Имеем
$$\Delta \theta=\frac{v\Delta t}{R}=\frac{v\Delta t}{p}\,eB(z).$$
При этом изменение глубины $z$ равно $\Delta z=v\Delta t\cos{\theta}$. 
Следовательно, 
$$\Delta \theta=\frac{\Delta z}{\cos{\theta}}\,\frac{eB(z)}{p}.$$  
Или, перейдя к дифференциалам,
\begin{equation}
\cos{\theta}\,d\theta=\frac{eB(z)}{p}\,dz.
\label{eq15_2}
\end{equation}
Так как при входе и выходе электрона в область поля $z=0$, интегрируя (\ref{eq15_2}),
получаем 
$$\int\limits_\alpha^{(pi-\beta)}\cos{\theta}\,d\theta=\int\limits_0^0
\frac{eB(z)}{p}\,dz=0.$$
Следовательно, $\sin{\alpha}=\sin{(\pi-\beta)}=\sin{\beta}$ и $\alpha=\beta$.
\begin{figure}[htb]
\centerline{\epsfig{figure=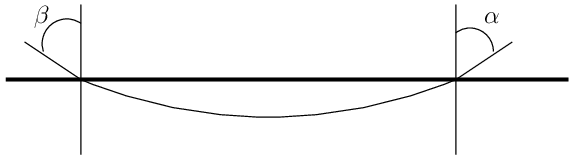,height=1.5cm}}
\end{figure}

\subsection*{Задача 3 {\usefont{T2A}{cmr}{b}{n}(5.21)}: Торможение в электрическом
поле}
Пройденный путь $L=2l$, где $l$ --- длина участка торможения. Работа электрических
сил на этом участке, $-e{\cal E}l$, по величине равна начальной кинетической энергии
электрона: $e{\cal E}l=mc^2(\gamma_0-1)$. Отсюда 
$$l=\frac{mc^2(\gamma_0-1)}{e{\cal E}}\;\;\;\mbox{и}\;\;\; 
L=\frac{2mc^2(\gamma_0-1)}{e{\cal E}}.$$

Из уравнения движения $$\frac{dp_x}{dt}=-e{\cal E}=\mathrm{const}$$
следует, что $p_x=p_0-e{\cal E}t$. В начальную точку электрон возвращается 
с противоположным импульсом, поэтому $-p_0=p_0-e{\cal E}t$ и получаем, что в  
начальную точку электрон возвращается через время
$$t=\frac{2p_0}{e{\cal E}}=\frac{2mV_0\gamma_0}{e{\cal E}}.$$

\subsection*{Задача 4 {\usefont{T2A}{cmr}{b}{n}(5.23)}: За какое время  сигнал
связи догонит ракету?}
Пусть $S^\prime$ --- мгновенно сопутствующая ракете система отсчета. Через время
$dt^\prime$, которому в Л-системе $S$ соответствует время $dt=\gamma dt^\prime$,
скорость ракеты в системе  $S^\prime$ станет $dV^\prime = a^\prime dt^\prime$.
По формуле сложения скоростей
$$V+dV=\frac{dV^\prime+V}{1+\frac{dV^\prime V}{c^2}}\approx (dV^\prime+V)
\left (1-\frac{dV^\prime V}{c^2}\right)\approx V+dV^\prime (1-\beta^2).$$
Следовательно, изменение скорости ракеты в системе $S$ за время $dt$ будет $dV=
dV^\prime (1-\beta^2)=dV^\prime/\gamma^2$. Поэтому
$$\frac{dV}{dt}=\frac{dV^\prime/\gamma^2}{\gamma dt^\prime}=\frac{1}
{\gamma^3}~\frac{dV^\prime}{dt^\prime}=\frac{a^\prime}{\gamma^3}.$$
Путь, пройденный ракетой за время $t$ после старта, равен
$$L(t)=\int\limits_0^tV dt= \int\limits_0^{V(t)}V\,\frac{\gamma^3}{a^\prime}
\,dV=\frac{c^2}{2a^\prime}\int\limits_0^{V^2(t)/c^2}\frac{d\beta^2}
{(1-\beta^2)\sqrt{1-\beta^2}}=\frac{c^2}{a^\prime}\,[\gamma(t)-1],$$
где мы учли, что
$$\gamma^3\beta d\beta=\frac{1}{2}\frac{d\beta^2}
{(1-\beta^2)\sqrt{1-\beta^2}}=d\left(\frac{1}{\sqrt{1-\beta^2}}\right)
=d\gamma.$$
С другой стороны,
$$\frac{d(V\gamma)}{dt}=\gamma\frac{dV}{dt}+V\frac{d\gamma}{dt}=
\frac{a^\prime}{\gamma^2}+\beta^2a^\prime=a^\prime.$$
Следовательно, $$\beta(t)\gamma(t)=\frac{a^\prime t}{c}$$ и 
$$\gamma(t)=\sqrt{1+[\beta(t)\gamma(t)]^2}=\sqrt{1+\frac{a^{\,\prime 2}t^2}
{c^2}}.$$
С учетом этого формула для $L(t)$ принимает вид
\begin{equation}
L(t)=\frac{c^2}{a^\prime}\,\left[\sqrt{1+\frac{a^{\,\prime 2}t^2}
{c^2}}-1\right ].
\label{eq15_4}
\end{equation}
Пусть сигнал догонит ракету за время $\tau$. Тогда $L(T+\tau)=c\tau$. Используя
(\ref{eq15_4}), это условие можно переписать как
$$\sqrt{1+\frac{a^{\,\prime 2}(T+\tau)^2}{c^2}}=1+\frac{a^\prime \tau}{c}$$
или
$$1+\frac{a^{\,\prime 2}T^2}{c^2}+\frac{a^{\,\prime 2}\tau^2}{c^2}+
2\frac{a^{\,\prime 2}T\tau}{c^2}=1+\frac{a^{\,\prime 2}\tau^2}{c^2}+
2\frac{a^\prime \tau}{c}.$$
Отсюда окончательно получаем $$\tau=\frac{a^\prime T^2}{2(c-a^\prime T)}.$$
Сигнал не догонит ракету, если $c-a^\prime T\le 0$, что для времени $T$ дает
$$T\ge\frac{c}{a^\prime}.$$

\subsection*{Задача 5 {\usefont{T2A}{cmr}{b}{n}(5.11)}: Магнитный фильтр}
\begin{figure}[htb]
\centerline{\epsfig{figure=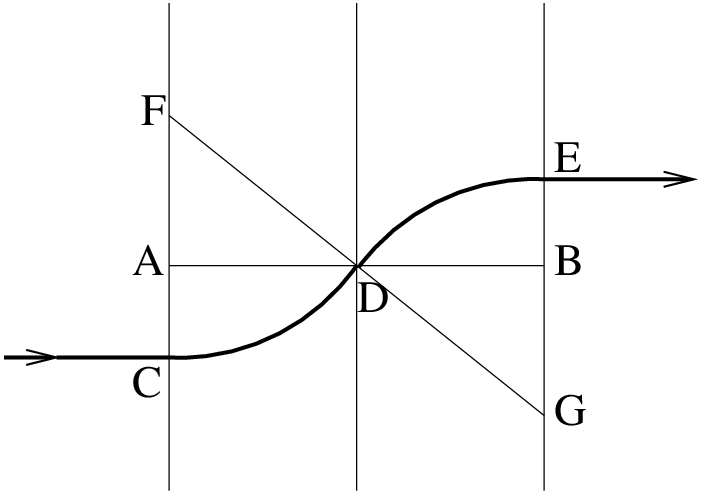,height=4cm}}
\end{figure}
Из рисунка видно, что $AC=BE=R-\sqrt{R^2-d^2}$, так как $FD=FC=R$, $GD=GE=R$ и
$AD=DB=d$. Поэтому расстояние между пучками протонов и электронов после прохождения
фильтра будет
$$z=2\left (R_p+R_e-\sqrt{R_p^2-d^2}-\sqrt{R_e^2-d^2}\right ).$$
При этом
$$R_e=\frac{p_e}{eB}\approx\frac{2\cdot 10^9~\mbox{эВ}/3\cdot 10^8~\mbox{м/с}}
{e\cdot 1~\mbox{Тл}}=\frac{2\cdot 10^9~\mbox{В}}{1~\mbox{Тл}\cdot
3\cdot 10^8~\mbox{м/с}}=\frac{20}{3}~\mbox{м}.$$
Имплульс протона
$$p_p=\sqrt{\frac{E^2}{c^2}-M^2c^2}\approx \sqrt{3}~\mbox{ГэВ/c}.$$ 
Следовательно,
$$R_p=\frac{\sqrt{3}\cdot 10^9}{1\cdot 3\cdot 10^8}~\mbox{м}=\frac{10}
{\sqrt{3}}~\mbox{м}.$$
Как видим, $R_e,R_p\gg d$ и, используя $\sqrt{1-x}\approx 1-x/2$ при $x\ll 1$,
получаем 
$$\sqrt{R_p^2-d^2}\approx R_p-\frac{d^2}{2R_p},\;\;\;
\sqrt{R_e^2-d^2}\approx R_e-\frac{d^2}{2R_e}.$$
Окончательно,
$$z=2\,\frac{d^2}{2}\left(\frac{1}{R_p}+\frac{1}{R_e}\right)=
0.2^2\left(\frac{\sqrt{3}}{10}+\frac{3}{20}\right)~\mbox{м}\approx
13~\mbox{мм}.$$

\subsection*{Задача 6 {\usefont{T2A}{cmr}{b}{n}(5.19)}:  Радиус кривизны 
траектории электрона в верхней точке}
{\usefont{T2A}{cmr}{m}{it} Первое решение.}
Во второй задаче этого семинара мы уже получили уравнение (\ref{eq15_2}), которое 
определяет изменение угла $\theta$ в зависимости от $z$. В начале $\theta=0$, 
а в верхней точке траектории $\theta=\pi/2$. Поэтому
$$\int\limits_0^{\pi/2}\cos{\theta}\,d\theta=\frac{eB_0}{p}\int\limits_0
^{z_0}\left (1+\frac{z}{a}\right )dz,$$
где $z_0$ --- координата верхней точки $O$. Следовательно,
\begin{equation}
1=\frac{eB_0}{p}\left (z_0+\frac{z_0^2}{2a}\right ),
\label{eq15_6}
\end{equation}
что для $z_0$ дает уравнение
$$z_0^2+2az_0-\frac{2ap}{eB_0}=0.$$
Учитывая, что $z_0>0$, получаем
$$z_0=-a+a\sqrt{1+\frac{2p}{eB_0a}}.$$
Следовательно, магнитное поле в точке $O$ равно
$$B(z_0)=B_0\left (1-1+\sqrt{1+\frac{2p}{eB_0a}}\right )=
B_0\sqrt{1+\frac{2p}{eB_0a}}.$$
Поэтому радиус кривизны траектории электрона в верхней точке траектории $O$ равен
$$R(z_0)=\frac{p}{eB(z+0)}=\frac{mV\gamma}{eB_0}\,\frac{1}{\sqrt{1+
\frac{2mV\gamma}{eB_0a}}}.$$

{\usefont{T2A}{cmr}{m}{it} Второе решение.}
Из уравнения движения электрона с зарядом $-e$,
$$\frac{d\vec{p}}{dt}=-e\vec{V}\times\vec{B},$$
с учетом того, что магнитное поле направлено вдоль оси $y$, получаем
$$\frac{dp_x}{dt}=eBV_z.$$
Но $p_x=m\gamma V_x$, а величина скорости, и следовательно $\gamma$, при движении
в магнитном поле не меняются, так как сила Лоренца не совершает работу. Поэтому 
получаем
$$\frac{dV_x}{dt}=\frac{eB}{m\gamma}\,V_z$$
или $$dV_x=\frac{eB}{m\gamma}\,dz=\frac{eB_0}{m\gamma}\left(1+\frac{z}{a}
\right)dz.$$
Проинтегрируем это уравнение с учетом того, что при $z=0$, когда электрон влетает 
в область магнитного поля, имеем $V_x=0$, а в наивысшей точке $O$ --- $V_x=V$ (так 
как в этой точке $V_z=0$). В результате получим
$$V=\frac{eB_0}{m\gamma}\left(z_0+\frac{z_0^2}{2a}\right).$$
Это эквивалентно уравнению (\ref{eq15_6}).

\subsection*{Задача 7 {\usefont{T2A}{cmr}{b}{n}(5.9)}: Какая часть мюонов выйдет
из области магнитного поля, не распавшись?}
\begin{figure}[htb]
\centerline{\epsfig{figure=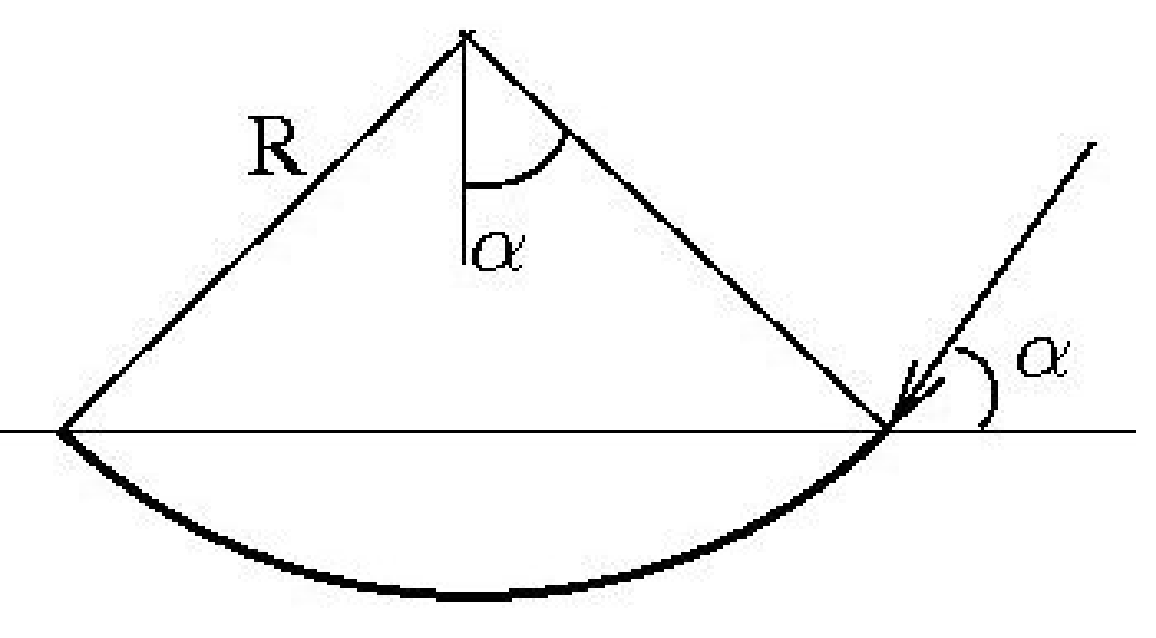,height=4cm}}
\end{figure}
Мюоны проходят путь (см. рисунок) $$s=2\alpha R=2\alpha\,\frac{p}{eB}$$
за время
$$t=\frac{s}{V}=2\alpha\,\frac{p}{eVB}.$$
Не распавшись, из области магнитного поля выйдет следующая часть мюонов: 
$$\frac{N}{N_0}=\exp{\left (-\frac{t}{\gamma \tau}\right )}.$$
Но
$$\frac{t}{\gamma \tau}=2\alpha\,\frac{p}{eV\gamma\tau B}=2\alpha\,\frac{m}
{eB\tau}$$
и
$$\frac{2m}{eB\tau}=\frac{2mc^2}{eB\tau c^2}=\frac{2\cdot 105\cdot 10^6}
{10^{-2}\cdot 2\cdot 10^{-6}\cdot 9\cdot 10^{16}}\approx 1,2\cdot 10^{-1}.$$
Следовательно,
$$\frac{N}{N_0}\approx \exp{(-1,2\cdot 10^{-1}\cdot\alpha)}\approx 1-1,2\cdot 
10^{-1}\cdot\alpha.$$

\section[Семинар 16]
{\centerline{Семинар 16}}

\subsection*{Задача 1 {\usefont{T2A}{cmr}{b}{n}(5.27)}: Столкновение протона и
электрона в электрическом поле}
Имеем следующие уравнения движения для протона ($e$ --- заряд протона)
$$\frac{d\vec{p}_p}{dt}=e\vec{\cal{E}}$$
и для электрона
$$\frac{d\vec{p}_e}{dt}=-e\vec{\cal{E}}.$$
Следовательно, за одно и то же время протон и электрон приобретут одинаковые 
по величине, но противоположные импульсы. Их энергии будут разными, но, согласно 
закону сохранения энергии, $E_p+E_e=m_pc^2+m_ec^2+e{\cal{E}}L$. Следовательно, 
в момент столкновения
$$p_p+p_e=\left (m_p+m_e+e{\cal{E}}L,\;\vec{0}\right ),$$
где $p_p,p_e$ --- 4-импульсы протона и электрона; мы временно положили $c=1$. Закон
сохранения 4-импульса в реакции $p+e^-\to p+e^-+\pi^++\pi^-$ дает
$$(p_p+p_e)^2=(p_p^\prime+p_e^\prime+p_{\pi^+}+p_{\pi^-})^2.$$
Левую часть вычисляем в Л-системе: $(p_p+p_e)^2=(m_p+m_e+e{\cal{E}}L)^2$. Правую
часть --- в системе центра инерции, где на пороге реакции частицы покоятся:
$(p_p^\prime+p_e^\prime+p_{\pi^+}+p_{\pi^-})^2=(m_p+m_e+2m_\pi)^2$. 
Следовательно, $(m_p+m_e+e{\cal{E}}L)^2=(m_p+m_e+2m_\pi)^2$ и (восстановили 
скорость света $c$)
$$L=\frac{2m_\pi c^2}{e{\cal{E}}}=\frac{2\cdot 140\cdot 10^6}{10^7}~\mbox{м}
\approx 28~\mbox{м}.$$
Если протон стартует на время $T$ раньше электрона, он приобретет на $e{\cal{E}}T$ 
больше импульса. Следовательно, в момент столкновения
$$p_p+p_e=\left (m_p+m_e+e{\cal{E}}L, \;e\vec{\cal{E}}T\right )$$
и условие порога реакции будет
$$(m_p+m_e+e{\cal{E}}L)^2-(e{\cal{E}}T)^2=(m_p+m_e+2m_\pi)^2.$$
Отсюда
$$L=\frac{1}{e{\cal{E}}}\left[\sqrt{(m_p+m_e+2m_\pi)^2+(e{\cal{E}}T)^2}-m_p-
m_e\right].$$
Или, если восстановить скорость света $c$,
$$L=\frac{c^2}{e{\cal{E}}}\left[\sqrt{(m_p+m_e+2m_\pi)^2+\left(\frac{e{\cal{E}
}T}{c}\right )^2}-m_p-m_e\right]\approx 31,5~\mbox{м}.$$

\subsection*{Задача 2 {\usefont{T2A}{cmr}{b}{n}(5.29)}: Предельные значения
компонент импульса и скорости}
Из уравнения движения
$$\frac{d\vec{p}}{dt}=\vec{F}$$
следует, что
$$\frac{dp_\parallel}{dt}=0,\;\;\;\frac{dp_\perp}{dt}=F.$$
Поэтому
$$p_\parallel=mV_0\gamma_0=\mathrm{const}\;\;\;\mbox{и}\;\;\;
p_\perp=Ft \to \infty.$$
Из равенств $\vec{p}^2=p_\parallel^2+p_\perp^2$ и $\vec{p}=m\vec{V}\gamma$ 
получаем $$m^2c^2\beta^2\gamma^2=m^2c^2\beta_0^2\gamma_0^2+(Ft)^2.$$
С учетом $\beta^2\gamma^2=\gamma^2-1$ это дает
$$\gamma=\gamma_0\sqrt{1+\left(\frac{Ft}{mc\gamma_0}\right)^2}.$$
Тогда из $p_\parallel=mV_\parallel\gamma=mV_0\gamma_0$ получаем
$$V_\parallel=\frac{V_0\gamma_0}{\gamma}=\frac{V_0}{\sqrt{1+\left(\frac{Ft}
{mc\gamma_0}\right)^2}}\to 0\;\;\mbox{при}\;\; t\to\infty.$$
Аналогично из  $p_\perp=mV_\perp\gamma=Ft$ получаем
$$V_\perp=\frac{Ft}{m\gamma}=\frac{Ft}{m\gamma_0}\,\frac{1}{\sqrt{1+\left(
\frac{Ft}{mc\gamma_0}\right)^2}}\to \frac{Ft}{m\gamma_0}\,\frac{mc\gamma_0}
{Ft}=c\;\;\mbox{при}\;\; t\to\infty.$$

\subsection*{Задача 3 {\usefont{T2A}{cmr}{b}{n}(5.30)}: Высота траектории и
минимальная скорость электрона}
\begin{figure}[htb]
\centerline{\epsfig{figure=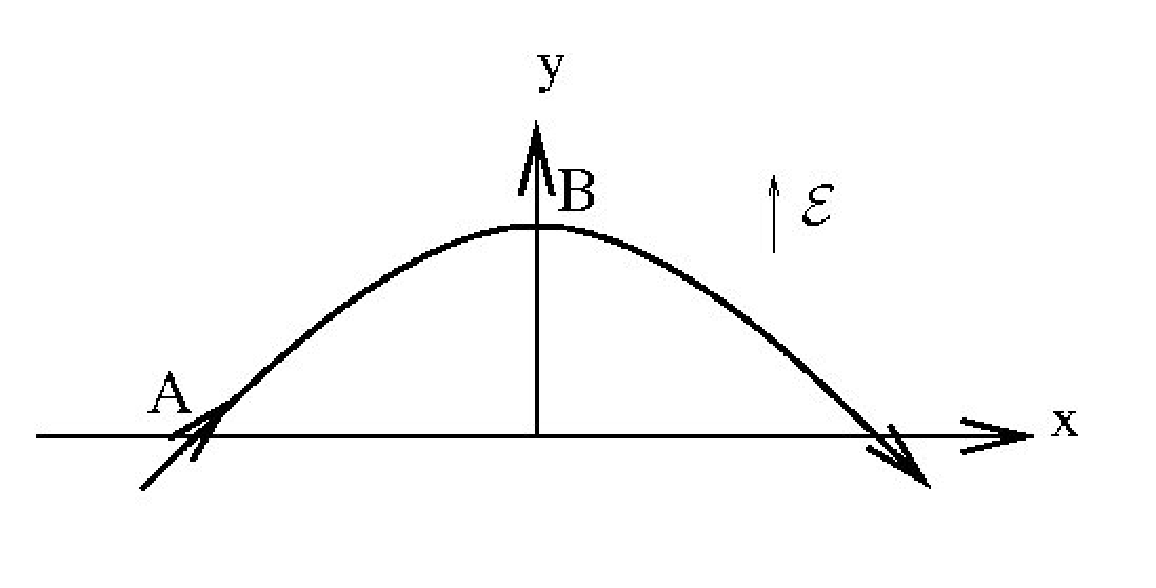,height=4cm}}
\end{figure}
Импульс электрона вдоль оси $x$ не меняется: $p_x=p_{0x}=p_0\cos{45^\circ}=
p_0/\sqrt{2}$. Закон сохранения энергии при переходе из точки $A$ в точку $B$ (см.
рисунок) дает
$$\sqrt{p_0^2+m_e^2}=\sqrt{\left(\frac{p_0}{\sqrt{2}}\right)^2+m_e^2}+
e{\cal{E}}H,$$
где $e$ --- заряд протона (т.е. заряд электрона с обратным знаком: $e>0$). Отсюда
(мы востановили скорость света $c$, которую в предыдущей формуле положили равной 
единице)
\begin{equation}
H=\frac{c}{e{\cal{E}}}\left [\sqrt{p_0^2+m_e^2c^2}-\sqrt{\frac{p_0^2}{2}+
m_e^2c^2}\right].
\label{eq16_3}
\end{equation}
Начальная кинетическая энергия электрона равна
$$T_0=c\sqrt{p_0^2+m_e^2c^2}-m_ec^2.$$
Отсюда 
$$c\sqrt{p_0^2+m_e^2c^2}=T_0+m_ec^2,\;\;\mbox{и}\;\;
c\sqrt{\frac{p_0^2}{2}+m_e^2c^2}=
\sqrt{\frac{(T_0+m_ec^2)^2+m_e^2c^4}{2}}.$$
Подставляя это в (\ref{eq16_3}), получаем окончательно
$$H=\frac{1}{e{\cal{E}}}\left[T_0+m_ec^2-\sqrt{\frac{(T_0+m_ec^2)^2+
m_e^2c^4}{2}}\right ]\approx 0,38\,\mbox{м}.$$
Минимальную скорость электрон имеет в точке $B$ и при этом его импульс равен $p_0/
\sqrt{2}$. Поэтому
$$\frac{m V_{min}}{\sqrt{1-V_{min}^2/c^2}}=\frac{p_0}{\sqrt{2}}.$$
Отсюда
$$V_{min}=c\sqrt{\frac{p_0^2}{2m_e^2c^2+p_0^2}}=c\sqrt{\frac{E_0^2-m_e^2c^4}
{E_0^2+m_e^2c^4}}=c\sqrt{\frac{(T_0+m_ec^2)^2-m_e^2c^4}
{(T_0+m_ec^2)^2+m_e^2c^4}}\approx 0,89\,c.$$

\subsection*{Задача 4: Минимальный угол разлета $\gamma$-квантов}
Закон сохранения 4-импульса $p_++p_-=k_1+k_2$ дает $(p_++p_-)^2=(k_1+k_2)^2$.
Но (положили $c=1$) $(p_++p_-)^2=2m_e^2+2m_eE$ и $(k_1+k_2)^2=2k_1\cdot k_2=
2E_1E_2(1-\cos{\theta})$. Следовательно, $m_e(m_e+E)=E_1E_2(1-\cos{\theta})$
и 
\begin{equation}
1-\cos{\theta}=\frac{m_e(m_e+E)}{E_1E_2}=\frac{m_e(m_e+E)}{E_1(E+m_e
-E_1)},
\label{eq16_4}
\end{equation}
так как $E_2=E+m_e-E_1$. Функция $f(E_1)=E_1(E+m_e-E_1)$ имеет максимум
при $E_1=(E+m_e)/2=E_2$. Этому значению, согласно (\ref{eq16_4}), соответствует
минимальное значение $1-\cos{\theta}$, т.~е. максимальное значение $\cos{\theta}$
и минимальное значение угла $\theta$. Поэтому
$$1-\cos{\theta_{min}}=2\sin^2{\frac{\theta_{min}}{2}}=4\frac{m_e}{m_e+E}$$
и окончательно (восстановили скорость света $c$)
$$\theta_{min}=2\arcsin{\sqrt{\frac{2m_ec^2}{m_ec^2+E}}}.$$

\subsection*{Задача 5 {\usefont{T2A}{cmr}{b}{n}(3.19)}: Скорость второго корабля
относительно Земли}
Пусть в системе первого корабля второй корабль имеет скорость $V$. С точки зрения 
второго корабля источник сигнала приближается к нему со скоростью $V$. Поэтому он
будет принимать сигнал с частотой $\omega$, которая определяется из эффекта Доплера:
$$\omega=\gamma_V\omega_1(1+\beta_V)=\omega_1\sqrt{\frac{1+\beta_V}{1-
\beta_V}}.$$
Второй корабль отражает такую же частоту $\omega$ (в своей системе отсчета). Поэтому
первый корабль будет принимать частоту
$$\omega_2=\gamma_V\omega(1+\beta_V)=\omega\sqrt{\frac{1+\beta_V}{1-
\beta_V}}=\omega_1\frac{1+\beta_V}{1-\beta_V}.$$
Но скорость второго корабля относительно Земли
$$V_2=\frac{V-V_1}{1-VV_1/c^2}$$ или
$$\beta_2=\frac{\beta_V-\beta_1}{1-\beta_V\beta_1}.$$
Отсюда
$$\beta_V=\frac{\beta_1+\beta_2}{1+\beta_1\beta_2}$$
и
$$\frac{\omega_2}{\omega_1}=\frac{1+\beta_V}{1-\beta_V}=\frac{1+\beta_1\beta_2
+\beta_1+\beta_2}{1+\beta_1\beta_2-\beta_1-\beta_2}.$$
Решая относительно $\beta_2$, получаем окончательно
$$\beta_2=\frac{\beta_1(\omega_2+\omega_1)-(\omega_2-\omega_1)}
{\beta_1(\omega_2-\omega_1)-(\omega_2+\omega_1)}.$$

\section[Семинар 17]
{\centerline{Семинар 17}}

\subsection*{Задача 1 {\usefont{T2A}{cmr}{b}{n}(\hspace*{-1.5mm}\cite{8}, 
2.14)}: Какое время покажут часы?}
{\usefont{T2A}{cmr}{m}{it} Первое решение.}
В Л-системе пройдет время $\gamma T$ и расстояние между кораблями будет $2V\cdot
\gamma T$. В Л-системе сигнал со скоростью $c$ догоняет корабль, который убегает
 со скоростью $V$. Поэтому сигнал догонит за время
$$\tau=\frac{2V\gamma T}{c-V}=\frac{2\beta\gamma}{1-\beta}\,T.$$
Следовательно, в этот момент часы в Л-системе покажут 
$$t=\gamma T+\tau=\gamma T\left (\frac{2\beta}{1-\beta}+1\right )=
\gamma T\,\frac{1+\beta}{1-\beta}.$$
Если ось $x$ направим по движению второго корабля, координата события ``сигнал догнал
второй корабль'' будет $x=Vt$. Поэтому в этот момент часы на втором корабле покажут
время $$t^\prime_2=\gamma\left(t-\frac{V}{c^2}\,x\right)=\gamma t(1-\beta^2)=
\frac{t}{\gamma}=T\,\frac{1+\beta}{1-\beta}=3T.$$
Какое время покажут в этот момент часы первого корабля, если ``этот момент'' (т. е. 
одновременность) понимается в смысле координатной системы, связанной с первым кораблем?
Если ось $x$ направим по движению первого корабля, то координата события ``сигнал 
догнал второй корабль'' будет $x=-Vt$. Поэтому, согласно преобразованию Лоренца, 
в этом случае будем иметь
$$t^\prime_1=\gamma t(1+\beta^2)=T\,\frac{1+\beta^2}{(1-\beta)^2}=5T.$$

{\usefont{T2A}{cmr}{m}{it} Второе решение.}
В системе первого корабля второй корабль удаляется со скоростью
$$u=\frac{V+V}{1+v^2/c^2}=\frac{2V}{1+\beta^2}.$$
Следовательно, в момент посылки сигнала расстояние до него было $2VT/(1+\beta^2)$.
Сигнал догонит второй корабль за время (по часам первого корабля)
$$\tau^\prime =\frac{2VT/(1+\beta^2)}{c-2V/(1+\beta^2)}=\frac{2\beta T}
{(1-\beta)^2}.$$
Поэтому, когда сигнал догонит второй корабль, часы первого корабля покажут время
$$t^\prime_1=\tau^\prime+T=T\,\frac{1+\beta^2}{(1-\beta)^2}=5T.$$
С точки зрения космонавтов первого корабля часы на втором корабле идут $\gamma_u$ раз
медленнее. Но 
$$\gamma_u=\gamma_{V\oplus V}=\gamma^2(1+\beta^2)=\frac{1+\beta^2}
{1-\beta^2}.$$ 
Поэтому в момент встречи сигнала со вторым кораблем часы второго корабля покажут время
$$t^\prime_2=\frac{t^\prime_1}{\gamma_u}=T\,\frac{1+\beta}{1-\beta}=3T.$$

\subsection*{Задача 2: Частота принимаемого сигнала при постоянном угле пеланга}
В системе корабля источник сигнала (маяк) движется со скоростью $V$. Так как угол
$\theta$ дан в системе маяка, используем следующую формулу для эффекта Доплера:
$\nu=\gamma\nu_0 (1+\beta\cos{\theta})$, где $\nu$ --- это частота принимаемого 
на корабле сигнала. Следовательно, $\nu<\nu_0$, если
$\gamma(1+\beta\cos{\theta})<1$, что дает
$$\cos{\theta}<\frac{1}{\beta}\left (\frac{1}{\gamma}-1\right).$$
По условию $\beta=0,8$, что означает $\gamma=1/0,6$. Поэтому получаем
$\cos{\theta}<-1/2$ и $\theta>120^\circ$.

\subsection*{Задача 3 {\usefont{T2A}{cmr}{b}{n}(\hspace*{-1.5mm}\cite{8},
2.53)}: Под каким углом выйдет луч света через боковой торец?}
\begin{figure}[htb]
\centerline{\epsfig{figure=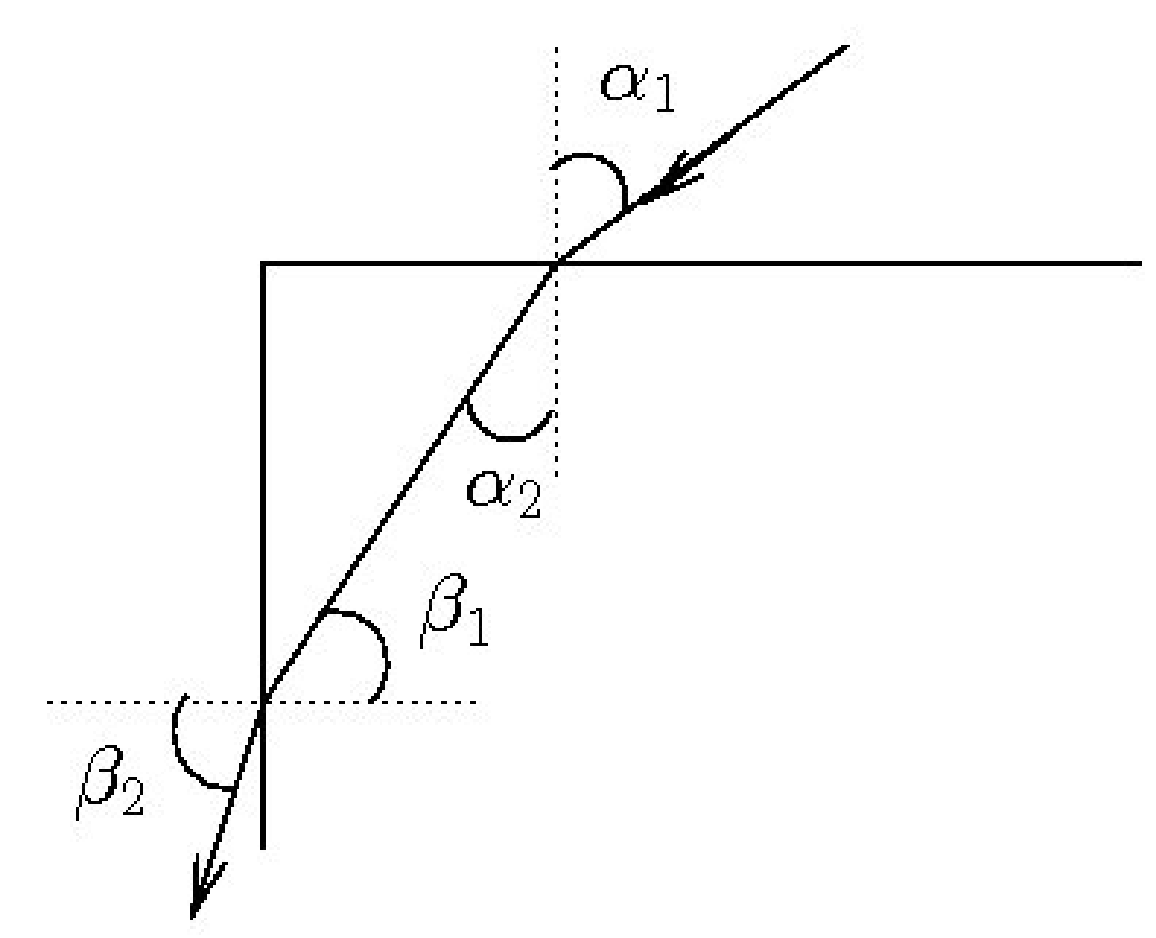,height=4cm}}
\end{figure}
В системе покоя бруска имеем картину, показанную на рисунке. Так как для воздуха 
показатель преломления близок к единице, закон преломления (закон Снеллиуса) будет 
иметь вид $\sin{\alpha_1}=n\,\sin{\alpha_2}$ и $n\,\sin{\beta_1}=\sin{\beta_2}$.
Но $\beta_1=\pi/2-\alpha_2$ и из этих уравнении получаем
$$\sin{\beta_2}=n\,\cos{\alpha_2}=n\sqrt{1-\frac{\sin^2{\alpha_1}}{n^2}}.$$
Следовательно, $\sin{\beta_2}=\sqrt{n^2-\sin^2{\alpha_1}}$. Сначала найдем 
$\alpha_1$. В Л-системе луч падает вертикально, т. е. $c_x=0,\,c_y=-c$. В системе
бруска ($S^\prime$-система) будем иметь
$$c_x^\prime=\frac{c_x-V}{1-\frac{c_x V}{c^2}}=-V,\;\;\;c_y^\prime=\frac{c_y}
{\gamma\left (1-\frac{c_x V}{c^2}\right )}=-\frac{c}{\gamma}.$$
При этом, разумеется,
$$c_x^{\prime\,2}+c_y^{\prime\,2}=V^2+\frac{c^2}{\gamma^2}=c^2.$$
Следовательно,
$$\sin{\alpha_1}=\frac{|c_x^\prime|}{c}=\beta.$$
Таким образом, $\sin{\beta_2}=\sqrt{n^2-\beta^2}$ и для луча, который вышел из 
бруска, будем иметь $c_x^{\prime\prime}=-c\,\cos{\beta_2}=-c\,\sqrt{1+
\beta^2-n^2}$ и $c_y^{\prime\prime}=-c\,\sin{\beta_2}=-c\,\sqrt{n^2-\beta^2}$. 
Перейдем обратно в Л-систему:
$$\tilde{c}_x=\frac{c_x^{\prime\prime}+V}{1+\frac{c_x^{\prime\prime} V}
{c^2}}.$$
Следовательно, угол к вертикали, под которым луч выйдет через боковой торец бруска 
после преломления в Л-системе, определяется из равенства
$$\tg{\theta}=\left|\frac{\tilde{c}_x}{\tilde{c}_y}
\right|\frac{c_x^{\prime\prime}+V}{c_y^{\prime\prime}}\,\gamma=
\frac{\beta-\sqrt{1+\beta^2-n^2}}{\sqrt{(n^2-\beta^2)(1-\beta^2)}}.$$
Так как $n^2>1$, то $\beta>\sqrt{1+\beta^2-n^2}$ и $\tilde{c}_x>0$. 
Следовательно, в Л-системе луч из боковой грани выходит вперед, а не назад. При этом
бусок успевает убежать от луча: пока фотон передвигается из точки $B$ в точку 
$B^\prime$, точка $A$ грани передвигается в точку $A^\prime$ и при этом $AA^\prime>
AB^\prime$ (см. рисунок).
\begin{figure}[htb]
\centerline{\epsfig{figure=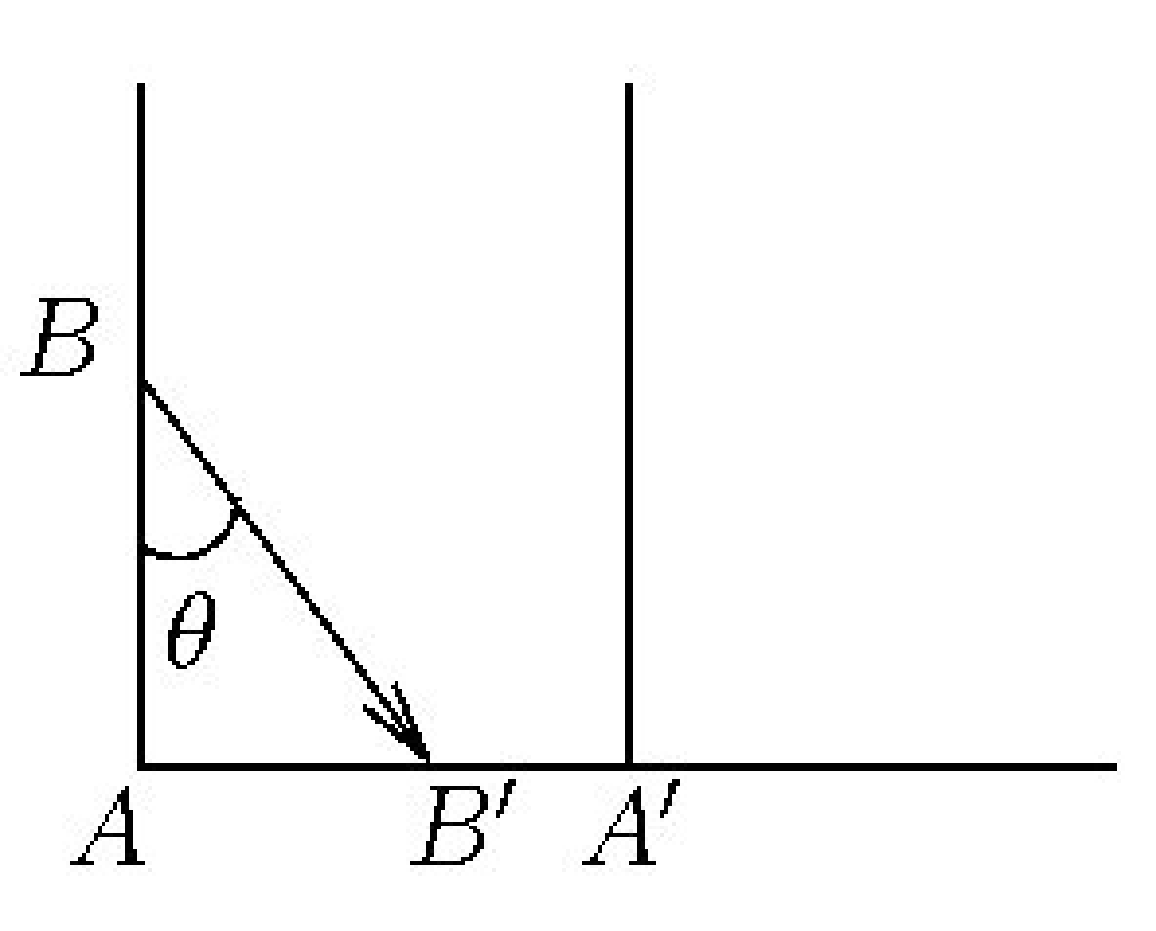,height=4cm}}
\end{figure}
Докажем это. $AA^\prime>AB^\prime$ эквивалентно неравенству $\beta>\sin{\theta}$ 
(скорость грани больше, чем горизонтальная проекция скорости фотона при движении от $B$
к $B^\prime$). Так как
$$1+\ctg^2{\theta}=1+\frac{(n^2-\beta^2)(1-\beta^2)}{(\beta-\sqrt{1+\beta^2-
n^2})^2}=\frac{(1-\beta\sqrt{1+\beta^2-n^2})^2}{(\beta-\sqrt{1+\beta^2-
n^2})^2},$$
то $$\sin{\theta}=\frac{1}{\sqrt{1+\ctg^2{\theta}}}=\frac{\beta-\sqrt{1+
\beta^2-n^2}}{1-\beta\sqrt{1+\beta^2-n^2}}$$ и надо доказать, что
\begin{equation}
\beta>\frac{\beta-\sqrt{1+\beta^2-n^2}}{1-\beta\sqrt{1+\beta^2-n^2}}.
\label{eq17_3}
\end{equation}
Так как $n>1>\beta$, то $\sqrt{1+\beta^2-n^2}<1$ и $1-\beta\sqrt{1+\beta^2-n^2}
>0$. Поэтому неравенство (\ref{eq17_3}) эквивалентно неравенству
$$\beta-\beta^2\sqrt{1+\beta^2-n^2}>\beta-\sqrt{1+\beta^2-n^2},$$
или $\beta^2<1$, что верно.

Луч не выйдет из бруска, если будет иметь место полное внутреннее отражение: $\beta_2=
\frac{\pi}{2}$. Но тогда $\sqrt{n^2-\beta^2}=\sin{\beta_2}=1$, что дает $n=
\sqrt{1+\beta^2}$. Следовательно, чтобы луч вышел из бруска, мы должны иметь
$n<\sqrt{1+\beta^2}$.

\subsection*{Задача 4: Энергия $\gamma$-кванта в реакции $e^++e^-\to
\omega+\gamma$}
Пишем закон сохранения 4-импульса $p_++p_-=p_\omega+k$, или $p_\omega=p-k$,
где $p=p_++p_-=(2E,\vec{0})$. Возведем обе части равенства в квадрат, 
$M_\omega^2=p_\omega^2=4E^2-4EE_\gamma$. Отсюда
$$E_\gamma=\frac{4E^2-M_\omega^2}{4E}=E-\frac{M_\omega^2}{4E}.$$
Восстановим по размерности скорость света $c$:
$$E_\gamma=E-\frac{M_\omega^2c^4}{4E}\approx 110~\mbox{МэВ}.$$

\subsection*{Задача 5 {\usefont{T2A}{cmr}{b}{n}(\hspace*{-1.5mm}\cite{8}, 
5.16)}: Распад каона в пузырьковой камере}
Пусть в системе покоя каона мюон имеет энергию $E^\prime$ и импульс $p^\prime$.
Тогда в Л-системе будем иметь (положили $c=1$) $p_x=\gamma(p^\prime_x+\beta_K
E^\prime)$, где $\beta_K$ --- скорость каона в Л-системе. Следовательно,
$$p_{max}=\gamma(p^\prime+\beta_K E^\prime),\;\;\;
p_{min}=\gamma|(-p^\prime+\beta_K E^\prime)|.$$
Так как $R=\frac{p}{qB}$, из условия задачи следует, что
$$N=\frac{p_{max}}{p_{min}}=\frac{\beta_K E^\prime+p^\prime}
{|\beta_K E^\prime-p^\prime|}.$$
Но
$$\frac{p^\prime}{E^\prime}=\beta^\prime_\mu$$
есть скорость мюона в системе покоя каона. Следовательно,
$$N=\frac{\beta_K+\beta^\prime_\mu}{|\beta_K-\beta^\prime_\mu|}$$
и
\begin{equation}
\beta^\prime_\mu=\frac{N\pm 1}{N\mp 1}\,\beta_K.
\label{eq17_5a}
\end{equation}
Здесь верхний знак берется, если $\beta_K<\beta^\prime_\mu$, и нижний --- если
$\beta_K>\beta^\prime_\mu$. 

Найдем теперь $\beta^\prime_\mu$ по-другому. В системе покоя каона имеем закон 
сохранения 4-импульса $p_K^\prime=p_\mu^\prime+p_\nu^\prime$, где $p_K^\prime=
(M_K,\vec{0})$. Следовательно, $0=p_\nu^{\prime\,2}=(p_K^\prime-p_\mu^\prime)^2
=M_K^2+m_\mu^2-2M_KE^\prime$. Отсюда
$$E^\prime=\frac{M_K^2+m_\mu^2}{2M_K}\;\;\mbox{и}\;\;
p^\prime=\sqrt{E^{\prime\,2}-m_\mu^2}=\frac{M_K^2-m_\mu^2}{2M_K}.$$
Поэтому
$$\beta^\prime_\mu=\frac{p^\prime}{E^\prime}=\frac{M_K^2-m_\mu^2}{M_K^2
+m_\mu^2}.$$
Отсюда, с учетом (\ref{eq17_5a}), получаем
\begin{equation}
\frac{M_K}{m_\mu}=\sqrt{\frac{1+\beta^\prime_\mu}{1-\beta^\prime_\mu}}=
\sqrt{\frac{N\mp 1+\beta_K\,(N\pm 1)}{N\mp 1-\beta_K\,(N\pm 1)}}.
\label{eq17_5b}
\end{equation}
Средняя длина трека каона до распада равна 
$$l=\beta_K\gamma_K\tau=\frac{\beta_K \tau}{\sqrt{1-\beta_K^2}}.$$
Отсюда (восстановили скорость света $c$)
$$\beta_K=\frac{1}{\sqrt{1+\left (\frac{c\tau}{l}\right)^2}}\approx 0,457.$$
Подставляя это значение $\beta_K$ в (\ref{eq17_5b}), получаем
$$\frac{M_K}{m_\mu}\approx 4,7,\;\;\mbox{или}\;\;1,3,$$
где первое число соответствует верхнему знаку в формуле (\ref{eq17_5b}), а второе 
число --- нижнему знаку. Реальным массам частиц соответствует первое решение.

\subsection*{Задача 6 {\usefont{T2A}{cmr}{b}{n}(\hspace*{-1.5mm}\cite{8}, 
5.41)}: Какую часть топлива израсходует ракета?}
Пусть в сопутствующей системе ракеты за короткое время излучилась энергия 
${\cal{E^\prime}}$. Законы сохранения энергии и импульса имеют вид
$$mc^2={\cal{E^\prime}}+(m+\Delta m)c^2\gamma^\prime,\;\;\;
0=-\frac{\cal{E^\prime}}{c}+(m+\Delta m)V^\prime\gamma^\prime,$$
где $V^\prime,\,\gamma^\prime$ --- скорость и $\gamma$-фактор, которые появятся 
у ракеты из-за отдачи. Исключая из этих уравнений ${\cal{E^\prime}}$, получаем
в первом порядке по малым величинам $\beta^\prime$ и $\Delta m$ (в этом порядке 
$\gamma^\prime\approx 1$):
\begin{equation}
\frac{\Delta m}{m}=-\beta^\prime.
\label{eq17_6}
\end{equation}
Согласно релятивистскому закону сложения скоростей, скорость ракеты после отдачи в 
Л-системе дается выражением $$\beta+\Delta \beta=\frac{\beta^\prime+\beta}
{1+\beta\beta^\prime}\approx \beta+(1-\beta^2)\,\beta^\prime.$$
Отсюда $$\beta^\prime\approx\frac{\Delta\beta}{1-\beta^2}$$
и (\ref{eq17_6}) принимает вид $$\frac{\Delta m}{m}=-\frac{\Delta\beta}
{1-\beta^2}.$$ Перейдем в этом уравнении от конечных разностей к дифференциалам и 
проинтегрируем:
$$\ln{\frac{M_1}{M_2}}=-\int\limits_{M_1}^{M_2}\frac{dm}{m}=
\int\limits_{\beta_1}^{\beta_2}\frac{d\beta}
{1-\beta^2}=\frac{1}{2} \int\limits_{\beta_1}^{\beta_2}d\beta\left[\frac{1}
{1-\beta}+\frac{1}{1+\beta}\right ]=\ln{\sqrt{\frac{(1+\beta_2)(1-\beta_1)}
{(1-\beta_2)(1+\beta_1)}}}.$$
Отсюда
$$\frac{M_2}{M_1}=\sqrt{\frac{(1-\beta_2)(1+\beta_1)}
{(1+\beta_2)(1-\beta_1)}}\approx \sqrt{\frac{1-\beta_2}
{1-\beta_1}}=\frac{1}{\sqrt{10}},$$
и окончательно
$$\frac{\Delta M}{M_1}=\frac{M_1-M_2}{M_1}=1-\frac{1}{\sqrt{10}}\approx
68\%.$$

\section[Семинар 18]
{\centerline{Семинар 18}}

\subsection*{Задача 1: За какое время волна от камня достигнет берега?}
В системе реки $S^\prime$ берег движется со скоростью $-V$ и точка $A$, откуда 
бросили камень, к моменту прихода волны сместится в точку $A^\prime$, которая имеет 
координату $x^\prime=-Vt^\prime$.
Причем $t^\prime$ определяется из условия (см. рисунок) $ut^\prime=\sqrt{a^2+
V^2t^{\prime\,2}}$.
\begin{figure}[htb]
\centerline{\epsfig{figure=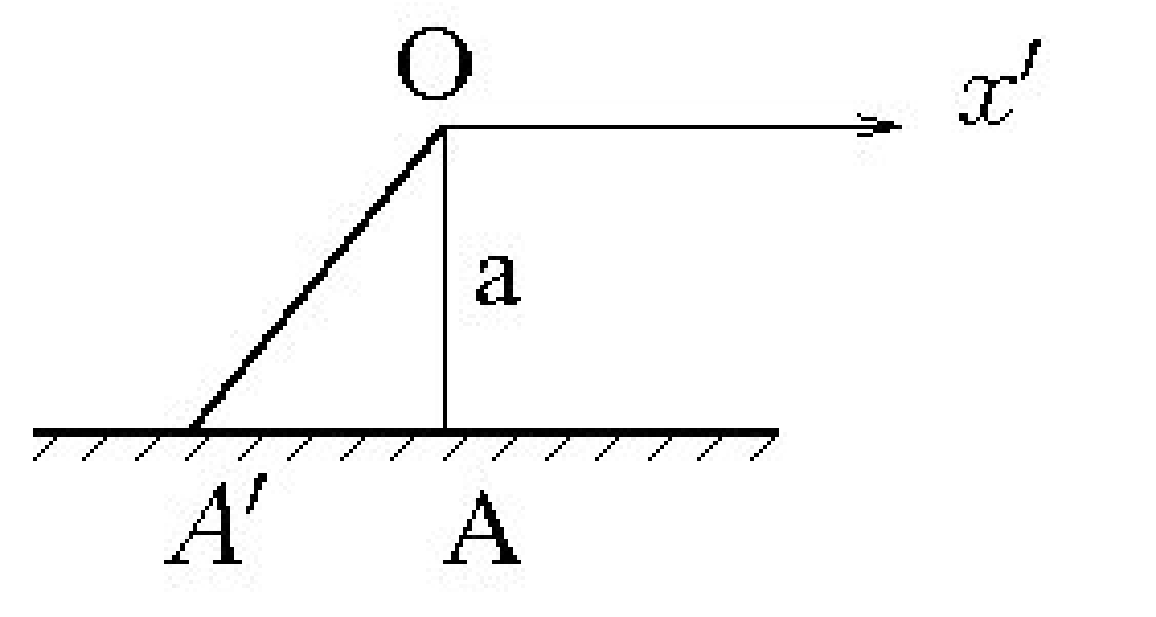,height=2.5cm}}
\end{figure}
Отсюда $$t^\prime=\frac{a}{\sqrt{u^2-v^2}}.$$
В системе берега волна придет в точку,  откуда бросили камень, через время
$$t_A=\gamma_V\left (t^\prime+\frac{V}{c^2}x^\prime\right)=\gamma_V t^\prime 
(1-\beta_V^2)=\frac{t^\prime}{\gamma_V}.$$
Подставляя $t^\prime$, получаем окончательно
$$t_A=\frac{a}{\gamma_V\sqrt{u^2-V^2}}=\frac{a}{c}\sqrt{\frac{c^2-V^2}
{u^2-V^2}}.$$

Чтобы найти, за какое время волна от камня достигнет берега в системе берега, проследим 
в системе $S^\prime$ за участком волны, который движется под углом $\theta^\prime$
к оси $x^\prime$ (см. рисунок).
\begin{figure}[htb]
\centerline{\epsfig{figure=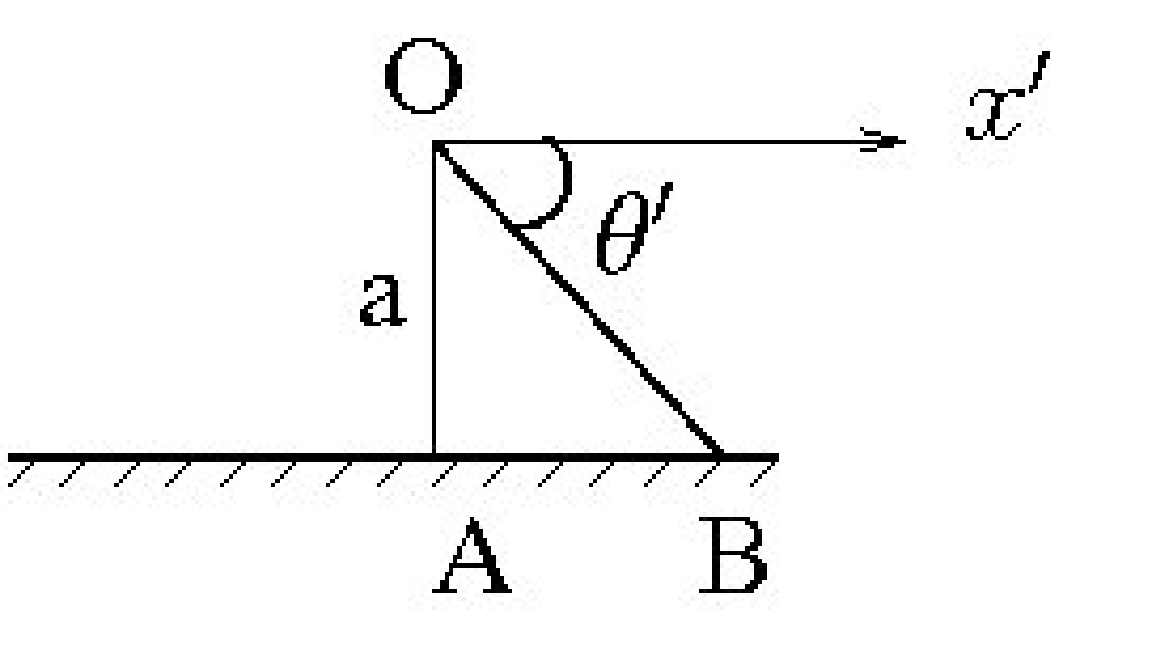,height=3cm}}
\end{figure}
Тогда в этой системе волна достигнет берега за время
$$t^\prime_B=\frac{|OB|}{u}=\frac{a}{u\sin{\theta^\prime}}$$
в точке с координатой $x_B^\prime=a\ctg{\theta^\prime}$. В системе берега этот 
участок волны достигнет берега через время
$$t_B=\gamma_V\left (t^\prime_B+\frac{V}{c^2}x_B^\prime\right)=
\gamma_V\,\frac{a}{u}\,\frac{1+\beta_V\beta_u\cos{\theta^\prime}}
{\sin{\theta^\prime}}.$$
Надо найти минимум функции
$$f(\theta^\prime)=\frac{1+\beta_V\beta_u\cos{\theta^\prime}}
{\sin{\theta^\prime}}.$$
Условие
$$\frac{df(\theta^\prime)}{d\theta^\prime}=-\frac{\cos{\theta^\prime}(1+
\beta_V\beta_u\cos{\theta^\prime})}{\sin^2{\theta^\prime}}-\beta_V\beta_u=
-\frac{\cos{\theta^\prime}+\beta_V\beta_u}{\sin^2{\theta^\prime}}=0$$
означает, что угол $\theta^\prime_m$, который соответствует минимальному значению 
$t_B$, определяется из уравнения $\cos{\theta^\prime_m}=-\beta_V\beta_u$. 
Следовательно,
$$t_{min}=\gamma_V\,\frac{a}{u}\,\frac{1-\beta^2_V\beta^2_u}{\sqrt{1-
\beta^2_V\beta^2_u}}=\frac{a}{u}\,\sqrt{\frac{1-\beta^2_V\beta^2_u}
{1-\beta^2_V}}.$$

\subsection*{Задача 2 {\usefont{T2A}{cmr}{b}{n}(\hspace*{-1.5mm}\cite{8}, 
2.47)}: Сколько вещества окажется внутри пенала?}
{\usefont{T2A}{cmr}{m}{it} Первое решение.}
В системе облака пенал имеет длину $\frac{L}{\gamma_V}$, а волна упругой деформации 
догоняет крышку $A$ со скоростью
$$u_1=\frac{u+V}{1+\frac{uV}{c^2}}=\frac{u+V}{1+\beta_u\beta_V}.$$
Крышка $A$, в свою очередь, движется со скоростью $V$. Поэтому волна догонит крышку 
$A$ за время $$t=\frac{L}{\gamma_V(u_1-V)}.$$
Но
$$u_1-V=\frac{u+V}{1+\beta_u\beta_V}-V=\frac{u(1-\beta_V^2)}{1+\beta_u\beta_V}
=\frac{u}{\gamma_V^2}\,\frac{1}{1+\beta_u\beta_V}.$$
Поэтому
$$t=\frac{L}{\gamma_V}\,\frac{\gamma_V^2}{u}(1+\beta_u\beta_V)=
\gamma_V\,\frac{L}{u}\,(1+\beta_u\beta_V).$$
За время $t$ крышка $A$ передвинется на расстояние $Vt$. Следовательно, внутри пенала
окажутся те пылинки, которые в момент удара первой пылинки о дно $B$ находились от $B$
на расстоянии
$$\frac{L}{\gamma_V}+Vt=\frac{L}{\gamma_V}+L\gamma_V\,\frac{V}{u}\,(1+\beta_u
\beta_V)=L\gamma_V\left (1+\frac{V}{u}\right ).$$
Поэтому масса вещества, собранного внутри пенала, равна
$$m=\rho LS\gamma_V\left (1+\frac{V}{u}\right ).$$

{\usefont{T2A}{cmr}{m}{it} Второе решение.}
В системе пенала облако пыли имеет плотность $\gamma_V \rho$. После удара первой 
пылинки о дно крышка $A$ в этой системе закроется через время $t^\prime=L/u$. За
это время в пенал успеют дополнительно  войти те пылинки,  которые в момент удара 
первой пылинки о дно находились на расстоянии $Vt^\prime$ от крышки $A$. 
Следовательно, внутри пенала окажутся пылинки с общей массой
$$m=\gamma_V \rho SL+\gamma_V \rho SVt^\prime=\rho LS\gamma_V\left 
(1+\frac{V}{u}\right ).$$

\subsection*{Задача 3 {\usefont{T2A}{cmr}{b}{n}(\hspace*{-1.5mm}\cite{8},
3.23)}: Какова частота сигналов, принимаемых на подводных лодках?}
Пусть радар первой лодки начинает излучать сигнал. Одно колебание в этом сигнале 
совершается за время $T_0=1/\nu_0$ по часам первой лодки. В системе воды пройдет
$\gamma_1$-раз больше времени: $T_1=\gamma_1 T_0=\gamma_1/\nu_0$, где
$$\gamma_1=\frac{1}{\sqrt{1-\frac{V_1^2}{c^2}}}.$$
За это время передний фронт сигнала передвинется на расстояние $(c/n)T_1$, а лодка 
--- на $V_1T_1$. Следовательно, в системе воды длина волны (длина сигнала, которая
соответствует одному колебанию) будет
$$\lambda_1=\left (\frac{c}{n}-V_1\right )T_1=\frac{\gamma_1}{\nu_0}
\left (\frac{c}{n}-V_1\right ).$$
Этот сигнал движется навстречу второй лодке со скоростью $c/n$. Следовательно, он 
будет принят второй лодкой за время 
$$\tau=\frac{\lambda_1}{(c/n)+V_2}=\frac{\gamma_1}{\nu_0}\,\frac{c-nV_1}
{c+nV_2}$$
по часам системы воды. По часам второй лодки время принятия сигнала будет в 
$\gamma_2$ раз меньше:
$$T_1^\prime=\frac{\tau}{\gamma_2}=\frac{1}{\nu_0}\,\frac{\gamma_1}{\gamma_2}
\,\frac{1-n\beta_1}{1+n\beta_2}.$$
Но $T_1^\prime=1/\nu_1$, где $\nu_1$ --- частота принимаемого второй лодкой сигнала 
от радара первой лодки. Следовательно,
$$\nu_1=\nu_0\,\frac{\gamma_2}{\gamma_1}\,\frac{1+n\beta_2}{1-n\beta_1}.$$
Аналогично, частота принимаемого первой лодкой сигнала от радара второй лодки равна
$$\nu_2=\nu_0\,\frac{\gamma_1}{\gamma_2}\,\frac{1+n\beta_1}{1-n\beta_2}.$$

\subsection*{Задача 4: Минимальный угол разлета $\gamma$-квантов}
{\usefont{T2A}{cmr}{m}{it} Первое решение.}
Так как энергия налетающего протона пороговая, $\pi^0$-мезон в системе центра масс
покоится. Следовательно, в Л-системе он движется со скоростью системы центра масс
$$V=\frac{pc^2}{E+m_pc^2},$$
где $p$ --- импульс налетающего протона, а $E$ --- его энергия. Но $pc=\sqrt{E^2-
m_p^2c^4}$. Поэтому скорость $\pi^0$-мезона в Л-системе равна
$$V=c\,\frac{\sqrt{E^2-m_p^2c^4}}{E+m_pc^2},$$
и его $\gamma$-фактор будет
$$\gamma_\pi=\frac{1}{\sqrt{1-\frac{V^2}{c^2}}}=\left[ 1-\frac{E^2-m_p^2c^4}
{(E+m_pc^2)^2}\right]^{-1/2}=\sqrt{\frac{1}{2}\left ( 1+\frac{E}{m_pc^2}
\right )}.$$
Зная $\gamma$-фактор, находим энергию $\pi^0$-мезона:
$$E_\pi=m_\pi c^2\gamma_\pi=m_\pi c^2\sqrt{\frac{1}{2}\left ( 1+
\frac{E}{m_pc^2}\right )}.$$
Рассмотрим распад этого $\pi^0$-мезона. Закон сохранения 4-импульса $p_\pi=k_1+k_2$ 
дает
$$m_\pi c^2=p_\pi^2=(k_1+k_2)^2=2k_1\cdot k_2=2\left (\frac{E_1E_2}{c^2}-
\frac{E_1E_2}{c^2}\,\cos{\theta}\right )=4\,\frac{E_1E_2}{c^2}\,\sin^2{
\frac{\theta}{2}}.$$
Следовательно, угол разлета $\gamma$-квантов определяется из уравнения
$$\sin{\frac{\theta}{2}}=\frac{m_\pi c^2}{2\sqrt{E_1E_2}}=
\frac{m_\pi c^2}{2\sqrt{E_1(E_\pi-E_1)}}.$$
Угол $\theta$ минимальный, когда $E_1(E_\pi-E_1)$ достигает максимального значения.
А это случается при симметричном распаде: $E_1=E_2=E_\pi/2$. Следовательно,
\begin{equation}
\sin{\frac{\theta_{min}}{2}}=\frac{m_\pi c^2}{E_\pi}=\sqrt{\frac{2}{1+
\frac{E}{m_pc^2}}}.
\label{eq18_4}
\end{equation}
Надо найти пороговую энергию протона $E$. Пусть $p_1,p_2$ --- 4-импульсы протонов
до реакции, а $p_1^\prime,p_2^\prime$ --- их 4-импульсы после реакции. Тогда
из закона сохранения 4-импульса имеем $(p_1+p_2)^2=(p_1^\prime + p_2^\prime+
p_\pi)^2$. Времeнно положим $c=1$, и правую сторону посчитаем в системе центра масс,
где при пороговой энергии продукты реакции покоятся:
$(p_1^\prime + p_2^\prime+p_\pi)^2=(2m_p+m_\pi)^2$, а левую часть --- 
в Л-системе: $(p_1+p_2)^2=2m_p^2+2m_pE$. Следовательно, $(2m_p+m_\pi)^2=2m_p^2+
2m_pE$ и (восстановили скорость света $c$)
$$E=\frac{2m_p^2+m_\pi^2+4m_pm_\pi}{2m_p}\,c^2.$$
Поэтому
$$\frac{E}{m_pc^2}=1+2\,\frac{m_\pi}{m_p}+\frac{1}{2}\,\frac{m_\pi^2}
{m_p^2}.$$
Подставляя это в (\ref{eq18_4}), получаем окончательно
$$\sin{\frac{\theta_{min}}{2}}=\sqrt{\frac{2}{2+2\,\frac{m_\pi}{m_p}+
\frac{1}{2}\,\frac{m_\pi^2}{m_p^2}}}=\sqrt{\frac{1}{1+\frac{m_\pi}{m_p}+
\frac{1}{4}\,\frac{m_\pi^2}{m_p^2}}}=\frac{1}{1+\frac{m_\pi}{2m_p}}.$$
Следовательно,
$$\theta_{min}=2\arcsin{\frac{2m_p}{2m_p+m_\pi}}\approx 138^\circ.$$

{\usefont{T2A}{cmr}{m}{it} Второе решение.}
Согласно (\ref{eq18_4}),
$$\sin{\frac{\theta_{min}}{2}}=\frac{m_\pi c^2}{E_\pi}=\frac{1}{\gamma_\pi},$$
где $$\gamma_\pi=\frac{1}{\sqrt{1-\frac{V^2}{c^2}}}$$
и $V$ есть скорость $\pi^0$-мезона в Л-системе. Перейдем в систему центра масс. 
В этой системе продукты реакции покоятся. Следовательно, система центра масс движется 
относительно Л-системы со скоростью $V$. Протон, который покоился в лабораторной 
системе, в системе центра масс имеет скорость $-V$. Так как в системе центра масс 
суммарный импульс равен нулю, приходим к выводу, что налетающий протон в этой системе 
имеет скорость $V$. Следовательно, в системе центра масс, $\gamma$-факторы обоих 
протонов до реакции равны $\gamma_\pi$ и закон сохранения энергии в этой системе 
принимает вид $2m_pc^2\gamma_\pi=(2m_p+m_\pi)c^2$. Отсюда
$$\gamma_\pi=\frac{2m_p+m_\pi}{2m_p}$$
и
$$\sin{\frac{\theta_{min}}{2}}=\frac{1}{\gamma_\pi}=\frac{2m_p}{2m_p+m_\pi}.$$

\subsection*{Задача 5: Скорость дрейфа в магнитном поле}
\begin{figure}[htb]
\centerline{\epsfig{figure=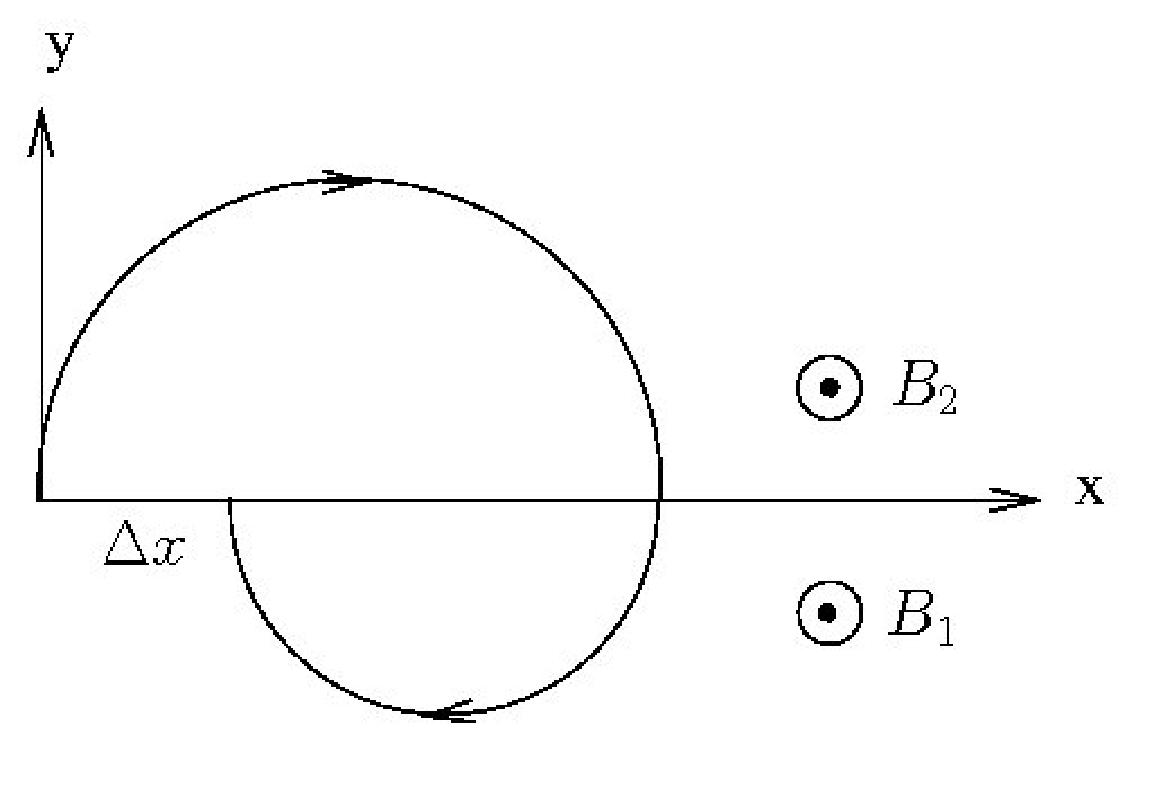,height=4cm}}
\end{figure}
Из рисунка видно, что 
$$\Delta x=2(R_2-R_1)=\frac{2p}{e}\left (\frac{1}{B_2}-\frac{1}{B_1}\right ),$$
где $p$ --- импульс электрона. С другой стороны, полуокружности электрон проходит за 
времена
$$t_2=\frac{\pi R_2}{V}=\frac{\pi p}{eVB_2} \;\;\;\mbox{и}\;\;\;
t_1=\frac{\pi R_1}{V}=\frac{\pi p}{eVB_1}.$$
Следовательно, перемещение $\Delta x$ вдоль оси $x$ совершается за время
$$\Delta t=t_2+t_1=\frac{\pi p}{eV} \left (\frac{1}{B_2}+\frac{1}{B_1}
\right ),$$ и средняя скорость перемещения вдоль оси $x$ (скорость дрейфа) равна
$$\bar{V}_{\mbox{др}}=\frac{\Delta x}{\Delta t}=\frac{2}{\pi}\,V\,\frac{B_1-B_2}
{B_1+B_2}.$$

\subsection*{Задача 6: Найти первоначальную энергию частицы}
Уравнения движения имеют вид
$$p_x=p_0,\;\;\;\frac{dp_y}{dt}=e{\cal{E}},$$
где ось $x$ направлена по линии первоначального движения частицы, а ось $y$ --- по
электрическому полю. Следовательно, 
$p_y=e{\cal{E}} t$ и $p^2=p_x^2+p_y^2=p_0^2+e^2{\cal{E}}^2 t^2$.
Но $p=mc\beta\gamma$ и $\beta^2\gamma^2=\gamma^2-1$. Поэтому
$$\gamma^2-1=\gamma_0^2-1+\left(\frac{e{\cal{E}}}{mc}\right)^2t^2$$
и
$$\gamma=\gamma_0\sqrt{1+\left(\frac{e{\cal{E}}}{mc\gamma_0}\right)^2t^2}.$$
В $x$-направлении перемещение $d$ определяется равенством
$$d=\int\limits_0^T V_x\,dt=\frac{p_0}{m}\int\limits_0^T\frac{dt}{\gamma},$$
так как $p_0=p_x=mV_x\gamma$. Но
$$\int\limits_0^T\frac{dt}{\gamma}=\frac{1}{\gamma_0}\int\limits_0^T\frac{dt}
{\sqrt{1+\left(\frac{e{\cal{E}}}{mc\gamma_0}\right)^2t^2}}=
\frac{mc}{e{\cal{E}}}\,\mathrm{arcsh}\,{\frac{e{\cal{E}}T}{mc\gamma_0}}.$$
Таким образом,
$$\frac{md}{p_0}=\frac{mc}{e{\cal{E}}}\,\mathrm{arcsh}\,{\frac{e{\cal{E}}cT}
{E_0}},$$
где $E_0=mc^2\gamma_0$ --- первоначальная энергия частицы. Из этого равенства 
можно определить $E_0$:
$$E_0=\frac{e{\cal{E}}cT}{\sh{\frac{e{\cal{E}}d}{p_0c}}}.$$

\section[Семинар 19]
{\centerline{Семинар 19}}

\subsection*{Задача 1: Какую задержку между вспышками зарегистрирует наблюдатель?}
Два события: наблюдатель $X$ увидел свет от вспышки $A$ и  наблюдатель $X$ увидел 
свет от вспышки $B$, в системе $X$ как одновременны, так и одноместны. Поэтому 
в Л-системе эти события тоже будут одновременны. С другой стороны, в Л-системе первое 
событие произойдет в момент времени
$$\tau_A=\frac{R}{\sqrt{c^2-V^2}}=\frac{R}{c}\,\gamma,$$
как это очевидно из рисунка.
\begin{figure}[!h]
\centerline{\epsfig{figure=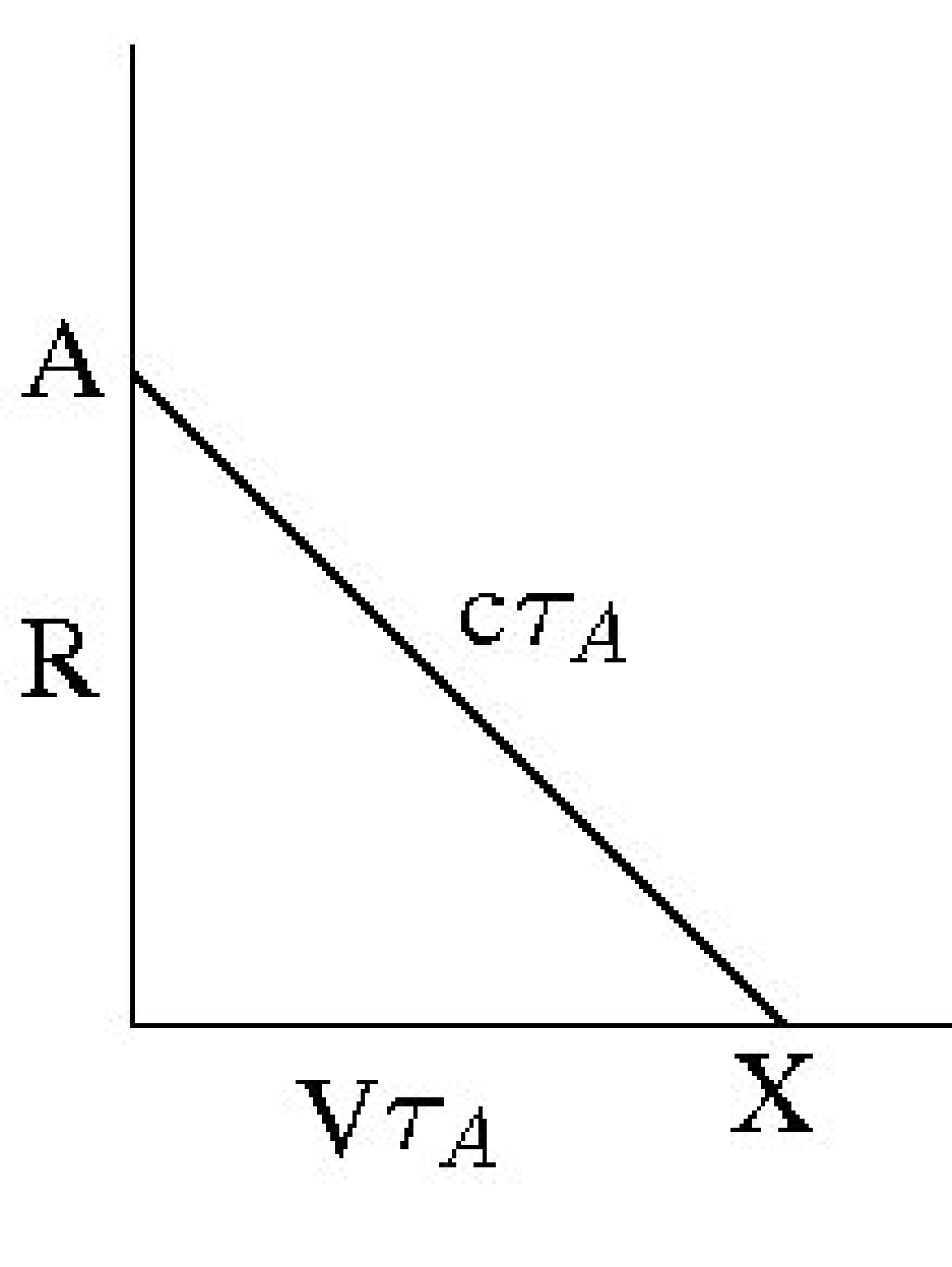,height=4cm}}
\end{figure}

Пусть в Л-системе вспышка в точке $B$ происходит в момент времени $t_B$. В этот момент
наблюдатель $X$ находится от точки $B$ на расстоянии $R-Vt_B$. Поэтому наблюдатель 
$X$ увидит свет от вспышки $B$ в момент времени
$$\tau_B=t_B+\frac{R-Vt_B}{c+V},$$
так как он движется навстречу световому сигналу от вспышки $B$. Но, как выяснили, 
$\tau_A=\tau_B$. Поэтому 
$$\frac{R}{c}\,\gamma=t_B+\frac{R-Vt_B}{c+V}.$$
Отсюда
$$t_B=\frac{R}{c}[\gamma(1+\beta)-1]=\frac{R}{c}\left[\frac{5}{4}\left(1+
\frac{3}{5}\right)-1\right]=\frac{R}{c}.$$
Теперь перейдем в систему $Y$ (ось $x$ всегда направлена по движению системы отсчета):
\begin{figure}[htb]
\centerline{\epsfig{figure=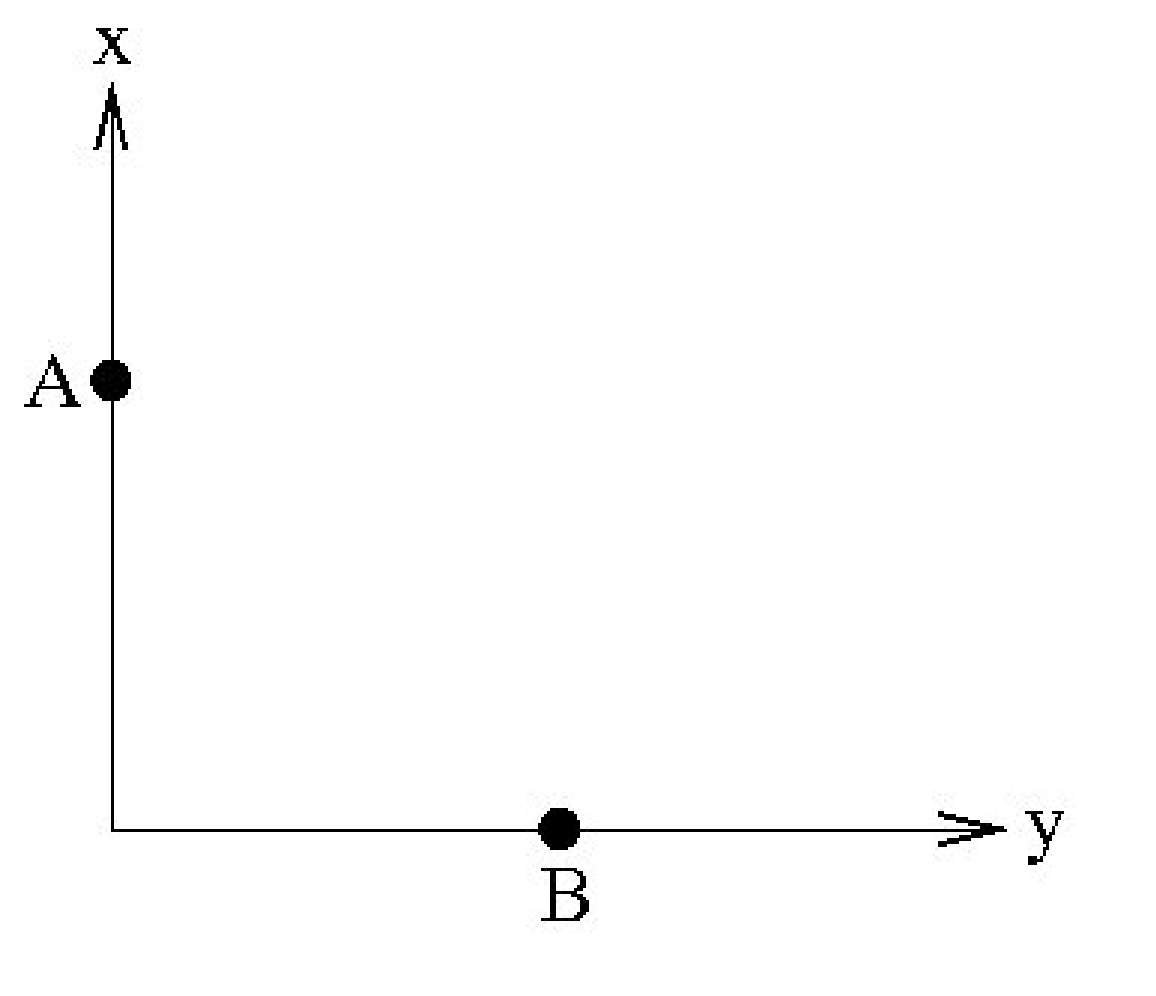,height=5cm}}
\end{figure}

Координаты вспышки $A$ в Л-системе равны $x_A=R,\,y_A=0,\,t_A=0$. Поэтому в системе 
$Y$ будем иметь
$$x_A^\prime=\gamma(x_A-Vt_A)=\gamma R,\;\;y_A^\prime=y_A=0,\;\;
t_A^\prime=\gamma\left(t_A-\frac{V}{c^2}\,x_A\right )=-\gamma\beta\,\frac{R}
{c}.$$
Следовательно, наблюдатель $Y$ увидит свет от вспышки $A$ в момент времени (по своим
часам)
$$\tau^\prime_A=t_A^\prime+\frac{\sqrt{x_A^{\prime\,2}+y_A^{\prime\,2}}}{c}=
\frac{\gamma R}{c}(1-\beta)=\frac{5}{4}\left(1-\frac{3}{5}\right)\frac{R}{c}=
\frac{1}{2}\,\frac{R}{c}.$$
Координаты вспышки $B$ в Л-системе равны $x_B=0,\,y_B=R,\,t_B=\frac{R}{c}$.
Поэтому в системе $Y$ будем иметь
$$x_B^\prime=\gamma(x_B-Vt_B)=-\gamma Vt_B=-\gamma\beta R,\;\;y_B^\prime=y_B=
R,\;\;t_B^\prime=\gamma\left(t_B-\frac{V}{c^2}\,x_B\right )=\gamma t_B=\gamma
\frac{R}{c}.$$
Следовательно, наблюдатель $Y$ увидит свет от вспышки $B$ в момент времени
$$\tau^\prime_B=t_B^\prime+\frac{\sqrt{x_B^{\prime\,2}+y_B^{\prime\,2}}}{c}=
\frac{R}{c}[\gamma+\sqrt{\gamma^2\beta^2+1}]=2\gamma\frac{R}{c}=\frac{5}{2}\,
\frac{R}{c}.$$
Искомая задержка $$\tau^\prime_B-\tau^\prime_A=\frac{5}{2}\,\frac{R}{c}-
\frac{1}{2}\,\frac{R}{c}=2\,\frac{R}{c}.$$
 
\subsection*{Задача 2: Распад $K_S$-мезона}
Сохранение 4-импульса дает $p_K=p_++p_-$. Отсюда $p_-^2=(p_K-p_+)^2=p_K^2-
2p_K\cdot p_++p_+^2$. Но (положили $c=1$) $p_K^2=M^2$, $p_+^2=p_-^2=m^2$ и
$p_K\cdot p_+=E_KE_+-\vec{p}_K\cdot\vec{p}_+=E_KE_+$. Следовательно, $M^2=
2E_KE_+$ и (восстановили скорость света $c$) 
$$E_K=\frac{M^2c^4}{2E_+}=\frac{500}{2(140+26,66)}\,500~\mbox{МэВ}\approx
750~\mbox{МэВ}.$$
Из закона сохранения импульса $\vec{p}_K=\vec{p}_++\vec{p}_-$, с учетом $\vec{p_+}
\perp \vec{p}_K$, Получаем (см. рисунок) $|\vec{p}_-|=\sqrt{\vec{p}_K^2+
\vec{p}_+^2}$.
\begin{figure}[htb]
\centerline{\epsfig{figure=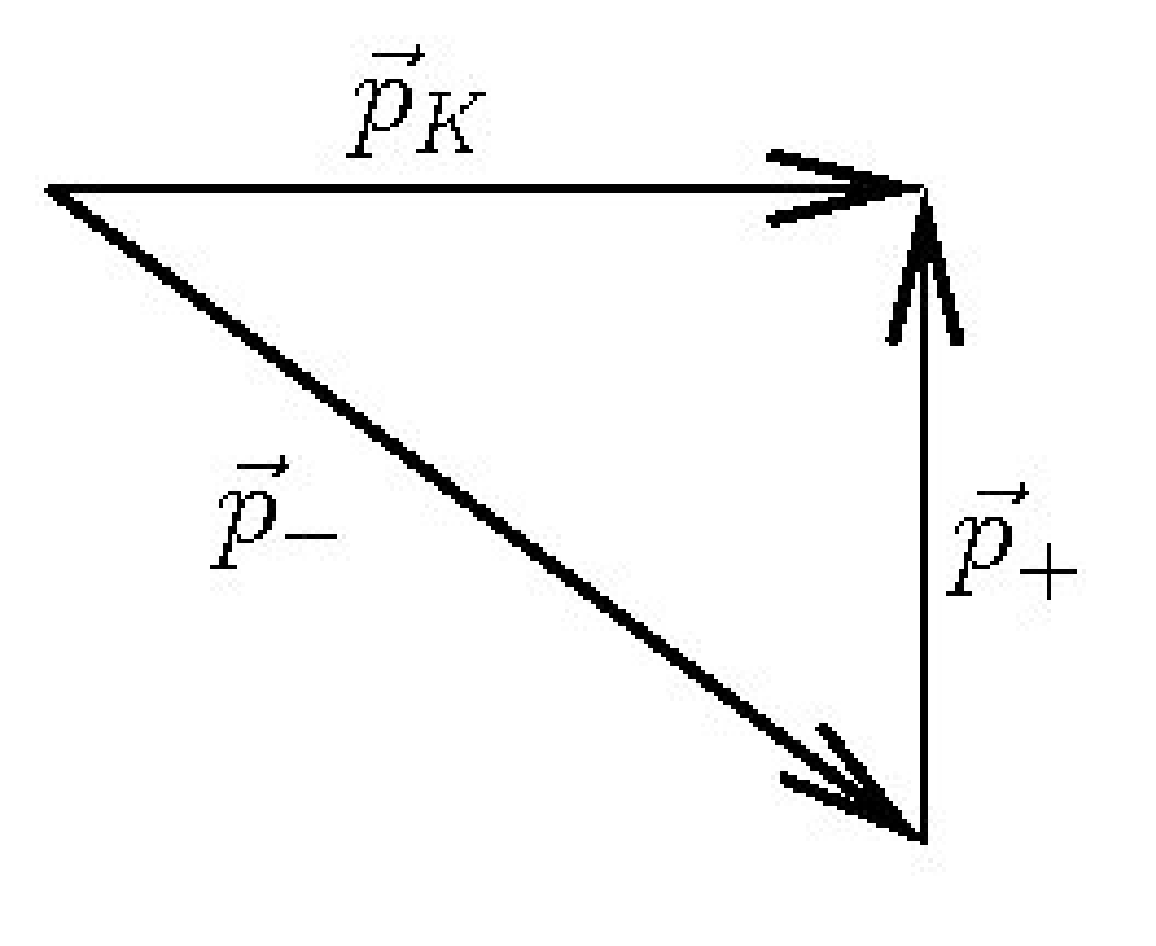,height=2cm}}
\end{figure}

\noindent Но $c^2\vec{p}^2=E^2-m^2c^4$. Поэтому
$$c|\vec{p}_-|=\sqrt{E_K^2+E_+^2-(Mc^2)^2-(mc^2)^2}\approx 566~\mbox{МэВ}$$
и $|\vec{p}_-|\approx 566~\mbox{МэВ}/c$.

\subsection*{Задача 3: Пучок мюонов в магнитном поле}
Пусть скорость мюонов $V$. Тогда 115 оборотов они сделают за время 
$t=115\frac{2\pi R}{V}$ (в Л-системе). В собственной системе мюонов пройдет время 
$$t^\prime=\frac{t}{\gamma}=230\pi\,\frac{R}{V\gamma}\,.$$
Но $$R=\frac{p}{eB}=\frac{mV\gamma}{eB}.$$
Поэтому $$t^\prime=230\pi\,\frac{m}{eB}.$$ По условию
$$\frac{1}{3}=\frac{N}{N_0}=e^{-\frac{t^\prime}{\tau}}.$$ Поэтому
$$\frac{t^\prime}{\tau}=230\pi\,\frac{m}{eB\tau}=\ln{3}.$$ Отсюда
$$B=230\pi\,\frac{mc^2}{e\tau c^2\ln{3}}\approx 0,35~\mbox{Т},$$
где при вычислении подставили
$$\frac{mc^2}{e}\approx 106\cdot10^6~\mbox{В}.$$
Найдем $\gamma$-фактор мюона. Из
$$R=\frac{mV\gamma}{eB}=\frac{mc}{eB}\,\beta\gamma$$
получаем 
$$\sqrt{\gamma^2-1}=\beta\gamma=\frac{eBR}{mc}=\frac{eBRc}{mc^2}$$
и $$\gamma=\sqrt{1+\left (\frac{eBRc}{mc^2}\right)^2}\approx 50.$$
Начальная плотность пучка в собственной системе отсчета
$$n_0^\prime=\frac{n_0}{\gamma}\approx 2\cdot 10^8~\mbox{см}^{-3}.$$

\subsection*{Задача 4 {\usefont{T2A}{cmr}{b}{n}(5.33)}: Монета на наклонной
плоскости}
В отличии от нерелятивистской версии задачи (семинар 2, задача 6), уравнение движения
монеты будет 
$$\frac{d\vec{p}}{dt}=\vec{F}-F\,\frac{\vec{p}}{p}.$$
Умножим обе части скалярно на $\vec{p}$ и учтем, что
$$\vec{p}\cdot \frac{d\vec{p}}{dt}=\frac{1}{2}\frac{d\vec{p}^2}{dt}=
\frac{1}{2}\frac{dp^2}{dt}=p\frac{dp}{dt},\;\;\vec{p}\cdot\vec{F}=p_yF.$$
Получаем $$p\frac{dp}{dt}=p_yF-Fp$$ или
$$\frac{dp}{dt}==\frac{p_y}{p}F-F.$$
Сравним это с уравнением для проекции $p_y$:
$$\frac{dp_y}{dt}=F_y-F\,\frac{p_y}{p}=F-F\,\frac{p_y}{p}.$$
Как видим,
$$\frac{dp}{dt}+\frac{dp_y}{dt}=\frac{d}{dt}(p+p_y)=0.$$
Это означает, что $p+p_y$ во время движения монеты не меняется. В начале 
$p=mV\gamma_V$ и $p_y=0$. В конце $p=p_y=mu\gamma_u$. Следовательно,
$mV\gamma_V=2mu\gamma_u$ и $\beta_V\gamma_V=2\beta_u\gamma_u$. С учетом
$\beta\gamma=\sqrt{\gamma^2-1}$ получаем $\gamma_V^2-1=4(\gamma_u^2-1)$.
Отсюда
$$\gamma_u^2=\frac{\gamma_V^2+3}{4}$$ и
$$\beta_u^2=1-\frac{1}{\gamma_u^2}=\frac{\gamma_V^2-1}{\gamma_V^2+3}=
\frac{1/(1-\beta_V^2)-1}{1/(1-\beta_V^2)+3}=
\frac{\beta_V^2}{4-3\beta_V^2}.$$
Окончательно
$$u=\frac{V}{\sqrt{4-3\,\frac{V^2}{c^2}}}.$$

\subsection*{Задача 5 {\usefont{T2A}{cmr}{b}{n}(2.39)}: Скорость частиц в системе
центра масс}
В Л-системе 4-скорости частиц имеют вид
$$u_1=(c\gamma_1,\,\vec{V}_1\gamma_1),\;\;
u_2=(c\gamma_2,\,\vec{V}_2\gamma_1),$$
а в системе центра масс
$$u_1^\prime=(c\gamma,\,\vec{V}\gamma),\;\;
u_2^\prime=(c\gamma,\,-\vec{V}\gamma).$$
Из инвариантности скалярного произведения 4-векторов $u_1\cdot u_2=u_1^\prime\cdot 
u_2^\prime$ следует, что
$$c^2\gamma_1\gamma_2(1-\vec{\beta}_1\cdot \vec{\beta}_2)=
c^2\gamma^2(1+\beta^2)=c^2\frac{1+\beta^2}{1-\beta^2}.$$
Решая это уравнение относительно $\beta$, получаем
$$\beta=\sqrt{\frac{\gamma_1\gamma_2(1-\vec{\beta}_1\cdot \vec{\beta}_2)-1}
{\gamma_1\gamma_2(1-\vec{\beta}_1\cdot \vec{\beta}_2)+1}}.$$

\subsection*{Задача 6 {\usefont{T2A}{cmr}{b}{n}(2.51)}: Ориентация ракеты связи}
Скорость ракеты в Л-системе находим по формулам релятивистского сложения скоростей
$$u_x=\frac{u_x^\prime+V}{1+\frac{u_x^\prime V}{c^2}},\;\;
u_y=\frac{u_y^\prime}{\gamma\left (1+\frac{u_x^\prime V}{c^2}\right )}.$$
Следовательно,
$$u_x=\frac{u^\prime\cos{\alpha^\prime}+V}{1+\beta^\prime\beta
\cos{\alpha^\prime}},\;\;
u_y=\frac{u^\prime\sin{\alpha^\prime}}{\gamma(1+\beta^\prime\beta
\cos{\alpha^\prime})},$$
где
$$\beta=\frac{V}{c},\;\;\beta^\prime=\frac{u^\prime}{c},\;\;\mbox{и}\;\;
\gamma=\frac{1}{\sqrt{1-\beta^2}}.$$
Угол $\theta$, который вектор $\vec{u}$ составляет с осью $x$ (угол между векторами
скорости ракеты и корабля в Л-системе), определяется из соотношения
$$\tg{\theta}=\frac{u_y}{u_x}=\frac{u^\prime\sin{\alpha^\prime}}
{\gamma(u^\prime\cos{\alpha^\prime}+V)}=\frac{\beta^\prime
\sin{\alpha^\prime}}{\gamma(\beta^\prime\cos{\alpha^\prime}+\beta)}.$$

Чтобы найти угол $\alpha$ между осью ракеты и осью $x$, надо знать координаты начала 
и конца ракеты в один и тот же момент времени в Л-системе, скажем, при $t=0$. Выберем 
начало отсчета времени так, что в момент $t=t^\prime=0$ начала систем $S$ и 
$S^\prime$ совпадают и происходит запуск ракеты. В системе $S^\prime$ начало 
ракеты $A$ в момент времени $t^\prime$ имеет координаты 
$x^\prime_A=u^\prime t^\prime\cos{\alpha^\prime},\;\;
y^\prime_A=u^\prime t^\prime\sin{\alpha^\prime},$
a конец ракеты $B$ имеет координаты
$$x^\prime_B=(l^\prime+u^\prime t^\prime)\cos{\alpha^\prime},\;\;
y^\prime_B=(l^\prime+u^\prime t^\prime)\sin{\alpha^\prime},$$
где $l^\prime$ --- длина ракеты в этой системе (заметим, что в системе корабля
$S^\prime$ скорость ракеты $\vec{u}^\prime$ направлена вдоль ее оси). Если $x_A,
\,y_A$ и $x_B,\,y_B$ являются координатами начала и конца ракеты в Л-системе в момент
времени $t=0$, то по преобразованиям Лоренца
$$x^\prime_B(t^\prime_B)-x^\prime_A(t^\prime_A)=\gamma (x_B-x_A),\;\;\mbox{и}
\;\;y^\prime_B(t^\prime_B)-y^\prime_A(t^\prime_A)=y_B-y_A.$$
Следовательно, 
$$\tg{\alpha}=\frac{y_B-y_A}{x_B-x_A}=\frac{\gamma\, [y^\prime_B(t^\prime_B)-
y^\prime_A(t^\prime_A)]}{x^\prime_B(t^\prime_B)-x^\prime_A(t^\prime_A)}=
\gamma\,\frac{\sin{\alpha^\prime}\,[l^\prime+u^\prime(t^\prime_B-t^\prime_A)]}
{\cos{\alpha^\prime}\,[l^\prime+u^\prime(t^\prime_B-t^\prime_A)]}=
\gamma\,\tg{\alpha^\prime}.$$
Заметим, что если $0<\alpha^\prime<\pi/2$,
$$\tg{\theta}=\frac{\tg{\alpha^\prime}}{\gamma\left (1+\frac{\beta}
{\beta^\prime\cos{\alpha^\prime}}\right )}<\tg{\alpha^\prime}<\tg{\alpha}.$$
Следовательно, $\theta<\alpha$ и в Л-системе ракета летит, как показано на рисунке.
\begin{figure}[htb]
\centerline{\epsfig{figure=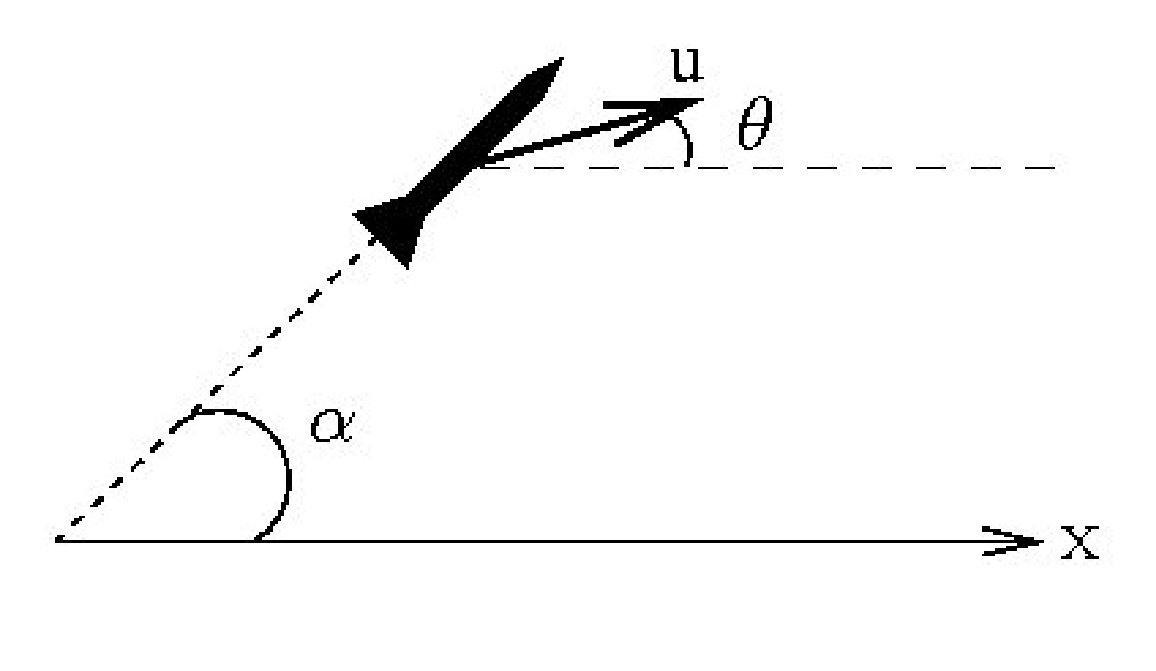,height=3cm}}
\end{figure}

{\usefont{T2A}{cmr}{m}{it} Вариант решения.}
Найдем угол $\alpha$, используя Лоренцево сокращение продольных размеров. Перейдем 
в систему ракеты. Ось $x$ направим по движению корабля в этой системе. Тогда в системе
корабля компоненты скорости лабораторного наблюдателя будут $V_x^\prime=V\,
\cos{\alpha^\prime}$, $V_y^\prime=V\,\sin{\alpha^\prime}$. По формуле
релятивистского сложения скоростей найдем компоненты скорости наблюдателя в системе 
ракеты:
$$\tilde{V}_x=\frac{V\,\cos{\alpha^\prime}+u^\prime}{1+\beta\beta^\prime\,
\cos{\alpha^\prime}},\;\;\tilde{V}_y=\frac{V\,\sin{\alpha^\prime}}
{\gamma^\prime(1+\beta\beta^\prime\,\cos{\alpha^\prime})}.$$
Пусть в системе ракеты линия относительного движения лабораторного наблюдателя
и ракеты составляет угол $\tilde{\beta}$ с осью $x$ и, следовательно, с осью ракеты, 
так как в этой системе ось ракеты ориентирована вдоль линии относительного движения
ракеты и корабля. Тогда
$$\tg{\tilde{\beta}}=\frac{\tilde{V}_y}{\tilde{V}_x}=
\frac{\beta\sin{\alpha^\prime}}{\gamma^\prime(\beta\cos{\alpha^\prime}+
\beta^\prime)}.$$
В Л-системе ракета движется со скоростью $\vec{u}=\vec{u}^\prime\oplus\vec{V}$.
Проекция ракеты $L\sin{\tilde{\beta}}$ ($L$ --- собственная длина ракеты),
перпендикулярная к линии относительного движения ракеты и лабораторного наблюдателя,
не изменится, а продольная проекция $L\cos{\tilde{\beta}}$ уменьшится $\gamma_u$
раз. Поэтому в Л-системе угол $\beta$, который ось ракеты составляет с линией
относительного движения ракеты и лабораторного наблюдателя, определяется из
$$\tg{\beta}=\frac{L\sin{\tilde{\beta}}}{L\cos{\tilde{\beta}}/\gamma_u}=
\gamma_u\tg{\tilde{\beta}}=\frac{\beta\sin{\alpha^\prime}}{\beta
\cos{\alpha^\prime}+\beta^\prime}\,\gamma(1+\beta\beta^\prime\,
\cos{\alpha^\prime}),$$
где мы воспользовались тем, что $\gamma_u=\gamma\gamma^\prime(1+\beta\beta^\prime
\,\cos{\alpha^\prime})$ согласно общей формуле
\begin{equation}
\gamma_{\vec{V}_1\oplus\vec{V}_2}=\gamma_1\gamma_2\left(1+\frac{\vec{V}_1\cdot
\vec{V}_2}{c^2}\right).
\label{eq19_6}
\end{equation}
Докажем (\ref{eq19_6}). Пусть некое тело движется относительно системы $S^\prime$
со скоростью $\vec{V}_1$, а сама система $S^\prime$ движется со скоростью 
$\vec{V}_2$ относительно системы $S$. Тогда относительно системы $S$ тело движется 
со скоростью $\vec{V}=\vec{V}_1\oplus\vec{V}_2$, и его 4-скорость будет 
$u_1=(\gamma_Vc,\gamma_V\vec{V})$. 
\newline 4-скорость начала отсчета системы $S$ в самой 
системе $S$ будет $u_2=(c,\vec{0})$. В системе $S^\prime$ начало отсчета системы 
$S$ движется со скоростью $-\vec{V}_2$. Поэтому в этой системе $u^\prime_1=
(\gamma_1c,\gamma_1\vec{V}_1)$ и $u^\prime_2=(\gamma_2c,-\gamma_2\vec{V}_2)$. 
Скалярное произведение 4-векторов инвариантно. Поэтому
$\gamma_Vc^2=u_1\cdot u_2=u_1^\prime\cdot u_2^\prime=\gamma_1\gamma_2 c^2+
\gamma_1\gamma_2\vec{V}_1\cdot\vec{V}_2$,
и отсюда следует (\ref{eq19_6}).

Так как в Л-системе линия относительного движения ракеты и лабораторного наблюдателя 
составляет угол $\theta$ с осью $x$, будем иметь $\alpha=\theta+\beta$ и, 
следовательно, 
$$\tg{\alpha}=\tg{(\theta+\beta)}=\frac{\tg{\theta}+\tg{\beta}}
{1-\tg{\theta}\,\tg{\beta}}.$$
Если подставить в этой формуле
$$\tg{\theta}=\frac{\tg{\alpha^\prime}}{\gamma\left(1+\frac{\beta}
{\beta^\prime\cos{\alpha^\prime}}\right)}$$
и
$$\tg{\beta}=\frac{\tg{\alpha^\prime}}{1+\frac{\beta^\prime}
{\beta\cos{\alpha^\prime}}}\,\gamma(1+\beta\beta^\prime\cos{\alpha^\prime}),$$
получим
$$\tg{\alpha}=\gamma\,\tg{\alpha^\prime}\;\frac{\frac{1}{\gamma^2}\left (
1+\frac{\beta^\prime}{\beta\cos{\alpha^\prime}}\right )+\left(1+\frac{\beta}
{\beta^\prime\cos{\alpha^\prime}}\right)(1+\beta\beta^\prime
\cos{\alpha^\prime})}{\left(1+\frac{\beta}{\beta^\prime\cos{\alpha^\prime}}
\right)\left (1+\frac{\beta^\prime}{\beta\cos{\alpha^\prime}}\right )-
\tg^2{\alpha^\prime}(1+\beta\beta^\prime\cos{\alpha^\prime})}.$$
Но
$$\frac{1}{\gamma^2}\left (1+\frac{\beta^\prime}{\beta\cos{\alpha^\prime}}
\right )+\left(1+\frac{\beta}{\beta^\prime\cos{\alpha^\prime}}\right)(1+
\beta\beta^\prime\cos{\alpha^\prime})=$$
$$=(1-\beta^2)\left (1+\frac{\beta^\prime}{\beta\cos{\alpha^\prime}}
\right )+\left(1+\frac{\beta}{\beta^\prime\cos{\alpha^\prime}}\right)(1+
\beta\beta^\prime\cos{\alpha^\prime})=$$
$$=2+\frac{\beta^{\prime\,2}-\beta^2\beta^{\prime\,2}+\beta^2}{\beta
\beta^\prime\cos{\alpha^\prime}}+\beta\beta^\prime\cos{\alpha^\prime}=
2+\frac{\beta^{\prime\,2}-\beta^2\beta^{\prime\,2}\sin^2{\alpha^\prime}+
\beta^2}{\beta\beta^\prime\cos{\alpha^\prime}},$$
и
$$\left(1+\frac{\beta}{\beta^\prime\cos{\alpha^\prime}}
\right)\left (1+\frac{\beta^\prime}{\beta\cos{\alpha^\prime}}\right )-
\tg^2{\alpha^\prime}(1+\beta\beta^\prime\cos{\alpha^\prime})=$$
$$=1+\frac{\beta}{\beta^\prime\cos{\alpha^\prime}}+\frac{\beta^\prime}
{\beta\cos{\alpha^\prime}}+\frac{1}{\cos^2{\alpha^\prime}}-\tg^2{\alpha^\prime}
-\frac{\beta\beta^\prime\sin^2{\alpha^\prime}}{\cos{\alpha^\prime}}=$$
$$=2+\frac{\beta^{\prime\,2}-\beta^2\beta^{\prime\,2}\sin^2{\alpha^\prime}+
\beta^2}{\beta\beta^\prime\cos{\alpha^\prime}}.$$
Следовательно, $\tg{\alpha}=\gamma\;\tg{\alpha^\prime}$.

\section[Семинар 20]
{\centerline{Семинар 20}}

\subsection*{Задача 1 {\usefont{T2A}{cmr}{b}{n}(6.5)}: Зависимость силы и
потенциала от координаты}
$x(t)$ --- непрерывная функция. Поэтому точная функциональная зависимость $x(t)$,
изображенная на рисунке к задаче, имеет вид
$$x(t)=\left\{\begin{array}{cc} \frac{a}{\ch{T}-1}\,\left[\ch{T}-\ch{(t-T)}
\right], & 0\le t\le 2T, \vspace*{2mm}\\  
\frac{-a}{\ch{T}-1}\,\left[\ch{T}-\ch{(t-3T)}\right], &
2T\le t\le 4T.\end{array}\right . $$
Вычислим скорость
$$\dot x(t)=\left\{\begin{array}{cc} \frac{-a}{\ch{T}-1}\,\sh{(t-T)}, 
& 0\le t\le 2T, \vspace*{2mm}\\  
\frac{a}{\ch{T}-1}\,\sh{(t-3T)}, &
2T\le t\le 4T,\end{array}\right . $$
и ускорение
$$\ddot x(t)=\left\{\begin{array}{cc} \frac{-a}{\ch{T}-1}\,\ch{(t-T)}, 
& 0\le t\le 2T, \vspace*{2mm}\\  
\frac{a}{\ch{T}-1}\,\ch{(t-3T)}, &
2T\le t\le 4T.\end{array}\right . $$
Как видим, если $0\le t\le 2T$, т.~е. если $0\le x\le a$, то
$$\ddot x(t)=x(t)-\frac{a\,\ch{T}}{\ch{T}-1},$$
а если $2T\le t\le 4T$, т.~е. если $-a\le x\le 0$, то
$$\ddot x(t)=x(t)+\frac{a\,\ch{T}}{\ch{T}-1}.$$
Поэтому
$$F(x)=\left\{\begin{array}{cc} m\left(x-\frac{a\,\ch{T}}{\ch{T}-1}\right),
& 0< x\le a,\vspace*{2mm}\\ m\left(x+\frac{a\,\ch{T}}{\ch{T}-1}\right),
& -a\le x<0.\end{array}\right . $$
Точку $x=0$ надо исключить, так как $F(x)$ имеет разрыв в этой точке.

В частном случае $\ch{T}=2$ будем иметь
$$x(t)=\left\{\begin{array}{cc} a\left[2-\ch{(t-T)}
\right], & 0\le t\le 2T, \vspace*{2mm}\\  
-a\left[2-\ch{(t-3T)}\right], &
2T\le t\le 4T.\end{array}\right . $$
и
$$F(x)=\left\{\begin{array}{cc} m\left(x-2a\right),
& 0< x\le a,\vspace*{2mm}\\ m\left(x+2a\right),
& -a\le x<0.\end{array}\right . $$
Потенциал находим так:
$$U(x)=-\int F(x)\,dx=\left\{\begin{array}{cc} -m\left(\frac{x^2}{2}-2ax
\right)+\mathrm{const},& 0< x\le a,\vspace*{2mm}\\ -m\left(\frac{x^2}{2}+
2ax\right)+\mathrm{const},& -a\le x<0.\end{array}\right . $$
Выберем $\mathrm{const}=0$ и восстановим точку $x=0$, так как $U(x)$ непрерывна
в этой точке. Окончательно
$$U(x)=\left\{\begin{array}{cc} -\frac{mx}{2}\,(x-4a),& 0\le x\le a,
\vspace*{2mm}\\ -\frac{mx}{2}\,(x+4a),& -a\le x\le 0.\end{array}\right . $$

\subsection*{Задача 2 {\usefont{T2A}{cmr}{b}{n}(6.6)}: Закон движения частицы
при нулевой полной энергии}
Из закона сохранения энергии
$$\frac{mV^2}{2}+U(x)=E=0$$
следует, что
$$V^2=-\frac{2U(x)}{m}=\frac{2\alpha}{m}\,x^4,$$
или
$$\frac{dx}{dt}=V=\pm\sqrt{\frac{2\alpha}{m}}\,x^2.$$
Разделяем переменные и интегрируем:
$$\pm\sqrt{\frac{2\alpha}{m}}\,t=\int\limits_{x_0}^x\frac{dx}{x^2}=
\frac{1}{x_0}-\frac{1}{x},$$
где $x_0=x(0)$ --- значение координаты при $t=0$. Окончательно
$$x(t)=\frac{x_0}{1\mp\sqrt{\frac{2\alpha}{m}}\,x_0t}.$$
Знак определяется из начальных условий --- из знака $V_0=V(0)$:
$$V_0=\pm\sqrt{\frac{2\alpha}{m}}\,x_0^2.$$

Фазовая траектория тоже определяется из закона сохранения энергии
$\frac{p^2}{2m}+U(x)=E=0$, что дает $p=\pm\sqrt{2\alpha m}\,x^2$. В точке
$x=0$ скорость равна нулю, т.~е. частица останавливается. Поэтому возможны только 
следующие фазовые траектории:
\begin{itemize}
\item $x_0>0,\,V_0>0$ --- фазовая траектория начинается с точки $(x_0,V_0)$ и 
уходит на бесконечность по параболе $p=\sqrt{2\alpha m}\,x^2$.
\item $x_0>0,\,V_0<0$ --- фазовая траектория есть кусочек параболы 
$p=-\sqrt{2\alpha m}\,x^2$ между начальной точкой $(x_0,V_0)$ и конечной
точкой $(0,0)$.
\item $x_0<0,\,V_0>0$ --- фазовая траектория есть кусочек параболы 
$p=\sqrt{2\alpha m}\,x^2$ между начальной точкой $(x_0,V_0)$ и конечной
точкой $(0,0)$.
\item $x_0<0,\,V_0<0$ --- фазовая траектория, начиная с точки $(x_0,V_0)$,
уходит на бесконечность по параболе $p=-\sqrt{2\alpha m}\,x^2$.
\item $x_0=0,\,V_0=0$ --- это точка неустойчивого равновесия и фазовая траектория
вырождается в точку.
\end{itemize}
 
\subsection*{Задача 3 {\usefont{T2A}{cmr}{b}{n}(6.8)}: Траектории частицы на
фазовой плоскости}
\begin{figure}[htb]
\begin{center}
\epsfig{figure=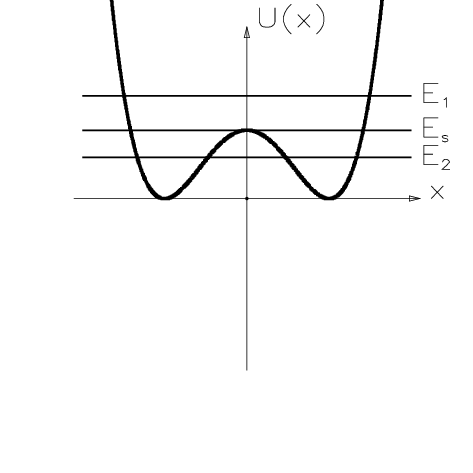,height=3cm}
\hfill
\epsfig{figure=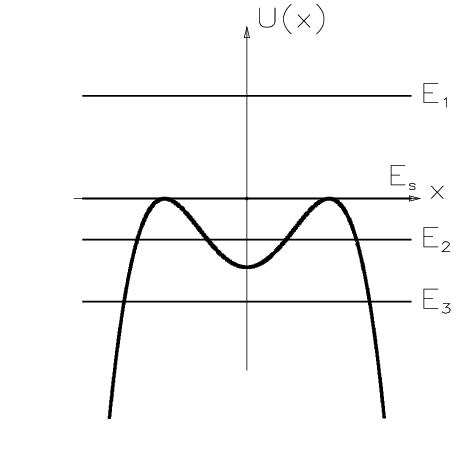,height=3cm}
\hfill
\epsfig{figure=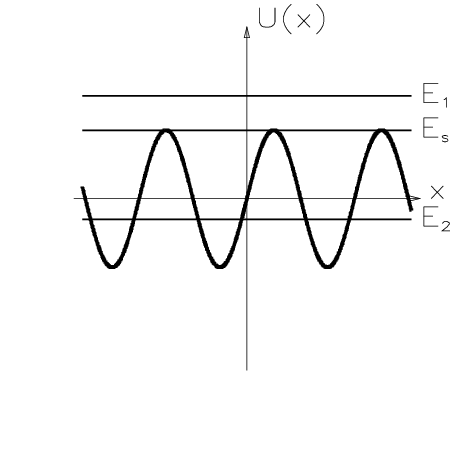,height=3cm}
\hfill
\epsfig{figure=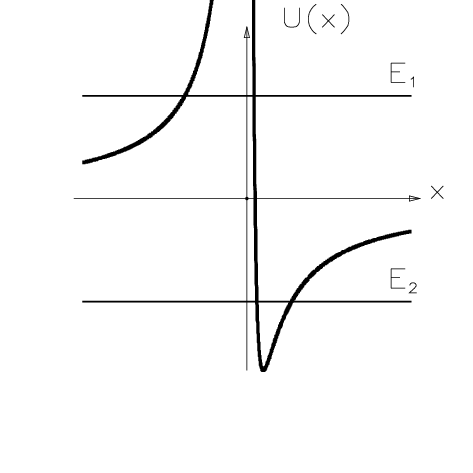,height=3cm}
\vfill
\epsfig{figure=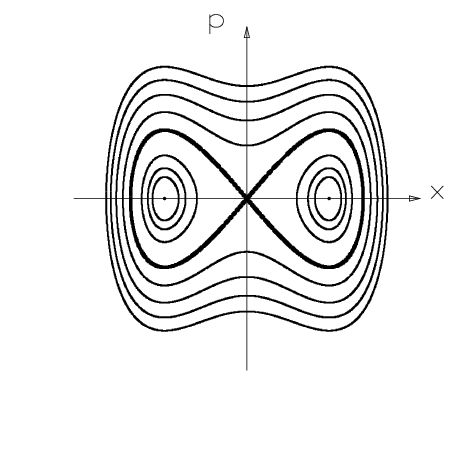,height=3cm}
\hfill
\epsfig{figure=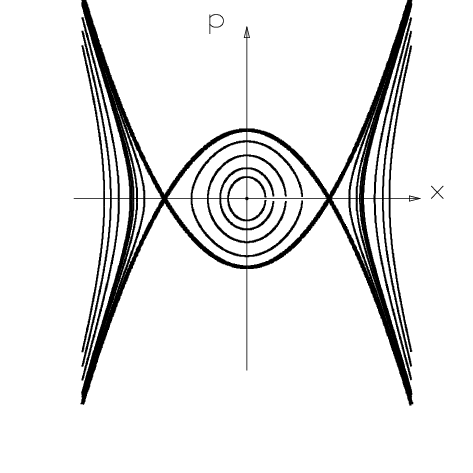,height=3cm}
\hfill
\epsfig{figure=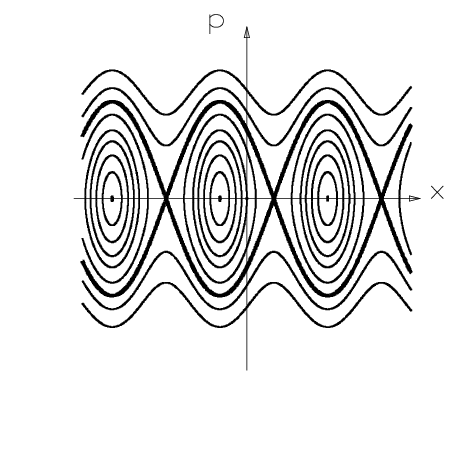,height=3cm}
\hfill
\epsfig{figure=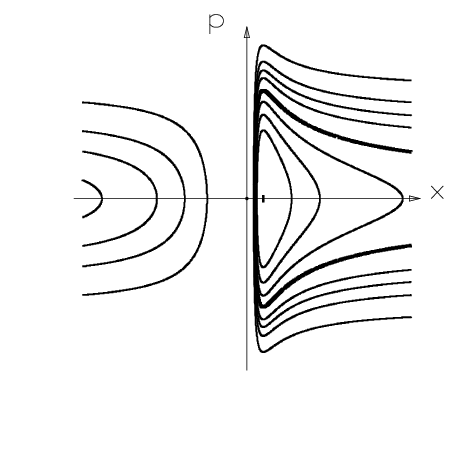,height=3cm}
\end{center}
\end{figure}
Закон сохранения энергии
$$\frac{p^2}{2m}+U(x)=E$$
дает уравнение фазовой траектории при одномерном движении. Движение возможно только в
области, где $E>U$. На рисунке приведены графики заданных потенциалов (верхняя часть 
рисунка) и соответствующие фазовые траектории.
 
Для потенциала $U(x)=\alpha^2(x^2-b^2)^2$ движение всегда финитное. Если энергия 
$E<\alpha^2b^4$ (энергия $E_2$ на рисунке), имеются две области финитного движения.  
Если энергия $E>\alpha^2b^4$ (энергия $E_1$ на рисунке) --- область финитного 
движения одна. Энергия $E_s=\alpha^2b^4$ соответствует 
сепаратрисе (жирная кривая на рисунке), которая разделяет эти два режима движения.

Для потенциала $U(x)=-\alpha^2(x^2-b^2)^2$ возможны как финитное движение (энергия 
$E_2$ на рисунке), так и инфинитное движение (энергии $E_1$ и $E_3$ на рисунке).
Сепаратрисе (жирная кривая на рисунке) соответствует энергия $E_s=0$.

Для потенциала $U(x)=U_0\,\sin{kx}$ движение возможно, только если $E>-U_0$.
Если $E>U_0$, движение инфинитное (энергия $E_1$ на рисунке). Если $-U_0<E<U_0$,
появляются бесконечно много областей финитного движения  (энергия $E_2$ на рисунке).
Сепаратрисе (жирная кривая на рисунке) соответствует энергия $E_s=U_0$.

Для потенциала $U(x)=\alpha^2(\frac{b^2}{x^2}-\frac{1}{x}$, движение возможно, 
только если $E>U_{min}$, где $U_{min}=-\frac{\alpha^2}{4b^2}$ соответствует
минимуму потенциала при $x=2b^2$ (в точке, где производная потенциала $U^\prime(x)$
равняется нулю). Если $E<0$ (энергия $E_2$ на рисунке), движение финитное. Если
$E>0$ (энергия $E_1$ на рисунке), появляются две области инфинитного движения
(соответственно для отрицательных и положительных $x$). Сепаратрисе (жирная кривая на 
рисунке) соответствует энергия $E_s=U_0$.   

\subsection*{Задача 4 {\usefont{T2A}{cmr}{b}{n}(6.9)}: Как изменится  форма и
объем области в фазовом пространстве?}
Если бы не было стенок, через некоторое время область, занятая частицами, имела бы форму
параллелограмма  $ACDF$ (см. рисунок). Но при достижении стенки при $x=L$ частицы 
упруго отражаются. Математически это отражение для $BCDE$ части параллелограмма (для 
которой $x>L$) означает преобразование $x\to L-(x-L)=2L-x,\;\; p\to -p$. 
В результате $BCDE$ переходит в $EBCD$. Но часть частиц из  $EBCD$ имеет $x<0$.
Это означает, что они испытали упругое отражение от стенки при $x=0$, что соответствует
преобразованию $x \to -x,\;\;p\to -p$. В результате $GHCD$ часть с $x<0$ перейдет
в закрашенную $GHCD$. Следовательно, в данном случае область в фазовом пространстве,
занятая частицами, состоит из трех закрашенных полосок $ABEF$, $EBHG$ и $GHCD$
(см. рисунок). 
\begin{figure}[htb]
\centerline{\epsfig{figure=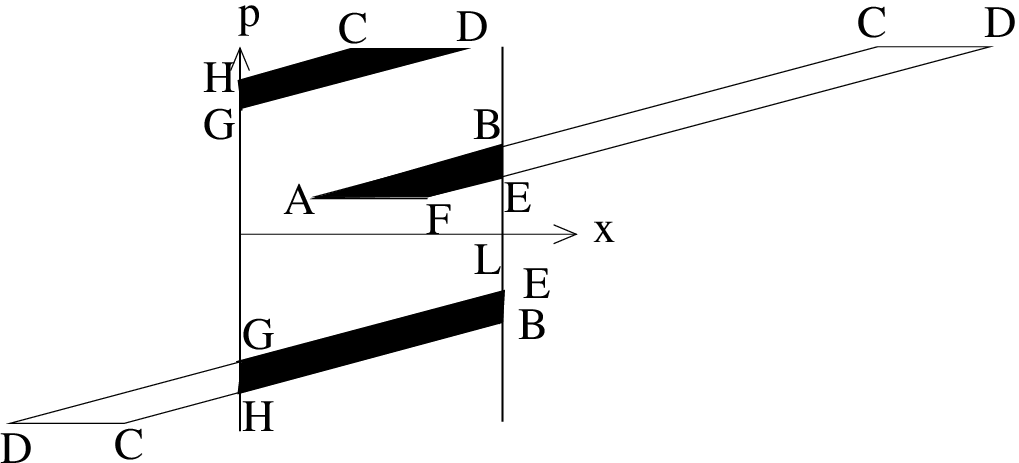,height=4cm}}
\end{figure}

В общем случае область, занятая частицами при времени $t$, находится по следующему 
алгоритму. Сначала находим область, которая была бы при отсутствии стенок (параллелограмм
$ACDF$). Координаты вершин этого параллелограмма равны
$$\begin{array}{ll}
A\left( x_0+\frac{p_0}{m}\,t,\,p_0\right ), & 
C\left( x_0+\frac{p_0+\Delta p}{m}\,t,\,p_0+\Delta p\right ),\\
D\left( x_0+\Delta x+\frac{p_0+\Delta p}{m}\,t,\,p_0+\Delta p\right ), &
F\left( x_0+\Delta x+\frac{p_0}{m}\,t,\,p_0\right ),
\end{array}$$
где $m$ --- масса частиц. Все частицы $(x,p)$ из этого параллелограмма с $x>L$ 
подверглись упругим отражениям от стенок при $x=L$ и $x=0$. Число отражений равно 
$n=[x/L]$, если $x>nL$, и $n-1$,  если $x=nL$ (так как в этом случае частица 
достигла стенки, но еще не успела отразиться). Здесь $[x/L]$ означает целую часть числа 
$x/L$. Так как $$m=\left [n+1-\frac{x}{L}\right]=\left [1-\left(\frac{x}{L}-
n\right)\right]=\left\{\begin{array}{c} 0,\;\;\mbox{если}\;\; x>nL, \\
1,\;\;\mbox{если}\;\; x=nL,\end{array} \right .$$
то число отражений равно $n-m$. Так как каждое отражение меняет знак импульса, после
всех отражений значение импульса будет $(-1)^{n-m}p$. Значение координаты после первого
отражения будет $2L-x$, после второго --- $x-2L$, после третьего --- $4L-x$, после
четвертого --- $x-4L$ и т.~д. В частности,  после $(2j-1)$-го отражения значение 
координаты будет $2jL-x$, а  после $2j$-го отражения --- $x-2jL$. Заметим, что 
$$2\left[\frac{n+1}{2}\right]=\left\{\begin{array}{l} 2j,\;\;\mbox{если}\;\; 
n=2j-1, \\ 2j,\;\;\mbox{если}\;\; n=2j.\end{array}\right .$$
Поэтому после всех отражений координата точки $(x,p)$ с $x>L$ из первоначального 
параллелограмма $ACDF$ станет
$$(-1)^{n-m}\left (x-2\left[\frac{n-m+1}{2}\right]L\right)=
(-1)^n\left (x-2\left[\frac{n+1}{2}\right]L\right).$$
То, что приведенное выражение не зависит от $m$, можно непосредственно проверить для
$x=nL$, когда $m=1$ (во всех остальных случаях $m=0$). Таким образом, отражения
приводят к преобразованию
$$(x,p)\to\left((-1)^n\left (x-2\left[\frac{n+1}{2}\right]L\right),\;
(-1)^{n-m}p\right)$$
для точек с $x>L$ из первоначального параллелограмма $ACDF$. В результате область 
фазового пространства, занятое частицами, будет состоять из $n_D-n_A+1$ полосок, если
только крайние точки $A$ и $D$ не достигли стенок. Здесь $n_A=[x_A/L]$ и 
$n_D=[x_D/L]$. Если крайне правая точка $D$ достигает стенок и еще не отразилась, число 
полосок будет на единицу меньше, т.~е. $n_D-n_A$. Если стенок достигает крайне левая 
точка $A$, то в области кроме $n_D-n_A+1$ полосок появится еще изолированная точка, 
которая соответствует еще не отразившейся частице $A$. Количество полосок со временем
неограниченно растет, так как в первоначальном параллелограмме $ACDF$ разность координат
крайних точек $x_D-x_A=\Delta x+\frac{\Delta p}{m}\,t$ растет со временем. При 
этом, так как отражения не меняют площадь, суммарная площадь области фазового 
пространства, занятого частицами, остается постоянной. Пример эволюции формы этой области 
показан ниже на рисунке.  
\begin{figure}[htb]
\begin{center}
\epsfig{figure=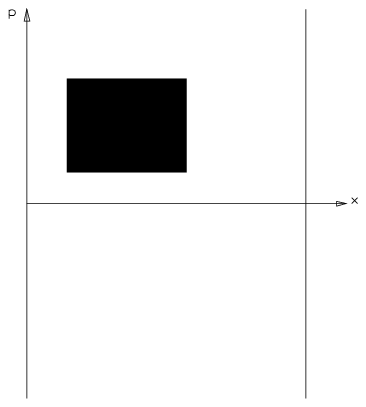,width=0.23\textwidth}
\hfill
\epsfig{figure=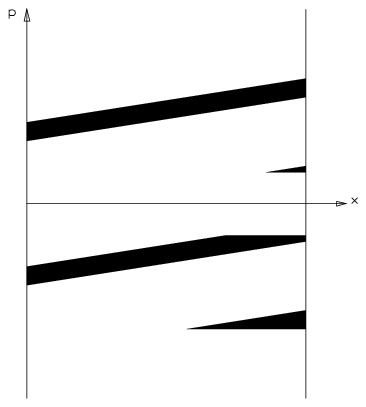,width=0.23\textwidth}
\hfill
\epsfig{figure=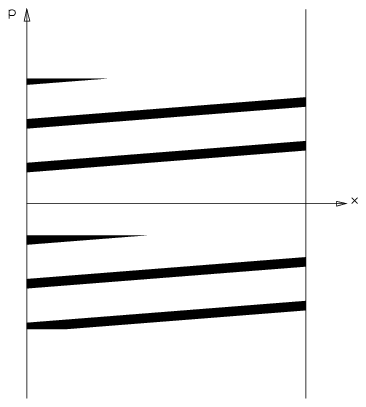,width=0.23\textwidth}
\hfill
\epsfig{figure=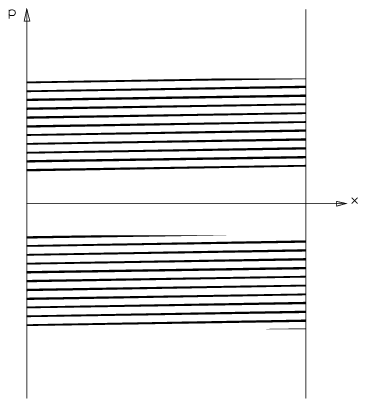,width=0.23\textwidth}
\end{center}
\end{figure}

\subsection*{Задача 5 {\usefont{T2A}{cmr}{b}{n}(6.11)}: Зависимость периода от
энергии}
Период движения
$$T=\sqrt{2m}\int\limits_{x_{min}}^{x_{max}}\frac{dx}{\sqrt{E-U(x)}}=
\sqrt{2m}\int\limits_{x_{min}}^{x_{max}}\frac{dx}{\sqrt{E-\alpha |x|^\beta}}.
$$
Здесь $x_{min}$ и $x_{max}$ --- решения уравнения $E=\alpha |x|^\beta$, т.~е.
$x_{min}=-x_0$, $x_{max}=x_0$, где $x_0=(E/\alpha)^{1/\beta}$. Преобразуем 
интеграл:
$$T=\sqrt{\frac{2m}{E}}\int\limits_{-x_0}^{x_0}\frac{dx}{\sqrt{1-|x/x_0|
^\beta}}=\sqrt{\frac{2m}{E}}\,x_0\,\int\limits_{-1}^1
\frac{dy}{\sqrt{1-|y|^\beta}},$$
где на последнем шаге мы сделали подстановку $y=x/x_0$. Следовательно,
$$T=\sqrt{\frac{2m}{E}}\,\left(\frac{E}{\alpha}\right)^\frac{1}{\beta}I,$$
где 
$$I=\int\limits_{-1}^1\frac{dy}{\sqrt{1-|y|^\beta}}$$
есть некоторое число. Заметим, что несобственный интеграл $I$ не расходится и дает 
конечное число. Например, особенность около верхнего предела интегрируема: после 
подстановки $y=1-x$ будем иметь ($\epsilon$ --- некоторое малое число)
$$\int\limits_{1-\epsilon}^1\frac{dy}{\sqrt{1-|y|^\beta}}=
\int\limits_0^\epsilon\frac{dx}{\sqrt{1-(1-x)^\beta}}\approx
\int\limits_0^\epsilon\frac{dx}{\sqrt{\beta x}}=\left .2\sqrt{\frac{x}
{\beta}}\right|_0^\epsilon=2\sqrt{\frac{\epsilon}{\beta}}.$$
Также интегрируема особенность около нижнего предела:
$$\int\limits_{-1}^{-1+\epsilon}\frac{dy}{\sqrt{1-|y|^\beta}}=
\int\limits_0^\epsilon\frac{dx}{\sqrt{1-|x-1|^\beta}}=
\int\limits_0^\epsilon\frac{dx}{\sqrt{1-(1-x)^\beta}}\approx
2\sqrt{\frac{\epsilon}{\beta}},$$
где мы сделали подстановку $y=x-1$.

Как видим, $$T\sim E^{\frac{1}{\beta}-\frac{1}{2}}.$$

\subsection*{Задача 6 {\usefont{T2A}{cmr}{b}{n}(6.19)}: Закон движения тела при
наличии силы трения}
Уравнение движения имеет вид
$$m\frac{dV}{dt}=mg-\alpha V^2.$$
Разделяем переменные
$$dt=\frac{dV}{g-\frac{\alpha}{m}\,V^2}=\frac{m}{\alpha}\,\frac{dV}
{V_m^2-V^2}$$
и интегрируем
$$t=\frac{m}{\alpha}
\int\limits_0^V\frac{dV}{V_m^2-V^2}=\frac{m}{2\alpha V_m}\int\limits_0^V
\left(\frac{1}{V_m-V}+\frac{1}{V_m+V}\right)\,dV=\frac{m}{2\alpha V_m}\,
\ln{\frac{V_m+V}{V_m-V}},$$
где $V_m^2=mg/\alpha$. Следовательно,
$$\frac{V_m+V}{V_m-V}=\exp{\left(\frac{2\alpha V_m}{m}\,t\right)}=
\exp{\left(2\sqrt{\frac{\alpha g}{m}}\;t\right)},$$
и
$$V(t)=V_m\,\frac{\exp{\left(2\sqrt{\frac{\alpha g}{m}}\,t\right)}-1}
{\exp{\left(2\sqrt{\frac{\alpha g}{m}}\,t\right)}+1}=V_m
\th{\left(\sqrt{\frac{\alpha g}{m}}\;t\right)}=\sqrt{\frac{mg}{\alpha}}
\th{\left(\sqrt{\frac{\alpha g}{m}}\;t\right)}.$$
$x(t)$ можно найти интегрированием этого выражения:
$$x(t)=\frac{m}{\alpha}\int\limits_0^t\;\frac{d\,\ch{\left(
\sqrt{\frac{\alpha g}{m}}\;t\right)}}{\ch{\left(
\sqrt{\frac{\alpha g}{m}}\;t\right)}}=\frac{m}{\alpha}\ln{\left[
\ch{\left(\sqrt{\frac{\alpha g}{m}}\;t\right)}\right]}.$$

Как вариант, $x(t)$ можно найти и по-другому. А именно, из сохранения энергии 
с учетом работы силы трения
$$\Delta\frac{mV^2}{2}=mg\Delta x-\alpha V^2\Delta x$$
получаем
$$\frac{m}{2}dV^2=(mg-\alpha V^2)\,dx,$$ 
где мы от конечных разностей перешли к дифференциалам. Разделяем переменные и 
интегрируем
$$x=\frac{m}{2}\int\limits_0^V\frac{dV^2}{mg-\alpha V^2}=
-\frac{m}{2\alpha}\ln{\left(1-\frac{\alpha V^2}{mg}\right)}=
-\frac{m}{2\alpha}\ln{\left(1-\th^2{\left(\sqrt{\frac{\alpha g}{m}}\;t\right)}
\right )}.$$
Но 
$$1-\th^2{\left(\sqrt{\frac{\alpha g}{m}}\;t\right)}=\frac{1}
{\ch^2{\left(\sqrt{\frac{\alpha g}{m}}\;t\right)}},$$
и окончательно получаем
$$x(t)=\frac{m}{\alpha}\ln{\left[
\ch{\left(\sqrt{\frac{\alpha g}{m}}\;t\right)}\right]}.$$

\subsection*{Задача 7: Отношение периодов движения частицы в левой и 
правой потенциальных ямах}
\noindent Период движения в левой яме при $E\to 0$ есть
$$T_1=\sqrt{\frac{2m}{k}}\int\limits_{-a}^a\frac{dx}{(b-x)\sqrt{1-\left(
\frac{x}{a}\right )^2}}.$$
Сделаем подстановку
$$\sqrt{1-\left(\frac{x}{a}\right )^2}=\left(1-\frac{x}{a}\right)t$$
или 
$$x=a\frac{t^2-1}{t^2+1}=a\left\{1-\frac{2}{t^2+1}\right \}.$$
Тогда
$$dx=\frac{4at\,dt}{(1+t^2)^2},\;\;1-\frac{x}{a}=\frac{2}{1+t^2},\;\;
x-b=\frac{(a-b)t^2-(a+b)}{1+t^2}$$
и
$$T_1=2a\sqrt{\frac{2m}{k}}\int\limits_0^\infty\frac{dt}{(b-a)t^2+(a+b)}=
2\pi\sqrt{\frac{m}{2k}}\left [\left(\frac{b}{a}\right)^2-1\right]^{-1/2}.$$
В правой яме имеем малые колебания. Поэтому
$$T_2=2\pi\sqrt{\frac{m}{k_{\mbox{эфф}}}},$$
где $k_{\mbox{эфф}}=U^{\prime\prime}(b).$
Но 
$$\frac{U^{\prime}}{k}=\frac{2}{a}\left(\frac{x}{a}\right)(x-b)^2+
2\left[\left(\frac{x}{a}\right)^2-1\right](x-b)$$
и
$$\frac{U^{\prime\prime}}{k}=2\left[\left(\frac{b}{a}\right)^2-1\right].$$
Следовательно,
$$T_2=2\pi\sqrt{\frac{m}{2k}}\left [\left(\frac{b}{a}\right)^2-1\right]
^{-1/2}=T_1.$$

Заметим, что равенство периодов движения в левой и правой потенциальных ямах имеет 
место даже тогда, когда энергия $E>0$ не малая величина \cite{28,29}. 

\section[Семинар 21]
{\centerline{Семинар 21}}

\subsection*{Задача 1 {\usefont{T2A}{cmr}{b}{n}(6.40)}: Падение капли при наличии
тумана}
{\usefont{T2A}{cmr}{m}{it} Первое решение.}
Неподвижные капельки тумана не привносят собой импульса. Следовательно, поток импульса 
отсутствует и уравнение движения капли имеет вид
\begin{equation}
\frac{dp}{dt}=\frac{d}{dt}(mv)=mg.
\label{eq21_1a}
\end{equation}
За короткое время $\Delta t$ капля радиусом $R$ сталкивается с теми капельками тумана,
которые находятся на ее пути, т.~е. находятся в объеме, который равен объему цилиндра
с поперечным сечением $\pi R^2$ и высотой $V\Delta t$. Следовательно, приращение
массы капли $\Delta m \sim \pi R^2 V\Delta t$ и $\frac{dm}{dt}=k\pi R^2 V$,
где $k=\mathrm{const}$ есть некий коэффициент пропорциональности. Поэтому имеем 
$$\frac{dm}{dt}=\frac{d}{dt}\left(\frac{4\pi}{3}\rho R^3\right)=4\pi\rho R^2
\frac{dR}{dt}=k\pi R^2 V$$
и, следовательно,
\begin{equation}
\frac{dR}{dt}=\frac{kV}{4\rho}=\alpha V,
\label{eq21_1b}
\end{equation}
где $\alpha=\frac{k}{4\rho}$ --- некая другая константа. Но $V=\frac{dx}{dt}$
и (\ref{eq21_1b}) дает $\frac{d}{dt}(R-\alpha x)=0$. Отсюда, так как при
$t=0$ имеем $x(0)=0$ и $R(0)\approx 0$, получаем $R(t)=\alpha\, x(t)$. Тогда
$$m=\frac{4\pi}{3}\rho R^3=\frac{4\pi}{3}\rho \alpha^3 x^3= \beta x^3,$$
где мы ввели еще одну константу $\beta$. Подставляя $m=\beta x^3$ в 
(\ref{eq21_1a}), получаем
$$\frac{d}{dt}(x^3\dot x)=gx^3.$$
Но $x^3\dot x=\frac{1}{4}\,\frac{dx^4}{dt}$ и уравнение принимает вид
$$\frac{d^2\,x^4}{dt^2}=4gx^3.$$
Решим это уравнение. Обозначая $y=x^4$, запишем его в виде $\ddot{y}=4g\, y^{3/4}$.
Умножим обе стороны этого уравнения на $\dot y$:
$$\dot y\,\ddot{y}=\frac{1}{2}\frac{d\,{\dot y}^2}{dt}=4g \,y^{3/4}\,\dot y=
\frac{16}{7}\,g\,\frac{d\,y^{7/4}}{dt}.$$
Следовательно, 
$${\dot y}^2=\frac{32g}{7}\,y^{7/4}+\mathrm{const}.$$
При $t=0$ имеем $y(0)=x^4(0)=0$ и $\dot y(0)=4x^3(0)\,\dot x(0)=0$, поэтому
$\mathrm{const}=0$ и получаем
$$\frac{dy}{dt}=\sqrt{\frac{32g}{7}}\,y^{7/8}.$$
Разделяем переменные и интегрируем:
$$\sqrt{\frac{7}{32g}}\,\int\limits_0^y y^{-7/8}dy=\sqrt{\frac{7}{32g}}\,8\,
y^{1/8}=\sqrt{\frac{7}{32g}}\,8\,\sqrt{x}.$$
Поэтому
$$x(t)=\frac{32g}{7\cdot 64}\,t^2=\frac{g}{14}\,t^2,$$
что показывает, что капля падает с ускорением $g/7$.

{\usefont{T2A}{cmr}{m}{it} Второе решение.}
При абсолютно неупругом ударе капли с капельками тумана теряется энергия и  
выделяется в виде тепла. Если капля массы $m$, имея скорость $V$, абсолютно неупругo 
сталкивается с неподвижными капельками общей массой $\Delta m$, то закон сохранения 
импульса позволяет найти выделившееся количество теплоты:
$$\Delta Q=\frac{(mV)^2}{2m}-\frac{(mV)^2}{2(m+\Delta m)}=\frac{mV^2}{2}
\left (1-\frac{1}{1+\Delta m/m}\right )\approx \frac{\Delta m\,V^2}{2}.$$
Поэтому, если капля упала с высоты $x$, закон сохранения энергии с учетом 
выделившегося тепла будет иметь вид
\begin{equation}
\int\limits_0^x mg\,dx=\frac{mV^2}{2}+\int\limits_0^m\frac{V^2}{2}\,dm.
\label{eq21_1c}
\end{equation}
Но, как видели, $m=\beta x^3$, и с учетом этого (\ref{eq21_1c}) принимает вид
$$\frac{1}{4}\,gx^4=\frac{1}{2}\,x^3V^2+\frac{3}{2}\int\limits_0^x
V^2x^2\,dx.$$
Продифференцируем это уравнение по $x$:
$$gx^3=\frac{3}{2}\,x^2V^2+\frac{1}{2}\,x^3\,\frac{dV^2}{dx}+\frac{3}{2}\,
V^2x^2=3\,x^2V^2+\frac{1}{2}\,x^3\,\frac{dV^2}{dx}.$$
Следовательно,
$$2g=\frac{dV^2}{dx}+\frac{6}{x}\,V^2=\frac{1}{x^6}\,\frac{d}{dx}\left (
x^6V^2\right ).$$
Поэтому
$$x^6V^2=2g\int\limits_0^x x^6\,dx=\frac{2}{7}\,gx^7$$
и
$$V^2=2\,\frac{g}{7}\,x.$$
Разумеется, мы можем решить это дифференциальное уравнение методом разделения 
переменных, но достаточно сравнить его  с уравнением свободного падения $V^2=2gx$, 
чтобы заключить, что капля падает с ускорением $g/7$.

\subsection*{Задача 2 {\usefont{T2A}{cmr}{b}{n}(6.34)}:  Закон движения цепи}
{\usefont{T2A}{cmr}{m}{it} Первое решение.}
Поток импульса отсутствует, так как к движущейся части цепи присоединяются 
первоначально неподвижные звенья, не привносящие собой импульса. Таким образом, 
уравнение движения будет 
$$\frac{dp}{dt}=\frac{d}{dt}(mv)=mg,$$
где $m$ --- масса движущейся части цепи. Но $m=\rho x$, где $\rho$ --- линейная 
плотность цепи, а $x$ --- длина движущейся части цепи. Следовательно, уравнение
движения принимает вид
$$\frac{d}{dt}(x\dot x)=gx,$$ или
$$\frac{d^2x^2}{dt^2}=2gx.$$
Пусть $y=x^2$. Тогда $\ddot y=2g\sqrt{y}$ и
$$\dot y\ddot y=\frac{1}{2}\frac{d\dot y^2}{dt}=2g\sqrt{y}\,\dot{y}=\frac{4}
{3}\,g\,\frac{dy^{3/2}}{dt}.$$
Поэтому $$\dot y^2=\frac{8g}{3}\,y^{3/2}+\mathrm{const}.$$
На самом деле $\mathrm{const}=0$, так как $y(0)=x^2(0)=0$ и $\dot t(0)=2\dot 
x(0)\,x(0)=0$. Следовательно,
$$\frac{dy}{dt}=\sqrt{\frac{8g}{3}}\,y^{3/4}.$$
Разделяем переменные и интегрируем:
$$t=\sqrt{\frac{3}{8g}}\int\limits_0^yy^{-3/4}\,dy=4\sqrt{\frac{3}{8g}}\,
y^{1/4}=\sqrt{\frac{6x}{g}}.$$
Таким образом $$x=\frac{g}{6}\,t^2.$$

{\usefont{T2A}{cmr}{m}{it} Второе решение.}
Так же как в предыдущей задаче о падении капли в тумане, можно использовать закон 
сохранения энергии с учетом выделившегося  тепла. Если подвижная часть цепи имеет 
длину $x$, то это означает, что ее центр масс опустился на величину $x/2$. Поэтому 
закон сохранения энергии будет иметь вид
\begin{equation}
mg\frac{x}{2}=\frac{mV^2}{2}+\int\limits_0^m\frac{V^2}{2}\,dm,
\label{eq21_2}
\end{equation}
где последний член и есть количество  выделившегося  тепла. Но $m=\rho x$ и 
(\ref{eq21_2}) перепишется как
$$g\frac{x^2}{2}=\frac{xV^2}{2}+\int\limits_0^x\frac{V^2}{2}\,dx.$$
Продифференцируем это уравнение по $x$:
$$gx=\frac{V^2}{2}+\frac{x}{2}\,\frac{dV^2}{dx}+\frac{V^2}{2}=V^2+\frac{x}{2}
\,\frac{dV^2}{dx}.$$
Следовательно,
$$2g=\frac{2}{x}\,V^2+\frac{dV^2}{dx}=\frac{1}{x^2}\,\frac{d}{dx}\left 
(x^2V^2\right ).$$
Отсюда
$$x^2V^2=2g\int\limits_0^xx^2\,dx=\frac{2g}{3}\,x^3,$$
и  $$V^2=2\,\frac{g}{3}\,x.$$
Последнее соотношение показывает, что цепь падает с ускорением $g/3$.

\subsection*{Задача 3 {\usefont{T2A}{cmr}{b}{n}(6.31)}:  Падение каната на чашку
весов}
На канат действуют сила тяжести $mg$ и сила реакции чашки $F$. Под действием этих 
двух сил $x$-компонента (ось $x$ направлена вниз) импульса каната меняется. Если 
$x$ --- длина не упавшей еще части каната, то $x=L-gt^2/2$. Следовательно, 
$\Delta x=-gt\,\Delta t$ и изменение $x$-компоненты импульса каната будет
$$\Delta p=\frac{m}{L}(x+\Delta x)(V+\Delta V)-\frac{m}{L}\,xV\approx
\frac{m}{L}(x\Delta V+V\Delta x).$$
Подставляя сюда выражения для $x$,  $\Delta x$ и $\Delta V=g\Delta t$, получаем
$$\Delta p=\frac{m}{L}\,g\Delta t\left (L-\frac{gt^2}{2}\right )-\frac{m}{L}
g^2t^2\,\Delta t=mg\,\Delta t-\frac{3m}{2L}\,g^2t^2\,\Delta t.$$
Поэтому можем написать 
$$mg\,\Delta t-\frac{3m}{2L}\,g^2t^2\,\Delta t=(mg-F)\,\Delta t,$$
что для силы $F$ дает
$$F=\frac{3m}{2L}\,g^2t^2.$$
Время падения каната $$t=\sqrt{\frac{2L}{g}}.$$
Поэтому
$$F_{max}=\frac{3m}{2L}\,g^2\,\frac{2L}{g}=3mg.$$

\subsection*{Задача 4 {\usefont{T2A}{cmr}{b}{n}(6.13)}: Столкновение металлических
шаров}
Пусть потенциальная энергия шара при деформации $x$ равна $U(x)=\gamma x^\delta$. 
Закон сохранения энергии имеет вид 
$$2m\,\frac{\dot x^2}{2}+2U(x)=2m\,\frac{V^2}{2},$$
где $V$ --- первоначальная скорость шаров. Поэтому
$$\frac{dx}{dt}=\sqrt{V^2-\frac{2}{m}\,U(x)}.$$
Максимальная деформация, $x_{max}$, определяется из условия
$$V^2-\frac{2}{m}\,U(x_{max})=0,$$
а время столкновения шаров $T$ равно
$$T=2\int\limits_0^{x_{max}}\frac{dx}{\sqrt{V^2-\frac{2}{m}\,U(x)}}.$$
В интеграле
$$T=2\int\limits_0^{x_{max}}\frac{dx}{\sqrt{V^2-\frac{2\gamma}{m}\,x^\delta}}=
\frac{2}{V}\int\limits_0^{x_{max}}\frac{dx}{\sqrt{1-\frac{2\gamma}{m}\,
x^\delta}}$$
сделаем замену переменных:
$$x=\left(\frac{mV^2}{2\gamma}\,y\right)^{1/\delta}.$$
Тогда
$$\frac{2\gamma}{m}\,x^\delta=y,\;\;\;
dx=\left(\frac{mV^2}{2\gamma}\right)^{1/\delta}\frac{1}{\delta}y^{\frac{1}
{\delta}-1}\,dy$$
и, кроме того,  $y_{max}=1$, так как
$$1-\frac{2\gamma}{m}\,x_{max}^\delta=0.$$
Поэтому получим 
$$T=\frac{2}{\delta}\left(\frac{m}{2\gamma}\right)^{\frac{1}{\delta}}\,
V^{\frac{2}{\delta}-1}\,I_\delta,$$
где $$I_\delta=\int\limits_0^1\frac{y^{\frac{1}{\delta}-1}}{\sqrt{1-y}}\,dy.$$
По условию задачи $T=\alpha\,V^{-1/5}$. Поэтому должны иметь
$$\alpha=\frac{2}{\delta}\left (\frac{m}{2\gamma}\right)^{\frac{1}{\delta}}
I_\delta$$
и
$$\frac{2}{\delta}-1=-\frac{1}{5}.$$
Следовательно, $\delta=5/2$ и
$$\alpha=\frac{4}{5}\,I_{5/2}\,\left (\frac{m}{2\gamma}\right)^{\frac{2}{5}},
\;\;\mbox{что означает}\;\;\gamma=\frac{m}{2}\left(\frac{5\alpha}{4I_{5/2}}
\right)^{-5/2}.$$
Окончательно
$$U(x)=\frac{m}{2}\left(\frac{4I_{5/2}}{5\alpha}\,x\right)^{5/2},\;\;
\mbox{и}\;\;F(x)=-\frac{dU}{dx}=-5m\left(\frac{4I_{5/2}}{5\alpha}\right)^{5/2}
\,x^{3/2}.$$
Заметим, что интеграл $$I_{5/2}=\int\limits_0^1\frac{dy}{y^{3/5}\sqrt{1-y}}=
B\left(\frac{1}{2},\frac{2}{5}\right)$$
конечен. Здесь $B(\alpha,\beta)$ --- это бета функцию Эйлера:
$$B(\alpha,\beta)=\int\limits_0^1 (1-t)^{\alpha-1}t^{\beta-1}\,dt.$$

Примерные траектории шаров на фазовой плоскости (в системе центра масс шаров, 
$X$ --- координата центра шара) показаны на рисунке.
\begin{figure}[htb]
\centerline{\epsfig{figure=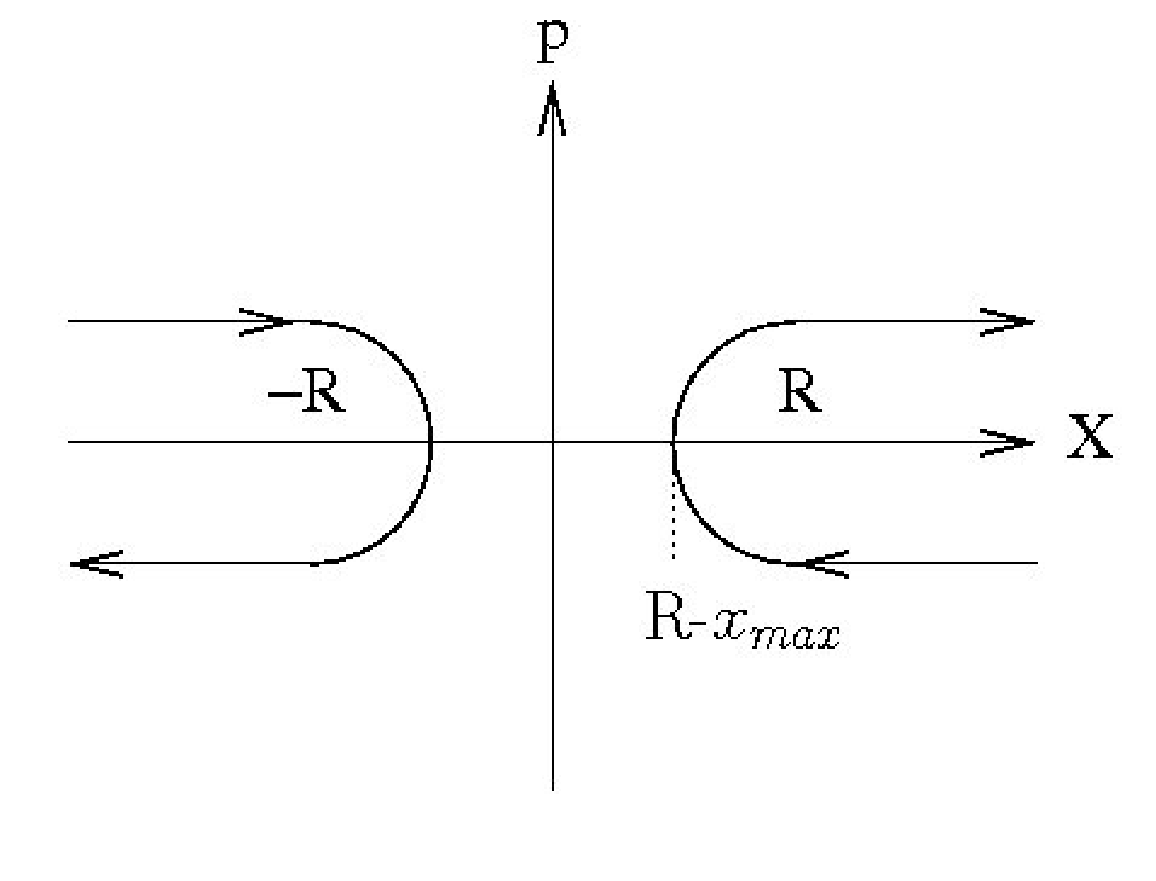,height=4cm}}
\end{figure} 

\subsection*{Задача 5 {\usefont{T2A}{cmr}{b}{n}(6.14)}: Зависимость периода 
колебаний частицы от энергии}
{\usefont{T2A}{cmr}{m}{it} Первое решение.}
Пусть $E\le ka^2/2$, тогда
$$T=\sqrt{2m}\int\limits_{x_1}^{x_2}\frac{dx}{\sqrt{E-\frac{kx^2}{2}}}=
\sqrt{\frac{2m}{E}}\int\limits_{x_1}^{x_2}\frac{dx}{\sqrt{1-\frac{kx^2}
{2E}}},$$
где точки поворота $$x_1=-\sqrt{\frac{2E}{k}},\;\;\;x_2=\sqrt{\frac{2E}{k}}$$
определяются из условия $E=kx^2/2$. Интеграл вычисляется с помощью подстановки
$$y=\sqrt{\frac{k}{2E}}\;x,$$
что дает
$$T=\sqrt{\frac{2m}{E}}\,\sqrt{\frac{2E}{k}}\int\limits_{-1}^{1}
\frac{dy}{\sqrt{1-y^2}}=2\pi\sqrt{\frac{m}{k}}.$$
Если $E> ka^2/2$, то
$$T=\sqrt{2m}\int\limits_{-a}^a\frac{dx}{\sqrt{E-\frac{kx^2}{2}}}=
2\sqrt{\frac{m}{k}}\int\limits_{y_{min}}^{y_{max}}\frac{dy}{\sqrt{1-y^2}},$$
где $$y_{min}=-\sqrt{\frac{k}{2E}}\;a,\;\;\;y_{max}=\sqrt{\frac{k}{2E}}\;a.$$
Следовательно,
$$T=\left . 4\sqrt{\frac{m}{k}}\,\arcsin{y}\right|_0^{\sqrt{\frac{k}{2E}}\,a}=
4\sqrt{\frac{m}{k}}\,\arcsin{\sqrt{\frac{ka^2}{2E}}}.$$
Окончательно
$$T=\left\{\begin{array}{cc} 2\pi\sqrt{\frac{m}{k}}, &\mbox{если}\;\; 
E\le\frac{ka^2}{2},\\  & \\ 4\sqrt{\frac{m}{k}}\,\arcsin{\sqrt{\frac{ka^2}
{2E}}}, &\mbox{если}\;\;  E>\frac{ka^2}{2}.
\end{array}\right . $$

{\usefont{T2A}{cmr}{m}{it} Второе решение.}
Пусть $E> ka^2/2$. Если бы не было отражающих стенок при $x=\pm a$, частица бы
колебалась с периодом $T_0=\frac{2\pi}{\omega}=2\pi\sqrt{\frac{m}{k}}$ и 
с амплитудой $A=\sqrt{\frac{2E}{k}}$. Наличие стенок приведет к тому, что период 
колебаний сократится на время $4\tau$, где $\tau$ --- это время, которое требуется, 
чтобы координата частицы при отсутствии отражающих стенок уменьшилась от $x_0=A$ при 
$t=0$, до $a=A\,\cos{\omega\tau}$ при $t=\tau$. Следовательно,
$$\tau=\frac{1}{\omega}\,\arccos{\frac{a}{A}}=\sqrt{\frac{m}{k}}\,\arccos{
\sqrt{\frac{ka^2}{2E}}}$$
и $$T=T_0-4\tau=4\sqrt{\frac{m}{k}}\left(\frac{\pi}{2}-\arccos{\sqrt{\frac
{ka^2}{2E}}}\right )=4\sqrt{\frac{m}{k}}\,\arcsin{\sqrt{\frac{ka^2}{2E}}}.$$

\subsection*{Задача 6 {\usefont{T2A}{cmr}{b}{n}(6.15)}:  По какому закону
обращается в бесконечность период движения частицы?}
Так как в точке $x=a$ имеем максимум, то $U^\prime(a)=0$ и около этой точки 
$$U(x)\approx U(a)+\frac{1}{2}U^{\prime\,\prime}(a)(x-a)^2,$$
причем $U^{\prime\,\prime}(a)<0$. Пусть $(U(a)-E)/E=\Delta$. Тогда около точки
$x=a$
$$1-\frac{U(x)}{E}\approx -\Delta +k\left(\frac{x}{a}-1\right)^2,$$ где 
$$k=-\frac{a^2U^{\prime\,\prime}(a)}{2E}>0.$$ 
Левая точка поворота $x_1$ находится из условия $E-U(x_1)=0$, что дает
$$\frac{x_1}{a}=1+\sqrt{\frac{\Delta}{k}}.$$
Период движения $$T=\sqrt{2m}\int\limits_{x_1}^{x_2}\frac{dx}{\sqrt{E-U(x)}}=
a\sqrt{\frac{2m}{E}}\int\limits_{x_1}^{x_2}\frac{d\left(\frac{x}{a}\right)}
{\sqrt{1-\frac{U(x)}{E}}}$$
набирается в основном около нижнего предела
$$y_1=\frac{x_1}{a}=1+\sqrt{\frac{\Delta}{k}}.$$
Поэтому когда $\Delta\to 0$ с точностью до членов порядка $\sqrt{\Delta}$
$$T= a\sqrt{\frac{2m}{E}}\int\limits_{y_1}\frac{dy}{\sqrt{k}(y-1)}\approx
-a\sqrt{\frac{2m}{Ek}}\;\ln{\sqrt{\frac{\Delta}{k}}}\approx
\sqrt{\frac{m}{-U^{\prime\,\prime}(a)}}\;\ln{\frac{-a^2U^{\prime\,\prime}(a)}
{U(a)-E}}.$$

\subsection*{Задача 7 {\usefont{T2A}{cmr}{b}{n}(6.24)}: Скорость движения 
кальмара}
{\usefont{T2A}{cmr}{m}{it} Первое решение.}
Пусть в единицу времени кальмар выбрасывает массу воды $\mu$ со скоростью $u$ 
относительно себя. Изменение импульса для системы кальмар плюс $\mu\,dt$ 
количество воды равно (ось $x$ направлена по движению кальмара, $m$ --- масса 
кальмара)
$$\Delta p_x=m(V+\Delta V)+\mu\,dt\,(V-u)-mV=F_{\mbox{тр}}\,dt=
-\alpha\,V\,dt.$$
Поэтому получаем следующее уравнение движения
\begin{equation}
\frac{dV}{dt}=\frac{\mu}{m}\,u-\frac{\alpha+\mu}{m}\,V.
\label{eq21_7}
\end{equation}
Разделяем переменные и интегрируем
$$t=\int\limits_0^V\frac{dV}{\frac{\mu}{m}\,u-\frac{\alpha+\mu}{m}\,V}=
-\frac{m}{\alpha+\mu}\ln{\frac{\frac{\mu}{m}\,u-\frac{\alpha+\mu}{m}\,V}
{\frac{\mu}{m}\,u}}.$$
Отсюда
$$V(t)=\frac{\mu\,u}{\alpha+\mu}\left [1-\exp{\left(-\frac{\alpha+\mu}{m}\,t
\right)}\right ].$$
В системе кальмара, когда кальмар выбрасывает воду, вся ее работа идет на увеличение 
кинетической энергии воды. Следовательно, $N\,dt=\mu\,dt\,u^2/2$, и
$$\mu=\frac{2N}{u^2}.$$

{\usefont{T2A}{cmr}{m}{it} Второе решение.}
Рассмотрим кальмара как систему переменного состава. Уравнение ее движения будет
$$\frac{dp_x}{dt}=F_{\mbox{тр}}+\Pi_x,$$
где $\Pi_x$ есть поток импульса в систему. Когда кальмар заглатывает массу 
неподвижной воды $\mu\,dt$, эта вода не привносит с собой импульса в систему. 
Когда же кальмар выбрасывает эту самую порцию воды со скоростью $\tilde{u}_x=V-u$ 
относительно лабораторной системы, вода с собой уносит $x$-компоненту импульса  
$\mu\,dt(V-u)$. Поэтому поток импульса в систему равен $\Pi_x=-\mu(V-u)=\mu(u-V)$ 
и уравнение движения принимает вид
$$\frac{dp_x}{dt}=-\alpha\,V+\mu(u-V),$$
что эквивалентно уравнению (\ref{eq21_7}).

\subsection*{Задача 8 {\usefont{T2A}{cmr}{b}{n}(6.27)}: Закон движения пробки в
горлышке бутылки с подогретым шампанским}
Сила трения пропорциональна площади той боковой поверхности пробки, которая все еще
соприкасается с горлышком бутылки. Поэтому $F_{\mbox{тр}}\sim (L-x)$, где $L$ ---
длина горлышка, $x$ --- смещение нижнего конца пробки в горлышке. Следовательно,
$F_{\mbox{тр}}=k(L-x)$, где $k$ это некоторый коэффициент пропорциональности.
Вместо него мы можем использовать максимальное значение силы трения $F_0=F_{\mbox{тр}}
(x=0)=kL$ и написать
$$F_{\mbox{тр}}(x)=F_0\left(1-\frac{x}{L}\right).$$
Уравнение движения будет иметь вид
$$m\ddot x=F_{\mbox{давл}}-F_{\mbox{тр}}=F_{\mbox{давл}}-F_0\left(1-\frac{x}{L}
\right).$$
Сила давления $F_{\mbox{давл}}$ практически не меняется, так как обьем газа в бутылке
меняется незначительно при движении пробки. Тогда общее решение уравнения движения
$$\ddot x-\frac{F_0}{mL}\,x=\frac{F_{\mbox{давл}}-F_0}{m}$$
имеет вид
$$x(t)=C_1\exp{\left(\sqrt{\frac{F_0}{mL}}\,t\right)}+
C_2\exp{\left(-\sqrt{\frac{F_0}{mL}}\,t\right)}+\left(1-\frac{F_{\mbox{давл}}}
{F_0}\right)L.$$
Начальные условия $x(0)=0$ и $\dot x(0)=0$ определяют константы $C_1$ и $C_2$
через систему уравнений
$$C_1+C_2=-\left(1-\frac{F_{\mbox{давл}}}{F_0}\right)L,\;\; C_1-C_2=0.$$
Решая эту систему, окончательно получаем
$$x(t)=L\left(\frac{F_{\mbox{давл}}}{F_0}-1\right)\left[\ch{\left(\sqrt{
\frac{F_0}{mL}}\,t\right)}-1\right].$$

\section[Семинар 22]
{\centerline{Семинар 22}}

\subsection*{Задача 1 {\usefont{T2A}{cmr}{b}{n}(6.25)}: Сколько времени тело
находилось в полете?}
Уравнение движения имеет вид
$$m\frac{dV_x}{dt}=-mg-\alpha V_x.$$
Разделяем переменные и интегрируем
$$T=-\int\limits_{V_1}^{-V_2}\frac{dV_x}{g+\frac{\alpha}{m}V_x}=\frac{m}
{\alpha}\,\ln{\frac{1+\frac{\alpha}{mg}\,V_1}{1-\frac{\alpha}{mg}\,V_2}}.$$
Используя $\ln{(1+x)}\approx x-x^2/2$, когда $x\ll 1$, получаем
$$T\approx \frac{m}{\alpha}\left\{\frac{\alpha V_1}{mg}-\frac{1}{2}\left(
\frac{\alpha V_1}{mg}\right)^2+\frac{\alpha V_2}{mg}+\frac{1}{2}\left(
\frac{\alpha V_2}{mg}\right)^2\right\}=\frac{V_1+V_2}{g}\left[1-\frac{\alpha}
{2mg}(V_1-V_2)\right].$$

\subsection*{Задача 2 {\usefont{T2A}{cmr}{b}{n}(7.1)}: Частота колебаний доски}
Пусть $P_1$ и $P_2$ --- силы реакции, действующие на доску со стороны валиков в точках
соприкосновения после того, как доска массой $M$ сместилась вправо на маленькую 
величину $x$ (см. рисунок).  
\begin{figure}[htb]
\centerline{\epsfig{figure=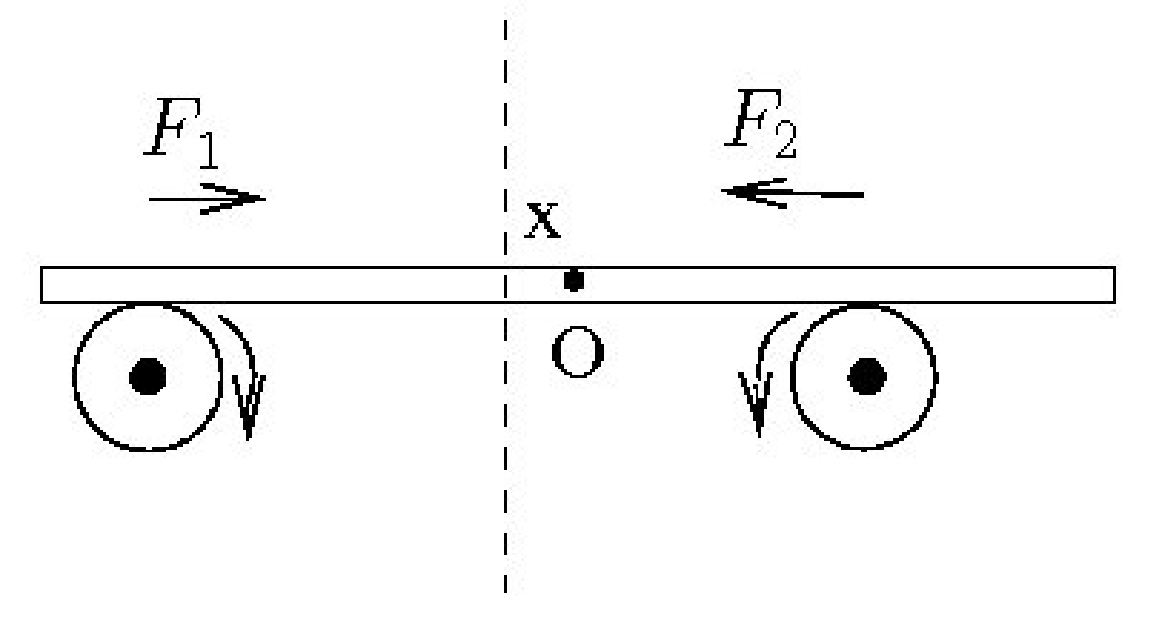,height=3cm}}
\end{figure}
Тогда $P_1+P_2=Mg$, так как в вертикальном направлении доска
находится в равновесии. Кроме того, доска не вращается около центра масс и, 
следовательно, моменты сил $P_1$ и $P_2$ относительно центра масс доски (точка $O$ 
на рисунке) должны быть равны:
$$P_1\left(\frac{L}{2}+x\right)=P_2\left(\frac{L}{2}-x\right).$$
Из этих двух уравнений получаем
$$P_1-P_2=-\frac{2}{L}(P_1+P_2)=-\frac{2Mg}{L}\,x.$$
Поэтому возвращающая сила трения (вернее, ее $x$-компонента) равна
$$F=F_1-F_2=\mu (P_1-P_2)=-\frac{2Mg\mu}{L}\,x,$$
и уравнение движения в горизонтальном направлении будет иметь вид
$$M\ddot x=-\frac{2Mg\mu}{L}\,x$$
или $$\ddot x+\omega^2\,x=0,$$ где
$$\omega^2=\frac{2g\mu}{L}.$$
Следовательно, частота колебаний доски
$$\omega=\sqrt{\frac{2g\mu}{L}}.$$

\subsection*{Задача 3 {\usefont{T2A}{cmr}{b}{n}(7.2)}: Частота малых колебаний
жидкости в трубке}
Пусть жидкость в трубке сместилась от равновесного уровня на маленькую величину $x$
вправо. Кинетическая энергия жидкости будет $T=m\dot x^2/2$, где $m$ --- масса 
жидкости, так как вся жидкость движется с одинаковой скоростью $\dot x$.
Потенциальная энергия жидкости будет
$$U=\frac{mg}{2L}\,(L-x)\,\frac{(L-x)\cos{\alpha}}{2}+
\frac{mg}{2L}\,(L+x)\,\frac{(L+x)\cos{\alpha}}{2},$$
где $L=h/\cos{\alpha}$ --- длина равновесного столбика жидкости в каждом из колен 
$V$-образной трубки. Действительно, например, в левом колене находится масса жидкости
$\frac{m}{2L}(L-X)$ и центр масс этой части жидкости находится на высоте
$\frac{1}{2}\,(L-x)\cos{\alpha}$. 

Следовательно, полная энергия жидкости
$$E=T+U=\frac{m\dot x^2}{2}+\frac{mg\cos{\alpha}}{4L}\,[(L-x)^2+(L+x)^2]=
\frac{m\dot x^2}{2}+\frac{mg\cos{\alpha}}{2L}\,x^2+\mathrm{const}.$$
Полная энергия сохраняется. Поэтому
$$\frac{dE}{dt}=m\dot x\ddot x+\frac{mg\cos{\alpha}}{L}\,x\dot x=0.$$
После сокращения на $m\dot x$ получаем уравнение гармонических колебаний с частотой
$$\omega=\sqrt{\frac{g\cos{\alpha}}{L}}=\sqrt{\frac{g}{h}}\,\cos{\alpha}.$$

\subsection*{Задача 4 {\usefont{T2A}{cmr}{b}{n}(7.3)}: При каком условии
колебания устойчивы?}
Пусть первоначальное натяжение пружин равно $a$ и пусть нить сместилась вправо на
маленькую величину $x$ так, что левая пружина удлинилась на $x$, а правая сжалась
на $x$ (см. рисунок).
\begin{figure}[htb]
\centerline{\epsfig{figure=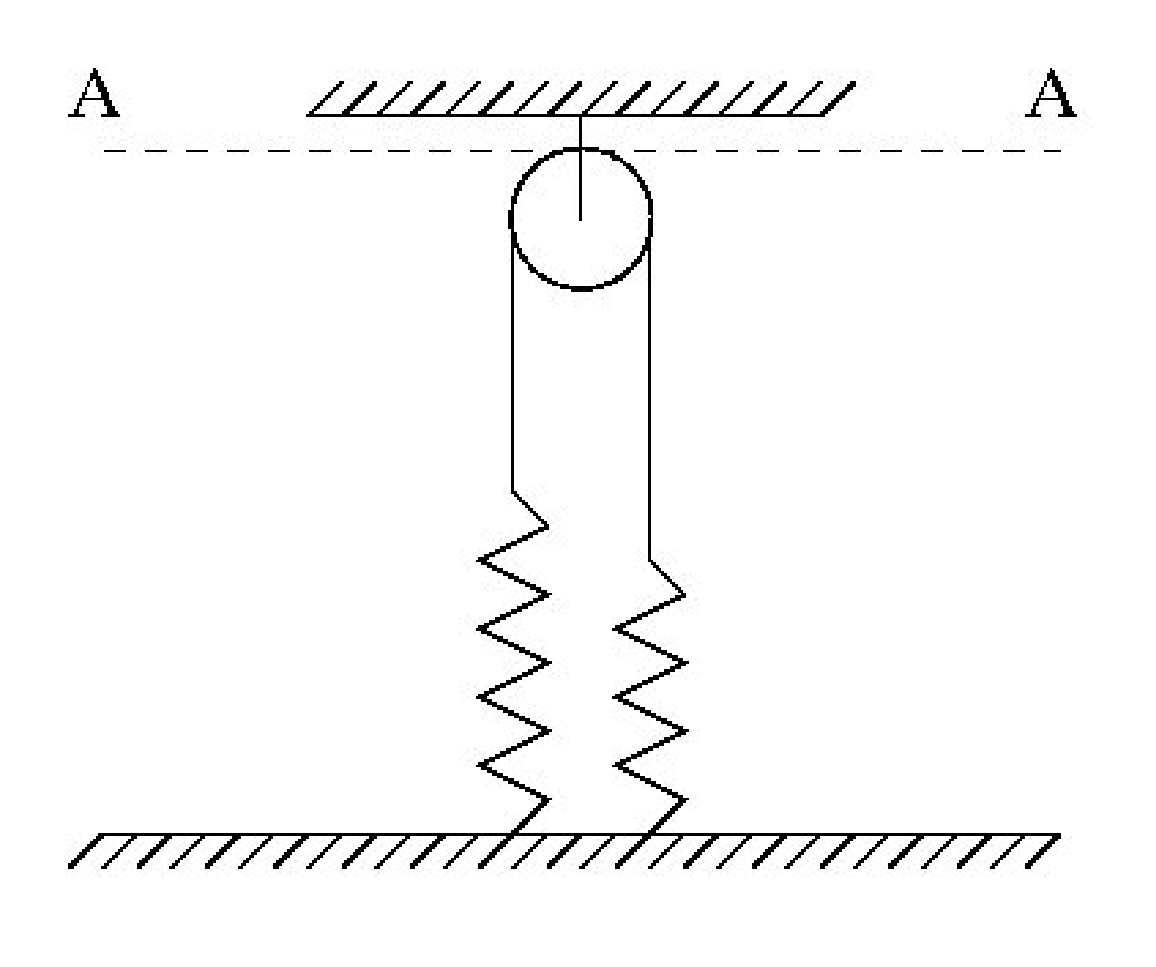,height=2.7cm}}
\end{figure}

\noindent Упругая потенциальная энергия пружин равна
$$U_1=\frac{k}{2}\,(a+x)^2+\frac{k}{2}\,(a-x)^2=k(a^2+x^2).$$
Гравитационную потенциальную энергию будем отсчитывать от уровня $AA$ (см. рисунок).
Правая часть нити имеет массу $$\frac{M}{L}\left(\frac{L}{2}+x\right ),$$
и ее центр масс находится ниже уровня $AA$ на
$$\frac{1}{2}\left(\frac{L}{2}+x\right ).$$
Здесь $M$ --- полная масса нити. Аналогичные величины для левой части находятся 
заменой $x\to -x$. Поэтому гравитационная потенциальная энергия нити равна
$$U_2=-\frac{Mg}{L}\left(\frac{L}{2}+x\right )\frac{1}{2}\left(\frac{L}
{2}+x\right )-\frac{Mg}{L}\left(\frac{L}{2}-x\right )\frac{1}{2}\left
(\frac{L}{2}-x\right )=-\frac{Mg}{L}\left(\frac{L^2}{4}+x^2\right).$$
Вся нить движется со скоростью $\dot x$. Поэтому кинетическая энергия $T=M\dot x^2/2$
и полная энергия $E=T+U_1+U_2$ равна
$$E=\frac{M\dot x^2}{2}+\left(k-\frac{Mg}{L}\right)\,x^2+\mathrm{const}.$$
Полная энергия сохраняется. Поэтому
$$\frac{dE}{dt}=M\,\dot x\,\ddot x+2\left(k-\frac{Mg}{L}\right)\,x\,\dot 
x=0.$$
После сокращения на $M\dot x$ получаем уравнение гармонических колебаний с частотой
$$\omega=\sqrt{2\left(\frac{k}{M}-\frac{g}{L}\right)}.$$
Колебания станут неустойчивыми при 
$$\frac{g}{L}>\frac{k}{M},$$
когда $\omega$ становится мнимым.

\subsection*{Задача 5 {\usefont{T2A}{cmr}{b}{n}(6.35)}: За какое время цепь
соскользнет со стола?}
{\usefont{T2A}{cmr}{m}{it} Первое решение.}
В точке соприкосновения цепи с полом на цепь действует сила реакции пола $T$. Найдем
ее. Пусть $\rho$ --- масса единицы длины цепи. Кусочек цепи длиной $\Delta x$ перед
моментом падения на пол имеет импульс $\rho\Delta x V$ и под действием силы $T$
останавливается за время $\Delta t=\Delta x/V$. Поэтому $\rho\Delta x V=T\Delta 
t=T\Delta x/V$ и $T=\rho\,V^2$.

Рассмотрим кусок цепи длиной $H+\Delta x$, где кусочек длиной $\Delta x$ вначале 
лежит неподвижно на столе. Через время $\Delta t$ этот кусочек присоединится к 
движущейся части цепи, но аналогичный фронтальный кусочек выбранного куска цепи будет 
лежать неподвижно на полу (см. рисунок). 
\begin{figure}[htb]
\centerline{\epsfig{figure=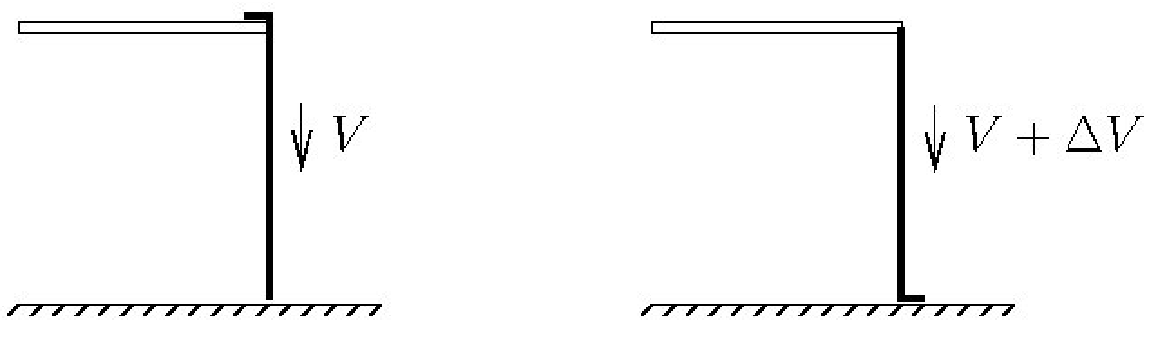,height=3cm}}
\end{figure}
Поэтому изменение импульса этого куска цепи длиной 
$H+\Delta x$ будет $\Delta p=m(V+\Delta V)-mV=m\Delta V=\rho\,H\,\Delta V$, 
где $m$ --- масса  движущейся части цепи длиной $H$. Это изменение импульса произошло 
под действием силы $mg-T=\rho(gH-V^2)$ и поэтому $\Delta p=(mg-T)\Delta t=
\rho(gH-V^2)\,\Delta t$, что дает следующее уравнение движения
\begin{equation}
\frac{dV}{dt}=g-\frac{V^2}{H}=g\left(1-\frac{V^2}{u^2}\right ),
\label{eq22_5}
\end{equation}
где $u=\sqrt{gH}$. Разделяем переменные и интегрируем
$$gt=\int\limits_0^V\frac{dV}{1-\frac{V^2}{u^2}}=\frac{1}{2}\int\limits_0^V
\left[\frac{1}{1-\frac{V}{u}}+\frac{1}{1+\frac{V}{u}}\right ]\,dV=
\frac{u}{2}\ln{\frac{u+V}{u-V}}.$$
Следовательно,
$$\frac{u+V}{u-V}=e^{2gt/u}$$
и $$V=u\;\frac{e^{2gt/u}-1}{e^{2gt/u}+1}=u\th{\left(\frac{gt}{u}\right)}=
\sqrt{gH}\,\th{\left(\sqrt{\frac{g}{H}}\,t\right )}.$$
Время соскальзывания $T$ определяется из условия
$$2H=\int\limits_0^T V(t)\,dt=\sqrt{gH}\int\limits_0^T\th{\left(\sqrt{\frac{g}
{H}}\,t\right )}dt=H\ln{\left(\ch{\sqrt{\frac{g}{H}}\,T}\right)},$$
где мы воспользовались интегралом
$$\int\th{x}\,dx=\int\frac{d(\ch{x})}{\ch{x}}=\ln{(\ch{x})}.$$ 
Поэтому
$$\ch{\left(\sqrt{\frac{g}{H}}\,T\right)}=e^2$$
и $$T=\sqrt{\frac{H}{g}}\,\arch{(e^2)}=\sqrt{\frac{H}{g}}\,\ln{(e^2+
\sqrt{e^4-1})}\approx 2,7\,\sqrt{\frac{H}{g}}.$$

{\usefont{T2A}{cmr}{m}{it} Второе решение.}
Рассмотрим движущуюся часть цепи с нижней границей чуть выше точки соприкосновения 
с полом как систему переменного состава. Неподвижные звенья цепи на столе, которые 
поочередно вовлекаются в движение, не привносят с собой импульса. Звенья цепи, которые
уходят из системы, а потом ударяются об пол и останавливаются, за время $\Delta t$ 
уносят с собой импульс $\rho\,V\,\Delta t\,V=\rho\,V^2\,\Delta t$. Поэтому поток
импульса в систему равен $\Pi=-\rho\,V^2$, и уравнение движения будет
$$\frac{dp}{dt}=mg+\Pi=mg-\rho\,V^2.$$
Но $p=mV=\rho\,H\,V$, и это уравнение совпадает с (\ref{eq22_5}).

Заметим, что если бы мы нижнюю границу системы провели чуть далее точки 
соприкосновения цепи с полом, то поток импульса в систему равнялся бы нулю, 
так как из системы поочередно уходили бы неподвижные звенья, лежащие на полу. Но в этом 
случае появилась бы сила реакции пола $T$, действующая на систему, и уравнение 
движения осталось бы прежним.

{\usefont{T2A}{cmr}{m}{it} Третье решение.}
Так же как в задаче о падении капли в тумане, можно использовать закон сохранения 
энергии с учетом выделившегося  тепла. Пусть $x$ --- длина упавшей на пол цепи.
Потеря энергии (переход ее в тепло) происходит, во-первых, когда новые звенья с общей
массой $\Delta m$ вовлекаются в движение, и, во-вторых, когда такой же массы  
$\Delta m$ часть цепи ударяется об пол и останавливается. В обоих случаях выделившееся
тепло равно $\Delta m\, V^2/2$. Поэтому закон  сохранения энергии с учетом 
выделившегося  тепла запишется в виде
$$\rho x g H=\rho H \frac{V^2}{2}+\int\limits_0^{\rho\,x}V^2\,dm.$$
Но $dm=\rho\,dx$, и после сокращения на $\rho$ это уравнение принимает вид
$$xgH=H\,\frac{V^2}{2}+\int\limits_0^x V^2\,dx.$$
Продифференцируем по $x$:
\begin{equation}
g\,H=\frac{H}{2}\,\frac{dV^2}{dx}+V^2.
\label{eq22_5a}
\end{equation}
Однако 
$$\frac{dV^2}{dx}=\frac{dV^2}{dt}\,\frac{dt}{dx}=\frac{dV^2}{dt}\,\frac{1}
{V}=2\,\frac{dV}{dt}$$
и легко видеть, что из (\ref{eq22_5a}) получится уравнение движения (\ref{eq22_5}).

\subsection*{Задача 6 {\usefont{T2A}{cmr}{b}{n}(6.42)}: Закон движения ледяного
метеорита}
За время $\Delta t$ от метеорита отделяется оболочка с массой $-\Delta M$, где
$\Delta M$ --- изменение массы самого метеорита. В системе отсчета метеорита испарение
изотропно. Поэтому в этой системе от метеорита уходит импульс $\Delta p^\prime_x=0$.
В лабораторной системе $\Delta p_x=\Delta p^\prime_x+(-\Delta M)V=-\Delta M\, V$,
где $V$ --- скорость метеорита. Так как этот импульс (вернее, его $x$-компонента,
ось $x$ направлена по движению метеорита) уходит из системы, поток импульса в систему
будет $$\Pi_x=-\frac{\Delta p_x}{\Delta t}=\frac{\Delta M}{\Delta t}\, V=
\dot M\,V.$$ Следовательно, уравнение движения будет иметь вид (мы  пренебрегли силой 
тяжести по сравнению с силой торможения $F=-\alpha S V$)
$$\frac{dp_x}{dt}=\dot M\,V+M\,\dot V=F+\Pi_x=F+\dot M\,V,$$
или $$M\frac{dV}{dt}=F=-\alpha S V.$$
Поделим это уравнение на $dM/dt=-\beta S$ и разделим переменные:
$$\frac{dV}{V}=\frac{\alpha}{\beta}\,\frac{dM}{M}.$$
Интегрируя, получаем $$\frac{V}{V_0}=\left(\frac{M}{M_0}\right)^{\alpha/\beta}=
\left(\frac{R}{R_0}\right)^{3\alpha/\beta},$$
где мы учли, что $M=\frac{4\pi}{3}\rho R^3$. Продифференцируем это последнее 
равенство по времени: $\dot M=4\pi\rho R^2\dot R=-\beta S=-\beta 4\pi R^2$.
Поэтому $\dot R=-\beta/\rho$, что дает $$R=R_0-\frac{\beta\,t}{\rho}$$ и
$$V=V_0\left (1-\frac{\beta\, t}{\rho R_0}\right)^{3\alpha/\beta}.$$

\subsection*{Задача 7 {\usefont{T2A}{cmr}{b}{n} \cite{30}}: Когда
исчезнет сила давления на блок?}
Пусть длина правой части веревки $z$, левой $2L-z$. $z$-Компонента 
импульса веревки будет (ось $z$ направлена вниз)
$$p_z=\frac{m}{2L}z\dot{z}-\frac{m}{2L}(2L-z)\dot{z}=
\frac{m}{L}(z-L)\dot{z}.$$
Поэтому можем написать
$$\frac{dp_z}{dt}=\frac{m}{L}[\dot{z}^2+(z-L)\ddot{z}]=mg-F,$$
где $F$ как раз равно силе давления на блок. Отсюда
\begin{equation}
F=mg-\frac{m}{L}[\dot{z}^2+(z-L)\ddot{z}].
\label{eq22_61}
\end{equation}
С другой стороны, уравнения движения отдельно правой части и отдельно левой 
части выглядят так (проекция на ось $z$):
\begin{equation}
\frac{d}{dt}\left (\frac{m}{2L}z\dot{z}\right )=\frac{m}{2L}zg-T+\Pi
\label{eq22_62}
\end{equation}
и
\begin{equation}
\frac{d}{dt}\left (-\frac{m}{2L}(2L-z)\dot{z}\right )=\frac{m}{2L}(2L-z)g-T
+\Pi,
\label{eq22_63}
\end{equation}
где $\Pi=\frac{m}{2L}\dot{z}^2$ есть поток импульса в правую часть веревки (сколько 
импульса приносят с собой новые части в единицу времени). В левую часть веревки 
проекция потока импульса на ось $z$ тоже положительна (уходящие части уносят импульс 
в противоположном к $z$ направлению, что эквивалентно приходу импульса по направлению
$z$ - величина $z$-проекции импульса у левой части при этом увеличивается).

Заметим, что (\ref{eq22_62}) и (\ref{eq22_63}) эквивалентны уравнениям
\begin{equation}
\frac{m}{2L}z\ddot{z}=\frac{m}{2L}zg-T,
\label{eq22_64}
\end{equation} 
\begin{equation}
\frac{m}{2L}(2L-z)\ddot{z}=T-\frac{m}{2L}(2L-z)g.
\label{eq22_65}
\end{equation}
Их сумма дает
\begin{equation}
\ddot{z}=\frac{g}{L}(z-L).
\label{eq22_66}
\end{equation} 
Умножая это уравнение на $\dot{z}$ и интегрируя (правую часть в пределах от $L$ до 
$z$), получаем
$$\frac{1}{2}\dot{z}^2=\frac{g}{L}\left(\frac{z^2}{2}-Lz\right )+
\frac{gL}{2},$$
или (это фактически закон сохранения энергии)
\begin{equation}
\dot{z}^2=\frac{g}{L}(z-L)^2.
\label{eq22_67}
\end{equation} 
Уравнения (\ref{eq22_61}),(\ref{eq22_66}) и (\ref{eq22_67}) дают для силы 
давления на блок
$$F=mg\left [ 1-2\frac{(z-L)^2}{L^2}\right ].$$
Следовательно, она исчезает (и дальше веревка слетит с блока), когда
$$z-L=\frac{L}{\sqrt{2}}.$$

Но самое интересное в этой задаче то, что $F\ne 2T$. Действительно, 
если сложить уравнения (\ref{eq22_62}) и (\ref{eq22_63}), видно что $F=2T-2\Pi$. 
Следовательно, сила натяжения веревки не исчезает в момент, когда давление на блок 
равно нулю:
$$2T=F+\frac{m}{L}\dot{z}^2= mg\left [ 1-\frac{(z-L)^2}{L^2}\right ],$$
и когда $$z-L=\frac{L}{\sqrt{2}},$$ сила натяжения веревки $$T=\frac{mg}{4}.$$

Чтобы понять, почему $F\ne 2T$, надо проследить за судьбой того бесконечно маленького 
участка веревки, которая отсоединяется от левой части и присоединяется к правой. Пусть
блок имеет конечный радиус $R$. Тогда участок левой части веревки длиной $\pi R$, 
который имел скорость $-\dot{z}$, исчезает из левой части за время 
$\Delta t= \frac{\pi R}{\dot{z}}$ и такой же участок, уже со скоростью $\dot{z}$,
появляется в правой части. То есть эффективное ускорение этого участка вдоль оси $z$
есть $a=\frac{2\dot{z}}{\Delta t}= \frac{2\dot{z}^2}{\pi R}$ и стремится к 
бесконечности, когда $R\to 0$. Но масса этого участка $\Delta m=\frac{m}{2L}\pi R$
стремится к нулю и $\Delta m \, a=\frac{m}{L}\dot{z}^2$ остается конечной. 
Следовательно, $F$ и $2T$ должны отличаться, чтобы обеспечить $\Delta m \, a=2T-F$.

\section[Семинар 23]
{\centerline{Семинар 23}}

\subsection*{Задача 1 {\usefont{T2A}{cmr}{b}{n}(7.19)}: Какой путь пройдет шайба
до остановки?}
Уравнение движения имеет вид $$M\ddot x=-kx-\alpha \dot x,$$
или $$\ddot x+\omega_0^2x+2\beta\dot x=0,$$ где
$$\omega_0^2=\frac{k}{M},\;\;\mbox{и}\;\;\beta=\frac{\alpha}{2M}.$$
Соответствующее характеристическое уравнение $\lambda^2+2\beta\lambda+\omega_0^2=0$
имеет решения $\lambda=-\beta\pm\sqrt{\beta^2-\omega_0^2}\approx -\beta\pm
i\,\omega_0$, так как $\beta^2\ll\omega_0^2$. Следовательно,
$$x(t)=(C_1e^{i\omega_0t}+C_2e^{-i\omega_0t})e^{-\beta t}.$$
Начальные условия 
$$x(0)=A=C_1+C_2,\;\; \dot x(0)=0=-\beta(C_1+C_2)+i\omega_0
(C_1-C_2)\approx i\omega_0 (C_1-C_2)$$ определяют константы $C_1$ и $C_2$:
$C_1=C_2=\frac{A}{2}$, что дает $$x(t)=Ae^{-\beta t}\cos{\omega_0 t}.$$
Это затухающие колебания с последовательными амплитудами $A_n=Ae^{-\beta t_n}$, где
$n$ целое число $0,1,2,\ldots$ и $t_n$ определяется из условия $\cos{\omega_0 t_n}
=\pm 1$, т.~е. $\omega_0 t_n=n\pi$ и $t_n=n\pi/\omega_0$. Пройденный шайбой путь 
равен
$$s=A+2A_1+2A_2+\cdots=2A\left (1+e^{-\frac{\pi\beta}{\omega_0}}+
e^{-2\frac{\pi\beta}{\omega_0}}+e^{-3\frac{\pi\beta}{\omega_0}}+\cdots 
\right)-A.$$
В скобках имеем сумму бесконечной геометрической прогрессии. Поэтому 
$$s=\frac{2A}{1-e^{-\frac{\pi\beta}{\omega_0}}}-A=
A\frac{1+e^{-\frac{\pi\beta}{\omega_0}}}{1-e^{-\frac{\pi\beta}{\omega_0}}}=
A\,\cth{\frac{\pi\beta}{2\omega_0}}.$$
Так как $$\frac{\pi\beta}{2\omega_0}=\frac{\alpha\pi}{4}\,\frac{1}
{\sqrt{Mk}}\ll 1,$$ то $$s\approx A\,\frac{4}{\alpha\pi}\,\sqrt{Mk}=
\frac{4A}{\pi}\,\sqrt{\frac{Mk}{\alpha^2}}\gg A.$$

\subsection*{Задача 2 {\usefont{T2A}{cmr}{b}{n}(6.44)}: Закон движения нити}
{\usefont{T2A}{cmr}{m}{it} Первое решение.}
Пусть конец нити, к которому приложена сила, сдвинулся от начального положения (верхний
рисунок) на $AC=x$ (см. рисунок). 
\begin{figure}[htb]
\centerline{\epsfig{figure=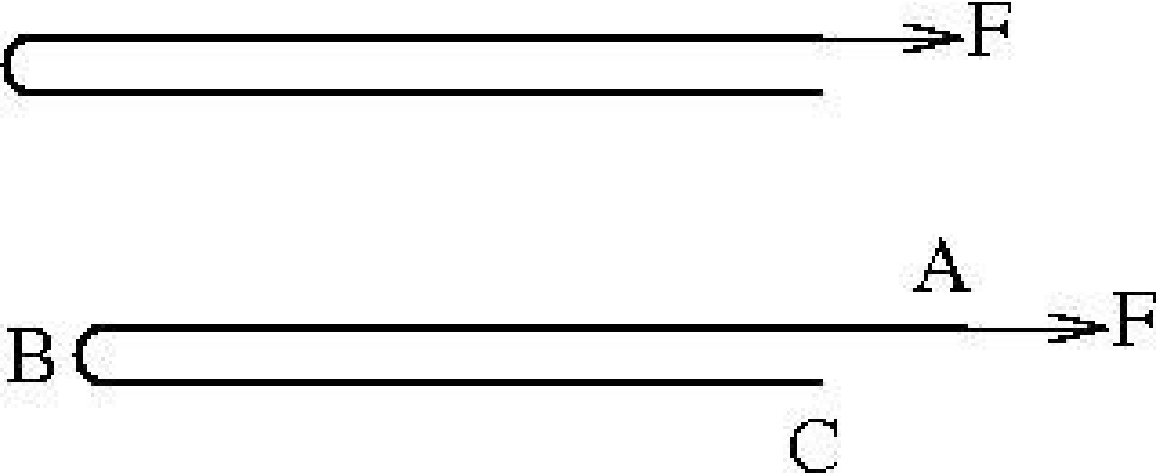,height=3cm}}
\end{figure}
Часть нити $AB$ движется со скоростью $V=dx/dt$, а часть $BC$ --- неподвижна. 
На рисунке видно, что $2\,BC+x=L$, где $L$ --- длина нити. Поэтому $BC=(L-x)/2$ и 
$AB=L-BC=(L+x)/2$. Следовательно, импульс нити $$p=\frac{M}{L}\,\frac{L+x}{2}\,
V,$$ где $M$ --- масса нити. Уравнение движения нити будет
\begin{equation}
\frac{dp}{dt}=\frac{M}{2L}\,\frac{d}{dt}[(L+x)V]=F.
\label{eq23_2}
\end{equation}
Отсюда
$$(L+x)V=(L+x)\frac{dx}{dt}=\frac{2FL}{M}\,t.$$
Разделяем переменные и интегрируем
$$\int\limits_0^x(L+x)\,dx=Lx+\frac{x^2}{2}=\frac{2FL}{M}\int\limits_0^t
t\,dt=\frac{FL}{M}\,t^2.$$ Для $x$ получаем квадратное уравнение
$$x^2+2Lx-\frac{2FL}{M}\,t^2=0.$$
Поэтому $$x=-L\pm\sqrt{L^2+\frac{2FL}{M}\,t^2}.$$ Знак минус не подходит, так как
$x$ --- величина положительная. Окончательно
$$x(t)=L\left(\sqrt{1+ \frac{2FLt^2}{ML}}-1\right).$$

{\usefont{T2A}{cmr}{m}{it} Второе решение.}
Кинетическая энергия движущейся части нити $AB=(L+x)/2$ равна
$$T=\frac{M}{4L}\,(L+x)\,V^2.$$
Когда очередная порция нити длиной $d(AB)=dx/2$ присоединяется к движущейся части
$AB$, выделяется тепло (см. решение задачи о падении капли в тумане)
$$dQ=\frac{1}{2}\,\frac{M}{L}\,d(AB)\,V^2=\frac{M}{4L}\,V^2\,dx.$$
Поэтому закон сохранения энергии с учетом выделившегося  тепла запишется как
$$Fx=\frac{M}{4L}\,(L+x)\,V^2+\frac{M}{4L}\int\limits_0^xV^2\,dx.$$
Продифференцируем это уравнение по $x$:
$$F=\frac{M}{2L}\left[\frac{1}{2}V^2+\frac{L+x}{2}\,\frac{dV^2}{dx}+
\frac{1}{2}V^2\right]=\frac{M}{2L}\left[V^2+\frac{L+x}{2}\,\frac{dV^2}{dx}
\right ].$$
Но $$\frac{dV^2}{dx}=\frac{dV^2}{dt}\,\frac{1}{dx/dt}=2\,\frac{dV}{dt},$$
и получаем уравнение
$$F=\frac{M}{2L}\left[V^2+(L+x)\frac{dV}{dt}\right]=
\frac{M}{2L}\,\frac{d}{dt}[(L+x)V],$$
которое совпадает с уравнением (\ref{eq23_2}).

\subsection*{Задача 3 {\usefont{T2A}{cmr}{b}{n}(7.4)}: Время столкновения сильно
накачанного мяча со стенкой}
Пусть деформация мяча равна $x$ (см. рисунок). Тогда площадь соприкосновения мяча со
стенкой будет $S=\pi r^2=\pi [R^2-(R-x)^2]\approx 2\pi R x$. Следовательно, сила,
которая тормозит мяч, равна $F=PS=2\pi R P x$.
\begin{figure}[htb]
\centerline{\epsfig{figure=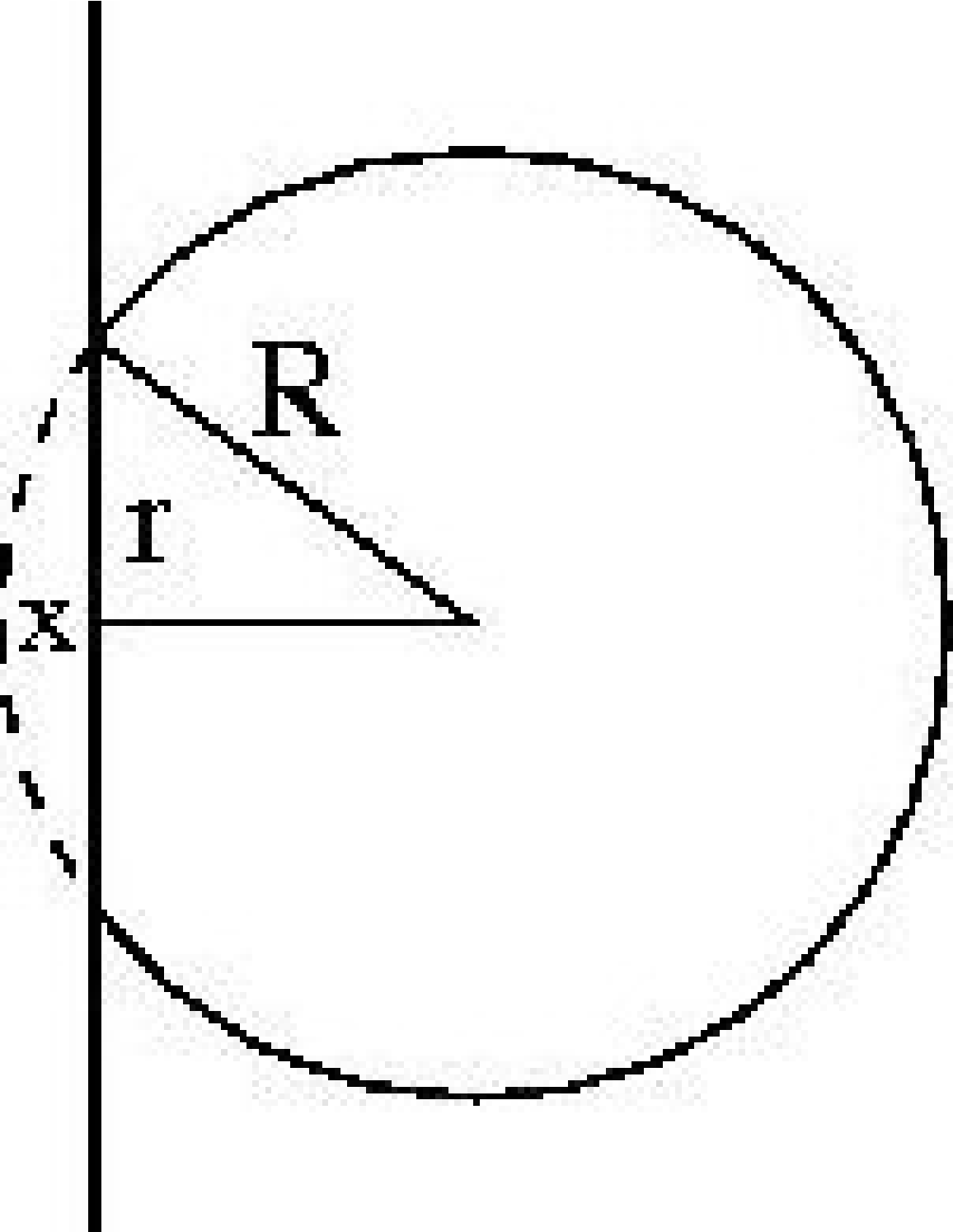,height=3cm}}
\end{figure}

Уравнение движения мяча $m\ddot x=-2\pi R P x$ есть уравнение гармонических колебаний
с частотой $$\omega=\sqrt{\frac{2\pi R P}{m}}.$$ Это соответствует периоду 
колебаний $$T=2\pi\sqrt{\frac{m}{2\pi R P}}=\sqrt{\frac{2\pi m}{R P}}.$$
Время столкновения
$$\tau=\frac{T}{2}=\sqrt{\frac{\pi m}{2RP}}\approx 10^{-2}~\mbox{с}.$$

\subsection*{Задача 4 {\usefont{T2A}{cmr}{b}{n}(7.11)}: Период колебаний грузика}
Пусть грузик отклонили на угол $\alpha$ от вертикали справа, тогда как стенка 
отклонена на угол $\phi$ слева. Если $\alpha<\phi$, наличие стенки никак не повляет 
на движение грузика, и он будет колебаться с частотой $\omega=\sqrt{g/L}$. Если же 
$\alpha>\phi$, то в момент времени $t_1$, которое определяется из условия 
$\alpha\cos{(\omega t_1)}=-\phi$, грузик столкнется со стенкой. Если бы
стенки не было, грузик бы продолжал движение слева до остановки в момент времени 
$t=T/2=\pi/\omega$. Потом, двигаясь уже справа, в момент времени $t_2=T-t_1$ снова 
бы пересек воображаемую стенку. Упругое столкновение со стенкой приводит к тому, что 
от полного периода свободного грузика $T=2\pi\sqrt{\frac{L}{g}}$ вырезается период
времени $\Delta t=t_2-t_1=T-2t_1$ между $t_1$ и $t_2$. Поэтому искомый новый 
период колебаний грузика будет $T^\prime=T-\Delta t=2t_1$. Но
$$t_1=\frac{1}{\omega}\arccos{\left(-\frac{\phi}{\alpha}\right)}=
\frac{1}{\omega}\left [\frac{\pi}{2}+\arcsin{\left(\frac{\phi}{\alpha}
\right)}\right ].$$
Поэтому, окончательно,
$$T^\prime=\frac{2}{\omega}\left [\frac{\pi}{2}+\arcsin{\left(\frac{\phi}
{\alpha}\right)}\right ]=\frac{\pi}{\omega}\left [1+\frac{2}{\pi}\,
\arcsin{\left(\frac{\phi}{\alpha}\right)}\right ].$$

Если удар неупругий, после удара об стенку грузик останавливается, а потом колебается
с периодом $T$ и амплитудой $\phi$.

\subsection*{Задача 5 {\usefont{T2A}{cmr}{b}{n}(7.13)}: Частота малых
колебаний бусинки}
{\usefont{T2A}{cmr}{m}{it} Первое решение.}
Бусинка движется по эллипсу (сумма расстояний от бусинки до точек крепления концов 
нити постоянна и равна длине нити $L$). Уравнение эллипса
$$\frac{x^2}{a^2}+\frac{y^2}{b^2}=1.$$
С другой стороны, фокусы эллипса (точки крепления концов нити) имеют координаты
$(-d/2,0)$ и $(d/2,0)$. Поэтому
$$\sqrt{\left(x+\frac{d}{2}\right)^2+y^2}+\sqrt{\left(x-\frac{d}{2}\right)^2+
y^2}=L.$$
Подставляя сюда верхнюю точку эллипса $x=0,y=b$, а потом крайне правую точку
$x=a,y=0$, получаем уравнения $2\sqrt{b^2+d^2/4}=L$ и $2a=L$. Следовательно,
$$a=\frac{L}{2},\;\;\;b=\frac{1}{2}\sqrt{L^2-d^2}.$$
Около нижней точки равновесия
$$y=-b\sqrt{1-\frac{x^2}{a^2}}\approx -b\left(1-\frac{x^2}{2a^2}\right).$$
Поэтому при малых колебаниях $\dot y =bx\dot x/a^2\ll \dot x$ и полная энергия
бусинки $$E\approx \frac{m}{2}\dot x^2+mgy\approx \frac{m}{2}\dot x^2-
mgb\left(1-\frac{x^2}{2a^2}\right)$$
сохраняется. Поэтому
$$\frac{dE}{dt}=m\dot x\,\ddot x+\frac{mgb}{a^2}\dot x\,x=0,$$
что дает уравнение гармонических колебаний $\ddot x+\omega^2\,x=0$ с частотой
$$\omega=\sqrt{\frac{gb}{a^2}}=\sqrt{2\,\frac{g}{L}\,\sqrt{1-\frac{d^2}
{L^2}}}.$$

{\usefont{T2A}{cmr}{m}{it} Второе решение.}
Частота малых колебаний $$\omega=\sqrt{\frac{g}{R}},$$
где $R$ есть радиус кривизны эллипса в нижней точке. Около этой точки
$$y=-b\sqrt{1-\frac{x^2}{a^2}}\approx -b\left(1-\frac{x^2}{2a^2}\right).$$
С другой стороны, с точностью до квадратичных членов, эти точки эллипса должны
лежать на окружности $x^2+(y-R+b)^2=R^2$, так как центр соприкасающейся окружности
имеет координаты $(x=0,y=R-b)$. Поэтому
$$x^2+\left(\frac{bx^2}{2a^2}-R\right)^2=R^2.$$
Оставляя только члены не выше квадратичных по $x$, отсюда получаем $R=a^2/b$ и, 
следовательно, $$\omega=\sqrt{\frac{gb}{a^2}}=\sqrt{2\,\frac{g}{L}\,
\sqrt{1-\frac{d^2}{L^2}}}.$$
 
\subsection*{Задача 6 {\usefont{T2A}{cmr}{b}{n}(7.14)}: Период колебаний точки
в циклоидальной чашке}
{\usefont{T2A}{cmr}{m}{it} Первое решение.}
Если $x=R(\phi+\sin{\phi})$ и $y=R(1-\cos{\phi})$, то $\dot x=R\dot\phi+
R\dot\phi\,\cos{\phi}$ и $\dot y=R\dot\phi\,\sin{\phi}$. Поэтому кинетическая
энергия равна
$$T=\frac{m}{2}(\dot x^2+\dot y^2)=mR^2\dot\phi^2(1+\cos{\phi}).$$
Учитывая выражение для потенциальной энергии $U=mgy=mgR(1-\cos{\phi})$, для полной
энергии получаем 
$$E=mR^2\dot\phi^2(1+\cos{\phi})+mgR(1-\cos{\phi})=2mR^2\dot\phi^2\cos^2{
\frac{\phi}{2}}+2mgR\sin^2{\frac{\phi}{2}}.$$
Введем новую переменную $q=\sin{(\phi/2)}$. Тогда $\dot q =(\dot\phi/2)
\cos{(\phi/2)}$ и выражение для энергии принимает вид
$E=8mR^2\dot q^2+2mgRq^2$. Полная энергия сохраняется, поэтому
$$\dot E=16MR^2\dot q\,\ddot q+4mgR\dot q\,q=0.$$
Сокращая на $16MR^2\dot q$, получаем уравнение гармонических колебаний с частотой
$$\omega=\sqrt{\frac{g}{4R}}.$$

{\usefont{T2A}{cmr}{m}{it} Второе решение.}
Этот способ решения подходит только для нахождения периода малых колебаний. Для таких 
колебаний $$\omega=\sqrt{\frac{g}{\rho}},$$
где $\rho$ есть радиус кривизны циклоидальной чашки в нижней точке $\phi=0$. Этот 
радиус кривизны можно найти так. Если $\phi$ --- маленькая величина, то с точностью
до квадратичных членов $x\approx 2R\phi$ и $y=2R\sin^2{(\phi/2)}\approx
R\phi^2/2$. С другой стороны, с этой точностью точка $(x,y)$ должна лежать на
окружности $x^2+(y-\rho)^2=\rho^2$. Подставляя сюда выражения для $x$ и $y$ и 
оставляя только члены не выше квадратичной, получаем $4R^2\phi^2-R\rho\,\phi^2=0$.
Следовательно, $\rho=4R$. 

\section[Семинар 24]
{\centerline{Семинар 24}}

\subsection*{Задача 1 {\usefont{T2A}{cmr}{b}{n}(7.18)}:  Движение стрелки
амперметра после разрыва цепи}
Пусть $\omega$ --- частота собственных колебании стрелки. Уравнение движения
имеет вид
$$I\ddot \varphi=-k\varphi-M\,\frac{\dot \varphi}{|\dot \varphi|},$$
где $M$ --- момент сил сухого трения, который направлен против угловой скорости 
(заметим, что $\frac{\dot \varphi}{|\dot \varphi|}=1$, если $\dot \varphi>0$
и $\frac{\dot \varphi}{|\dot \varphi|}=-1$, если $\dot \varphi<0$). 
Следовательно,
$$\ddot \varphi+\omega^2\,\varphi=-\frac{M}{I}\,\frac{\dot \varphi}{|\dot 
\varphi|}.$$
Обозначим $$\frac{M}{I}=\omega^2\,\varphi_1=\frac{k}{I}\,\varphi_1.$$
Т.~е. если $\varphi=\varphi_1$, то возникающий момент пружинного механизма стрелки
$k\,\varphi_1$ в точности равен моменту сил сухого трения $M$. Поэтому если 
$\varphi<\varphi_1$, стрелка не будет поворачиваться, так как трение сможет 
уравновесить возвращающий момент $k\,\varphi$ (при этом момент сил трения будет
меньше его максимального значения $M$). Следовательно, имеем следующее уравнение
движения стрелки:
$$\ddot \varphi+\omega^2\,\varphi=\left \{ \begin{array}{c} -\omega^2\,
\varphi_1,\;\;\mbox{если}\;\;\dot \varphi>0, \\ \quad\omega^2\,\varphi_1,
\;\;\mbox{если}\;\;\dot \varphi<0. \end{array} \right . $$
Сразу после разрыва цепи $\varphi$ уменьшается, т.~e. $\dot \varphi<0$ и 
уравнение движения есть $\ddot \varphi+\omega^2\,\varphi=\omega^2\,\varphi_1$.
Общее решение этого уравнения имеет вид 
$$\varphi(t)=C_1e^{i\omega t}+C_2e^{-i\omega t}+\varphi_1.$$
Начальные условия $\varphi(0)=C_1+C_2+\varphi_1=\varphi_0$ и
$\dot\varphi(0)=i\omega(C_1-C_2)=0$ определяют константы $C_1$ и $C_2$:
$$C_1=C_2=\frac{\varphi_0-\varphi_1}{2},$$ и, следовательно,
$\varphi(t)=(\varphi_0-\varphi_1)\,\cos{\omega\,t}+\varphi_1$.
Тогда $\dot \varphi(t)=-\omega\,(\varphi_0-\varphi_1)\,\sin{\omega\,t}$ и
фазовая траектория есть половина эллипса (см. рисунок) с центром в точке $\varphi=
\varphi_1$:
$$\frac{(\varphi-\varphi_1)^2}{(\varphi_0-\varphi_1)^2}+\frac{\dot \varphi^2}
{\omega^2(\varphi_0-\varphi_1)^2}=1.$$

\begin{figure}[htb]
\centerline{\epsfig{figure=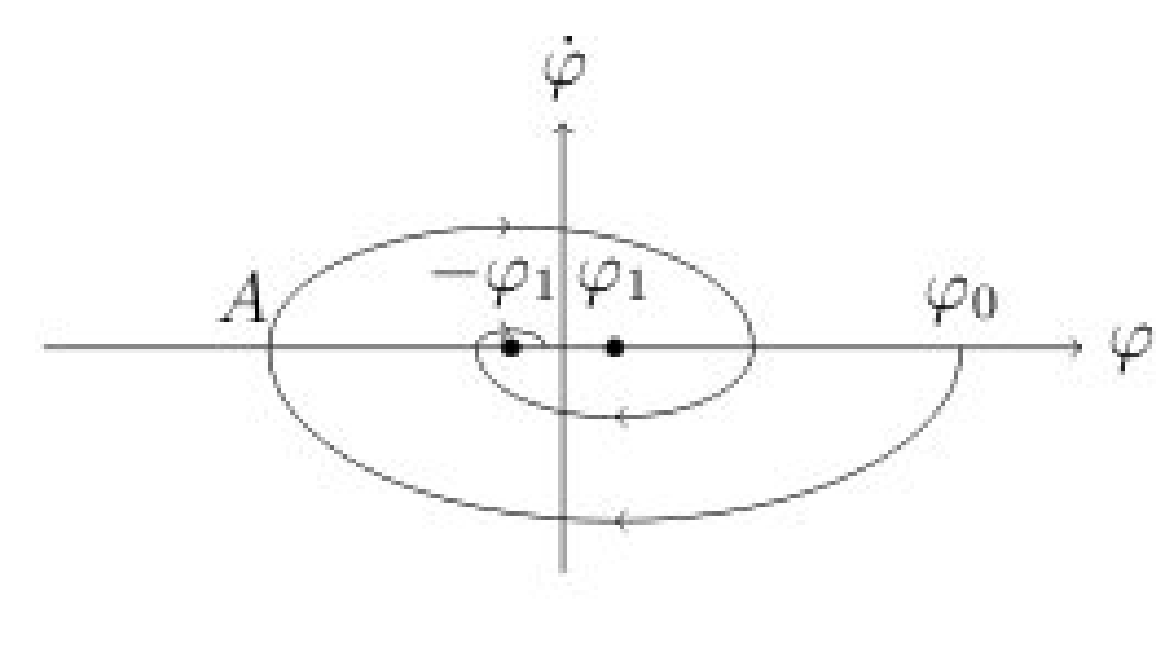,height=4cm}}
\end{figure}

В точке $A$ стрелка остановится и пойдет в другую сторону. При этом $\dot \varphi>0$ 
и уравнение движения будет $\ddot \varphi+\omega^2\,\varphi=\omega^2\,\varphi_1$. 
Оно отличается от предыдущего только знаком $\varphi_1$. Поэтому соответствующая
фазовая траектория тоже будет половина эллипса, только с центром в точке $\varphi=
-\varphi_1$ и с полуосями $\varphi_0-3\varphi_1$ и $\omega^2(\varphi_0-
3\varphi_1)$. Дальше все повторится по аналогичной схеме: следующая половинка эллипса
будет иметь центр в  в точке $\varphi=\varphi_1$ и полуоси $\varphi_0-5\varphi_1$ 
и $\omega^2(\varphi_0-5\varphi_1)$, и так далее. Когда стрелка попадет между
точками $\varphi=-\varphi_1$ и $\varphi=\varphi_1$, она остановится и колебания
прекратятся.

\subsection*{Задача 2 {\usefont{T2A}{cmr}{b}{n}(7.20)}: Оптимальный режим
демпфирования колебаний}
Пусть напряжение на конденсаторе $U$, а ток в цепи $I$. В катушке возникает Э.Д.С. 
самоиндукции ${\cal E}=-L\frac{dI}{dt}$. По закону Кирхгофа
\begin{equation}
U+IR={\cal E}=-L\frac{dI}{dt}.
\label{eq24_2a}
\end{equation}
Получим это уравнение по-другому. Энергия системы
$$E=\frac{CU^2}{2}+\frac{LI^2}{2}$$
уменьшается из-за выделения тепла в сопротивлении $R$. Поэтому $\Delta E=-I^2R\,
\Delta t$, или $$CU\Delta U+LI\,\Delta I=-I^2R\,\Delta t.$$ Но $C\Delta U=
\Delta Q=I\,\Delta t$. Поэтому получаем уравнение $IU\Delta t+LI\Delta I=
-I^2R\,\Delta t$, эквивалентное к уравнению (\ref{eq24_2a}). Используя $I=\dot Q$
и $U=\frac{Q}{C}$, из (\ref{eq24_2a}) получаем 
\begin{equation}
\ddot{Q}+\frac{R}{L}\,\dot Q+\frac{1}{LC}\,Q=0.
\label{eq24_2b}
\end{equation}
Характеристическое уравнение 
$$\lambda^2+\frac{R}{L}\,\lambda+\frac{1}{LC}=0$$
имеет корни
$$\lambda=-\frac{R}{2L}\pm\sqrt{\frac{R^2}{4L^2}-\frac{1}{LC}}.$$
Если $$\frac{R^2}{4L^2}<\frac{1}{LC}\;\;\left(\mbox{т.~е.}\;\;R^2<4\frac{L}{C}
\right),$$
будем иметь затухающие колебания (при начальных значениях $Q(0)=Q_0,\,\dot Q(0)=0$):
$$Q(t)=Q_0e^{-\beta t}\left [\cos{\omega t}+\frac{\beta}{\omega}\,
\sin{\omega t}\right],$$
где
$$\beta=\frac{R}{2L}\;\;\mbox{и}\;\;\omega=\sqrt{\frac{1}{LC}-\frac{R^2}
{4L^2}}.$$
Такой режим не выгоден. Поэтому возьмем 
$$\frac{R^2}{4L^2}\ge\frac{1}{LC}\;\;\left(\mbox{т.~е.}\;\;R^2\ge 4\frac{L}{C}
\right)).$$
Тогда будем иметь чистое затухание
$$Q(t)=e^{-\beta t}(C_1e^{\gamma t}+C_2e^{-\gamma t}),$$
где $$\gamma=\sqrt{\frac{R^2}{4L^2}-\frac{1}{LC}},$$
а константы $C_1$ и $C_2$ определяются из начальных условий $Q(0)=Q_0=C_1+C_2$ и 
$\dot Q(0)=0=-\beta(C_1+C_2)+\gamma(C_1-C_2)$, что дает
$$C_1=\frac{Q_0}{2}\left (1+\frac{\beta}{\gamma}\right),\;\;
C_2=\frac{Q_0}{2}\left (1-\frac{\beta}{\gamma}\right).$$
Следовательно, 
\begin{equation}
Q(t)=Q_0e^{-\beta t}\left [\ch{\gamma t}+\frac{\beta}{\gamma}\,
\sh{\gamma t}\right ].
\label{eq24_2c}
\end{equation}
В пределе $t\to\infty$ множитель
$$\ch{\gamma t}+\frac{\beta}{\gamma}\,\sh{\gamma t}\to
\frac{1}{2}\left(1+\frac{\beta}{\gamma}\right)\,e^{\gamma t}$$
мешает быстрому затуханию, если $\gamma\ne 0$. Поэтому наиболее оптимальный режим
затухания получается при $$R^2=4\,\frac{L}{C},$$
когда $\gamma=0$. Вычисляя предел $\gamma\to 0$ выражения (\ref{eq24_2c}), 
получаем, что в этом случае $$Q(t)=Q_0e^{-\beta t}(1+\beta\,t).$$

\subsection*{Задача 3 {\usefont{T2A}{cmr}{b}{n}(7.22)}: Добротность и затухание
колебаний стрелки амперметра}
Если уравнение движения стрелки амперметра имеет вид $\ddot x+2\beta\dot x+
\omega_0^2x=0$, то добротность определяется как отношение $$Q=\frac{\pi}{\beta T}=
\frac{\omega}{2\beta},$$
где $$\omega=\frac{2\pi}{T}=\sqrt{\omega_0^2-\beta^2}$$
есть частота колебаний. Отсюда $4Q^2\beta^2=\omega^2=\omega_0^2-\beta^2$ и
$$\beta=\frac{\omega_0}{\sqrt{1+4Q^2}}.$$
Так как $Q\gg 1$, то $\beta/\omega\ll 1$ и колебания стрелки амперметра, при
начальных условиях $x(0)=A,\,\dot x(0)=0$, имеют вид
$$x(t)=Ae^{-\beta t}\left [\cos{\omega t}+\frac{\beta}{\omega}\,\sin{\omega t}
\right ]\approx Ae^{-\beta t}\cos{\omega t}.$$
Амплитуда уменьшится в 20 раз за время $t$ такое, что $e^{-\beta t}=1/20$. Поэтому
$$t=\frac{\ln{20}}{\beta}\approx \frac{3}{\beta}=\frac{3}{\omega_0}
\sqrt{1+4Q^2}\approx \frac{6Q}{\omega_0}\approx 19~\mbox{с}.$$

\subsection*{Задача 4 {\usefont{T2A}{cmr}{b}{n}(7.34)}: Амплитуда
установившихся колебаний при осциллирующей точке подвеса}
Пусть $l+x$ --- координата шарика, $z=a\cos{\Omega t}$ --- координата точки подвеса
пружинки. Здесь $l$ --- длина нерастянутой пружины. Растяжение пружины равно
$\Delta l=x-z$. Поэтому на шарик действует сила упругости $F_x=-k\Delta l=-k(x-z)$
и уравнение движения будет $m\ddot x=-k(x-a\cos{\Omega t})$, или
$$\ddot x+\omega^2 x=a\omega^2\cos{\Omega t},$$
где $\omega^2=k/m$. Вынужденные колебания будут иметь вид $x=A\cos{\Omega t}$.
После подстановки в уравнение движения для амплитуды колебаний $A$ получаем уравнение
$A(\omega^2-\Omega^2)=a\omega^2$. Следовательно,
$$A=\frac{a\omega^2}{\omega^2-\Omega^2}.$$

\subsection*{Задача 5 {\usefont{T2A}{cmr}{b}{n}(7.29)}: Амплитуда
колебаний после воздействия внешней силы}
Будем использовать формулу \cite{31}:
\begin{equation}
z(t)=e^{i\omega t}\left [z_0+\int\limits_0^t\frac{F(\tau)}{m}\,
e^{-i\omega \tau}d\tau \right ],
\label{eq24_5}
\end{equation}
где $z(t)=\dot x(t)+i\omega\,x(t)$ и $z_0=z(0)$. В нашем случае $z_0=0$ и 
$F(t)=0$, если $t>T/2$. Поэтому
$$z(t)=\frac{F_0}{m}\,e^{i\omega t}\int\limits_0^{T/2}\sin{\omega\tau}\,
e^{-i\omega \tau}d\tau.$$ Но
$$\int\limits_0^{T/2}\sin{\omega\tau}\,e^{-i\omega \tau}d\tau=\frac{1}{2i}
\int\limits_0^{T/2} (1-e^{-2i\omega\tau})d\tau=\frac{T}{4i},$$
так как $\omega T/2=\pi$ и легко увидеть, что интеграл от второго члена равен нулю.
Следовательно,
$$z(t)=\frac{F_0T}{4im}\,e^{i\omega t}=\frac{F_0T}{4im}(\cos{\omega t}+
i\sin{\omega t})=\frac{F_0T}{4m}\sin{\omega t}-i\,\frac{F_0T}{4m}
\cos{\omega t}.$$
Это означает, что $$x=-\frac{F_0T}{4m}\cos{\omega t}.$$
Амплитуда колебаний равна $$A=\frac{F_0T}{4m}=\frac{2\pi F_0}{4m\omega^2}=
\frac{\pi}{2}\,\frac{F_0}{k}.$$
Из приведенного решения видно, что если внешняя сила действовала в течение $N$ 
полупериодов синусоиды, амплитуда увеличится $N$ раз.

\subsection*{Задача 6 {\usefont{T2A}{cmr}{b}{n}(7.24)}: Амплитуда
колебаний после действия прямоугольного импульса}
{\usefont{T2A}{cmr}{m}{it} Первое решение.}
По формуле (\ref{eq24_5}), так как $z_0=0$, получаем
$$z(\tau)=e^{i\omega \tau}\int\limits_0^\tau\frac{F}{m}\, e^{-i\omega t} dt=
\frac{F}{im\omega}(e^{i\omega \tau}-1)=\frac{F}{im\omega}(\cos{\omega\tau}
-1+i\sin{\omega\tau}).$$
Энергия колебаний осциллятора (после выключения силы в момент $t=\tau$ она не будет
меняться) равна
$$E=\frac{m}{2}|z|^2=\frac{F^2}{2m\omega^2}[(\cos{\omega\tau}-1)^2+
\sin^2{\omega\tau}]=\frac{F^2}{m\omega^2}(1-\cos{\omega\tau})=\frac{2F^2}
{k}\sin^2{\frac{\omega\tau}{2}},$$
где мы учли, что $m\omega^2=k$. С другой стороны, $E=kA^2/2$, где $A$ есть амплитуда
колебаний. Следовательно,
$$\frac{kA^2}{2}=\frac{2F^2}{k}\sin^2{\frac{\omega\tau}{2}}\;\;\mbox{и}\;\;
A=\frac{2F}{k}\left|\sin{\frac{\omega\tau}{2}}\right|.$$

{\usefont{T2A}{cmr}{m}{it} Второе решение.}
При $t\le\tau$ уравнение движения имеет вид ($x$ есть растяжение пружины) 
$m\ddot x=-kx+F+mg$, или
$$\ddot x+\omega^2x=\frac{F+mg}{m},\;\;\mbox{где}\;\;\omega^2=\frac{k}{m}.$$
Его общее решение имеет вид
$$x(t)=\frac{F+mg}{m\omega^2}+C_1\cos{\omega t}+C_2\sin{\omega t}.$$
Условие $\dot x(0)=0$ означает $C_2=0$. В начале груз был в равновесии, т.~е.
$x(0)=mg/k$. Поэтому должны иметь $$\frac{F+mg}{k}+C_1=\frac{mg}{k}.$$
Это определяет $C_1=-F/k$. Следовательно, при $t\le\tau$ закон движения груза есть
\begin{equation}
x(t)=\frac{F}{k}(1-\cos{\omega t})+\frac{mg}{k}.
\label{eq24_6a}
\end{equation} 
При $t>\tau$ груз колебается около положения с амплитудой $A$:
\begin{equation}
x(t)=\frac{mg}{k}+A\sin{[\omega(t-\tau)+\alpha]}.
\label{eq24_6b}
\end{equation}
Тогда 
$$x(\tau)=\frac{mg}{k}+A\sin{\alpha}\;\;\mbox{и}\;\;
\dot x(\tau)=A\omega\cos{\alpha}.$$
Поэтому
\begin{equation}
A^2=\left(x(\tau)-\frac{mg}{k}\right)^2+\frac{(\dot x(\tau))^2}{\omega^2}.
\label{eq24_6c}
\end{equation}
С другой стороны, при $t=\tau$ значения $x(\tau)$ и $\dot x(\tau)$, вычисленные по
формулам (\ref{eq24_6a}) и (\ref{eq24_6b}), должны совпадать. Поэтому, согласно
(\ref{eq24_6a}), будем иметь
$$x(\tau)=\frac{F}{k}(1-\cos{\omega \tau})+\frac{mg}{k},\;\;
\dot x(\tau)=\frac{F\omega}{k}\sin{\omega\tau}. $$
Подставляя эти значения в (\ref{eq24_6c}), получим
$$A^2=\frac{F^2}{k^2}\left[ (1-\cos{\omega \tau})^2+\sin^2{\omega\tau}\right ]
=\frac{2F^2}{k^2}(1-\cos{\omega \tau})=\frac{4F^2}{k^2}\sin^2{\frac{\omega
\tau}{2}}.$$
Следовательно, $$A=\frac{2F}{k}\left|\sin{\frac{\omega\tau}{2}}\right|.$$

\section[Семинар 25]
{\centerline{Семинар 25}}

\subsection*{Задача 1 {\usefont{T2A}{cmr}{b}{n}(6.43)}: Обезьянка на веревке}
Пусть длина свешивающейся части веревки равна $x$. На эту часть веревки действуют 
следующие силы: сила тяжести $\frac{m}{L}xg$, сила натяжения веревки $T$ в точке
$A$ (см. рисунок) и сила $F$ от обезьянки, которая равна $Mg$, так как обезьянка
неподвижна.

Уравнение движения для свешивающейся части веревки $AB$ имеет вид
\begin{equation}
\frac{d}{dt}\left (\frac{m}{L}x\dot x\right )=\frac{m}{L}xg+Mg-T+\pi_x,
\label{eq25_1a}
\end{equation}
где $\pi_x$ есть поток импульса в свешивающийся части веревки. Так как за время
$\Delta t$ к свешивающейся части веревки добавляется кусочек длиной $\Delta x=\dot x
\Delta t$, который уже имел скорость $\dot x$, то 
$$\pi_x=\frac{\frac{m}{L}\Delta x \dot x}{\Delta t}=\frac{m}{L}\dot x^2.$$
Поэтому уравнение движения (\ref{eq25_1a}) принимает вид
\begin{equation}
\frac{m}{L}x\ddot x=\frac{m}{L}xg+Mg-T.
\label{eq25_1b}
\end{equation}
\begin{figure}[htb]
\centerline{\epsfig{figure=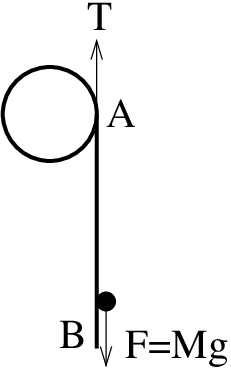,height=4cm}}
\end{figure}

\noindent Вращение блока ускоряется силой натяжения веревки, поэтому 
$I\ddot \varphi=TR$.
Но момент инерции $I$ создается той частью веревки, которая на блоке. Поэтому
$$I=\frac{m}{L}(L-x)R^2,$$
и для силы натяжения веревки получаем
$$T=\frac{m}{L}(L-x)R\ddot \varphi.$$
Веревка не проскальзывает. Следовательно, $\dot x=R\dot \varphi$ и 
$\ddot x=R\ddot \varphi$. Подставляя
$$T=\frac{m}{L}(L-x)\ddot x$$
в (\ref{eq25_1b}), получаем
$$x\ddot x=xg-(L-x)\ddot x+\frac{ML}{m}g,$$
или
$$\ddot x-\frac{g}{L}=\frac{M}{m}g.$$
Решение этого уравнения, которое соответствует начальным условиям $x(0)=l_0$,
$\dot x(0)=0$, есть
$$x(t)=A\,\ch{\left(\sqrt{\frac{g}{L}}\,t\right)}-\frac{ML}{m},$$
где константа $A$ определяется из условия
$$A-\frac{ML}{m}=l_0.$$
Скорость веревки (и, следовательно, скорость обезянки относительно веревки) равна
$$V(t)=\dot x(t)=\sqrt{\frac{g}{L}}\left(l_0+\frac{ML}{m}\right)\,
\sh{\left(\sqrt{\frac{g}{L}}\,t\right)}=\sqrt{gL}\left(\frac{l_0}{L}+
\frac{M}{m}\right)\,\sh{\left(\sqrt{\frac{g}{L}}\,t\right)}.$$

\subsection*{Задача 2 {\usefont{T2A}{cmr}{b}{n}(7.35)}: Амплитуда установившихся
колебаний маятника при горизонтальных осцилляциях точки подвеса}
{\usefont{T2A}{cmr}{m}{it} Первое решение.}
\begin{figure}[htb]
\centerline{\epsfig{figure=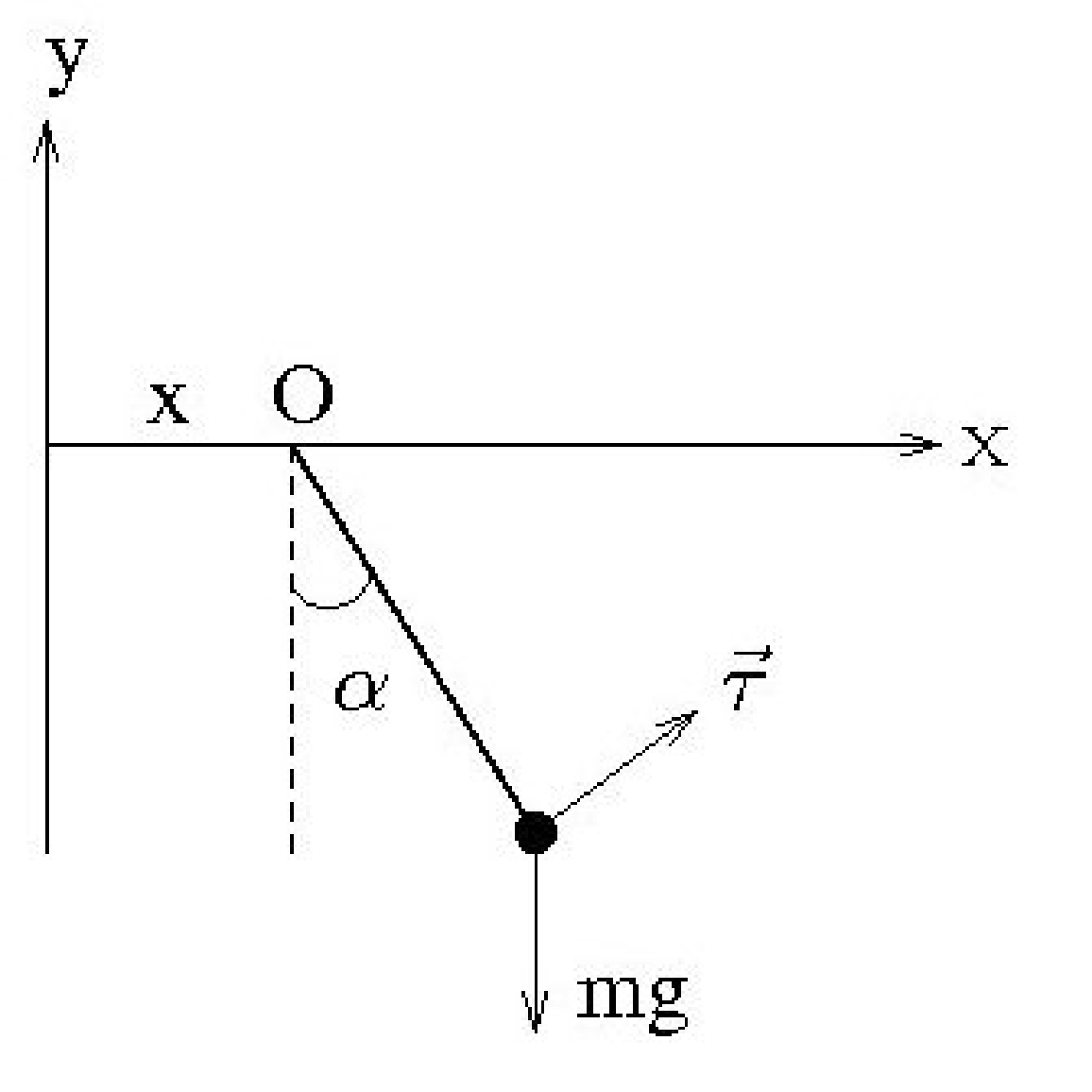,height=5cm}}
\end{figure}
Скорость грузика $\vec{V}$ относительно лабораторной системы складывается из его 
скорости $l\dot \alpha \vec{\tau}$ относительно точки подвеса $O$ и скорости 
$\dot x\,\vec{i}$ точки $O$ относительно лабораторной системы:
$$\vec{V}=l\dot \alpha \vec{\tau}+\dot x\,\vec{i},$$
здесь $\tau$ есть единичный вектор в перпендикулярном к нити направлении (см. 
рисунок).

Следовательно, ускорение грузика будет
$$\vec{a}=\ddot x\,\vec{i}+l\ddot \alpha \vec{\tau}+l\dot \alpha \dot{ 
\vec{\tau}}.$$
Так как $\tau$ единичный вектор, $$\frac{d\vec{\tau}^2}{dt}=2\vec{\tau}\cdot
\dot{\vec{\tau}}=0.$$
Следовательно, $\dot{\vec{\tau}}$ и $\vec{\tau}$ --- взаимно перпендекулярные 
вектора, и, проектируя $m\vec{a}=\vec{F}$ на направление $\vec{\tau}$, получаем
$$m(\ddot x \cos{\alpha}+l\ddot \alpha)\approx m(\ddot x +l\ddot \alpha)=
\vec{F}\cdot\vec{\tau}=-mg\sin{\alpha}\approx -mg\alpha.$$
Таким образом, имеем следующее уравнение движения
\begin{equation}
\ddot \alpha+\frac{g}{l}\,\alpha=-\frac{\ddot x}{l}=\frac{b}{l}\,\Omega^2\,
\cos{\Omega t}.
\label{eq25_2}
\end{equation}
Вынужденные колебания ищем в виде $\alpha(t)=A\,\cos{\Omega t}$. Подставляя в 
(\ref{eq25_2}), получаем
$$A\left(-\Omega^2+\frac{g}{l}\right )\cos{\Omega t}=\frac{b\Omega^2}{l}
\cos{\Omega t}.$$
Следовательно,
$$A=\frac{b}{l}\,\frac{\Omega^2}{\omega^2-\Omega^2},$$
где $\omega^2=g/l$.

{\usefont{T2A}{cmr}{m}{it} Второе решение.}
Координаты грузика равны $$x=b\cos{\Omega t}+l\sin{\alpha}\approx
b\cos{\Omega t}+l\alpha,\;y=-l\cos{\alpha}.$$ 
Поэтому
$$\dot x\approx -b\Omega\sin{\Omega t}+l\dot\alpha,\;\;
\ddot x\approx -b\Omega^2\cos{\Omega t}+l\ddot\alpha,\;\;
\dot y=l\dot\alpha\sin{\alpha}\ll \dot x.$$
Следовательно, энергия грузика
$$E\approx  \frac{m}{2}(-b\Omega\sin{\Omega t}+l\dot\alpha)^2-mgl\cos{\alpha}
\approx  \frac{m}{2}(-b\Omega\sin{\Omega t}+l\dot\alpha)^2-mgl\left ( 1-
\frac{\alpha^2}{2}\right).$$
Энергия не сохраняется, так как на точку подвеса $O$ действует внешняя сила 
$F=m\ddot x$, которая в единицу времени совершает работу 
$$N=FV_O=F\frac{d(b\cos{\Omega t})}{dt}=
m(l\ddot\alpha-b\Omega^2\cos{\Omega t})(-b\Omega\sin{\Omega t}).$$
Поэтому $\dot E=N$, что дает
$$m(l\dot\alpha-b\Omega\sin{\Omega t})(l\ddot\alpha-b\Omega^2\cos{\Omega t})+
mgl\alpha\dot\alpha=m(l\ddot\alpha-b\Omega^2\cos{\Omega t})(-b\Omega
\sin{\Omega t}).$$
После сокращений (и деления на $\dot\alpha$), это уравнение приводится к виду 
(\ref{eq25_2}).

\subsection*{Задача 3 {\usefont{T2A}{cmr}{b}{n}(7.36)}: Резонанс при движении
поезда}
Заменим все рессоры одной эффективной пружиной с жесткостью $k$. Тогда $k\Delta x=mg$,
где $\Delta x$ --- прогиб, $m$ --- масса вагона. Частота свободных колебаний вагона
будет $$\omega=\sqrt{\frac{k}{m}}=\sqrt{\frac{g}{\Delta x}}.$$
Амплитуда колебаний максимальна при резонансе, когда время между прохождения стыков
$t=L/V$ станет равным периоду свободных колебаний $T=2\pi/\omega$. Из
$$\frac{L}{V}=2\pi\sqrt{\frac{\Delta x}{g}}$$
получаем
$$V=\frac{L}{2\pi}\sqrt{\frac{g}{\Delta x}}\approx 62,9~\mbox{м}/\mbox{с}
\approx 226~\mbox{км}/\mbox{ч}.$$

\subsection*{Задача 4 {\usefont{T2A}{cmr}{b}{n}(7.30)}: Амплитуда колебаний при
резонансе}
Пусть показания амперметра $I$. В амперметре магнито-электрической системы крутящий 
момент стрелки создаётся благодаря взаимодействию между полем постоянного магнита и 
током, который проходит через обмотку рамки и, следовательно, этот вращающий момент
пропорционален току $i(t)$, проходящему через обмотку рамки. Поэтому уравнение 
колебательных движений стрелки амперметра будет иметь вид $\ddot I+\omega_0^2I=
C\,i(t)$, где $C$ --- некоторая константа. Но если ток $i(t)=i_0$ не меняется, то
показания амперметра $I$ должен совпасть с $i_0$. Отсюда $C=\omega_0^2$ и
окончательно будем иметь уравнение движения
$$\ddot I+\omega_0^2I=\omega_0^2\, i(t)=\omega_0^2\, i_0\,\sin{\omega t}.$$
Решение ищем в виде $I=I_0\,\sin{\omega t}$. Подставляя это в уравнение движения,
мы находим $I_0(\omega_0^2-\omega^2)\sin{\omega t}=\omega_0^2\, i_0\,
\sin{\omega t}$. Следовательно, амплитуда вынужденных колебаний
$$I_0=i_0\,\frac{\omega_0^2}{\omega_0^2-\omega^2}=i_0\,\frac{\nu_0^2}
{\nu_0^2-\nu^2}.$$
Ее величина, при условиях задачи, равна $|I_0|=(1/99)~A\approx 10~mA$.

При резонансе надо учесть затухание, и уравнение движения будет
$$\ddot I+2\beta\dot I+\omega_0^2I=\omega_0^2\, i_0\,\sin{\omega t}.$$
Удобно ввести комплексную переменную $z$ такую, что $I=Re\,z$ и которая удовлетворяет
уравнению, из которого вытекает уравнение движения для $I$:
$$\ddot z+2\beta\,\dot z+\omega_0^2\,z=-i\,\omega_0^2\,i_0\,e^{i\omega t}.$$
Ищем решение в виде $z=z_0e^{i\omega t}$ и получаем
$$z_0(-\omega^2+2\beta\,i\omega+\omega_0^2)e^{i\omega t}=-i\,\omega_0^2\,
i_0\,e^{i\omega t},$$
что дает
$$z_0=\frac{-i\,\omega_0^2\,i_0}{\omega_0^2-\omega^2+2i\,\beta\,\omega}.$$
При резонансе $\omega=\omega_0$, и будем иметь $z_0=-\omega_0\,i_0/2\beta$.
Поэтому
$$I(t)=Re\left(-\frac{\omega_0\,i_0}{2\beta}\right )\,e^{i\omega t}=
-\frac{\omega_0\,i_0}{2\beta}\,\cos{\omega t}.$$
Амплитуда колебаний  $I_0=\omega_0\,i_0/2\beta$. Но, как нашли в задаче 7.22,
$$\beta=\frac{\omega_0}{\sqrt{1+4Q^2}}.$$
Следовательно,
$$I_0=\frac{\sqrt{1+4Q^2}}{2}\,i_0\approx Q\,i_0=10~A.$$

\subsection*{Задача 5 {\usefont{T2A}{cmr}{b}{n}(7.26)}: Найти энергию,
приобретенную осциллятором}
{\usefont{T2A}{cmr}{m}{it} Первое решение.}
Энергия осциллятора $E=\frac{m}{2}|z|^2$, где 
$$z=e^{i\omega t}\left [z_0+\int\limits_0^t\frac{F(s)}{m}\,e^{-i\omega s}ds
\right ].$$
Но в нашем случае
$$\int\limits_0^t\frac{F(s)}{m}\,e^{-i\omega s}ds=\frac{F_0}{m}\int\limits_0^t
\left (e^{-i\omega s}-e^{-s(i\omega+1/\tau)}\right )=\frac{F_0}{m}\left [
\frac{1-e^{-i\omega t}}{i\omega}+\frac{-e^{-t(i\omega+1/\tau)}-1}
{i\omega+1/\tau}\right ],$$
что в пределе $t\to\infty$ принимает вид
$$z=\frac{F_0}{m}\left [
\frac{1-e^{-i\omega t}}{i\omega}-\frac{1}{i\omega+1/\tau}\right].$$
Следовательно, когда $t\gg\tau$,
$$z=e^{i\omega t}\left\{z_0+\frac{F_0}{m}\left [
\frac{1-e^{-i\omega t}}{i\omega}-\frac{1}{i\omega+1/\tau}\right]\right\}$$
и
$$E=\frac{m}{2}\left |z_0+\frac{F_0}{m}\left [
\frac{1-e^{-i\omega t}}{i\omega}-\frac{1}{i\omega+1/\tau}\right]\right |^2.$$
Но 
$$\frac{1-e^{-i\omega t}}{i\omega}=-\frac{i}{\omega}\left (1-\cos{\omega t}+
i\,\sin{\omega t}\right)=\frac{1}{\omega}\left [\sin{\omega t}-i(1-
\cos{\omega t})\right ]$$
и
$$\frac{1}{i\omega+1/\tau}=\frac{\tau}{1+i\omega\tau}=\frac{\tau(1-i\omega
\tau)}{1+\omega^2\tau^2}.$$
Поэтому, так как $z_0=V_0+i\omega x_0$, будем иметь
$$E=\frac{m}{2}\left |V_0+\frac{F_0}{m}\left (\frac{\sin{\omega t}}{\omega}
-\frac{\tau}{1+\omega^2\tau^2}\right )+i\left [\omega x_0+\frac{F_0}{m}\left(
\frac{\omega\tau^2}{1+\omega^2\tau^2}-\frac{1-\cos{\omega t}}{\omega}\right )
\right ]\right |^2.$$
Присутствие $\cos{\omega t}$ и $\sin{\omega t}$ членов соответствует тому, при 
$t\gg\tau$ осциллятор колебается около нового положения равновесия, которое 
определяется силой $F_0$, и во время этих колебаний меняет свою энергию за счет 
работы этой внешней силы. Пример подобной ситуации --- колебания груза на пружинке 
в силе тяжести. В среднем работа  внешней силы равна нулю, и средняя энергия 
осциллятора не меняется. Поэтому, чтобы найти, какую энергию приобрел осциллятор, 
$E$ надо усреднить по периоду колебаний. При этом 
$$<\cos{\omega t}>=<\sin{\omega t}>=0,\;\;\;\mbox{и}\;\;\;
<\cos^2{\omega t}>=<\sin^2{\omega t}>=\frac{1}{2}.$$ Поэтому
\begin{eqnarray} &&
<E>=\frac{m}{2}\left[V_0^2-2\frac{V_0F_0}{m}\,\frac{\tau}{1+\omega^2\tau^2}+
\frac{F_0^2}{m^2}\left(\frac{1}{2\omega^2}+\frac{\tau^2}{(1+\omega^2\tau^2)^2}
\right )+\omega^2x_0^2+\right . \nonumber \\ && \left . 
+2\omega x_0\frac{F_0}{m}\left (\frac{\omega\tau^2}{1+\omega^2\tau^2}-
\frac{1}{\omega}\right )+\frac{F_0^2}{m^2}\left (\frac{\omega^2\tau^4}
{(1+\omega^2\tau^2)^2}+\frac{3}{2\omega^2}-\frac{2\tau^2}{1+\omega^2\tau^2}
\right )\right]= \nonumber \\ && = E_0-F_0\,\frac{x_0+V_0\tau}
{1+\omega^2\tau^2}+\frac{F_0^2}{2m\omega^2}\,\frac{2+\omega^2\tau^2}
{1+\omega^2\tau^2}.\nonumber
\end{eqnarray}
Следовательно, переданная осциллятору энергия равна
$$\Delta E=\frac{F_0^2}{2m\omega^2}\,\frac{2+\omega^2\tau^2}{1+\omega^2\tau^2}
-F_0\,\frac{x_0+V_0\tau}{1+\omega^2\tau^2}.$$
Если $\omega\tau\gg 1$ (медленное включение), то
$$\Delta E\approx \frac{F_0^2}{2m\omega^2}=\frac{F_0^2}{2k},$$
что есть энергия упругой деформации пружины (с жесткостью $k=m\omega^2$) в новом
положении равновесия. Если $\omega\tau\ll 1$ (быстрое включение), то
$$\Delta E\approx \frac{F_0^2}{m\omega^2}-F_0x_0=F_0\left (\frac{F_0}{k}-
x_0\right )$$
есть работа силы $F_0$ при перемещении из $x_0$ в  новое положение равновесия.

{\usefont{T2A}{cmr}{m}{it} Второе решение.}
Уравнение движения осциллятора имеет вид
$$\ddot x+\omega^2\,x=\frac{F_0}{m}(1-e^{-t/\tau}).$$
Ищем его частное решение в виде $x=a+be^{-t/\tau}$. Подставляя в уравнение движения,
находим 
$$a=\frac{F_0}{m\omega^2},\;\;\;b=-\frac{F_0}{m}\,\frac{\tau^2}{1+\omega^2
\tau^2}.$$
Следовательно, общее решение уравнения движения имеет вид
$$x(t)=A\cos{\omega t}+B\sin{\omega t}+\frac{F_0}{m\omega^2}-\frac{F_0}{m}\,
\frac{\tau^2}{1+\omega^2\tau^2}\,e^{-t/\tau}.$$
Неизвестные константы $A$ и $B$ находим из начальных условий $x_0=x(0)$, $V_0=
\dot x(0)$, что дает
$$A=x_0-\frac{F_0}{m\omega^2}+\frac{F_0}{m}\,\frac{\tau^2}{1+\omega^2\tau^2},
\;\;\;B=\frac{V_0}{\omega}-\frac{F_0}{m\omega}\,\frac{\tau}{1+\omega^2
\tau^2}.$$
Энергия осциллятора
$$E=\frac{m}{2}\left (\dot{x}^2+\omega^2\,x^2\right).$$
Когда $t\gg\tau$, 
$$x(t)\approx A\cos{\omega t}+B\sin{\omega t}+\frac{F_0}{m\omega^2}.$$
Поэтому после усреднения получаем (заметим, что $<\sin{\omega t}\cos{\omega t}>\sim
<\nolinebreak \sin{2\omega t}>=0$)
$$<x^2>=\frac{1}{2}(A^2+B^2)+\left(\frac{F_0}{m\omega^2}\right)^2,\;\;\;
<\dot{x}^2>=\frac{1}{2}\omega^2(A^2+B^2).$$
Следовательно,
$$<E>=\frac{m}{2}\left[\frac{F_0^2}{m^2\omega^2}+\omega^2(A^2+
B^2)\right].$$
Но
\begin{eqnarray} &&
\omega^2(A^2+B^2)=\omega^2\left(x_0-\frac{F_0}{m\omega^2}\,\frac{1}
{1+\omega^2\tau^2}\right)^2+\left(V_0-\frac{F_0}{m}\,\frac{\tau}{1+\omega^2
\tau^2}\right)^2=\nonumber \\ && =
V_0^2+\omega^2\,x_0^2-\frac{2F_0}{m}\,\frac{x_0+V_0\tau}
{1+\omega^2\tau^2}+\frac{F_0^2}{m^2\omega^2}\,\frac{1}{1+\omega^2\tau^2},
\nonumber
\end{eqnarray}
и получаем
$$<E>=E_0-F_0\,\frac{x_0+V_0\tau}{1+\omega^2\tau^2}+\frac{F_0^2}{2m\omega^2}
\left(1+\frac{1}{1+\omega^2\tau^2}\right).$$

\subsection*{Задача 6 {\usefont{T2A}{cmr}{b}{n}(7.39)}: Как меняется высота
подскока шарика над плитой?}
{\usefont{T2A}{cmr}{m}{it} Первое решение.}
При постоянном поле тяжести $g$ фазовая траектория определяется из закона сохранения 
энергии
$$E=\frac{p^2}{2m}+mgx$$
и имеет вид
$$x=\frac{1}{mg}\left(E-\frac{p^2}{2m}\right)=h-\frac{p^2}{2m^2g},$$
так как $E=mgh$. При этом движение циклическое и во время одного периода цикла импульс
шарика меняется от $p_{max}=m\sqrt{2gh}$ до $-p_{max}$. Поэтому площадь фазовой 
траектории равна
$$S=\int\limits_{-p_{max}}^{p_{max}} x(p)\,dp=2\int\limits_0^{p_{max}}
\left (h-\frac{p^2}{2m^2g}\right)dp=2\left (p_{max}h-\frac{p_{max}^3}{6m^2g}
\right )=\frac{4}{3}mh\sqrt{2gh}.$$
При медленном изменении $g$ площадь фазовой траектории сохраняется. Следовательно,
$gh^3=\mathrm{const}$ и $h\sim g^{-1/3}$.

{\usefont{T2A}{cmr}{m}{it} Второе решение.}
Из выражения для энергии следует, что
$$gh=\frac{\dot{x}^2}{2}+gx.$$
Возьмем производную и воспользуемся уравнением движения шарика $\ddot x=-g$. 
В результате получим 
\begin{equation}
\dot g\,h+g\dot h=\dot{x}(\ddot x+g)+\dot g\,x=\dot g\,x.
\label{eq25_6a}
\end{equation}
Во время нескольких подскоков шарика можно считать, что $g$ и $h$ не меняются,
период движения равен $T=2V_0/g$ и внутри одного периода шарик движется по закону 
$x(t)=V_0t-gt^2/2$, где $V_0=\sqrt{2gh}$. Усредним уравнение (\ref{eq25_6a}) по 
одному периоду:
\begin{equation}
\dot g\,h+g\dot h=\dot g\, <x>.
\label{eq25_6b}
\end{equation}
Но
$$<x>=\frac{1}{T}\int\limits_0^T x(t)\,dt=
\frac{1}{T}\int\limits_0^T\left(V_0t-gt^2/2\right)dt=\frac{V_0^2}{3g}=
\frac{2}{3}\,h.$$
Подставляя это в (\ref{eq25_6b}), получаем $g\,\dot{h}=-h\,\dot{g}/3$, или
$$\frac{\dot{g}}{g}+3\frac{\dot{h}}{h}=\frac{d}{dt}\left(\ln{g}+3\ln{h}
\right )=\frac{d}{dt}\ln{(gh^3)}=0.$$
Отсюда следует, что $gh^3=\mathrm{const}$.

\section[Семинар 26]
{\centerline{Семинар 26}}

\subsection*{Задача 1 {\usefont{T2A}{cmr}{b}{n}(8.3)}: Мягкая посадка космического
аппарата}
Пусть $m$ --- масса космического аппарата, $M$ --- масса планеты. При мягкой посадке
энергия аппарата на поверхности планеты равна его потенциальной энергии $-GmM/R$.
Поэтому закон сохранения энергии запишется как
$$\frac{mV_\infty^2}{2}=-G\frac{mM}{R}+Fh=-\frac{mV_2^2}{2}+Fh.$$
Отсюда
$$h=\frac{m(V_\infty^2+V_2^2)}{2F}.$$
Здесь $V_2$ --- это вторая космическая скорость, которая определяется из условия
$$\frac{mV_2^2}{2}-G\frac{mM}{R}=0.$$

\subsection*{Задача 2 {\usefont{T2A}{cmr}{b}{n}(8.10)}: Период движения частицы
в центральном поле}
$x$-Проекция силы равна
$$F_x=-\frac{\partial U}{\partial x}=-2\alpha\,r\,\frac{\partial r}
{\partial x}.$$
Но $$\frac{\partial r}{\partial x}=\frac{\partial}{\partial x}\sqrt{x^2+y^2+
z^2}=\frac{x}{\sqrt{x^2+y^2+z^2}}=\frac{x}{r}.$$
Следовательно, $F_x=-2\alpha\,x$ и уравнение движения будет $m\ddot x=-2\alpha\,x$.
Это уравнение гармонических колебаний с частотой $\omega^2=2\alpha/m$. Следовательно,
период колебаний равен $$T=\frac{2\pi}{\omega}=\pi\sqrt{\frac{2m}{\alpha}}.$$
Из симметрии потенциала ясно, что в $y$- и $z$-направлениях тоже будут гармонические 
колебания с той же частотой $\omega$.

\subsection*{Задача 3 {\usefont{T2A}{cmr}{b}{n}(8.2)}: Давление в центре жидкой
планеты}
Выделим маленький цилиндрический объем жидкости с поперечным сечением $S$ и высотой
$\Delta x$ на расстоянии $x$ от центра планеты. Условие равновесия выделенного 
элемента имеет вид 
$$pS-(p+\Delta p)S-\Delta m\,g^*(x)=0,$$ где $\Delta m=\rho S\,\Delta x$ 
и
$$g^*(x)=\frac{G}{x^2}\,\frac{4\pi}{3}\,\rho\,x^3=G\,\frac{M}{R^3}\,x=
g\,\frac{x}{R}$$
есть ускорение свободного падения на расстоянии $x$ от центра планеты массой $M$ и 
радиуса $R$. При этом $g=GM/R^2$ --- ускорение свободного падения на поверхности 
планеты. Следовательно, $$-\Delta p\,S-\rho\,S\,\Delta x\,g\,\frac{x}{R}=0,$$
и получаем уравнение
$$\frac{dp}{dx}=-\rho\,g\,\frac{x}{R}.$$
Разделяем переменные и интегрируем:
$$\int\limits_{p_c}^0dp=-\frac{\rho\,g}{R}\int\limits_0^R x\,dx.$$
Здесь $p_c$ --- искомое давление в центре планеты (на поверхносты планеты $p=0$).
В результате получаем
$$p_c=\frac{1}{2}\,\rho\,g\,R\approx 1,8\cdot 10^{11}~\mbox{н}/\mbox{м}^2
\approx 2\cdot 10^6~\mbox{атм}.$$
Если не пренебрегать сжимаемостью жидкости при увеличении давления и считать, что 
$\rho(x)=\rho_c (1-\alpha\,x)$, то для массы планеты внутри сферы радиуса $x$ 
получим
$$M(x)=\int \rho(r) dV=4\pi\rho_c\int\limits_0^x (1-\alpha\, r)r^2\,dr=
\frac{4\pi}{3}\,\rho_c\,x^3\left[1-\frac{3}{4}\,\alpha\,x\right ].$$
Следовательно, 
$$g^*(x)=G\,\frac{M(x)}{x^2}=\frac{4\pi}{3}\,\rho_cGx\left[1-\frac{3}{4}\,
\alpha\,x\right ],$$
и в этом случае условие равновесия примет вид
$$\frac{dp}{dx}=-\rho(x)\,g^*(x)=-\frac{4\pi\rho_c^2G}{3}\,x\,(1-\alpha x)
\left[1-\frac{3}{4}\,\alpha\,x\right ].$$
Поэтому
$$p_c=\frac{4\pi\rho_c^2G}{3}\int\limits_0^R x\,(1-\alpha x)
\left[1-\frac{3}{4}\,\alpha\,x\right ]dx=\frac{4\pi\rho_c^2G}{6}R^2\left[
1-\frac{7}{6}\,\alpha\,R+\frac{3}{8}\,\alpha^2\,R^2\right].$$
Вводя ускорение свободного падения на поверхности планеты
$$g=\frac{4\pi}{3}\,\rho_cG\,R\left[1-\frac{3}{4}\,\alpha\,R\right ]$$
и средную плотность планеты
$$\bar \rho=\frac{M(R)}{\frac{4\pi}{3}\,R^3}=\rho_c\left[1-\frac{3}{4}\,
\alpha\,R\right ],$$
ответ для давления $p_c$ можно переписать так:
$$p_c=\frac{1}{2}\bar \rho g R\frac{1-\frac{7}{6}\,\alpha\,R+\frac{3}{8}\,
\alpha^2\,R^2}{\left[1-\frac{3}{4}\,\alpha\,R\right ]^2}.$$

\subsection*{Задача 4 {\usefont{T2A}{cmr}{b}{n}(8.12)}: Покинет ли планета
звезду?}
Найдем скорость планеты $V$ до взрыва из второго закона Ньютона
$$m\frac{V^2}{r}=G\frac{mM}{r^2},$$
где $r$ --- радиус орбиты планеты. Следовательно,
$$V^2=G\frac{M}{r}.$$
Кинетическая энергия планеты
$$E=\frac{mV^2}{2}=\frac{1}{2}G\frac{mM}{r}$$
не изменится сразу после взрыва. Поэтому полная энергия планеты после взрыва будет
$$E=\frac{1}{2}G\frac{mM}{r}-G\frac{m(1-\alpha)M}{r}=G\frac{mM}{r}\left(
\alpha-\frac{1}{2}\right).$$
Планета покинет звезду, если $E\ge 0$, т.~е. если $\alpha\ge 1/2$.

\subsection*{Задача 5 {\usefont{T2A}{cmr}{b}{n}(8.23)}: При какой скорости
частица не провалится в воронку?}
В цилиндрических координатах (ось $z$ направлена вдоль оси воронки) энергия частицы
равна
$$E=\frac{m}{2}\left (\dot\rho^2+\rho^2\dot\varphi^2+\dot z^2\right )+
mgz.$$
Но $\rho=\rho_0+z\,\ctg{\alpha}$. Отсюда $\dot z=\tg{\alpha}\,\dot\rho$ и 
получаем
\begin{equation}
E=\frac{m}{2}\left (\dot\rho^2[1+\tg^2{\alpha}]+\rho^2\dot\varphi^2\right )+
mgz.
\label{eq26_5}
\end{equation}
Из-за цилиндрической симметрии $z$-компонента момента импульса $L_z=m(x\dot y-
y\dot x)$ сохраняется. Но из $x=\rho\cos{\varphi},\,y=\rho\sin{\varphi}$ следует,
что $$x\dot y-y\dot x=\rho\cos{\varphi}(\dot\rho\sin{\varphi}+\rho\dot\varphi
\cos{\varphi})-\rho\sin{\varphi}(\dot\rho\cos{\varphi}-\rho\dot\varphi
\sin{\varphi})=\rho^2\dot\varphi.$$
Таким образом, $L_z=m\rho^2\dot\varphi=mV(\rho_0+h\,\ctg{\alpha}),$
так как в начальный момент $z=h$, $\rho=\rho_0+h\,\ctg{\alpha}$ и $V=\rho\dot
\varphi$. Отсюда $$\dot \varphi=\frac{V(\rho_0+h\,\ctg{\alpha})}{\rho^2}$$ и
(\ref{eq26_5}), с учетом начального значения энергии, примет вид
$$\frac{mV^2}{2}+mgh=E=\frac{m}{2}\,\dot\rho^2\,\frac{1}{\cos^2{\alpha}}+
\frac{mV^2R^2}{2\rho^2}+mgz,$$
где обозначили $R=\rho_0+h\,\ctg{\alpha}$. Следовательно,
$$g(h-z)\ge\frac{V^2}{2}\left(\frac{R^2}{\rho^2}-1\right)=\frac{V^2}{2}\,
\frac{(R-\rho)(R+\rho)}{\rho^2}.$$
Но $R-\rho=(h-z)\,\ctg{\alpha}$ и получаем неравенство
$$(h-z)\left(g-\frac{V^2}{2}\,\ctg{\alpha}\,
\frac{R+\rho}{\rho^2}\right)\ge 0.$$
Чтобы частица не провалилась в воронку, нам надо $z> 0$. Если $z\ge h$, то все уже 
хорошо. Поэтому возьмем  $h>z$. Тогда должны иметь
$$g-\frac{V^2}{2}\,\ctg{\alpha}\,\frac{R+\rho}{\rho^2}\ge 0,$$ или
$$\rho^2-\frac{V^2\,\ctg{\alpha}}{2g}\,\rho-\frac{V^2\,\ctg{\alpha\,R}}{2g}
\ge 0.$$
Так как $\rho$ --- положительная величина, это возможно, только если
$$\rho\ge\rho_+=\frac{V^2\,\ctg{\alpha}}{4g}+\sqrt{\frac{V^4\,\ctg{\alpha}^2}
{16g^2+\frac{V^2\,\ctg{\alpha\,R}}{2g}}}.$$
Условие $z> 0$ означает $\rho>\rho_0$, так как  $\rho=\rho_0+z\,\ctg{\alpha}$.
Таким образом, частица не провалится в воронку, если $\rho_+>\rho_0$, что эквивалентно
неравенству
$$\sqrt{\frac{V^4\,\ctg{\alpha}^2}{16g^2+\frac{V^2\,\ctg{\alpha\,R}}{2g}}}>
\rho_0-\frac{V^2\,\ctg{\alpha}}{4g}.$$
После возведения обеих сторон этого неравенства в квадрат получим
$$\frac{V^2}{2g}\,\ctg{\alpha}\,(R+\rho_0)>\rho_0^2.$$
Но $R+\rho_0=2\rho_0+h\,\ctg{\alpha}$ и окончательно получаем условие
$$V^2>\frac{2g\rho_0^2}{(2\rho_0+h\,\ctg{\alpha})\ctg{\alpha}}=
\frac{g\rho_0\,\tg{\alpha}}{1+\frac{h}{2\rho_0}\,\ctg{\alpha}}.$$

\subsection*{Задача 6 {\usefont{T2A}{cmr}{b}{n}(8.26)}: Сечение падения потока
метеоритов на Землю}
Эффективный потенциал имеет вид
$$U_{\mbox{эфф}}=-G\frac{mM}{r}+\frac{L^2}{2mr^2}=-G\frac{mM}{r}+
E\,\frac{b^2}{r^2},$$
где мы учли, что момент импульса метеорита $L=mV_\infty b$. Здесь $b$ --- 
прицельный параметр метеорита, $m$ --- его масса, $M$ --- масса Земли и
$E=mV_\infty^2/2$ --- энергия метеорита.

Минимальное расстояние метеорита от центра Земли определяется из условия
\begin{equation}
E=U_{\mbox{эфф}}(r_{min})=-G\frac{mM}{r_{min}}+E\,\frac{b^2}{r_{min}^2}.
\label{eq26_6}
\end{equation}
Но 
$$G\frac{mM}{r_{min}}=G\frac{mM}{R}\,\frac{R}{r_{min}}=\frac{mV_2^2}{2}\,
\frac{R}{r_{min}},$$
где $R$ --- радиус Земли, а $V_2$ есть вторая космическая скорость. Следовательно, 
(\ref{eq26_6}) примет вид
$$\frac{mV_\infty^2}{2}=-\frac{mV_2^2}{2}\,\frac{R}{r_{min}}+
\frac{mV_\infty^2}{2}\,\frac{b^2}{r_{min}^2},$$
или
$$r_{min}^2+\left(\frac{V_2}{V_\infty}\right)^2R\,r_{min}-b^2=0.$$
Отсюда
$$r_{min}=-\frac{1}{2}\left(\frac{V_2}{V_\infty}\right)^2R+
\sqrt{\frac{1}{4}\left(\frac{V_2}{V_\infty}\right)^4R^2+b^2}.$$
Метеорит падает на Землю, если $r_{min}\le R$. Поэтому получаем условие
$$\sqrt{\frac{1}{4}\left(\frac{V_2}{V_\infty}\right)^4R^2+b^2}\le R+
\frac{1}{2}\left(\frac{V_2}{V_\infty}\right)^2R.$$
Возводим обе части неравенства в квадрат и получаем
$$b^2\le R^2+\left(\frac{V_2}{V_\infty}\right)^2R^2.$$
Сечение падения потока метеоритов
$$\sigma=\pi b_{max}^2=\pi R^2\left(1+\frac{V_2^2}{V_\infty^2}\right).$$

\subsection*{Задача 7 {\usefont{T2A}{cmr}{b}{n}(7.42)}: Адиабатическое изменение
радиуса круговой орбиты}
В центральном поле момент импульса $L$ сохраняется. Для круговой орбиты $L=mVr$.
Следовательно, $w=V^2r^2$ --- сохраняющаяся величина. С другой стороны, на круговой 
орбите $$m\frac{V^2}{r}=G\frac{mM}{r^2}.$$
Отсюда $$V^2=G\frac{M}{r}$$ и $w=GMr$. Следовательно, $Mr$ сохраняется, и чтобы
радиус увеличился вдвое, надо, чтобы масса уменьшилась вдвое. Но закон уменьшения массы
имеет вид $$\frac{dM}{dt}=-10^{-9}M.$$ Разделяем переменные и интегрируем:
$$\int\limits_M^{M/2}\frac{dM}{M}=-10^{-9}\int\limits_o^T dt,$$
что дает $\ln{2}=10^{-9}T$. Следовательно, $T=10^9\ln{2}~\mbox{лет}$.

\subsection*{Задача 8 {\usefont{T2A}{cmr}{b}{n}(7.40)}: Осциллятор с медленно
возрастающей массой}
{\usefont{T2A}{cmr}{m}{it} Первое решение.}
Для осциллятора адиабатическим инвариантом является $E/\omega$. В случае маятника
$\omega=\sqrt{g/l}$ --- не зависит от массы и период $T$ не будет менятся. 
Из адиабатического инварианта следует, что энергия тоже не будет меняться. Но 
$E\sim mA^2$, поэтому амплитуда $A\sim m^{-1/2}$.

В случае груза на пружинке $\omega=\sqrt{k/m}\sim m^{-1/2}$ и $T\sim 1/\omega\sim
m^{1/2}$.  Следовательно, $E\sim \omega\sim m^{-1/2}$ и, так как $E=kA^2/2$, 
будем иметь $A\sim E^{1/2}\sim m^{-1/4}$.

{\usefont{T2A}{cmr}{m}{it} Второе решение.} 
Уравнение движения осциллятора имеет вид
\begin{equation}
\frac{dp}{dt}=\dot m \dot x+m\ddot x =-m\omega^2 x.
\label{eq26_8a}
\end{equation}
Энергия осциллятора
$$E=\frac{m\dot x^2}{2}+\frac{m\omega^2x^2}{2}$$
меняется со скоростью
$$\dot E=\frac{\dot m \dot x^2}{2}+\frac{\dot m \omega^2 x^2}{2}+
m\omega\dot\omega x^2+\dot x\,(m\ddot x+m\omega^2 x),$$
что после применения уравнения движения (\ref{eq26_8a}) перепишется так:
\begin{equation}
\dot E=-\frac{\dot m \dot x^2}{2}+\frac{\dot m \omega^2 x^2}{2}+
m\omega\dot\omega x^2.
\label{eq26_8b}
\end{equation}
Усредним (\ref{eq26_8b}) по периоду колебаний осциллятора. За период колебаний
масса осциллятора $m$, частота $\omega$ и амплитуда колебаний $A$ меняются 
незначительно. Поэтому эти величины можно считать постоянными во время усреднения. 
Тогда из $x(t)=A\cos{(\omega t+\phi)}$ следует, что среднее значение $x^2$ равно
$A^2/2$, а среднее значение  $\dot x^2$ равно $A^2\omega^2/2$. Следовательно,
после усреднения будем иметь
\begin{equation}
\dot E= -\frac{\dot m \omega^2 A^2}{4}+\frac{\dot m \omega^2 A^2}{4}+
\frac{m\omega\dot\omega A^2}{2}=\frac{m\omega\dot\omega A^2}{2}.
\label{eq26_8c}
\end{equation}
Но среднее за период колебаний значение энергии осциллятора равно
$$E=\frac{m\omega^2 A^2}{4}+\frac{m\omega^2 A^2}{4}=\frac{m\omega^2 A^2}{2},$$
и (\ref{eq26_8b}) можно переписать так:
$$\frac{\dot E}{E}=\frac{\dot \omega}{\omega}.$$
Это последнее равенство как раз показывает, что $E/\omega$ сохраняется:
$$\frac{d}{dt}\left (\frac{E}{\omega}\right )=\frac{\dot E}{\omega}-
\frac{E\dot\omega}{\omega^2}=\frac{E}{\omega}\,\left(\frac{\dot E}{E}-
\frac{\dot \omega}{\omega}\right )=0.$$

\subsection*{Задача 9 {\usefont{T2A}{cmr}{b}{n}(7.41)}: Осциллятор в процессе
таяния}
Пусть масса осциллятора за время $\Delta t$ меняется на $\Delta m$. Испарившаяся 
оболочка с массой $-\Delta m$ уносит $x$-компоненту импульса $\Delta p_x=(-\Delta 
m)\dot x$. Таким образом, имеется поток импульса в систему 
$$\pi_x=-\frac{\Delta p_x}{\Delta t}=\dot m\,\dot x.$$
Знак минус потому, что поток импульса показывает, сколько импульса приходит в систему
за единицу времени из-за изменения состава системы.  Следовательно, уравнение движения
осциллятора имеет вид
$$\frac{dp_x}{dt}=-m\omega^2 x+\pi_x=-m\omega^2 x+\dot m\,\dot x,$$
что эквивалентно уравнению движения Ньютоновского типа, только с переменной массой:
\begin{equation}
m\ddot x=-m\omega^2 x.
\label{eq26_9a}
\end{equation}
Так как уравнение движения не имеет каноническую форму $dp_x/dt=F_x$, площадь фазовой 
траектории не является адиабатическим инвариантом и $E/\omega$ в данном случае не 
сохраняется. Найдем, какой вид имеет истинный адиабатический инвариант для данной 
системы.

Энергия осциллятора
$$E=\frac{m\dot x^2}{2}+\frac{m\omega^2x^2}{2}$$
меняется со скоростью
$$\dot E=\frac{\dot m \dot x^2}{2}+\frac{\dot m \omega^2 x^2}{2}+
m\omega\dot\omega x^2+\dot x\,(m\ddot x+m\omega^2 x),$$
что после применения уравнения движения (\ref{eq26_9a}) перепишется так:
\begin{equation}
\dot E=\frac{\dot m \dot x^2}{2}+\frac{\dot m \omega^2 x^2}{2}+
m\omega\dot\omega x^2.
\label{eq26_9b}
\end{equation}
Усредним (\ref{eq26_9b}) по периоду колебаний осциллятора. При этом $x^2$ заменится
на $A^2/2$, а $\dot x^2$ --- на $A^2\omega^2/2$, где $A$ --- медленно меняющаяся  
амплитуда колебаний. В результате усреднения получим
\begin{equation}
\dot E=\frac{\dot m \omega^2 A^2}{2}+\frac{m\omega\dot\omega A^2}{2}=
\frac{\dot m}{m}\,E+\frac{\dot \omega}{\omega}\,E,
\label{eq26_9c}
\end{equation}
где $E=m\omega^2 A^2/2$ есть среднее за период колебаний значение энергии 
осциллятора. Полученное равенство (\ref{eq26_9c}) показывает, что $E/(m\omega)$ 
сохраняется:
$$\frac{d}{dt}\left (\frac{E}{m\omega}\right )=\frac{\dot E}{m\omega}-
\frac{E\dot\omega}{m\omega^2}-\frac{E\dot m}{m^2\omega}=\frac{E}{m\omega}\,
\left(\frac{\dot E}{E}-\frac{\dot \omega}{\omega}-\frac{\dot m}{m}\right )
=0.$$

Применим найденный адиабатический инвариант $E/(m\omega)$ к конкретным осцилляторам 
задачи. В случае маятника $\omega=\sqrt{g/l}$ --- не зависит от массы и период 
$T$ не будет меняться. Из адиабатического инварианта следует, что $E/m\sim A^2$ тоже 
не будет меняться. Таким образом, в этом случае амплитуда остается постоянной.

В случае груза на пружинке $\omega=\sqrt{k/m}\sim m^{-1/2}$ и $T\sim 1/\omega\sim
m^{1/2}$.  Следовательно, $E\sim m\omega\sim m^{1/2}$ и, так как $E=kA^2/2$, 
амплитуда будет уменьшаться по закону  $A\sim E^{1/2}\sim m^{1/4}$.

\subsection*{Задача 10 {\usefont{T2A}{cmr}{b}{n}(7.43)}: Частица в ящике,
который медленно поднимают за один конец}
{\usefont{T2A}{cmr}{m}{it} Первое решение.} 
Пусть $\alpha$ --- угол наклона, $x$ --- координата частицы вдоль дна ящика.
Закон сохранения энергии при фиксированном угле наклона $\alpha$
$$E=\frac{p^2}{2m}+mgx\sin{\alpha}$$
определяет фазовую траекторию. Когда ящик медленно поднимают за один конец, площадь
фазовой траектории не меняется (является адиабатическим инвариантом). В начале эта
площадь равна (см. рисунок) $S=2mVL$.
\begin{figure}[htb]
\centerline{\epsfig{figure=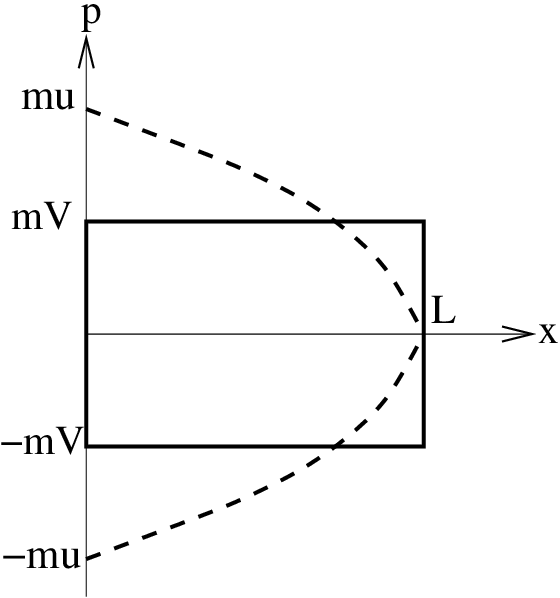,height=4cm}}
\end{figure}
При предельном угле наклона фазовая траектория показана пунктирной линией и ее площадь
равна
$$S=\int\limits_{-p_{max}}^{p_{max}}x\,dp=\frac{1}{mg\sin{\alpha}}
\int\limits_{-p_{max}}^{p_{max}}\left ( E-\frac{p^2}{2m}\right )dp=
\frac{2}{mg\sin{\alpha}}\left (Ep_{max}-\frac{p_{max}^3}{6m}
\right ).$$
Так как $E=p_{max}^2/2m=mgL\sin{\alpha}$, то получаем
$$S=\frac{2p_{max}^3}{3m^2g\sin{\alpha}}=\frac{4mL}{3}\sqrt{2gL
\sin{\alpha}}=2mVL,$$
где последнее равенство следует из сохранения площади фазовой траектории. Решая 
полученное уравнение относительно $\sin{\alpha}$, получаем, что предельный угол 
наклона определяется из условия
$$\sin{\alpha}=\frac{9V^2}{8gL}.$$

{\usefont{T2A}{cmr}{m}{it} Второе решение.} 
Уравнение движения частицы имеет вид
$$m\ddot x =-mg\sin{\alpha}.$$
Энергия частицы
$$E=\frac{m\dot x^2}{2}+mgx\sin{\alpha},$$
с учетом уравнения движения, меняется со скоростью
\begin{equation}
\dot E=\dot x (m\ddot x+mg\sin{\alpha})+mgx\dot\alpha\cos{\alpha}=
mgx\dot\alpha\cos{\alpha}.
\label{eq26_10a}
\end{equation}
Усредним это уравнение по периоду движения частицы. Так как угол $\alpha$ меняется 
очень медленно, во время усреднения его можно считать постоянным. Пусть $\alpha$ 
меньше предельного угла. Когда частица движется вверх по дну ящика, его координата 
меняется как
$$x=ut-\frac{1}{2}g\sin{\alpha}\,t^2,$$
где $u$ --- начальная скорость частицы около нижней стенки ящика. Через время 
$T/2=(u-u_1)/(g\sin{\alpha})$ частица, имея ненулевую скорость $u_1$, упруго 
отразится от верхней стенки ящика и закон ее движения примет форму
$$x=L-u_1\left(t-\frac{T}{2}\right)-\frac{1}{2}g\sin{\alpha}\left(t-\frac{T}
{2}\right)^2.$$
Поэтому среднее за период значение координаты равно
$$<x>=\frac{1}{T}(I_1+I_2),$$
где
$$I_1=\int\limits_0^{T/2}\left (ut-\frac{1}{2}g\sin{\alpha}\,t^2\right )dt$$
и
$$I_2=\int \limits_{T/2}^T\left (L-u_1\left(t-\frac{T}{2}\right)-\frac{1}{2}
g\sin{\alpha}\,\left(t-\frac{T}{2}\right )^2\right )dt.$$
Вычисляя интегралы, получаем
$$<x>=\frac{L}{2}+\frac{(u-u_1)^2}{12g\sin{\alpha}}=
\frac{L}{2}+\frac{(\sqrt{E}-\sqrt{E-mgL\sin{\alpha}})^2}{6mg\sin{\alpha}},$$
где на последнем этапе мы учли, что
$$E=\frac{mu^2}{2}=\frac{mu_1^2}{2}+mgL\sin{\alpha}.$$
Следовательно, после усреднения уравнение (\ref{eq26_10a}) принимает вид
\begin{equation}
\frac{dE}{d\sin{\alpha}}=mg\left(\frac{L}{2}+\frac{(\sqrt{E}-
\sqrt{E-mgL\sin{\alpha}})^2}{6mg\sin{\alpha}}\right ).
\label{eq26_10b}
\end{equation}
Вместо энергии $E$ удобно ввести новую безразмерную переменную $\tau$ через 
$E=mg\sin{\alpha}L\tau$. Тогда (\ref{eq26_10b}) примет вид
$$\tau+\sin{\alpha}\frac{d\tau}{d\sin{\alpha}}=\frac{1}{2}+\frac{
(\sqrt{\tau}-\sqrt{\tau-1})^2}{6}.$$
Разделяем переменные и интегрируем
$$\int\limits_\epsilon^{\sin{\alpha}}\frac{d\sin{\alpha^\prime}}
{\sin{\alpha^\prime}}=\int\limits_{\tau_0}^1\frac{3d\tau}{1-2\tau-
\sqrt{\tau(\tau-1)}}.$$
Здесь $\alpha$ --- предельный угол, при котором $u_1=0$, т.~е. $E=mgL\sin{\alpha}$
и соответственно $\tau=1$. В качестве нижнего предела взяли не ноль, так как интеграл 
расходится, а некую маленькую величину $\epsilon$. Соответствующее значение
$\tau$ находится из $mV^2/2=mgL\epsilon\tau$ и равно
$$\tau_0=\frac{V^2}{2gL\epsilon}\gg 1.$$
В интеграле по $\tau$ сделаем подстановку Эйлера $\sqrt{\tau(\tau-1)}=\tau t$,
что дает
$$\tau=\frac{1}{1-t^2},\;\;\;d\tau=\frac{2t\,dt}{(1-t^2)^2}.$$
В результате получим
$$\ln{\frac{\sin{\alpha}}{\epsilon}}=\int\limits_0^{t_0}\frac{6t\,dt}
{(1-t^2)(1+t+t^2)},$$
где
\begin{equation}
t_0=\sqrt{1-\frac{1}{\tau_0}}=\sqrt{1-\frac{2gL\epsilon}{V^2}}\approx
1-\frac{gL\epsilon}{V^2}.
\label{eq26_10c}
\end{equation}
Разложим подынтегральное выражение на элементарные дроби:
$$\frac{6t}{(1-t^2)(1+t+t^2)}=\frac{at+b}{1+t+t^2}+\frac{c}{1+t}+
\frac{d}{1-t},$$
где неизвестные коэффициенты $a,b,c,d$ удовлетворяют системе уравнений
$$d=a+c,\;\;b=2d,\;\;b+c+d=0,\;\;a+2d=6.$$
Эта система легко решается с результатом
$$d=1,\;\;a=4d=4,\;\;b=2d=2,\;\;c=-3d=-3.$$
Таким образом
$$\frac{6t}{(1-t^2)(1+t+t^2)}=\frac{2(2t+1)}{1+t+t^2}-\frac{3}{1+t}+
\frac{1}{1-t}.$$
После данного разложения интегрирование становится тривиальным, получаем
$$\ln{\frac{\sin{\alpha}}{\epsilon}}=\ln{\frac{(1+t_0+t_0^2)^2}{(1-t_0)
(1+t_0)^3}}\approx \ln{\frac{9V^2}{8gL\epsilon}},$$
где для $t_0$ воспользовались выражением (\ref{eq26_10b}) и оставили только
ведущие по $\epsilon$ члены. Отсюда
$$\sin{\alpha}=\frac{9V^2}{8gL}.$$
Как видим, вспомогательная переменная $\epsilon$ благополучно сокращается.

\section[Семинар 27]
{\centerline{Семинар 27}}

\subsection*{Задача 1: Уравнение орбиты в задаче Кеплера}
В центральном поле движение плоское: так как $\vec{r}\cdot\vec{L}=0$, где 
$\vec{L}=\vec{r}\times\vec{p}$ есть момент импульса, $\vec{r}$ лежит в плоскости, 
перпендикулярной к постоянному вектору $\vec{L}$. Пусть $r,\varphi$ --- полярные 
координаты в этой плоскости. Тогда $L=mr^2\dot\varphi$ и, следовательно,
$$\dot\varphi=\frac{L}{mr^2}.$$
Подставим это в выражение для энергии:
$$E=\frac{mV^2}{2}-\frac{\alpha}{r}=\frac{m}{2}(\dot r^2+r^2\,\dot\varphi^2)
-\frac{\alpha}{r}.$$
В результате получим
$$E=\frac{m\dot r^2}{2}+\frac{L^2}{2mr^2}-\frac{\alpha}{r},$$
или
$$\frac{E}{L^2}=\frac{m\dot r^2}{2L^2}+\frac{1}{2mr^2}-\frac{\alpha}{L^2r}.$$
Но
$$\frac{m\dot r}{L}=\frac{m\dot r}{mr^2\dot\varphi}=\frac{1}{r^2}\,\frac{dr}
{d\varphi}=-\frac{d\chi}{d\varphi},$$
где $\chi=1/r$. Поэтому
$$\frac{E}{L^2}=\frac{1}{2m}\left(\frac{d\chi}{d\varphi}\right)^2+
\frac{\chi^2}{2m}-\frac{\alpha}{L^2}\,\chi.$$
Умножим это равенство на $2m$ и потом продифференцируем по $\varphi$:
$$2\,\frac{d^2\chi}{d\varphi^2}\, \frac{d\chi}{d\varphi}+2\chi\,\frac{d\chi}
{d\varphi}-\frac{2m\alpha}{L^2}\,\frac{d\chi}{d\varphi}=0.$$
Отсюда или $\frac{d\chi}{d\varphi}=0$, что соответствует движению по окружности $r=
\mathrm{const}$, или
\begin{equation}
\frac{d^2\chi}{d\varphi^2}+\chi=\frac{m\alpha}{L^2}.
\label{eq27_1}
\end{equation}
Общее решение уравнения (\ref{eq27_1}) имеет вид
$$\chi=A\,\cos{\varphi}+B\,\sin{\varphi}+\frac{m\alpha}{L^2}.$$
Пусть при $\varphi=0$, $r=r_{min}$. Тогда должны иметь
$$A+\frac{m\alpha}{L^2}=\frac{1}{r_{min}}.$$
Кроме того, так как $\chi=1/r$ при $\varphi=0$ будет иметь максимальное значение, то
$$\left .\frac{d\chi}{d\varphi}\right |_{\varphi=0}=0,$$
что дает $B=0$. Следовательно,
\begin{equation}
\frac{1}{r}=\left(\frac{1}{r_{min}}-\frac{m\alpha}{L^2}\right)
\cos{\varphi}+\frac{m\alpha}{L^2}.
\label{eq27_1a}
\end{equation}
Обозначим
$$p=\frac{L^2}{m\alpha},\;\;\;e=\frac{p}{r_{min}}-1.$$
Тогда уравнение орбиты (\ref{eq27_1a}) примет вид
$$\frac{p}{r}=1+e\,\cos{\varphi}.$$
Это уравнение эллипса с параметром $p$ и эксцентриситетом $e$.

\subsection*{Задача 2 {\usefont{T2A}{cmr}{b}{n}(8.39)}: Третья космическая
скорость}
Пусть $V$ --- скорость ракеты относительно Земли при запуске после выхода из 
атмосферы. Скорость $V^\prime$, которая останется у ракеты, когда она преодолеет 
тяготение Земли и отойдет от нее далеко, находим из закона сохранения энергии: 
$$\frac{mV^2}{2}-G\,\frac{mM}{R}=\frac{mV^{\prime\,2}}{2}.$$
Заметим, что на этом этапе гравитационный потенциал Солнца можно не принимать во 
внимание, так как данное изменение положения ракеты, хотя много больше радиуса Земли 
$R$, много меньше расстояния $r$ от Земли до Солнца и, следовательно, гравитационный 
потенциал Солнца при этом практически не меняется. Таким образом,
$$V^{\prime\,2}=V^2-V_2^2,$$
где $V_2$ --- вторая космическая скорость, которая определяется из условия
$$\frac{mV_2^2}{2}=G\,\frac{mM}{R}.$$
Относительно Солнца скорость будет $\tilde V=V^{\prime}+V_{\mbox{орб}}$, если
запускаем вдоль траектории Земли, чтобы использовать ее орбитальную скорость 
$V_{\mbox{орб}}$. Чтобы ракета покинула солнечную систему, должны иметь
$$\frac{m\tilde V^2}{2}-G\,\frac{mM_\odot}{r}=0.$$
Отсюда $\tilde V=\sqrt{2}\,V_{\mbox{орб}}$, где орбитальная скорость Земли
определяется из условия
$$M\,\frac{V_{\mbox{орб}}^2}{r}=\frac{MM_\odot}{r^2}.$$
Следовательно, для третьей космической скорости $V_3$ получаем уравнение
$$\sqrt{2}\,V_{\mbox{орб}}=\sqrt{V_3^2-V_2^2}+V_{\mbox{орб}}.$$
Отсюда
$$V_3=\sqrt{(\sqrt{2}-1)^2\,V_{\mbox{орб}}^2+V_2^2}=
\sqrt{(3-2\sqrt{2})\,V_{\mbox{орб}}^2+V_2^2}\approx 17~\mbox{км}/\mbox{с}.$$

\subsection*{Задача 3 {\usefont{T2A}{cmr}{b}{n}(8.60)}: Комета Галлея}
Сравним орбиты кометы и Земли. Для кометы $2a=(36+0.6)~\mbox{а.е.}$, для Земли
$a_\oplus=1~\mbox{а.е.}$ По второму закону Кеплера $T^2\sim a^3$. Для Земли
$T_\oplus=1~\mbox{год}$. Поэтому
$$T=T_\oplus\left(\frac{a}{a_\oplus}\right)^{3/2}\approx 78~\mbox{лет}.$$

\subsection*{Задача 4 {\usefont{T2A}{cmr}{b}{n}(8.52)}: Изменение орбиты спутника}
Для эллиптических орбит
$$r_{min}=\frac{p}{1+e},$$
где параметр орбиты $p$ и эксцентриситет $e$ равны
$$p=\frac{L^2}{m\alpha},\;\;\;e=\sqrt{1+\frac{2EL^2}{m\alpha^2}}.$$
В начале $r_{min}=r$ и $e=0$. Следовательно, $p=r$. При радиальной добавке 
к скорости момент импульса $L$ не меняется. Поэтому должны иметь
$$r_1=\frac{p}{1+e}=\frac{r}{1+e}.$$
Отсюда
$$e=\frac{r}{r_1}-1.$$
Так как радиальная добавка $V_r$ перпендикулярна к первоначальной скорости спутника,
энергия спутника после изменения скорости будет
$$E^\prime=E+\frac{mV_r^2}{2},$$
где $E$ --- начальная энергия спутника. Поэтому
$$e^2=1+\frac{2E^\prime L^2}{m\alpha^2}=1+\frac{2EL^2}{m\alpha^2}+
\frac{mV_r^2L^2}{m\alpha^2}=\frac{V_r^2L^2}{\alpha^2},$$
так как (вспомним, что первоначально $e=0$)
$$1+\frac{2EL^2}{m\alpha^2}=0.$$
Приравняв два выражения для эксцентриситета $e$, получаем
$$\frac{r}{r_1}-1=\frac{|V_r|L}{\alpha}.$$
Но $$L=mVr=\frac{\alpha}{V},$$ где $V$ есть скорость спутника на круговой орбите,
которая определяется из второго закона Ньютона:
$$m\,\frac{V^2}{r}=\frac{\alpha}{r^2}.$$
Таким образом, получаем 
$$\frac{r}{r_1}-1=\frac{|V_r|}{\alpha}\,\frac{\alpha}{V}=\frac{|V_r|}
{V}.$$
Окончательно
$$\frac{V_r}{V}=\pm\left(\frac{r}{r_1}-1\right).$$
То, что одну и ту же орбиту можно получить как при $V_r>0$ (точка $A$), так и при
$V_r<0$ (точка $B$), видно на рисунке.
\begin{figure}[htb]
\centerline{\epsfig{figure=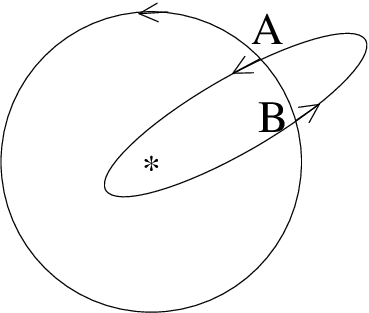,height=4cm}}
\end{figure}

{\usefont{T2A}{cmr}{m}{it} Вариант решения.}
По формуле
$$a=-\frac{\alpha}{2E^\prime},$$
для большой полуоси орбиты $a$ получаем
$$a=-\frac{\alpha}{2E\left(1+\frac{mV_r^2}{2E}\right)}=\frac{r}{1+
\frac{mV_r^2}{2E}},$$
так как $r=-\frac{\alpha}{2E}$. На круговой орбите
$$E=\frac{mV^2}{2}-\frac{\alpha}{r}=-\frac{mV^2}{2},$$
так как $mV^2/r=\alpha/r^2$. Таким образом,
$$a=\frac{r}{1-\frac{V_r^2}{V^2}}.$$
С другой стороны,
$$a=\frac{1}{2}(r_{min}+r_{max})=\frac{1}{2}\left (\frac{p}{1+e}+\frac{p}
{1-e}\right )=\frac{p}{1-e^2}.$$
Приравняв эти выражения для $a$ друг к другу, получаем 
$$e=\frac{|V_r|}{V}.$$

\subsection*{Задача 5 {\usefont{T2A}{cmr}{b}{n}(8.4)}: Поле тяготения внутри
полости}
\begin{figure}[htb]
\centerline{\epsfig{figure=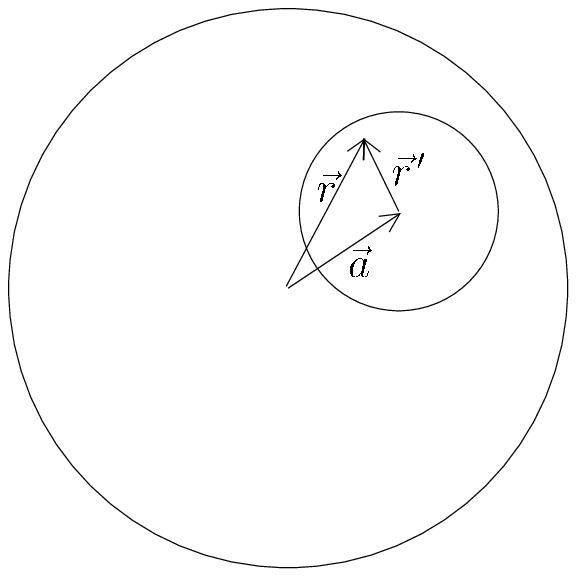,height=4cm}}
\end{figure}
В сплошном шаре напряженность гравитационного поля в точке с радиус-вектором $\vec{r}$
равна
$$\vec{E}=-G\,\frac{M(r)}{r^2}\,\frac{\vec{r}}{r}=-G\,\frac{4\pi}{3}\,\rho r^3
\,\frac{\vec{r}}{r^3}=-\frac{4\pi}{3}\,G\rho\,\vec{r},$$
где $\rho$ --- плотность вещества, $M(r)$ --- масса вещества внутри сферы радиуса 
$r$. Шар с плотностью вещества $\rho$ и полостью внутри можно заменить на сплошной 
шар плюс другой сплошной шар на месте полости, но с плотностью вещества $-\rho$. 
Поэтому
$$\vec{E}=-\frac{4\pi}{3}\,G\rho\,\vec{r}+\frac{4\pi}{3}\,G\rho\,\vec{r}^
{\,\prime}=-\frac{4\pi}{3}\,G\rho\,\vec{a},$$
так как $\vec{r}=\vec{a}+\vec{r}^{\,\prime}$. Как видим, гравитационное поле внутри
полости однородно и антипараллельно вектору $\vec{a}$, соединяющему центр шара 
с центром полости.

\subsection*{Задача 6 {\usefont{T2A}{cmr}{b}{n}(8.28)}: Сечение падения частиц
на сферу}
Эффективный потенциал 
$$U_{\mbox{эфф}}=\frac{\alpha}{r}+\frac{L^2}{2mr^2}.$$
Но, если $b$ --- прицельный параметр, $L=mV_\infty b$, $E=mV_\infty^2/2$  и 
$$\frac{L^2}{2mr^2}=E\,\frac{b^2}{r^2}.$$
Следовательно,
$$U_{\mbox{эфф}}=\frac{\alpha}{r}+E\,\frac{b^2}{r^2}.$$
Минимальное расстояние от центра $r_{min}$ является решением уравнения $E=
U_{\mbox{эфф}}(r_{min})$, или
$$E=\frac{\alpha}{r_{min}}+E\,\frac{b^2}{r_{min}^2}.$$
Это дает
$$r_{min}^2-\frac{\alpha}{E}\,r_{min}-b^2=0$$
и $$r_{min}=\frac{\alpha}{2E}+\sqrt{\frac{\alpha^2}{4E^2}+b^2}.$$
Нам нужно $r_{min}\le R$, т.~е.
$$\sqrt{\frac{\alpha^2}{4E^2}+b^2}\le R-\frac{\alpha}{2E}.$$
После возведения обеих частей неравенства в квадрат получаем
\begin{equation}
b^2\le R^2\left(1-\frac{\alpha}{RE}\right).
\label{eq27_6}
\end{equation}
Следовательно, сечение падения
$$\sigma=\pi b_{max}^2=\pi R^2\left(1-\frac{\alpha}{RE}\right).$$
Видно, что $\sigma=0$, если $E=\alpha/R$. При дальнейшем уменьшении $E$,
$\sigma=0$, так как неравенство (\ref{eq27_6}) не будет иметь решения, т.~е.
$r_{min}> R$ при любом прицельном параметре. Окончательно
$$\sigma=\left\{\begin{array}{cc} \pi R^2\left(1-\frac{\alpha}{RE}\right), 
& E\ge \frac{\alpha}{R}, \\ 0, & E< \frac{\alpha}{R}. \end{array}
\right . $$

\subsection*{Задача 7 {\usefont{T2A}{cmr}{b}{n}(8.24)}: Границы движения частицы}
В цилиндрической системе координат
$$E=\frac{m}{2}(\dot\rho^2+\rho^2\dot\varphi^2+\dot z^2)+mgz.$$
Из-за симметрии $z$-компонента момента импульса сохраняется:
$$L_z=m\rho^2\dot\varphi=mV\sqrt{\frac{H}{\alpha}}.$$
Поэтому
$$E=\frac{m}{2}(\dot\rho^2+\dot z^2)+\frac{mV^2H}{2\alpha\rho^2}+mgz.$$
Отсюда, если учесть, что энергия сохраняется и в начале
$$E=\frac{mV^2}{2}+mgH,$$
получаем
$$\frac{m}{2}(\dot\rho^2+\dot z^2)+\frac{mV^2H}{2\alpha\rho^2}+mgz=
\frac{mV^2}{2}+mgH.$$
Следовательно, так как $\alpha\rho^2=z$,
$$\dot\rho^2+\dot z^2=V^2+2gH-V^2\,\frac{H}{z}-2gz=(z-H)\left(\frac{V^2}{z}-
2g\right)\ge 0.$$
Поэтому или
$$\begin{array}{c} z-H\ge 0, \\ \frac{V^2}{z}-2g\ge 0,\end{array}$$
или
$$\begin{array}{c} z-H\le 0, \\ \frac{V^2}{z}-2g\le 0.\end{array}$$
В первом случае
$$H\le z\le\frac{V^2}{2g},$$
и такие границы движения реализуются при $H\le\frac{V^2}{2g}$. Во втором случае
$$\frac{V^2}{2g}\le z\le H,$$
и такие границы движения реализуются при $H\ge\frac{V^2}{2g}$.

\subsection*{Задача 8 {\usefont{T2A}{cmr}{b}{n}(8.15)}: Сечение отрыва планеты
быстрой звездой}
\begin{figure}[htb]
\centerline{\epsfig{figure=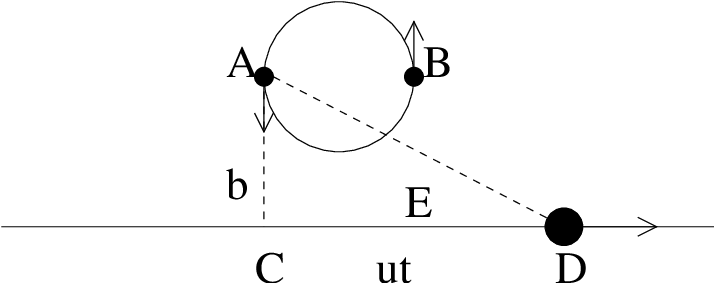,height=3cm}}
\end{figure}
Оценим, какую дополнительную скорость приобретает планета при прохождении звезды. 
Так как звезда проходит быстро, планета не успевает сдвинуться --- приобретет 
только дополнительный импульс. Пусть при $t=0$ планета находится в точке $A$, 
а звезда --- в точке $C$ под ней (см. рисунок). В произвольный момент времени $t$ 
поперечная составляющая силы, действующая на планету со стороны звезды, равна ($b$ ---
прицельный параметр звезды, $u$ --- ее скорость)
$$F_{y}=G\,\frac{mM}{b^2+u^2t^2}\,\cos{\angle CAD}=G\,\frac{mM}{b^2+u^2t^2}\,
\frac{b}{\sqrt{b^2+u^2t^2}}.$$
Поэтому, согласно второму закону Ньютона,
$$\frac{dV_y}{dt}=\frac{F_y}{m}=\frac{GMb}{(b^2+u^2t^2)^{3/2}}.$$
Следовательно, изменение поперечной скорости планеты при прохождении звезды будет
$$\Delta V_y=\int\limits_{-\infty}^\infty  \frac{GMb\,dt}{(b^2+u^2t^2)^{3/2}}=
2GMb\int\limits_0^\infty\frac{dt}{(b^2+u^2t^2)^{3/2}}=
\frac{2GM}{bu}\int\limits_0^\infty \frac{dx}{(1+x^2)^{3/2}}.$$
Но
$$\int\limits_0^\infty \frac{dx}{(1+x^2)^{3/2}}=\left .\frac{x}{\sqrt{1+x^2}}
\right|_0^\infty=1,$$
и получаем
\begin{equation}
\Delta V_y=\frac{2GM}{bu}.
\label{eq27_8}
\end{equation}
Орбитальная скорость планеты $V_{\mbox{орб}}$ на круговой орбите радиусом $R$ 
определяется из условия $$m\,\frac{V_{\mbox{орб}}^2}{R}=G\,\frac{mM_\oplus}
{R^2},$$ что дает
$$G=\frac{RV_{\mbox{орб}}^2}{M_\oplus}.$$
Поэтому (\ref{eq27_8}) можно переписать так
$$\Delta V_y=2\,\frac{M}{M_\oplus}\,\frac{R}{b}\,\frac{V_{\mbox{орб}}^2}{u}.$$
Отрыв планеты происходит, если $V_{\mbox{орб}}+\Delta V_y\ge V_2=\sqrt{2}\,
V_{\mbox{орб}}$, где $V_2=\sqrt{2}\,V_{\mbox{орб}}$ --- вторая космическая 
скорость для планеты. Это дает
$$b\le\frac{2}{\sqrt{2}-1}\,\frac{M}{M_\oplus}\,\frac{V_{\mbox{орб}}}{u}\,R.$$
Сечение ионизации будет
\begin{equation}
\sigma=\pi b_{max}^2=\pi R^2\frac{4}{(\sqrt{2}-1)^2}\,\left(\frac{M}
{M_\oplus}\right)^2\left(\frac{V_{\mbox{орб}}}{u}\right)^2.
\label{eq27_8a}
\end{equation}
На самом деле мы рассмотрели самый благоприятный случай, когда орбитальная скорость
планеты способствует отрыву. Поэтому (\ref{eq27_8a}) дает верхний предел на 
сечение ионизации --- сечение ионизации не может быть больше этого значения.
Сечение гарантированной ионизации будет меньше: для этого должно быть $\Delta V_y-
V_{\mbox{орб}}\ge \sqrt{2}\,V_{\mbox{орб}}$, и оно соответствует случаю, когда при 
$t=0$ планета находится в точке $B$, а звезда --- в точке $E$ под ней (см. рисунок).

\subsection*{Задача 9 {\usefont{T2A}{cmr}{b}{n}(8.17)}: Время жизни атома
водорода}
Энергия электрона меняется со скоростью (подразумевается CGSE система единиц)
$$\frac{dE}{dt}=-\frac{2e^2a^2}{3c^3}=-\frac{2e^2V^4}{3c^3r^2}=
-\frac{2e^6}{3m^2c^3r^4},$$
так как ускорение электрона на круговой орбите $a=V^2/r$ и, согласно второму закону
Ньютона $mV^2/r=e^2/r^2$, квадрат скорости $V^2=e^2/(mr)$. С другой стороны, на 
круговой орбите
$$E=\frac{mV^2}{2}-\frac{e^2}{r}=-\frac{e^2}{2r}.$$
Поэтому
$$dE=\frac{e^2}{2r^2}\,dr,$$
и получаем
$$\frac{e^2}{2r^2}\,\frac{dr}{dt}=-\frac{2e^6}{3m^2c^3r^4}.$$
Разделяем переменные и интегрируем,
$$\int\limits_{r_0}^0r^2\,dr=-\frac{4e^4}{3m^2c^3}\int_0^Tdt.$$
Поэтому
$$T=\frac{m^2c^3r_0^3}{4e^4}\approx 1,6\cdot 10^{-11}~\mbox{с}.$$

\section[Семинар 28]
{\centerline{Семинар 28}}

\subsection*{Задача 1: Вектор Рунге--Ленца}
Введем полярные координаты $r$ и $\varphi$ в плоскости движения (которая 
перпендикулярна постоянному вектору $\vec{L}=\vec{r}\times\vec{p}=\sgn{(\dot
\varphi)}L\,\vec{k}$ момента
импульса). Тогда $\dot{\vec{e}}_r=\dot\varphi\,\vec{e}_\varphi$ и 
$\dot{\vec{e}}_\varphi=-\dot\varphi\,\vec{e}_r$, где в терминах ортов $\vec{i},
\vec{j}, \vec{k}$ декартовой системы, полярные орты $\vec{e}_r$, $\vec{e}_\varphi$
имеют вид
$$\vec{e}_r=\frac{\vec{r}}{r}=\cos{\varphi}\,\vec{i}+\sin{\varphi}\,\vec{j},
\;\;\vec{e}_\varphi=\vec{k}\times \vec{e}_r=-\sin{\varphi}\,\vec{i}+
\cos{\varphi}\,\vec{j}.$$
В задаче Кеплера уравнение движения имеет вид
$$m\frac{d\vec{V}}{dt}=-\frac{\alpha}{r^2}\,\vec{e}_r=\frac{\alpha}
{r^2\dot\varphi}\,\dot{\vec{e}}_\varphi.$$
Но $$\vec{L}=m\,\vec{r}\times\dot{\vec{r}}=mr\,\vec{e}_r\times(\dot r\vec{e}_r+
r\dot\varphi\,\vec{e}_\varphi)=mr^2\dot\varphi\,\vec{e}_r\times\vec{e}_\varphi
=mr^2\dot\varphi\,\vec{k}.$$
Поэтому получаем
$$\frac{d\vec{V}}{dt}=\frac{\alpha}{L\,\sgn{(\dot\varphi)}}\,
\dot{\vec{e}}_\varphi$$
и 
$$\frac{d}{dt}\left(\vec{V}-\frac{\alpha}{L\,\sgn{(\dot\varphi)}}\,
\vec{e}_\varphi\right)=0,$$
так как при движении остаются постоянными как величина момента импульса $L$, так и 
знак $\sgn{(\dot\varphi)}=|\dot\varphi|/\dot\varphi$ орбитальной скорости
$\dot\varphi$.

Как видим, в задаче Кеплера с потенциалом $U(r)=-\alpha/r$ сохраняется вектор
$$\vec{h}=\vec{V}-\sgn{(\dot\varphi)}\,\frac{\alpha}{L}\,\vec{e}_\varphi.$$
Этот вектор называется вектором Гамильтона \cite{33,34}. Следовательно, вектор
$$\vec{e}=\frac{1}{\alpha}\,\vec{h}\times\vec{L}=\frac{1}{\alpha}\,\vec{V}
\times\vec{L}-\vec{e}_r.$$
тоже сохраняется. Это и есть вектор Рунге--Ленца.

Уравнение орбиты можно получить следующим образом. Пусть ось $x$ направлена по вектору
$\vec{e}$. Тогда
$$\vec{e}\cdot\vec{r}=er\,\cos{\varphi}=\frac{1}{\alpha}\,\vec{r}\cdot(\vec{V}
\times\vec{L})-r=\frac{1}{\alpha}\,\vec{L}\cdot(\vec{r}\times\vec{V})-r=
\frac{L^2}{m\alpha}-r.$$
Следовательно, $$\frac{p}{r}=1+e\cos{\varphi},$$
где $p=L^2/(m\alpha)$. Из этого вывода следует, что величина вектора Рунге--Ленца
равна эксцентриситету орбиты. В этом можно убедиться и прямым вычислением:
$$e^2=\frac{(\vec{V}\times \vec{L})^2}{\alpha^2}-\frac{2}{\alpha r}\vec{r}
\cdot(\vec{V}\times \vec{L})+1=\frac{V^2L^2}{\alpha^2}-\frac{2L^2}
{m\alpha r}+1,$$
что можно переписать так
$$e^2=1+\frac{2L^2}{m\alpha^2}\left(\frac{mV^2}{2}-\frac{\alpha}{r}\right)=
1+\frac{2L^2E}{m\alpha^2}.$$

{\usefont{T2A}{cmr}{m}{it} Уравнение орбиты из вектора Гамильтона.} 
Уравнение орбиты можно получить еще таким способом. Пусть ось $x$ выбрана так, что при 
$\varphi=0$ длина радиус-вектора $\vec{r}$ принимает минимальное значение $r=r_m$.
Пусть $\dot\varphi >0$ (см. рисунок).
\begin{figure}[htb]
\centerline{\epsfig{figure=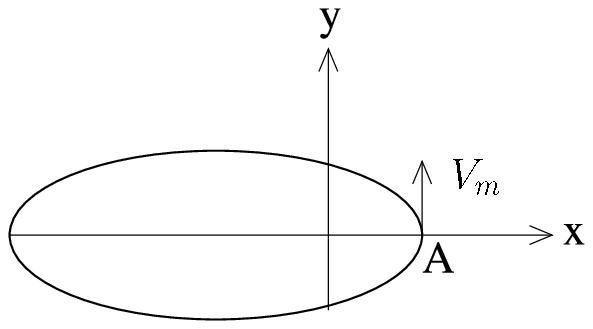,height=2cm}}
\end{figure}
Вычисляя вектор Гамильтона в точке $A$, получим
$$\vec{h}=\left(V_m-\frac{\alpha}{L}\right )\vec{j}=\left(V_m-\frac{\alpha}{L}
\right )(\sin{\varphi}\,\vec{e}_r+\cos{\varphi}\,\vec{e}_\varphi),$$
где $V_m$ --- скорость частицы при $r=r_m$ (в точке $A$). С другой стороны,
$$\vec{h}=\vec{V}-\frac{\alpha}{L}\,\vec{e}_\varphi=\dot r\,\vec{e}_r+
\left(r\dot\varphi-\frac{\alpha}{L}\right )\vec{e}_\varphi,$$
так как $\vec{V}=\dot r \,\vec{e}_r+r\dot\varphi\,\vec{e}_\varphi$. Приравняв
эти два выражения для $\vec{h}$, получаем
\begin{eqnarray} &&
\dot r=\left(V_m-\frac{\alpha}{L}\right )\sin{\varphi},\nonumber\\&&
r\dot\varphi=\frac{\alpha}{L}+\left(V_m-\frac{\alpha}{L}\right )\cos{\varphi}.
\nonumber
\end{eqnarray}
Отсюда, если поделить первое уравнение на второе, будем иметь
$$\frac{1}{r}\,\frac{dr}{d\varphi}=\frac{\left(V_m-\frac{\alpha}{L}\right )
\sin{\varphi}}{\frac{\alpha}{L}+\left(V_m-\frac{\alpha}{L}\right )
\cos{\varphi}}.$$
Разделяем переменные и интегрируем, что дает
$$\ln{\frac{r}{r_m}}=-\ln{\frac{V_m}{\frac{\alpha}{L}+\left(V_m-\frac{\alpha}
{L}\right )\cos{\varphi}}}.$$
Следовательно,
$$\frac{p}{r}=1+e\,\cos{\varphi},$$
где
$$p=\frac{Lr_mV_m}{\alpha}=\frac{L^2}{m\alpha},\;\;\;(\mbox{так как}\;\;
L=mV_mr_m),$$
и
$$e=\frac{LV_m}{\alpha}-1.$$
Заметим, что
$$e^2=1-2\,\frac{LV_m}{\alpha}+\frac{L^2V_m^2}{\alpha^2},$$
и если учесть $L=mV_mr_m$, получаем
$$e^2=1+\frac{2L^2}{m\alpha^2}\left (\frac{mV_m^2}{2}-\frac{\alpha}{r_m}
\right)=1+\frac{2L^2E}{m\alpha^2}.$$

\subsection*{Задача 2 {\usefont{T2A}{cmr}{b}{n}(8.65)}: Падение Земли на Солнце}
{\usefont{T2A}{cmr}{m}{it} Первое решение.} 
Падение можно рассмотреть как предельный случай движения по траектории с большой
полуосью $a=R/2$, где $R$ --- расстояние от Земли до Солнца (см. рисунок).
\begin{figure}[htb]
\centerline{\epsfig{figure=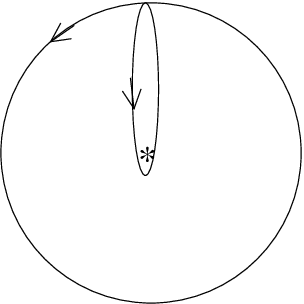,height=2cm}}
\end{figure}

По второму закону Кеплера период такого движения $T$ удовлетворяет соотношению
$$\frac{T^2}{T_\oplus^2}=\frac{(R/2)^3}{R^3}=\frac{1}{8},$$
где $T_\oplus$ --- 1~год. Таким образом,
$$T=\frac{T_\oplus}{2\sqrt{2}}$$ и время падения
$$t=\frac{T}{2}=\frac{T_\oplus}{4\sqrt{2}}\approx 2,1~\mbox{месяца}.$$
 
{\usefont{T2A}{cmr}{m}{it} Второе решение.} 
Если Землю остановить на орбите, ее энергия будет $E=-\alpha/R$. Тогда закон 
сохранения энергии $$\frac{m\dot r^2}{2}-\frac{\alpha}{r}=E,$$
дает
\begin{equation}
\frac{dr}{dt}=-\sqrt{\frac{2}{m}\left(\frac{\alpha}{r}-\frac{\alpha}{R}
\right)}=-\sqrt{\frac{2\alpha}{mR}\left(\frac{R}{r}-1\right)}.
\label{eq28_2}
\end{equation}
Знак минус потому, что при падении $r$ уменьшается. Введем вместо $r$ новую переменную
$\xi$ с помощью соотношения 
$$r=\frac{R}{2}(1-\cos{\xi}).$$
Тогда
$$\sqrt{\frac{R}{r}-1}=\sqrt{\frac{1+\cos{\xi}}{1-\cos{\xi}}}=\ctg{\frac{\xi}
{2}}.$$
Кроме того, 
$$dr=\frac{R}{2}\sin{\xi}\,d\xi=R\,\sin{\frac{\xi}{2}}\cos{\frac{\xi}
{2}}\,d\xi.$$
Поэтому (\ref{eq28_2}) примет вид
$$\frac{d\xi}{dt}=-\frac{1}{R}\,\sqrt{\frac{2\alpha}{mR}}\,\frac{1}
{\sin^2{\frac{\xi}{2}}}=-\frac{1}{R}\,\sqrt{\frac{2\alpha}{mR}}\,\frac{2}
{1-\cos{\xi}}.$$
Разделяем переменные и интегрируем:
$$t=-\frac{R}{2}\,\sqrt{\frac{mR}{2\alpha}}\int\limits_\pi^0(1-\cos{\xi})
d\xi=\frac{\pi R}{2}\,\sqrt{\frac{mR}{2\alpha}}.$$
При этом мы учли, что вначале при $t=0$ имеем $r=R$ и, следовательно, $\xi=\pi$,
а в конце, когда падение закончится, $r=0$ и, следовательно, $\xi=0$. Но
$$T_\oplus=\frac{2\pi R}{V_\oplus}=\frac{2\pi R}{\sqrt{\frac{\alpha}{mR}}}=
2\pi R\sqrt{\frac{mR}{\alpha}},$$
и окончательно
$$t=\frac{T_\oplus}{4\sqrt{2}}.$$

\subsection*{Задача 3 {\usefont{T2A}{cmr}{b}{n}(8.43)}: Возвращение с Луны}
\begin{figure}[htb]
\centerline{\epsfig{figure=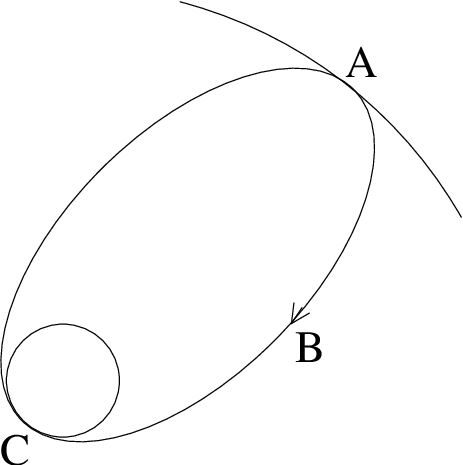,height=3cm}}
\end{figure}
Траектория возвращения --- это часть эллипса $ABC$, причем большая полуось этого эллипса
удовлетворяет уравнению $2a\approx R+r$, где $R$ --- радиус Земли, $r\gg R$ --- 
расстояние Земля--Луна (см. рисунок). Следовательно, в системе Земли, энергия 
космического аппарата должна быть 
$$E=-\frac{\alpha}{2a}\approx -\frac{\alpha}{R+r}.$$
С другой стороны, около Луны (в точке $A$) энергия аппарата
$$E=\frac{mV^2}{2}-\frac{\alpha}{r}.$$ 
Приравняв эти два выражения для энергии, находим кинетическую энергию аппарата
$$\frac{mV^2}{2}=\frac{\alpha}{r}-\frac{\alpha}{R+r}=\frac{\alpha R}{r(r+R)}
\approx \frac{\alpha R}{r^2}=\frac{GmM_\oplus R}{r^2}.$$
Отсюда, учитывая, что орбитальная скорость Луны $V_{\mbox{л}}$ определяется из
$$V^2_{\mbox{л}}=\frac{GM_\oplus}{R},$$
получаем
$$V=\sqrt{\frac{2R}{r}}\,V_{\mbox{л}}.$$
Это скорость космического аппарата относительно Земли. Скорость относительно Луны равна
$$\tilde V=V-V_{\mbox{л}}=-\left (1-\sqrt{\frac{2R}{r}}\right)\,V_{\mbox{л}}
<0.$$
Отрицательный знак показывает, что надо стартовать против движения Луны. Но $\tilde V$
--- это скорость аппарата после преодоления тяготения Луны. Стартовая скорость $V_0$
с поверхности Луны определяется из
$$\frac{mV_0^2}{2}-G\,\frac{mM_{\mbox{л}}}{R_{\mbox{л}}}=\frac{m\tilde V^2}
{2}.$$
Но $$G\,\frac{M_{\mbox{л}}}{R_{\mbox{л}}}=g_{\mbox{л}}R_{\mbox{л}}$$
и окончательно получаем
$$V_0=\sqrt{2g_{\mbox{л}}R_{\mbox{л}}+\tilde V^2}=\sqrt{2g_{\mbox{л}}
R_{\mbox{л}}+\left (1-\sqrt{\frac{2R}{r}}\right)^2V^2_{\mbox{л}}}\approx
2,6~\mbox{км}/\mbox{с}.$$

\subsection*{Задача 4 {\usefont{T2A}{cmr}{b}{n}(8.51)}: Минимальная начальная
скорость при полете на Марс и на Венеру}
Траектория аппарата должна быть эллипсом, который с одной стороны касается орбиты
Земли (в точке $A$), а с другой стороны --- орбиты Марса (в точке $B$). Для такой
траектории (см. рисунок) $2a=r+r_m=2,52~\mbox{а.е.}$ 
\begin{figure}[htb]
\centerline{\epsfig{figure=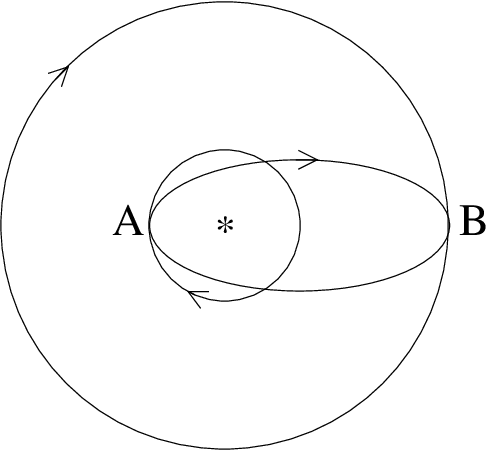,height=4cm}}
\end{figure}

Полная энергия аппарата на такой траектории будет
$$E=-\frac{\alpha}{2a}=-\frac{\alpha}{r+r_m},$$
где $\alpha=GmM_\odot$, $m$ --- масса аппарата, $M_\odot$ --- масса Солнца.
Если скорость аппарата в точке $A$ равна $V$, то
$$\frac{mV^2}{2}-\frac{\alpha}{r}=E=-\frac{\alpha}{r+r_m}.$$
Поэтому
$$\frac{mV^2}{2}=\frac{\alpha}{r}-\frac{\alpha}{r+r_m}=
m\,\frac{GM_\odot}{r}\,\frac{r_m}{r+r_m}=mV_\oplus^2\,\frac{r_m}{r+r_m},$$
где $$V_\oplus=\sqrt{\frac{GM_\odot}{r}}$$
есть орбитальная скорость Земли. Следовательно,
$$V=V_\oplus\sqrt{\frac{2r_m}{r+r_m}}.$$
Это скорость относительно Солнца. Скорость аппарата относительно Земли будет 
$$\tilde V=V-V_\oplus=V_\oplus\left (\sqrt{\frac{2r_m}{r+r_m}}-1\right).$$
Такая скорость должна остаться после преодоления притяжения Земли. Поэтому стартовая
скорость определяется из
$$\frac{mV_0^2}{2}-\frac{GmM_\oplus}{R}=\frac{m\tilde V^2}{2},$$
где $R$ --- радиус Земли. Но $$G\,\frac{M_\oplus}{R}=\frac{V_2^2}{2},$$
где $V_2\approx 11,2~\mbox{км}/\mbox{с}$ --- вторая космическая скорость. 
Следовательно, скорость космического аппарата при выходе из атмосферы Земли для полета 
на Марс должна быть
$$V_0=\sqrt{\tilde V^2+V_2^2}=\sqrt{V_2^2+V_\oplus^2\left (\sqrt{\frac{2r_m}
{r+r_m}}-1\right)^2}\approx 11,6~\mbox{км}/\mbox{с}.$$
\begin{figure}[htb]
\centerline{\epsfig{figure=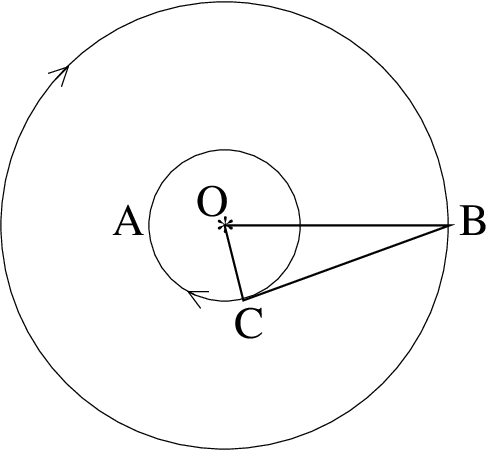,height=3cm}}
\end{figure}

\noindent Продолжительность полета находим по второму закону Кеплера:
$$t=\frac{T}{2}=\frac{T_\oplus}{2}\left(\frac{r+r_m}{2r}\right)^{3/2}\approx
0,7~\mbox{лет},$$
где $T_\oplus=1~\mbox{год}$. За это время Земля переместится из точки $A$ в точку 
$C$ (см. рисунок), причем $\angle COB=0,7\cdot 360^\circ-180^\circ=0,2\cdot 
360^\circ=72^\circ$. Тогда расстояние от Земли до Марса
$$CB=\sqrt{r^2+r_m^2-2rr_m\cos{72^\circ}}\approx 1,5~\mbox{а.е.}$$

При полете на Венеру $2a=r+r_v$ и скорость аппарата относительно Солнца в точке
$A$ должна быть
$$V=V_\oplus\sqrt{\frac{2r_v}{r+r_v}}.$$
Скорость относительно Земли будет $\tilde V=V-V_\oplus<0$, т.~е. аппарат надо
запускать против движения Земли. Стартовая скорость
$$V_0=\sqrt{\tilde V^2+V_2^2}=\sqrt{V_2^2+V_\oplus^2\left (\sqrt{\frac{2r_v}
{r+r_v}}-1\right)^2}\approx 11,5~\mbox{км}/\mbox{с}.$$

\subsection*{Задача 5 {\usefont{T2A}{cmr}{b}{n}(8.42)}: Радиус оптимальной орбиты}
Пусть скорость на круговую орбиту равна $V$. По закону сохранения энергии
$$\frac{m(V+\Delta V)^2}{2}-G\,\frac{mM}{r}=\frac{mV_\infty^2}{2}.$$
Но на круговой орбите
$$m\,\frac{V^2}{r}=G\,\frac{mM}{r^2},$$
т.~е. $$G\,\frac{mM}{r}=mV^2,$$
и получаем
$$(V+\Delta V)^2-2V^2=V_\infty^2,$$
или
$$(\Delta V)^2+2V\Delta V-V^2-V_\infty^2=0.$$
Решаем относительно $\Delta V$: $$\Delta V=-V\pm\sqrt{2V^2+V_\infty^2}.$$ 
Знак минус не подходит, так как при этом $|\Delta V|$ будет слишком большим. 
Следовательно,
$$\Delta V=\sqrt{2V^2+V_\infty^2}-V.$$
Найдем, при каком $V$ имеем минимум:
$$\frac{d(\Delta V)}{dV}=\frac{2V}{\sqrt{2V^2+V_\infty^2}}-1=0$$
дает $2V^2=V_\infty^2$. Поэтому
$$(\Delta V)_{min}=\sqrt{2V_\infty^2}-\frac{V_\infty}{\sqrt{2}}=
\frac{V_\infty}{\sqrt{2}}.$$
Соответствующий оптимальный радиус орбиты определяется из 
$$r=\frac{GM}{V^2}=\frac{GM}{R}\,\frac{R}{V^2}=R\,\frac{V_2^2}{2V^2}=
R\,\frac{V_2^2}{V_\infty^2}.$$
Здесь $R$ --- радиус Земли, $V_2$ --- вторая космическая скорость.

\subsection*{Задача 6 {\usefont{T2A}{cmr}{b}{n}(8.66)}: Падение баллистической
ракеты}
Сравним воображаемые орбиты ракеты и околоземного спутника с такой же массой в 
предположении, что вся масса Земли сосредоточена в ее центре (см. рисунок).
\begin{figure}[htb]
\centerline{\epsfig{figure=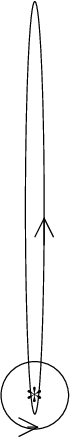,height=4cm}}
\end{figure}

Пусть большая полуось орбиты ракеты $a^\prime$. Для спутника $a=R$, где $R$ --- 
радиус Земли. Но $a=-\alpha/(2E)$. Поэтому
$$\frac{a^\prime}{a}=\frac{E}{E^\prime}.$$
По второму закону Кеплера,
$$T^\prime=T\left(\frac{a^\prime}{a}\right)^{3/2}=T\left(\frac{E}{E^\prime}
\right)^{3/2}.$$
Для спутника имеем
$$m\frac{V_1^2}{R}=G\frac{mM}{R^2}=mg.$$
Т.~е. $V_1=\sqrt{gR}$ и $$T=\frac{2\pi R}{V_1}=2\pi\sqrt{\frac{R}{g}}.$$
Его полная энергия
$$E=\frac{mV_1^2}{2}-G\frac{mM}{R}=\frac{mgR}{2}-mgR=-\frac{1}{2}\,mgR.$$
Для ракеты
$$E^\prime=\frac{mV^2}{2}-G\frac{mM}{R}=\frac{mV^2}{2}-mgR=-\frac{1}{2}\,mgR
\left(2-\frac{V^2}{gR}\right).$$
Следовательно,
$$\frac{E^\prime}{E}=2-\frac{V^2}{gR}$$
и
$$T^\prime=2\pi\sqrt{\frac{R}{g}}\left(2-\frac{V^2}{gR}\right)^{-3/2}\approx
5,5~\mbox{час}.$$
Как ясно из рисунка, время, через которое возвратится баллистическая ракета, 
запущенная с поверхности Земли, мало будет отличаться от $T^\prime$.

\section[Семинар 29]
{\centerline{Семинар 29}}

\subsection*{Задача 1 {\usefont{T2A}{cmr}{b}{n}(8.18)}: Эффект
Пойнтинга--Робертсона}
{\usefont{T2A}{cmr}{m}{it} Первое решение.} 
В единицу времени частица поглощает энергию
$$\Delta E=N\,\frac{\pi a^2}{4\pi r^2}=N\,\frac{a^2}{4r^2},$$
где $a$ --- радиус частицы, $r$ --- радиус ее орбиты. При этом в тангенциальном
направлении импульс не поглощается: $\Delta p_x=0$. В системе частицы  поглощенная
энергия равна
$$\Delta E^\prime=\gamma(\Delta E-V\Delta p_x)\approx \Delta E.$$
Эта энергия в системе частицы переизлучается изотропно: $\Delta p^\prime_x=0$. 
В системе Солнца излучение не изотропно и уносит $x$-компоненту импульса 
$$\Delta \tilde p_x=\gamma\left(\Delta p^\prime_x+\frac{V}{c^2}\Delta 
E^\prime\right)=
\gamma\,\frac{V}{c^2}\,\Delta E^\prime\approx \frac{V}{c^2} \Delta E.$$
Следовательно, в системе Солнца частица в единицу времени теряет $x$-компоненту 
импульса
$$\Delta \tilde p_x=\frac{V}{c^2}\,N\,\frac{a^2}{4r^2}.$$
Поэтому эффективная сила торможения равна
$$F=\frac{V}{c^2}\,N\,\frac{a^2}{4r^2}.$$
Скорость частицы находим из 
$$V^2=G\,\frac{M_\odot}{r},$$
что дает $V\approx 3\cdot 10^4~\mbox{м}/\mbox{с}$ и $F\approx 1,5\cdot 10^{-12}
~\mbox{н}$. 

{\usefont{T2A}{cmr}{m}{it} Второе решение {\usefont{T2A}{cmr}{b}{n}\cite{35}}.}
Будем считать, что падающее на частицу излучение состоит из моноэнергетических фотонов
с частотой $\omega$. Тогда в мгновенно сопутствующей частице системе $S^\prime$ в 
единицу времени частица поглощает импульс 
$$\Delta \vec{p}^{\,\prime}=\frac{\hbar\omega^\prime}{c}\;(n^\prime c 
S^\prime)\,\vec{m}^{\,\prime}= \hbar\omega^\prime n^\prime A^\prime\,
\vec{m}^{\,\prime}.$$
Это и есть изменение импульса частицы в единицу времени, так как из-за изотропности
переизлучения частица не теряет импульс во время переизлучения. Здесь $n^\prime$ ---
концентрация фотонов в падающем потоке, $A^\prime$ --- эффективное поперечное сечение 
частицы, $m^\prime$ --- единичный вектор в направлении движения падающих фотонов.
Следовательно, уравнение движения частицы в системе  $S^\prime$ имеет вид
$$\frac{d\vec{p}^{\,\prime}}{d\tau}=\hbar\omega^\prime n^\prime A^\prime\,
\vec{m}^{\,\prime}.$$
В этой системе энергия частицы не меняется (переизлучается столько же энергии, 
сколько-же падает). Таким образом, 4-сила в системе $S^\prime$ имеет вид
$$F^{\prime\,\mu}=(0,\, \hbar\omega^\prime n^\prime A^\prime\,
\vec{m}^{\,\prime}).$$
В системе Солнца $S$ 4-силу $F^\mu$ находим по преобразованиям Лоренца:
\begin{eqnarray} &&
F^0= \gamma\left(F^{\prime\,0}+\frac{\vec{V}\cdot\vec{F^{\,\prime}}}{c}
\right ), \nonumber \\ &&
\vec{F}=\vec{F}^{\,\prime}+\left [(\gamma-1)\frac{\vec{V}\cdot
\vec{F^{\,\prime}}}{V^2}+\gamma\,\frac{F^0}{c}\right]\vec{V}.
\nonumber
\end{eqnarray}
Это дает
\begin{equation}
\vec{F}=\hbar\omega^\prime n^\prime A^\prime \left [\vec{m}^{\,\prime}+
(\gamma-1)\,\frac{\vec{V}\cdot\vec{m}^{\,\prime}}{V^2}\,\vec{V}\right ].
\label{eq29_1}
\end{equation}
Надо $n^\prime$, $\omega^\prime$ и $\vec{m}^{\,\prime}$ выразить через 
соответствующие величины в системе Солнца $S$. Так как 
$$\left(\frac{\hbar\omega}{c},\, \frac{\hbar\omega}{c}\,\vec{m}^{\,\prime}
\right )$$
есть 4-вектор (4-вектор энергии-импульса фотона), то преобразования Лоренца дают
\begin{eqnarray} &&
\omega^\prime=\gamma\omega\left(1-\frac{\vec{V}\cdot\vec{m}}{c}\right),
\nonumber \\ &&
\omega^\prime \vec{m}^{\,\prime}=\omega \left[\vec{m}-\left (\frac{\gamma}
{c}-(\gamma-1)\frac{\vec{V}\cdot\vec{m}}{V^2}\right )\vec{V}\right ].
\nonumber
\end{eqnarray}
Подставляя второе равенство в (\ref{eq29_1}), получаем после некоторой алгебры
$$\vec{F}=\hbar\omega n^\prime A^\prime\left [\vec{m}-\gamma^2\left(
1-\frac{\vec{V}\cdot\vec{m}}{c}\right)\frac{\vec{V}}{c}\right].$$
$j^\mu=(cn,\,cn\vec{m})$ --- тоже 4-вектор (4-вектор потока фотонов). Поэтому
по преобразованиям Лоренца получаем
$$n^\prime=\gamma n \left(1-\frac{\vec{V}\cdot\vec{m}}{c}\right),$$
и окончательно
\begin{equation}
\vec{F}=\hbar\omega nA^\prime\gamma\left(1-\frac{\vec{V}\cdot\vec{m}}{c}\right)
\left [\vec{m}-\gamma^2\left(
1-\frac{\vec{V}\cdot\vec{m}}{c}\right)\frac{\vec{V}}{c}\right].
\label{eq29_1a}
\end{equation}
В условиях задачи $V/c\ll 1$ поэтому можно положить
$\gamma\approx 1,\;\; A^\prime\approx A=\pi a^2$.  Кроме того,
$$\hbar\omega n c=\frac{N}{4\pi r^2},\;\;\;\vec{V}\perp \vec{m},$$
и (\ref{eq29_1a}) примет вид
$$\vec{F}\approx \frac{Na^2}{4cr^2}\left [\vec{m}-\frac{\vec{V}}{c}\right].$$
Первый член радиальный и соответствует давлению света на частицу. Второй член направлен
против скорости частицы и соответствует силе торможения Пойнтинга--Робертсона. 

\subsection*{Задача 2 {\usefont{T2A}{cmr}{b}{n}(8.50)}: Изменение орбиты
космической станции}
Пусть $H_1=200~\mbox{км}$, $H_2=210~\mbox{км}$ и соответствующие скорости 
космической станции на этих высотах равны $V_1$ и $V_2$. Закон сохранения момента
импульса дает $mV_1(R+H_1)=mV_2(R+H_2)$, где $R$ --- радиус Земли. Поэтому
$$V_2=\frac{R+H_1}{R+H_2}\,V_1.$$
По закону сохранения энергии
$$\frac{mV_1^2}{2}-\frac{\alpha}{R+H_1}=\frac{mV_2^2}{2}-\frac{\alpha}{R+H_2}=
\frac{mV_1^2}{2}\left(\frac{R+H_1}{R+H_2}\right)^2-\frac{\alpha}{R+H_2},$$
где $\alpha=GmM_\oplus$. Отсюда получаем
\begin{equation}
\frac{mV_1^2}{2}=\frac{\alpha(R+H_2)}{(R+H_1)(2R+H_1+H_2)}.
\label{eq29_2}
\end{equation}
Если $V$ --- первоначальная скорость на круговой орбите, то
$$\frac{mV^2}{R+H_1}=\frac{\alpha}{(R+H_1)^2}.$$
Из этого соотношения можно выразить $\alpha$ через $V$ и подставить в 
(\ref{eq29_2}). В результате получим 
$$V_1=\sqrt{\frac{2(R+H_2)}{2R+H_1+H_2}}\;V$$ и
$$\Delta V=V_1-V=\left(\sqrt{\frac{2(R+H_2)}{2R+H_1+H_2}}-1\right)V.$$
Так как $H_1,H_2\ll R$, то $V\approx 7,9~\mbox{км}/\mbox{с}$ --- первая 
космическая скорость,
$$\frac{2(R+H_2)}{2R+H_1+H_2}\approx \left(1+\frac{H_2}{R}\right)
\left(1-\frac{H_1+H_2}{2R}\right)\approx 1+\frac{H_2-H_1}{2R}$$
и
$$\Delta V\approx \left(\sqrt{1+\frac{H_2-H_1}{2R}}-1\right )V\approx
\frac{H_2-H_1}{4R}\,V\approx 3~\mbox{м}/\mbox{с}.$$

\subsection*{Задача 3 {\usefont{T2A}{cmr}{b}{n}(8.49)}: Прирост скорости
в перигее и высота апогея}
{\usefont{T2A}{cmr}{m}{it} Первое решение.} 
Пусть $V$ --- скорость в перигее. Из формул 
$$r_{min}=\frac{p}{1+e},\;\;\;p=\frac{L^2}{m\alpha}=\frac{mV^2r_{min}^2}
{\alpha},\;\;\;\frac{mV_2^2}{2}=\frac{\alpha}{R},$$
где $V_2$ --- вторая космическая скорость, $R$ --- радиус Земли, получаем для 
эксцентриситета $e$:
$$1+e=\frac{mV^2r_{min}}{\alpha}=\frac{2r_{min}}{R}\,\frac{V^2}{V_2^2}.$$
Поэтому, так как $r_{max}=p/(1-e)$, будем иметь
$$\frac{r_{min}}{r_{max}}=\frac{1-e}{1+e}=\frac{2}{1+e}-1=\frac{V_2^2}{V^2}\,
\frac{R}{r_{min}}-1.$$
Отсюда
$$\frac{r_{min}\,\Delta r_{max}}{r_{max}^2}=2\frac{V_2^2\,\Delta V}{V^3}\,
\frac{R}{r_{min}}$$ и 
$$\frac{\Delta r_{max}}{r_{max}}=\frac{r_{max}}{r_{min}}\,\frac{2\Delta V}
{V}\,\frac{V_2^2\,R}{V^2\,r_{min}}=\frac{2\Delta V}{V}\,\frac{r_{max}}
{r_{min}}\,\left(\frac{r_{min}}{r_{max}}+1\right)=\frac{2\Delta V}{V}\left(
1+\frac{r_{max}}{r_{mim}}\right).$$
Поэтому
$$\frac{\Delta V}{V}=\frac{1}{2}\,\frac{\Delta r_{max}}{r_{max}}\left(
1+\frac{r_{max}}{r_{mim}}\right)^{-1}.$$
Но
$$\left(1+\frac{r_{max}}{r_{mim}}\right)^{-1}=1-\left(1+\frac{r_{min}}
{r_{max}}\right)^{-1}=1-\frac{V^2\,r_{min}}{V_2^2\,R},$$
и окончательно получаем
\begin{equation}
\frac{\Delta V}{V}=\frac{1}{2}\,\frac{\Delta r_{max}}{r_{max}}\left(
1-\frac{V^2\,r_{min}}{V_2^2\,R}\right).
\label{eq29_3}
\end{equation}

{\usefont{T2A}{cmr}{m}{it} Второе решение.}
Из $r_{min}+r_{max}=2a=-\alpha/E$ имеем
$$\Delta r_{max}=\frac{\alpha}{E^2}\,\Delta E=\frac{\alpha}{E^2}\,mV\,
\Delta V=\frac{\alpha m\,V^2}{E^2}\,\frac{\Delta V}{V}=\frac{\alpha}{E}\,
\frac{m\,V^2}{E}\,\frac{\Delta V}{V},$$
так как $$E=\frac{mV^2}{2}-\frac{\alpha}{r_{min}}.$$ 
Следовательно,
$$\frac{\Delta V}{V}=\frac{E}{\alpha}\,\frac{E}{mV^2}\,\Delta r_{max}.$$
Но
$$\frac{E}{\alpha}=-\frac{1}{r_{min}}\left (1-\frac{mV^2\,r_{min}}{2\alpha}
\right)=-\frac{1}{r_{min}}\left (1-\frac{V^2\,r_{min}}{V_2^2\,R}\right ),$$
и из
$$r_{max}=2a-r_{min}=-\frac{\alpha}{E}-r_{min}=-\frac{r_{min}}{E}\left(
\frac{\alpha}{r_{min}}+E\right)=-\frac{r_{min}}{E}\,\frac{MV^2}{2}$$
получаем
$$\frac{E}{mV^2}=-\frac{r_{min}}{2r_{max}}.$$
Таким образом, приходим к формуле (\ref{eq29_3}).

\subsection*{Задача 4 {\usefont{T2A}{cmr}{b}{n}(8.53)}: Баллистическая ракета}
{\usefont{T2A}{cmr}{m}{it} Первое решение.} 
Траектория ракеты --- это часть эллипса $ACB$ (см. рисунок). 
\begin{figure}[htb]
\centerline{\epsfig{figure=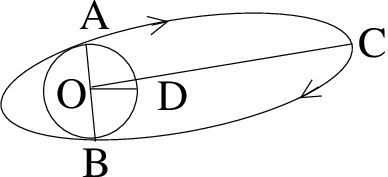,height=3cm}}
\end{figure}

\noindent Пусть уравнение эллипса имеет вид $$\frac{p}{r}=1-e\cos{\varphi},$$
где угол $\varphi$ отсчитывается от оси эллипса $OC$ и эксцентриситет $e\ne 0$. 
В точках $A$ и $B$, $r=R$, где $R$ --- радиус Земли. При этом в точке $A$ 
$\varphi=90^\circ-\beta$, а в точке $B$  $\varphi=270^\circ-\beta$, где 
$\beta$ --- возможный угол наклона оси эллипса $OC$ по отношению к экваториальной 
плоскости. Поэтому должны иметь
$$\frac{p}{R}=1-e\cos{(90^\circ-\beta)}=1+e\sin{\beta},\;\;\;
\mbox{и}\;\;\;\frac{p}{R}=1-e\cos{(270^\circ-\beta)}=1-e\sin{\beta}.$$
Следовательно, $\beta=0$ и $p=R$. С другой стороны,
$$p=\frac{L^2}{m\alpha}=\frac{V_0^2\cos^2{\theta}\,R^2}{GM_\oplus},$$
так как момент импульса ракеты $L=mV_0\cos{\theta}$. Таким образом, должны иметь
$$\frac{V_0^2\cos^2{\theta}\,R^2}{GM_\oplus}=R$$ и $$V_0^2\cos^2{\theta}=
\frac{GM_\oplus}{R}=gR.$$
Окончательно $V_0\cos{\theta}=\sqrt{gR}$.

{\usefont{T2A}{cmr}{m}{it} Второе решение.}
Согласно закону сохранения энергии, величина скорости ракеты $V=V_0$ одна и та же
в точках $A$ и $B$. Сохраняющийся вектор Гамильтона, так как $\dot\varphi<0$,
имеет вид 
$$\vec{h}=\vec{V}+\frac{\alpha}{L}\,\vec{e}_\varphi.$$
В точках $A$ и $B$ имеем следующую картину (см. рисунок). 
\begin{figure}[htb]
\centerline{\epsfig{figure=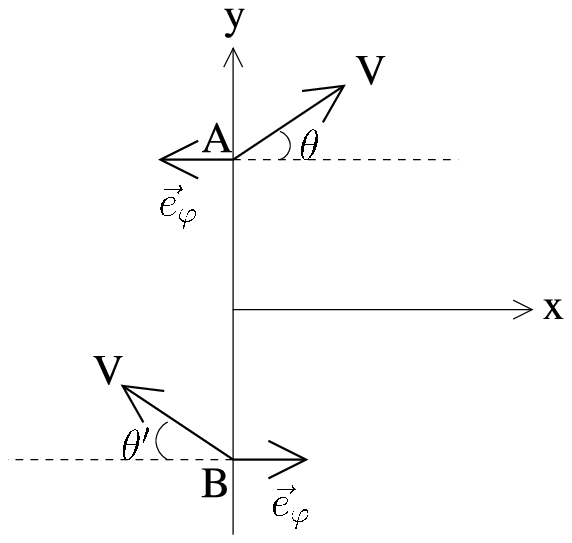,height=4cm}}
\end{figure}

\noindent Сохранения $y$-компонент вектора Гамильтона $h_{A\,y}=h_{B\,y}$ дает 
$V\sin{\theta}=V\sin{\theta^\prime}$. Следовательно, $\theta=\theta^\prime$. 
Что касается $x$-компонент вектора Гамильтона, из рисунка ясно, что $$h_{A\,x}=
V\cos{\theta}-\frac{\alpha}{L}$$ и $$h_{B\,x}=-V\cos{\theta}+\frac{\alpha}{L}
=-h_{A\,x}.$$ 
Поэтому закон сохранения $x$-компонент вектора Гамильтона $h_{B\,x}=h_{A\,x}$
в данном случае означает, что $h_{B\,x}=h_{A\,x}=0$. Следовательно,
$$V\cos{\theta}=\frac{\alpha}{L}=\frac{GM_\oplus}{RV\cos{\theta}}$$
и $$V\cos{\theta}=\sqrt{\frac{GM_\oplus}{R}}=\sqrt{gR}.$$

\subsection*{Задача 5 {\usefont{T2A}{cmr}{b}{n}(8.74)}: Частота малых колебаний
грузиков}
Направим ось $x$ внизу (см. рисунок).
\begin{figure}[htb]
\centerline{\epsfig{figure=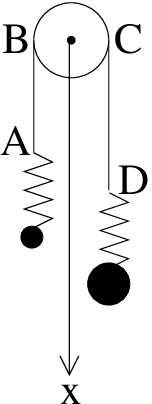,height=4cm}}
\end{figure}

\noindent Уравнения движения грузиков имеют вид
$$m\,\ddot x_1=mg-k(x_1-AB-l_1),\;\;\;M\,\ddot x_2=Mg-k(x_2-CD-l_2),$$
где $l_1,l_2$ --- длины недеформированных пружин. Так как нить предполагается 
безмассовой, силы, приложенные со стороны пружин к ее концам $A$ и $D$, должны быть 
равны друг другу (иначе нить начнет двигаться с бесконечным ускорением): 
$k(x_1-AB-l_1)=k(x_2-CD-l_2)$. Следовательно, 
\begin{equation}
AB-CD=x_1-x_2+l_2-l_1.
\label{eq29_5a}
\end{equation}
С другой стороны, нерастяжимость нити означает
\begin{equation}
AB+CD=L=\mathrm{const},
\label{eq29_5b}
\end{equation}
так как участок нити $BC$ имеет постоянную длину $\pi R$. Из (\ref{eq29_5a}) и
(\ref{eq29_5b}) находим
$$AB=\frac{1}{2}(x_1-x_2+L+l_2-l_1).$$
Поэтому
$$x_1-AB-l_1=\frac{x_2+x_2}{2}-\frac{L+l_1+l_2}{2},$$
и уравнения движения грузиков примут вид 
$$m\,\ddot x_1=mg-\frac{k}{2}(x_1+x_2-L-l_1-l_2),\;\;\;
M\,\ddot x_1=Mg-\frac{k}{2}(x_1+x_2-L-l_1-l_2).$$
Если первое из этих уравнений умножить на $M$, а второе --- на $m$, а потом взять 
сумму и разность полученных уравнений, получим систему:
\begin{eqnarray} &&
mM\,\ddot X=2mMg-\frac{k}{2}(m+M)(X-L-l_1-l_2),\nonumber \\ &&
mM\,\ddot x=-\frac{k}{2}(M-m)(X-L-l_1-l_2),
\label{eq29_5c}
\end{eqnarray}
где $X=x_1+x_2$ и $x=x_1-x_2$. Вводя приведенную массу $\mu=mM/(m+M)$, первое
уравнение можно переписать так:
$$\ddot X+\frac{k}{2\mu}\,X=2g+\frac{k}{2\mu}\,(L+l_1+l_2).$$
Решение этого уравнения с начальными условиями $X(0)=L+l_1+l_2$, $\dot X(0)=0$
имеет вид
\begin{equation}
X(t)=L+l_1+l_2+\frac{4\mu g}{k}(1-\cos{\omega t}),
\label{eq29_5d}
\end{equation}
где 
$$\omega=\sqrt{\frac{k}{2\mu}}.$$ 
С учетом (\ref{eq29_5d}) второе уравнение из системы (\ref{eq29_5c}) примет вид
$$\ddot x=-2g\,\frac{M-m}{M+m}\,(1-\cos{\omega t}).$$
Решение этого уравнения с начальными условиями $x(0)=x_0$, $\dot x(0)=0$
имеет вид
\begin{equation}
x(t)=x_0-at^2-\frac{2a}{\omega^2}\,\cos{\omega t},
\label{eq29_5e}
\end{equation}
где $$a=\frac{M-m}{M+m}\,g.$$
Интересно рассмотреть пределы (\ref{eq29_5d}) и (\ref{eq29_5e}) при $k\to\infty$,
когда пружины переходят в нерастяжимую нить, и $\omega\to\infty$:
\begin{eqnarray} &&
X(t)=x_1(t)+x_2(t)=L+l_1+l+2=x_{10}+x_{20},\nonumber \\ &&
x(t)=x_1(t)-x_2(t)=x_0-at^2=x_{10}-x_{20}-at^2, \nonumber
\end{eqnarray}
что соответствует ожидаемому
$$x_1(t)=x_{10}-\frac{at^2}{2},\;\;\; x_2(t)=x_{20}+\frac{at^2}{2}.$$ 

\section[Семинар 30]
{\centerline{Семинар 30}}

\subsection*{Задача 1 {\usefont{T2A}{cmr}{b}{n}(8.73)}: Как изменится скорость
хода часов?}
Пусть $I_1$ --- момент инерции маховика, $I_2$ --- момент инерции корпуса часов
относительно оси вращения маховика, а $\phi_1$ и $\phi_2$ --- углы поворота маховика
и корпуса часов соответственно. Тогда уравнения движения имеют вид
$$I_1\ddot \phi_1=-k(\phi_1-\phi_2),\;\;I_2\ddot \phi_2=-k(\phi_2-\phi_1).$$
Вводя относительный угол поворота $\phi=\phi_2-\phi_1$, из этих уравнений можно 
получить $I_1I_2\ddot\phi=-k(I_1+I_2)\phi$, или
$$\ddot\phi+\omega^2\phi=0,$$
где
$$\omega^2=k\,\frac{I_1+I_2}{I_1I_2},$$
и соответственно,
$$T=\frac{2\pi}{\omega}=2\pi\sqrt{\frac{1}{k}\,\frac{I_1I_2}{I_1+I_2}}.$$
Когда часы закреплены, можно считать $I_1/I_2\approx 0$ и
$$T_0\approx 2\pi\sqrt{\frac{I_1}{k}}.$$
Когда часы не закреплены,
$$\frac{T^2}{T_0^2}\approx\frac{I_2}{I_1+I_2}<1.$$
Следовательно, $T<T_0$ и незакрепленные часы идут быстрее. Если $I_1\ll I_2$,
$$\frac{\Delta T}{T_0}=\frac{T-T_0}{T_0}\approx \sqrt{\frac{1}{1+I_1/I_2}}-1
\approx -\frac{I_1}{2I_2}.$$

\subsection*{Задача 2 {\usefont{T2A}{cmr}{b}{n}(8.72)}: Период малых продольных
колебаний осциллятора}
Пусть $x_1$ --- координата массы $m$, $x_2$ --- координата массы $M$. Уравнения
движения будут иметь вид
$$m\ddot x_1=k(x_2-x_1-L),\;\;\;M\ddot x_1=-k(x_2-x_1-L),$$
где $L$ --- длина недеформированной пружины. Вводя относительную координату 
$x=x_2-x_1$, из этих уравнений можно получить $mM\ddot x=-k(m+M)(x-L)$, или
$$\ddot x+\omega^2 x=\frac{kL}{\mu},$$
где $\mu=mM/(m+M)$ --- приведенная масса и $\omega^2=k/\mu$. Следовательно, период
колебаний 
$$T=\frac{2\pi}{\omega}=2\pi\sqrt{\frac{\mu}{k}}=2\pi\sqrt{\frac{1}{k}\,
\frac{mM}{m+M}}.$$

\subsection*{Задача 3 {\usefont{T2A}{cmr}{b}{n}(8.72)}: Через какое время
столкнутся две точки?}
Задача эквивалентна с задачей падения точки с приведенной массой $\mu=mM/(m+M)$ на 
центр поля притяжения $U(r)=-G\frac{mM}{r}$. А эта последняя решается также, как 
задача о падении Земли на Солнце (семинар 28, задача 2). Следовательно, две точки 
столкнутся через время
$$t=\frac{T_0}{4\sqrt{2}},$$
где $T_0$ --- период обращения точки с массой $\mu$ на круговой орбите радиуса $R$.
Но $T_0=2\pi R/V$, где скорость точки $V$ определяется из условия (второй закон 
Ньютона)
$$\mu\,\frac{V^2}{R}=G\,\frac{mM}{R^2}.$$
Отсюда
$$V^2=\frac{GmM}{\mu R}=\frac{G}{R}(m+M).$$
Таким образом,
$$T_0=\frac{2\pi R}{\sqrt{\frac{G}{R}(m+M)}}$$ и
$$t=\frac{\pi R}{2\sqrt{2}}\,\sqrt{\frac{R}{G(m+M)}}.$$
Если $m=M=1~\mbox{кг}$, $R=1~\mbox{м}$, то численно получаем
$t\approx 0,96\cdot 10^5~\mbox{с}\approx 27~\mbox{час}$.

\subsection*{Задача 4 {\usefont{T2A}{cmr}{b}{n}(8.85)}: Полная масса системы
"двойная звезда"}
Вектор $\vec{r}=\vec{r}_2-\vec{r}_1$ описывает эллипс, который соответствует
движению точки с массой $\mu=m_1m_2/(m_1+m_2)$ в поле $U(r)=-Gm_1m_2/r$.
$r_{min}$ и $r_{max}$ этого эллипса и будут минимальным и максимальным расстоянием
между звездами. Но для эллипса $2a=r_{min}+r_{max}$, а, согласно второму закону
Кеплера, 
$$\frac{a\sqrt{a}}{T}=\frac{R\sqrt{R}}{T_0},$$
где $T_0$ --- период обращения точки на круговой орбите радиуса $R$. Определяя
скорость точки из второго закона Ньютона
$$\mu\,\frac{V^2}{R}=G\,\frac{m_1m_2}{R^2},$$
находим
$$T_0=\frac{2\pi R}{V}=\frac{2\pi R}{\sqrt{\frac{G}{R}(m_1+m_2)}}.$$
Поэтому
$$\frac{R\sqrt{R}}{T_0}=\frac{1}{2\pi}\sqrt{G(m_1+m_2)},$$
и второй закон Кеплера перепишется так:
$$\frac{a\sqrt{a}}{T}=\frac{1}{2\pi}\sqrt{G(m_1+m_2)}.$$
Отсюда
$$m_1+m_2=\frac{4\pi^2}{G}\,\frac{a^3}{T^2}=\frac{\pi^2}{2G}\,\frac{(r_{min}
+r_{max})^3}{T^2}.$$

\subsection*{Задача 5 {\usefont{T2A}{cmr}{b}{n}(8.96)}: Сечение рассеяния
электрона на протоне}
{\usefont{T2A}{cmr}{m}{it} Первое решение.} 
На рисунке видно, что угол рассеяния $\theta=2\varphi_0-\pi$. 
\begin{figure}[htb]
\centerline{\epsfig{figure=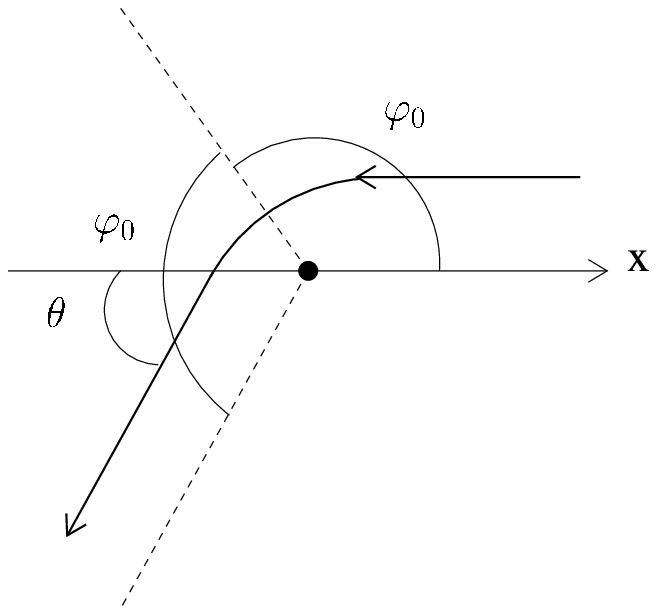,height=5cm}}
\end{figure}

\noindent Чтобы найти $\varphi_0$, найдем уравнение траектории. Момент импульса
$L=mV_\infty\rho$, где $\rho$ --- прицельный параметр. Поэтому закон сохранения 
энергии запишется как 
$$\frac{mV_\infty^2}{2}=\frac{m\dot{r}^2}{2}+\frac{L^2}{2mr^2}-\frac{e^2}{r}=
\frac{m\dot{r}^2}{2}+\frac{mV_\infty^2\rho^2}{2r^2}-\frac{e^2}{r}.$$
Поделим это на $L^2=m^2\dot{\varphi}^2r^4$ и учтем, что 
$$\frac{\dot{r}}{\dot{\varphi}}=\frac{dr}{d\varphi}.$$
В результате получим
$$\frac{mV_\infty^2}{2L^2}=\frac{1}{2mr^4}\left(\frac{dr}{d\varphi}\right)^2+
\frac{mV_\infty^2\rho^2}{2L^2r^2}-\frac{e^2}{L^2r}.$$
Далее учтем, что $L=mV_\infty\rho$ и $$\frac{1}{r^2}\,\frac{dr}{d\varphi}=
\frac{ds}{d\varphi},$$
где $s=1/r$. Это дает
$$\frac{1}{2m\rho^2}=\frac{1}{2m}\left(\frac{ds}{d\varphi}\right)^2+\frac{s^2}
{2m}-\frac{e^2}{m^2V_\infty^2\rho^2}\,s.$$
Продифференцируем это выражение по $\varphi$:
$$0=\frac{1}{m}\left(\frac{ds}{d\varphi}\right)\frac{d^2s}{d\varphi^2}+
\frac{1}{m}\,s\,\frac{ds}{d\varphi}-\frac{e^2}{m^2V_\infty^2\rho^2}\,
\frac{ds}{d\varphi}.$$
Но $ds/d\varphi\ne 0$, поэтому получаем
$$\frac{d^2s}{d\varphi^2}+s=\frac{e^2}{m^2V_\infty^2\rho^2}.$$
Общее решение этого уравнения имеет вид
$$s(\varphi)=A\cos{\varphi}+B\sin{\varphi}+\frac{e^2}{m^2V_\infty^2\rho^2}.$$
Когда $\varphi=0$, $r=\infty$ и $s=0$. Поэтому
$$A=-\frac{e^2}{m^2V_\infty^2\rho^2}.$$
Когда электрон находится очень далеко, он движется практически по прямой с прицельным 
параметром $\rho$. Поэтому для маленьких углов $\varphi$, $\rho\approx 
r\sin{\varphi}\approx r\varphi$. Или $s\approx\varphi/\rho$. Отсюда
$ds/d\varphi=1/\rho$ при $\varphi=0$. Это условие определяет константу $B$:
$$B=\frac{1}{\rho}.$$
Окончательно уравнение траектории получается такой:
$$s(\varphi)=A(\cos{\varphi}-1)+\frac{1}{\rho}\,\sin{\varphi}.$$
Когда $r=r_{min}$, тогда $s=s_{max}$ и $ds/d\varphi=0$. Таким образом, 
$\varphi_0$ определяется из условия
$$-A\sin{\varphi_0}+\frac{1}{\rho}\,\cos{\varphi_0}=0.$$
Отсюда $$\ctg{\varphi_0}=A\rho=-\frac{e^2}{mV_\infty^2\rho}.$$
Поэтому
\begin{equation}
\tg{\frac{\theta}{2}}=\tg{\left(\varphi_0-\frac{\pi}{2}\right)}=-
\ctg{\varphi_0}=\frac{e^2}{mV_\infty^2\rho}
\label{eq30_5a}
\end{equation}
и $\theta\ge 90^\circ$, только если прицельный параметр 
$$\rho\le \frac{e^2}{mV_\infty^2}.$$
Следовательно, сечение рассеяния
$$\sigma=\pi\rho_{max}^2=\frac{\pi e^4}{(mV_\infty^2)^2}.$$
Но $mV_\infty^2=2T$, где $T$ --- начальная кинетическая энергия электрона.
Поэтому окончательно
$$\sigma=\frac{\pi e^4}{4T^2}.$$

Теперь рассмотрим ситуацию, когда протон не закреплен. В системе центра инерции
$$\vec{r}_e=\frac{m_p}{m_p+m_e}\,\vec{r} \;\;\mbox{и}\;\;
\vec{r}_p=-\frac{m_e}{m_p+m_e}\,\vec{r},$$
где $\vec{r}=\vec{r}_e-\vec{r}_p$. Это показывает, что $\vec{r}_e$ поворачивается
на такой же угол, как $\vec{r}$ (в системе центра инерции). Но угол ``рассеяния'' для
$\vec{r}$  определяется как угол рассеяния эффективной частицы массы $\mu=\frac{m_e
m_p}{m_p+m_e}$ в поле $U=e^2/r$. Если этот угол рассеяния $\theta$, то, как мы уже
нашли,
\begin{equation}
\tg{\frac{\theta}{2}}=\frac{e^2}{\mu V_\infty^2\rho}=\frac{m_p+m_e}{m_p}\,
\frac{e^2}{2\rho T},
\label{eq30_5b}
\end{equation}
так как до рассеяния $\vec{V}_\infty=\vec{V}_{e\infty}-\vec{V}_{p\infty}=
\vec{V}_{e\infty}$.
Но здесь угол рассеяния $\theta$ дан в системе центра инерции, а нас интересует 
угол рассеяния в лабораторной системе $\theta_1$. Какая связь между ними?

Скорость системы центра масс 
$$\vec{V}=\frac{m_e\vec{V}_{e\infty}+m_p\vec{V}_{p\infty}}{m_p+m_e}=
\frac{m_e\vec{V}_{e\infty}}{m_p+m_e}.$$
Поэтому скорость электрона в системе центра инерции до рассеяния будет
$$\vec{u}_e=\vec{V}_{e\infty}-\vec{V}=\frac{m_p\vec{V}_{e\infty}}{m_p+m_e}.$$
В системе центра инерции скорость электрона не меняется по величине, только 
поворачивается на угол $\theta$. Поэтому скорость электрона после рассеяния в этой 
системе будет
$$\vec{u}_e^\prime= \frac{m_pV_{e\infty}}{m_p+m_e}\,\vec{n},$$
где $n_x=-\cos{\theta},\;n_y=-\sin{\theta}$. В лабораторной системе скорость 
электрона после рассеяния  будет
$$\vec{V}_e^\prime=\vec{u}_e^\prime+\vec{V}=\vec{V}+\frac{m_pV_{e\infty}}
{m_p+m_e}\,\vec{n}.$$
Если $\theta_1$ --- угол рассеяния в лабораторной системе, то
$$\tg{\theta_1}=\frac{-\vec{V}_{ey}^\prime}{-\vec{V}_{ex}^\prime}=
\frac{\frac{m_pV_{e\infty}}{m_p+m_e}\,\sin{\theta}}{\frac{m_eV_{e\infty}}
{m_p+m_e}+\frac{m_pV_{e\infty}}{m_p+m_e}\,\cos{\theta}}=\frac{m_p\sin{\theta}}
{m_e+m_p\cos{\theta}},$$
так как 
$$V_x=-\frac{m_eV_{e\infty}}{m_p+m_e},\;\;V_y=0.$$
Следовательно,
\begin{equation}
\tg{\theta_1}=\frac{m_p\sin{\theta}}{m_e+m_p\cos{\theta}}.
\label{eq30_5c}
\end{equation}
Нас интересует рассеяние на угол $\theta_1>90^\circ$. Максимальному прицельному 
параметру соответствует $\theta_1=90^\circ$. Тогда $\tg{\theta_1}=\infty$, и из
(\ref{eq30_5c}) получаем
$$\cos{\theta}=-\frac{m_e}{m_p}.$$
Но 
$$\cos{\theta}=\cos^2{\frac{\theta}{2}}-\sin^2{\frac{\theta}{2}}=
\frac{\cos^2{\frac{\theta}{2}}-\sin^2{\frac{\theta}{2}}}
{\cos^2{\frac{\theta}{2}}-\sin^2{\frac{\theta}{2}}}=
\frac{1-\tg^2{\frac{\theta}{2}}}{1+\tg^2{\frac{\theta}{2}}},$$
и из
$$\frac{1-\tg^2{\frac{\theta}{2}}}{1+\tg^2{\frac{\theta}{2}}}=-\frac{m_e}
{m_p}$$ получаем
$$\tg^2{\frac{\theta}{2}}=\frac{m_p+m_e}{m_p-m_e}.$$
С другой стороны, из (\ref{eq30_5b}),
$$\tg^2{\frac{\theta}{2}}=\frac{e^4}{4T^2\rho_{max}^2}\,\frac{(m_p+m_e)^2}
{m_p^2}.$$
Следовательно,
$$\rho_{max}^2=\frac{e^4}{4T^2}\,\frac{(m_p+m_e)^2}{m_p^2}\,
\frac{m_p-m_e}{m_p+m_e}=\frac{e^4}{4T^2}\left(1-\frac{m_e^2}{m_p^2}\right).$$
Поэтому сечение
$$\sigma=\pi\rho_{max}^2=\frac{\pi e^4}{4T^2}\left(1-\frac{m_e^2}{m_p^2}
\right)=\sigma_0\left(1-\frac{m_e^2}{m_p^2}\right),$$
где $\sigma_0$ --- сечение рассеяния для закрепленного протона. Формула для 
$\sigma_0$ дана в системе СГСЭ. Поэтому, чтобы найти численное значение, надо 
подставить $e=4,8\cdot 10^{-10}~\mbox{Фр}$ и $T=10~\mbox{кЭв}=1,6\cdot 
10^{-8}~\mbox{эрг}$. В результате получим $ \sigma_0\approx 1,6\cdot 
10^{-22}~\mbox{см}^2=160~\mbox{б}$.

{\usefont{T2A}{cmr}{m}{it} Второе решение.} 
Уравнение (\ref{eq30_5a}), который играет центральную роль при рассмотрении рассеяния
в кулоновском поле, можно получить из закона сохранения вектора Гамильтона. При 
рассеянии, изображенной на рисунке, $\dot\varphi>0$ и вектор Гамильтона имеет вид
$$\vec{h}=\vec{V}-\frac{\alpha}{L}\,\vec{e}_\varphi,$$ где $\alpha=e^2$.
До рассеяния  $h_x(-\infty)=-V$, так как $\vec{e}_{\varphi x}
(-\infty)=0$. После рассеяния $h_x(\infty)=-V\cos{\theta}-
\frac{\alpha}{L}\,\sin{\theta}$, так как $\vec{e}_{\varphi x}(\infty)=
\sin{\theta}$. Но $h_x$ не меняется, так как вектор Гамильтона сохраняется 
в кулоновском поле. Следовательно, $h_x(\infty)=h_x(-\infty)$, что дает
$$V(1-\cos{\theta})=\frac{\alpha}{L}\,\sin{\theta}.$$
Отсюда
$$\tg{\frac{\theta}{2}}=\frac{1-\cos{\theta}}{\sin{\theta}}=\frac{\alpha}
{LV}=\frac{\alpha}{mV^2\rho}=\frac{\alpha}{2T\rho}.$$

\subsection*{Задача 6 {\usefont{T2A}{cmr}{b}{n}(8.95)}: Рассеяние быстрых
электронов}
Пусть угол рассеяния $\theta$ (см. рисунок) маленький. 
\begin{figure}[htb]
\centerline{\epsfig{figure=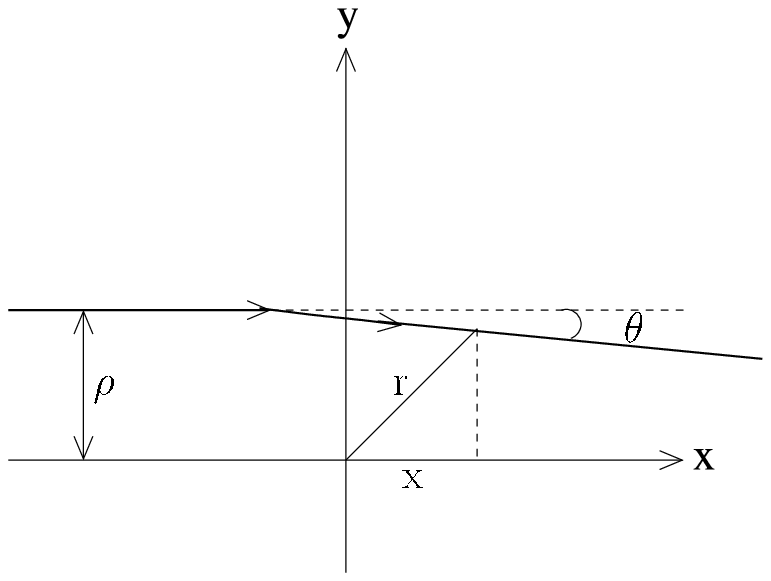,height=5cm}}
\end{figure}

В лабораторной системе
$$\tg{\theta}=\frac{p_{1y}}{p_1}\approx \theta,$$
где $\vec{p}_1$ --- импульс частицы после рассеяния. При этом $p_1\approx p_0$,
где $\vec{p}_0$ --- импульс частицы до рассеяния. Так как до рассеяния $p_{0y}=0$,
будем иметь 
$$p_{1y}=\Delta p_y=\int\limits_{-\infty}^\infty F_y\,dt=-\int\limits_{-\infty}
^\infty\frac{y}{r}\frac{\alpha}{r}\,dt.$$
При маленьких углах рассеяния в первом приближении $y\approx \rho$ все время, а
проекция скорости частицы на ось $x$ не меняется и равна первоначальной скорости 
$V_0$. Поэтому $dt\approx dx/V_0$ и 
$$p_{1y}\approx -\int\limits_{-\infty}^\infty\frac{\alpha\rho}{r^2}\,\frac{dx}
{V_0}\approx -\frac{\alpha\rho}{V_0}\int\limits_{-\infty}^\infty\frac{dx}{x^2+
\rho^2}=-\frac{\alpha}{V_0}\left .\arctg\frac{x}{\rho}\right|_{-\infty}
^\infty=-\frac{\alpha\pi}{V_0}.$$
Следовательно, угол рассеяния
$$\theta\approx\frac{p_{1y}}{p_1}\approx -\frac{\alpha\pi}{p_0V_0}$$
не зависит от прицельного параметра $\rho$.

\section[Семинар 31]
{\centerline{Семинар 31}}

\subsection*{Задача 1 {\usefont{T2A}{cmr}{b}{n}(8.90)}: Рассеяние на абсолютно
упругой сфере}
\begin{figure}[htb]
\centerline{\epsfig{figure=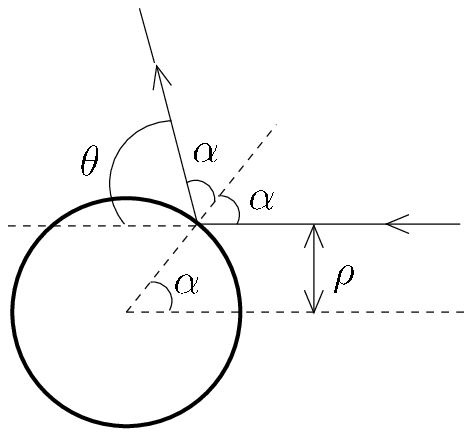,height=4cm}}
\end{figure}
\noindent Угол рассеяния (см. рисунок) $\theta=\pi-2\alpha$. Следовательно,
$$\cos{\frac{\theta}{2}}=\cos{\left(\pi-\alpha\right)}=\sin{\alpha}=
\frac{\rho}{R}.$$
Поэтому, если $\rho\le R$,
$$\theta=2\arccos{\frac{\rho}{R}}.$$
Если $\rho> R$, рассеяние не происходит.

\subsection*{Задача 2 {\usefont{T2A}{cmr}{b}{n}(8.91)}: Сечение рассеяния при
упругом столкновении частицы со сферой}
Угол рассеяния в системе центра масс равен углу рассеяния эффективной частицы с массой
$\mu=\frac{m_1m_2}{m_1+m_2}$ на абсолютно упругой сфере радиусом $R$.
Согласно предыдущей задаче,
$$\theta=2\arccos{\frac{\rho}{R}}.$$
Рассеянию в лабораторной системе на угол $\theta_1=\frac{\pi}{2}$ соответствует
угол рассеяния $\theta$ в системе центра масс такой, что (см. решение задачи 5
предыдущего семинара)
$$\cos{\theta}=-\frac{m_1}{m_2}.$$
Следовательно,
$$\frac{\rho}{R}=\cos{\frac{\theta}{2}}=\sqrt{\frac{1}{2}(1+\cos{\theta})}=
\sqrt{\frac{1}{2}\left (1-\frac{m_1}{m_2}\right )}.$$
Поэтому
$$\rho_{max}^2=\frac{R^2}{2}\left (1-\frac{m_1}{m_2}\right )$$ и 
$$\sigma=\pi\rho_{max}^2=\frac{\pi R^2}{2}\left (1-\frac{m_1}{m_2}
\right )=\frac{\pi R^2}{4},$$
так как $m_1=m$ и $m_2=2m$.

\subsection*{Задача 3 {\usefont{T2A}{cmr}{b}{n}(8.102)}: Зависимость угла
рассеяния от прицельного параметра}
Закон сохранения энергии позволяет написать
$$\frac{mV_\infty^2}{2}=\frac{m\dot{r}^2}{2}+\frac{mV_\infty^2\rho^2}{2r^2}+
\frac{\alpha}{r}+\frac{\beta}{r^2}.$$
Поделим это на $L^2=m^2\dot{\varphi}^2r^4$ и учтем 
$$\frac{\dot{r}}{\dot{\varphi}}=\frac{dr}{d\varphi}.$$
В результате получим
$$\frac{mV_\infty^2}{2L^2}=\frac{1}{2mr^4}\left(\frac{dr}{d\varphi}\right)^2+
\frac{mV_\infty^2\rho^2}{2L^2r^2}+\frac{\alpha}{rL^2}+\frac{\beta}{L^2r^2}.$$
Так как $L=mV_\infty\rho$, это выражение можно переписать как
$$\frac{1}{2m\rho^2}=\frac{1}{2m}\left(\frac{ds}{d\varphi}\right)^2+\frac{s^2}
{2m}+\frac{\alpha s}{m^2V_\infty^2\rho^2}+\frac{\beta s^2}{m^2V_\infty^2
\rho^2},$$
где $s=1/r$. Продифференцируем полученное равенство по $\varphi$:
$$0=\frac{1}{m}\left(\frac{ds}{d\varphi}\right)\frac{d^2s}{d\varphi^2}+
\frac{s}{m}\frac{ds}{d\varphi}+\frac{\alpha}{m^2V_\infty^2\rho^2}\frac{ds}
{d\varphi}+\frac{2\beta s}{m^2V_\infty^2\rho^2}\frac{ds}{d\varphi}.$$
так как $\frac{ds}{d\varphi}\ne 0$, отсюда получаем дифференциальное уравнение
$$\frac{d^2s}{d\varphi^2}+\left(1+\frac{2\beta}{mV_\infty^2\rho^2}\right )s=
-\frac{\alpha}{mV_\infty^2\rho^2}.$$
Общее решение этого уравнения имеет вид
$$s(\varphi)=A\cos{(\epsilon\varphi)}+B\sin{(\epsilon\varphi)}-
\frac{\alpha}{mV_\infty^2\rho^2\epsilon^2},$$
где
$$\epsilon=\sqrt{1+\frac{2\beta}{mV_\infty^2\rho^2}}=
\sqrt{1+\frac{\beta}{T\rho^2}}.$$
В этой формуле $T$ --- кинетическая энергия падающей частицы.

Начальное условие $s(0)=1/r(0)=0$ дает
$$A=\frac{\alpha}{mV_\infty^2\rho^2\epsilon^2}=\frac{\alpha}{2T\rho^2
\epsilon^2}.$$
Кроме того (см. решение   задачи 5 предыдущего семинара) 
$$\frac{ds}{d\varphi}(0)=\frac{1}{\rho}.$$
Поэтому получаем $\epsilon B=1/\rho$ и $B=\frac{1}{\epsilon\rho}$.
Окончательно
$$s(\varphi)=A\left[\cos{(\epsilon\varphi)}-1\right]+\frac{1}{\epsilon\rho}
\sin{(\epsilon\varphi)}.$$
Угол рассеяния $\theta=\pi-2\varphi_0$, где $\varphi_0$ определяется из условия
(см. решение   задачи 5 предыдущего семинара)
$$\frac{ds}{d\varphi}(\varphi_0)=0.$$
Это дает
$$-A\epsilon\sin{(\epsilon\varphi_0)}+\frac{1}{\rho}\cos{(\epsilon\varphi_0)}
=0$$
и
$$\tg{(\epsilon\varphi_0)}=\frac{1}{A\rho\epsilon}=\frac{2T\rho\epsilon}
{\alpha}.$$
Следовательно,
$$\theta=\pi-\frac{2}{\epsilon}\,\arctg{\left(\frac{2T\rho\epsilon}
{\alpha}\right)}.$$

\subsection*{Задача 4 {\usefont{T2A}{cmr}{b}{n}(8.98)}: Теневая область при
кулоновском рассеянии}
{\usefont{T2A}{cmr}{m}{it} Первое решение.}
Частицы не смогут попасть за огибающую семейства их траекторий. Если семейство 
определяется уравнением $F(r,\varphi,\varphi_0)=0$, где $\varphi_0$ --- параметр, 
то огибающая определяется из системы
\begin{eqnarray} 
F(r,\varphi,\varphi_0)&=&0, \nonumber \\  \frac{\partial}{\partial 
\varphi_0}F(r,\varphi,\varphi_0)&=&0.
\label{eq31_4a}
\end{eqnarray}
Из этой системы надо исключить $\varphi_0$ и получим уравнение огибающей в виде
$r=r(\varphi)$. В нашем случае траектории из семейства --- это гиперболы
(считаем, что поток падает справа)
$$\frac{p}{r}=-1+e\cos{(\varphi-\varphi_0)}.$$
Надо параметр орбиты $p$ и эксцентриситет $e$ выразить через $\varphi_0$. Через 
прицельный параметр $\rho$ эксцентриситет выражается так:
$$e=\sqrt{1+\frac{2EL^2}{m\alpha^2}}=\sqrt{1+\frac{m^2V_\infty^4\rho^2}
{\alpha^2}}.$$
Но (см. решение   задачи 5 предыдущего семинара)
$$\tg{\varphi_0}=\frac{mV_\infty^2}{\alpha}\,\rho.$$
Следовательно, 
$$e=\sqrt{1+\tg^2{\varphi_0}}=\frac{1}{|\cos{\varphi_0}|}=
\frac{1}{\cos{\varphi_0}},$$
так как $\varphi_0\le \pi/2$. 

Для параметра орбиты имеем
$$p=\frac{L^2}{m\alpha}=\frac{mV_\infty^2\rho^2}{\alpha}=\frac{\alpha}
{mV_\infty^2}\,\tg^2{\varphi_0}=\frac{\alpha}{2E}\,\tg^2{\varphi_0}.$$
Следовательно, уравнение гиперболической траектории имеет вид
\begin{equation}
\frac{\alpha}{2E}\,\frac{\tg^2{\varphi_0}}{r}=-1+\frac{\cos{(\varphi-
\varphi_0)}}{\cos{\varphi_0}},
\label{eq31_4b}
\end{equation}
или
$$F(r,\varphi,\varphi_0)=\frac{\alpha}{2E}\,\frac{\tg^2{\varphi_0}}{r}+
1-\frac{\cos{(\varphi-\varphi_0)}}{\cos{\varphi_0}}=0.$$
Тогда второе уравнение системы (\ref{eq31_4a}) будет
\begin{equation}
\frac{\partial F(r,\varphi,\varphi_0)}{\partial \varphi_0}=
\frac{\alpha}{rE}\,\frac{\tg{\varphi_0}}{\cos^2{\varphi_0}}-
\frac{\sin{(\varphi-\varphi_0)}}{\cos{\varphi_0}}-
\frac{\cos{(\varphi-\varphi_0)\sin{\varphi_0}}}{\cos^2{\varphi_0}}=0.
\label{eq31_4c}
\end{equation}
Используя $$\sin{(\varphi-\varphi_0)}=\sin{\varphi}\cos{\varphi_0}-
\cos{\varphi}\sin{\varphi_0}$$ и $$\cos{(\varphi-\varphi_0)}=\cos{\varphi}
\cos{\varphi_0}+\sin{\varphi}\sin{\varphi_0},$$ можно упростить
$$\frac{\sin{(\varphi-\varphi_0)}}{\cos{\varphi_0}}+
\frac{\cos{(\varphi-\varphi_0)\sin{\varphi_0}}}{\cos^2{\varphi_0}}=
\sin{\varphi}+\sin{\varphi}\tg^2{\varphi_0}=\frac{\sin{\varphi}}
{\cos^2{\varphi_0}},$$
и из (\ref{eq31_4c}) получим
\begin{equation}
\frac{\alpha}{E}\,\frac{\tg{\varphi_0}}{r}=\sin{\varphi}.
\label{eq31_4d}
\end{equation}
Это позволяет исключить $ \varphi_0$ из системы (\ref{eq31_4a}). В частности,
\begin{equation}
\frac{\cos{(\varphi-\varphi_0)}}{\cos{\varphi_0}}=\cos{\varphi}+
\sin{\varphi}\tg{\varphi_0}=\cos{\varphi}+\frac{E}{\alpha}\,r\,\sin^2{\varphi}.
\label{eq31_4e}
\end{equation}
С учетом (\ref{eq31_4d}) и (\ref{eq31_4e}), из (\ref{eq31_4b}) получим уравнение 
огибающей:
$$\frac{E}{2\alpha}\,r\sin^2{\varphi}=-1+\cos{\varphi}+\frac{E}{\alpha}\,r
\sin^2{\varphi},$$
или
$$1-\cos{\varphi}=\frac{E}{2\alpha}\,r\,\sin^2{\varphi}=\frac{E}{2\alpha}\,r
(1+\cos{\varphi})(1-\cos{\varphi}).$$
Сокращая на $1-\cos{\varphi}$, получаем уравнение параболы
\begin{equation}
\frac{2\alpha}{E}\,\frac{1}{r}=1+\cos{\varphi}.
\label{eq31_4f}
\end{equation}
Если поток падает не справа, а слева, уравнение огибающей будет
$$\frac{2\alpha}{E}\,\frac{1}{r}=1-\cos{\varphi}.$$

{\usefont{T2A}{cmr}{m}{it} Второе решение.}
Зафиксируем угол $\varphi$ и из семейства (\ref{eq31_4b}) найдем ту траекторию 
(т.~е. соответствующее значение параметра $\varphi_0$), которая дает минимальное
значение $r$ для данного угла. Для этого функция
$$f(\varphi_0)=\frac{1}{\tg^2{\varphi_0}}\left (-1+\frac{\cos{(\varphi-
\varphi_0)}}{\cos{\varphi_0}}\right )=\frac{\sin{\varphi}}{\tg{\varphi_0}}-
\frac{1-\cos{\varphi}}{\tg^2{\varphi_0}}$$
должна быть максимальной. Удобно ввести новую переменную $\chi=1/\tg{\varphi_0}$ и
искать максимум функции $f(\chi)=\sin{\varphi}\,\chi-(1-\cos{\varphi})\chi^2$.
Тогда условие $df(\chi)/d\chi=0$ сразу дает
$$\chi=\frac{\sin{\varphi}}{2(1-\cos{\varphi})}.$$
Подставляя соответствующее $\tg{\varphi_0}=1/\chi$ в
$$\frac{\alpha}{2E}\,\frac{\tg^2{\varphi_0}}{r}=-1+\frac{\cos{(\varphi-
\varphi_0)}}{\cos{\varphi_0}}=-1+\cos{\varphi}+\sin{\varphi}\,\tg{\varphi_0},$$
получаем
$$\frac{2\alpha}{E}\,\frac{(1-\cos{\varphi})^2}{\sin^2{\varphi}}=-1+
\cos{\varphi}+2(1-\cos{\varphi})=1-\cos{\varphi}.$$
Отсюда с учетом  $\sin^2{\varphi}=(1+\cos{\varphi})(1-\cos{\varphi})$ и после 
соответствующих сокращений получается (\ref{eq31_4f}).

{\usefont{T2A}{cmr}{m}{it} Третье решение.}
Воспользуемся сохранением вектора Гамильтона (см. решение задачи 1 семинара 28. 
Заметим, что там рассматривался притягивающий потенциал $U(r)=-\alpha/r$. Поэтому 
в приведенной там формуле надо изменить знак $\alpha$, так как у нас сейчас 
отталкивающий потенциал $U(r)=\alpha/r$)
$$\vec{h}=\vec{V}+\frac{\alpha}{L}\,\vec{e}_\varphi.$$
\begin{figure}[htb]
\centerline{\epsfig{figure=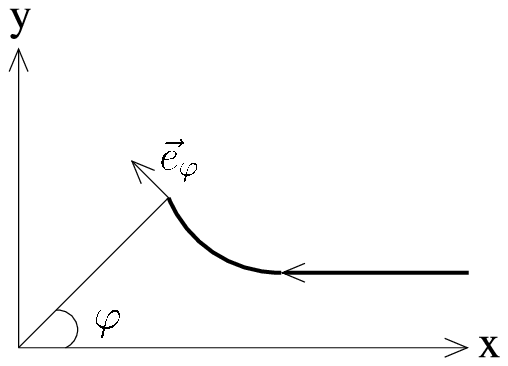,height=4cm}}
\end{figure}
Для данного угла $\varphi$ рассмотрим соответствующую точку на траектории, которая 
характеризуется моментом импульса $L$. В $x$- и $y$-проекциях закон сохранения вектора 
Гамильтона запишется так (см. рисунок):
\begin{eqnarray} &&
-V_\infty=V_x-\frac{\alpha}{L}\sin{\varphi}, \nonumber \\ &&
\frac{\alpha}{L}=V_y+\frac{\alpha}{L}\cos{\varphi}.
\nonumber
\end{eqnarray}
Поэтому для $V^2=V_x^2+V_y^2$ получаем
\begin{equation}
V^2=\left(\frac{\alpha}{L}\sin{\varphi}-V_\infty\right)^2+
\frac{\alpha^2}{L^2}(1-\cos{\varphi})^2=2\frac{\alpha^2}{L^2}+V_\infty^2-
2\frac{\alpha}{L}\,V_\infty\sin{\varphi}-2\frac{\alpha^2}{L^2}\cos{\varphi}.
\label{eq31_4aa}
\end{equation}
С другой стороны, $V^2$ можно найти из закона сохранения энергиии
$$\frac{mV_\infty^2}{2}=\frac{mV^2}{2}+\frac{\alpha}{r},$$
что дает
$$V^2=V_\infty^2-2\frac{\alpha}{mr}.$$
Подставляя это в левую часть (\ref{eq31_4aa}), получаем после тривиальной алгебры
\begin{equation}
\frac{1}{mr}=\alpha z^2\cos{\varphi}-\alpha z^2+V_\infty z\,\sin{\varphi},
\label{eq31_4bb}
\end{equation}
где $z=1/L$. Надо выбрать $z$ (момент импульса $L$) таким образом, чтобы $r$ был
минимальным. Поэтому требуем
$$\frac{d}{dz}\left(\alpha z^2\cos{\varphi}-\alpha z^2+V_\infty z\,
\sin{\varphi}\right )=2\alpha z\cos{\varphi}-2\alpha z+V_\infty \sin{\varphi}
=0,$$
что дает
$$z=\frac{V_\infty \sin{\varphi}}{2\alpha (1-\cos{\varphi})}.$$
Согласно (\ref{eq31_4bb}), такому $z$ соответствует
$$\frac{1}{mr}=\frac{V_\infty^2\sin^2{\varphi}}{4\alpha(1-\cos{\varphi})}=
\frac{V_\infty^2}{4\alpha}(1+\cos{\varphi}).$$
Следовательно,
$$\frac{1}{r}=\frac{E}{2\alpha}(1+\cos{\varphi}),$$
что эквивалентно (\ref{eq31_4f}).

\subsection*{Задача 5 {\usefont{T2A}{cmr}{b}{n}(9.4)}: Сила упругого сжатия 
обруча}
Представим, что обруч сместился вниз на $\Delta x$ (см. рисунок). 
\begin{figure}[htb]
\centerline{\epsfig{figure=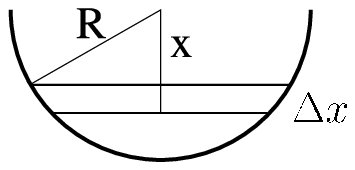,height=1.5cm}}
\end{figure}

\noindent  Изменение его радиуса при этом равно
$$\Delta r=\sqrt{R^2-(x+\Delta x)^2}-r=\sqrt{r^2+x^2-(x+\Delta x)^2}-r\approx
\sqrt{r^2-2x\,\Delta x}-r\approx -\frac{x}{r}\Delta x,$$
так как
$$\sqrt{r^2-2x\,\Delta x}\approx  r\left(1-\frac{x}{r^2}\Delta x\right).$$
При перемещении обруча сила тяжести совершает виртуальную работу $A_1=mg\Delta x$,
а сила упругости обруча $T$ (которая стремится увеличить радиус обруча $r$) 
совершает виртуальную работу $$A_2=2\pi\Delta r T=-2\pi\frac{x}{r}\,T\Delta x.$$ 
Согласно принципу возможных перемещений, $A_1+A_2=0$. Это условие определяет $T$:
$$T=\frac{mgr}{2\pi x}=\frac{mgr}{2\pi\sqrt{R^2-r^2}}.$$

\section[Семинар 32]
{\centerline{Семинар 32}}

\subsection*{Задача 1 {\usefont{T2A}{cmr}{b}{n}(9.1)}: Натяжение цепочки, надетой
на гладкий конус}
Сместим виртуально цепочку из $AB$ и $CD$ (см. левый рисунок). Работа силы натяжения
будет $A_1=-T\Delta l$, где $\Delta l$ --- удлинение цепочки. Работа силы тяжести
равна $mg\Delta h$. Согласно принципу возможных перемещений, $A_1+A_2=mg\Delta h
-T\Delta l=0$. Следовательно, $$T=mg\,\frac{\Delta h}{\Delta l}.$$
\begin{figure}[htb]
\centerline{\epsfig{figure=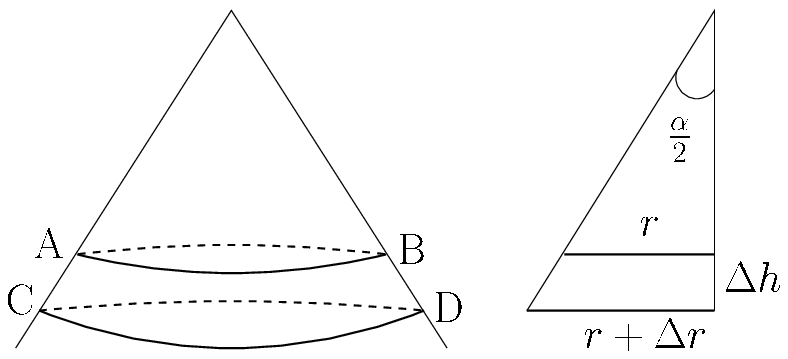,height=4cm}}
\end{figure}

На правом рисунке $\Delta r=\Delta h\,\tg{\frac{\alpha}{2}}$. Поэтому
$\Delta l=2\pi (r+\Delta r)- 2\pi r=2\pi\Delta r=2\pi\Delta h\,
\tg{\frac{\alpha}{2}}$. Окончательно
$$T=\frac{mg}{2\pi}\,\ctg{\frac{\alpha}{2}}.$$

\subsection*{Задача 2 {\usefont{T2A}{cmr}{b}{n}(9.2)}: Натяжение цепочки, надетой
на гладкую сферу}
Так же как в предыдущей задаче получаем 
$$T=mg\,\frac{\Delta h}{\Delta l}.$$
Но (см. рисунок) $r^2=R^2-(R-h)^2$. Дифференцируя это равенство, получаем
$r\Delta r=(R-h)\Delta h=\sqrt{R^2-r^2}\Delta h$. Поэтому
$$\frac{\Delta h}{\Delta l}=\frac{\Delta h}{2\pi\Delta r}=\frac{1}{2\pi}\,
\frac{r}{\sqrt{R^2-r^2}}.$$
Окончательно 
$$T=\frac{mg}{2\pi}\,\frac{r}{\sqrt{R^2-r^2}}=\frac{P}{2\pi}\,\frac{r}
{\sqrt{R^2-r^2}}.$$
\begin{figure}[htb]
\centerline{\epsfig{figure=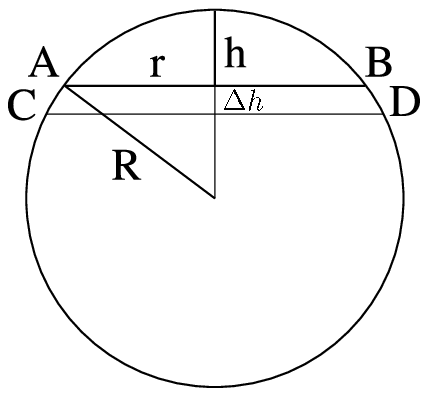,height=3cm}}
\end{figure}

\subsection*{Задача 3 {\usefont{T2A}{cmr}{b}{n}(9.8)}: Найти натяжение нити}
Надо, чтобы центр тяжести конструкции лежал на вертикали, иначе относительно точки 
$A$ будем иметь ненулевой момент вращения. Центр тяжести $K$ находится в середине 
средней линии $MN$ (см. рисунок).
\begin{figure}[htb]
\centerline{\epsfig{figure=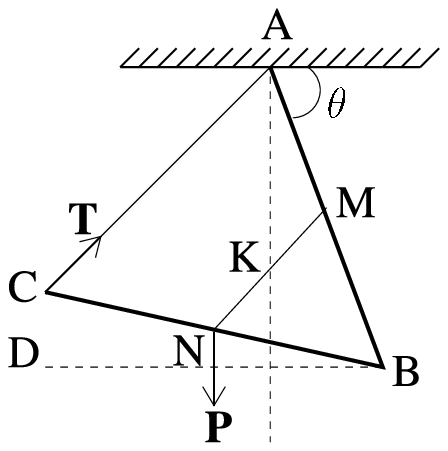,height=4cm}}
\end{figure}

\noindent Следовательно,
$$KM=\frac{1}{2}MN=\frac{1}{4}AC=\frac{L}{4}.$$
Рассмотрим треугольник $AMK$. $\angle KMA=120^\circ$, так как $\angle NMB=
60^\circ$ ($MB=BN=NM=L/2$, т.~е. треугольник $MNB$ равносторонний). По теореме
синусов, $$\frac{AK}{\sin{120^\circ}}=\frac{KM}{\sin{(90^\circ-\theta)}}.$$
Следовательно,
$$\cos{\theta}=\frac{KM}{AK}\,\sin{120^\circ}=\frac{KM}{AK}\,\sin{60^\circ}.$$
Но, по теореме косинусов,
$$AK^2=KM^2+AM^2-2KM\cdot AM\cos{120^\circ}=KM^2+AM^2+2KM\cdot AM
\cos{60^\circ}=\frac{7L^2}{16}.$$
Поэтому
$$\cos{\theta}=\frac{L/4}{\sqrt{7}(L/4)}\,\frac{\sqrt{3}}{2}=\frac{\sqrt{3}}
{2\sqrt{7}},$$ и 
$$\tg^2{\theta}=\frac{1}{\cos^2{\theta}}-1=\frac{28}{3}-1=\frac{25}{3}.$$
Окончательно получаем
$$\tg{\theta}=\frac{5}{\sqrt{3}} \;\;\;\mbox{и}\;\;\;\theta\approx 71^\circ.$$

Условие равновесия действующих на стержень $CB$ моментов вращения относительно точки
$B$ имеет вид $$P\,\frac{L}{2}\,\cos{(\theta-60^\circ)}=Th,$$
где $h=L\,\sin{60^\circ}=L\sqrt{3}/2$ --- высота треугольника $ABC$. Но
$$\cos{(\theta-60^\circ)}=\cos{\theta}\cos{60^\circ}+\sin{\theta}
\sin{60^\circ}=\frac{3\sqrt{3}}{2\sqrt{7}},$$
так как $$\sin{\theta}=\sqrt{1-\frac{3}{28}}=\frac{5}{2\sqrt{7}}.$$
Следовательно,
$$PL\,\frac{3\sqrt{3}}{4\sqrt{7}}=TL\,\frac{\sqrt{3}}{2}.$$
Отсюда
$$T=\frac{3P}{2\sqrt{7}}.$$

\subsection*{Задача 4 {\usefont{T2A}{cmr}{b}{n}(9.19)}: Период вращения
нейтронной звезды}
Сохранение момента импульса $I_0\omega_0=I\omega$ дает $R_0^2\omega_0=R^2\omega$,
так как момент инерции $I\sim mR^2$ (для шара $I=(2/5)mR^2$). 
Поэтому для периода вращения $T=2\pi/\omega$ получаем
$$T=\left(\frac{R}{R_0}\right)^2T_0.$$
Но масса звезды сохраняется:
$$\frac{4}{3}\,\pi\rho_0R_0^3=\frac{4}{3}\,\pi\rho R^3.$$
Следовательно, $R/R_0=(\rho_0/\rho)^{1/3}$, и окончательно
$$T=\left(\frac{\rho_0}{\rho}\right)^{2/3}T_0\approx 10^{-3}~\mbox{с}.$$ 

\subsection*{Задача 5 {\usefont{T2A}{cmr}{b}{n}(9.25)}: Сколько оборотов сделает
цилиндр?}
Энергия цилиндра $E=\frac{1}{2}I\omega^2$ расходится на работу против сил трения.
Для цилиндра $I=\frac{1}{2}mR^2$. Следовательно, $E=\frac{1}{4}m\omega^2R^2$.
На основание цилиндра выделим маленький элемент с площадью $dS$ на расстоянии $r$ от
оси цилиндра. На него приходится сила давления $mg\frac{dS}{\pi R^2}$. Поэтому сила
трения, которая действует на этот элемент, равна $dF=\mu mg\frac{dS}
{\pi R^2}$. При повороте цилиндра на угол $\alpha$ эта сила трения совершает работу
$$dA=-dF\,r\alpha=-\frac{\mu mg\alpha}{\pi R^2}\,rdS.$$
Но $dS=rdrd\varphi$, где $\varphi$ есть полярная координата элемента площади $dS$.
Поэтому суммарная работа сил трения равна
$$A=-\frac{\mu mg\alpha}{\pi R^2}\int\limits_0^R r\,2\pi r\,dr=
-\frac{2}{3}\mu mgR\alpha.$$
Следовательно, цилиндр остановится после поворота на угол $\alpha$ такого, что
$$\frac{2}{3}\mu mgR\alpha=\frac{1}{4}m\omega^2R^2.$$
Это дает
$$\alpha=\frac{3}{8}\,\frac{\omega^2 R}{\mu g}.$$
Число оборотов
$$N=\frac{\alpha}{2\pi}=\frac{3}{16 \pi}\,\frac{\omega^2 R}{\mu g}.$$

\subsection*{Задача 6 {\usefont{T2A}{cmr}{b}{n}(9.28)}: Цилиндрическая банка
с жидкостью}
Пусть момент инерции банки равен $I$. Когда жидкость полностью раскрутится, момент
инерции системы будет  $2I$. По закону сохранения момента импульса $I\omega_0=
2I\omega$, находим $\omega=\omega_0/2$. Первоначальная энергия вращения равна
$E_0=\frac{1}{2}I\omega_0^2$, конечная --- $E=\frac{1}{2}2I\omega^2=\frac{1}
{4}I\omega_0^2$. Следовательно, в тепло перешла энергия 
$$Q=E_0-E=\frac{1}{4}I\omega_0^2=\frac{1}{2}E_0.$$

\subsection*{Задача 7 {\usefont{T2A}{cmr}{b}{n}(9.26)}: Установившаяся скорость
вращения дисков}
При соприкосновении дисков возникают силы трения $\vec{F}_1=-\vec{F}_2$ (см. 
рисунок).
\begin{figure}[htb]
\centerline{\epsfig{figure=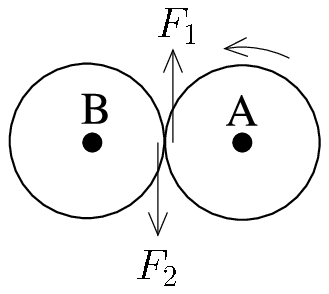,height=3cm}}
\end{figure}
Сила трения $\vec{F}_1$ создает тормозящий момент $M=FR$ относительно оси $A$ (здесь
$F=|\vec{F}_1|=|\vec{F}_2|$). На второй диск будет действовать такой же момент
относительно оси $B$ (созданный силой $\vec{F}_2$). Следовательно, уравнения движения
дисков имеют вид
$$I\dot\omega_1=-M,\;\;\;I\dot\omega_2=-M.$$
При этом положительное направление вращения --- это направление вращения 
первоначального диска. Сила трения $\vec{F}_2$ раскручивает второй диск 
в отрицательном направлении. Вычитая из первого уравнение второе, получаем 
$$I\frac{d}{dt}(\omega_1-\omega_2)=0.$$
Следовательно, $\omega_1-\omega_2=\mathrm{const}$. Первоначально $\omega_1=
\omega$, $\omega_2=0$. В конце угловые скорости дисков по величине равны, но второй 
диск вращается в отрицательном направлении: $\omega_1^\prime=-\omega_2^\prime=
\omega^\prime$. Поэтому $\omega=\omega^\prime_1-\omega^\prime_2=2\omega^\prime$,
и $\omega^\prime=\omega/2$.

Первоначальная энергия равна $E_1=I\omega^2/2$, конечная --- $E_2=2(I\omega^
{\prime\,2}/2)=I\omega^2/4$. Следовательно, в тепло перешла энергия
$Q=E_1-E_2=I\omega^2/4=E_1/2$.

\subsection*{Задача 8 {\usefont{T2A}{cmr}{b}{n}(9.32)}: Частота колебаний обруча}
Найдем момент инерции обруча относительно точки подвеса $A$ (см. левый рисунок).
\begin{figure}[htb]
\centerline{\epsfig{figure=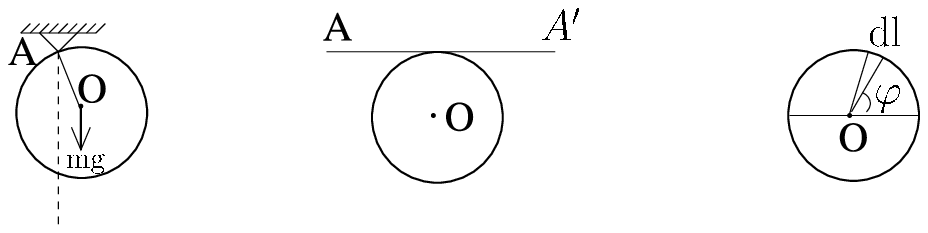,height=2.5cm}}
\end{figure}
По теореме Штейнера, $I_A=I_0+mR^2$. Но момент инерции обруча относительно центра 
есть $I_0=mR^2$. Следовательно, $I_A=2mR^2$. При малом отклонении $AO$ от 
вертикали на угол $\varphi$ сила тяжести $mg$ создает возвращающий момент 
$mgR\sin{\varphi}\approx mgR\varphi$. Поэтому уравнение движения будет 
$I_A\ddot \varphi=-mgR\varphi$. Отсюда получаем частоту малых колебаний
$$\omega=\sqrt{\frac{mgR}{I_A}}=\sqrt{\frac{g}{2R}}.$$
Если колебания происходят перпендикулярно к плоскости обруча, то в уравнение движения
будет присутствовать другой момент инерции обруча: относительно оси $AA^\prime$
(см. средний рисунок). Найдем сначала момент инерции относительно оси, проходящей через 
центр обруча и параллельной $AA^\prime$. Момент инерции элемента обруча длиной $dl$
(см. правый рисунок) равен $dI_0=(R\cos{\varphi})^2\,dm=\frac{mR^2}{2\pi}\,
\cos^2{\varphi}\,d\varphi$, так как $dm=\frac{m}{2\pi R}\,R\,d\varphi$.
Поэтому 
$$I_0=\int\limits_0^{2\pi}\frac{mR^2}{2\pi}\,\cos^2{\varphi}\,d\varphi=
\frac{mR^2}{4\pi}\int\limits_0^{2\pi}(1+\cos{2\varphi})\,d\varphi=\frac{1}{2}
\,mR^2.$$
Более просто этот результат можно получить, если учесть, что для плоского тела
$I_x+I_y=I_z$ и что для обруча $I_x=I_y$, $I_z=mR^2$.

По теореме Штейнера находим момент инерции относительно оси $AA^\prime$:
$$I_{AA^\prime}=\frac{1}{2}\,mR^2+mR^2=\frac{3}{2}\,mR^2.$$ 
Поэтому частота малых колебаний будет
$$\omega^\prime=\sqrt{\frac{mgR}{I_{AA^\prime}}}=\sqrt{\frac{2g}{3R}}.$$

\subsection*{Задача 9 {\usefont{T2A}{cmr}{b}{n}(9.40)}: Частота колебаний обруча
внутри неподвижной гладкой сферы}
Обруч $AB$ вращается относительно точки $O$ - центра сферы (см. левый рисунок).
\begin{figure}[!h]
\centerline{\epsfig{figure=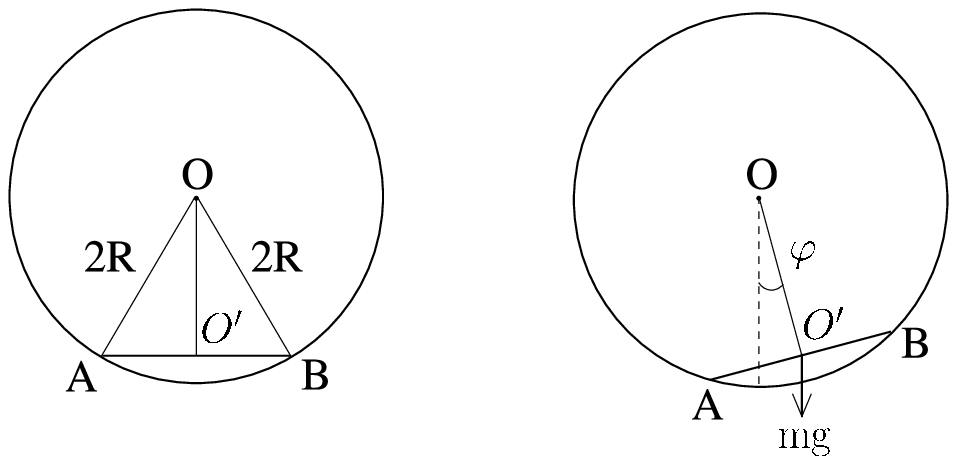,height=3cm}}
\end{figure}
Поэтому надо найти момент инерции обруча относительно оси проходящей через $O$ 
перепендикулярно к плоскости рисунка (параллельно к плоскости обруча). По теореме 
Штейнера $I_O=I_{O^\prime}+m|OO^\prime|^2$. Но $I_{O^\prime}=\frac{1}{2}\,mR^2$
и $|OO^\prime|=2R\,\sin{60^\circ}=\sqrt{3}\,R$. Следовательно,
$$I_O=\frac{1}{2}\,mR^2+3\,mR^2=\frac{7}{2}\,mR^2.$$
Возращающий момент силы тяжести при отклонении на маленький угол $\varphi$ равен (см.
правый рисунок) $M=mg\,|OO^\prime|\sin{\varphi}\approx \sqrt{3}\,mgR\,\varphi$.
Таким образом, уравнение движения обруча имеет вид
$$\frac{7}{2}\,\,mR^2\,\ddot\varphi=-\sqrt{3}\,mgR\,\varphi$$
и $$\omega^2=\frac{\sqrt{3}\,mgR}{7R/2}=\frac{2\sqrt{3}\,g}{7R}.$$

\section[Семинар 33]
{\centerline{Семинар 33}}

\subsection*{Задача 1 {\usefont{T2A}{cmr}{b}{n}(9.13)}: Моменты инерции однородных
тел}
{\usefont{T2A}{cmr}{m}{it} 1. Кольцо.} Пусть $R_1$, $R_2$ --- внутренний и 
внешний радиус кольца, $S=\pi(R_2^2-R_1^2)$ --- его площадь. Тогда
$$I_z=\int\limits_{R_1}^{R_2}\frac{m}{S}\,r^2(2\pi r)dr=\frac{m}{2}\,
\frac{R_2^4-R_1^4}{R_2^2-R_1^2}=\frac{1}{2}\,m(R_1^2+R_2^2).$$
Здесь $m$ --- масса кольца, $2\pi r\,dr$ --- площадь очень тонкого (шириной $dr$)
кольца радиуса $r$.

Для плоских тел $I_x+I_y=I_z$. Из симметрии кольца следует, что $I_x=I_y$. Поэтому
$$I_x=I_y=\frac{1}{2}\,I_z=\frac{1}{4}\,m(R_1^2+R_2^2).$$

{\usefont{T2A}{cmr}{m}{it} 2. Диск.} Это частный случай кольца, когда $R_1=0,\,
R_2=R$. Поэтому
$$I_z=\frac{1}{2}\,mR^2,\;\;\;I_x=I_y=\frac{1}{4}\,mR^2.$$

{\usefont{T2A}{cmr}{m}{it} 3. Цилиндр.} Его можно представить как совокупность 
бесконечно тонких дисков. Для каждого диска $dI_z=\frac{1}{2}dm\,R^2$. Поэтому
$$I_z=\frac{1}{2}\,R^2\int dm=\frac{1}{2}\,mR^2,$$
где $m$ --- масса цилиндра. 

Из симметрии цилиндра очевидно, что $I_x=I_y$. По теореме Штейнера, для элементарного 
диска,
$dI_x=\frac{1}{4}dm\,R^2+dm\,z^2$, где $z$ --- это расстояние плоскости диска от
медианной плоскости цилиндра (проходящий через его центр). Но 
$$dm=\frac{m}{\pi R^2 H}\,\pi R^2 dz=\frac{m}{H}\,dz.$$
Поэтому
$$I_x=I_y=\frac{1}{4}\,m\,R^2+\int\limits_{-H/2}^{H/2}\frac{m}{H}z^2\,dz=
\frac{1}{4}\,m\,R^2+\frac{1}{12}\,m\,H^2.$$

{\usefont{T2A}{cmr}{m}{it} 4. Палочка.} Это частный случай цилиндра, когда $R=0,\,
H=l$. Поэтому
$$I_z=0,\;\;\;I_x=I_y=\frac{1}{12}\,ml^2.$$

{\usefont{T2A}{cmr}{m}{it} 5. Сфера.} Из симметрии сферы следует, что $I_x=I_y=
I_z=I$. Но
$$I_x=\int dm\,(y^2+z^2),\;\; I_y=\int dm\,(x^2+z^2),\;\;I_z=\int dm\,(x^2+
y^2).$$
Складывая их, получаем $3I=2R^2\int dm=2mR^2$, так как для всех точек сферы
$x^2+y^2+z^2=R^2$. Поэтому
$$I=\frac{2}{3}\,mR^2.$$

{\usefont{T2A}{cmr}{m}{it} 5. Шар.} Его можно представить как совокупность 
бесконечно тонких концентрических сфер. Для каждой сферы (радиусом $r$)
$$dm=\frac{m}{\frac{4}{3}\pi R^3}\,4\pi r^2\,dr=\frac{3m}{R^3}\,r^2\,dr.$$
Поэтому
$$I=\frac{2}{3}\int r^2 dm=2\frac{m}{R^3}\int\limits_0^R r^4\,dr=\frac{2}{5}
\,mR^2.$$

\subsection*{Задача 2 {\usefont{T2A}{cmr}{b}{n}(9.18)}: Метеорная пыль и
продолжительность суток}
Пусть весь поток метеорной пыли в единицу времени равен $\mu$. За время $t$ масса 
Земли увеличится на $\mu t$ и при этом метеорное вещество образует тонкую сферу 
радиусом $R$ --- радиусом Земли. Момент инерции будет
$$I=\frac{2}{5}\,mR^2+\frac{2}{3}\,\mu t R^2.$$
Из сохранения углового момента $I_0\omega_0=I\omega$ получаем для периода 
$$T=\frac{2\pi}{\omega}=\frac{I}{I_0}\,T_0=\left(1+\frac{5}{3}\,\frac{\mu}{m}
\,t\right )T_0,$$
где $m$ --- масса Земли, $I_0=(2/5)mR^2$ --- ее первоначальный момент инерции и
$T_0=24~\mbox{ч}$. 

\subsection*{Задача 3 {\usefont{T2A}{cmr}{b}{n}(9.27)}: Раскручивание неподвижных
дисков}
Направление первоначального вращения примем за положительное. При соприкосновении 
дисков возникнают одинаковые по величине силы трения  (так как все диски одинаковые), 
которые замедляют вращение первого диска и раскручивают первоначально неподвижные 
диски (см. рисунок).
 
Уравнения движения дисков имеют вид (здесь $M=FR$)
$I\dot\omega_1=-2M,\;\;I\dot\omega_2=-M,\;\;I\dot\omega_3=-M$.
Поэтому $I(\dot\omega_1-\dot\omega_2-\dot\omega_3)=0$ и, следовательно, 
$\omega_1-\omega_2-\omega_3=\mathrm{const}$. В начале $\omega_1=\omega,\,
\omega_2=\omega_3=0$. В конце $\omega_1=\omega^\prime,\,
\omega_2=\omega_3=-\omega^\prime$. Следовательно, $\omega=\omega^\prime-
(-\omega^\prime)-(-\omega^\prime)=3\omega^\prime$ и $\omega^\prime=
\omega/3$.
\begin{figure}[htb]
\centerline{\epsfig{figure=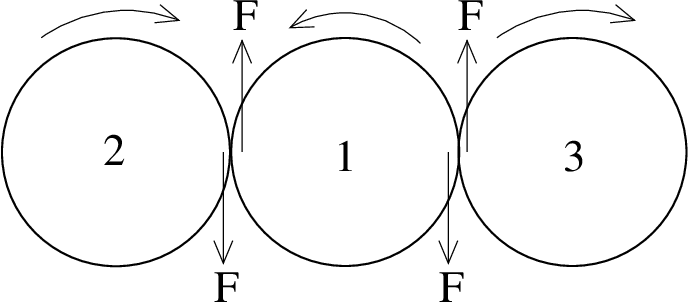,height=2cm}}
\end{figure}

\subsection*{Задача 4 {\usefont{T2A}{cmr}{b}{n}(9.35)}: Частота колебаний
проволочного равностороннего треугольника}
Момент силы тяжести относительно точки $A$ будет (см. рисунок) $M=mg|AO|
\sin{\varphi}\approx mg\varphi |AO|$.
\begin{figure}[htb]
\centerline{\epsfig{figure=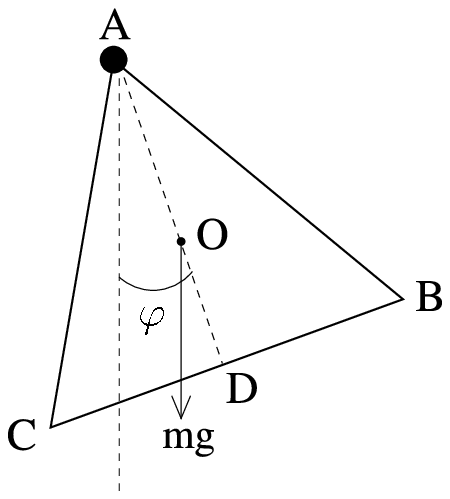,height=3cm}}
\end{figure}
Здесь $O$ --- это центр тяжести треугольника (точка пересечения медиан). Поэтому 
$|AO|$ равен двум третям медианы $AD$. Но в равностороннем треугольнике высота 
совпадает с медианой. Следовательно,
$$|AO|=\frac{2}{3}\,L\sin{60^\circ}=\frac{L}{\sqrt{3}},$$
и уравнение движения будет
$$I\ddot\varphi=-\frac{mgL}{\sqrt{3}}\,\varphi.$$
Соответственно, для частоты получаем
$$\omega^2=\frac{mgL}{\sqrt{3}I}.$$
Надо найти момент инерции $I=I_{AB}+I_{BC}+I_{AC}$. По теореме Штейнера,
$$I_{AB}=I_{AC}=\frac{m}{3}\,L^2\,\frac{1}{12}+\frac{m}{3}\left(\frac{L}{2}
\right)^2=\frac{mL^2}{9},$$
где $m$ --- полная масса треугольника. Аналогично
$$I_{BC}=\frac{m}{3}\,L^2\,\frac{1}{12}+\frac{m}{3}|AD|^2=\frac{m}{3}\,L^2\,
\frac{1}{12}+\frac{m}{3}\,L^2\,\frac{3}{4}=\frac{5mL^2}{18},$$
так как $|AD|=|AC|\,\sin{60^\circ}=\sqrt{3}\,L/2$. Полный момент инерции 
треугольника относительно точки $A$ равен
$$I=2\,\frac{mL^2}{9}+\frac{5mL^2}{18}=\frac{1}{2}\,mL^2.$$
Поэтому
$$\omega^2=\frac{mgL}{\sqrt{3}\frac{1}{2}mL^2}=\frac{2g}{\sqrt{3}L}.$$

\subsection*{Задача 5 {\usefont{T2A}{cmr}{b}{n}(9.38)}: Частота колебаний
номерка из гардероба}
Найдем центр тяжести номерка. У сплошного диска центр тяжести совпадает 
с геометрическим центром, т.~е. имеет координату $x=R$ (см. правый рисунок).
\begin{figure}[htb]
\centerline{\epsfig{figure=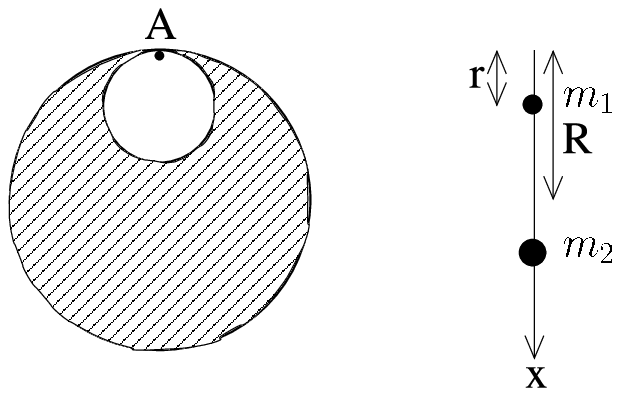,height=3cm}}
\end{figure}
Но сплошной диск можно представить как совокупность двух тел: маленького диска 
радиусом $r$ и номерка из гардероба. Пусть искомая координата центра тяжести номерка 
равна $x=X$. Так как координата центра тяжести маленького диска есть $x=r$, можем 
написать
\begin{equation}
R=\frac{m_1r+m_2X}{m},
\label{eq33_5a}
\end{equation}
где $m$ --- масса, которую бы имел сплошной диск радиусом $R$, $m_1=\frac{r^2}{R^2}\,
m$ --- масса маленького диска, $m_2=\frac{R^2-r^2}{R^2}\,m$ --- масса номерка.
Здесь мы учли, что массы частей однородного диска пропорциональны их площадям.  
Решая (\ref{eq33_5a}) относительно $X$, получаем
$$X=\frac{R^3-r^3}{R^2-r^2}.$$
Для физического маятника
\begin{equation}
\omega^2=\frac{m_2gl}{I}=\frac{mg}{I}\,\frac{R^3-r^3}{R^2},
\label{eq33_5b}
\end{equation}
где $l=X$ есть расстояние от точки подвеса до центра тяжести номерка, $I$ ---  
момент инерции номерка относительно точки $A$ (см. левый рисунок). Момент инерции 
сплошного диска относительно точки $A$ был бы
$$I_R=\frac{1}{2}\,mR^2+mR^2=\frac{3}{2}\,mR^2.$$
Но $I_R=I+I_r$, где
$$I_r=\frac{1}{2}\,m_1r^2+m_1r^2=\frac{3}{2}\,m_1r^2=\frac{3}{2}\,m\,
\frac{r^4}{R^2}$$
есть момент инерции маленького диска  относительно точки $A$. Поэтому
$$I=I_R-I_r=\frac{3}{2}\,m\,\frac{R^4-r^4}{R^2}.$$
Подставляя это в (\ref{eq33_5b}), получаем окончательно
$$\omega^2=\frac{2g}{3}\,\frac{R^3-r^3}{R^4-r^4}.$$

\subsection*{Задача 6 {\usefont{T2A}{cmr}{b}{n}(9.43)}: Частота малых колебаний
полушара}
{\usefont{T2A}{cmr}{m}{it} Первое решение.}
Найдем момент инерции относительно оси, которая проходит через точку $A$ (см. левый 
рисунок) перпендикулярно к плоскости рисунка (ось $x$).
\begin{figure}[htb]
\centerline{\epsfig{figure=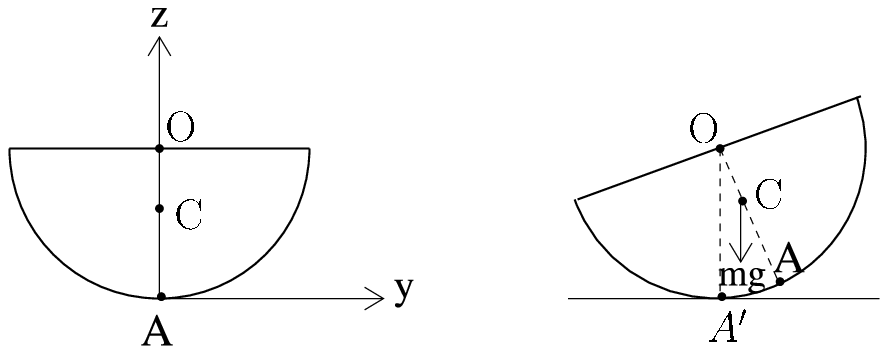,height=3cm}}
\end{figure}

\noindent Выделим маленький диск между $z$ и $z+dz$. Его масса будет
$$dm=\frac{m}{\frac{4}{6}\,\pi R^3}\,\pi r^2\,dz=\frac{3m}{2R^3}\,z(2R-z)\,
dz,$$
где $r^2=R^2-(R-z)^2=z(2R-z)$. Момент инерции этого диска относительно горизонтальной
оси, проходяшей через его центр, равен $dI_x=\frac{1}{4}\,dm\,r^2$, а относительно 
параллельной оси, проходяшей через $A$:
$$dI_A=\frac{1}{4}\,dm\,r^2+dm\,z^2=dm\left[\frac{z}{4}\,(2R-z)+z^2\right]=
\frac{dm}{4}\,z(2R+3z).$$
Поэтому момент инерции полушара будет
$$I_A=\frac{3m}{8R^3}\int\limits_0^R z^2(2R+3z)(2R-z)\,dz=\frac{13}{20}\,
mR^2.$$
Надо еще найти координату центра масс полушара (точка $C$):
$$z_c=|AC|=\frac{1}{m}\int\limits_0^R z\,dm=\frac{3}{2R^3}\int\limits_0^R 
z^2(2R-z)\,dz=\frac{5R}{8}.$$
Пусть полушар наклонили на угол $\varphi$ (см. правый рисунок). Точка соприкосновения
полушара с поверхностью $A^\prime$ лежит на вертикали, так как поверхность эта 
касается полушара и, следовательно, перпендикулярна к радиусу. Сила тяжести 
относительно мгновенной оси вращения, проходящей через  $A^\prime$, создает 
возвращающий момент
$$M=mg|OC|\sin{\varphi}\approx \frac{3}{8}\,mgR\varphi,$$
так как $|OC|=R-|AC|=3R/8$. При малых колебаниях можно считать, что момент инерции 
полушара относительно мгновенной оси вращения не меняется: $I_{A^\prime}\approx I_A$.
Поэтому уравнение движения полушара будет
$$I_A\ddot\varphi=-\frac{3}{8}\,mgR\varphi,$$
и для частоты малых колебаний получаем
$$\omega^2=\frac{1}{I_A}\,\frac{3mgR}{8}=\frac{15g}{26R}.$$

{\usefont{T2A}{cmr}{m}{it} Второе решение.}
Так как полушар катится без проскальзывания, $V=\omega R$ и, следовательно,
$|AB|=V\Delta t=R\omega\Delta t=R\varphi$ (см. рисунок).
\begin{figure}[htb]
\centerline{\epsfig{figure=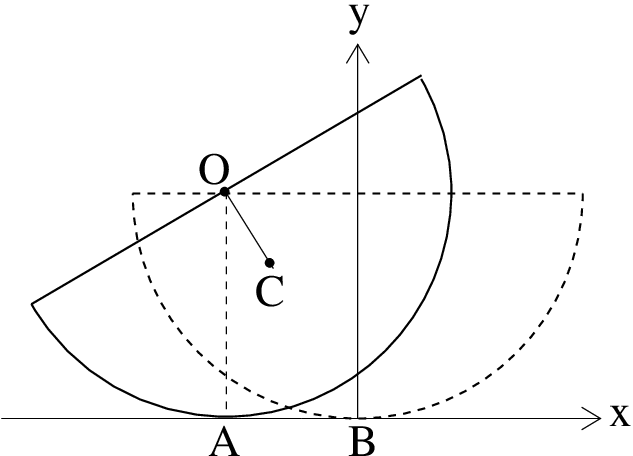,height=4cm}}
\end{figure}
Поэтому координаты центра масс полушара $C$ будут (для $\varphi\ll 1$)
$$x=-R\varphi+\frac{3}{8}\,R\sin{\varphi}\approx -\frac{5}{8}\,R\varphi,\;\;\;
y=R-\frac{3}{8}\,R\cos{\varphi}\approx \frac{5}{8}\,R+\frac{3}{16}\,R
\varphi^2.$$
При этом $\dot y\sim R\varphi\dot\varphi \ll \dot x\sim R\dot\varphi$. Поэтому
с точностью до квадратичных членов энергия полушара будет 
$$E\approx \frac{1}{2}m\dot x^2+\frac{1}{2}I_C\dot \varphi^2+mgy\approx
\frac{1}{2}\,m\,\frac{25}{64}\,R^2\dot\varphi^2+\frac{1}{2}I_C\dot \varphi^2+
mg\left(\frac{5}{8}\,R+\frac{3}{16}\,R \varphi^2\right).$$
Здесь $I_C$ --- это момент инерции полушара. Но (см. предыдущее решение)
$$I_C=I_A-m(R-|OC|)^2=\frac{13}{20}\,mR^2-\frac{25}{64}\,mR^2,$$
и для энергии получаем
$$E\approx \frac{13}{40}\,mR^2\dot\varphi^2+mg\left(\frac{5}{8}\,R+\frac{3}{16}
\,R \varphi^2\right).$$
Энергия сохраняется. Поэтому
$$0=\frac{dE}{dt}=m\dot\varphi\left(\frac{13}{20}\,R^2\ddot\varphi+
\frac{3}{8}\,gR \varphi \right),$$
что дает уравнение гармонических колебаний
$$\ddot\varphi+\frac{15}{26}\,\frac{g}{R}\,\varphi=0.$$

\subsection*{Задача 7 {\usefont{T2A}{cmr}{b}{n}(8.97)}: Рассеяние протона на
протоне}
В системе центра масс (а в данном случае лабораторная система совпадает с системой 
центра масс), угол рассеяния первого протона совпадат с углом рассеяния фиктивной 
частицы с приведенной массой $\mu=m_1m_2/(m_1+m_2)=m_p/2$ и радиус-вектором 
$\vec{r}=\vec{r}_1-\vec{r}_2$ в том же потенциале, что описывает взаимодействие 
сталкивающихся протонов. Это следует из соотношений
$$\vec{r}_1=\frac{m_2}{m_1+m_2}\,\vec{r}=\frac{1}{2}\,\vec{r},\;\;\;
\vec{r}_2=-\frac{m_1}{m_1+m_2}\,\vec{r}=-\frac{1}{2}\,\vec{r},$$
где $\vec{r}=\vec{r}_1-\vec{r}_2$ есть относительный радиус-вектор. Таким образом,
в системе центра масс, где выполняются эти соотношения, радиус-вектор первого протона
$\vec{r}_1$ в процессе рассеяния поворачивается на такой же угол, на какой 
поворачивается относительный радиус-вектор $\vec{r}$. Следовательно, дифференциальное
сечение рассеяния дается формулой Резерфорда 
$$d\sigma=\left(\frac{e^2}{2\mu V_\infty^2}\right)^2\frac{d\Omega_1}{\sin^4{
\frac{\theta_1}{2}}},$$
где $d\Omega_1=\sin{\theta_1}\,d\theta_1\,d\varphi_1$ --- телесный угол, 
$e$ --- заряд протона (в системе CGSE), и $V_\infty=|\vec{V}_{1\infty}-\vec{V}_
{2\infty}|=2V_{p\infty}$. Но тогда $2\mu V_\infty^2=4m_pV_{p\infty}^2=8E$. 
По $\varphi_1$ сразу проинтегрируем:
$$d\sigma(\theta_1)=\frac{\pi e^4}{32E^2}\,\frac{\sin{\theta_1}}{\sin^4{\frac{
\theta_1}{2}}}=\frac{\pi e^4}{16E^2}\,\frac{\cos{\frac{\theta_1}{2}}}{\sin^3
{\frac{\theta_1}{2}}}\,d\theta_1.$$
Протон, который летит слева, ничем не отличается от протона, который летит справа.
Поэтому, если мы не отслеживаем траектории сталкивающихся частиц, а детектор просто 
регистрирует протон, летящий после рассеяния под углом $\theta$, это может быть и 
второй протон (первый протон при этом рассеивается на угол $\pi-\theta$). 
Следовательно, в этом случае дифференциальным сечением рассеяния будет сумма
\begin{equation}
d\sigma=d\sigma(\theta)+d\sigma(\pi-\theta)=\frac{\pi e^4}{16E^2}\left(
\frac{\cos{\frac{\theta}{2}}}{\sin^3{\frac{\theta}{2}}}+
\frac{\sin{\frac{\theta}{2}}}{\cos^3{\frac{\theta}{2}}}\right)\,d\theta.
\label{eq33_7}
\end{equation}
В этой формуле $\theta=\theta_1$ --- это по существу есть угол рассеяния первого 
протона. Детектор зарегистрирует протон, рассеянный на угол больше, чем $\theta_0$, 
если $\theta_1$ меняется в пределах от $\theta_0$ до $\pi-\theta_0$ (если 
$\theta_1>\pi-\theta_0$, то $\theta_2<\theta_0$ и мы зарегистрируем протон, 
летящий под углом $\theta<\theta_0$ --- это будет второй протон). Поэтому, чтобы 
найти полное сечение рассеяния, мы должны проинтегрировать (\ref{eq33_7}) в этих 
пределах. При этом каждую конфигурацию мы посчитаем дважды, так как $\theta_1$ и 
$\pi-\theta_1$ дают неразличимые конфигурации. Поэтому результат интегрирования мы 
должны поделить на два, чтобы исключить двойной счет, и окончательно
$$\sigma(\theta>\theta_0)=\frac{1}{2}\,\frac{\pi e^4}{16E^2}\int\limits_
{\theta_0}^{\pi-\theta_0}\left(
\frac{\cos{\frac{\theta_1}{2}}}{\sin^3{\frac{\theta_1}{2}}}+
\frac{\sin{\frac{\theta_1}{2}}}{\cos^3{\frac{\theta_1}{2}}}\right)\,d\theta_1=
\frac{\pi e^4}{16E^2}\left(\frac{1}{\sin^2{\frac{\theta_0}{2}}}-
\frac{1}{\cos^2{\frac{\theta_0}{2}}}\right).$$
Если $\theta_0=90^\circ$, то $\sigma(\theta>90^\circ)=0$.
 
\subsection*{Задача 8 {\usefont{T2A}{cmr}{b}{n}(8.107)}: Интенсивность пучка
$\alpha$-частиц}
Пусть сечение рассеяния  на угол больше $90^\circ$ равно $\sigma$. 
Данная $\alpha$-частица расеется на угол больше $90^\circ$, если на ее пути 
в цилиндрике вокруг ее траектории с поперечным сечением $\sigma$ и длиной 
$d=10~\mbox{мкм}=10^{-5}~\mbox{м}$ окажется ядро. Но вероятность того, что ядро 
окажется там, равна отношению объемов $(\sigma d)/(S d)=\sigma/S$, где $S$ --- 
поперечное сечение пучка. Это справедливо только в том случае, если имеется всего 
лишь одно ядро. На пути пучка на самом деле много ядер. Их число равно $N=nSd$, где 
$n=\rho/M$ --- число ядер в единицу объема, $M$ --- масса одного ядра, $\rho $ ---
плотность вещества мишени. Поэтому вероятность рассеяния будет в $N$ раз больше: 
$$p=N\,\frac{\sigma}{S}=n\sigma d.$$
За время $t$ на фольгу падает $Jt$ $\alpha$-частиц, где $J$ --- это интенсивность 
пучка $\alpha$-частиц. Следовательно, число рассеяний в фольге за время $t$ равно
$K=Jtn\sigma d$. Отсюда $J=K/(tn\sigma d)$. Найдем сечение рассеяния на угол 
больше $90^\circ$. Так как масса $\alpha$-частицы много меньше массы ядра, можно
использовать формулу Резерфорда 
$$d\sigma=\left(\frac{2Ze^2}{4E}\right)^2\frac{d\Omega}
{\sin^4{\frac{\theta}{2}}},\;\;\;d\Omega=2\pi\sin{\theta}\,d\theta,$$
где $\theta$ --- это угол рассеяния в лабораторной системе. Здесь мы учли, что заряд
$\alpha$-частицы равен $2e$, а заряд ядра --- $Ze$, где для золота $Z=79$. Поэтому
$$\sigma=\int\limits_{\pi/2}^\pi d\sigma=\pi Z^2\frac{e^4}{E^2}.$$
Выражая все в единицах CGSE: $e=4,8\cdot10^{-10}~\mbox{Фр}$, $E=10~\mbox{МэВ}=
10^7\cdot 1,6\cdot 10^{-12}~\mbox{эрг}$, получаем $\sigma\approx 4,1\cdot
10^{-24}~\mbox{см}^2=4,1\cdot 10^{-28}~\mbox{см}^2$. Кроме того,
$$n=\frac{\rho}{M}=\frac{19,3\cdot 10^3~\mbox{кг}/~\mbox{м}^3}
{197\cdot 1,66\cdot 10^{-27}~\mbox{кг}}\approx 5,9\cdot 10^{28}~
\mbox{м}^{-3},$$
так как для золота $M\approx 197~\mbox{а.е.м}$ и $\mbox{а.е.м}\approx 1,66\cdot 
10^{-27}~\mbox{кг}$. Поэтому окончательно
$$J\approx \frac{1}{3600\cdot 5,9\cdot 10^{28}\cdot 4,1\cdot 10^{-28}\cdot
10^{-5}}~\mbox{с}^{-1}\approx 1,1~\mbox{с}^{-1}.$$

\section[Семинар 34]
{\centerline{Семинар 34}}

\subsection*{Задача 1 {\usefont{T2A}{cmr}{b}{n}(9.55)}: Каким участком сабли
следует рубить лозу?}
Пусть саблю держим в точке $A$ и опускаем поступательно со скоростью $V$ (см. 
рисунок).
\begin{figure}[htb]
\centerline{\epsfig{figure=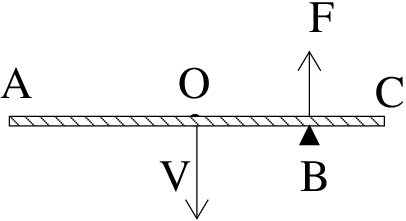,height=2.0cm}}
\end{figure}
Во время удара в точке $B$ на саблю действует сила $F$. В результате удара импульс
сабли уменьшится на $\int Fdt$. Поэтому скорость центра масс сабли после удара будет
$$u=V-\frac{1}{m}\int Fdt.$$
Кроме того, сила $F$ создает вращательный момент $F\left(\frac{L}{2}-x\right )$
относительно центра масс $O$. Здесь $x$ --- это расстояние от точки удара $B$ до
конца сабли $C$. Поэтому в результате удара сабля приобретет угловую скорость
$$\omega=\frac{1}{I}\int F\left(\frac{L}{2}-x\right )dt=
\frac{1}{I}\left(\frac{L}{2}-x\right )\int Fdt,$$
где $I=\frac{mL^2}{12}$ --- момент инерции сабли относительно ее центра масс.
Следовательно, после удара точка $A$ участвует как в поступательном, так и во 
вращательном движении, и ее скорость будет
$$V^\prime=u+\omega\,\frac{L}{2}=V-\frac{1}{m}\int Fdt+\frac{L}{2I}
\left(\frac{L}{2}-x\right )\int Fdt.$$
Удара не почувствуем, если $V^\prime=V$. Следовательно,
$$-\frac{1}{m}\int Fdt+\frac{L}{2I}\left(\frac{L}{2}-x\right )\int Fdt=0,$$
что дает
$$x=\frac{L}{2}-\frac{2I}{mL}=\frac{L}{2}-\frac{L}{6}=\frac{L}{3}.$$

\subsection*{Задача 2 {\usefont{T2A}{cmr}{b}{n}(9.80)}: С каким ускорением надо
тянуть вверх нить?}
Когда нить тянем, сила натяжение нити $T$ создает вращающий момент $M=TR$ относительно
центра тяжести $O$ (см. рисунок).
\begin{figure}[htb]
\centerline{\epsfig{figure=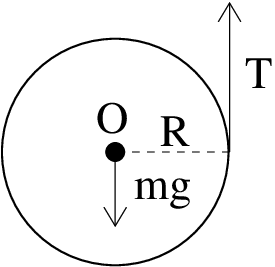,height=2.0cm}}
\end{figure}
Следовательно, $I\dot\omega=TR$. Так как катушка не падает, то $T=mg$. Поэтому 
получаем $\dot \omega=mgR/I$. С другой стороны, если $x$ --- это длина нити, которую
мы уже размотали, то $\dot x=R\dot\varphi =R\omega$ и ускорение $a=\ddot x=R\dot
\omega$. Следовательно, 
\begin{equation}
a=\frac{mgR^2}{I}.
\label{eq34_2}
\end{equation}
Надо найти момент инерции катушки $I$. Для начала найдем момент инерции усеченного 
конуса с радиусами оснований $R_1=R$ и $R_2$ и длиной $L$. Выделим бесконечно тонкий 
диск на расстоянии $x$ от меньшего основания (см. рисунок).
\begin{figure}[htb]
\centerline{\epsfig{figure=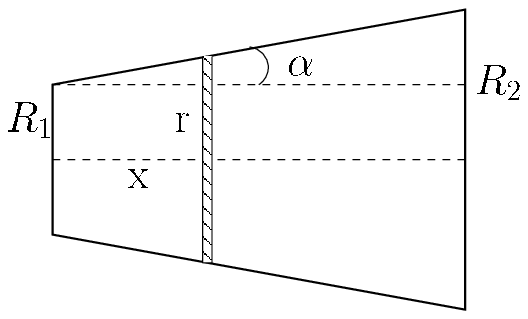,height=3.0cm}}
\end{figure}
Масса этого диска будет $dm_1=\rho\pi r^2 dx$, где $\rho$ --- плотность материала
усеченного конуса, $r=R_1+x\tg{\alpha}$ --- радиус диска. Но $\tg{\alpha}=
(R_2-R_1)/L$. Поэтому 
$$r=R_1+\frac{R_2-R_1}{L}\,x$$
и $$dx=\frac{L}{R_2-R_1}\,dr.$$
Для момента инерции $I_1$ усеченного конуса получаем
$$I_1=\int \frac{1}{2}\,dm_1\,r^2=\frac{\pi\rho L}{2(R_2-R_1)}\int\limits_{R_1}
^{R_2}r^4\,dr=\frac{\pi\rho L}{10}\,\frac{R_2^5-R_1^5}{R_2-R_1}=
\frac{31}{10}\,\pi\rho,$$
так как, согласно условию задачи, $R_2=2$, $R_1=1$ и $L=1$. 

Момент инерции центрального цилиндра длины $L^\prime=5$ и массы $m_2=\rho\pi R_1^2
L^\prime=5\,\pi\rho$ равен
$$I_2=\frac{1}{2}m_2R_1^2=\frac{1}{2}\rho\pi L^\prime R_1^4=\frac{5}{2}\,
\pi\rho.$$
Следовательно, полный момент инерции катушки будет
\begin{equation}
I=2I_1+I_2=\frac{87}{10}\,\pi\rho.
\label{eq34_2a}
\end{equation}
Нужна еще масса катушки $m$. Масса усеченного конуса равна
$$m_1=\int dm_1=\int\limits_0^L \rho\pi r^2 dx=
\frac{\rho\pi L}{R_2-R_1}\int\limits_{R_1}^{R_2}r^2\,dr=\frac{\rho\pi L}
{3(R_2-R_1)}\,(R_2^3-R_1^3)=\frac{7}{3}\,\pi\rho.$$
Поэтому полная масса катушки равна
\begin{equation}
m=2m_1+m_2=\frac{29}{3}\,\pi\rho.
\label{eq34_2b}
\end{equation}
Подставляя (\ref{eq34_2a}) и (\ref{eq34_2b}) в  (\ref{eq34_2}), с учетом
$R=R_1=1$, получаем окончательно
$$a=\frac{10}{9}\,g.$$

\subsection*{Задача 3 {\usefont{T2A}{cmr}{b}{n}(9.76)}: На какой высоте цилиндр
оторвется от горки? }
\begin{figure}[htb]
\centerline{\epsfig{figure=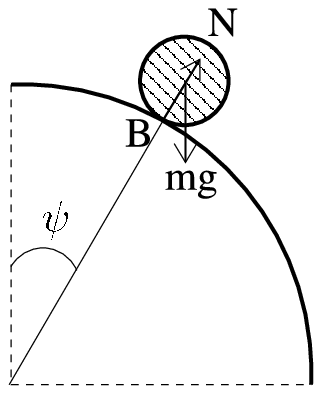,height=3.0cm}}
\end{figure}
Кинетическая энергия маленького цилиндра с моментом инерции $I=mr^2/2$ равна
$$T=\frac{1}{2}\,mV^2+\frac{1}{2}\,I\omega^2=\frac{3}{4}\,mV^2.$$
Последнее равенство следует из того, что при отсутствии проскальзывания $V=\omega r$
(точка $B$ маленького цилиндра, которая участвует как в поступательном движении 
цилиндра со скоростью $V$, так и в его вращательном движении с противоположно 
направленной линейной скоростью $\omega r$, неподвижна). С другой стороны, закон 
сохранения энергии позволяет написать $T=mg(R+r)(1-\cos{\psi})$ (кинетическая 
энергия цилиндра возникает из-за уменьшения его потенциальной энергии). Поэтому 
получаем
$$V^2=\frac{4}{3}\,g(R+r)(1-\cos{\psi}).$$
В точке отрыва сила реакции $N$ зануляется и центростремительное ускорение 
$$a_c=\frac{V^2}{r+R}=\frac{4}{3}\,g(1-\cos{\psi})$$
создается только радиальной компонентой силы тяжести $mg\cos{\psi}$. Поэтому в этой
точке $ma_c=mg\cos{\psi}$, что дает $$\frac{4}{3}(1-\cos{\psi})=\cos{\psi}.$$
Отсюда $\cos{\psi}=4/7$, и высота отрыва будет
$$H=(R+r)\cos{\psi}=\frac{4}{7}\,(R+r).$$

\subsection*{Задача 4 {\usefont{T2A}{cmr}{b}{n}(9.62)}: Максимальное число
оборотов футбольного мяча}
Скорость мяча после удара будет $V=\frac{1}{m}\int F dt$. Чтобы максимально закрутить
мяч, точка удара должна быть на максимальном удалении от центра мяча, равного радиусу
мяча $R$. Поэтому угловая скорость мяча после удара будет $\omega=\frac{1}{I}\int 
FR dt$, где $I=2mR^2/3$ --- момент инерции мяча. Следовательно, 
$$\omega=\frac{mVR}{I}=\frac{3V}{2R}.$$
За время полета $t=L/V$ мяч повернется на угол $\varphi =\omega t=\frac{3L}{2R}$
и, следовательно, сделает
$$N=\frac{\varphi}{2\pi}=\frac{3L}{4\pi R}\approx 24$$
оборотов.

\subsection*{Задача 5 {\usefont{T2A}{cmr}{b}{n}(9.53)}: Столкновение шарика со
стержнем}
Пусть после удара шарик движется в прежнем направлении со скоростью $V^\prime$,
а скорость центра масс стержня и его угловая скорость равны соответственно $u$ и 
$\omega$ (см. рисунок). 
\begin{figure}[htb]
\centerline{\epsfig{figure=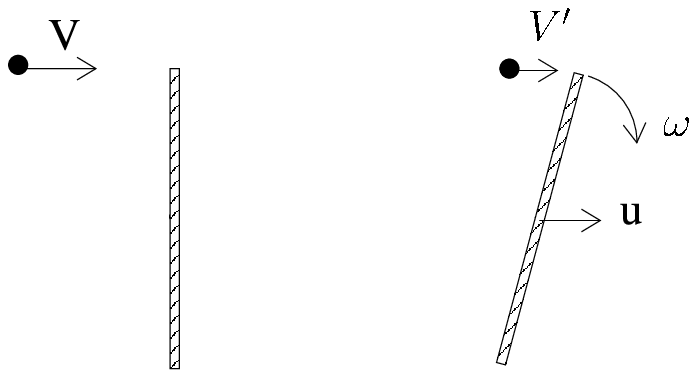,height=3.0cm}}
\end{figure}
Сохранение момента импульса относительно центра стержня запишется как
$$mV\,\frac{L}{2}=mV^\prime\,\frac{L}{2}+I\omega,$$
где $I=ML^2/12$ --- момент инерции стержня. Поэтому
$$I\omega^2=\frac{1}{I}(I\omega)^2=\frac{m^2L^2}{4I}(V-V^\prime)^2=
\frac{3m^2}{M}(V-V^\prime)^2,$$
и из закона сохранения энергии
$$\frac{mV^2}{2}=\frac{mV^{\prime\,2}}{2}+\frac{Mu^2}{2}+\frac{I\omega^2}{2}$$
получаем
$$Mu^2=m(V^2-V^{\prime\,2})-I\omega^2=m(V-V^\prime)\left[V+V^\prime-
\frac{3m}{M}\,V+\frac{3m}{M}\,V^\prime\right]=$$ $$=Mu\left[V+V^\prime-
\frac{3m}{M}\,V+\frac{3m}{M}\,V^\prime\right],$$
где последний шаг следует из закона сохранения импульса $mV=mV^\prime+Mu$.
Следовательно,
$$u=V+V^\prime-\frac{3m}{M}\,V+\frac{3m}{M}\,V^\prime.$$
Сдругой стороны, из закона сохранения импульса
$$u=\frac{m}{M}(V-V^\prime).$$ 
Поэтому должны иметь
$$V+V^\prime-\frac{3m}{M}\,V+\frac{3m}{M}\,V^\prime=\frac{m}{M}\,V-
\frac{m}{M}\,V^\prime.$$
Отсюда
$$V^\prime\left(1+4\frac{m}{M}\right)=V\left(4\frac{m}{M}-1\right)$$
и
$$V^\prime=\frac{4m-M}{4m+M}\,V.$$
Видно, что если $4m<M$, шарик изменит направление своего движения на 
противоположное.

\subsection*{Задача 6 {\usefont{T2A}{cmr}{b}{n}(9.68)}: С каким ускорением
скатывается бочка?}
\begin{figure}[htb]
\centerline{\epsfig{figure=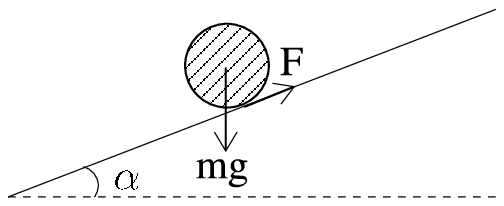,height=2.0cm}}
\end{figure}
Уравнение движения вдоль наклонной плоскости будет $(m+M)a=(m+M)g\sin{\alpha}-F$.
Сила трения $F$ раскручивает только бочку. Поэтому $I\dot\omega=FR$, где $I=mR^2$ 
--- момент инерции бочки. Но, так как проскальзывания нет, $a=R\dot\omega$. Поэтому
$$F=\frac{I\dot\omega}{R}=\frac{Ia}{R^2}=ma.$$
Подставляя это в уравнение движения и решая относительно $a$, получаем
$$a=\frac{m+M}{2m+M}\,g\sin{\alpha}.$$

\subsection*{Задача 7: частота малых колебаний стержня}
Пусть $\theta$ -- угол, который стержень образует с осью $x$, $x_m,y_m$ 
-- координаты центра масс, $x_1,y_1$ и $x_2,y_2$ -- координаты левого и правого 
конца. Тогда $x_1=x_m-\frac{l}{2}\cos{\theta}$, $x_2=x_m+\frac{l}{2}
\cos{\theta}$, $y_1=y_m-\frac{l}{2}\sin{\theta}$ и $y_2=y_m+\frac{l}{2}
\sin{\theta}$. Но $y_1=\alpha\,x_1^2$ и $y_2=\alpha\,x_2^2$. T.e.
$$y_m\pm\frac{l}{2}\sin{\theta}=\alpha\left(x_m\pm\frac{l}{2}\cos{\theta}
\right)^2.$$
Отсюда
$$y_m=\alpha \left (x_m^2+\frac{l^2}{4}\cos^2{\theta}\right ).$$
Подставляя это, например, в $y_2=\alpha\,x_2^2$, находим
$$x_m=\frac{1}{2\alpha}\tg{\theta}.$$
Потенциальная энергия стержня
$$U=mgy_m=\frac{mg\alpha}{4}\left [ l^2\cos^2{\theta}+\frac{1}{\alpha^2}
\tg^2{\theta}\right ].$$
Кинетическая энергия 
$$T=\frac{1}{2}m\left (\dot{x}_m^2+\dot{y}_m^2\right )+\frac{1}{2}\,
\frac{1}{12}\,ml^2\dot{\theta}^2.$$
Но
$$\dot{x}_m=\frac{1}{2\alpha}\,\frac{\dot \theta}{\cos^2{\theta}}\;\;
\mbox{и}\;\;\dot{y}_m=2\alpha[x_m\dot{x}_m-\frac{l^2}{4}\dot{\theta}
\sin{\theta}\cos{\theta}]=\frac{\alpha}{2}\dot{\theta}\sin{\theta}\left [
\frac{1}{\alpha^2}\,\frac{1}{\cos^3{\theta}}-l^2\cos{\theta}\right ].$$
Поэтому полная энергия
$$E=T+U=\frac{m}{2}f(\theta)\dot{\theta}^2+U(\theta),$$
где
$$f(\theta)=\frac{1}{4\alpha^2}\,\frac{1}{\cos^4{\theta}}+
\frac{\alpha^2}{4}\sin^2{\theta}\cos^2{\theta}\left [\frac{1}{\alpha^2}\,
\frac{1}{\cos^4{\theta}}-l^2\right ]^2+\frac{1}{12}l^2.$$
Положение равновесия определяется из $\frac{dU}{d\theta}=0$, что дает
$$-l^2\sin{\theta}\cos{\theta}+\frac{1}{\alpha^2}\,\frac{1}{\cos^2{\theta}}
\,\frac{\sin{\theta}}{\cos{\theta}}=0.$$
Т.е. или $\sin{\theta}=0$, или
$$\cos^4{\theta}=\frac{1}{\alpha^2l^2}.$$
Если $\alpha l<1$, второго решения нет.
$$\ddot U(\theta)=\frac{mg\alpha}{2}\left [ -l^2\cos^2{\theta}+l^2\sin^2{
\theta}+\frac{1}{\alpha^2\cos^2{\theta}}+\frac{3}{\alpha^2}\,\frac{\sin^2{
\theta}}{cos^4{\theta}}\right ].$$
При $\theta=0$ и $\alpha l<1$,  
$$\ddot U(0)=\frac{mg}{2\alpha}(1-\alpha^2 l^2)>0, \;\;
f(0)=\frac{1}{12\alpha^2}[3+\alpha^2l^2].$$
Поэтому с точностью до квадратичных членов
$$E=\frac{m}{2}\,\frac{3+\alpha^2l^2}{12\alpha^2}\dot{\theta}^2+\frac{1}{2}\,
\frac{mg}{2\alpha}(1-\alpha^2l^2)\theta^2+U(0).$$
При $E=A\dot{\theta}^2+B\theta^2$ сохранение энергии $\frac{dE}{dt}=0$ дает
уравнение движения $\ddot{\theta}+\frac{B}{A}\theta=0$ и, следовательно,
$\omega^2=\frac{B}{A}$. Таким образом, если $\alpha l<1$, есть только одно 
положение равновесия при $\theta=0$ и
$$\omega^2=6g\alpha\,\frac{1-\alpha^2l^2}{3+\alpha^2l^2}.$$

Если $\alpha l>1$, то $\ddot U(0)<0$ и положение равновесия при 
$\theta=0$ становится неустойчивым. Устойчивому равновесию отвечает
$$\cos{\theta_m}=\pm\frac{1}{\sqrt{\alpha l}}.$$
При этом
$$\ddot U(\theta_m)=2mgl(\alpha l-1)>0,\;\;f(\theta_m)=\frac{1}{3}l^2$$
и, следовательно,
$$\omega^2=6\,\frac{g}{l}\,(\alpha l-1).$$

\section[Семинар 35]
{\centerline{Семинар 35}}

\subsection*{Задача 1 {\usefont{T2A}{cmr}{b}{n}(9.66)}: Минимальное значение
угловой скорости}
На обруч будет действовать сила трения $F=\mu mg$, которая будет его тормозить 
с ускорением $a=F/m=\mu g$ и перекручивать в другую сторону с угловым ускорением
$\epsilon=FR/I=\mu g/R$ (см. рисунок).
\begin{figure}[htb]
\centerline{\epsfig{figure=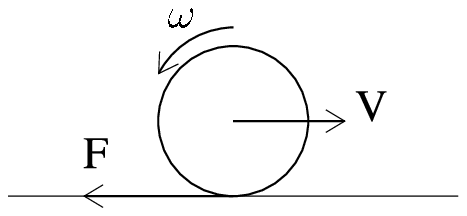,height=2.0cm}}
\end{figure} 

\noindent Следовательно,
$$V=V_0-\mu gt,\;\;\;\mbox{и}\;\;\;\omega=-\omega_0+\frac{\mu g}{R}\,t,$$
где мы за положительное направление вращения взяли направление, противоположное 
первоначальной закрутке. Если $\omega_0$ маленькое, через некоторое время будем иметь
$V=\omega R$ и дальше обруч покатится без проскальзывания, т.~е. сила трения исчезнет
и скорость обруча перестанет меняться. Найдем, через какое время это случится. Условие
$$V_0-\mu gt=\left (-\omega_0+\frac{\mu g}{R}\,t\right )R$$
дает
$$t=\frac{V_0+\omega_0R}{2\mu g}.$$
При этом установившаяся скорость будет
$$V=V_0-\mu g\,\frac{V_0+\omega_0R}{2\mu g}=\frac{V_0-\omega_0R}{2}.$$
Эта скорость неотрицательна, если $\omega_0\le V_0/R$. Если же $\omega_0> V_0/R$,
установившаяся скорость будет отрицательной, т.~е. обруч покатится назад. В этом 
случае до того, как обруч перекрутится и изменит направление вращения, скорость 
обнулится и потом станет отрицательной. Условие $V=\omega R$ наступит при $V<0$, 
$\omega<0$ и обруч покатится  назад со скоростью (по величине)
$$|V|=\frac{|V_0-\omega_0R|}{2}=\frac{\omega_0R-V_0}{2}.$$  

\subsection*{Задача 2 {\usefont{T2A}{cmr}{b}{n}(9.67)}:  Ускорение скатывания
с наклонной плоскости}
Пусть сила трения равна $F$ и цилиндр (или шар) скатывается без скольжения (см. 
рисунок).
\begin{figure}[htb]
\centerline{\epsfig{figure=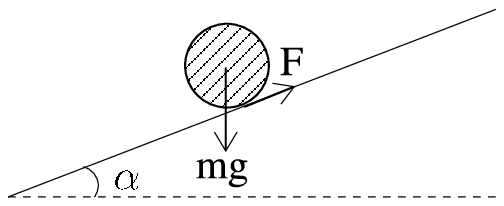,height=2.0cm}}
\end{figure} 

\noindent Тогда его ускорение определяется из 
\begin{equation}
ma=mg\sin{\alpha}-F,
\label{eq35_2}
\end{equation}
а угловое ускорение из
$\dot\omega=FR/I$. Так как проскальзывания нет, $R\dot\omega=a$. Поэтому для силы
трения получаем $$F=\frac{I\dot\omega}{R}=\frac{I}{R^2}\,a.$$
Подставляя это в (\ref{eq35_2}),  получаем
$$a=\frac{g\sin{\alpha}}{1+\frac{I}{mR^2}}.$$
Для полого цилиндра $I=mR^2$ и $a=\frac{1}{2}\,g\sin{\alpha}$. Для сплошного 
цилиндра $I=\frac{1}{2}\,mR^2$ и $a=\frac{2}{3}\,g\sin{\alpha}$. Для шара
$I=\frac{2}{5}\,mR^2$ и $a=\frac{5}{7}\,g\sin{\alpha}$.

\subsection*{Задача 3 {\usefont{T2A}{cmr}{b}{n}(10.16)}: Устойчивость
равновесия пузырька}
Выталкивающая сила, действующая на пузырек, равна весу воды с объемом пузырька. Но 
вес воды есть векторная сумма силы тяжести $mg$ и центробежной силы $m\Omega^2R\sin
{\alpha}$, где $R=D/2$ есть радиус стеклянной трубки. Следовательно, при малом 
отклонении, на пузырек действуют сила $F_1=mg$, направленная вверх, и сила 
$F_2=m\Omega^2R\sin{\alpha}\approx m\Omega^2R\alpha$, направленная к оси 
вращения (см. рисунок). Равновесие будет устойчивым, если тангенциальная компонента
силы $F_2$ больше, чем  тангенциальная компонента силы $F_1$. Т.~е. если
$F_2\cos{\alpha}>F_1\sin{\alpha}$. Так как $\alpha\ll 1$, то $\cos{\alpha}
\approx 1$, $\sin{\alpha}\approx \alpha$ и условие устойчивости принимает вид 
$m\Omega^2R\alpha>mg\alpha$, что означает
$$\Omega^2>\frac{g}{R}=\frac{2g}{D}.$$
\begin{figure}[htb]
\centerline{\epsfig{figure=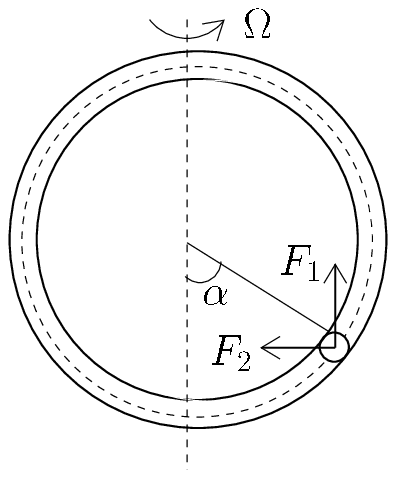,height=4.0cm}}
\end{figure}

\subsection*{Задача 4 {\usefont{T2A}{cmr}{b}{n}(9.88)}: Устойчивость свободного
вращения спичечного коробка}
Пусть $I_1$, $I_2$, $I_3$ --- моменты инерции спичечной коробки. Свободное вращение
коробки описывается уравнениями Эйлера
$$I_1\dot\omega_1-(I_2-I_3)\omega_2\omega_3=0,\;\;
I_2\dot\omega_2-(I_3-I_1)\omega_3\omega_1=0,\;\;
I_3\dot\omega_3-(I_1-I_2)\omega_1\omega_2=0.$$
Пусть первоначально коробку раскрутили вокруг первой оси с угловой скоростью 
$\omega_1$. Рассмотрим малое отклонение от такого движения, т.~е. будем предполагать,
что $\omega_2$, $\omega_3$, а также $\dot\omega_1$, $\dot\omega_2$ и 
$\dot\omega_3$ --- маленькие величины. Тогда с точностью до членов первого порядка 
будем иметь
$$I_2\ddot\omega_2\approx (I_3-I_1)\dot \omega_3\omega_1\approx
\frac{(I_3-I_1)(I_1-I_2)}{I_3}\,\omega_1^2\omega_2.$$
Следовательно, с этой точностью
$$\ddot\omega_2+\frac{(I_1-I_3)(I_1-I_2)}{I_2I_3}\,\omega_1^2\omega_2=0.$$
Если $(I_1-I_3)(I_1-I_2)>0$, это будет уравнением гармонических колебаний, 
$\omega_2$ останется все время маленькой и будет колебаться около нуля. Если же 
$(I_1-I_3)(I_1-I_2)<0$, то возмущение $\omega_2$ будет экпоненциально расти и
не останется маленькой.

Следовательно, вращение спичечной коробки вокруг главной оси устойчиво, если 
соответствующий момент инерции максимален или минимален. Вращение же вокруг оси с 
промежуточным значением момента инерции неустойчиво. Например, если $I_2>I_1>I_3$,
вращение вокруг первой оси будет неустойчивым.

\subsection*{Задача 5: Длина висящего каната}
Можно считать, что сила тяжести $mg$ приложена в центре масс каната, т.~е. в середине
$B$ (см. рисунок).
Жесткость половины каната $AB$ в два раза больше. Поэтому она растянется до длины 
$l_1^\prime$ такой, что $mg=2k\left (l_1^\prime-l/2\right)$. Отсюда 
$$l_1^\prime=\frac{l}{2}+\frac{mg}{2k}.$$
Нижняя половина каната $BC$ не растянется, так как на ней не действует ``эффективная''
сила тяжести. Следовательно, длина каната, когда он висит вертикально, равна
$$l^\prime=l_1^\prime+\frac{l}{2}=l+\frac{mg}{2k}.$$
\begin{figure}[htb]
\centerline{\epsfig{figure=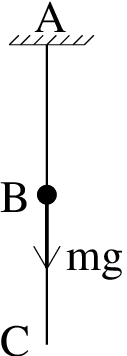,height=3.0cm}}
\end{figure}

\noindent Это решение простое, но не совсем строгое, так как растягивается, конечно, 
каждый участок каната.

{\usefont{T2A}{cmr}{m}{it} Более строгое решение.}
Заменим канат на систему из $N$ грузиков массы $m/N$ каждый и $N$ невесомых пружин
длины $l/N$ каждый  (см. рисунок). 
\begin{figure}[htb]
\centerline{\epsfig{figure=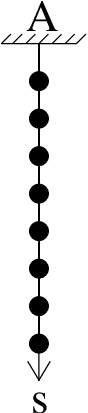,height=3.0cm}}
\end{figure}
Жесткость маленькой пружинки будет в $N$ раз больше, чем жесткость каната, так как у нее
длина в $N$ раз меньше. Поэтому условие равновесия $i$-го грузика будет
\begin{equation}
\frac{m}{N}\,g=Nk\left(s_i-s_{i-1}-\frac{l}{N}\right )-
Nk\left(s_{i+1}-s_i-\frac{l}{N}\right )=Nk\left[s_i-s_{i-1}+s_i-s_{i+1}
\right],
\label{eq35_5a}
\end{equation}
где $s_i$ --- координата $i$-го грузика. Но это только если $i<N$. Для последнего 
грузика будем иметь 
\begin{equation}
\frac{m}{N}\,g=Nk\left(s_N-s_{N-1}-\frac{l}{N}\right ).
\label{eq35_5b}
\end{equation}
Нас интересует предел $N\to\infty$. Поэтому введем непрерывную переменную $x$ через
соотношения $x=i\,\frac{l}{N}$ и будем рассматривать $s_i$ как функцию $x$. Тогда
$$s_{i+1}-s_i=s(x+dx)-s(x)\approx \frac{ds}{dx}\,dx,$$
где $dx=l/N$. Кроме того,
$$s_{i+1}-s_i-(s_i-s_{i-1})\approx \left [s^\prime (x)-s^\prime (x-dx)\right]
dx\approx s^{\prime\prime}(x-dx)(dx)^2\approx s^{\prime\prime}(x)(dx)^2,$$
где $s^\prime (x)=\frac{ds(x)}{dx}$ и $s^{\prime\prime} (x)=\frac{d^2s(x)}
{dx^2}$. Но $N^2k(dx)^2=kl^2$ и уравнение (\ref{eq35_5a}) в пределе $N\to\infty$ 
принимает вид
\begin{equation}
mg=-kl^2\,\frac{d^2s(x)}{dx^2},
\label{eq35_5c}
\end{equation}  
тогда как уравнение (\ref{eq35_5b}) в пределе $N\to\infty$ дает (после сокращения на
kl) следующее условие:
\begin{equation}
\left . \frac{ds(x)}{dx}\right|_{x=l}-1=0.
\label{eq35_5d}
\end{equation} 
Кроме того, будем иметь граничное условие $s(0)=0$, так как чем ближе точка подвеса, 
тем меньше растягивается соответствующий участок каната. 

Решение уравнения (\ref{eq35_5c}) имеет вид
$$s(x)=-\frac{1}{2}\,\frac{mg}{kl^2}\,x^2+A_1x+A_2.$$
Условие $s(0)=0$ дает $A_2=0$, тогда как условие (\ref{eq35_5d}) определяет в этом 
случае $A_1$ через уравнение $$A_1=\frac{mg}{kl}+1.$$ 
Следовательно,
$$s(x)=x+\frac{mgx}{kl}\left(1-\frac{x}{2l}\right ).$$
Координата конца каната (его длина) равна
$$l^\prime=s(l)=l+\frac{mg}{2k}.$$
   
\subsection*{Задача 6:  Колебания полушара на гладкой горизонтальной плоскости}
{\usefont{T2A}{cmr}{m}{it} Первое решение.}
Так как в горизонтальном направлении сила не действует, центр тяжести полушара движется
только по вертикали. При малых колебаниях этим смещением по вертикали тоже можно 
пренебречь и считать, что полушар вращается вокруг центра масс. При наклоне полушара
на угол $\phi\ll 1$ сила реакции опоры $N\approx mg$ создает возвращающий момент
$Nb\sin{\phi}\approx (mgb)\phi$, где $b$ -- расстояние от центра масс $C$ до 
геометрического центра шара $O$. Следовательно, уравнение движения будет
$I_C\ddot{\phi}=-mgb\,\phi$ и частота малых колебаний $$\omega^2=\frac{mgb}
{I_C}.$$
Момент инерции полушара относительно $O$ есть половина момента инерции шара (масссы 
$2m$): $I_O=\frac{1}{2}\frac{2}{5}(2m)R^2=\frac{2}{5}mR^2$. По теореме Штейнера
$I_O=I_C+mb^2$. Но $b=\frac{3}{8}R$ (см. задачу 6 семинара 33). Поэтому
$I_C=I_O-mb^2=\frac{83}{320}mR^2$ и
$$\omega^2=\frac{120}{83}\,\frac{g}{R}.$$

{\usefont{T2A}{cmr}{m}{it} Второе решение.}
Так как трения нет, горизонтальное положение центра масс полушара $C$ не меняется и 
для маленьких углов поворота $\varphi$ координаты центра масс будут (см. рисунок)
\begin{figure}[htb]
\centerline{\epsfig{figure=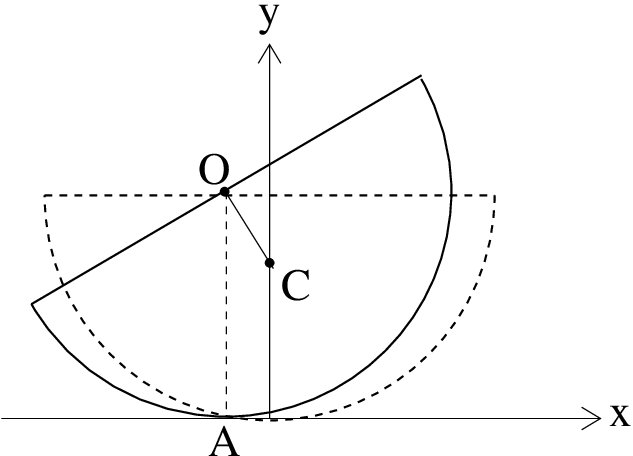,height=4cm}}
\end{figure}
$$x=0,\;\;\;y=R-\frac{3}{8}\,R\cos{\varphi}\approx \frac{5}{8}\,R+
\frac{3}{16}\,R\varphi^2.$$
При этом $\dot y\sim R\varphi\dot\varphi$ второго порядка по малости. Поэтому
с точностью до квадратичных членов энергия полушара будет 
$$E\approx \frac{1}{2}I_C\dot \varphi^2+mgy\approx
\frac{83}{640}\,\dot \varphi^2+mg\left(\frac{5}{8}\,R+\frac{3}{16}\,R 
\varphi^2\right).$$
Здесь $I_C=83/320$ --- это момент инерции полушара (см. предыдушее решение).
Энергия сохраняется. Поэтому
$$0=\frac{dE}{dt}=m\dot\varphi\left(\frac{83}{320}\,R^2\ddot\varphi+
\frac{3}{8}\,gR \varphi \right),$$
что дает уравнение гармонических колебаний
$$\ddot\varphi+\frac{120}{83}\,\frac{g}{R}\,\varphi=0.$$

\section[Семинар 36]
{\centerline{Семинар 36}}

\subsection*{Задача 1 {\usefont{T2A}{cmr}{b}{n}(10.4)}: Отклонение тела при
падении}
Угол между векторами $\vec{V}$ и $\vec{\omega}$ есть $\alpha=\varphi+90^\circ$
(см. рисунок). 
\begin{figure}[htb]
\centerline{\epsfig{figure=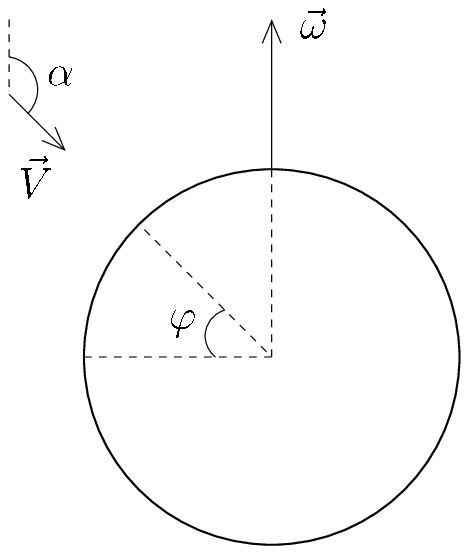,height=4cm}}
\end{figure}

\noindent  Векторное произведение $\vec{V}\times\vec{\omega}$ направлено от 
плоскости рисунка к нам, т.~е. отклонение за счет действия силы Кориолиса $2m\vec{V}
\times\vec{\omega}$ будет на восток. Оценим  $(\vec{V}\times\vec{\omega})_{max}$.
Максимальная скорость 
$V_{max}\approx\sqrt{2gh}\approx 10^2~\mbox{м}/\mbox{с}$. Угловая скорость 
вращения Земли $\omega\approx 7,3\cdot 10^{-5}~\mbox{с}^{-1}$. Поэтому 
$(\vec{V}\times\vec{\omega})_{max}\sim 10^2\cdot 7,3\cdot 10^{-5}\approx 
10^{-2}\mbox{м}/\mbox{с}^2\ll g$. Центробежное ускорение $\omega^2 R\sim 3\cdot 
10^{-2}\mbox{м}/\mbox{с}^2$ тоже много меньше $g$. Следовательно, при вычислении 
времени падения можно считать, что тело свободно падает по вертикали с ускорением $g$ 
и время падения $\tau=\sqrt{2h/g}\approx 10~\mbox{с}$. 

Направим ось $y$ (в системе Земли) к востоку. Тогда $\dot{V}_y=2V\omega\sin{\alpha}
=2V\omega\cos{\varphi}=V\omega$, так как $\varphi=60^\circ$ и $\cos{\varphi}=
1/2$. Но $V\approx gt$, поэтому будем иметь
$\dot{V}_y\approx g\omega\,t$, что дает $V_y(t)=\dot{y}= g\omega\,t^2/2$, так 
как $V_y(0)=0$. Интегрируем еще раз: $y(t)=g\omega\,t^3/6$, так как $y(0)=0$.
Окончательно
$$y(\tau)=\frac{1}{6}g\omega\tau^3\approx 0,12~\mbox{м}=12~\mbox{см}.$$

\subsection*{Задача 2 {\usefont{T2A}{cmr}{b}{n}(10.5)}: Наклон поверхности воды
в реке}
Поверхность воды перпендикулярна равнодействующей силы тяжести и силы Кориолиса. 
Центробежной силой можно пренебречь, так как ее радиальная компонента незначительно 
уменьшает силу тяжести, а тангенциальная компонента направлена вдоль течения реки и
наклон поверхности воды не вызывает. Угол между векторами  $\vec{V}$ и 
$\vec{\omega}$ есть $\alpha=180^\circ-\varphi$ (см. левый рисунок).  
\begin{figure}[htb]
\centerline{\epsfig{figure=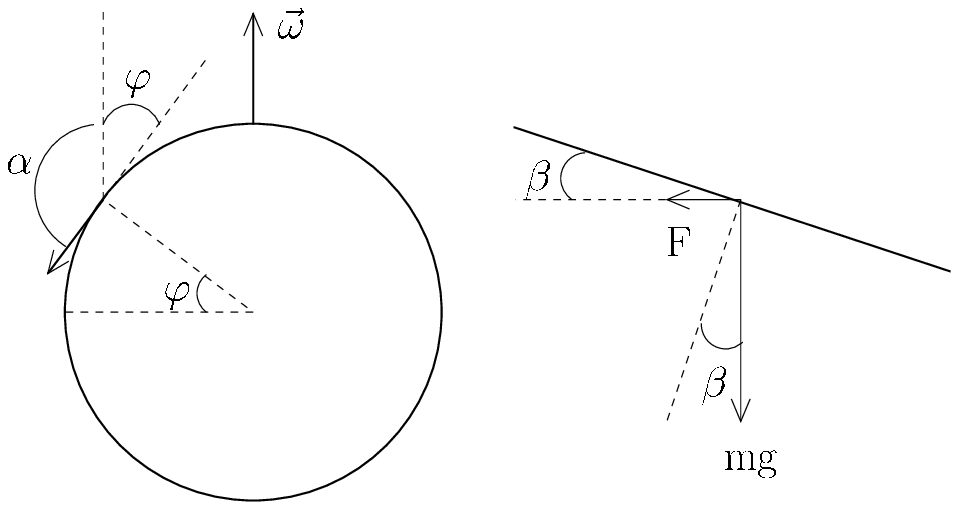,height=4.5cm}}
\end{figure}

\noindent Векторное произведение $\vec{V}\times\vec{\omega}$ направлено от нас, 
т.~е. к западу. Поэтому на западном берегу уровень воды будет выше. Угол наклона 
$\beta$ определяется из соотношения (см. правый рисунок) $\tg{\beta}=\frac{F}{mg}$,
где модуль силы Кориолиса равен $F=2mV\omega\sin{(180^\circ-\varphi)}=2mV\omega
\sin{\varphi}$. Так как $V\omega\ll g$, можно считать $\tg{\beta} \approx \beta$,
и окончательно $$\beta\approx \frac{2V\omega\sin{\varphi}}{g}.$$

\subsection*{Задача 3 {\usefont{T2A}{cmr}{b}{n}(10.12)}: Скорость пробки
относительно трубки в момент вылета}
Когда пробка находится на расстоянии $x$ от оси вращения, на нее действует центробежная
сила (в системе трубки) $m\omega^2 x$. Поэтому уравнение движения будет $m\ddot x=
m\omega^2 x$, или $\ddot x-\omega^2 x=0$. Общее решение имеет вид
$x(t)=A_1e^{\omega t}+A_2e^{-\omega t}$. Константы $A_1$ и $A_2$ определим из 
начальных условий: $x(0)=x_0$ дает $A_1+A_2=x_0$, тогда как $\dot{x}(0)=0$ дает
$A_1-A_2=0$. Следовательно, $A_1=A_2=x_0/2$, и $x(t)=x_0\ch{\omega t}$. Время 
движения $T$ определяется из соотношения $x(T)=L$. Поэтому
$$T=\frac{1}{\omega}\,\arch{\frac{L}{x_0}}=\frac{1}{\omega}\,\ln{\left(
\frac{L}{x_0}+\sqrt{\frac{L^2}{x_0^2}-1}\right)}.$$
Скорость вылета $V=\dot{x}(T)=\omega x_0\sh{\omega T}$.

\subsection*{Задача 4 {\usefont{T2A}{cmr}{b}{n}(10.13)}: Уравнение движения
бусинки}
Центробежная сила будет $F_c=m\omega^2\rho=m\omega^2x\sin{\alpha}$, где 
$\rho$ --- расстояние от бусинки до оси вращения, $x$ --- расстояние от бусинки до
вершины конуса (конца палочки; см. рисунок).
\begin{figure}[htb]
\centerline{\epsfig{figure=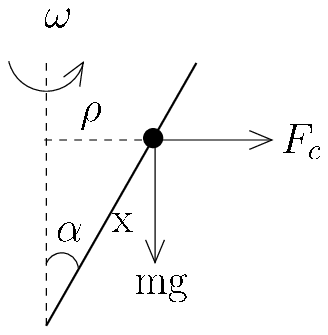,height=3cm}}
\end{figure}

\noindent Уравнение движения вдоль палочки будет $m\ddot x=m\omega^2x\sin^2{\alpha}
-F_{\mbox{Тр}}$. Сила трения $|F_{\mbox{Тр}}|=\mu N$, где сила реакции $\vec{N}$
уравновешивает векторную сумму силы Кориолиса $2m\dot{x}\omega\sin{\alpha}$ и 
нормальной (относительно палочки) компоненты центробежной силы $m\omega^2x\sin{
\alpha}\cos{\alpha}$. Поэтому 
$$N^2=4m^2\dot{x}^2\omega^2\sin^2{\alpha}+m^2\omega^4x^2\sin^2{\alpha}
\cos^2{\alpha},$$
и
$$|F_{\mbox{Тр}}|=\mu N=\mu m\omega\sin{\alpha}\sqrt{4\dot{x}^2+x^2\omega^2
\cos^2{\alpha}}.$$
Уравнение движения будет иметь вид
$$\ddot x=\omega^2x\sin^2{\alpha}-\mu\omega\,\frac{\dot{x}}{|\dot{x}|}\,
\sin{\alpha}\sqrt{4\dot{x}^2+x^2\omega^2\cos^2{\alpha}}.$$
Мы ввели множитель $\dot{x}/|\dot{x}|$, чтобы учесть тот факт, что сила трения 
$\vec{F}_{\mbox{Тр}}$ всегда направлена против скорости. 

\subsection*{Задача 5 {\usefont{T2A}{cmr}{b}{n}(9.22)}: Угловая скорость вращения
желоба после соскальзывания тела}
Пусть после соскальзывания тела желоб вращается с угловой скоростью $\omega$, 
а скорость тела относительно желоба равна $V$. При этом скорость $\vec{V}$ 
составляет угол $30^\circ$ с горизонталью и, следовательно, угол $180^\circ-
30^\circ$ со скоростью точки $A$ желоба относительно земли, которая по величине равна 
$\omega R$ 
(см. рисунок).
\begin{figure}[htb]
\centerline{\epsfig{figure=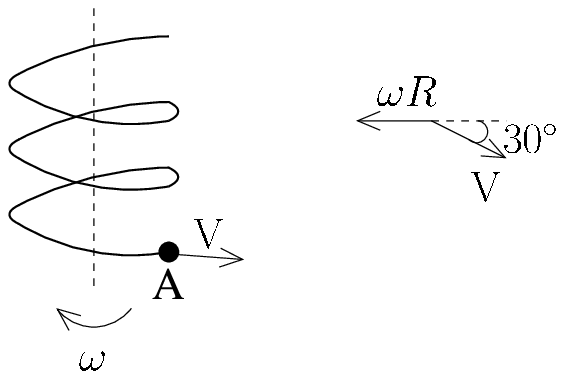,height=4cm}}
\end{figure}

\noindent Поэтому скорость тела относительно земли будет
$$\sqrt{V^2+(\omega r)^2-2V\omega R\cos{30^\circ}}=\sqrt{V^2+(\omega r)^2-
\sqrt{3}V\omega R},$$
и, так как момент инерции желоба равен $I=MR^2$, закон сохранения энергии запишется 
как
$$mgH=\frac{m}{2}\left (V^2+\omega^2 r^2-\sqrt{3}V\omega R\right)+
\frac{1}{2}MR^2\omega^2.$$
С другой стороны, так как горизонтальная компонента скорости тела относительно земли 
равна $V\cos{30^\circ}-\omega R=\sqrt{3}V/2-\omega R$, закон сохранения 
вертикальной $z$-компоненты момента импульса относительно оси желоба имеет вид
$$m\left(\frac{\sqrt{3}V}{2}-\omega R\right)R-MR^2\omega=0.$$
Отсюда получаем
$$V=\frac{2(m+M)\omega R}{\sqrt{3}m}.$$
Это можно подставить в законе сохранения энергии и после некоторой алгебры получим
$$mgH=\frac{4M+m}{6m}(m+M)R^2\omega^2.$$
Поэтому окончательно
$$\omega=\frac{m\sqrt{6gH}}{R{\sqrt{(m+M)(4M+m)}}}= 
\frac{m\sqrt{6gH}}{R{\sqrt{m^2+5mM+4M^2}}}.$$

\subsection*{Задача 6 {\usefont{T2A}{cmr}{b}{n}\cite{36}}: Отклонение 
отвеса от линии к центру Земли}
$\vec{F}=-\nabla U$. Поэтому сферические компоненты силы будут
$F_r=-(\vec{e}_r\cdot\nabla)U$, $F_\theta=-(\vec{e}_\theta\cdot\nabla)U$ и
$F_\phi=-(\vec{e}_\phi\cdot\nabla)U$. Орты сферической системы имеют вид
$$\vec{e}_r=\sin{\theta}\cos{\phi}\,\vec{i}+\sin{\theta}\sin{\phi}\,\vec{j}
+\cos{\theta}\,\vec{k}=\frac{\partial x}{\partial r}\,\vec{i}+
\frac{\partial y}{\partial r}\,\vec{j}+\frac{\partial z}{\partial r}\,
\vec{k},$$
$$\vec{e}_\theta=\cos{\theta}\cos{\phi}\,\vec{i}+\cos{\theta}\sin{\phi}\,
\vec{j}-\sin{\theta}\,\vec{k}=\frac{1}{r}\left [\frac{\partial x}
{\partial \theta}\,\vec{i}+\frac{\partial y}{\partial \theta}\,\vec{j}+
\frac{\partial z}{\partial \theta}\,\vec{k}\right ],$$
$$\vec{e}_\phi=-\sin{\phi}\,\vec{i}+\cos{\phi}\,\vec{i}=\frac{1}
{r\sin{\theta}}\left [\frac{\partial x}{\partial \phi}\,
\vec{i}+\frac{\partial y}{\partial \phi}\,\vec{j}+\frac{\partial z}
{\partial \phi}\,\vec{k}\right ].$$
Поэтому получаем
$$F_r=-\frac{\partial U}{\partial r}=-\frac{GM}{r^2}\left [1-\frac{3}{2}
\,a_2\,\frac{R^2}{r^2}(3\cos^2{\theta}-1)\right ],$$ $$
F_\theta=-\frac{1}{r}\frac{\partial U}{\partial \theta}=-\frac{3}{2}\,a_2
\,\frac{GM}{r^2}\frac{R^2}{r^2}\sin{2\theta}, $$ $$
F_\phi=-\frac{1}{r\sin{\theta}}\frac{\partial U}{\partial \phi}=0.$$
Если бы Земля не вращалась, угол отклонения отвеса был бы
$$\alpha_1\approx \tg{\alpha_1}=\left |\frac{F_\theta}{F_r}\right |\approx
\frac{3}{2}\,a_2\,\sin{2\theta}.$$
Из-за вращения Земли возникает дополнительное отклонение. Проекция центробежной силы на
$\vec{e}_\theta$ будет 
$$m\omega^2R\sin{\theta}\cos{\theta}=mg\,\frac{\omega^2R}{2g}\sin{2\theta}.$$
Поэтому дополнительное отклонение равно
$$\alpha_0\approx \frac{\omega^2R}{2g}\sin{2\theta}.$$
Полное отклонение
$$\alpha\approx \left [\frac{3}{2}\,a_2+ \frac{\omega^2R}{2g}\right]
\sin{2\theta}.$$
Здесь $\theta$ не широта, а полярный угол сферической системы координат, ось $z$ 
которий проходит через северный полюс. Заметим, что $\frac{\omega^2R}{g}
\approx 3,45\cdot 10^{-3}$ такого же порядка, как $3a_2\approx 3,3\cdot 10^{-3}$.

\subsection*{Задача 7 {\usefont{T2A}{cmr}{b}{n}(\hspace*{-1.5mm}\cite{37}, 
487)}: Орбита геостационарного спутника после радиального возмущения}
{\usefont{T2A}{cmr}{m}{it} Первое решение.}
Перейдем в неинерциальную систему отсчета (с координатами $x^{\,\prime}$ и  
$y^{\,\prime}$) земного наблюдателя. Уравнение движения спутника в этой системе 
имеет вид
$$m\ddot{\vec{r}}^{\,\prime}=-\frac{GmM}{r^3}\,\vec{r}+m\omega^2\,\vec{r}+
2m\,\dot{\vec{r}}^{\,\prime}\times\vec{\omega},$$
где $\vec{r}=\vec{R}+\vec{r}^{\,\prime}$ (см. рисунок) и $\omega$ --- угловая 
скорость вращения Земли (что совпадает с угловой скоростью геостационарного спутника 
до возмущения). 
\begin{figure}[htb]
\centerline{\epsfig{figure=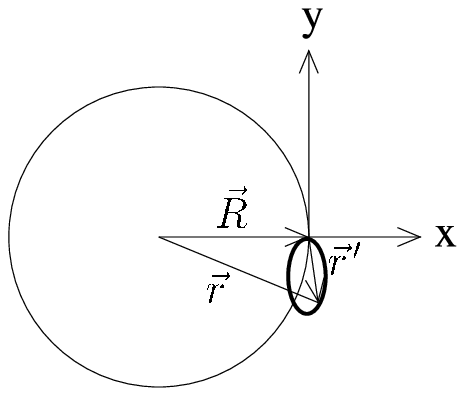,height=4cm}}
\end{figure}

\noindent Проектируя уравнение движения спутника на координатные оси, получаем
\begin{equation}
m\ddot{x}^{\,\prime}=-\frac{GmM}{(R+x^{\,\prime})^2}+m\omega^2(R+x^{\,\prime})+
2m\omega\dot{y}^{\,\prime}
\label{eq36_7a}
\end{equation}
и
\begin{equation}
m\ddot{y}^{\,\prime}=-\frac{GmM}{R^3}\,y^{\,\prime}+m\omega^2\,y^{\,\prime}-
2m\omega\dot{x}^{\,\prime}.
\label{eq36_7b}
\end{equation}
Здесь мы учли, что $r=\sqrt{(R+x^{\,\prime})^2+y^{\prime\,2}}\approx R+
x^{\,\prime}$,
и отбросили члены второго порядка малости по $x^{\,\prime}$ и  $y^{\,\prime}$. На 
геостационарной орбите
\begin{equation}
\frac{GmM}{R^2}=m\omega^2R.
\label{eq36_7c}
\end{equation} 
Поэтому из (\ref{eq36_7b}) получаем $\ddot{y}^{\,\prime}=-2\omega\dot{x}^{\,
\prime}$,
и, так как $x^{\,\prime}(0)=0$ и $\dot{y}^{\,\prime}(0)=0$, 
\begin{equation}
\dot{y}^{\,\prime}=-2\omega\, x^{\,\prime}.
\label{eq36_7d}
\end{equation}
С другой стороны, уравнение (\ref{eq36_7a}), с учетом
$$\frac{1}{(R+x^{\,\prime})^2}\approx\frac{1}{R^2}\left(1-2\,\frac{x^{\,
\prime}}{R}
\right)$$
и (\ref{eq36_7c}), примет вид
$$m\ddot{x}^{\,\prime}=-3m\omega^2x^{\,\prime}+2m\omega\dot{y}^{\,\prime}=-m
\omega^2 x^{\,\prime},$$
где на последнем этапе мы учли (\ref{eq36_7d}). Следовательно, $x^{\,\prime}=A\,
\sin{\omega t}+B\,\cos{\omega t}$. Начальные условия $x^{\,\prime}(0)=0,\,
\dot{x}^{\,\prime}(0)=\Delta V$ определяют константы $A$ и $B$ как $A=\Delta V/
\omega$ и $B=0$. Поэтому
$$x^{\,\prime}=\frac{\Delta V}{\omega}\,\sin{\omega t}.$$
Тогда (\ref{eq36_7d}) дает $\dot{y}^{\,\prime}=-2\Delta V\,\sin{\omega t}$, что
с учетом $y^{\,\prime}(0)=0$ означает
$$y^{\,\prime}=2\frac{\Delta V}{\omega}(\cos{\omega t}-1).$$
Заметим, что
$$\frac{x^{\prime\,2}}{(\Delta V/\omega)^2}+\frac{(y^{\,\prime}+2\Delta V/
\omega)}{(2\Delta V/\omega)^2}=1.$$
Следовательно, возмущенная траектория в системе земного наблюдателя --- это эллипс.

{\usefont{T2A}{cmr}{m}{it} Второе решение.}
Так как добавка к скорости радиальная, момент импульса $L$ не меняется. Следовательно,
параметр орбиты $p=\frac{L^2}{m\alpha}$ тоже не меняется и равен $R$. Уравнение 
новой орбиты будет
$$\frac{p}{r}=1-e\,\cos{(\varphi-\varphi_m)}.$$
В точке $A$, $r=R=p$ и $\varphi=0$ (см. рисунок). Поэтому должны иметь 
$\cos{\varphi_m}=0$, т.~е. $\varphi_m=\pi/2$, и уравнение орбиты будет
$$\frac{R}{r}=1-e\,\sin{\varphi}.$$  
\begin{figure}[htb]
\centerline{\epsfig{figure=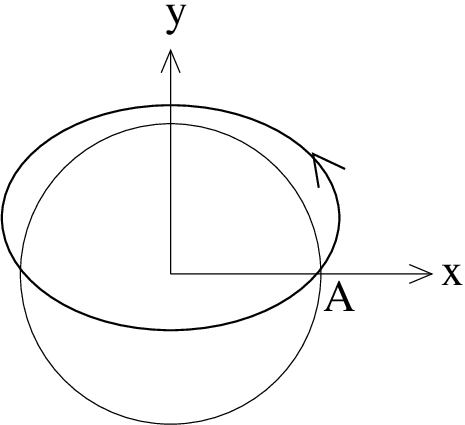,height=3cm}}
\end{figure}

\noindent Для круговой орбиты эксцентриситет равен нулю, т.~е.
$$1+\frac{2L^2E}{m\alpha^2}=0.$$
После возмущения будем иметь ($V$ --- это скорость спутника на круговой орбите):
$$e=\sqrt{1+\frac{2L^2}{m\alpha^2}\left (E+\frac{1}{2}m\,\Delta V^2\right)}=
=\frac{L\Delta V}{\alpha}=\frac{\Delta V}{V}\ll 1.$$
Последнее равенство следует из $\frac{L^2}{m\alpha}=R$ и $L=mVR$. Следовательно,
$$r=\frac{R}{1-e\,\sin{\varphi}}\approx R(1+e\,\sin{\varphi})\approx
R(1+e\,\sin{\omega t}),$$
так как в нулевом порядке $\varphi\approx \omega t$. Чтобы найти поправку первого
порядка по $e$ к $\varphi(t)$, воспользуемся законом сохранения момента импульса. Из
$L=mr^2\dot\varphi$ следует, что
$$\frac{d\varphi}{dt}=\frac{L}{mr^2}=\frac{L}{mR^2}(1-e\,\sin{\varphi})^2=
\omega(1-e\,\sin{\varphi})^2,$$
так как 
$$\frac{L}{mr^2}=\frac{mVR}{mR^2}=\frac{V}{R}=\omega.$$
Разделяя переменные, с учетом того, что
$$\frac{d\varphi}{(1-e\,\sin{\varphi})^2}\approx (1+2e\,\sin{\varphi})\,
d\varphi,$$
получаем уравнение $(1+2e\,\sin{\varphi})\,d\varphi=\omega dt$. Интегрируем, 
предполагая, что $\varphi(0)=0$ (т.~е. в начальный момент времени спутник находился
в точке $A$):
$$\varphi-2e(\cos{\varphi}-1)=\omega t.$$
Поэтому
$$\varphi=\omega t+2e(\cos{\varphi}-1)\approx \omega t+2e(\cos{\omega t}-1).$$
Теперь можно найти координаты спутника $x=r\cos{\varphi}$ и $y=r\sin{\varphi}$
в инерциальной системе отсчета. Используя
$$\cos{\left[\omega t+2e(\cos{\omega t}-1)\right]}\approx
\cos{\omega t}-2e\sin{\omega t}(\cos{\omega t}-1),$$
будем иметь
$$x\approx R(1+e\sin{\omega t})\left[\cos{\omega t}-2e\sin{\omega t}
(\cos{\omega t}-1)\right]\approx R\left[\cos{\omega t}-e\sin{\omega t}
\cos{\omega t}+2e\sin{\omega t}\right].$$
Аналогично, так как
$$\sin{\left[\omega t+2e(\cos{\omega t}-1)\right]}\approx
\sin{\omega t}+2e\cos{\omega t}(\cos{\omega t}-1),$$
будем иметь
$$y\approx R(1+e\sin{\omega t})\left[\sin{\omega t}+2e\cos{\omega t}
(\cos{\omega t}-1)\right]\approx  R\left[\sin{\omega t}+e(1-\cos{\omega t})^2
\right].$$
\begin{figure}[htb]
\centerline{\epsfig{figure=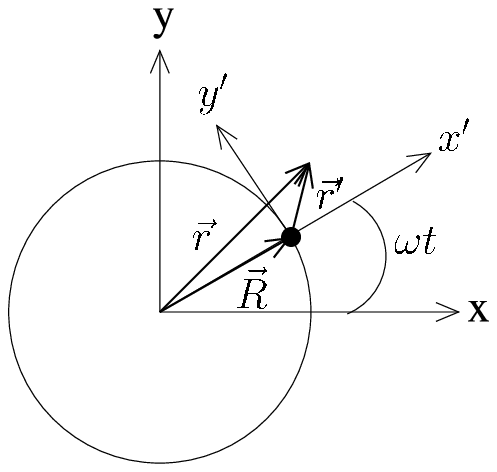,height=4cm}}
\end{figure}
Чтобы перейти в неинерциальную систему отсчета земного наблюдателя, надо учесть, что 
начало этой системы сдвинуто на вектор $\vec{R}=R(\cos{\omega t}\,\vec{i}+
\sin{\omega t}\,\vec{j})$, а координатные оси повернуты на угол $\omega t$ (см. 
рисунок). Поэтому координаты вектора $\vec{r}^{\,\prime}=\vec{r}-\vec{R}$ в 
неинерциальной системе отсчета будут
$$x^{\,\prime}=(x-R\cos{\omega t})\cos{\omega t}+(y-R\sin{\omega t})
\sin{\omega t}=Re\sin{\omega t}$$
и
$$y^{\,\prime}=-(x-R\cos{\omega t})\sin{\omega t}+(y-R\sin{\omega t})
\cos{\omega t}=-2Re(1-\cos{\omega t}).$$
Заметим, что
$$Re=R\,\frac{\Delta V}{V}=\frac{\Delta V}{\omega}.$$

%\clearpage
\addtocontents{toc}{\cftpagenumberson{section}}
\section[Список литературы]{\centerline{Список литературы}}
%\addcontentsline{toc}{section}{Список литературы}
%\begin{thebibliography}{99}

\end{document}